\documentclass[12pt,oneside]{book}
\usepackage{latexsym,amsfonts,amsmath,amscd,amssymb,epsf}
\usepackage[latin1]{inputenc}
\usepackage[american]{babel}
\usepackage[dvips]{graphicx}
\usepackage{bbm}
\usepackage{epsfig}
\usepackage{graphics,color}
\usepackage{mathrsfs}
\usepackage{color}
\usepackage{graphicx}
\usepackage{caption}
\usepackage{subcaption}
\usepackage{cancel}
\usepackage{mathtools}
\usepackage{slashed}
\usepackage[toc,page]{appendix}

\advance \topmargin by -\headheight
\advance \topmargin by -\headsep

\evensidemargin \oddsidemargin
\def\th{\theta}

  \def\tr{{\hbox{\rm Tr}}}

\def\ie{{\em i.e.}}
\def\ie{\hbox{\it i.e.}}

\def\CC{{\mathchoice
{\rm C\mkern-8mu\vrule height1.45ex depth-.05ex
width.05em\mkern9mu\kern-.05em}
{\rm C\mkern-8mu\vrule height1.45ex depth-.05ex
width.05em\mkern9mu\kern-.05em}
{\rm C\mkern-8mu\vrule height1ex depth-.07ex
width.035em\mkern9mu\kern-.035em}
{\rm C\mkern-8mu\vrule height.65ex depth-.1ex
width.025em\mkern8mu\kern-.025em}}}

\def\RR{{\rm I\kern-1.6pt {\rm R}}}

\def\ZZ{{\rm Z}\kern-3.8pt {\rm Z} \kern2pt}
\def\IB{\relax{\rm I\kern-.18em B}}
\def\ID{\relax{\rm I\kern-.18em D}}
\def\II{\relax{\rm I\kern-.18em I}}
\def\IP{\relax{\rm I\kern-.18em P}}

\newcommand{\beq}{\begin{equation}}
\newcommand{\eeq}{\end{equation}}
\newcommand{\rc}{\nonumber\\}
\newcommand{\bear}{\begin{eqnarray}}
\newcommand{\eear}{\end{eqnarray}}

\def\to{\rightarrow}

\def\tr{{\rm Tr}}
\def\to{\rightarrow}

\newfont{\namefont}{cmr10}
\newfont{\addfont}{cmti7 scaled 1440}
\newfont{\boldmathfont}{cmbx10}
\newfont{\headfontb}{cmbx10 scaled 1728}
\def\ie{{\it i.e.}}

\def\revise#1       {\raisebox{-0em}{\rule{3pt}{1em}}%
                     \marginpar{\raisebox{.5em}{\vrule width3pt\
                     \vrule width0pt height 0pt depth0.5em
                     \hbox to 0cm{\hspace{0cm}{%
                     \parbox[t]{4em}{\raggedright\footnotesize{#1}}}\hss}}}}

\def\ee           {{\rm e}}

\def\tr           {\mathop{\rm Tr}}

\def\sqr#1#2{{\vcenter{\vbox{\hrule height.#2pt
 \hbox{\vrule width.#2pt height#1pt \kern#1pt
 \vrule width.#2pt}\hrule height.#2pt}}}}

\def\a{\alpha}

\def\r{\rho}

\def\be{\begin{equation}}
\def\ee{\end{equation}}

\def\m{\mu}

\def\l{\lambda}
\def\n{\nu}

\def\tr{{\tilde\rho}}
\newcommand{\eel}[1]{\label{#1}\end{equation}}
\newcommand{\bea}{\begin{eqnarray}}
\newcommand{\eea}{\end{eqnarray}}
\newcommand{\eeal}[1]{\label{#1}\end{eqnarray}}

\def\m{\mu}
\def\n{\nu}
\def\a{\alpha}

\def\r{\rho}

\def\s{\sigma}

\def\l{\lambda}

\def\be{\begin{equation}}
\def\ee{\end{equation}}

\def \l {\lambda}

\def \m {\mu}
\def \n {\nu}

\def\l{\lambda}

\def\CC{\mathcal{C}}

\def\tr{{\rm Tr}}

\def\lbldef#1#2{\expandafter\gdef\csname #1\endcsname {#2}}

\def\href#1#2{#2}

\definecolor{M_Beige}         {rgb}{0.96 , 0.96 , 0.86}

\definecolor{M_Brown}         {rgb}{0.65 , 0.16 , 0.16}
\definecolor{M_Gold}          {rgb}{0.12 , 0.84 , 0.30}

\definecolor{M_LemonChiffon}  {rgb}{1.00 , 0.98 , 0.80}

\definecolor{M_Orange}        {rgb}{1.00 , 0.60 , 0.00}

\definecolor{M_Pink}          {rgb}{1.00 , 0.75 , 0.80}

\definecolor{M_Gre}          {rgb}{0.05 ,0.46 , 0.00}

\begin{document}


\begin{titlepage}
\begin{center} \Large \bf Holographic duality and applications
\end{center}
\vskip 0.3truein
\begin{center}
\ Yago Bea 
\footnote{PhD Thesis, Universidade de Santiago de
Compostela, Spain (September, 2016). \\
Advisor: Alfonso V. Ramallo}\\
\vspace{0.3in}
Departamento de F\'\i sica de Part\'\i culas, Universidade de
Santiago de Compostela \\and\\
Instituto Galego de F\'\i sica de Altas Enerx\'\i as (IGFAE)\\
E-15782 Santiago de Compostela, Spain\\
\vspace{0.15in}
yago.bea@fpaxp1.es
\vspace{0.3in}
\end{center}
\vskip 1truein
\begin{center}
\bf ABSTRACT
\end{center}

In this thesis we review some results on the generalization of the gauge/gravity duality to new cases by using T-duality and by including fundamental matter, finding applications to condensed matter physics. First, we construct new supersymmetric solutions of type IIA/B and eleven-dimensional supergravity by using non-abelian T-duality. Second, we construct a type IIA supergravity solution with D6-brane sources, dual to an unquenched massive flavored version of the ABJM theory. Third, we study a probe D6-brane with worldvolume gauge fields in the ABJM background, obtaining the dual description of a quantum Hall system. Moreover, we consider a system of a probe D6-brane in the ABJM background and study quantum phase transitions of its dual theory.

\smallskip
\end{titlepage}
\setcounter{footnote}{0}
\setcounter{page}{1}


\tableofcontents


\chapter*{Motivation}
\addcontentsline{toc}{chapter}{Motivation}

This thesis is based on the gauge/gravity duality, which emerged from string theory in the nineties.
We give a historical introduction and contextualize the work that we will present. \newline

In the twentieth century it took place the consolidation of two fundamental developments in physics. On the one hand, the theory of general relativity was constructed to explain gravity, one of the four fundamental forces of nature, and became very successful describing the macroscopic physics. On the other hand, the birth of quantum field theory gave a natural framework for elementary particle physics, which explains the other three fundamental forces of nature, weak, strong and electromagnetic, obtaining a very accurate description of the subatomic physics. However, these two developments are incompatible, and a desirable unification in a quantum theory of gravity remained as an open problem. In the seventies the string theory was constructed, and it turned out to be a promising candidate to perform such unification. Thus, string theory can be considered as a `theory of everything', in the sense that it aims to explain the four fundamental forces of nature in a unified and common way.

String theory was originally formulated in the context of the strong interactions. In 1968 Veneziano proposed an amplitude \cite{veneziano} to explain some characteristics of the new hadronic experimental data. Soon after Veneziano's proposal, a theory which reproduced the amplitude was found in the form of a quantized string.  In 1970 the Nambu-Goto action for the bosonic string was formulated \cite{nambugoto} and in 1971 Ramond, Neveu and Schwarz constructed the fermionic string \cite{fermionicstring}. Nevertheless, at the beginning of the seventies the Veneziano amplitude became unable to explain the new experimental results, and a new theory, quantum chromodynamics, revealed itself as a much more promising candidate for the description of strong interactions. Therefore, string theory ceased to be a good candidate for the strong interactions. In 1974, Scherk and Schwarz \cite{scherkschwarz} reinterpreted the string theory as a much more fundamental theory: a quantum theory of gravity. The scale of the theory was reduced from the strong interaction scale to the Planck scale. Previously, as a theory of strong interactions, the massless spin 2 particle present in the spectrum of the theory was absent in the experimental results, giving rise to an unsolved puzzle, but now, as a theory of quantum gravity, this particle had a natural interpretation as the graviton. String theory became a good candidate for the unification of the four fundamental forces, as in the low energy limit only the massless modes are left, including the graviton and lower spin particles, which could account for the matter and gauge bosons. Also, Einstein gravity is correctly recovered in the classical limit.

 In 1976 Gliozzi Scherk and Olive \cite{GSO} imposed the so-called GSO projection which removed some inconsistencies present in string theory. They imposed spacetime supersymmetry, property that became of fundamental importance. Supersymmetry was developed coetaneously with string theory during the seventies, and also its local version, supergravity.

 String theory was not unique, and by 1984 there were 3 different and consistent versions: type I, type IIA and type IIB string theories. In 1985 it took place the `first superstring revolution' with the discovery by Gross, Harvey, Martinec and Rohm \cite{heterotic} of two new string theories: heterotic $SO(32)$ and heterotic $E_8 \times E_8$. These two theories turned out to be particularly interesting because they have room enough for the standard model gauge group $U(1) \times SU(2) \times SU(3)$ and there was some hope that the standard model could be derived from them. But the possibilities of compactification were proven to be larger than $10^{500}$ and the particular way in which the standard model arises from string theory remained unclarified. 

     The presence of five different and consistent string theories was a difficulty in finding a unique fundamental theory. This difficulty vanished in 1995 when the `second superstring revolution' took place with the discovery (mostly by Witten \cite{wittendualities}) of a large web of dualities relating all the string theories. Besides, these dualities also related the string theories to eleven-dimensional supergravity, a theory predicted by Nahm \cite{nahm} and constructed soon thereafter by  Cremmer, Julia and Scherk \cite{cremmer}. 


  By that time, new objects called branes were being studied in string theory. D-branes are non perturbative
solutions consisting of hyperplanes where open strings can end. These open strings describe
the dynamics of the branes, and its massless modes realize a gauge theory
on the worldvolume of the brane. Alternatively, D-branes can be obtained as solitonic objects in the description of closed strings. The closed string interact with the branes, as they are massive objects. This open/closed dual description points towards the
existence of a duality between gauge theories and string theory.

In the context of the recently discovered branes, the study of D3-branes in type IIB supergravity led Maldacena to conjecture in 1997 the AdS/CFT correspondence \cite{Maldacena:1997re}. This proposal states that the D3-branes in flat space, after taking the decoupling limit, admit two equivalent descriptions, dual to each other, one being the closed string description, corresponding to string theory on $AdS_5 \times S^5$, and the other being the open string description, corresponding to $3+1$ dimensional ${\cal N}=4$ $SU(N)$ SYM conformal field theory. Soon after this proposal, a lot of new examples appeared generalizing the conjecture to less supersymmetric and non-conformal cases. 

The AdS/CFT duality can be formulated in different levels of generality. The statement above is the `strong version' of the conjecture. Taking the low energy limit, the conjecture states that type IIB supergravity plus stringy corrections in $AdS_5 \times S^5$ is dual to $3+1$ ${\cal N}=4$ SYM conformal field theory in the t'Hooft limit. Finally, neglecting the $\alpha'$ corrections we obtain the `weak formulation' of the duality, which states that type IIB supergravity on  $AdS_5 \times S^5$ is dual to  $3+1$ dimensional ${\cal N}=4$ $SU(N)$ SYM conformal field theory in the large $N$ limit and strong t'Hooft coupling $\lambda$. This weak version of the conjecture is the most useful, and will be the one used in this thesis. This is why we will focus on the study of supergravity, in particular in type IIA/B and eleven-dimensional  supergravity. 


The usefulness of the conjecture lies on the fact that it is a strong/weak duality, \ie, if on one side of the duality the coupling constant is weak, on the other side it is strong and vice versa. This means that if we want to compute observables in a strong coupling region of a quantum field theory, we can consider its supergravity description, were the coupling constant is weak, and perform the computations. Then, performing an easy computation in classical supergravity we can obtain results in the strong coupling regime of a quantum field theory, where other tools are limited or inexistent. The drawback of this powerful method is that the quantum field theories with known gravity dual are very limited, and still far from realistic theories. The hope is that we can extract universal properties that are valid for a set of similar theories, and then extrapolate to the realistic theories. Another drawback of the gauge/gravity duality is that even if it has undergone many non trivial tests, there is no proof of the conjecture. \newline






T-duality in string theory was discovered in the eighties. It relates two seemingly different theories that are compactified on a circle. These two theories are
two different descriptions of the same physics, and they are said to be T-dual to each other. For example, type IIA and B string theories on Minkowski spacetime compactified on a circle are T-dual to each other. Locally, T-duality is given by the Buscher rules \cite{buscher} \cite{Bergshoeff:1995as}, and globally, it corresponds to an exchange of the Chern number of a $U(1)$-fibration and the $H_3$ flux on its base \cite{Bouwknegt:2003vb}. 
On the other hand, non-abelian T-duality (NATD) was introduced at the end of the eighties \cite{de la Ossa:1992vc} as a generalization of T-duality to non-abelian isometry groups. Buscher rules were known for the NSNS sector, and it was applied in the context of sigma models. Nevertheless, the transformation rules for the RR sector remained unclarified until 2010 \cite{Sfetsos:2010uq}. With the complete Buscher rules, the NATD was used to obtain new supergravity solutions of type IIA/B and, then, new examples of the AdS/CFT duality. In particular, new $AdS$ solutions were found by this method. Nevertheless, the global properties of the transformation are not known, and then the dual solution is only known locally (sometimes stated as ``the range of the dual coordinates is not known''). As a consequence, the dual field theory to this new NATD background can not be understood with full detail. In this thesis we construct new supersymmetric solutions of type IIA/B and eleven-dimensional supergravity using NATD. \newline

In the original formulation of the gauge/gravity duality, all the fields of the field theory are in the adjoint representation of the gauge group. In order to make contact with phenomenological theories, it is convenient to introduce fundamental matter (recall that fundamental matter corresponds to quarks in QCD or electrons in condensed matter physics). The fundamental matter was introduced in the correspondence in 2002 by Karch and Katz \cite{Karch:2002sh}, using D7 probe branes in the $AdS_5 \times S^5$ solution. Using probe branes in the gravity side corresponds to having quenched fundamental matter in the field theory side. In order to take into account the full effect of the fundamental matter, one has to solve the equations of motion of supergravity coupled to the D-brane sources. In general, it is a difficult problem to obtain such backgrounds. For a stack of localized flavor branes, the problem has reduced symmetries, Dirac deltas and dependence on several variables. In order to simplify it, one can consider a continuous distribution of sources (as in electromagnetism a continuous source of charges), recovering the symmetries, the dependence in one variable and avoiding Dirac deltas. This method is known in the literature as the smearing technique.  In this thesis, the smearing technique is used to obtain a new supergravity background that generalizes the ABJM model by including massive fundamental matter. \newline

The holographic duality has found a lot of applications, mostly in QCD and condensed matter physics.  In particular, after the conjecture was formulated, the applications to condensed matter physics have increased continuously, and we can find applications to the description of superconductors, superfluids, insulators, metals, strange metals, topological insulators, Hall effect, Kondo model, etc. In this thesis we present a new model for the fractional quantum Hall effect based on the ABJM model. Besides, we also present a new model exhibiting quantum phase transitions.

\subsection*{About this thesis}

This thesis is based on the papers \cite{Bea:2013jxa, Bea:2014yda, Bea:2015fja, Bea:2016fcj} and has the following structure:

\begin{itemize}
\item The first chapter is a general introduction to the gauge/gravity duality. First, we review the fundamental aspects of supergravity. Second, we review T-duality in the context of supergravity, and consider its non-abelian generalization. Third,  we introduce the AdS/CFT correspondence and analyze the addition of flavor to it. Finally, we comment on some applications of the AdS/CFT conjecture.

\item The second chapter reviews some applications of non-abelian T-duality. In the context of type IIA/B supergravity, the non-abelian T-duality is used to obtain new solutions from previously know solutions. The aim is to obtain new interesting cases of the gauge/gravity duality. In particular, new $AdS_3$ fixed points are constructed. Also, we analyze the field theory duals of the new supergravity solutions, via the computation of some observables. 

\item In the third chapter we construct a new supergravity IIA background with D6-branes sources, whose field theory dual corresponds to a generalization of the ABJM theory that includes unquenched massive flavors. This construction uses the smearing technique. Several observables are analyzed, confirming a RG flow between two fixed points at the IR and UV. 

\item In the fourth chapter we construct a supergravity solution based on the ABJM model, dual to a quantum Hall system. The fundamental matter is introduced in the supergravity dual via probe D6-branes, in which non-trivial woldvolume gauge fields are turned on. This set up allows to have Hall states, and the filling fraction is computed. For particular values of the parameters, supersymmetric solutions are found.

\item In the fifth chapter, we construct a supergravity solution based on the ABJM model, dual to a condensed matter system exhibiting quantum phase transitions. The fundamental matter is introduced via D6-branes, and charge density is turned on. In the ABJM background, the probe brane undergoes a second order phase transition at zero charge density. In the partially backreacted ABJM solution, the phase transition becomes first order, and takes place at a finite value of the charge density.

\end{itemize} \newpage


\chapter{Introduction}

We start this chapter by a review of the fundamental aspects of supergravity, which is the basic ground where this thesis is formulated. Besides, the elementary aspects of T-duality and non-abelian T-duality in the context of supergravity are reviewed. We also give a brief introduction to the AdS/CFT conjecture, and analyze the addition of flavor to it. Finally, we comment on the applications of the duality.

\section{Supergravity}
\label{supergravity1}

Supergravity is obtained by gauging supersymmetry, \ie, by considering local supersymmetry, instead of global. In a natural way, supergravity theories contain Einstein gravity, and due to its supersymmetric character, they are well behaved theories and thus good candidates for a quantum theory of gravity.

The string theories at low energies are described by supergravity theories. For example, type IIA(B) string theory at low energies is described by IIA(B) supergravity, and M-theory at low energies is described by eleven-dimensional supergravity. These supergravity theories are particularly interesting in the context of AdS/CFT. The AdS/CFT conjecture was born from string theory, and it relates a string theory background with a quantum field theory. But the duality is actually useful in a particular regime, precisely in the low energy limit of the string theory side, when it is well described by classical supergravity.

In general, supergravity theories can be constructed in different dimensions, with different gauge groups and with different amounts of supersymmetry. 
The different supergravity theories are very heterogeneous, and it is difficult to consider a common treatment. In order to define what a supergravity theory is, let us consider the eleven-dimensional supergravity. This theory was predicted by Nahm \cite{nahm} and constructed soon thereafter by  Cremmer, Julia and Scherk \cite{cremmer}. 


\subsubsection{Eleven-dimensional supergravity}

The bosonic geometrical data of eleven-dimensional supergravity \cite{Figueroa-O'FarrillMaximalSupergravity} consists of $(M,g,F)$, where $(M,g)$ is an eleven-dimensional lorentzian manifold with a Spin structure and $F \in \Omega^4(M)$, $dF=0$, is a closed four-form. There is also fermionic data, but we will restrict to bosonic solutions. The bosonic equations of motion of eleven-dimensional supergravity are a generalization of the Einstein and Maxwell equations in four dimensions. The generalization of the Einstein equation is:
\begin{subequations}\label{bothequations0}
\begin{equation}
\text{Ric}(X,Y)=\frac{1}{2} \langle i_X F ,i_Y F \rangle - \frac{1}{6} g(X,Y) \langle F , F \rangle ~~ , 
\label{Einsteinequation1}
\end{equation}
where the left hand side is the Ricci tensor and the right hand side is related to the energy momentum tensor of the field $F$, and $X,Y \in  \mathfrak{X} (M)$ are vector fields in $M$. The scalar product on forms $\langle - , -  \rangle$ is the natural one inherited from the metric. Rewriting the same equation in local coordinates:
\begin{equation}
R_{\mu \nu}= \frac{1}{12} F_{\mu \sigma_1 \sigma_2 \sigma_3} F_{\nu}^{ ~ \sigma_1 \sigma_2 \sigma_3} - \frac{1}{144} g_{\mu \nu}  F_{\sigma_1 \sigma_2 \sigma_3 \sigma_4} F^{\sigma_1 \sigma_2 \sigma_3 \sigma_4 } ~~ .
\label{Einsteinequation2}
\end{equation}
\end{subequations}
The generalization of the Maxwell equation is:
\begin{equation}
d * F = \frac{1}{2} F \wedge F ~~ .
\label{Maxwellequation}
\end{equation}
A set $(M,g,F)$ verifying equations (\ref{bothequations0}) and (\ref{Maxwellequation}) is a classical bosonic solution to eleven-dimensional supergravity. 

Equations  (\ref{bothequations0}) and (\ref{Maxwellequation}) can be obtained from an effective action:
\begin{equation}
S=\frac{1}{2 \kappa^2} \int_M   \left(*1 \  R - \frac{1}{2} F \wedge *F + \frac{1}{6} A \wedge F \wedge F \right) ~~ ,
\label{11sugraaction}
\end{equation}
where $A$ is a three form verifying $F=dA$, assuming that $F$ is exact. The same expression in local coordinates reads (let $U$ be a local chart of $M$):
\begin{equation}
S= \frac{1}{2 \kappa^2} \int_{U \subset M} d\xi^{11} \sqrt{-g}  \left( R -\frac{1}{48}   F_{\sigma_1 \sigma_2 \sigma_3 \sigma_4} F^{\sigma_1 \sigma_2 \sigma_3 \sigma_4 }\right)  +   \frac{1}{2 \kappa^2}\int_{U \subset M} \frac{1}{6} A \wedge F \wedge F   ~~ .
\label{11sugraaction2}
\end{equation}

We have chosen a Lorentz manifold with a Spin structure. A time-oriented and space-oriented Lorentz manifold $(M,g)$ has a Spin structure if the frame bundle can be lifted according to:
\begin{equation}
Spin_+(1,10) \mapsto SO_+(1,10) ~~.
\label{Spinlift}
\end{equation}
Not every manifold admits a Spin structure, and a topological obstruction can exist. A characterization for a Lorentz manifold to have a Spin structure is given in terms of Stiefel-Whitney classes \cite{karoubi}.

An equivalent and useful way of describing  the Spin structure is to consider a Clifford bundle on $M$, and then take the subbundle that is obtained by restriction to the Spin group inside the Clifford algebra at each point $p \in M$ (the obstruction to this reduction is the same as for the Spin structure). The Clifford algebra $Cl(1,10)$ can be read off from the classification of Clifford algebras \cite{Atiyah:1964zz}:

\begin{equation}
Cl(1,10) \cong Mat_{32} (\mathbb{R}) \oplus Mat_{32} (\mathbb{R}) ~~ .
\label{cliffordalgebra110}
\end{equation}

Let $\$ $ be the Spin bundle and $\varepsilon \in \Gamma (\$)$ a section, usually called spinor. There is a natural action of the exterior algebra on the Spin bundle:

\begin{equation}
c : \Lambda T^*M \cong  Cl(T^*M)  \rightarrow End \ \$ ~~ ,
\label{actioncliffordspinbundle}
\end{equation}
where the fist map is the bundle isomorphism induced by the vector space isomorphism between the exterior and Clifford algebras and the second map is induced from the action of the Clifford algebra $Cl(1,10)$ on the spinor representation $S$ of $Spin(1,10)$.

The supersymmetric character of the theory is given by a list of rules to transform bosons and fermions that leave invariant the equations of motion, the so called supersymmetry transformations. For bosonic solutions, where the fermionic fields are set to zero, the supersymmetric variation of bosonic fields is automatically zero, and the supersymmetric variation of the fermionic field is given by:
\begin{equation}
\delta_{\varepsilon} \Psi = D \varepsilon ~~ ,
\label{FermionicSusyVariation}
\end{equation}
where $D$, the supercovariant connection, is defined as:
\begin{subequations}\label{subequations1}
\begin{equation}
D_X \varepsilon \coloneqq \nabla_X \varepsilon + \frac{1}{6} c \left( \imath_X F \right) \varepsilon  - \frac{1}{12} c \left( X^\flat \wedge F \right) \varepsilon ~~~~~~~ \forall X\in \mathfrak{X}(M) ~~ ,
\label{Supersymmetrytransformation11dimsugra}
\end{equation}
where $\nabla$ is the spin connection and $X^\flat$ is the one-form dual to $X$. In components:
\begin{equation}
D_{\mu} \varepsilon  \coloneqq \nabla_{\mu} \varepsilon  + \frac{1}{36} \Gamma^{\sigma_1 \sigma_2 \sigma_3} F_{\mu_{} \sigma_1 \sigma_2 \sigma_3}  \varepsilon  - \frac{1}{288}  \Gamma_{\mu}^{~ \sigma_1 \sigma_2 \sigma_3 \sigma_4 }  F_{\sigma_1 \sigma_2 \sigma_3 \sigma_4}  \varepsilon ~~.
\label{Supersymmetrytransformation11dimsugra2}
\end{equation}
\end{subequations}

We say that a classical bosonic solution of eleven-dimensional supergravity $(M,g,F)$ is supersymmetric if there is a nonzero spinor $\varepsilon \in \Gamma (\$)$ which is parallel with respect to the supercovariant connection $D$, \ie, $D_X \varepsilon=0 \ $ $\forall X \in \mathfrak{X}(M)$, $\varepsilon \neq 0$.

A nonzero spinor satisfying $D \varepsilon=0$ is called a Killing spinor. As equation ($\ref{subequations1}$) is linear, the solutions form a vector space. The dimension of this vector space can range from 0 to 32, and it is common to use the parameter $\nu \coloneqq \text{dim} \{ \text{Killing spinors}\}/32$ to refer to the supersymmetry of a given background.


It has been proven that the algebraic structure generated by the infinitesimal isometries and the Killing spinors form a superalgebra for any bosonic supersymmetric solution of eleven-dimensional supergravity \cite{FigueroaO'Farrill:2004mx}. This superalgebra is known as `Killing superalgebra'. For example, for $AdS_4 \times S^7$ the Killing superalgebra is $\mathfrak{osp}(1|32)$. It has been also proven that this superalgebra is a filtered deformation of a subsuperalgebra of the Poincare superalgebra \cite{Figueroa-O'Farrill:2015efc}, and it is thought that the classification of the filtered subalgebras could allow a classification of the supersymmetric backgrounds (using this technique, it has been already rederived the classification of maximally supersymmetric backgrounds \cite{Figueroa-O'Farrill:2015efc}). \newline


The definition of Killing spinor could be generalized. We have considered a Spin structure on the manifold, and this can be generalized to Pin structures, or Clifford structures, restricting to the Pin group or Clifford group inside the Clifford algebra. Besides, complex Clifford algebras can be considered and Spin$^c$ structures instead of Spin structures, or more generally Pin$^c$ and Clifford$^c$ \cite{Moroianu}. The conditions for the existence of these structures can be also found in $\cite{karoubi}$. This kind of generalizations correspond to a generalization of the definition of supergravity as usually considered in the literature, and are still under investigation. It could be interesting to consider the potential applications to AdS/CFT. 
~~ \newline

\subsubsection{Supersymmetric solutions}

A general classification of bosonic supersymmetric solutions of eleven-dimensional supergravity is an open and difficult problem. Nevertheless, there are some partial classifications for some particular regions of the theory.

Let us consider the classification attending to the amount of supersymmetry. For eleven-dimensional supergravity it has been classified the solutions of maximal supersymmetry $\nu=1$ \cite{FigueroaO'Farrill:2002ft}, corresponding to symmetric spaces (basically, they are Minkowski, $AdS_4 \times S^7$, $AdS_7 \times S^4$ and the common Penrose limit of the last two cases). It has been proven that for $\nu=31/32$ \cite{Gran:2006cn, FigueroaO'Farrill:2007ica}  and $\nu=30/32$ \cite{Gran:2010tj} there are no solutions. Besides, it has been proven that for $\nu > 1/2$ the solutions are locally homogeneous \cite{FigueroaO'Farrill:2012fp}. For the following values there are known solutions:  1/32, 1/16, 3/32, 1/8, 5/32, 3/16, 1/4, 3/8, 1/2, 9/16, 5/8, 11/16, 3/4, 1, but for the other fractions of supersymmetry it is not known if there exist solutions.

Let us consider the classification for particular values of the four-form $F \in \Omega^4(M)$. First, for $F=0$, the superconnection $D$ reduces to the spin connection $\nabla$, and it suffices to classify the spin manifolds admitting parallel spinors with respect to the spin connection. In order to simplify this problem, it is commonly assumed that the eleven-dimensional manifold $M$ is a product of a Lorentz and a Riemann manifolds, $M=L \times R$. Choosing $L$ as the Minkowski space, then the problem reduces to find a riemannian manifold $R$ admitting parallel spinors.  The classification of simply connected, complete (in this subsection all manifolds will be simply connected and complete) Riemann manifolds admitting parallel spinors was obtained in \cite{Wang}. The idea is that the Riemann manifolds admitting parallel spinors are manifolds of special holonomy groups. The special holonomy groups where classified by Berger in 1957, and Wang \cite{Wang} used this result to classify those admitting parallel spinors. The result of Wang is that manifolds admitting parallel spinors are in one of the classes presented in Table \ref{Manifoldsadmittinparallelspinors}. To be precise, what is classified is not the whole set of manifolds admitting parallel spinors, but only the set of possible holonomy groups that the manifolds admitting parallel spinors can have. Knowing the special holonomy group allows to know the structure of the manifold.

\begin{table}[]
\centering
\begin{tabular}{ccccl}
\cline{1-4}
\multicolumn{1}{|c|}{dim}  & \multicolumn{1}{c|}{Holonomy}    & \multicolumn{1}{c|}{Geometry}    & \multicolumn{1}{c|}{n}       &  \\ \cline{1-4}
                           &                                  &                                  &                              &  \\ \cline{1-4}
\multicolumn{1}{|c|}{4k+2} & \multicolumn{1}{c|}{$SU_{2k+1}$} & \multicolumn{1}{c|}{Calabi-Yau}  & \multicolumn{1}{c|}{(1,1)}   &  \\ \cline{1-4}
\multicolumn{1}{|c|}{4k}   & \multicolumn{1}{c|}{$SU_{2k}$}   & \multicolumn{1}{c|}{Calabi-Yau}  & \multicolumn{1}{c|}{(2,0)}   &  \\ \cline{1-4}
\multicolumn{1}{|c|}{4k}   & \multicolumn{1}{c|}{$Sp_{k}$}    & \multicolumn{1}{c|}{hyperk\"ahler} & \multicolumn{1}{c|}{(k+1,0)} &  \\ \cline{1-4}
\multicolumn{1}{|c|}{7}    & \multicolumn{1}{c|}{$G_2$}       & \multicolumn{1}{c|}{exceptional} & \multicolumn{1}{c|}{1}       &  \\ \cline{1-4}
\multicolumn{1}{|c|}{8}    & \multicolumn{1}{c|}{$Spin_7$}    & \multicolumn{1}{c|}{exceptional} & \multicolumn{1}{c|}{(1,0)}   &  \\ \cline{1-4}
\end{tabular}
\caption{Manifolds admitting parallel spinors}
\label{Manifoldsadmittinparallelspinors}
\end{table}

We have considered an eleven-dimensional manifold as the product of a Minkowski space and a Riemann manifold $M=L \times R$, so we reduced the problem to the riemannian case. This has been the usual point of view adopted in the literature of string theory since its birth. Nevertheless, it is also interesting to consider parallel spinors in Lorentz manifolds. This problem was recently solved by Thomas Leistner. He obtained the classification of special holonomy manifolds for the lorentzian case, and as a corollary, the classification of the lorentzian manifolds admitting parallel spinors. In summary, the lorentzian manifolds admitting parallel spinors are the symmetric lorentzian spaces (in the riemannian case, symmetric spaces do not admit parallel spinors) and lorentzian manifolds with special holonomy group equal to $G \rtimes \mathbb{R}$, with $G$ a group of the riemannian case of Table \ref{Manifoldsadmittinparallelspinors}.

Let us now consider $F\neq 0$, for the particular case of $F$ being related to the volume form of the manifolds involved. Then the Killing spinor equation reduces to the geometrical Killing spinor equation \footnote{We use ``Killing spinor'' for a spinor satisfying the general equation (\ref{subequations1}) and ``geometrical Killing spinor'' for a spinor satisfying the particular equation (\ref{geometricalkillingspinor}) }, namely:

\begin{equation}
 \nabla_X \varepsilon= \lambda X \varepsilon ~~~~~~~ \forall X\in \mathfrak{X}(M) ~~ ,
\label{geometricalkillingspinor}
\end{equation}
with $\lambda \in \mathbb{C}$. As in the previous case, let us consider the simplification of considering the eleven-dimensional manifold as the product of a lorentzian manifold and a riemannian manifold. It turns out that the classification of riemannian manifolds admitting geometrical Killing spinors can be reduced to the classification of riemannian manifolds admitting parallel spinors with respect to the spin connection. This result was obtained by B\"ar \cite{Bar}, who used the cone construction. Given a manifold $X$ admitting geometrical Killing spinors, on the product manifold $\mathbb{R}^+ \times X$ the following metric can be constructed:
\begin{equation}
 ds_{\mathbb{R}^+ \times X}^2=dr^2+ r^2 ds_X^2~.
\label{conemetric}
\end{equation}
The geometrical Killing spinors on the base manifold $X$ are in 1 to 1 correspondence with the parallel spinors of the cone manifold $C=\mathbb{R}^+ \times X$. The corresponding geometries admitting geometrical Killing spinors are shown in Table \ref{GeometricalKillingSpinors}. 

An interesting question is if this cone construction can be extended to the lorentzian case. The answer is that it has not been done yet, because the classification in this case reduces to the classification of the manifolds with signature $(2,n-2)$ admitting parallel spinors, and the classification of special holonomy manifolds in signature $(2,n-2)$ is still unfinished.

\begin{table}[]
\centering
\begin{tabular}{ccccl}
\cline{1-4}
\multicolumn{1}{|c|}{dim X} & \multicolumn{1}{c|}{Holonomy of C} & \multicolumn{1}{c|}{Geometry of X}       & \multicolumn{1}{c|}{($n_+$, $n_-$)}                        &  \\ \cline{1-4}
                            &                                    &                                          &                                                            &  \\ \cline{1-4}
\multicolumn{1}{|c|}{k}     & \multicolumn{1}{c|}{\{1\}}         & \multicolumn{1}{c|}{round sphere}        & \multicolumn{1}{c|}{($2^{\lfloor k/2 \rfloor},2^{\lfloor k/2 \rfloor}$)} &  \\ \cline{1-4}
\multicolumn{1}{|c|}{4k-1}  & \multicolumn{1}{c|}{$Sp_{k}$}   & \multicolumn{1}{c|}{3-Sasaki}            & \multicolumn{1}{c|}{(k+1,0)}                               &  \\ \cline{1-4}
\multicolumn{1}{|c|}{4k-1}    & \multicolumn{1}{c|}{$SU_{2k}$}     & \multicolumn{1}{c|}{Sasaki-Einstein}     & \multicolumn{1}{c|}{(2,0)}                                 &  \\ \cline{1-4}
\multicolumn{1}{|c|}{4k+1}    & \multicolumn{1}{c|}{$SU_{2k+1}$}      & \multicolumn{1}{c|}{Sasaki-Einstein}     & \multicolumn{1}{c|}{(1,1)}                                 &  \\ \cline{1-4}
\multicolumn{1}{|c|}{6}     & \multicolumn{1}{c|}{$G_2$}         & \multicolumn{1}{c|}{nearly K\"ahler}     & \multicolumn{1}{c|}{(1,1)}                                 &  \\ \cline{1-4}
\multicolumn{1}{|c|}{7}     & \multicolumn{1}{c|}{$Spin_7$}      & \multicolumn{1}{c|}{weak $G_2$ holonomy} & \multicolumn{1}{c|}{(1,0)}                                 &  \\ \cline{1-4}
\end{tabular}
\caption{Manifolds admitting geometrical Killing spinors}
\label{GeometricalKillingSpinors}
\end{table}

For more general cases, with $F$ arbitrary, there are no known classifications for the manifolds admitting Killing spinors. For these backgrounds, the study must be done case by case, checking the existence of globally defined Killing spinors. 

Finally, it is also interesting to mention that there are some classifications for the more general case of supergravity defined using Spin$^c$ manifolds. For example, it has been classified the riemannian Spin$^c$ manifolds admitting parallel spinors \cite{Moroianu}.

These partial classifications have important applications to AdS/CFT. For example, it has allowed to extend the conjecture from the initial case of D3-branes in flat space to cone D3-branes, at the first stages of the formulation. One of the most remarkable examples is the Klebanov-Witten solution \cite{Klebanov:1998hh}.\newline

We have described in detail the basic aspects of eleven-dimensional supergravity. An analog study can be done for type IIA/B supergravity, but we will only outline the basic expressions that we will use. 
Along this thesis, we will mostly use type IIA and B supergravities.

\subsubsection{Type IIA supergravity}

It is a non-chiral maximally supersymmetric supergravity in ten dimensions. Type IIA supergravity can be obtained by dimensional reduction from eleven-dimensional supergravity. 

The bosonic geometrical data consists of $(M, g, \phi, B_2, A_1, A_3)$, where $(M, g)$ is a ten-dimensional Lorentz manifold with Spin structure, $\phi  \in \Omega^0 (M)$, $B_2 \in  \Omega^2 (M)$, $A_1 \in \Omega^1 (M)$, $A_3 \in \Omega^3 (M)$. Let us define:
\begin{eqnarray}
 H_3 &=&dB_2~, \\
 F_2 &=&dA_1 ~,   \\
 F_4 &=&dA_3 + H_3 \wedge A_1 ~.
\label{definitionsformsIIA}
\end{eqnarray}

The equations of motion are obtained from the action (in string frame):
$$S_{IIA}= \frac{1}{2 \kappa_{10}^2} \int d\xi^{10} \sqrt{-g} \Big\{ e^{-2 \phi} \left( R + 4 \partial_{\mu} \phi  \partial^{\mu} \phi - \frac{1}{2 \cdot 3!} H_3^2 \right) - \frac{1}{2 \cdot 2!} F_2^2 - \frac{1}{2\cdot 4!} F_4^2 \Big\}  $$
\begin{equation}
-\frac{1}{4 \kappa_{10}^2} \int B_2 \wedge dA_3 \wedge dA_3  ~,
\label{actiontypeIIBcomponents}
\end{equation}
where $2 \kappa_{10}^2=(2 \pi)^7 \alpha'^4 g_s^2$, and $g_s$ is the string coupling constant.

 The SUSY transformations for the dilatino $\lambda$ and the gravitino $\psi_m$ for type IIA supergravity in string frame are \cite{3},
\begin{eqnarray}
 \delta_{\epsilon}\lambda &=& \left[ \frac{1}{2}\Gamma^m \partial_m \Phi + \frac{1}{4\cdot3!}H_{mnp}\Gamma^{mnp} \Gamma_{11} + \frac{e^{\Phi} }{8} \left( \frac{3}{2!} F_{mn} \Gamma^{mn} \Gamma_{11} -\frac{1}{4!} F_{mnpq}\Gamma^{mnpq} \right)  \right] \epsilon \ ,  \nonumber  \\
 \delta_{\epsilon}\psi_m &=& \left[  \nabla_m+ \frac{1}{4\cdot2!}H_{mnp}\Gamma^{np} \Gamma_{11}  - \frac{e^{\Phi} }{8} \left( \frac{1}{2} F_{np} \Gamma^{np} \Gamma_{11} +\frac{1}{4!} F_{npqr}\Gamma^{npqr} \right) \Gamma_m  \right] \epsilon \ ,
 \label{susy0000AA}
\end{eqnarray}
where $\Gamma_{11}$ is the product of all gamma matrices, and $m,n,p,q,r \in \{1,...,10\}$.

\subsubsection{Type IIB supergravity}

There exists also a chiral maximal supergravity in ten dimensions, type IIB supergravity. It is related to type IIA supergravity by T-duality, and it can not be obtained by dimensional reduction from eleven-dimensional supergravity.

The bosonic geometrical data consists of $(M, g, \phi, B_2, \chi,A_2, A_4)$, where $(M, g)$ is a ten-dimensional Lorentz manifold with Spin structure, $\chi \in \Omega^0 (M)$, $B_2, A_2 \in \Omega^2 (M)$, $A_4 \in \Omega^4 (M)$. Let us define:
\begin{eqnarray}
 H_3 &=&dB_2~, \\
 F_1 &=&d\chi ~,   \\
 F_3 &=&dA_2+ \chi H_3  ~, \\
F_5 &=&dA_4 + H_3 \wedge A_2 ~.
\label{definitionsformsIIA}
\end{eqnarray} 

The equations of motion are obtained from the action (in string frame):


$$S_{IIB}= \frac{1}{2 \kappa_{10}^2} \int d\xi^{10} \sqrt{-g} \Big\{ e^{-2 \phi} \left( R + 4 \partial_{\mu} \phi  \partial^{\mu} \phi - \frac{1}{2 \cdot 3!} H_3^2 \right) - \frac{1}{2} \partial_{\mu} \chi  \partial^{\mu} \chi - \frac{1}{2 \cdot 3!} F_3^2- \frac{1}{4 \cdot 5!} F_5^2 \Big\}  $$
\begin{equation}
+\frac{1}{4 \kappa_{10}^2} \int dA_2 \wedge H_3 \wedge ( A_4 + \frac{1}{2} B_2 \wedge A_2)  ~,
\label{actiontypeIIBcomponents}
\end{equation}
where $2 \kappa_{10}^2=(2 \pi)^7 \alpha'^4 g_s^2$. The equations of motion obtained from this action are supplemented by a further condition, which states that the $F_5$ form is self-dual:
\begin{equation}
F_5=*F_5 ~.
\label{selfdualityofF5}
\end{equation}

 The SUSY transformations for the dilatino $\lambda$ and the gravitino $\psi_m$ for type IIB supergravity in string frame are \cite{3},
\begin{eqnarray}
 \delta_{\epsilon}\lambda &=& \left[ \frac{1}{2}\Gamma^m \partial_m \Phi + \frac{1}{4\cdot3!}H_{mnp}\Gamma^{mnp} \tau_3 - \frac{e^{\Phi} }{2}  F_m \Gamma^m(i\tau_2) -\frac{e^{\Phi}}{4\cdot3!} F_{mnp}\Gamma^{mnp} \tau_1 \right] \epsilon \ ,  \\
 \delta_{\epsilon}\psi_m &=& \left[  \nabla_m+ \frac{1}{4\cdot2!}H_{mnp}\Gamma^{np} \tau_3 + \frac{e^{\Phi}}{8} \left( F_n \Gamma^n (i\tau_2)+\frac{1}{3!}F_{npq}\Gamma^{npq} \tau_1 + \frac{1}{2\cdot5!}F_{npqrt}\Gamma^{npqrt} (i\tau_2)  \right) \Gamma_m \right] \epsilon \ ,
 \nonumber
 \label{susy0000}
\end{eqnarray}
where $\tau_i \ , \; i = 1,2,3$, are the Pauli matrices.


\section{T-duality}
\label{tduality}

T-duality was introduced in string theory in the eighties, and it has been applied in different contexts. Here we will review T-duality in the particular context of type IIA/B supergravity. We will first explain abelian T-duality and then non-abelian T-duality.

\subsection{Abelian T-duality}

Let us start from a type IIA/B supergravity background with a global $U(1)$-isometry, which leaves invariant not only the metric but all the fields of the solution.
Abelian T-duality is properly defined only for  backgrounds with such isometry. What T-duality does is to start from a background with a global $U(1)$-isometry in type IIA (IIB) supergravity and obtain a background with a global $U(1)$-isometry in type IIB (IIA) supergravity. The map is an involution, and performing T-duality again we recover the original background. It is conjectured that T-dual backgrounds are actually dual to each other, \ie, they are two different descriptions of the same physics. \newline

Locally, T-duality is given by the Buscher rules \cite{buscher} \cite{Bergshoeff:1995as}. Given a local expression for the metric and fields in terms of coordinates, and denoting by $\theta$ an adapted coordinate for the isometry, 
$(x^0,x^{\alpha})=(\theta, x^{\alpha})$, $\alpha=1,2,...,9$ , then the Buscher rules for the NSNS sector are:
\begin{eqnarray}
&  & \hat{g}_{00}=1/g_{00}~, ~~~~~~~ \hat{g}_{0\alpha}=B_{0 \alpha} / g_{00}  ~,   \nonumber \\
& & \hat{g}_{\alpha \beta}= g_{\alpha \beta} -(g_{0\alpha}g_{0\beta} - B_{0\alpha} B_{0 \beta})/g_{00} ~, \nonumber \\
& & \hat{B}_{0 \alpha} =g_{0 \alpha} /g_{00} ~,   \nonumber \\
& & \hat{B}_{\alpha \beta}=B_{\alpha \beta}  - (g_{0\alpha} B_{0 \beta} - g_{0\beta} B_{0 \alpha})/g_{00}    ~,    \nonumber \\
& & \hat{\phi}=\phi - \frac{1}{2} \log g_{00}~.
\label{abelianbuscherrules}
\end{eqnarray}
For the RR sector let us define:
\begin{equation}
\text{IIB}  ~~~~  P:=\frac{e^{\Phi}}{2} \sum_{n=0}^4 \slashed{F}_{2n+1} ~, ~~~~ \text{IIA}  ~~~~ P:=\frac{e^{\Phi}}{2} \sum_{n=0}^5 \slashed{F}_{2n}  ~,
\label{bispinortypeIIBIIA0}
\end{equation}
where $\slashed{F}_{i}:=1/i! \Gamma_{\mu_1...\mu_i} F_i^{\mu_1...\mu_i} $. Then, the transformation is given by:
\begin{equation}
\hat{P}=P \ \Omega^{-1}~,
\label{PhatPold}
\end{equation}
where:
\begin{equation}
\Omega= \frac{1}{\sqrt{g_{00}}} \Gamma_{11} \Gamma_0 ~.
\label{abelianbuscherrulesRR2}
\end{equation}

~\newline
Globally, the initial 10-dim background that has a $U(1)$-isometry, can be equivalently described by a $U(1)$-principal fibration over a 9-manifold. On the one hand, the $U(1)$-fibrations are classified by the Chern number (an integer positive number). On the other hand, the projection of the $H_3$ field over the base of the fibration gives a 2-form field, that integrated over the base manifold gives an integer number. Under T-duality, these two integer numbers are exchanged \cite{Bouwknegt:2003vb}, and the global properties of the transformation are cleanly specified.\newline

One interesting question is if supersymmetry is preserved along the procedure of T-duality.  Locally the supersymmetry is preserved and is the same before and after the T-duality. Nevertheless, in topologically non-trivial manifolds where the existence of global Killing spinors is non-trivial (as in the interesting AdS/CFT solutions), the question is not easy to answer. Following the conjecture, both theories are two different descriptions of the same physics, so the supersymmetry must be the same. But there are simple examples where the supersymmetry after the T-duality is lost, and the phenomenon of `supersymmetry without supersymmetry' appears. For example, in \cite{Duff:1998us} it is performed T-duality along the $S^1$ fibre of the `$S^5=\mathbb{CP}^2 \times S^1$' fibration in the $AdS_5 \times S^5$ solution, obtaining an $AdS_5 \times \mathbb{CP}^2 \times S^1$ solution in type IIA supergravity. This solution is not supersymmetric in the usual sense because $\mathbb{CP}^2$ is not Spin. But if the theory is T-dual to $AdS_5 \times S^5$ it is expected to have the same supersymmetry. The solution to this puzzle turns out to be that we have to consider Spin$^c$ Killing spinors, and $\mathbb{CP}^2$ is a Spin$^c$ manifold (for the definition of Spin$^c$ see section \ref{supergravity1}).\newline

The Buscher rules are derived from the T-dual action, via the following procedure. The idea is to start from the sigma model action for the NSNS fields, and gauge the $U(1)$ isometry $\partial \rightarrow D= \partial +A$ introducing a gauge field $A$.  Then, a Lagrange multiplier term is introduced, ensuring that  the gauge field is not dynamical. Finally, integrating out the gauge field we are left with the dual action. The Lagrange multiplier acts as the new coordinate. In the next section the computation is preformed for the more general case of non-abelian T-duality. \newline


We have considered the definition of abelian T-duality for a background with a global $U(1)$-isometry, or equivalently, with a $U(1)$-principal fibration. This definition can be extended to more general cases, for example, when the fibration is not a principal bundle. 
\newline

\subsection{Non-abelian T-duality}

Abelian T-duality can be generalized to non-abelian T-duality (NATD). The idea is to consider a background with a global $G$-isometry, where $G$ is a Lie group, which leaves invariant not only the metric but all the fields of the theory. Along this thesis, we will consider the particular case of $G=SU(2)$.

The non-abelian T-duality was
originally presented in 
\cite{de la Ossa:1992vc}
and was further developed 
and carefully inspected in \cite{Giveon:1993ai}-\cite{Borlaf:1996na}. See the lectures \cite{Quevedo:1997jb} for 
a nice account of some of the dualities that follow from a Buscher procedure. \\

Locally, we can consider the generalization of the Buscher rules to the $SU(2)$ case (for more details on notation and conventions see \cite{Itsios:2013wd}). If $L^i$, $(i=1,2,3)$ are the $SU(2)$ Maurer-Cartan one-forms, we can write the metric and NSNS $B_2$ form as:
\begin{align}
ds^2&=G_{\mu\nu} dx^{\mu} dx^{\nu} + 2 G_{ \mu i} dx^{\mu} L^i + g_{ij} L^i L^j ~, \\
B_2&=\frac{1}{2} B_{\mu \nu} dx^{\mu} \wedge dx^{\nu} + B_{\mu i} dx^{\mu}\wedge L^i + \frac{1}{2} b_{ij} L^i \wedge L^j ~,
\label{metricB2maurercartan}
\end{align}
where $\mu, \nu \in \{1,2,...,7\}$. Let us define:
\begin{equation}
Q_{\mu \nu}:=G_{\mu \nu}+B_{\mu \nu} ~, ~~ Q_{\mu i}:=G_{\mu i} + B_{\mu i} ~, ~~  Q_{i \mu}:=G_{i \mu} + B_{i \mu}      ~, ~~  E_{ij}:=g_{ij} + b_{ij}   ~,
\label{definitionsQAB}
\end{equation}
and from them the following block matrix:
\begin{equation}
Q_{AB}:=\left( \begin{array}{ccc}
Q_{\mu \nu} &   Q_{\mu i}  \\
 Q_{\nu j} &  E_{ij}  \end{array} \right) ~,
\label{definitionsQAB2}
\end{equation}
where $A,B \in \{1,2,...,10\}$. We can identify the dual metric and $B_2$ field as the symmetric and antisymmetric components, respectively,  of:

\begin{equation}
\hat{Q}_{AB}:=\left( \begin{array}{ccc}
Q_{\mu \nu}- Q_{\mu i } M_{ij}^{-1} Q_{j \nu} ~~ &   Q_{\mu j}M_{ji}^{-1}  \\
 -M_{ij}^{-1} Q_{j \mu}  ~~ &  M_{ij}^{-1}  \end{array} \right) ~,
\label{definitionsQAB2}
\end{equation}
where:
\begin{equation}
M_{ij}:=E_{ij}+ \epsilon_{ij}^k v_k ~,
\label{matrixMij}
\end{equation}
and $v_k$, $(k=1,2,3)$ are the new dual coordinates. Moreover, one finds that the dilaton receives a contribution at the quantum level
just as in abelian case:

\begin{equation}
\hat{\Phi}=\Phi - \frac{1}{2} \ln \left( \det M \right) ~.
\label{dualdilaton}
\end{equation}

In order to transform the RR fluxes, one must construct the following quantity out of the RR forms:
\begin{equation}
\text{IIB}  ~~~~  P:=\frac{e^{\Phi}}{2} \sum_{n=0}^4 \slashed{F}_{2n+1} ~, ~~~~ \text{IIA}  ~~~~ P:=\frac{e^{\Phi}}{2} \sum_{n=0}^5 \slashed{F}_{2n}~,
\label{bispinortypeIIBIIA}
\end{equation}
where $\slashed{F}_{i}:= \frac{1}{i!} \Gamma_{\mu_1...\mu_i} F_i^{\mu_1...\mu_i} $. Then, the dual RR fluxes are obtained from:
\begin{equation}
\hat{P}=P \ \Omega^{-1} ~,
\label{dualfluxesRR}
\end{equation}
where $\Omega=( \Gamma^1 \Gamma^2 \Gamma^3 + \zeta_{a} \Gamma^{a}) \Gamma_{11} / \sqrt{1+ \zeta^2}$, $\zeta^2=\zeta_a \zeta^a$,  $\zeta^a= \kappa^a_i z^i$, $\kappa^a_i \kappa^a_j=g_{ij}$ and $z_i=(b_i+v_i)/\det \kappa$, $b_{ij}=\epsilon_{ijk} b_k$. \newline

Let us review how to obtain the NSNS Buscher rules above from the T-dual  action. The lagrangian density for the NSNS sector of the initial solution is:

\begin{equation}
{\cal L}=Q_{AB} \partial_+ X^A \partial_- X^B  ~,
\label{initiallagrangian}
\end{equation}
where $\partial_{\pm} X^A=(\partial_{\pm} X^{\mu}, L^i_\pm)$. We omit the dilaton contribution. We then gauge the $SU(2)$ isometry by replacing the derivatives by covariant derivatives $\partial_{\pm} g \rightarrow D_{\pm} g=\partial_{\pm} g - A_{\pm} g$. The next step is to add a Lagrange multiplier term to ensure that the gauge fields are non-dynamical:
\begin{equation}
-i \tr (v F_{\pm}) ~, ~~~~ F_{\pm}=\partial_+ A_- - \partial_- A_+ - \left[ A_+, A_- \right]~.
\label{lagrangemultipliertermNATD}
\end{equation}
Now a gauge fixing choice must be done. We choose the 3 Lagrange multipliers $v_i$ as the new coordinates. Finally, in the last step we integrate out the gauge fields, obtaining the dual lagrangian density:
\begin{equation}
\hat{{\cal L}}=\hat{Q}_{AB} \partial_+ \hat{X}^A \partial_- \hat{X}^B  ~,
\label{finallagrangian}
\end{equation}
where $\partial_{\pm} \hat{X}^B = ( \partial_{\pm} X^{\mu} , \partial_{\pm} v^i)$. \newline

Globally, the rules for NATD are not known. This is a major difficulty, as without the global properties of the dual manifold, a lot of information is lost. For example, it is not known if NATD is in fact a duality, in the sense that the dual background is an equivalent description of the same theory, or it is a solution generation technique, \ie, a systematic method to obtain new backgrounds. 
\newline

The amount of supersymmetry preserved by a supergravity solution after a NATD transformation follows from the argument of \cite{Sfetsos:2010uq}, which has been proven in \cite{Kelekci:2014ima}. According to this, one just has to check the vanishing of the Lie-Lorentz (or Kosmann) derivative \cite{Kosmann} of the Killing spinor along the Killing vector that generates the isometry of the NATD transformation. More concretely, suppose that we want to transform a supergravity solution by performing a NATD transformation with respect to some isometry of the background that is generated by the Killing vector $k^\mu$. Then there is a simple criterion which states that if the Lie-Lorentz derivative of the Killing spinor along $k^\mu$ vanishes, then the transformed solution preserves the same amount of SUSY as the original solution. In the opposite scenario one has to impose more projection conditions on the Killing spinor in order to make the Lie-Lorentz derivative vanish. Thus in that case the dual background preserves less supersymmetry than the original one. Nevertheless, this is only a local description of the spinors, and for the analysis of globally defined spinors it is needed the global properties of the NATD.\newline

Let us summarize the basic differences between abelian T-duality and NATD: 
\begin{itemize}
\item The abelian T-duality is an involution, \ie, performing T-duality twice, the original background is recovered. Nevertheless, NATD is not an involution, as in the NATD background the isometries are no longer present (if there is no $SU(2)$ isometry it has no sense to apply NATD again).
\item   For compact commuting isometries one may argue that T-duality is actually a true
symmetry of string theory. There is no analogous statement for the non-abelian cases.
\item The global properties of the abelian T-duality are understood. Nevertheless, for the NATD they are not known yet. This problem is commonly stated as `the variables of the T-dual background are generically non-compact'.
\item In the abelian case we flip between type IIA and type IIB, but in the
non-abelian case we might change or stay within the same theory. If the dimension of the isometry group $G$ is odd, we flip from type IIA and B, and if the dimension is even, we stay within the same theory.
\end{itemize}


In chapter 2, we will perform NATD on backgrounds with $SU(2)$ isometry, to obtain new local solutions, and in particular, new $AdS_3$ solutions, which may be interesting for AdS/CFT applications.

\section{The AdS/CFT correspondence}

The AdS/CFT duality was originally formulated in 1997 by Juan Maldacena \cite{Maldacena:1997re}, who conjectured that type IIB string theory on $AdS_5 \times S^5$ is dual to $3+1$ ${\cal N}=4$  $SU(N)$  superconformal Yang-Mills theory.

The conjecture is based on two alternative descriptions of a D3-brane. Let us start by considering type IIB string theory on 9+1 Minkowski spacetime, and a stack of $N$ parallel D3-branes extended along 3+1 dimensions. String theory in this background contains two kinds
of perturbative excitations, open strings and closed strings. The open strings end on the D3-branes, and describe their excitations, while the closed strings are excitations of empty space. If we consider the system at energies lower than
the string scale $1/l_s$, then only the massless string states can be excited
. On the one hand, the closed string massless
states give a gravity supermultiplet in ten dimensions, whose low energy effective
action is the one of type IIB supergravity. On the other hand, the open string massless states give an
${\cal N} = 4$ vector supermultiplet in 3+1 dimensions, whose low energy effective
action is that of ${\cal N} = 4$ $SU(N)$ super-Yang-Mills theory. To be precise, the complete action of the massless excitations is:

\begin{equation}
S=S_{bulk}+S_{brane}+S_{int} ~~,
\label{actionbulkbraneinteraction}
\end{equation}
where $S_{bulk}$ is the action of ten dimensional supergravity plus higher derivative corrections,  $S_{brane}$ is defined on
the brane worldvolume and it contains the ${\cal N} =  4$ $SU(N)$ SYM action plus some higher derivative corrections, and $S_{int}$ describes the interactions between the brane and bulk excitations. 

Now we can take the low energy limit. In fact, the so-called decoupling limit or Maldacena limit consists in keeping the energy fixed and $\alpha' \rightarrow 0$, keeping also fixed $N$ and $g_s$ (the string coupling constant). In this limit the interaction action $S_{int}$ vanishes. In addition, all the higher derivative terms in the brane action vanish,
leaving only the pure 3+1 ${\cal N} = 4$ $SU(N)$ SYM conformal field theory, and the supergravity theory in the bulk becomes free.
So, in the low energy limit we have two decoupled systems:
free gravity in the bulk and  the 3+1  gauge
theory.

Let us now consider a different description of the same stack of D3-branes. Recall that in the low energy limit, type IIB string theory is well described by type IIB supergravity. The D3-brane solution in type IIB supergravity has the following form:
\begin{align}
 ds^2 & =h(r)^{-\frac{1}{2}} (-dt^2+dx^2+dy^2+dz^2) + h(r)^{\frac{1}{2}} \left( dr^2 + r^2 d\Omega^2_{S^5} \right) ~~, \nonumber \\
 F_5  & = (1+ * ) dt \wedge dx \wedge dy \wedge dz \wedge d(h(r)^{-1})~~,    ~~~~ F_3=0~~, ~~~~ F_1=0~~,  ~~~~ e^{\phi}=1  ~~,
\label{d3brane}
\end{align}
where,
\begin{equation}
h(r) :=1+\frac{R^4}{r^4}~~,  ~~~~ R^4:=4 \pi g_s \alpha'^2 N~.
\label{d3brane2}
\end{equation}

The D3-brane is charged under the $F_5$ self-dual form, and it is a massive object since it has some tension. Now notice that the $g_{tt}$ component of the metric is not constant, and depends on $r$. Then, the energy $E_r$ of an object as measured by an observer at a position $r$ and the energy $E_{\infty}$ measured by an observer at infinity are related by a redshift factor:
 \begin{equation}
E_r \ h(r)^{-\frac{1}{4}}=  E_{\infty} ~~,
\label{redshiftfactor}
\end{equation}

This means that a object moving towards $r=0$ will have less and less energy from the point of view of the observer at infinity.

Let us now take the low energy limit of the D3-brane solution \ref{d3brane}. In this low energy limit there are two kind of excitations, from the point of view of the observer at infinity. On the one hand, there are massless particles propagating in the bulk with increasing wavelength, and, on the other hand, there are any kind of excitation getting closer to $r=0$. In the low energy limit, these two kinds of modes decouple from each other. So, finally we end with two decoupled sectors, one is free bulk supergravity and the other is the near horizon region of the geometry. This near horizon geometry is obtained from \ref{d3brane} by approximating $h(r)=R^4/r^4$, obtaining a new exact solution to the supergravity equations of motion:
\begin{align}
 ds^2 & =\frac{r^2}{R^2} (-dt^2+dx^2+dy^2+dz^2) +  \frac{R^2}{r^2} dr^2 + R^2 d\Omega^2_{S^5}  ~~,  \nonumber \\
 F_5  & =  (1+ * )4 \frac{r^3}{R^4} dt \wedge dx \wedge dy \wedge dz \wedge dr~~,    ~~~~ F_3=0~~, ~~~~ F_1=0~~,  ~~~~ e^{\phi}=1  ~~,
\label{ads5timess5solution}
\end{align}
which is the geometry of  $AdS_5 \times S^5$.

In summary, we have described a D3-brane as a string theory object where the open strings can end, and alternatively, as a solution to supergravity. In both cases we have two decoupled sectors in the low energy limit. In each case, one of the two decoupled sectors is supergravity in flat space. Thus, it is natural to identify the second system
which appears in both descriptions. This leads to conjecture that 3+1 ${\cal N}=4$ $SU(N)$ SMY theory is dual to type IIB string theory on $AdS_5 \times S^5$ \newline

If the theories are conjectured to be the same, then the global symmetries must match.  On the one hand, the isometry group of $AdS_5$ 
\footnote{In the AdS/CFT context, by `$AdS_d$ space' it is meant `universal covering of $AdS_d$ space'. The reason is because in $AdS_d$ cyclic causal curves are allowed, but not in its universal covering. The topology of the first is $S^1 \times \mathbb{R}^{d-1}$ and the topology of the second is $\mathbb{R}^{d}$. The isometries of the $AdS_d$ space are $O(2,d-1)$ (the connected component of the neutral element is $SO_+(2,d-1)$), and the isometries of the universal covering of $AdS_d$ space are $SO(2,d-1)$.}
is $SO(2,4)$, and it matches with the conformal group of 3+1 Minkowski space, which is also  $SO(2,4)$. On the other hand, the isometry group of $S^5$ is $SO(6)$, and the R-symmetry of the field theory is $SO(6)_R$. Also these bosonic symmetries match when extended to the full supergroups on both sides of the duality. This supergroup is $SU(2,2|4)$, and its maximal bosonic subgroup is precisely $SO(2,4) \times SO(6)$.

Moreover, if the theories are conjectured to be dual, also the supersymmetry preserved by them must match. In both cases the supersymmetry is maximal, namely $\nu=1$. Notice that the initial configuration of D3-branes have only $\nu=1/2$ supersymmetries, but after the decoupling limit, the supersymmetry is enhanced to $\nu=1$. 

Besides, we can write down a dictionary for the parameters on both sides of the duality. The string coupling $g_s$ and the Yang-Mills coupling  $g_{YM}$ are related by: 
\begin{equation}
g_{YM}^2=4 \pi g_s ~.
\label{couplingconstants}
\end{equation} 
Moreover, the $\theta$ parameter on the gauge theory is identified with the expectation value of the RR scalar $\chi$ on the gravity side. These relations can be put together: 
\begin{equation}
\tau:=\frac{4 \pi i}{g_{YM}^2}+ \frac{\theta}{2\pi} = \frac{i}{g_s}+ \frac{\chi}{2 \pi} ~,
\label{couplingconstants2}
\end{equation} 
We have written the couplings in this way because both the string theory and gauge theory have an $Sl(2,\mathbb{Z})$ self duality symmetry, under which $\tau \rightarrow (a \tau + b)/( c \tau +d)$, where $a,b,c,d$ are integer numbers satisfying $ad-bc=1$. This transformation is related to S-duality, another important duality in string theory. 
The number of initial D3-branes $N$, after the low energy limit, corresponds to the flux of the $F_5$ form through the $S^5$ sphere:
\begin{equation}
N=\int_{S^5} F_5 ~.
\label{fluxF5}
\end{equation} 
 and on the field theory side, it is the range of the gauge group $SU(N)$.\newline

The AdS/CFT conjecture can be formulated at three different levels of generality. The correspondence as formulated above is the strong statement of the duality, which assumes that both theories are dual for any values of the parameters $g_s$ and $N$. The conjecture can also be formulated in a mild version, that states that 3+1 ${\cal N}=4$ $SU(N)$ SYM theory in the large $N$ limit is dual to type IIB classical string theory ($g_s \rightarrow 0$) on $AdS_5 \times S^5$. The large $N$ limit in the field theory side is also known as the t'Hooft limit. Recall that the t'Hooft limit is obtained by taking $g_{YM} \rightarrow 0$ and $N \rightarrow \infty$ while keeping fixed $\lambda:=g_{YM}^2 N$, the t'Hooft coupling constant. It corresponds to a topological expansion of the Feynman diagrams of the field theory where the quantity $1/N$ acts as the coupling constant, and at leading order in $1/N$ only the planar diagrams contribute.
Corrections $1/N$ on the field side corresponds to $g_s$ corrections on the string side.

We can also consider a weak version of the duality, by taking the large t'Hooft coupling limit $\lambda \rightarrow \infty$. On the string theory side this corresponds to take the limit $\alpha' \rightarrow 0$, where classical string theory reduces to classical supergravity. So, the duality in its weak form states that ${\cal N}=4$ $SU(N)$ SYM theory in the large $N$ limit and large t'Hooft coupling $\lambda$ is dual to classical type IIB supergravity on $AdS_5 \times S^5$.
This weak formulation of the duality is the most useful, because using classical supergravity we can access to strong coupling results in field theory. This will be the formulation that we will use along this thesis.\newline

In fact, the duality is a weak/strong duality. This means that if on one side of the duality the coupling constant  is strong, then on the other side is weak, and vice versa. This makes the duality very useful, when we consider a field theory at strong coupling, where the perturbative techniques are not available, and other tools are very limited (like lattice field theory, localization or integrability). In this case, we can consider the gravity dual, which is weakly coupled, and compute the observables using classical supergravity. For the particular case of ${\cal N}=4$ $SU(N)$ SYM theory in the large $N$ limit and large $\lambda$ limit, the theory is strongly coupled. Taking into account that $R^4/\alpha'^2=\lambda$, then the supergravity theory has a large $AdS_5$ and $S^5$ radius, and is classical supergravity. \newline

We have stated that the AdS/CFT duality establishes a relation between a gravity theory and a field theory. Let us now be more precise and write down the dictionary relating observables on both sides of the duality.

A conformal gauge theory is specified by a complete set of conformal operators. We are interested in the gauge invariant conformal operators polynomial in the canonical fields since these operators will have a definite conformal dimension and, moreover, will form finite multiplets of increasing dimension.  In order to compute correlators for a given operator ${\cal O}$, its generating functional $\Gamma({\cal O})$  is constructed by adding a source term to the lagrangian of the conformal field theory:
\begin{equation}
e^{\Gamma({\cal O})}=\langle e^{ \int h {\cal O}} \rangle ~,
\label{operatorgeneratingfunctional}
\end{equation}
where $h$ is a source for ${\cal O}$.

On the gravity side of the duality, the closed string excitations
can be described by fields living on $AdS_5 \times S^5$. When compactifying along the $S^5$, these
fields can be decomposed in terms of spherical harmonics, thus
giving rise to an infinite tower of modes.
Therefore, the
observables of the string theory are a set of five-dimensional fields living on $AdS_5$, and, according to the correspondence, they have to be matched to the observables of the gauge theory.

Recall, the metric of $AdS_5$ space with radius $R$ can be written as:
\begin{equation}
ds^2=\frac{r^2}{R^2}dx^2_{1,3}+ \frac{R^2}{r^2}dr^2 ~,
\label{adsmetric000}
\end{equation}
 In these variables the boundary of $AdS_5$ is at
$r \rightarrow \infty$ and is isomorphic to four-dimensional Minkowski space, spanned by the coordinates
$x_0, ..., x_3$.  Now let us consider a scalar field $\hat{h}(x,r)$ on $AdS_5$, to be associated to the operator ${\cal O}$ of the gauge
theory. 
The asymptotic dependence of $\hat{h}$ on the radial coordinate is given by the two independent solutions:
\begin{equation}
\hat{h}  \sim r^{\Delta-4}~,  ~~~~  \hat{h} \sim r^{-\Delta}~,  ~~~~ (r \rightarrow \infty)~, 
\label{asymptoticbehaviorhhat}
\end{equation}
where $\Delta=2+\sqrt{4+m^2 R^2}$, and $m$ is the five-dimensional mass. Then, the first solution dominates near the boundary, and we can write:
\begin{equation}
\hat{h}(x,r)  \rightarrow r^{\Delta-4} \hat{h}_{\infty} (x)~,  ~~~~ (r \rightarrow \infty)~, 
\label{asymptoticbehaviorhhat2}
\end{equation}
 where $\hat{h}_{\infty} (x)$ is a four-dimensional function living on the boundary and is defined only up
to conformal transformations, under which it has conformal dimension $\Delta-4$.  Finally, we can establish a relation between the field $\hat{h}$ and the operator ${\cal O}$, by identifying the generating functional of correlators of ${\cal O}$ with the five dimensional action $S_{AdS_5}(\hat{h})$ restricted to the solutions of $\hat{h}$ with boundary value $ \hat{h}_{\infty} (x)$:

\begin{equation}
\langle e^{ \int h {\cal O}} \rangle = e^{-S_{AdS_5} (\hat{h})} ~.
\label{operatorgeneratingfunctional}
\end{equation}

This relation was proposed in \cite{Gubser:1998bc}, \cite{Witten:1998qj}. Here we have considered the case of a scalar field, and a similar relation can be obtained for fields of other spins.\newline



After the original formulation in 1997 of the duality by Maldacena \cite{Maldacena:1997re}, the conjecture was extended to other cases, with less supersymmetry and without conformal invariance. Let us mention some of the most important solutions that appeared soon afterwards. In 1998 it was published the Klebanov-Witten solution \cite{Klebanov:1998hh}, a solution with reduced supersymmetry based on the conifold geometry. The conifold is a concrete case of the special holonomy manifolds revisited in section \ref{supergravity1}. This solution was followed by several generalizations, like the Klebanov-Tseytlin solution \cite{Klebanov:2000nc}, which use fractional D3-branes on the conifold, and the Klebanov-Strassler solution \cite{Klebanov:2000hb}, which use fractional D3-branes on the deformed conifold, both solutions being non-conformal. Another non-conformal solution with reduced supersymmetry, formulated in 2000, is the Pilch-Warner solution \cite{Pilch:2000ue}, a massive deformation of the ${\cal N}=4$ $SU(N)$ SYM theory, usually known as ${\cal N}=2^*$. Also, in 2000, it was published the Maldacena-Nu\~nez  solution \cite{Maldacena:2000mw}, in which D5-branes are wrapped in a calibrated cycle of a special holonomy manifold. 

A huge amount new examples of the correspondence were constructed until these days. 
One of the most important examples of the duality was formulated in 2008 by Aharony, Bergman, Jafferis and Maldacena \cite{Aharony:2008ug}, and it is known as the ABJM theory. This thesis is mostly based on this solution, so we devote the next subsection to explain this example in full detail.

\subsection{The ABJM theory}
\label{pureABJM}

The ABJM theory is one of the most important examples of the AdS/CFT duality. In the same way that the  ${\cal N}=4$ $SU(N)$ SYM theory is the best understood example of the AdS$_5$/CFT$_4$ correspondence, the ABJM theory is the best understood example of the AdS$_4$/CFT$_3$ correspondence. 

In the seminal paper by Maldacena in 1997 \cite{Maldacena:1997re}, it was conjectured the duality between the maximally supersymmetric $AdS_5 \times S^5$  type IIB supergravity solution and the ${\cal N}=4$ $SU(N)$ SYM field theory based on the decoupling limit of D3-branes in flat space. Besides, it is also noticed that there are two other maximally supersymmetric cases,  $AdS_4 \times S^7$ and $AdS_7 \times S^4$, in eleven-dimensional supergravity, corresponding to the decoupling limit of the M2- and M5-branes on flat space, respectively. Nevertheless, the concrete field theories living inside the M-branes were not known. 

The gauge theory living on the M5-brane in flat space was known to be a $5+1$ ${\cal N}=(2,0)$ superconformal theory with number of degrees of freedom scaling as $N^{3}$ in the large $N$ limit. Nowadays, the details of the theory still remain unknown. 

The gauge theory living on the M2-branes in flat space was known to be a conformal $2+1$ ${\cal N}=8$  supersymmetric Chern-Simons-matter theory, with R-symmetry $SO(8)$ and a specific matter content, and that should scale as $N^{3/2}$ in the large $N$ limit. But the details of the theory where not known.  An important step towards the understanding of this theory was obtained in \cite{Schwarz:2004yj}, where $2+1$ ${\cal N}=1$ and ${\cal N}=2$ superconformal $U(N)$ Chern-Simons-matter theories where constructed \footnote{Interestingly, the gravity duals of these theories where recently found in \cite{Guarino:2015jca}.}. Nevertheless, it seemed not possible to obtain interacting theories with more supersymmetry and, besides, these theories were parity violating (the theory on the M2-branes must be parity preserving). Another important step was obtained by Bagger and Lambert \cite{Bagger:2007vi}  \cite{Bagger:2007jr} \cite{Bagger:2006sk} and Gustavsson \cite{Gustavsson:2007vu} \cite{Gustavsson:2008dy} who constructed a maximally supersymmetric ${\cal N}=8$ Chern-Simons-matter theory which was proven to be equivalent \cite{VanRaamsdonk:2008ft} to a theory with the particular gauge group $SU(2) \times SU(2)$ and opposite levels $k$ and $-k$ \footnote{Notice that this theory is different from the ABJM for $N=2$ because the gauge group is $SU(2) \times SU(2)$ and not $U(2) \times U(2)$.}.

 Inspired by the previous results, in 2008 Aharony, Bergman, Jafferis and Maldacena (ABJM) \cite{Aharony:2008ug} constructed a $2+1$ dimensional ${\cal N}=6$ superconformal Chern-Simons-matter theory with gauge group $U(N) \times U(N)$ and opposite levels $k$ and $-k$. The bosonic matter content of the theory is four complex scalar fields $C_I$ $(I = 1, 2, 3, 4)$ transforming in the bifundamental representation $(\textbf{N}, \bar{\textbf{N}} )$ of the gauge group and their corresponding complex conjugate fields in the anti-bifundamental representation $( \bar{\textbf{N}}, \textbf{N} )$. The fermionic matter content is 4 Majorana fermions,    $\psi_I$ $(I = 1, 2, 3, 4)$.

The action of the ABJM theory is:

$$S_{ABJM}=\int_{{\cal M}_3} d^3x \  \tr \left[ \frac{k}{4\pi}\left(A\wedge dA + \frac{2}{3} A\wedge A \wedge A - \hat{A}\wedge d\hat{A} - \frac{2}{3} \hat{A}\wedge \hat{A} \wedge \hat{A}\right) \right.$$
$$-D_{\mu} C_I^{\dagger}  D^{\mu} C^I - i \psi^{I  \dagger}  \gamma^{\mu} D_{\mu} \psi_I$$
$$+\frac{4 \pi^2}{3 k^2} \left(C^IC_I^{\dagger}C^JC_J^{\dagger}C^KC_K^{\dagger} + C_I^{\dagger}C^IC_J^{\dagger}C^JC_K^{\dagger}C^K + 4 C^IC_J^{\dagger}C^KC_I^{\dagger}C^JC_K^{\dagger} - 6 C^IC_J^{\dagger}C^JC_I^{\dagger}C^KC_K^{\dagger} \right)$$

$$+\frac{2 \pi i}{k} \left( C_I^{\dagger}C^I {\psi}^{J \dagger} \psi_J -\psi^{\dagger J} C^I C_I^{\dagger} \psi_J - 2 C_I^{\dagger} C^J \psi^{\dagger I} \psi_j+2 \psi ^{\dagger J} C^I C_J^{\dagger} \psi_I\right.$$

\begin{equation}
\left.+ \epsilon^{IJKL}\psi_I C_J^{\dagger} \psi_K C_L^{\dagger} -  \epsilon_{IJKL} \psi^{\dagger I} C^J \psi^{\dagger K} C^L  \right) ~,
\label{abjmaction}
\end{equation}
where the covariant derivative acts as:
\begin{equation}
D_{\mu} C^I=\partial_{\mu} C^I + i(A_{\mu} C^I - C^I \hat{A}_{\mu})~.
\label{covariantderivative}
\end{equation}
The first line contains the Chern-Simons terms of the two gauge fields $A$ and $\hat{A}$ corresponding to the two gauge groups $U(N) \times U(N)$, with levels $k$ and $-k$ respectively. The second line contains the kinetic terms for the complex scalar fields and the Majorana spinors. The rest of the action is the potential. The potential is particularly chosen in order to preserve ${\cal N}=6$ supersymmetry and it is very constrained.

\begin{figure}[!ht]
  \centering
     \includegraphics[scale=0.4]{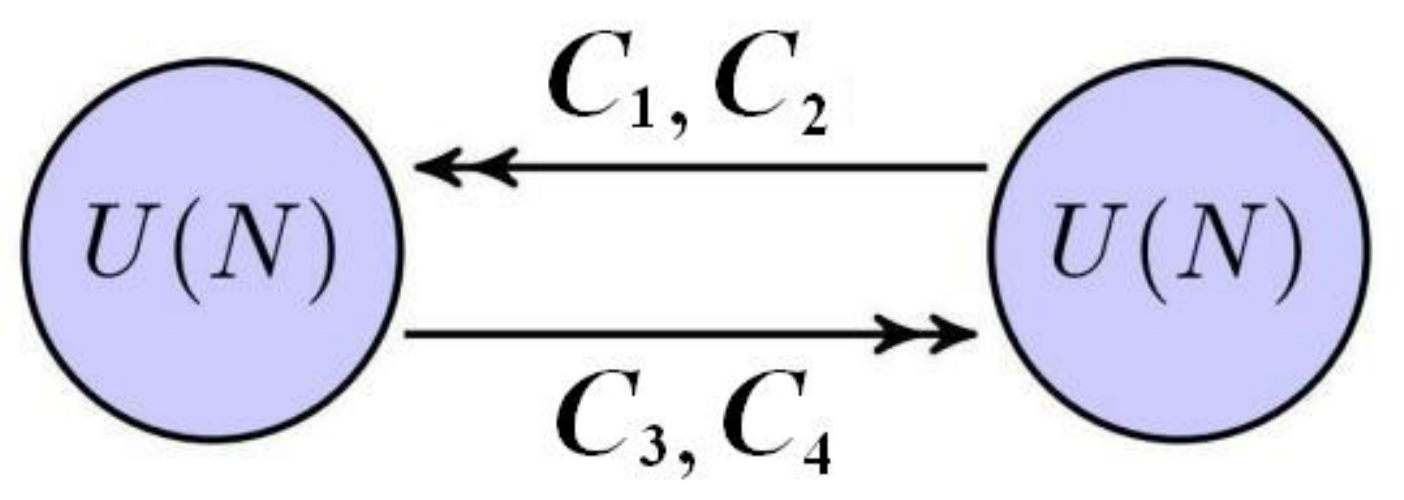}
  \caption{ABJM quiver diagram. The nodes correspond to the gauge groups, and the arrows represent the bosonic matter content $C_I$ ($I=1,2,3,4$) transforming in the bifundamental and anti-bifundamental representations.}
\end{figure}

The field content of the $2+1$ ABJM theory is similar to the $3+1$ Klebanov-Witten theory \cite{Klebanov:1998hh}, and also has a similar superpotential for the chiral fields. If we relabel $\{C_1, C_2, C_3, C_4\}=\{A_1,A_2,B_1^{\dagger},B_2^{\dagger}\}$ the superpotential is:
\begin{equation}
W=\frac{2 \pi}{k} \epsilon^{ij} \epsilon^{kl} \ \tr [A_i B_k A_j B_l] ~.
\label{ABJMsuperpotential}
\end{equation}

The ABJM theory is formulated on a 3 dimensional manifold ${\cal M}_3$. For the purpose of holography, this manifold will be chosen to be $\mathbb{R} \times S^2$ with the natural lorentzian metric, as this is the boundary of $AdS_4$. For the purpose of localization, the manifold will be chosen to be the euclidean manifold $S^3$ with the usual round metric.

The $R$ symmetry of the theory is $U(4)_R=U(1)_R \times SU(4)_R$. The scalars $C_I$ and the fermions
$\psi_I$ are in the fundamental $\textbf{4}$ representation of $SU(4)_R$, and have charge $+1$ under $U(1)_R$.

One can take the t'Hooft limit in the ABJM theory, which is given by:
\begin{equation}
N , \ k \rightarrow \infty ~~~ , ~~~~~~ \lambda := \frac{N}{k}=fixed ~,
\label{tHooftlimitABJM}
\end{equation}
where $\lambda$ is the t'Hooft coupling. It turns out that in the large $N$ limit, the degrees of freedom scale as $N^{3/2}$, a different behavior than in ${\cal N}=4$ $SU(N)$ SYM, which scales as $N^2$.

In general, the ABJM theory is not maximally supersymmetric, but for specific values of the parameters $N$ and $k$ the supersymmetry is enhanced from ${\cal N}=6$ to ${\cal N}=8$. It happens when $N=2$ or $k=1,2$. In particular, the case $k=1$ is interesting, as it corresponds to $N$ M2-branes in flat space, which in the decoupling limit give rise to the maximally supersymmetric solution $AdS_4 \times S^7$.

The ABJM theory is a conformal highly supersymmetric field theory that has nice mathematical properties. 
Using localization and matrix model techniques it has been obtained the exact interpolating function
for the free energy of ABJM theory on the three-sphere in the large $N$ limit, which implies in particular the $N^{3/2}$
behavior at strong coupling \cite{Drukker:2010nc}. This constitutes a non trivial check of the AdS/CFT correspondence.
Using localization also the Wilson Loops can be obtained. Besides, the ABJM theory has many integrability properties. \newline

Let us now consider the gravity dual description to the ABJM field theory. In order to motivate the conjecture, the authors of \cite{Aharony:2008ug} constructed a specific brane set up in type IIB string theory, involving D3-branes, D5-branes and NS5-branes, where one of the coordinates is compactified on $S^1$. Next, they performed a T-duality along this compact coordinate and obtained a solution in type IIA string theory. Afterwards, this solution is lifted to M theory, where it is shown to be the theory of M2-branes on a $\mathbb{C}^4/\mathbb{Z}_n$ singularity. 

So, the conjecture in its strong version states that the ABJM field theory is dual to M2-branes on a $\mathbb{C}^4/\mathbb{Z}_n$ singularity in M theory. 

In the large $N$ limit, the gravitational background becomes $AdS_4 \times S^7/\mathbb{Z}_k$ in eleven-dimensional supergravity. The $S^7/\mathbb{Z}_k$ space is a lens space, \ie, a fibre bundle with base $\mathbb{CP}^3$ and fiber $S^1$, and the possible fibrations are classified by an integer, in this case $k$. When the Chern-Simons level $k$ is large ($k^5 \gg N$) the size of the fiber is small and the system is better described in terms of type IIA   supergravity, after performing a dimensional reduction to ten dimensions along the Hopf fiber $S^1$. In the type IIA description, the geometry is $AdS_4 \times \mathbb{CP}^3$, where $\mathbb{CP}^3$ inherits the natural metric from the total space $S^7/\mathbb{Z}_k$, known as the Fubini-Study metric. The solution also has a constant dilaton, and $F_2$ and $F_4$ forms.

The $\mathbb{CP}^3$ manifold with the Funibi-Study metric has an isometry group $U(4)$. This precisely corresponds with the $R$-symmetry of the ABJM field theory $U(4)_R=U(1)\times SU(4)$. The manifold $\mathbb{CP}^3$ has a spin structure, and in particular a K\"ahler structure with K\"ahler form $J$. The dilaton $\phi$ and the $F_2$ and $F_4$ forms are:
\begin{equation}
e^{\phi}=\frac{2 k}{L}~, ~~~~ F_2=2 k J~, ~~~~ F_4=\frac{3 k}{2} L^2 \Omega_{AdS_4}~, 
\label{dilatonandformsF2F4}
\end{equation}
where $\Omega_{AdS_4}$ is the volume form of $AdS_4$ and $L$ is the radius of $AdS_4$, $L^4=2 \pi^2 N/k$. 

This type IIA description of the ABJM theory is the one that we will use in chapters 3, 4 and 5 of this thesis. \newline

The ABJM theory has been generalized in several directions. By adding fractional M2-branes it can be generalized (ABJ) to the case in which the ranges of the two gauge groups are different, $U(N) \times U(N+M)$, and in the type IIA description this corresponds to adding a topologically non trivial $B_2$ field proportional to $M$ \cite{Aharony:2008gk}. It can be generalized to the case where the two levels are not opposite, $k$ and $-k'$ with $k-k'\neq0$ and in the type IIA description this corresponds to a generalization to massive type IIA with mass parameter proportional to $k-k'$ \cite{Gaiotto:2009mv} \cite{Gaiotto:2009yz}. Moreover, the ABJM model admits a mass deformation (mABJM) that preserves the whole ${\cal N} = 6$ supersymmetry \cite{Gomis:2008vc}. The ABJM theory has been extended to different quivers and groups with different amount os supersymmetry in \cite{Imamura:2008nn} \cite{Jafferis:2008qz} \cite{Imamura:2008ji} \cite{Martelli:2008si}  \cite{Hanany:2008cd} \cite{Ooguri:2008dk} \cite{Benna:2008zy}.

    Another generalization is obtained by including fundamental matter. One of the purposes of this thesis is to consider the ABJM theory with unquenched massive fundamental matter, and this question will be addressed in chapter 3. In chapters 4 and 5 we will use the ABJM background and we will add quenched fundamental matter with gauge fields turned on and we will study its properties.

\section{Flavor in the gauge/gravity correspondence}

In the original formulation of the gauge/gravity correspondence by Maldacena \cite{Maldacena:1997re}, all the fields of the gauge theory are in the adjoint representation of the gauge group $SU(N)$. In the subsequent years, new examples of the correspondence where constructed, either with one gauge group and adjoint matter or quiver groups and adjoint or bifundamental matter (as in the ABJM theory). Soon, the need for fundamental matter appeared, and the way of how to introduce it was clarified in \cite{Karch:2002sh}.

The main motivation for introducing fundamental matter is to get closer to the phenomenological models, like QCD or those of condensed matter physics. In QCD the fundamental matter are the quarks, and in condensed matter physics the fundamental matter are the electrons.

\subsection{Brane construction}

 Let us start with the theory of ${\cal N}=4$ $SU(N)$ SYM and add flavor to it, as was first explained in \cite{Karch:2002sh}.

Let us consider a stack of $N_c$ D3-branes and a stack of $N_f$ D7-branes in type IIB string theory, in 10-dim Minkowski spacetime. The D3-branes are extended along 3+1 dimensions, and the D7-branes are extended along the same 3+1 dimensions and  4 transverse spatial directions. In this configuration, there are closed strings and open strings, and the open strings can have both ends on the D3-brane, both ends on the D7-brane or one end on the D3-brane and the other in the D7-brane. The lightest modes of these branes after quantization give rise to field theories living inside the D3 and the D7, with adjoint matter coming from the strings with both ends on the same brane, and bifundamental matter with one end in each brane. To be precise, after quantization we will have a $SU(N_c)$ gauge theory on the D3-branes, with adjoint matter ($\textbf{N}_c \otimes \bar{\textbf{N}}_c = \textbf{1}\oplus (\textbf{N}_c^2-1)$) coming from the strings with both ends on the D3-brane (3-3), we will have a $SU(N_f)$ gauge theory on the D7-branes, with adjoint matter ($\textbf{N}_f \otimes \bar{\textbf{N}}_f = \textbf{1}\oplus (\textbf{N}_f^2-1)$) coming from the strings with both ends on the D7-brane (7-7), and bifundamental matter  $\textbf{N}_c \otimes \bar{\textbf{N}}_f$ and $\textbf{N}_f \otimes \bar{\textbf{N}}_c$ coming from the strings (3-7) and (7-3) \footnote{Recall that for $SU(N)$ with $N\ge 3$ there are two fundamental representations, $\textbf{N}$ and $\bar{\textbf{N}}$, for $SU(2)$ there is only one, $\textbf{2}$, and $U(1)$ is not semisimple, so there is no definition of fundamental representation. In the text we assume $N\ge 3$ for simplicity.}. 

Starting from this configuration, let us consider the near horizon limit for the D3-branes. The lightest modes of the 3-3 string give rise to the ${\cal N}=4$ $SU(N)$ SYM with adjoint matter, as in the original case of Maldacena. Now, there is also the contribution from the 3-7 and 7-3 strings lightest modes, which give rise to a ${\cal N}=2$ hypermultiplet of matter transforming in the bifundamental representation of $SU(N_c) \times SU(N_f)$. The gauge group of the D7-brane becomes a global group in the gauge side, precisely the flavor group. The (7-7) strings then correspond to matter in the adjoint of the global group $SU(N_f)$ (it is common to abuse of language and refer to these bound states as mesons, but the theory is not confining, so they are just bound states).
On the gravity side, we will have some geometry, and in general it will be difficult to find. If we reduce to the particular case $N_f \ll N_c$, \ie, we take the t'Hooft limit with $N_f$ finite, then on the gravity side the geometry is not modified by the presence of the D7-branes, so we have $AdS_5 \times S^5$ where the D7-branes are treated as probe branes. On the gauge theory side, this corresponds to the `quenched approximation', a terminology used in lattice QCD, which means that the quark determinant is set to one.
Under the quenched approximation, many features of the fundamental matter can be captured, and it has turned out to be a very successful regime where many computations in the gauge/gravity duality have been carried out.

\subsubsection{Unquenched flavor}

 If the full effect of the fundamental matter has to be taken into account, then we can consider a number of $N_f$ D7-branes comparable to the number $N_c$ of D3-branes. To be precise, this corresponds in the large $N_c$ limit to take  $ N_c, N_f \rightarrow \infty $ and $ N_f / N_c = fixed $, and this is known as the Veneziano limit \cite{Veneziano:1976wm}. In this case, the low energy theory corresponds to type IIB supergravity coupled to the dynamics of the D7-branes, which is dictated by the DBI +WZ action. The total action of the problem is:

\begin{equation}
S=S_{IIB}+S_{DBI+WZ}~,
\label{totalactionforbackreaction}
\end{equation}
In practice, to find a backreacted solution one starts with a solution of pure supergravity where the gauge/gravity duality has been well understood. Then, one considers an ansatz that generalizes the solution and allow the backreaction of the flavor. Next, one introduces this ansatz in the general equations of motion obtained from \ref{totalactionforbackreaction}. Usually finding solutions to this equations is a difficult task, and in order to simplify the problem it is common to restrict to supersymmetric solutions. In that case, under an appropriate ansatz, it is enough to solve the dilatino and gravitino variations of type IIB, and the problem simplifies considerably. 

A concrete supersymmetric solution to the unquenched D3-D7 problem was found in \cite{Nunez:2010sf}. This solution is no longer conformal, and presents a Landau pole in the UV. For an example of a non-supersymmetric backreacted solution, see \cite{Faedo:2015urf}. \newline

An interesting question is what are the new features of the flavor that are captured when considering backreaction. For example, the beta function can get corrections, and the originally flavorless conformal background becomes non-conformal after backreaction \cite{Nunez:2010sf}. Another interesting feature is that some conformal dimensions become anomalous after backreaction (even if the backreacted background is still conformal) \cite{Conde:2011sw}. Also, the meson spectrum, after backreaction, will mix with the glueball spectrum. Besides, new dualities, like Seiberg-like dualities, may appear.  \newline

We can classify the approaches to the backreaction problem in two cases: the case in which the flavor branes are localized, and the case where the flavor branes are smeared. Let us consider the case in which the flavor branes are localized. This means that we have a stack of D7-branes at the same point. Then, the flavor symmetry is $SU(N_f)$, the source terms for the supergravity equations will have a Dirac delta at the location of the stack. Furthermore, the symmetries of the original (unflavored) background will be reduced, and dependence in more variables is introduced. 
Solving a problem with these features is a difficult task. In some cases exact solutions are found, for an example see  \cite{Gauntlett:1997pk}.

\subsubsection{The smearing technique}
\label{smearingtechnique}

The smearing technique consists in delocalizing the brane sources. The idea is the same as in electromagnetism, when a point charge is substituted by a continuous volume charge. In this way, no Dirac deltas are present, the symmetries of the original theory are restored, and the dependence on several variables disappear. As the branes are at different locations, the flavor group is $U(1)^{N_f}$ instead of $SU(N_f)$. 

This approach of the smearing technique was proposed in \cite{Bigazzi:2005md} in the context of non-critical holography. In practice, the smearing technique has been very useful and has lead to the discovery of a large amount of new backgrounds with backreacted flavor (for a review see \cite{Nunez:2010sf}). In general, obtaining a solution using the smearing technique is easier that with localized sources, and in many cases, even analytical solutions are obtained.\newline

In chapter 3 we will construct a model that generalizes the ABJM theory by including massive unquenched fundamental matter, using the smearing technique. In the next subsection we review a more elementary case where the fundamental matter is included in the ABJM theory in the quenched approximation.

\subsection{The ABJM theory with quenched flavor}
\label{ABJMquenchedflavor}

The ABJM theory can be generalized by including fundamental matter. One of the main interests of including fundamental matter are the applications in phenomenological models like QCD or condensed matter physics.

The easiest way of including fundamental matter is using the quenched approximation. Starting from the gravity description of ABJM theory in the type IIA regime, this can be obtained by including $N_f$ flavor probe D6-branes \cite{Hohenegger:2009as}, \cite{Gaiotto:2009tk}, with $N_f$ fixed when $N_c \rightarrow \infty$.

\begin{figure}[!ht]
  \centering
     \includegraphics[scale=1]{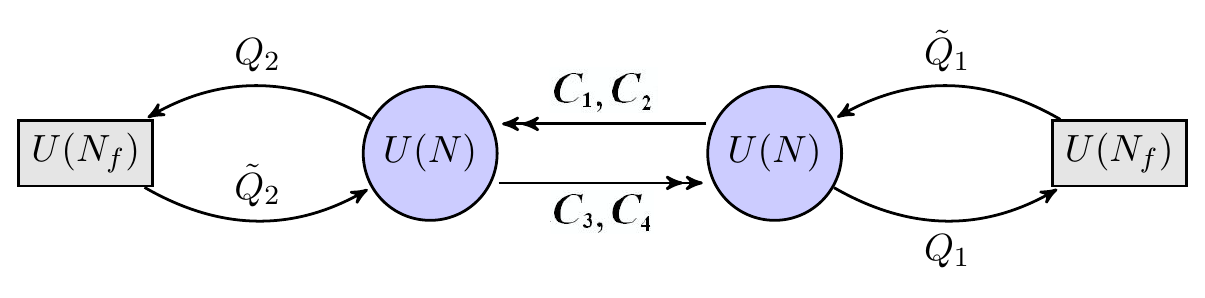}
  \caption{ABJM quiver diagram with flavor hypermultiplets}
\label{ABJMquiverflavor}
\end{figure}

The resulting field theory is the original ABJM plus a fundamental hypermultiplet in each node of the quiver diagram, as shown in Fig. \ref{ABJMquiverflavor}, preserving ${\cal N}=3$ supersymmetry of the original ${\cal N}=6$.

Let us comment on the precise brane construction. In order to introduce flavor in the ABJM theory, one can start by introducing two sets of $N_f$ D5-branes in the initial type IIB configuration. Each stack of D5-branes is situated at a specific position at the $x_6$ compact direction, in opposite places, and not coincident with the NS5 and NS5'. The open strings with ends 3-5 and 5-3 give rise to fields transforming in the $(\textbf{N}_c, \bar{\textbf{N}}_f,)$ and $(\bar{\textbf{N}}_c, \textbf{N}_f)$ representations. After T-dualizing to type IIA and lifting to M-theory, the D5-branes become KK-monopoles. In the type IIA description, they become D6-branes extended along $AdS_4$ and wrapping a lagrangian cycle $\mathbb{RP}^3$ inside $\mathbb{CP}^3$, which preserve half of the supersymmetries (${\cal N}=3$). 

This construction allows to conjecture that these probe D6-branes are dual to fundamental hypermultiplets in the ABJM theory, transforming in the fundamental  $(\textbf{N},1)$, $(1, \textbf{N})$ and anti-fundamental $(\bar{\textbf{N}},1)$, $(1, \bar{\textbf{N}})$ representations. The action of the ABJM field theory with this fundamental matter is the ABJM action plus the kinetic terms of the fundamental matter and the superpotential:

\begin{equation}
W=\frac{2 \pi}{k} Tr[(A_a B_a + Q_1 \tilde{Q}_1)^2 - (B_a A_a - Q_2 \tilde{Q}_2)^2 ]~.
\label{superpotentialABJMwithflavor}
\end{equation}

If the $N_f$ flavor branes are located at the same point,  the flavor group is $SU(N_f)$. \newline

Here we have reviewed the addition of flavor in the ABJM theory using probe D6-branes. This construction of adding flavor to ABJM using D6-branes will be the one that we will use along this thesis. Nevertheless, there are other brane set-ups for adding different kinds of fundamental matter to ABJM. For a list of supersymmetric flavor branes that allows to add different kinds of fundamental matter to ABJM see \cite{Ammon:2009wc}. \newline

For the reader who is familiar with the flavor D7-brane in $AdS_5 \times S^5$, we can compare it with the  flavor D6-brane in $AdS_4 \times \mathbb{CP}^3$ :

\begin{itemize}
\item  In both cases the codimension of the flavor brane is zero, \ie, the D6-brane fills the 2+1 dimensions of the ABJM theory and the D7-brane is extended along the 3+1 dimensions of the ${\cal N}=4$ $SU(N)$ SYM theory.

\item The D7-brane adds a fundamental hypermultiplet to the only $SU(N)$ node of the quiver diagram, while the D6-brane adds a fundamental hypermultiplet to each node $U(N) \times U(N)$ quiver. 

\item For the D7-brane embedding only the DBI part of the action contributes, while for the D6-brane embedding both DBI+WZ contribute.

\item  The D7-brane worldvolume is $AdS_5 \times S^3 \subset AdS_5 \times S^5$, while the D6-brane worldvolume is $AdS_4 \times \mathbb{RP}^3 \subset AdS_4 \times \mathbb{CP}^3$. 

\item For massive embeddings, where there is some brane profile, the brane terminates at some point before reaching the IR. At this point, in the D7 case the $S^3$ collapses to a point, and in the D6 case the $\mathbb{RP}^3$  becomes a $S^2$ (the $S^1$ fibre of $\mathbb{RP}^3$ collapses to a point but the base $S^2$ remains unaffected).

\end{itemize}


\section{The D-brane action}
\label{cp1sscbraneaction}
\setcounter{equation}{0}

The D-branes play a fundamental role in the derivation of the AdS/CFT correspondence. After taking the decoupling limit, the D-branes disappear, and we are left with the field theory and the supergravity theory. 
Nevertheless, in the previous section it has been shown how to add flavor to the AdS/CFT duality via D-branes. This flavor branes remain after the decoupling limit, so it is necessary to deal with its dynamics. If we restrict to the quenched approximation for the fundamental matter in the field theory, this corresponds to consider a probe Dp-brane in the supergravity background. 
In this section we will review the action of a D-brane in a generic supergravity background (for a review see \cite{Polchinski:1996na}), which will be used in chapters 3, 4 and 5.

D-branes are extended objects where open strings can end. The open strings ending on the brane must satisfy Dirichlet boundary conditions in the directions orthogonal to the hyperplane. Let us first recall that in the absence of any such object the massless bosonic sector of the open string consists of a ten-dimensional vector and, accordingly, the low energy effective field theory for the open superstring in flat space with Neumann boundary conditions is ${\cal
N}=1$ SYM. However, the zero modes of the open string along the directions with Dirichlet boundary conditions are not dynamical and so the
low energy fields corresponding to the massless modes of the  open string sector do not depend on these directions. Therefore, the low energy
effective theory one gets is the dimensional reduction of the ten-dimensional one, with the components of the gauge field along
directions with Dirichlet boundary conditions becoming scalars representing the fluctuations of the D-brane. The hyperplane is then a dynamical object whose fluctuations correspond to states of the open string spectrum.

The bosonic action of a Dp-brane can be computed by requiring that the
non-linear sigma model describing the propagation of an open string with
Dirichlet boundary conditions in a general supergravity background is
conformally invariant. The constraints arising for the bosonic fields are
the same as the equations of motion resulting from the following action
(in string frame):
\beq
S_{DBI}=-T_p\int_{\Sigma_{p+1}} d^{p+1}\xi\;e^{-\phi}\sqrt{\,-{\rm
det}\,(\,g+{\cal F}\,)}\,,
\label{cp1DBI}
\eeq
which is known as the Dirac-Born-Infeld (DBI) action. The presence of the
dilaton $\phi$ is due to the fact that the Dp-brane is an object of the open
string spectrum, hence the coupling $g_s^{-1}$ inside $T_p=(2 \pi)^{-p} g_s^{-1} \alpha'^{- \frac{p+1}{2}}$ , where $T_p$ is the tension of the brane. In (\ref{cp1DBI})  $\Sigma_{p+1}$ denotes the Dp-brane
worldvolume, and
$g$ is the induced metric on the Dp-brane resulting from the pullback to
the worldvolume of the background metric:
\beq
g_{ab}={\partial\,X^M\over \partial\xi^a}\,{\partial X^N\over\partial
\xi^b}\,G_{MN}\,,
\label{cp1indmetric}
\eeq
where $\xi^a$, $(a=0,\cdots,p)$ are the worldvolume coordinates,
and $X^M$ and $G_{MN}$ are the coordinates and metric of
the background. By making use of the worldvolume and spacetime
diffeomorphism invariance one can go to the so-called `static gauge'
where the worldvolume of the Dp-brane is aligned with the first $p+1$
spacetime coordinates. In this gauge the embedding of the Dp-brane is
given by the functions $X^m(\xi^a)$, where $X^m$, $(m=1,\cdots,9-p)$ are
the spacetime coordinates transverse to the Dp-brane.
Actually, in this `static gauge' the induced metric for a Dp-brane in flat
space takes the form:
\beq
g_{ab}=\eta_{ab}+{\partial\,X^m\over \partial\xi^a}\,{\partial X^n\over
\partial\xi^b}\,\delta_{mn}\,.
\label{cp1stindmetric}
\eeq
The two-form ${\cal F}$ present in eq. (\ref{cp1DBI}) is given by:
\beq
{\cal F}_{ab}=P[B]_{ab}+2\pi\,\alpha'\,F_{ab}\,,
\label{cp1calf}
\eeq
where $F=dA$ is the field strength of the gauge field living on the
worldvolume of the Dp-brane, $P\big[\cdots]$ denotes
the pullback to the worldvolume and $B$ is
the NSNS two-form potential. 


The DBI action (\ref{cp1DBI}) not only describes
the massless open string modes given by the worldvolume gauge field $A_a$
and the scalars $X^m$, but also couples the Dp-brane to the massless
closed string modes of the NSNS sector, namely $\phi$, $G_{MN}$ and
$B_{MN}$. Moreover,  this action is an abelian $U(1)$
gauge theory and when the target space is flat,
it reduces (at leading order in $\alpha'$) to YM in $p+1$ dimensions with
$9-p$ scalar fields plus higher derivative terms. Indeed, the quadratic
expansion of (\ref{cp1DBI}) for a Dp-brane in flat space is:
\beq
S_{DBI}\approx-\int_{\Sigma_{p+1}} d^{p+1}\xi\,e^{-\phi}
\left[T_p+{1\over
g_{YM}^2}\left({1\over2}F_{ab}F^{ab}+\partial_a\Phi^m\,
\partial^a\Phi^m\right)\right]\,,
\label{cp1dpYM}
\eeq
where we have applied the usual relation between coordinates and 
fields: $\Phi^m=1/(2\pi\,\alpha')\,X^m$, and that $g_{YM}^2= 2 (2\pi)^{p-2} g_s (\alpha')^{\frac{p-3}{2}}$.
After taking into account the
fermionic superpartners coming from the fermionic completion of the
action for the super D-brane, the low energy effective theory living on
the worldvolume becomes the  maximally supersymmetric SYM theory in $p+1$
dimensions (for a D-brane in flat space). In particular, for $p=3$ it is obtained the ${\cal N}=4$ $SU(N)$ SYM theory of the original Maldacena conjecture.


The worldvolume of a Dp-brane naturally couples to a $(p+1)$-form
RR potential
$C_{(p+1)}$, whose field strength is a $(p+2)$-form. 
Apart from the natural coupling term $\int_{\Sigma_{p+1}} C_{p+1}$, the
action coupling a Dp-brane to the RR potentials must contain additional
terms involving the worldvolume gauge field. By making use of T-duality
one can show that the part of the action of a Dp-brane describing its
coupling to the RR potentials is given by the following Wess-Zumino (WZ)
term:
\beq
S_{\,WZ}=T_p\,\int_{\Sigma_{p+1}}\,\sum_{r=0}^{p+1}P\left[C_{(r)}\right]
\wedge e^{\cal F}\, .
\label{cp1WZ}
\eeq
Although the integrand above involves forms of various ranks, the integral only picks out those proportional to the volume form of the Dp-brane worldvolume.

Finally, the action of a Dp-brane is given by the sum of the DBI and WZ
terms, (\ref{cp1DBI}) and (\ref{cp1WZ}) respectively,
which reads (in string frame):
\beq
S_{Dp}=-T_p\int_{\Sigma_{p+1}}d^{p+1}\xi\,e^{-\phi}\sqrt{\,-{\rm
det}\,(\,g+{\cal F}\,)}\,\pm\, T_p\int_{\Sigma_{p+1}}\,\sum_{r=0}^{p+1}\,
P\left[C_{(r)}\right]\wedge e^{\cal F}\,,
\label{cp1Dbraneact}
\eeq
where the + (-) in front of the second term corresponds to a Dp-brane (anti-Dp-brane).


\subsection{Kappa symmetry}
\label{cp1sskappasym}

The kappa symmetry is a fermionic gauge
symmetry of the worldvolume theory which is an important ingredient in the
covariant formulation of superstrings and
supermembranes \cite{Simon:2011rw}. It turns out to be a useful tool to find
supersymmetric embeddings of probe D-branes in a given background, and it will be used in chapters 3 and 4.

The bosonic Dp-brane is described by a map $X^m\left(\Sigma_{p+1}\right)$,
where $\Sigma_{p+1}$ is the worldvolume of the brane and $X^m$ are
the coordinates of the ten-dimensional target space. In order to construct the action
of the super Dp-brane, this map is replaced with a supermap
$\{Z^M\}=(X^m,\theta^\alpha)$ and, accordingly, the bosonic
fields with the corresponding superfields. In this subsection we will use the indices $M,N,\cdots,$ for the
superspace coordinates and the indices $m,n,\cdots,$ for the
bosonic coordinates of the ten-dimensional spacetime. Superspace forms can be
expanded in the coordinate basis $dZ^M$, or alternatively, in the one-form
frame $E^{\underline M}=E^{\underline M}_N\,dZ^N$, where the underlined
indices are flat and $E^{\underline M}_N$ is the supervielbein. Under the
action of the Lorentz group $E^{\underline M}$ decomposes into a vector
$E^{\underline m}$ and a spinor $E^{\underline\alpha}$, which is a
32-component Majorana spinor for IIA superspace and a doublet of chiral
Majorana spinors for IIB superspace. In particular, for flat superspace
one can write:
\beq
E^m=dX^m+\bar\theta\Gamma^m\,d\theta\,,\qquad\quad
E^\alpha=d\theta^\alpha\,,
\label{kpflatbs}
\eeq
where the indices are already flat. The supersymmetric generalization of
the DBI plus WZ terms of the action  (\ref{cp1Dbraneact}) for a
super Dp-brane reads:
\beq
S=-T_p\int_{\Sigma_{p+1}} d^{p+1}\xi\,\sqrt{-\det (g+\cal{F})}+T_p\int_{\Sigma_{p+1}}
\,\sum_{r=0}^{p+1}\, P\left[C_{(r)}\right]\wedge e^{\cal F}\,,
\label{kpsdpact}
\eeq
with:
\bear
&&g_{ab}=E^{\underline m}_a\,E^{\underline n}_b\,\eta_{\underline m
\underline n}\,,\rc\rc &&{\cal F}_{ab}=F_{ab}-E^{\underline M}_a
E^{\underline N}_b B_{\underline M\underline N}\,,\rc\rc
&&C_{(r)}={1\over r!}\,dZ^{M_1}\wedge\cdots\wedge dZ^{M_r}\,C_{M_1\cdots
M_r}\,,
\label{kpdpfields}
\eear
where the indices $a,b$ run over the worldvolume, $F$ is the field
strength of the worldvolume gauge field $A$ ($F=dA$)\footnote{With
respect to eq. (\ref{cp1Dbraneact}) $F$ has been made dimensionless by
means of the redefinition $2\pi\,\alpha'\,F\to F$.}, $B$ is the NSNS
two-form potential superfield, $C_{(r)}$ is the RR $r$-form gauge potential
superfield and $g_{ab}$ is the induced metric on the worldvolume expressed
in terms of the pullback to the worldvolume of the supervielbein, which is
given by:
\beq
E^{\underline m}_a=\partial_a Z^M\,E^{\underline m}_M\,.
\label{kpvielbpb}
\eeq
In ref. \cite{severaldaniel} the action (\ref{kpsdpact}) was shown to be
invariant under the local fermionic transformations:
\bear
&&\delta_\kappa\,E^{\underline m}=0\,,\rc\rc
&&\delta_\kappa\,E^{\underline \alpha}=(1+\Gamma_\kappa)\kappa\,,
\label{kpkpvar}
\eear
where $\delta_\kappa\,E^{\underline M}=(\delta_\kappa Z^N)
E^{\underline M}_N$. This generalizes to curved backgrounds the kappa
symmetry variations of the flat superspace coordinates:
\beq
\delta_\kappa\,\theta=(1+\Gamma_\kappa)\kappa\,,\qquad\quad\delta_\kappa\,
X^m=\bar\theta\Gamma^m\delta_\kappa\theta\,.
\eeq
One can easily check, reading the supervielbein from (\ref{kpflatbs}),
that these transformations fulfill eq. (\ref{kpkpvar}).

In terms of the induced gamma matrices
$\gamma_{a}=E^{\underline m}_a\,\Gamma_{\underline
m}$, the matrix $\Gamma_\kappa$ takes the form \cite{severaldaniel}:
\beq
\Gamma_{\kappa}={1\over \sqrt{-\det(g+{\cal F})}}\,
\sum_{n=0}^{\infty}\,{(-1)^n\over 2^n n!}\,\gamma^{a_1b_1
\cdots a_n b_n}\,{\cal F}_{a_1 b_1}\,\cdots\,{\cal F}_{a_n b_n}\,
J^{(n)}_{(p)}\,,
\label{kpgammakpdef}
\eeq
where $g$ and ${\cal F}$ are those defined in (\ref{kpdpfields}), and
$J^{(n)}_{(p)}$ is the following matrix:
\beq
J^{(n)}_{(p)} = \begin{cases}(\Gamma_{11})^{n+(p-2)/2}\,\,\Gamma_{(0)} & (\text{IIA}) ~, \cr\cr
(-1)^n (\sigma_3)^{n+(p-3)/2}\,\,i\sigma_2 \otimes \Gamma_{(0)} & (\text{IIB}) ~, \end{cases}
\label{kpmtj}
\eeq
with $\Gamma_{(0)}$ being:
\beq
\Gamma_{(0)}={1\over (p+1)!}\,\,\epsilon^{a_1\cdots a_{p+1}}\,\,
\gamma_{a_1\cdots a_{p+1}} ~,
\label{kpGammazero}
\eeq
and $\gamma_{a_1\cdots a_{p+1}}$ denoting the antisymmetrized product of
the induced gamma matrices. In eq. (\ref{kpmtj}) $\sigma_2$ and $\sigma_3$
are Pauli  matrices that act on the two Majorana-Weyl components 
(arranged as a two-dimensional vector) of  the type IIB spinors. 

What makes this fermionic symmetry a key ingredient in the formulation of
the super Dp-brane is that, upon gauge fixing, it eliminates the extra
fermionic degrees of freedom of the worldvolume theory, guaranteeing the
equality of fermionic and bosonic degrees of freedom on the worldvolume.
Recall that the number of bosonic degrees of freedom coming
from the transverse scalars is $9-p$. After adding up the $p-1$ physical
degrees of freedom corresponding to the worldvolume gauge field, the total
number of bosonic degrees of freedom is 8. On the other hand, one has 32
spinors $\theta$ which are cut in half by the equation of motion. We will
see that by gauge fixing the local kappa symmetry the spinorial degrees of
freedom are also halved, thus resulting the expected 8 physical spinors.

Kappa symmetry is a useful tool for finding embeddings of Dp-branes preserving some
supersymmetry \cite{severaldaniel2}. In particular,  we are interested in bosonic
configurations where the fermionic degrees of freedom vanish, \ie \
$\theta =0$, so we only need the variations of
$\theta$ up to linear terms in $\theta$. The supersymmetry plus kappa
symmetry transformations of $\theta $ is:
\beq
\delta\theta=\epsilon+(1+\Gamma_\kappa)\kappa ~,
\label{kpthetatransf}
\eeq
where $\epsilon$ is the supersymmetric variation parameter. Therefore,
although generically these transformations do not leave $\theta$ invariant,
one can choose an appropriate value of $\kappa$ such that
$\delta\theta=0$. Of course, this amounts to gauge fixing the kappa
symmetry. Indeed, let us impose the following gauge fixing condition:
\beq
{\cal P}\,\theta=0 ~,
\label{kpkpgaugefix}
\eeq
where ${\cal P}$ is a field independent projector, ${\cal P}^2=1$, so
the non-vanishing components of $\theta$ are given by $(1-{\cal
P})\,\theta$. The condition for preserving the gauge fixing condition
${\cal P}\delta\theta=0$ results in:
\beq
{\cal P}\delta\theta = {\cal P}\epsilon + {\cal P}(1 + \Gamma_{\kappa})\kappa =
0 ~, \label{kpunbrsusy}
\eeq
which determines $\kappa=\kappa(\epsilon)$. Then, after gauge fixing the
kappa symmetry, the transformation (\ref{kpthetatransf}) becomes a global
supersymmetry transformation. The condition of unbroken supersymmetry for the non-vanishing
components of $\theta$, namely $(1-{\cal P})\,\theta$, reads:
\beq
(1-{\cal P})\,\delta\theta = (1-{\cal P})\,\epsilon + 
(1-{\cal P})(1 + \Gamma_{\kappa})\,\kappa(\epsilon)= \epsilon + 
(1 + \Gamma_{\kappa})\,\kappa(\epsilon)=0 ~,
\label{kpunbrsusy2}
\eeq 
where in the last equality we have used eq. (\ref{kpunbrsusy}).
Multiplying the last equality by $(1-\Gamma_\kappa) $ one gets:
\beq
(1-\Gamma_\kappa)\,\epsilon=0 ~.
\label{kpfnalcond}
\eeq  
Then, the fraction of supersymmetry preserved by the brane is given
by the number of solutions to this equation. Notice that $\Gamma_\kappa$,
defined in eq. (\ref{kpgammakpdef}) depends on the first derivatives of
the embedding trough the induced metric and the pullback of the $B$
 field.  Finally, for backgrounds of reduced supersymmetry, $\epsilon$
(the target space supersymmetry parameter) must be substituted by the Killing
spinor  of the ten-dimensional geometry where the Dp-brane is embedded.

\section{Applications of the gauge/gravity duality}

The gauge/gravity duality can be used to explore  quantum field theories using gravity theories or vice versa. In most of its applications, classical supergravity is used to explore quantum field theories at strong coupling. The traditional method to explore quantum field theories was to use a perturbative approach, which is only valid at weak coupling regime. In general, few tools existed to address quantum field theories at strong coupling: lattice quantum field theory, integrability and localization, basically. The gauge/gravity duality has become a new powerful tool to address strongly coupled quantum field theories. 

Given a quantum field theory at strong coupling, the hope would be to find the exact gravity dual. Nevertheless, it is still not known how to obtain the gravity dual of any phenomenological theory of nature. Starting from highly supersymmetric and conformal theories, there has been large efforts to introduce new elements to get closer to the phenomenological theories, but even the more sophisticated models are still far from the real theories.  So, a natural question is: why is the duality useful if it is not known the gravity dual? What is expected from the duality is that we can extract universal properties. Indeed, we can consider several field  theories with known gravity dual, that share some features with a given phenomenological theory, and then extract some universal behavior from them. If a result holds for all of them, then we can conjecture that the result will be also valid for the phenomenological theory. So, the hope is that from the duality we can predict general behaviors, but not concrete numbers. \newline

Let us consider a list of fields were the AdS/CFT conjecture has found applications:

\begin{itemize}

\item Applications to QCD. Even if there have been large efforts devoted to find a theory as close as possible to the gravity dual of QCD (see, for example, Sakai-Sugimoto-Witten model \cite{Sakai:2004cn}), it seems that finding the exact gravity dual of QCD is a very challenging problem. Using models that share some properties with QCD, there have been efforts to understand different aspects of the theory, like the physics of the collisions happening at the particle accelerators (see for example \cite{Attems:2016tby}),  the phases of the QCD phase diagram (for example \cite{Faedo:2015urf}),  the physics of QCD with non vanishing theta angle (for example \cite{Bigazzi:2015bna}), the glueball spectrum and other particles of the spectrum (for example \cite{Pons:2004dk}),  the behavior of the jet events (for example \cite{Casalderrey-Solana:2016hrm} ),etc.


\item Applications to condensed matter physics. Condensed matter physics is one of the fields where the gauge/gravity duality has found more applications. Models have been constructed to describe holographic superconductors, fractional quantum Hall systems, fractional topological insulators, Kondo models, Weyl semimetals, holographic superfluids, dirty materials, etc. In condensed matter physics applications it is common to use the `bottom-up' approach, in which the gravity systems are not necessarily embedded in string theory. In these cases, the fact that the gravity model has a dual field theory is less clear. The models that are embedded in string theory are usually called `top-down' models.

\item Applications to astrophysics. For example, in \cite{Hoyos:2016zke} it is studied the cold
quark matter just above the deconfinement transition, using a D3-D7 system, in order to describe some phases of neutron stars.

\item Applications to cosmology.  Cosmological models can be studied from its field theory dual, see for example \cite{McFadden:2009fg}.

\item Applications to hydrodynamics. This concrete case of the duality is known as fluid/gravity duality, see \cite{Hubeny:2011hd} for a review.

\item Applications to non-equilibrium systems. In the last years real time evolution has been considered in gravitational systems, and a lot of numerical efforts in solving them have lead to a new branch of the duality, known as numerical holography. See for example \cite{Chesler:2010bi} \cite{Chesler:2008hg}.

\item Applications to entanglement entropy and quantum information theory. The famous Ryu-Takayanagi prescription offers a simple geometrical method to compute the entanglement entropy in a quantum field theory with known gravity dual.  For a review see \cite{Ryu}.

\item Applications to black hole physics. The black hole physics can be analyzed via its field theory dual, for an example see \cite{Berkowitz:2016muc}.

\end{itemize}

Here we have mentioned some applications of the duality, and a complete list would probably involve much more items. The subject has vastly extended in the last years, and the correspondence has found applications in many areas.

\subsubsection{The KSS bound}

Let us mention one of the first and most famous successes of the AdS/CFT conjecture. The duality can be used to compute transport properties in strongly coupled quantum field theories. In particular, in 2001 Policastro, Son and Starinets \cite{Policastro:2001yc} computed the shear viscosity in a holographic model, and the result, expressed in terms of the entropy density, is very simple:

\begin{equation}
\frac{\eta}{s}=\frac{1}{4 \pi} ~.
\label{kssbound}
\end{equation}

This value turned out to be the same for different holographic models, and in 2004 Kotvun, Son y Starinets (KSS)  \cite{Kovtun:2004de} conjectured that this is a lower bound for a set of theories (there are counterexamples where the bound is not satisfied, like in Gauss-Bonnet theories). 

The ratio of shear viscosity to entropy density  can be used to characterize how close a given fluid is to being perfect. Ordinary fluids, like water, have shear viscosities much higher than this value. Even the liquid helium has much higher viscosity. Nevertheless, there is a fluid whose shear viscosity is close to this ratio. It is the quark-gluon plasma, one of the phases of the phase diagram of QCD. This state of matter was first observed in the Relativistic Heavy Ion Collider~(RHIC) at Brookhaven National Laboratory in 2005.  After 10 years since this first measurement, the recent values obtained in the experiments are still compatible with the KSS bound (for a review of recent results see \cite{Shen:2015msa} ). Also, there are condensed matter systems whose shear viscosity is close to the conjectured bound, like for example, in graphene \cite{graphene}.


\subsubsection{Applications in this thesis}

In this thesis we will use the AdS/CFT correspondence to obtain new results in condensed matter physics. In chapter 4 we construct a gravity dual of a fractional quantum Hall system and in chapter 5 we construct a gravity dual of a system exhibiting quantum phase transitions.

The quantum Hall effect is a phenomenon present in gapped (2+1)-dimensional systems with broken parity symmetry. When electrons are confined in a heterojunction at low temperature and strong magnetic fields, the response to an applied electric field displays a striking behavior: the conductivity in the direction of the electric field vanishes, while the transverse conductivity is quantized and given by $(e^2/h)\nu$, where $\nu$ is the filling fraction, defined as the ratio of the charge density to the magnetic flux. In the integer quantum Hall effect (IQHE) $\nu\in {\mathbb Z}$, whereas $\nu$ is a rational number  in the fractional quantum Hall effect (FQHE). The integer quantum Hall effect is well explained by field theories at weak coupling. However, in a natural way the fractional quantum Hall effect is described by field theories at strong coupling, and then the AdS/CFT is a good arena to explore its properties. In chapter 4 a holographic model for a fractional quantum Hall system is constructed. In particular, supersymmetric Hall states are found.

Quantum phase transitions are phase transitions that happen at zero temperature, and the phase transition parameter is not the temperature but other parameter like the doping of the material, chemical potential, etc. In general, the approach to computing observables in quantum phase transitions is a theoretical challenge, as lowering the temperature near zero turns out to be a difficult process. Nevertheless, the AdS/CFT correspondence offers a very good scenario for this set up, as in its original formulation the temperature is already vanishing. Besides, the quantum phase transitions happen in systems at strong coupling. Then, the gauge/gravity duality is an ideal tool for studying these systems, and in chapter 5 we will construct a gravity dual of a system exhibiting quantum phase transition as the chemical potential is tuned.

Both models are based in the ABJM theory. This theory is particularly interesting to use for condensed matter physics purposes, as it is a Chern-Simmons matter theory in 2+1 dimensions, and many condensed matter systems are based in Chern-Simmons theories. In the original paper of Aharony, Bergman, Jafferis and Madacena \cite{Aharony:2008ug}, it is stated that one of the main motivations to construct the theory is to obtain new condensed matter physics applications. Nevertheless, after almost ten years, the theory has not been yet fully exploited, and in this thesis we give a further step in this direction.


\chapter{Non-abelian T duality and new $AdS_3$ backgrounds}




\section{Introduction}

\def\magenta{\textcolor{magenta}}
\def\Vol{\textrm{Vol}}

The aim of this chapter is to construct new solutions of type IIA/B and eleven-dimensional supergravity via non-abelian T-duality (NATD) and interpret the dual field theory of these new solutions according to the gauge/gravity correspondence.

In particular, we are interested in new $AdS_3 \times M_7$ supersymmetric solutions, but non-supersymmetric non-$AdS$ solutions are also obtained. As explained in section \ref{tduality}, the NATD is performed locally, and the global properties of the manifold $M_7$ are not known. Consequently, we will be able to obtain only some details of the dual field theories, as the full information requires the global details of the manifold\footnote{For example, the solutions $AdS_5 \times S^5$ and $AdS_5 \times S^5/\mathbb{Z}_2$ are locally equivalent, but globally different, and the field theories are different. The first one has gauge group $SU(N)$ and the second one $SU(N) \times SU(N)$}.





The NATD was used for the first time in the context of the gauge/gravity duality in the
paper \cite{Sfetsos:2010uq}. Indeed, Sfetsos and Thompson applied NATD to the
maximally supersymmetric example of $AdS_5\times S^5$, finding a 
metric and RR-fields that preserved ${\cal N}=2$ SUSY. When lifted 
to eleven dimensions, the background fits (not surprisingly) into 
the classification of
\cite{Lin:2004nb}. What is interesting 
is that Sfetsos and Thompson \cite{Sfetsos:2010uq} 
{\it generated} a new solution to the 
Gaiotto-Maldacena differential equation
\cite{Gaiotto:2009gz}, describing
${\cal N}=2$ SUSY CFTs of the Gaiotto-type \cite{Gaiotto:2009we}, \cite{Tachikawa:2013kta}.
This logic was profusely applied to less supersymmetric cases in 
\cite{Itsios:2012zv}-\cite{Itsios:2012dc}; finding new metrics and defining new QFTs by
the calculation of their observables.

Nevertheless, various puzzles associated with NATD remain. As explained above, the global properties of the new manifold are not known. In more concrete terms, the periodicity (if any) of the dual coordinates (the Lagrange multipliers in the sigma-model) is not known.
In the same vein, the precise field theory dual to the 
backgrounds generated by NATD is not clear.

In this chapter we present possible solutions to these puzzles---
at least in particular examples.
Our study begins with type IIB backgrounds dual to a compactification of the
Klebanov-Witten CFT \cite{Klebanov:1998hh} to a 2-d CFT. We will present
these backgrounds for different compactifications and perform a NATD
transformation on them, hence generating new smooth and SUSY 
solutions with $AdS_3$-factors
in type IIA and M-theory. Application of a further T-duality
generates new backgrounds in type IIB with an $AdS_3$-factor which are also smooth and preserve the same amount of SUSY. We will
make a proposal for the dual QFT and interpret the range of the 
dual coordinates in terms of a field theoretic operation.

The picture that emerges is that our geometries describe QFTs 
that become conformal 
at low energies. These CFTs live on the intersection of D2- and D6-branes 
suspended
between NS5-branes. While crossing the NS5-branes, 
charge for D4-branes is induced and new nodes of the quiver appear.

We will present the calculation of different observables of the 
associated QFTs that support the proposal made above. 
These calculations are performed in smooth 
supergravity solutions, hence they are trustable and capture the strong
dynamics of the associated 2-d CFTs.

The work performed in this chapter is a continuation of  the previous works of Sfetsos-Thompson
\cite{Sfetsos:2010uq} and Itsios-Nunez-Sfetsos-Thompson (INST) \cite{Itsios:2012zv},
\cite{Itsios:2013wd}. The connection between the material in this work
and those papers is depicted in Fig. \ref{RoadMap} below.
\begin{figure}[h]
	\centering
	\includegraphics[scale=0.6]{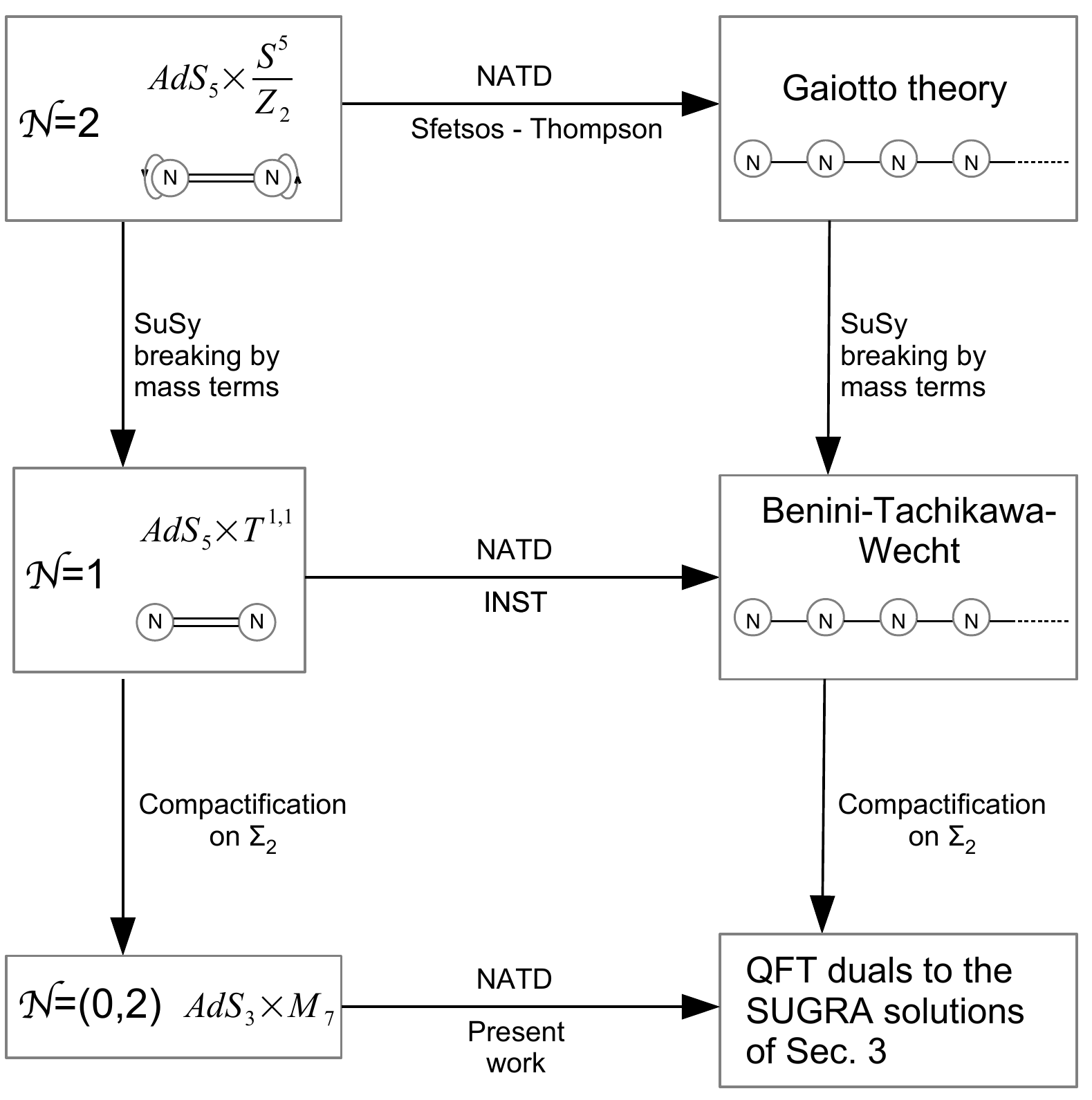}
	\caption{On the left: known solutions on which NATD is performed. On the right: QFT's that correspond to the NATD SUGRA solutions.}
\label{RoadMap}
\end{figure}

The organization of this chapter is as follows. In the first part
of this work, covered in sections \ref{section1.1} to \ref{seccionlift}, we present  type IIB backgrounds that are already 
known and the new ones (in type IIA, IIB and M-theory)
that we construct. Table \ref{tabla1} summarizes these solutions. In the second part of this work, starting in section \ref{comentarios},
 we begin the study of  the
field theoretical aspects encoded by the backgrounds presented in the first part.
Section \ref{chargessectionxx}  
deals with the quantized charges, defining ranks of the gauge groups.
Section \ref{centralchargessectionxx} 
studies the central charge computed holographically, either at the 
fixed points or along the anisotropic flows 
(where a proposal for a c-function is analyzed).
This observable  presents many clues
towards the understanding of the associated  QFTs. 
Section \ref{sectionEEWilson}
presents a detailed study of Wilson loops and entanglement entropy of the QFT
at the fixed points and along the flow. We discuss the results in section \ref{conclusiones}. The computations of the supersymmetry preserved by the solutions is relegated to appendix \ref{appendixsusy2}.


\vskip 20pt

\begin{table}[h]
\begin{tabular}{|c|c|c|c|c|}
\hline
\textrm{\textbf{Solutions}} & \textrm{\textbf{IIB}} &  \textrm{\textbf{NATD}} & \textrm{\textbf{NATD-T}} &\textrm{\textbf{Uplift}}
 \\
\hline
$
\begin{array}{l}
            \textrm{Flow from} \; AdS_5 \times T^{1,1} \; \textrm{to} \; AdS_3 \times H_2 \times T^{1,1}  \; \big ( \mathcal{N} = 1\big)
            \\[5pt]
            \textrm{Fixed points} \; AdS_3 \times \Sigma_2 \times T^{1,1}:  \Sigma_2 = S^2, T^2, H_2 
\end{array}
$
                                          &  \ref{section211xx} & \ref{s2h2natd} & \ref{NATD-T} & \ref{uplifteds2h2natd} 
\\
\hline
\textrm{The Donos-Gauntlett solution} &  \ref{sectionDG} &  \ref{natdDG} & \textrm{-}  &  \ref{upliftnatdDG}\\
\hline
\end{tabular}
\caption{Summary of solutions contained in sections \ref{section1.1} to \ref{seccionlift}. The numbers indicate  the section where they are discussed.}
\label{tabla1}
\end{table}



To be more precise, in the first part of this chapter (sections \ref{section1.1} to \ref{seccionlift}) we will exhaustively present a large set  of backgrounds solving
the type IIB  or type IIA supergravity equations. Most of them are new, but some are already present in the bibliography.
New solutions in eleven-dimensional
supergravity will also 
be discussed. These geometries, for the most part,
preserve some amount of SUSY.

The common denominator of these backgrounds will be the presence of 
an $AdS_3$ sub-manifold in the
ten or eleven-dimensional metric. This will be interpreted as 
the dual description of strongly coupled two dimensional 
conformal dynamics. In most of the cases, there is also a flow, 
connecting from an $AdS_3$ fixed point to an $AdS_5$, 
with boundary $R^{1,1}\times \Sigma_2$.
The manifold $\Sigma_2$ will be a constant curvature Riemann surface. 
As a consequence
we conjecture that the full geometry is describing the strongly 
coupled dynamics of a four dimensional QFT,
that is conformal at high energies and gets 
compactified on $\Sigma_2$ (typically preserving some amount of SUSY).
The QFT flows at low energies to a  2-d CFT that is also strongly coupled.

The solutions {that we are going to present} in this section, can be found by inspection of 
the type II equations. This requires a quite inspired ansatz. 
More practical is to search for solutions of this kind in 
five dimensional gauged supergravity, see the papers \cite{Buchel:2006gb,Gauntlett:2009zw}
for a detailed account of the lagrangians. 
Some other solutions are efficiently obtained by the use 
of generating techniques, for example a combination of
abelian and  non-abelian T-duality, 
that are applied to known (or new) backgrounds, as we show below. 

We now present in detail,  the solutions we will work with in this chapter. 
 
\section{Simple flows from $AdS_5 \times T^{1,1}$ to $AdS_3\times M_7$ in type IIB}
\label{section1.1}

We start this section by proposing a 
simple background in type IIB. In the sense of the Maldacena duality,
this describes the strongly coupled dynamics of an   
${\cal N}=1$ SUSY QFT in four dimensions, that 
is compactified to two dimensions on a manifold $\Sigma_2$.
In order to allow such a compactification we turn on a 1-form field, $A_1$, on the Riemann surface
$\Sigma_2$. Motivated by the works \cite{Buchel:2006gb,Gauntlett:2009zw}, where the authors consider dimensional reductions to five dimensions of type IIB supergravity backgrounds on any Sasaki-Einstein manifold, we propose the following ansatz,
\begin{equation}
 \begin{aligned}
 & \frac{ds^{2}}{L^2}=e^{2 A} \left(- dy_0^2+d y_1^2\right)+e^{2 B} ds^2_{\Sigma_{{2}}}+ dr^2+e^{2U}ds_{KE} ^2+e^{2V} \left( \eta + z A_1 \right)^2, 
 \\[5pt]
 & \frac{F_5}{L^4}=4 e^{-4U-V} \textrm{Vol}_5+2 J \wedge J \wedge \left( \eta + z A_1 \right) - z \textrm{Vol}_{\Sigma_2} \wedge J \wedge \left( \eta + z A_1 \right)
 \\[5pt]
&   ~~~~~ - ze^{-2B-V} \textrm{Vol}_{AdS_3} \wedge J, \;\;\; 
 \\[5pt]
&    \Phi=0, ~~~~~~ C_0=0, ~~~~~~ F_3=0, ~~~~~~ B_2=0.
 \end{aligned}
\label{NN02}
\end{equation}
%
%
 We will focus on the case in which the Sasaki-Einstein space is $T^{1,1}$, hence the K\"ahler-Einstein manifold is
 \begin{equation}
 ds_{KE}^2=\frac{1}{6}(\sigma_{1}^2+\sigma_{2}^2+\omega_1^2+\omega_2^2) \ ,
 \label{zazar}
\end{equation}
where we have defined,
\begin{equation}
\label{gaga}
 \begin{array}{lll}
  \sigma_{1}= d \theta_1,  &\hspace{-3mm} \sigma_{2}= \sin \theta_1 d\phi_1,   &\hspace{-6mm} \sigma_{3}= \cos \theta_1 d\phi_1,
 \\[10pt]
  \omega_1=\cos\psi \sin\theta_2 d\phi_2 -\sin\psi d\theta_2  , &\hspace{-3mm} \omega_2=\sin\psi \sin\theta_2 d\phi_2  + \cos\psi d\theta_2 , &\hspace{-6mm} \omega_3=d\psi +\cos\theta_2 d\phi_2  , 
  \\[10pt]
  \textrm{Vol}_{AdS_3}=e^{2A} dy_0 \wedge dy_1 \wedge dr, &\hspace{-3mm} \textrm{Vol}_5= e^{2B} \textrm{Vol}_{AdS_3} \wedge \textrm{Vol}_{\Sigma_{{2}}} , &\hspace{-6mm} z \in \mathbb{R} ,
  \\[10pt]
  \eta =\frac{1}{3}\left( d\psi + \cos \theta_1 d \phi_1 + \cos \theta_2 d \phi_2 \right), & \hspace{-3mm} J=\frac{-1}{6}\left( \sin \theta_1 d \theta_1 \wedge d \phi_1 + \sin \theta_2 d \theta_2\wedge d \phi_2  \right). & \textrm{}
 \end{array}
\end{equation}
%
%
The forms $\eta$ and $J$ verify the relation $d\eta = 2 \ J$.  The 
range of the angles in the $\sigma_{i}'s$ and the $\omega_i\textrm{'}s$ 
---the left invariant forms of $SU(2)$---  
is given by $0\leq\theta_{1,2}<\pi$, 
$0\leq\phi_{1,2}<2\pi$ and $0\leq\psi<4\pi$. {The $\omega_i$
satisfy $d\omega_i = \frac{1}{2} \ \epsilon_{ijk} \ \omega_j \wedge \omega_k$.}
 We also defined a one form $A_1$, that verifies $dA_1= \textrm{Vol}_{\Sigma_2}$. As usual $ds_{\Sigma_2}^2$ is the metric of the two dimensional surface of curvature $\kappa=(1,-1,0)$, denoting a sphere, hyperbolic plane\footnote{To be precise, we do not consider the hyperbolic plane $H_2$, as it has infinite volume. What we consider is a compact space $H_2/\Gamma$ obtained by quotient by a proper Fuchsian group \cite{kehagiasrusso}, and its volume is given by $4\pi (g-1)$, where $g$ is the genus of $H_2/\Gamma$.} or a torus 
respectively. In local coordinates these read,
\begin{equation}
\label{NN0}
 \begin{array}{llll}
  A_1= - \cos \alpha \ d \beta,  &  \textrm{Vol}_{\Sigma_{{2}}}= \sin \alpha \ d\alpha \wedge d \beta,     &  ds^2_{\Sigma_{{2}}}= d\alpha^2 + \sin^2 \alpha d \beta^2, & (\kappa=1) \ ,
 \\[10pt]
 A_1= \cosh \alpha \ d \beta,  & \textrm{Vol}_{\Sigma_{{2}}}= \sinh \alpha \ d\alpha 
\wedge d \beta,  &  ds^2_{\Sigma_{{2}}}= d\alpha^2 + \sinh^2 \alpha 
d \beta^2, & (\kappa=-1) \ ,
 \\[10pt]
   A_1 =  \alpha \ d \beta,  & \textrm{Vol}_{\Sigma_{{2}}}= d\alpha \wedge d \beta,  &  ds^2_{\Sigma_{{2}}}= d\alpha^2 + d \beta^2, & (\kappa=0) \ .
 \end{array}
\end{equation}
%
%
A natural vielbein for the metric (\ref{NN02}) is,
\begin{equation}
\begin{aligned}
& e^{y_0}=L e^A dy_0 \ ,   ~~~~~  e^{y_1}=L e^A dy_1 \ ,  ~~~~~  e^{\alpha}=L e^B d\alpha \ ,~~~~~~~  e^{\beta}=L e^B A_0 d\beta \ ,  ~~~~~~~  e^{r}=L dr \ , ~~~~~~~~~~~~
\\[5pt]
& e^{\sigma_{1}}= L \frac{e^{U}}{\sqrt{6}} \sigma_{1} \ ,  ~~~~  e^{\sigma_{2}}= L \frac{e^{U}}{\sqrt{6}} \sigma_{2} \ ,    ~~~~  e^{1}= L \frac{e^{U}}{\sqrt{6}} \omega_{1} \ ,  ~~~~ e^{2}=  L \frac{e^{U}}{\sqrt{6}} \omega_{2} \ ,   ~~~~ e^{3}=L e^V (\eta + z A_1) \ ,
\label{vielbein00}
\end{aligned}
\end{equation}
with $A_0=\sinh \alpha$ for $H_2$, $A_0=\sin \alpha$ for $S^2$  and $A_0=1$ for $T^2$.

As anticipated, the background above describes 
the strong dynamics for a compactification of a 
four dimensional QFT to two dimensions. 
{In the case that we are interested in this work} --- in which the 
K\"ahler-Einstein manifold is the one in eq. \eqref{zazar}--- 
the four dimensional QFT at high energy asymptotes to  
the Klebanov-Witten quiver 
\cite{Klebanov:1998hh} on $R^{1,1}\times \Sigma_2$. 
As it will be clear, most of our results will be valid
for the case of a general $Y^{p,q}$ or any other Sasaki-Einstein 
manifold and their associated QFT. Indeed, these solutions 
can be obtained by lifting to type IIB, simpler backgrounds of
the five-dimensional supergravity in \cite{Gauntlett:2009zw}.
In fact, the 5-d supergravity lagrangian was written for any Sasaki-Einstein 
internal space.

Assuming that the functions $A,B,U,V$ depend only on the radial 
coordinate $r$, we can calculate the BPS equations describing 
the SUSY preserving flow from $AdS_5\times T^{1,1}$ 
at large values of the radial coordinate to $AdS_3\times M_7$. 
The end-point of the flow will be dual to a  
2-d CFT obtained after taking the low energy limit of
a twisted KK compactification of the Klebanov-Witten QFT
on $\Sigma_2$. Imposing a set of projections on the 
SUSY spinors of type IIB---see appendix \ref{appendixsusy2} for details--- 
we find,
\begin{eqnarray}
& & A'-e^{-V-4U}\pm \frac{z}{2}e^{-2B-2U-V}=0,\nonumber\\[5pt]
& & B'-e^{-V-4U}\mp \frac{z}{2}e^{-2B-2U-V} \mp \frac{z}{2}e^{-2B+V}=0,\nonumber\\[5pt]
& & U'+e^{-V-4U}-e^{V-2U}=0,\label{O17}
\\[5pt]
& & V'-3e^{-V}+2e^{V-2U}+e^{-V-4U} \mp \frac{z}{2}e^{-2B-2U-V} \pm \frac{z}{2}e^{-2B+V}=0 \ ,\nonumber
\end{eqnarray}
where the upper signs are for $H_2$, the lower signs for $S^2$,  and $z=-\frac{1}{3}$ for both cases.
In the case of the torus the variation of the gravitino will force $z=0$, obtaining $A'=B'$, which does not permit an $AdS_3$ solution.
We now attempt {to} find simple solutions to the eqs.(\ref{O17}).
\subsection{Solution of the form $AdS_3\times H_2$, $AdS_3\times S^2$ and $AdS_3\times T^2$ with 'twisting'}
\label{section211xx}
{At this point we are going to} construct a flow between $AdS_5\times T^{1,1}$ 
and $AdS_3\times \Sigma_2\times M_5$.
To simplify the task, we  propose that the functions 
$U,V$ are constant, then the BPS equations imply that $U=V=0$, 
leaving us---in the case 
of $H_2$ with,
\begin{equation}
A'= 1 + \frac{e^{-2B}}{6},\;\;\; B'=1-\frac{e^{-2B}}{3} \ ,
\label{BPSH2}
\end{equation}
that can be immediately integrated,
 \begin{equation}
A=\frac{3}{2}r-\frac{1}{4} \ln \left( 1+e^{2r} \right) +a_0 \ ,    ~~~~~~ B=\ln\frac{1}{\sqrt{3}} + \frac{1}{2} \ln \left( 1+e^{2r} \right).
 \label{O19}
 \end{equation}
One of the  integration constants associated with these solutions 
corresponds to a choice of the origin 
of the holographic variable and the other constant, 
$e^{2a_0}$, sets the size of the 
three dimensional space ($y_0,y_1,r$). We will choose $a_0=0$ 
in what follows.

In the limit $r\rightarrow \infty$ (capturing the UV-dynamics of the QFT) we recover a Klebanov-Witten metric
     \begin{equation}
   A\sim r -\frac{1}{4}e^{-2r} +\frac{1}{8}e^{-4r} \ ,   ~~~~~ 
B \sim r -\frac{1}{2}\ln 3 +\frac{1}{2}e^{-2r}-\frac{1}{4}e^{-4r} \ ,
     \label{O20x}
     \end{equation}
whilst in the limit $r\rightarrow -\infty$ (that is dual to  the IR 
in the dual QFT) we obtain a supersymmetric solution of 
the form $AdS_3 \times H_2$ ,
     \begin{equation}
   A\sim\frac{3}{2}r -\frac{1}{4}e^{2r}+\frac{1}{8}e^{4r} \ ,  ~~~~~ 
B\sim\ln\frac{1}{\sqrt{3}} +\frac{1}{2}e^{2r} -\frac{1}{4}e^{4r}.
     \label{O21}
     \end{equation}
Let us write explicitly this background using eq. \eqref{zazar},
\begin{eqnarray}
& &      \frac{ds_{10}^2}{L^2}=\frac{e^{3r}}{\sqrt{1+e^{2r}}}\left( -dy_0^2+dy_1^2 \right)+\frac{1+e^{2r}}{3}\left( d\alpha^2+\sinh^2\alpha d\beta^2 \right)+dr^2+ {ds^2_{KE}} +\left( \eta - \frac{1}{3} A_1 \right)^2 \ ,\nonumber\\[5pt]
& & \frac{F_5}{L^4}=\frac{4}{3}e^{3r}\sqrt{1+e^{2r}}\sinh \alpha \ dy_0 \wedge dy_1 \wedge d\alpha \wedge d\beta \wedge dr +2 J \wedge J \wedge \left( \eta  - \frac{1}{3} A_1 \right) \nonumber\\[5pt]
& &  
     \qquad +\frac{1}{3} \sinh \alpha \ d \alpha \wedge d \beta \wedge J \wedge \left( \eta - \frac{1}{3}  A_1 \right) + \frac{e^{3r}}{\left(1+e^{2r}\right)^{\frac{3}{2}}} dy_0 \wedge dy_1 \wedge dr  \wedge J \ ,
     \\[5pt]
& &       A_1= \cosh \alpha \ d \beta,~~~~~~\Phi=0, ~~~~~~ C_0=0 ,~~~~~~ F_3=0, ~~~~~~ B_2=0 \ .
\nonumber
     \label{O20}
\end{eqnarray}
The solution above was originally presented in \cite{Gauntlett:2006af}.

We consider now the case of $S^2$. With the same assumptions 
about the functions $U,V$, the BPS equations (\ref{O17}) read,
\begin{equation}
A'=1-\frac{e^{-2B}}{6},\;\;\;     B'=1+\frac{e^{-2B}}{3}.
     \label{O25}
     \end{equation}
These can also be immediately integrated (with a suitable 
choice of integration constants), 
     \begin{equation}
   A=\frac{3}{2}r-\frac{1}{4} \ln \left( e^{2r} -1 \right) \ ,   ~~~~~~ B=\ln\frac{1}{\sqrt{3}} + \frac{1}{2} \ln \left( e^{2r}-1 \right) \ , ~~~~~~ r>0 \ .
     \label{O26}
     \end{equation}
This solution seems to be problematic close to $r=0$. 
Indeed, if we compute the Ricci scalar we obtain $R=0$, 
nevertheless, $R_{\mu\nu}R^{\mu\nu}\sim \frac{3}{32 L^4 r^4}+.... $ 
close to $r=0$. The solution is singular and we will not
study it further. 

It is interesting to notice that  a family of {\it non-SUSY} 
fixed point solutions exists. Indeed, we can consider 
the situation where $B,U,V$  and $A'(r)=a_1$ are constant. 
For $S^2$ and $H_2$, we find then that the Einstein equations impose $U=V=0$ and 
 \begin{equation}
     8+e^{-4B}z^2-4 \ a_1^2=0 \ ,  ~~~~~~ 4-z^2 e^{-4B}+\kappa \ e^{-2B}=0.
     \label{O12}
     \end{equation}
Where $\kappa=+1$ for $S^2$ and $\kappa=-1$ for $H_2$. The solution is,
  \begin{equation}
     z^2= e^{2B} \left( 4 e^{2B}  +  \kappa \right) \ ,   ~~~~~~  a_1^2=3  +  \kappa \ \frac{e^{-2B}}{4}~.
     \label{O13}
     \end{equation}
%
 For the $S^2$ case the range of parameters is,
     \begin{equation}
     B \in \mathbb{R} \ , ~~~~~~ z \in \mathbb{R}-\{0\} \ , ~~~~~~ a_1 \in \left( -\infty,\sqrt{3}\right) \cup \left( \sqrt{3}, +\infty \right) \ ,
     \label{O14}
     \end{equation}
while for ${H_2}$ we find,
  \begin{equation}
     B \in \Big[-\ln 2, +\infty \Big) \ , ~~~~~~ z \in \mathbb{R} \ , ~~~~~~ a_1 \in \Big( -\sqrt{3},-\sqrt{2} \ \Big] \cup \Big[ \ \sqrt{2}, \sqrt{3} \ \Big) \ .
     \label{O15}
     \end{equation}
Notice that in the case of $H_2$, there is a non-SUSY solution with $z=0$, hence no fibration
between the hyperbolic plane and the Reeb 
vector $\eta$. The SUSY fixed point  
in eq. \eqref{O20} is part of the family 
in eq. \eqref{O15}, with $z=-\frac{1}{3}$.

For the $T^2$ we also find an $AdS_3$ solution,
\begin{equation}
     a_1^2=3 \ , ~~~~~~ z ^2=4e^{4B} \ , ~~~~~~ U=0 \ , ~~~~~~ V=0 \ .
     \label{T7}
\end{equation}


%
     Notice that $z \in \mathbb{R}-\{0\} $.
This completes our presentation of what we will refer as 'twisted' solutions. 
By twisted we mean solutions where a gauge field is switched on the Riemann surface 
generating a fibration  between $\Sigma_2$ and the R-symmetry 
direction. In the following, we will present a background that 
also contains an $AdS_3$ factor, it 
flows in the UV to an $AdS_5$, 
but the field content and the mechanism of SUSY preservation are different
from the ones above.

\subsection{The Donos-Gauntlett-(Kim) background}
\label{sectionDG}

In this section we revisit a beautiful solution  written in 
\cite{Donos:2014eua}---this type of solution
was first studied in \cite{Donos:2008ug}. Due to the more
detailed study of \cite{Donos:2014eua}, we will refer to it as
the Donos-Gauntlett solution in the rest of this chapter.  

The background in \cite{Donos:2014eua} 
describes a flow  in the radial coordinate, from 
$AdS_5\times T^{1,1}$ to $AdS_3\times M_7$. 
The solution is very original. While the boundary of $AdS_5$ 
is of the form $R^{1,1}\times T^2$, the compactification on the
Riemann surface (a torus) does not use a 'twist'
of the 4d-QFT. This is reflected by the absence of a 
fibration of the Riemann surface on the R-symmetry direction $\eta$. 
Still, the background  preserves SUSY 
\footnote{The solutions in Eqs.(63)-(80) of the paper \cite{Nunez:2001pt}, 
can be thought
as an 'ugly' ancestor of the Donos-Gauntlett background.}. 
The {solution} contains 
an active NS-three form $H_3$ that together with 
the RR five form $F_5$ implies the presence of a 
RR-three form $F_3$. The authors of \cite{Donos:2014eua} found this configuration
by using a very inspired ansatz. 
We review this solution below, adding new information to complement that in
\cite{Donos:2014eua}.

The metric ansatz is given by,
\begin{align}
& &  \frac{ds^2}{L^2}   = e^{2A}\left(-dy_0^2+dy_1^2\right)+e^{2B}\left(d\alpha^2+d\beta^2\right)+dr^2+ e^{2U}ds_{KE}^{2}+e^{2V} \eta^2 \ , 
\label{metric-bef}
\end{align}
where $A,B,U,V$ are functions of the radial coordinate $r$ only. 
The line element
 $ds_{KE}^{2}$ is defined in eq. \eqref{zazar} 
and $\sigma_{i}$, $\omega_{i}$, $\eta$ are given in eq. \eqref{gaga}.
%
The natural vielbein is,
\begin{equation}
\begin{aligned}
& e^{y_0}=L e^{A} dy_0, \quad e^{y_1}=L e^{A} dy_1, \quad e^{\alpha}= Le^{B} d\alpha, \quad e^{\beta}= L e^{B} d\beta, \quad e^{r} = L dr,		
\\[5pt]
& e^{\sigma_{1}}= L \frac{e^{U}}{\sqrt{6}}\sigma_{1},  
\quad e^{\sigma_{2}}=L \frac{e^{U}}{\sqrt{6}}\sigma_{2},   \quad e^{1}= L \frac{e^{U}}{\sqrt{6}} \omega_1, \quad e^{2}= L \frac{e^{U}}{\sqrt{6}} \omega_2, \quad e^{3}=L e^V \eta\ .
\end{aligned}
\label{veil-bef}
\end{equation}
%
Note that compared to \cite{Donos:2014eua} 
we have relabelled $\phi_i \rightarrow -\phi_i$.
To complete the definition of the background, we also need 
\begin{align}
 {{\textrm{Vol}}_1= - \sigma_{1} \wedge \sigma_{2}, 
\qquad  {\textrm{Vol}}_2=\omega_1\wedge\omega_2}, 
\label{bdefn-bef}
\end{align}
and the fluxes,
\begin{equation}
\begin{aligned}
& \frac{1}{L^{4}}F_{5} =  4e^{2A+2B-V-4U}dy_0 \wedge dy_1 \wedge d\alpha \wedge d\beta \wedge d r+
\frac{1}{9}\eta \wedge {\textrm{Vol}}_1 \wedge {\textrm{Vol}}_2		
 \\[5pt]
& \qquad \quad  {+\frac{\lambda^{2}}{12} \Big[ d\alpha \wedge d\beta \wedge \eta \wedge ( \textrm{Vol}_1+{\textrm{Vol}}_2) + e^{2A-2B-V} dy_0 \wedge dy_1 \wedge d r \wedge ({\textrm{Vol}}_1+{\textrm{Vol}}_2)\Big]}, 
\\[5pt]
&  {\frac{1}{L^{2}} F_3=\frac{\lambda}{6} d\beta \wedge ({\textrm{Vol}}_1-{\textrm{Vol}}_2), \qquad \frac{1}{L^{2}}  B_2 = -\frac{\lambda }{6}\alpha\ ({\textrm{Vol}}_1-{\textrm{Vol}}_2)},
\end{aligned}
\label{fluxes-bef}
\end{equation}
where $\lambda$ is a constant that encodes the deformation of the space
(and the corresponding operator in the 4d CFT). 
The BPS equations for the above system are given by,
\begin{equation}
\label{BPS}
\begin{aligned}
& A' =	\frac{1}{4} \lambda ^2 e^{-2 B-2 U-V}+e^{-4 U-V}, 
\qquad B' =		e^{-4 U-V}-\frac{1}{4} \lambda ^2 e^{-2 B-2 U-V},	 
\\[5pt]
& U' =	e^{V-2 U}-e^{-4 U-V}, \qquad V' = 	-\frac{1}{4} \lambda ^2 e^{-2 B-2 U-V}-e^{-4 U-V}-2 e^{V-2 U}+3 e^{-V}.
\end{aligned}
\end{equation}
It is possible to recover the $AdS_5 \times T^{1,1}$ solution by setting $\lambda=0$ and
\begin{equation}
A=B=r, \qquad U=V=0.
\label{AdST11sol}
\end{equation}

Further we can recover the $AdS_3$ solution 
by setting $\lambda=2$ (this is just a conventional value adopted in
\cite{Donos:2014eua}) {and}
\begin{equation}
A=\frac{3^{3/4}}{\sqrt{2}}r, \qquad B=\frac{1}{4}\ln\left(\frac{4}{3}\right), \qquad U=\frac{1}{4}\ln\left(\frac{4}{3}\right), \qquad V=-\frac{1}{4}\ln\left(\frac{4}{3}\right) .
\label{AdS3sol}
\end{equation}
{A flow
(triggered by the deformation parameterized by  $\lambda$) between the asymptotic $AdS_5$ and
the $AdS_3$ fixed point}
can be {found numerically}.  \\

%


Up to this point, we have set the stage for our study, 
but most of the solutions discussed above have already been stated in the literature. {Here} we just added some new backgrounds and technical elaborations on the known ones.
Below, we will present genuinely new type IIA/B solutions. 
The technical tool {that we have used is} non-abelian T-duality. 
This technique, when applied to an $SU(2)$
isometry of the previously discussed solutions will generate new type IIA configurations. {In our examples these new solutions are nonsingular}. Their lift to eleven dimensions will 
produce new, smooth, $AdS_3$ configurations in M-theory. {Moreover, performing an additional abelian T-duality transformation
we generate new type IIB  backgrounds} with all fluxes turned on and an $AdS_3$ fixed point at the IR.
We move on to describe these.

\section{New backgrounds in type IIA: use of  non-abelian T-duality}

In this section we apply the technique of non-abelian T-duality on the type IIB backgrounds that we presented above. As a result we obtain new solutions of the type IIA supergravity.

\subsection{The non-abelian T-dual of the twisted solutions}
\label{s2h2natd}

We will start by applying non-abelian T-duality (NATD) to the 
background obtained via a twisted compactification
in section \ref{section1.1}. The configuration we will focus on
 is a particular case of that in eq. \eqref{NN02}. 
Specifically, in what follows we consider $U = V = 0$. These values for the functions $U$ and $V$ are compatible with the BPS system \eqref{O17}. In this case the background \eqref{NN02} simplifies to, 
\begin{equation}
 \label{NN01}
 \begin{aligned}
   & \frac{ds^{2}}{L^2}=e^{2 A} \left(- dy_0^2+d y_1^2\right)+e^{2 B} ds^2_{\Sigma_{{2}}}+ dr^2+\frac{1}{6} \left(\sigma_{1}^2+\sigma_{2}^2\right)+\frac{1}{6} \left(\omega_1^2+\omega_2^2\right) + \left( \eta + z A_1 \right)^2 \ ,
\\[5pt]
   & \frac{F_5}{L^4}=4 \textrm{Vol}_5+2 J \wedge J \wedge \left( \eta + z A_1 \right) - z \textrm{Vol}_{\Sigma_2} \wedge J \wedge \left( \eta + z A_1 \right) - ze^{-2B} \textrm{Vol}_{AdS_3} \wedge J \ .
 \end{aligned}
\end{equation}
%
%
As before, all other RR and NS fields are taken to vanish. Also, the 1-form $A_1$, the line element of the Riemann surface $\Sigma_2$ and the corresponding volume form, for each of the three cases that we consider here, are given in eq. \eqref{NN0}.

We will now present the {details} for the background after NATD
has been applied on the $SU(2)$ {isometry} described by the coordinates $(\theta_2,\phi_2,\psi)$. As is well-known
a gauge fixing has to be implemented {during the NATD procedure}. This leads to a  choice of three 'new coordinates' among the 
Lagrange multipliers {$(x_1,x_2,x_3)$} used in the NATD procedure  
and the 'old coordinates' ($\theta_2,\phi_2,\psi$) 
\footnote{The process of NATD and the needed gauge fixing was 
described in detail in \cite{Itsios:2013wd}, 
\cite{Macpherson:2014eza}.}. 
In all of our examples we consider a gauge fixing of the form,
\begin{equation}
 \theta_2 = \phi_2 = \psi = 0.
\label{gaugefixing012}
\end{equation}
 As a result, the Lagrange multipliers $x_1, x_2$ and $x_3$ play the r\^ole of the dual coordinates in the new background.
To display the natural symmetries of the background, 
we will quote the results using  spherical coordinates (the expressions in cylindrical and cartesian coordinates are written in appendix C of \cite{Bea:2015fja}). 
We define,
\begin{equation}
 \begin{aligned}
  & x_1=\rho \cos \xi \sin \chi \ , ~~~~~~ x_2=\rho \sin \xi \sin \chi \ , ~~~~~~ x_3=\rho \cos \chi \ ,
  \\[5pt]
  & \Delta =L^4 +54 \alpha'^2 \rho^2 \sin^2 \chi +36 \alpha'^2 \rho^2 \cos^2 \chi \ ,  ~~~~~ \tilde{\sigma}_{3}= \cos \theta_1 d \phi_1 + 3 z A_1.
    \label{NN10}
 \end{aligned}
\end{equation}
%
%
The NSNS sector of the transformed IIA background reads,
\begin{equation}
\label{NN11}
 \begin{aligned}
  & e^{-2 \widehat{\Phi}}=\frac{L^2}{324 \alpha'^3} \Delta, 
  \\
  & \frac{d\hat{s}^{2}}{L^2}=e^{2 A} \left(- dy_0^2+d y_1^2\right)+e^{2 B} ds^2_{\Sigma_2}+ dr^2+\frac{1}{6} \left(\sigma_{1}^2+\sigma_{2}^2\right) + \frac{\alpha'^2}{\Delta} \Bigg[ 6 \Big( \sin^2 \chi \big( d \rho^2 + \rho^2 (d \xi + \tilde{\sigma}_{3})^2 \big) 
  \\
   & \qquad + \rho \sin (2 \chi) d\rho d\chi+\rho^2 \cos^2 \chi d \chi^2 \Big) + 9 \Big( \cos \chi d \rho - \rho \sin \chi d \chi \Big)^2 +\frac{324 \alpha'^2}{L^4} \rho^2 d\rho^2 \Bigg] ,
  \\
  & \widehat{B}_2=\frac{\alpha'^3}{\Delta} \Bigg[ 36 \rho \cos \chi \Big(\rho \tilde{\sigma}_{3}\wedge d\rho +\rho \sin \chi d\xi \wedge \big(\sin \chi d\rho+\rho \cos \chi d \chi \big)\Big)
  \\
  & \qquad + \Big(  \frac{L^4}{\alpha'^2} \tilde{\sigma}_{3} - 54 \rho^2 \sin^2 \chi d \xi  \Big)\wedge \big( \cos \chi d \rho - \rho \sin \chi d \chi \big) \Bigg] \ ,
 \end{aligned}
\end{equation}
%
 %
while the RR sector is, 
\footnote{
 According to the democratic formalism, the higher rank RR forms are related to those of lower rank through the relation $ F_p = (-1)^{[\frac{p}{2}]} * F_{10-p}$.
}
 \begin{align}
 \label{NN12}
  \widehat{F}_0 &=0, \qquad \widehat{F}_2=\frac{L^4}{54\alpha'^{\frac{3}{2}}} \left( 2 {\sigma_{1}\wedge \sigma_{2}} + 3 z \textrm{Vol}_{\Sigma_2} \right) \ ,
  \nonumber\\[5pt]
  \widehat{F}_4 &=\frac{L^4}{18 \sqrt{\alpha'}} \Bigg[   3 z e^{-2B}  \big(d\rho \cos \chi - \rho \sin \chi d\chi \big)  \wedge \textrm{Vol}_{AdS_3} + z \ \rho \ \cos \chi \textrm{Vol}_{\Sigma_2} \wedge {\sigma_{1} \wedge \sigma_{2}}
  \nonumber\\[5pt]
   & -\frac{18 \alpha'^2 }{\Delta} \rho^2 \sin  \chi \Big( z \textrm{Vol}_{\Sigma_2}  + \frac{2}{3} {\sigma_{1} \wedge \sigma_{2}} \Big) \wedge \Big( 2  \cos \chi \big(\sin \chi d\rho+\rho \cos \chi d \chi \big)
  \\
  &  +3   \sin  \chi \big(\rho \sin \chi d \chi -\cos \chi d \rho \big) \Big)\wedge \big(d \xi +  \tilde{\sigma}_{3} \big) \Bigg]. 
 \nonumber
\nonumber
 \end{align}
%
%
%
The SUSY preserved by this background is discussed in 
appendix \ref{appendixsusy2}. We have checked that the equations of motion are solved by this background.

\subsection{The non-abelian T-dual of the Donos-Gauntlett solution}
\label{natdDG}

{In this section } we will briefly present the result of applying NATD to the Donos-Gauntlett solution \cite{Donos:2014eua}
that we described in section \ref{sectionDG}.

Like above, we will perform the NATD choosing a gauge such that the new coordinates are $(x_1,x_2,x_3)$.
We will quote the result in spherical coordinates (the expressions in cylindrical and cartesian coordinates are written in appendix C of \cite{Bea:2015fja}).
The expressions below are naturally more involved than those in section \ref{s2h2natd}.
{This is due to the fact that now there is a non-trivial NS 2-form that enters in the procedure (explicitly, in the string sigma model) 
which makes things more complicated.}
 As above, we start with some definitions,
\begin{equation}
  \mathcal{B}_{\pm}=\rho \cos \chi\pm\frac{L^2 \lambda}{6 \alpha'} \alpha \ , \qquad \mathcal{B}=\mathcal{B}_{+} \ ,  \qquad
  \Delta =L^4 e^{4 U+2 V}+54 \alpha'^2 e^{2 U} \rho^2 \sin^2 \chi +36 \alpha'^2 \mathcal{B}^2 e^{2V} \ .
    \label{s1}
\end{equation}
The NSNS sector is given by,
%
 \begin{align}
  e^{-2 \widehat{\Phi}} & =\frac{L^2}{324 \ \alpha'^3} \Delta,   \label{s5} 
\nonumber\\[5pt]
  \frac{d\hat{s}^{2}}{L^2} & =e^{2 A} \left(- dy_0^2+d y_1^2\right)+e^{2 B} \left( d \alpha^2+ d \beta^2\right)+ dr^2+\frac{e^{2 U}}{6} \left(\sigma_{1}^2+\sigma_{2}^2\right)+
\nonumber\\[5pt]
  &  +\frac{\alpha'^2}{\Delta} \Bigg[ 6 e^{2 U+2 V} \Bigg( \sin^2 \chi \Big( d \rho^2 + \rho^2 \big(d \xi + \sigma_{3} \big)^2 \Big)+ \rho \sin (2 \chi) d\rho d\chi+\rho^2 \cos^2 \chi d \chi^2 \Bigg) 
\nonumber\\[5pt]
  &  + 9 e^{4 U} \big( \cos \chi d \rho - \rho \sin \chi d \chi \big)^2 + \frac{324 \alpha'^2}{L^4} \Big( \big(\mathcal{B} \cos \chi+ \rho \sin^2 \chi \big) d\rho+ \rho \big(\rho \cos \chi - \mathcal{B} \big)\sin \chi d\chi \Big)^2 \Bigg],
\\[5pt]
  \widehat{B}_2 & =L^2\frac{\lambda}{6} \ \alpha \ {\sigma_{1}\wedge \sigma_{2}}+\frac{\alpha'^3}{\Delta} \Bigg[ 36 \ \mathcal{B} \  e^{2V} \Bigg(\sigma_{3}\wedge \Big( \big(\mathcal{B} \cos \chi+ \rho \sin^2 \chi \big) d\rho+ \rho \big(\rho \cos \chi - \mathcal{B} \big) \sin \chi d\chi \Big) 
\nonumber\\[5pt]
  &\hspace{-6mm}  +\rho \sin \chi d\xi \wedge \big(\sin \chi d\rho+\rho \cos \chi d \chi \big) \Bigg)+ e^{2U} \Big( e^{2V+2U} \frac{L^4}{\alpha'^2} \sigma_{3} - 54 \rho^2 \sin^2 \chi d \xi  \Big)\wedge \big( \cos \chi d \rho - \rho \sin \chi d \chi \big) \Bigg].
\nonumber
 \end{align}
%
%
The RR sector reads,
\begin{equation}
\label{s11}
 \begin{aligned}
  & \widehat{F}_0=0, \qquad
\widehat{F}_2=\frac{L^2}{6\alpha'^{\frac{3}{2}}} \Big( \lambda \alpha' d\mathcal{B}_{-}\wedge d\beta + \frac{2}{9} L^2 {\sigma_{1}\wedge \sigma_{2}} \Big), 
\\[5pt]
  & \widehat{F}_4=\frac{L^4 \lambda}{36 \sqrt{\alpha'}} \Bigg[ e^{V}\Big( \frac{2 L^2}{\alpha'}d\alpha -3 e^{-2B-2V} \lambda \big( \cos \chi d \rho - \rho \sin \chi d \chi \big) \Big) \wedge \textrm{Vol}_{AdS_3}
\\[5pt]
  &  \quad \; -\frac{6 \alpha'}{L^2} \Big( \rho \sin \chi \big( \sin \chi d\rho + \rho \cos \chi d \chi \big) \mathcal{B} d \mathcal{B}\Big) \wedge d \beta \wedge {\sigma_{1} \wedge \sigma_{2}}
\\[5pt]
  & \quad \; +\frac{36 \alpha'^2 }{\Delta} \rho \sin  \chi \Big( \frac{3 \alpha'}{L^2} d \beta \wedge d\mathcal{B}_{-}  -\frac{2}{3 \lambda} {\sigma_{1} \wedge \sigma_{2}} \Big) \wedge \Big( 2 e^{2V}\mathcal{B}  \big(\sin \chi d\rho+\rho \cos \chi d \chi \big)+ 
\\[5pt]
  & \quad \; +3 e^{2U}  \rho \sin  \chi \big(\rho \sin \chi d \chi -\cos \chi d \rho \big) \Big)\wedge \big(d \xi +  \sigma_{3} \big) \Bigg] \ .
 \end{aligned}
\end{equation}
%
Here again, aspects of the SUSY preserved by this background are
relegated to appendix \ref{appendixsusy2}. We have checked that the equations of motion are solved by this background.

\section{T-dualizing back from type IIA to IIB }
\label{NATD-T}
In this section, we will construct {\it new} type IIB 
Supergravity backgrounds with an $AdS_3$ factor at the IR. The idea is to obtain
these new solutions by performing an (abelian) T-duality
on the configurations  described 
by eqs. \eqref{NN11}-\eqref{NN12} which, in turn,
were obtained by performing NATD on the backgrounds 
of eq. \eqref{NN01}.
The full chain of dualities is NATD -T. 
The new solutions present an $AdS_3$ fixed point and
all RR and NS fields are switched on. 

It should be interesting to study if the $AdS_3$ fixed point of this geometry falls  within known classifications \cite{Gauntlett:2006ux}. If not,
to use them as inspiring ansatz to extend these taxonomic efforts.

In order to perform the T-duality, we will choose a Killing vector that has no fixed points, in such a way that the dual geometry has no singularities. An adapted system of coordinates for that Killing vector is obtained through the change of variables,
\begin{equation}
\alpha = \textrm{arccosh} \left( \cosh a \cosh b \right) \ , \qquad \beta = \arctan \left(  \frac{\sinh b}{\tanh a}  \right) \ ,
\label{NN2TTwewewe1}
\end{equation}
obtaining \footnote{
In fact using the coordinate transformation eq. \eqref{NN2TTwewewe1} we obtain $A_1 = \sinh a \ db + \textrm{total derivative}. $
},
\begin{equation}
 A_1= \sinh a \ db \ ,  \qquad  \textrm{Vol}_{\Sigma_2}= \cosh a\  da \wedge db \ , \qquad  ds^2_{\Sigma_2}=da^2 + \cosh^2 a \ db^2 \ .
\label{NN0000021}
\end{equation}

The Killing vector that we choose is the one given by translations along the $b$ direction. Its modulus is proportional to the quantity,
\begin{equation}
    \Delta_T = 54 \alpha'^2 z^2 \rho^2 \sin^2 \chi  \sinh^2 a + e^{2B} \cosh^2 a \  \Delta \ ,
\label{moduluskillingvector} 
\end{equation}
where $\Delta$ is defined in eq.(\ref{NN10}). Since $\Delta_T$  is never vanishing the isometry has no fixed points. 

To describe these new configurations, we define,
\begin{equation}
 A_3=3 z \sinh a \ \big( \cos \chi d\rho - 
\rho \sin \chi d \chi \big) - db \ .
  \label{NN1TT}
\end{equation}
 Then, we will have a NSNS sector,
%
\begin{align}
\label{NN2TT}
    e^{-2 \widetilde{\Phi}} & =\frac{L^4}{324 \alpha'^4} \Delta_T \ , 
\nonumber\\[5pt]
    \frac{d\tilde{s}^{2}}{L^2} & =e^{2 A} \left(- dy_0^2+d y_1^2\right)+e^{2 B} da^2+ dr^2+\frac{1}{6} \left(\sigma_{1}^2+\sigma_{2}^2\right)
\nonumber\\[5pt]
    & +\frac{\alpha'^2}{\Delta_T} \Bigg[  \frac{\Delta}{L^4} db^2 + 6 e^{2B} \cosh^2 a  \Big(  \rho^2 \sin^2 \chi      \big(  \sigma_{3}+ d\xi \big)^2 + \big(d \rho \  \sin \chi +\rho \ \cos \chi \  d \chi \big)^2 \Big)
\nonumber\\[5pt]
    &  +9 \big( z^2 \sinh^2 a+e^{2B} \cosh^2 a  \big) \Big( \big(d\rho \ \cos \chi -\rho \  \sin \chi \ d\chi \big)^2 + \frac{36 \alpha'^2}{L^4} \rho^2 d\rho^2 \Big) 
\nonumber\\[5pt]
    &   - 6 z \sinh a \ d b  \Bigg(  \Big(\frac{36 \alpha'^2}{L^4} \rho^2+1\Big)  \cos \chi \  d\rho - \rho \ \sin \chi \ d \chi \Bigg) \Bigg] \ ,
\\[5pt]
    \widetilde{B}_2 & =\frac{\alpha'^3}{\Delta_T}     \Bigg[  e^{2B} \cosh^2 a   \Bigg( 36 \rho \cos \chi \Big[ \rho \sigma_{3}\wedge d\rho + \rho  \sin \chi  d\xi \wedge \Big(\sin \chi   d\rho +\rho  \cos \chi   d\chi \Big)\Big]
\nonumber\\[5pt]
    &\hspace{-6mm} + \Big( \frac{L^4}{\alpha'^2} \sigma_{3} - 54 \rho ^2 d\xi  \sin ^2 \chi  \Big)\wedge \Big(d \rho  \cos \chi - \rho  \sin \chi d \chi  \Big) \Bigg)  
 - 18 \ z \ \rho^2 \ \sinh a \sin ^2 \chi \Big(      d \xi  \wedge A_3 + d b \wedge \sigma_{3}   \Big) \Bigg] \ ,
\nonumber
\end{align}
%
and a RR sector that reads,
%
\begin{align}
\label{NN3TT}
& \widetilde{F}_1 = \frac{z L^4}{18 \alpha'^2} \cosh a \ d a \ ,
\nonumber\\[5pt]
& \widetilde{F}_3 = \frac{L^4}{54\alpha'} \big(  2 A_3 + 3 z \rho \cos \chi \cosh a \ d a  \big)    \wedge         \sigma_{1}\wedge \sigma_{2}
\nonumber\\[5pt]
& \qquad +\frac{ z L^4 \alpha'  \cosh a }{\Delta_T} \rho^2 \sin \chi \big(  \sigma_{3} +  d \xi \big) \wedge \Big[    e^{2B} \cosh^2 a  \Big( \big ( 2 + \sin^2 \chi \big) d\chi - \sin \chi \cos \chi d \rho \Big) 
\nonumber\\[5pt]
& \qquad - z \sinh a \sin \chi A_3 \Big] \wedge d a \ ,
\nonumber\\[5pt]
& \widetilde{F}_5 = \frac{L^4}{6} e^{-2B}  \Big( 24 e^{4B} \rho \cosh a  da \wedge d \rho + z \big( d\rho  \cos \chi -\rho d\chi \sin \chi \big) \wedge d b \Big) \wedge \textrm{Vol}_{AdS_3} 
\\[5pt]
&  + \frac{L^4 \alpha'^2 \cosh a}{18 \Delta_T} \Bigg[   e^{2B} \cosh a \rho  \sin \chi  d \xi \wedge  \Bigg(  \big( d\rho  \sin \chi +\rho d \chi  \cos \chi \big)  \wedge \Big[ 24 \rho  \cos \chi A_3
\nonumber\\[5pt]
&  - z \Big( \frac{L^4}{\alpha'^2} + 54 \rho ^2 \sin ^2 \chi  \Big) \cosh a da \Big]
 + 18 \ \rho \  \sin  \chi  \big( 3 z \rho \cos \chi \cosh a d a -2 d\beta \big) \wedge \big(d \rho  \cos \chi -\rho  d \chi  \sin \chi \big) \Bigg)
\nonumber\\[5pt]
&  - 18 z^2 \  \rho ^3 \sinh a \ \sin ^2 \chi \  d \xi \wedge \Big( 3 z \sinh a \big(\sin^2 \chi d \rho + \rho \sin \chi \cos \chi d \chi \big) + \cos \chi A_3 \Big)  \wedge d a  \Bigg] \wedge  \sigma_{1}\wedge \sigma_{2} \ .    
\nonumber
\end{align}
%
We have checked that the equations of motion are solved by this background, either assuming the BPS equations (\ref{BPSH2}) or the non-SUSY solution (\ref{O15}). We have also checked that the Kosmann derivative vanishes without the need to impose further projections. These facts point to the conclusion that this new and smooth solution is also SUSY preserving.
 
 We will now present new backgrounds of eleven-dimensional supergravity.


\section{ Lift to M - theory }\label{seccionlift}

Here, we lift the solutions of  
sections \ref{s2h2natd} and \ref{natdDG} to eleven dimensions,
presenting new and smooth backgrounds of M-theory that describe 
the strong dynamics of a SUSY 2d CFT.

It is well known that given a solution of the 
type IIA supergravity the metric of the uplifted to eleven dimensions 
solution, has the following form,
\begin{equation}
ds^2_{11} = e^{-\frac{2}{3} \widehat{\Phi}} \ ds^2_{IIA} \ + \ e^{\frac{4}{3} \widehat{\Phi}} \ \big(  dx_{11} + \widehat{C}_1  \big)^2 \ ,
\end{equation}
where $\widehat{\Phi}$ is the dilaton of the 10-dimensional 
solution of the Type IIA SUGRA and $\widehat{C}_1$ is 
the 1-form potential that corresponds to the 
RR 2-form of the Type IIA background. Also, by
$x_{11}$ we denote the $11^{\textrm{th}}$ coordinate which corresponds 
to a $U(1)$ isometry as neither the metric tensor or flux  
explicitly depend on it.

The 11-dimensional geometry is supported by a 3-form potential $C^M_3$ 
which gives rise to a 4-form $F^M_4 = dC^M_3$. 
This 3-form potential can be written in terms 
of the 10-dimensional forms and the differential 
of the $11^{\textrm{th}}$ coordinate as,
\begin{equation}
 C^M_3 = \widehat{C}_3 + \widehat{B}_2 \wedge dx_{11} \ .
\end{equation}
The 3-form $\widehat{C}_3$ corresponds to the closed part of 
the 10-dimensional RR form 
$\widehat{F}_4 = d\widehat{C}_3 -\widehat{H}_3 \wedge \widehat{C}_1$.
Here, $\widehat{B}_2$ is the NS 2-form of the 10-dimensional 
type-IIA theory and $\widehat{H}_3 = d\widehat{B}_2$.
Hence we see that in order to describe the 11-dimensional 
solution we need the following ingredients,
\begin{equation}
 ds^2_{IIA} \ , \quad \widehat{\Phi} \ , \quad \widehat{B}_2 \ , \quad \widehat{C}_1 \ , \quad\widehat{C}_3 \,\,\,\, \textrm{or} \,\,\,\, \widehat{F}_4 \ .
\end{equation}
Let us now present these quantities for the cases of interest.
\subsection{Uplift of the NATD of the twisted solutions}
\label{uplifteds2h2natd}
As we mentioned above, in order to specify 
the M-theory background we need to read the field 
content of the 10-dimensional solution. For the case at hand we 
wrote the metric $ds^2_{IIA}$, 
the dilaton $\widehat{\Phi}$
and the NS 2-form $\widehat{B}_2$ in eq. \eqref{NN11}. 
Moreover, from the expression of the RR 2-form 
in eq. \eqref{NN12} we can immediately extract the 
1-form potential $\widehat{C}_1$, 
\begin{equation}
 \widehat{C}_1 = \frac{L^4}{54 \ a'^{\frac{3}{2}}} \ 
\big(  3 \ z \ A_1 - 2 \ \s_3  \big) \ .
\end{equation}
The 3-form potential--$\widehat{C}_3$--can be obtained 
from the RR 4-form of eq. \eqref{NN12}. After some algebra one finds,
\begin{equation}
\begin{aligned}
  \widehat{C}_3 & = \frac{L^4 \ \ e^{-2B}}{6 \ a'^{\frac{1}{2}}} \ z \ \r \ \cos\chi \ \textrm{Vol}_{AdS_3} + \frac{L^4 \ z \ \big(L^4 + 36 a'^2 \r^2\big) \ \cos\chi}{6 \ \a'^{\frac{1}{2}} \ \Delta} A_1 \wedge \sigma_{3} \wedge d\r
  \\[5pt]
                         & + \frac{L^4 \ \a'^{\frac{3}{2}} \ \r^2 \sin\chi}{\Delta} \Big( z \ A_1 - \frac{2}{3} \sigma_{3}  \Big) \wedge d\xi \wedge \Big( \r \ \big( 2 + \sin^2\chi  \big) \ d\chi -\sin\chi \ \cos\chi \ d\r  \Big)
   \\[5pt]
                         & + \frac{L^4 \ z \ \r}{18 \ \a'^{\frac{1}{2}} \ \Delta} \Big( \Delta \ \cos\chi \ \big(  \sigma_{3} \wedge \textrm{Vol}_{\Sigma_2} - 2 A_1 \wedge d \sigma_{3} \big) - 3 L^4 \sin\chi \ A_1 \wedge \sigma_{3} \wedge d\chi   \Big) \ .
 \end{aligned}
\end{equation}
We close this section by observing that, if the coordinate $\rho$
takes values in a finite interval, then the radius of the 
M-theory circle, $e^{\frac{4}{3} \widehat{\Phi}}$
never blows up, because the function $\Delta$ that appears in the expression 
of the dilaton is positive definite. We have checked that the equations of motion of the 11-dimensional Supergravity are solved by this background.
\subsection{Uplift of the NATD of the Donos-Gauntlett solution}
\label{upliftnatdDG}
Here, the NSNS fields of the 10-dimensional theory have 
been written in detail in eq. \eqref{s5}. 
In order to complete the description of the M-theory background 
we need also to consider the potentials 
$\widehat{C}_1$ and $\widehat{C}_3$ that are encoded in the RR fields of 
eq. \eqref{s11}. Hence from the RR 2-form potential 
we can easily read $\widehat{C}_1$,
\begin{equation}
  \widehat{C}_1 = \frac{L^2 \l \, \mathcal{B}_{-}}{6 \a'^{\frac{1}{2}}} \, d\beta - \frac{L^4}{27 \a'^{\frac{3}{2}}} \, \s_3 \ ,
\end{equation}
where the function $\mathcal{B}_{-}$ has been defined in 
eq. \eqref{s1}. Also, from the RR 4-form we can obtain 
the potential $\widehat{C}_3$ which in this case is,
\begin{equation}
 \begin{aligned}
  \widehat{C}_3 & = \frac{e^{-2B-V} L^4 \l}{36 \a'^{\frac{3}{2}}} \Big( 2 e^{2B + 2V} L^2 \alpha - 3 \l \a' \r \cos\chi \Big) \textrm{Vol}_{AdS_3} 
\\[5pt]
      & - \frac{\a'^{\frac{1}{2}} L^2 \r \sin\chi}{18 \Delta} \Big( 18 \a' \l \mathcal{B}_{-} \big( \s_3 + d\xi \big) \wedge d\beta + 4 L^2 \s_{3} \wedge d\xi \Big) \wedge \Sigma_1
\\[5pt]
      & + \frac{L^2 \l \a'^{\frac{1}{2}}}{12 }  \Big(\mathcal{B}_{-}^2 + \mathcal{B}^2 + \r^2 \sin^2\chi \Big) d\beta \wedge d\s_{3} \ .
 \end{aligned}
\end{equation}
Here for brevity we have defined the 1-form $\Sigma_1$ in the following way:
\begin{equation}
\Sigma_1 = 6 \a' e^{2V} \mathcal{B} \ \big(  \sin\chi d\r +\r \cos\chi d\chi  \big) - 9 \a' e^{2U} \r \sin\chi \ \big(  \cos\chi d\r -\r \sin\chi d\chi  \big)\ .
\end{equation}
Finally, we observe that the radius of the 
M-theory circle is finite, for reasons similar to those discussed in
the  previous example. We have checked that the equations of motion of the 11-dimensional Supergravity are solved by this background.\\

This completes our presentation of this set of new and exact solutions. A summary of all the solutions can be found in Table \ref{tabla1}. The expressions for these backgrounds
in cartesian and cylindrical coordinates are written in appendix C of \cite{Bea:2015fja} .

We will now move on to the second part of this chapter. We will study 
aspects of the field theories that our new and smooth backgrounds 
are defining.



\section{General comments on the Quantum Field Theory}\label{comentarios}
Let us start our study of the correspondence between our new metrics 
with their respective field theory  dual. We will state some general points that these field theories will fulfill.

In the case of the backgrounds corresponding to the compactifications described in section \ref{section211xx}, our field theories are obtained by a twisted KK-compactification on a two dimensional manifold---that can be $H_2, \  S^2$ or $T^2$. The original field theory is, as we mentioned, the Klebanov-Witten quiver, that controls the  high energy dynamics of our system. The bosonic part of the global symmetries for this  QFT in the UV are 
\begin{equation}
SO(1,3)\times SU(2)\times SU(2)\times U(1)_R\times U(1)_B ,
\end{equation}
where, as we know the $SO(1,3)$ is enhanced to $SO(2,4)$.
The theory contains two vector multiplets ${\cal W}^i=(\lambda^i, A_\mu^i)$, for $i=1,2$, together with four chiral multiplets 
${\cal A}_\alpha=(A_\alpha,\psi_{\alpha})$ for $\alpha=1,2$ and ${\cal B}_{\dot{\alpha}} =(B_{\dot{\alpha} } 
,\chi_{\dot{\alpha} }   )$ with $\dot{\alpha}=1,2$.

These fields transform as vector, spinors and scalars under $SO(1,3)$---that is $A_\alpha, B_{\dot{\alpha}}$ are singlets, the fermions transform in the $\bf(\frac{1}{2}, 0) \oplus (0,\frac{1}{2})$ and the 
vectors in the $\bf (\frac{1}{2},\frac{1}{2})$. The transformation  under the 'flavor' quantum numbers 
$SU(2)\times SU(2)\times U(1)_R\times U(1)_B $ is,
\begin{eqnarray}
& & A_\alpha=(2,1,\frac{1}{2}, 1),\;\;\;~~  B_{\dot{\alpha}}=(1,2,\frac{1}{2},-1),\nonumber\\
& & \psi_\alpha=(2,1,-\frac{1}{2},1),\;\;\; \chi_{\dot{\alpha}}=(1,2,-\frac{1}{2},-1),\\
& & \lambda^i=(1,1,1,0),\;\;\; ~~~~ A_\mu^i=(1,1,0,0).
\nonumber
\label{transflawsxx}
\end{eqnarray}

The backgrounds in section \ref{section211xx},
are describing the strong coupling regime of the field theory above, in the case in which we compactify the  D3-branes
on $\Sigma_2$ twisting the theory. This means, mixing the R-symmetry $U(1)_R$ with the $SO(2)$ isometry of $\Sigma_2$.
This twisting is reflected in the metric fibration between 
the $\eta$-direction (the Reeb vector) and the $\Sigma_2$.
The fibration is implemented by a vector field $A_1$ 
in eq. \eqref{NN02}. The twisting mixes the R-symmetry of the QFT, 
represented by $A_1$ in the dual description,  
with (part of) the Lorentz group.
In purely field theoretical terms, we are modifying the covariant derivative of different fields that under the  combined action of 
the spin connection and the R-symmetry (on the curved part of the space) will read $D_\mu\sim \partial_\mu+\omega_\mu+ A_\mu$.

In performing this twisting, the fields decompose under 
$SO(1,3)\to SO(1,1)\times SO(2)$. 
The decomposition is straightforward for the bosonic fields. 
For the fermions, we have that $\bf(\frac{1}{2}, 0) $ 
decomposes as ${\bf (+,\pm)}$ and similarly 
for the $(0,\frac{1}{2})$ spinors. 

The twisting itself is the 'mixing' between the $\pm$ charges of the spinor
and its R-symmetry charge. There is an analog operation for the vector and 
scalar fields. Some fields are scalars under the diagonal group in
$U(1)_R \times SO(2)_{\Sigma_2}$. Some are spinors and some are vectors.
Only the scalars under the diagonal group are massless. These determine the 
SUSY content of the QFT.
This particular example 
amounts to preserving two supercharges. 
There are two massless vector multiplets and two massless matter multiplets. The rest of the fields get a mass whose set by the inverse size of the compact manifold. In other words, the field theory at low energies is a two dimensional  CFT (as indicated by the $AdS_3$ factor), preserving $(0,2)$ SUSY and  obtained by a twisted compactification
of the Klebanov-Witten CFT. The QFT is deformed in the UV by a relevant operator of dimension two, as we can read from eq. \eqref{O20x}. 

An alternative way to think about this QFT is as the one describing
the low energy excitations of a stack of $N_c$ D3-branes wrapping  
a calibrated space $\Sigma_2$
inside a Calabi-Yau 4-fold.

The situation with the metrics in section \ref{sectionDG} is 
more subtle. In that case there is also a flow from the Klebanov-Witten quiver to a two-dimensional CFT preserving (0,2) SUSY. The difference is that this second QFT is not apparently obtained via a twisting procedure. As emphasized by the authors of \cite{Donos:2014eua}, 
the partial breaking of SUSY is due (from a five-dimensional 
supergravity perspective) to 'axion' fields depending on the torus directions. These axion fields are proportional to a  
deformation parameter---that we called $\lambda$ in eq. \eqref{fluxes-bef}. 
The deformation in the UV QFT is driven 
by an operator of dimension four that was identified to be $Tr(W_1^2-W_2^2)$ and a  dimension six operator that acquires a VEV, as discussed in \cite{Donos:2014eua}. 

To understand the dual field theory to the  IIA backgrounds obtained after non-abelian T-duality and presented in section \ref{s2h2natd} involves more intricacy. Indeed, it is at present unclear what is the analog field theoretical operation of non-abelian T-duality. There are, nevertheless, important hints. 
Indeed, the foundational paper of Sfetsos and Thompson \cite{Sfetsos:2010uq}, that sparked the interest of the uses of non-abelian T-duality in quantum field theory duals, showed that if one starts with a background of the form $AdS_5\times S^5/Z_2$, a particular solution of the Gaiotto-Maldacena system (after lifting to M-theory) is generated \cite{Gaiotto:2009gz}.
This is hardly surprising, as the backgrounds of eleven-dimensional supergravity with an $AdS_5$ factor and preserving $\mathcal{N}=2$ SUSY in four dimensions, have been classified. What is interesting is that the solution generated by Sfetsos and Thompson
appears as a zoom-in on the particular class of solutions in \cite{Maldacena:2000mw}. This was extended in 
\cite{Itsios:2013wd} that computed the action of non-abelian T-duality on the end-point of  the flow from the $\mathcal{N}=2$ conformal quiver with adjoints to the Klebanov-Witten CFT. Again, not surprisingly, the  backgrounds obtained correspond to the $\mathcal{N}=1$ version of the Gaiotto $T_N$ theories---
these were called {\it Sicilian} field theories by Benini, Tachikawa and Wecht
in  \cite{Benini:2009mz}, see Fig. \ref{RoadMap}.
It is noticeable, that while the Sicilian theories can be 
obtained by a twisted compactification of M5-branes on $H_2, S^2, T^2$, 
the case obtained using non-abelian T-duality corresponds only 
to M5-branes compactified on $S^2$ and preserving minimal SUSY in four dimensions. What we propose in this paper is that the twisted compactification on $\Sigma_2$ of a Sicilian gauge theory
 can be studied by using the backgrounds we discussed in section \ref{s2h2natd} and their M-theory counterparts.
We will elaborate more about the 2-d CFTs and 
their flows in the coming sections.
 
 In the following, we will calculate different observables of  these QFT's by using the backgrounds as a 'definition' of the 2d SCFT. The backgrounds are smooth and thus the observables have trustable results. 
Hence, we are defining a QFT by its observables, 
calculated in a consistent way by the dual solutions. 
The hope is that these calculations together with 
other efforts can help map the space of 
these new families of CFTs. To the study of these observables we turn now.

\section{Quantized charges}\label{chargessectionxx}
In this section, we will study the quantized charges on the string side.
This analysis appears in the field theory part of the chapter because these charges will, as in the
canonical case of $AdS_5\times S^5$, translate into the ranks of the gauge theory local symmetry groups.

The NATD produced local solutions to the 10-dim SUGRA equations of motion. Nevertheless, it is still not known how to obtain the global properties of these new geometries. 
Some quantities associated to a particular solution, like the Page charges below, are only well-defined when the global properties of the background are known. Since we have only a local description of our solution, we will propose very reasonable global results for the Page charges, mostly based on physical intuition.

Let us start by analyzing a quantity that is proposed to be periodic in the string theory.
We follow the ideas introduced in \cite{Lozano:2014ata} and further elaborated in \cite{Macpherson:2014eza}. 
To begin with, we focus on the NATD version of the twisted solutions; described in section \ref{s2h2natd}. Let us define the quantity,
 \begin{equation}
b_0= \frac{1}{4 \pi^2 \alpha'} \int_{\Pi_2} B_2  ~~ \in  \ [0,1] \ ,
\label{page9b}
\end{equation}
where the cycle $\Pi_2$ is defined as,
 \begin{equation}
\Pi_2=S^2   ~~~  \Big\{  \chi , \xi  , \alpha=0, \rho=\textrm{const} , d\beta=-\frac{1}{3z} d\xi    \Big\} \ .
\label{page9ceeett}
\end{equation}
As the topology of the NATD theory is not known, we propose that this cycle is present in the geometry. This cycle will have a globally defined volume form, which in a local description can be written as  $\textrm{Vol}_{\Pi_2}=\sin \chi \ d \xi \wedge d\chi$. 
We then find,
 \begin{equation}
b_0=\frac{1}{4 \pi^2 \alpha'} \int_{\Pi_2} \alpha' \rho \sin \chi d\xi \wedge d \chi= \frac{\rho}{\pi}  ~~ \in  \ [0,1].
\label{page9deee}
\end{equation}
Again, we emphasize that this is a {\it proposal} made in \cite{Lozano:2014ata} and used in \cite{Macpherson:2014eza}. Moving further than $\pi$ along the variable $\rho$ can be 'compensated' by performing a large gauge transformation on the $B_2$-field,
\begin{equation}
B_2 \rightarrow B_2'=B_2 - \alpha '  n \pi \sin \chi d \xi \wedge d\chi.
\label{page10scsc}
\end{equation}
We will make extensive use of this below.

 Let us now focus on the conserved magnetic charges defined for our backgrounds. We will start the analysis for the case of the solutions in section \ref{section211xx}.

\subsection{Page charges for the twisted solutions}
We will perform this study for the solutions before and 
after the NATD, and we will obtain how the Page charges transform under the NATD process.
Page charges  (in contrast to Maxwell charges) have
 proven fundamental in understanding aspects of the dynamics of 
field theories---see for example \cite{Benini:2007gx}.

In the following, we use as definition of Page charge,
\begin{equation}
\begin{aligned}
&  N_{Dp} \big{|}_{\Pi_{8-p}}=\frac{1}{2\kappa_{10}^2 T_{Dp}} \int_{\Pi_{8-p}}  \left( \sum\limits_{i} F_i  \right) \wedge  e^{-B_2},      ~~~~~~         \left. N_{NS5} \right|_{\Pi_3} =  \frac{1}{2\kappa_{10}^2 T_{NS5}} \int_{\Pi_3} H_3 \ ,
\\[5pt]
& 2\kappa_{10}^2 =(2\pi)^{7} \alpha'^{4},      ~~~~~~   T_{Dp} =\frac{1}{(2\pi)^{p} \alpha'^{\frac{p+1}{2}}  },  ~~~~~~ T_{NS5} = T_{D5} \ .
\end{aligned}
\label{Page5g}
\end{equation}
In particular, for D3-branes we have,
\begin{equation}
N_{D3} \big{|}_{\Pi_5}=\frac{1}{2\kappa_{10}^2 T_{D3}} \int_{\Pi_5} \left( F_5 - B_2 \wedge F_3 + \frac{1}{2}B_2 \wedge B_2 \wedge F_1 \right).\;\;
\nonumber
\label{page201tgrtg}
\end{equation}

The topology of the internal space is 
$\Sigma_2 \times S^2 \times S^3$.  
First, we consider the cycle $S^2 \times S^3$. 
Second, we can consider some cycles given by the product of $\Sigma_2$ with a generator of $H_3 \left( S^2 \times S^3 , \mathbb{Z}  \right)$. Notice that the $S^2 \times S^3$ is realized as a $U(1)$ fibration over an $S_1^2 \times S_2^2$ base. A smooth 3-manifold, $S_1^3$,  that can be used to generate $H_3 \left( S^2 \times S^3 , \mathbb{Z}  \right)$ is provided by the circle bundle restricted to the $S_1^2$ factor. We can also choose $S_2^3$, defined to be the circle bundle restricted to the $S_2^2$ factor of the base space. In summary, the relevant cycles are,
\begin{equation}
\begin{aligned}
\Pi_5^{(1)}&=&S^2 \times S^3  ~\Big\{\theta_1,\phi_1,\theta_2,\phi_2, \psi \Big\},~~~~  \Pi_5^{(2)}&=\Sigma_2 \times S_1^3                ~  \Big\{\alpha,\beta, \theta_1,\phi_1, \psi \Big\}, \\
\Pi_5^{(3)}&=& \Sigma_2 \times S_2^3   ~ \Big\{ \alpha,\beta ,\theta_2,\phi_2, \psi \Big\}\ .  ~~~~~~~~~
\label{page6tt}
\end{aligned}
\end{equation}
The background fields $B_2$, $F_1$ and $F_3$ are vanishing, and only $F_5$ contributes. 
The specific components of $F_5$ in eq. \eqref{NN01} that have non vanishing pullback on these cycles are,
\begin{equation}
F_5= \frac{L^4}{9} \ \textrm{Vol}_1 \wedge \textrm{Vol}_2 \wedge \eta + \frac{L^4}{6} z \textrm{Vol}_{\Sigma_2} \wedge \left( \textrm{Vol}_1 + \textrm{Vol}_2 \right) \wedge \eta + ...
\label{page6a}
\end{equation}
We explicitly see that it is a globally defined form, as all the involved quantities ($\eta$, $\textrm{Vol}_{\Sigma_2}$, $\textrm{Vol}_1$, $\textrm{Vol}_2$) are well defined globally.
The associated Page charges of D3-branes for the background  around 
eq. \eqref{NN01} are,
\begin{equation}
N_{D3} \big{|}_{\Pi_5^{(1)}}=\frac{4 L^4}{27 \alpha'^2 \pi} \ ,~~~~~~~~~~~~~~ \widehat{N}_{D3} \big{|}_{\Pi_5^{(2)}}=\widetilde{N}_{D3} \big{|}_{\Pi_5^{(3)}}=\frac{L^4}{\alpha'^2} \frac{z \  vol \left( \Sigma_2 \right)}{18 \pi^2} \ ,
\label{page9dvdfv}
\end{equation}
where $vol\left( \Sigma_2 \right)$ is the total volume of the two-manifold $\Sigma_2$.\footnote{Notice that we use Vol for volume elements (differential forms) and $vol$ for the actual volumes of the manifolds (real numbers).}
As usual, the first relation quantizes the size of the space,
\begin{equation}
L^4=\frac{27 \pi}{4} \alpha'^2 N_{D3}.
\label{page10xdf}
\end{equation}
We have then defined three D3-charges. The one associated with $N_{D3}$ is the usual one appearing
also in the $AdS_5\times S^5$ case. The other two can be thought as charges  'induced'  by the wrapping of the D3-branes on the Riemann surface. The reader may wonder  whether these charges are present in the backgrounds found after NATD. We turn to this now.

The particular  expressions for Page charges in type IIA are,
\begin{equation}
\begin{aligned}
N_{D6} \big{|}_{\Pi_2}&=\frac{1}{2\kappa_{10}^2 T_{D6}} \int_{\Pi_2} \left( \widehat{F}_2-\widehat{F}_0 \ \widehat{B}_2\right), 
\\[5pt]
N_{D4} \big{|}_{\Pi_4}&=\frac{1}{2\kappa_{10}^2 T_{D4}} \int_{\Pi_4} \left( \widehat{F}_4- \widehat{B}_2 \wedge \widehat{F}_2 + \frac{1}{2} \widehat{F}_0 \  \widehat{B}_2 \wedge \widehat{B}_2 \right), 
\\[5pt]
N_{D2} \big{|}_{\Pi_6}&=\frac{1}{2\kappa_{10}^2 T_{D2}} \int_{\Pi_6} \left( \widehat{F}_6- \widehat{B}_2 \wedge \widehat{F}_4+\frac{1}{2} \widehat{B}_2 \wedge \widehat{B}_2 \wedge \widehat{F}_2- \frac{1}{6} \widehat{F}_0 \ \widehat{B}_2 \wedge \widehat{B}_2 \wedge \widehat{B}_2  \right).
\label{page201uu}
\end{aligned}
\end{equation}
We label the radius of the space of the geometry in eq. \eqref{NN11} by $\widehat{L}$, to distinguish it from $L$, the quantized radius before the NATD.

In order to properly define the cycles to be considered, we should know the topology of this NATD solution. However, we have obtained only a local expression for this solution, and we do not know the global properties. As we explained above, we will  present a proposal to define the Page charges that would explain the transmutation of branes through the NATD. We propose the relevant cycles to be, \footnote{Intuitively, we can think that the branes transform under NATD as 3 consecutive T-dualities. For example, in the first 5-cycle of eq. (\ref{page6tt}), the NATD is performed along 3 of the coordinates, ($\theta_2, \phi_2, \psi$), in such a way that we end up with a 2-cycle, the first cycle in eq.(\ref{page7r}), associated with D6-branes. In the second 5-cycle of eq.(\ref{page6tt}) the NATD only affects the $\psi$-direction, so it disappears, and two more directions are added in order to complete the 3 T-dualities, ending up with a 6-cycle in eq.(\ref{page7r}), associated with D2-branes.}
\begin{equation}
\Pi_2^{(1)}  \ \  \Big\{\theta_1,\phi_1 \Big\} ,  ~~~~ 
      \Pi_6^{(2)}= \Big\{\alpha,\beta,\theta_1, \phi_1, \rho, \xi,\chi=\frac{\pi}{2} \Big\} ,  ~~~~  \Pi_2^{(3)} \  \  \Big\{\alpha,\beta \Big\} .
\label{page7r}
\end{equation}
The associated charges are,
\begin{equation}
\label{page8zr}
 \begin{aligned}
  & N_{D6} \big{|}_{\Pi_2^{(1)}}=\frac{2 \widehat{L}^4}{27 \alpha'^2 } \ ,    \qquad \quad  \widetilde{N}_{D6}\big{|}_{\Pi_2^{(3)}}=\frac{\widehat{L}^4}{\alpha'^2} \frac{z \ vol \left( \Sigma_2 \right)}{36 \pi} \ ,
  \\[10pt]
  & \widehat{N}_{D2} \big{|}_{\Pi_6^{(2)}}=\frac{\widehat{L}^4}{\alpha'^2} \frac{z \  vol \left( \Sigma_2 \right)  vol \left(\rho, \xi \right)}{ 144 \pi^4}=\frac{n^2}{4}\frac{\widehat{L}^4}{\alpha'^2} \frac{z \  vol \left( \Sigma_2 \right)}{36 \pi}.
 \end{aligned}
\end{equation}
%
%
In the last expression, we performed the integral over the $\rho$-coordinate in the interval $[0, n\pi]$.
These three charges are in correspondence with the ones before the NATD in eq. \eqref{page9dvdfv}.
Indeed, we can compute the quotients,
\begin{align}
& & \frac{\widehat{N}_{D3} }{N_{D3}}=\frac{\widetilde{N}_{D3}}{N_{D3}}=\frac{3}{8\pi}z \  vol \left( \Sigma_2 \right), \qquad \frac{4 \widehat{N}_{D2} }{n^2 N_{D6}}=\frac{\widetilde{N}_{D6}}{N_{D6} }=\frac{3}{8\pi}z \  vol \left( \Sigma_2 \right).
\label{manaxx}
\end{align}
These quotients indicate a nice correspondence between charges before
and after the duality.

Using the first relation in eq. \eqref{page8zr}  we quantize the size $\widehat{L}$ of the space after NATD to be
$\widehat{L}^4=\frac{27}{2} \alpha'^2 N_{D6}$.

A small puzzle is presented by the possible existence of charge for D4-branes, as there would be no quantized number before the NATD to make them correspond to. To solve this puzzle, we propose that there should be a globally defined closed non-exact form that allows us to perform a large gauge transformation for the $\widehat{B}_2$, in such a way that all the D4 Page charges vanish. In local coordinates, we have a gauge transformation,
\begin{equation}
\widehat{B}_2 \rightarrow \widehat{B}'_2 = \widehat{B}_2 + \alpha' d\left[ \rho \cos \chi \ \tilde{\sigma}_3 \right],
\label{page9ss9}
\end{equation}
written in such a way that the integrand has at least one leg along a non-compact
coordinate,
\begin{equation}
\widehat{F}_4-\widehat{B}'_2 \wedge \widehat{F}_2=\frac{z \widehat{L}^4}{6 \sqrt{\alpha'}} e^{-2B} \left(\cos \chi d\rho - \rho \sin \chi d \chi \right) \wedge \textrm{Vol}_{AdS_3}.
\label{page9adtrtg}
\end{equation}
Hence, any Page charge for D4-branes is vanishing. To be precise, the  D2 charge $\widehat{N}_{D2}$ must be computed after choosing this gauge, as it depends on $\widehat{B}_2$, but it turns out to be the same as calculated in eq. \eqref{page8zr}.

The motion in the $\rho$-coordinate, as we discussed above, can be related to large gauge transformations of the $\widehat{B}_2$-field.
The  large transformation that 'compensates' for motions in $\rho$, namely
$
 \widehat{B}_2 \rightarrow \widehat{B}'_2=\widehat{B}_2 - \alpha '  n \pi \sin \chi d \xi \wedge d\chi,
$
has the effect of changing the Page charges associated with D4-branes, that were initially vanishing. Indeed, if we calculate  for the following four cycles,
\begin{equation}
\Pi_4^{(1)}  ~~  \Big\{\theta_1,\phi_1, \chi, \xi \Big\} \ , ~~~~~~ ~~
\Pi_4^{(2)} ~~  \Big\{\alpha,\beta , \chi, \xi \Big\} \ , ~~~~
\label{page12y}
\end{equation}
the Page charge of D4-branes varies according to,
\begin{equation}
\begin{aligned}
\Delta N_{D4} \big{|}_{\Pi_4^{(1)}}&=\frac{1}{2\kappa_{10}^2 T_{D4}} \int_{\Pi_4^{(1)}} \left(- \Delta  \widehat{B}_2 \wedge \widehat{F}_2  \right) =- n \frac{L'^4}{\alpha'^2}\frac{2}{27} =-n N_{D6},   
\\[10pt]
\Delta N_{D4} \big{|}_{\Pi_4^{(2)}}&=\frac{1}{2\kappa_{10}^2 T_{D4}} \int_{\Pi_4^{(2)}} \left(- \Delta  \widehat{B}_2 \wedge \widehat{F}_2  \right) = -n \frac{L'^4}{\alpha'^2}\frac{z}{36 \pi}  vol \left( \Sigma_2 \right)=-n \widetilde{N}_{D6}  .
\label{page11n}
\end{aligned}
\end{equation}
We can interpret these findings in the following way. Our QFT  (after the NATD) can be thought as living on the world-volume of a superposition of D2- and D6-branes. Motions in the $\rho$-coordinate induce charge of D4-branes, which can be interpreted as new gauge groups appearing. This suggest that we are working with a linear quiver, with many gauge groups. Moving $n \pi$-units in $\rho$ generates or 'un-higgses'  new gauge groups of rank $ n N_{D6}$ and  $n \widetilde{N}_{D6}$. Computing volumes or other observables that involve integration on the $\rho$-coordinate amounts to working with a  QFT with different gauge group, depending on the range in $\rho$ we decide to integrate over. Notice that $\rho$ is not a holographic coordinate. Motions in $\rho$ are not changing the energy in the dual QFT. For the $AdS$-fixed points the theory is conformal and movements in $\rho$ do not change that.

In the paper \cite{Macpherson:2014eza} , the motion in the $\rho$-coordinate was argued to be related to a form of 'duality' (a Seiberg-type of duality was argued to take place, in analogy with the mechanism of the Klebanov-Strassler duality cascade, but in a CFT context). That can not be the whole story as other observables, like for example the number of degrees of freedom in the QFT, change according to the range of the $\rho$-integration. Hence the motion in $\rho$ can not be just a duality. 
 We are proposing here that moving in the $\rho$-coordinate amounts to changing the quiver, adding gauge groups, represented by the increasing D4 charge.

We will now present the same analysis we have performed above, but for the case of the background in section \ref{sectionDG}.

\subsection{Page charges for the Donos-Gauntlett solution}
Part of the analysis that follows was carefully done  in the original work of \cite{Donos:2014eua} and here we will extend the study for the solution after the NATD that we presented in section \ref{natdDG}. We will give an outline of the results as the general structure is similar to the one displayed by the twisted solutions of the previous section.

We focus on the original Donos-Gauntlett background first. We denote by $d_{\alpha}$, $d_{\beta}$ the two radii of the torus (the cycles of the torus are then of size $2\pi d_\alpha$ and $2\pi d_\beta$ respectively) and consider
five different cycles in the geometry,

\begin{equation}
\label{page20ab}
 \begin{aligned}
& \Pi_5^{(1)}=S^2 \times S^3  ~~  \Big\{\theta_1,\phi_1,\theta_2,\phi_2, \psi \Big\} ,  ~~
\Pi_5^{(2)}=T^2 \times S_1^3                ~~   \Big\{\alpha,\beta, \theta_1,\phi_1, \psi \Big\},     ~~\\
&\Pi_5^{(3)}=T^2 \times S_2^3   ~~ \Big\{ \alpha,\beta ,\theta_2,\phi_2, \psi \Big\} , \\
& \Pi_3^{(4)}=S_1^1 \times s(S)  ~~ \Big\{ \alpha, \theta_1= \theta_2, \phi_1= - \phi_2 , \psi=\textrm{const} \Big\},   ~~ \\
&\Pi_3^{(5)}=S_2^1 \times s(S)  ~~ \Big\{ \beta, \theta_1= \theta_2, \phi_1= - \phi_2 , \psi=\textrm{const} \Big\} .
 \end{aligned}
\end{equation}
%
%
The  Page charges associated with D3-, D5- and NS5-branes are,
%
%
\begin{equation}
\label{page21a}
 \begin{aligned}
  & N_{D3} \big{|}_{\Pi_5^{(1)}}=\frac{4 L^4}{27 \alpha'^2 \pi} \ ,  \qquad \widehat{N}_{D3} \big{|}_{\Pi_5^{(2)}}= - \widetilde{N}_{D3} \big{|}_{\Pi_5^{(3)}}=  2 \ \frac{L^4}{\alpha'^2} \frac{\lambda^2 \ d_{\alpha} d_{\beta}}{9} \ ,
  \\[10pt]
 & N_{NS5} \big{|}_{\Pi_3^{(4)}}=- 2 \ \frac{L^2 \lambda \ d_{\alpha}}{3 \alpha'} \ , \qquad 
     N'_{D5} \big{|}_{\Pi_3^{(5)}}=2 \ \frac{L^2 \lambda \ d_{\beta}}{3 \alpha'} \ .
 \end{aligned}
\end{equation}
After the NATD, we focus on the background around eqs. (\ref{s5}) (\ref{s11}). 
We consider the  following cycles in the geometry,
%
\begin{equation}
\label{page273}
 \begin{aligned}
  & \Pi_2^{(1)}  ~~   \Big\{ \theta_1,\phi_1 \Big\}, ~~~~
\Pi_6^{(2)}         ~~   \Big\{ \alpha, \beta, \theta_1,\phi_1, \rho, \xi , \chi=\frac{\pi}{2}  \Big\}, ~~~~
\Pi_2^{(3)}    ~~ \Big\{ \alpha,\beta \Big\}, ~~~~
\\[10pt]
  & \Pi_6^{(4)} ~~ \Big\{ \beta, \theta_1, \phi_1, \rho, \chi, \xi \Big\}, ~~~~
\Pi_2^{(5)}  ~~ \Big\{ \beta, \chi, \rho=\rho_0 \Big \}.   ~~~~~~~~~~~~~~~~~~~~~~~~~~~~~~~
 \end{aligned}
\end{equation}
The  correspondent Page charges defined on them (the $\rho$-coordinate is taken in the $[0, n\pi]$ interval),
%
%
\begin{equation}
\label{page8khg}
 \begin{aligned}
   & N_{D6} \big{|}_{\Pi_2^{(1)}}=\frac{2 \widehat{L}^4}{27 \alpha'^2 } \ ,    \qquad     \widetilde{N}_{D6}\big{|}_{\Pi_2^{(3)}}=  - \frac{\widehat{L}^4}{\alpha'^2} \frac{ \lambda^2  \pi}{18} d_{\alpha} d_{\beta} \ , \qquad  N'_{D6} \big{|}_{\Pi_2^{(5)}}=\frac{ L'^2 }{ \alpha' } \frac{\lambda}{3} \rho_0 \ d_{\beta} \ ,
   \\[10pt]
   & \widehat{N}_{D2} \big{|}_{\Pi_6^{(2)}}=-\frac{\widehat{L}^4}{\alpha'^2} \frac{  \lambda^2 \   vol \left( T^2 \right)  vol  \left(\rho , \xi \right)}{288 \pi^4}  =- \frac{\widehat{L}^4}{\alpha'^2} \lambda^2 \frac{n^2 \pi}{72} d_{\alpha} d_{\beta} ,
   \\[10pt]
   &  N_{D2} \big{|}_{\Pi_6^{(4)}}= \frac{\widehat{L}^2}{\alpha'} \frac{\lambda}{24 \pi^3} d_{\beta} \  vol \left( \rho, \chi, \xi \right)= \frac{\widehat{L}^2}{\alpha'} \lambda \frac{n^3 \pi}{18} d_{\beta}.  ~~~~~~~~~~~
 \end{aligned}
\end{equation}
From the first relation we obtain,
$
\widehat{L}^4=\frac{27}{2} \alpha'^2 N_{D6}$.  Like in the case of the twisted solutions, we can choose a gauge for the $\widehat{B}_2$ field
%
%
\begin{equation}
\label{page9joijo}
 \begin{aligned}
  & \widehat{B}_2 \rightarrow \widehat{B}'_2=\widehat{B}_2 + \delta \widehat{B}_2 \ ,
\\[5pt]
  & \delta \widehat{B}_2=\frac{9 \alpha'^2 \lambda}{2 L^2} \ d \beta \wedge \Big( \rho \sin \chi \ d \big( \rho \sin \chi \big)+ \mathcal{B} \ d \mathcal{B} \Big)+ L^2 \frac{ \lambda}{6} \big( d \alpha \wedge \sigma_{3} - \alpha \ \sigma_{1}\wedge \sigma_{2} \big), 
 \end{aligned}
\end{equation}
such that the Page charge of D4-branes, when computed on  every possible compact  4-cycle, is vanishing. 
Indeed, after the gauge transformation, we have
\begin{equation}
\widehat{F}_4-\widehat{B}'_2 \wedge \widehat{F}_2  =\frac{L^4 \lambda}{36 \sqrt{\alpha'}} e^{V}\left( \frac{2 L^2}{\alpha'}d\alpha -3 e^{-2B-2V} \lambda \ d(\rho \cos \chi) \right) \wedge \textrm{Vol}_{AdS_3}.
\label{page9hu}
\end{equation}
Any integral over compact manifolds is vanishing. Like in the case of the twisted solutions, $\widehat{N}_{D2}$ and $N_{D2}$ should be recalculated after choosing this gauge;  but their value turns out to be unchanged.

We can apply the same  string theory considerations on the quantity $b_0$ that now is defined as an integral over the cycle,
\begin{equation}
\Pi_2=S^2 \ ,  \qquad \{  \chi , \xi , \alpha = \textrm{const},\rho= \textrm{const}    \}.
\label{page9c}
\end{equation}
We then calculate,
 \begin{equation}
b_0=\frac{1}{4 \pi^2 \alpha'} \int_{\Pi_2} \alpha' \rho \sin \chi d\xi \wedge d \chi= \frac{\rho}{\pi}  ~ \in  \ [0,1].
\label{page9d0}
\end{equation}
If we move further than $\pi$ along the variable $\rho$, we can compensate this by performing the large gauge transformation
$
\widehat{B}_2 \rightarrow \widehat{B}'_2=\widehat{B}_2 - \alpha '  n \pi \sin \chi d \xi \wedge d\chi .
$
We now consider the correspondent variation of Page charges for D4-branes, that can be calculated using the following cycles,
\begin{equation}
\Pi_4^{(1)} ~~ \{ \theta_1,\phi_1, \chi, \xi \} \ , ~~~~~~~~ 
\Pi_4^{(2)} ~~ \{\alpha,\beta , \chi, \xi \}, ~~~~
\label{page1200}
\end{equation}
to be,
\begin{equation}
\begin{aligned}
\Delta Q^P_{D4} \big{|}_{\Pi_4^{(1)}}&=\frac{1}{2\kappa_{10}^2 T_{D4}} \int_{\Pi_4^{(1)}} \left(- \Delta  \widehat{B}_2 \wedge \widehat{F}_2  \right) =- n \frac{L^4}{\alpha'^2}\frac{2}{27}   =- n N_{D6} \ ,
\\[5pt]
\Delta Q^P_{D4} \big{|}_{\Pi_4^{(2)}}&=\frac{1}{2\kappa_{10}^2 T_{D4}} \int_{\Pi_4^{(2)}} \left(- \Delta  \widehat{B}_2 \wedge \widehat{F}_2  \right) =-  n \frac{L^4}{\alpha'^2}\frac{\pi \lambda^2}{18 }  d_{\alpha} d_{\beta}=n \widetilde{N}_{D6} \ .
\label{page1100}
\end{aligned}
\end{equation}
The variation of the Page charges of D2-branes vanishes under this large gauge transformation. For the Donos-Gauntlett solution we observe a structure very similar to the one discussed for the twisted solutions. 
Again, here we would propose that the NATD background 'un-higgses'
gauge groups of rank $n N_{D6}$ as we move in units of $n\pi$ in the $\rho$-coordinate.

We move now to the study of another important observable of 
our dual 2-d and 4-d CFTs.

\section{Central charges and c-theorem}
\label{centralchargessectionxx}
In this section, we will study the central charge, an important observable of the different strongly coupled QFTs that our backgrounds in sections \ref{section1.1} to \ref{seccionlift} are defining.

Let us start with a brief summary of the ideas behind this observable. 
The RG-flow can be understood as the motion of the different couplings of the QFT $\lambda_i$  in terms of a parameter $t=-\ln \mu$, such that for a given operator $\widehat{O}$,
\begin{equation}
\frac{d\widehat{O}}{dt}=-\beta_i(\lambda) \frac{\partial \widehat{O}}{\partial\lambda_i}.
\end{equation}
Hence, the beta functions $\beta_i(\lambda)$ are the 'velocities' of the motion towards the IR.
It is interesting to define a 'c-function' $c(t)$ with the property that it decreases when flowing to low energies,
\begin{equation}
\frac{dc(t)}{dt} \leq 0.
\end{equation}
Such that at stationary points $\frac{dc(t)}{dt}=0$ implies that $\beta_i(\lambda)=0$ and vice versa.
The intuition behind this quantity is that massless degrees of freedom are lifted by relevant deformations, the flow to low energies then coarse-grains away these lifted modes. This intuition is realized in different situation: the two dimensional case, with Zamolodchikov's definition of $c(g)$ 
\cite{Zamolodchikov:1986gt} or Cardy's conjecture for the 'a-theorem' 
\cite{Cardy:1988cwa},  proven by Komargodski and Schwimmer in
\cite{Komargodski:2011vj}. There are 
different versions of the c-theorem, varying in 'strength' and generality. See the paper 
\cite{Barnes:2004jj} for a summary.

As we discussed, the  c-function is properly defined
only at the conformal points of a QFT. 
Hence, we can define it properly in
all of our backgrounds only at the $AdS_5$ and $AdS_3$ fixed points. In those cases the central charge is basically proportional to the volume of the 'internal manifold' $M_d$ (the complement-space of $AdS_5$ or $AdS_3$). 
 Indeed, there exists a well-established formalism to calculate central charges
 that uses the relation between this quantity and the Weyl anomaly $\mathcal{A}$ in the QFT when placed on a generic curved space. This was first discovered in \cite{Brown:1996ata}  (before the Maldacena conjecture was formulated!) and a  complete understanding was  developed in \cite{Skenderis:1998ata}.
 Indeed, for conformal field theories in two and four dimensions, associated with $AdS_3$
and $AdS_5$ geometries respectively, we have \footnote{The central charges $a$ and $c$ are equal
at the leading order in an  $N_c$-expansion. This is the result captured by the Supergravity approximation used in this paper. Also notice that the L's entering in this formulas are relative to an $AdS$ space expressed in the canonical form. }
\begin{equation}
\label{skenderishenningson}
\hskip -4.2cm \textrm{2 dimensions:}\  \qquad \mathcal{A}=-\frac{L \bar{R}}{16 \pi G_N^{(3)}}=-\frac{c}{24 \pi}\bar{R} , ~~~~~ ~~~ c=\frac{3}{2} \frac{L}{G_N^{(3)}} \ ,
\end{equation}
and
\begin{equation}
\begin{aligned}
& \textrm{4 dimensions:} \, \quad   \mathcal{A}=-\frac{L^3 }{8 \pi G_N^{(5)}}  \left( - \frac{1}{8}\bar{R}^{ij}\bar{R}_{ij} +\frac{1}{24} \bar{R}^2 \right)  =\frac{1 }{16 \pi^2 }\Bigg[c \Big( \bar{R}^{ijkl}\bar{R}_{ijkl} -2 \bar{R}^{ij}\bar{R}_{ij} +\frac{1}{3}\bar{R}^2 \Big) 
\\[5pt]
& \qquad\qquad\qquad\qquad\quad \; - a \Big( \bar{R}^{ijkl}\bar{R}_{ijkl} -4 \bar{R}^{ij}\bar{R}_{ij} + \bar{R}^2 \Big)  \Bigg] \ ,   ~~~~~~~    c=a=\frac{\pi}{8}\frac{L^3}{G_N^{(5)}} \ , 
\end{aligned}
\end{equation}
where  $\bar{R}_{ijkl}$, $\bar{R}_{ij}$ and $\bar{R}$ 
are the Riemann and Ricci tensors and scalar of the boundary metric. 
The Newton constant in different dimensions is calculated according to
(we take $g_s=1$ as in the rest of this paper),
 \begin{equation}
G_N^{(10)}=8 \pi^6  \alpha'^4,  ~~~~~ G_N^{(10-d)}=\frac{G_N^{(10)}}{vol (M_d)} . 
\label{Central1a}
\end{equation}
For solutions presenting a flow between these fixed points (or generically, an RG flow), a quantity that 
 gives an idea of the number of degrees of freedom can also be defined. This quantity measures an 'effective volume' of the internal space, that is a volume that is weighted together with the dilaton
and  other factors in the metric. To define such a quantity one goes back to a proposal by Freedman, Gubser, Pilch and Warner \cite{Freedman:1999gp}---see also the paper \cite{Alvarez:1998wr} for earlier attempts. Indeed, for any background (that
should be considered to be a solution of a D-dimensional supergravity, possibly connected with string theory) of the form,
\begin{equation}
ds_{D}^2= e^{2A(r)}dx_{1,D-2}^2 + dr^2 \ ,
\label{manaza}
\end{equation}
and assuming that the matter fields satisfy certain Energy conditions
\cite{Freedman:1999gp}, it was proven  using the Einstein equations that the quantity,
\begin{equation}
c\sim \frac{1}{(A')^{D-2}},
\label{zaza}
\end{equation}
is monotonically increasing towards the UV \cite{Freedman:1999gp}. This proposal was extended by 
Klebanov, Kutasov and Murugan in the paper \cite{Klebanov:2007ws}, to account for an RG-flow in a $d+1$ dimensional QFT, dual to a generic metric and dilaton of the form,
\begin{equation}
ds^2= \alpha_0(r)\Big[ dx_{1,d}^2+ \beta_0(r) dr^2  \Big]
+ g_{ij}(r,\vec{\theta}) d\theta^id\theta^j,\;\;\;\; \Phi(r).
\label{ppeerroo}
\end{equation}
In these cases and in cases where the functions $\alpha_0, \Phi$ depend also on the internal coordinates $\alpha_0(r,\vec{\theta}), \Phi(r,\vec{\theta})$, the formulas of \cite{Klebanov:2007ws} were extended in \cite{Macpherson:2014eza} 
 to be,
\begin{equation}
c= d^d \frac{ \beta_0(r)^\frac{d}{2}  \widehat{H}^{\frac{2d+1}{2}}}{ \pi G_N^{(10)} (\widehat{H}')^d}, \;\;\;\; \widehat{H}= \Big( \int d\vec{\theta} \sqrt{e^{-4\Phi} \det[g_{int} ]\alpha_0^d} \Big)^2.
\label{kkmcentral}
\end{equation}
At conformal points, i.e. when calculated in $AdS$ backgrounds, this gives a constant result, in agreement with eq. \eqref{skenderishenningson}. For backgrounds
with a flow, the quantity in eq. \eqref{kkmcentral} gives an idea of the number of degrees of freedom that participate in the dynamics of the QFT at a given energy.

In the following sections, 
we will quote the results for central charges 
according to eq. \eqref{skenderishenningson} 
for the conformal field theories in two and four dimensions. Following that,
we will write the result that eq. \eqref{kkmcentral} gives for the flows between theories.

\subsection{Central charge at conformal points}

As anticipated, we will quote here the results for eq. \eqref{skenderishenningson} for the different $AdS_3$ and $AdS_5$ fixed points. We start with the twisted solutions of section \ref{section211xx}. We will use that the volume of the $T^{1,1}$ space is $ vol(T^{1,1})=16\pi^3/27$.

\noindent{\underline{Twisted geometries}}

 For the IR $AdS_3$ fixed point, the volume of the internal compact manifold is $
vol (M_7)= \frac{1}{3}   L^7  vol (\Sigma_2) vol (T^{1,1})$, and the central charge is,
\begin{equation}
 c= 9 \big{|} N_{D3} \widehat{N}_{D3} \big{|}  =  \frac{L^8}{\alpha'^4} \frac{ vol (\Sigma_2) vol (T^{1,1})}{24 \pi^6}  .
\label{Central42}
\end{equation}
At the UV $AdS_5$ fixed point, the volume of the internal compact manifold is $vol (M_5)= L^5  vol (T^{1,1})$, and the result for the central charge is,
\begin{equation}
c=\frac{27}{64} N_{D3}^2=\frac{L^8}{\alpha'^4} \frac{ vol (T^{1,1})}{64 \ \pi^5} \ .
\label{Central3aoee}
\end{equation}
After the NATD, we must  consider the new radius of the space $\widehat{L}$ and the volume of the new 5-dim compact space $4 \pi^2 vol (\rho, \chi, \xi)$. The computations turn out to be similar as the previous ones, and we obtain that the central charges before and after NATD, for both the two and four dimensional CFTs, are related by,
\begin{equation}
\frac{\hat{c}}{c}=\frac{\widehat{L}^8}{L^8} \frac{4 \pi^2  vol (\rho, \chi, \xi)}{ vol (T^{1,1})}.
\label{Central5aa}
\end{equation}
Let us comment on the quantity $vol (\rho, \chi, \xi)$, that  appears 
in the calculation of the Page charges in section \ref{chargessectionxx}---see for example, eq. \eqref{page8khg} and also in the computation of the 
entanglement entropy of section \ref{sectionEEWilson}.
Indeed, if we calculate,  \footnote{We identify the integral with the volume of the manifold spanned by the new coordinates. This becomes more apparent if we use the expressions in the appendix, in different coordinate systems. }
\begin{equation}
\begin{aligned}
& vol (\rho, \chi, \xi)=\int_{0}^{n\pi} \rho^2 d\rho \int_{0}^{2\pi} d\xi \int_{0}^{\pi}\sin\chi d\chi= \frac{4\pi^4}{3}n^3 \ ,
\\[10pt]
& \frac{\hat{c}}{c}=\frac{\widehat{L}^8}{L^8} \frac{4 \pi^2  vol (\rho, \chi, \xi)}{ vol (T^{1,1})}= \frac{36 \pi N_{D6}^2}{N_{D3}^2} n^3 \ .
\end{aligned}
\label{xxyz}
\end{equation}
We have performed the $\rho$-integral in the interval $[0,n\pi]$. 
The logic behind this choice was spelt out 
in section \ref{chargessectionxx}, see below eq. \eqref{page11n}. 
The proposal is that moving in units of $\pi$ in the $\rho$-coordinate implies 'un-higgsing' a gauge group, hence we would have a linear quiver gauge theory. The central charge captures this un-higgsing procedure, increasing according to how many groups we 'create'. What is interesting is the $n^3$
behavior in eq. \eqref{xxyz}. Indeed, if $n$ were associated with the rank 
of a gauge group, this scaling would be precisely the one obtained 
in Gaiotto-like CFTs (also valid for the $\mathcal{N}=1$ 'Sicilian' theories 
of \cite{Benini:2009mz}). Indeed,   
the NATD procedure when applied to the $AdS_5\times T^{1,1}$ 
background creates  metric and fluxes  similar to those characterizing 
the Sicilian CFTs. The backgrounds in sections \ref{section1.1} to \ref{seccionlift} are dual to a compactification of the Klebanov Witten CFT  
and (using the NATD background) the Sicilian CFT 
on a two space $\Sigma_2$. The two-dimensional 
IR fixed point of these flows is described by our 'twisted $AdS_3$' and its NATD. The central charge  of the Sicilian CFT and its compactified version is presenting a behavior that goes like
a certain rank to a third power $c\sim n^3$. 
This suggest that crossing $\rho=\pi$ amounts to adding D4-branes and 
Neveu-Schwarz five branes and $n$ is the number of  branes 
that were added ---see eqs. (\ref{page11n}) and (\ref{page1100})---  or crossed\footnote{Thanks to Daniel Thompson for a 
discussion about this.}. Had we integrated on the interval $[n\pi, (n+1)\pi]$, we would have obtained a scaling like
$c\sim N_{D_4}^2$ at leading order in $n$.

\noindent {\underline{Donos-Gauntlett geometry}}

We study here the central charge for the Donos-Gauntlett background in section \ref{sectionDG}.
For the $AdS_3$ fixed point, the volume of the internal compact manifold is

 $$vol (M_7)= L^7 \left(4/3 \right)^{5/4} 4 \pi^2 d_{\alpha} d_{\beta} \ vol (T^{1,1}) ~~,$$ 

and the central charge is,
\begin{equation}
c= 3 \ \big{|} N_{D3} \widehat{N}_{D3} \big{|}=\frac{2}{3} \frac{L^8}{\alpha'^4} \frac{d_{\alpha} d_{\beta} vol (T^{1,1})}{\pi^4} \ . 
\label{Central3ao1}
\end{equation}
For the $AdS_5$ fixed point, the volume of the internal compact manifold $vol (M_5) =  L^5  vol (T^{1,1})$, and the central charge results in,
\begin{equation}
c=\frac{27}{64} N_{D3}^2=\frac{L^8}{\alpha'^4} \frac{ vol (T^{1,1})}{64 \ \pi^5} \ .
\label{Central3ao}
\end{equation}

The quotient of central charges before and after the NATD for the Donos-Gauntlett QFT, are given by a similar expression to that in eq. \eqref{Central5aa}.

We move now to study a quantity that gives an idea 
of the degrees of freedom along a flow.

\subsection{Central charge for flows across dimensions}
\label{CCdifdims}
In the previous section, we calculated the central charge for two and four dimensional CFTs dual to the $AdS_3$ and $AdS_5$ fixed points of the flow. In this section, we will use the developments in  \cite{Freedman:1999gp} and \cite{Klebanov:2007ws}, to write a c-function along the flows between these fixed points. We will find various subtleties,
\begin{itemize}
\item{When considered as a low energy two-dimensional CFT, the definition of the c-function evaluated on the flows will not detect the presence of the four dimensional CFT in the far UV.}
\item{We attempt to generalize the formula of \cite{Klebanov:2007ws}
for anisotropic cases (that is for field theories that undergo a spontaneous compactification on $\Sigma_2$). This new definition will detect both the two dimensional and four dimensional conformal points, but will not necessarily be decreasing towards the IR. This is not in contradiction with 'c-theorems' that assume Lorentz invariance.}
\end{itemize}
We move into discussing these different points in our particular examples.
To start, we emphasize that the formulas in eqs.(\ref{manaza})-(\ref{kkmcentral}), contain the same information. Indeed, the authors of 
\cite{Klebanov:2007ws} present a 'spontaneous compactification' of a higher dimensional Supergravity (or String theory) to $d+2$ dimensions, see eq. \eqref{ppeerroo}. Moving the reduced system to Einstein frame and observing that the $T_{\mu\nu}$ of the  matter in the lower dimension satisfies  certain positivity conditions imposed in \cite{Freedman:1999gp}, use of  eq. \eqref{zaza} implies eq. \eqref{kkmcentral}. Hence, we will apply eqs. \eqref{ppeerroo}-\eqref{kkmcentral} to our different compactifications in sections \ref{section1.1} to \ref{seccionlift}.

{\underline{Twisted and Donos-Gauntlett solutions.}}
For the purpose of the flows both twisted and Donos-Gauntlett solutions present
a similar qualitative behavior.
We start by considering the family of backgrounds in eq. \eqref{NN02} as dual to field theories in $1+1$ dimensions.
In this case the quantities relevant for the calculation of the central charge are,
\begin{equation}
\begin{aligned}
& d=1,\;\; \alpha_0 =L^2 e^{2A},\;\;\; \beta_0=e^{-2A},\;\;\; e^{\Phi}=1,
\\[5pt]
& \frac{ds_{int}^2}{L^2}=e^{2 B} ds^2_{\Sigma_{{2}}}+e^{2U}ds_{KE} ^2+e^{2V} \left( \eta + z A_1 \right)^2.
\end{aligned}
\end{equation}
We calculate the quantity
\begin{equation}
\widehat{H}= {\cal N}^2 e^{2(B+4U+V+A)},     \qquad      {\cal N}=\frac{(4\pi)^3 vol (\Sigma_2) L^8}{108}.
\label{ccc}
\end{equation}
Then, we obtain
\begin{equation}
c=\frac{{\cal N} e^{2B+4U+V}}{2 \pi G_N^{(10)}(2B'+4U'+V'+A')}.
\end{equation}
Using the BPS equations describing these flows in eq. \eqref{O17} we can get an expression without derivatives. Specializing for the solution with an $H_2$ in section \ref{section211xx}, we find
\begin{equation}
c=\frac{{\cal N}}{9 \pi G_N^{(10)}}\frac{(1+e^{2r})^2}{1+2e^{2r}}.
\label{vava}
\end{equation}
We can calculate this for the background we obtained in 
section \ref{s2h2natd}, by application of NATD. 
The result and procedure will be straightforward, 
but we will pick a factor of the volume of the space 
parametrized by the new coordinates,
$
vol (\rho, \chi, \xi)$.

For the purposes of the RG-flow, the quotient of the central charges will 
be the same as the quotient in eq. \eqref{xxyz}. This was indeed observed 
in the past \cite{Itsios:2013wd}, \cite{Macpherson:2014eza}
and is just a consequence of the invariance of the quantity 
$\big(e^{-2\Phi}\sqrt{\det[g]}\big)$ under NATD.

Coming back to eq. \eqref{vava}, we find that in the far IR, represented by $r\to -\infty$, 
the central charge is constant. But in the far UV ($r\to \infty$), 
we obtain a result that is not characteristic of a CFT. 
Hence, this suggest that the definition is only capturing the behavior 
of a 2-dim QFT. In other words, the four dimensional QFT may 
be thought as a two dimensional QFT, but  with an infinite number of fields. 

The absence of the four dimensional fixed point in our eq. \eqref{vava} can be accounted if we generalize the prescription to calculate central charges for an anisotropic 4-dim QFT. Holographically
this implies working with a background of the form,\footnote{A natural generalization of eq. \eqref{olaqase} is $
ds^2 =  -\alpha_0 dt^2 +\alpha_1 dy_1^2+....+\alpha_d dy_d^2
 +\Pi_{i=1}^{d}\alpha_i^{\frac{1}{d}}
\beta dr^2 + g_{ij}d\theta^i d\theta^j.$}
\begin{equation}
ds_{10}^2 =  -\alpha_0 dy_0^2 +\alpha_1 dy_1^2+\alpha_2 ds_{\Sigma_2}^2
 +\left( \alpha_1 \alpha_2^2 \right)^{\frac{1}{3}}
\beta_0 dr^2 + g_{ij}d\theta^i d\theta^j.
\label{olaqase}
\end{equation}
In this case we define,
\begin{equation}
\begin{aligned}
 & G_{ij}d\xi_i d\xi_j=\alpha_1 dy_1^2 + \alpha_2 ds_{\Sigma_2}^2
 + g_{ij}d\theta^i d\theta^j,  
\\[5pt]
 & \widehat{H}= \left( \int d\theta^i \sqrt{e^{-4\Phi} \det[G_{ij}] }  \right)^2, 
\\[5pt]
 & c=d^d\frac{\beta_0^{\frac{d}{2}} \widehat{H}^{\frac{2d+1}{2}}}{\pi G_N^{(10)}(\widehat{H}')^d}.
\end{aligned}
\label{centralanisotropic}
\end{equation}

We can apply this generalized definition to the flow for the twisted 
$H_2$ background of section \ref{section211xx}, and Donos-Gauntlett background of section \ref{sectionDG} (for more examples the reader is referred to appendix D of \cite{Bea:2015fja} ).
In this case, we consider them as dual to a field theory in $3+1$  
anisotropic dimensions (two of the dimensions are compactified on a $\Sigma_2$).
The quantities relevant for the calculation of the central charge are,
\begin{equation}
\begin{aligned}
& & d=3,\;\; \alpha_1 =L^2 e^{2A},\;\;\; \alpha_2=L^2 e^{2B} , \;\;\; \beta_0=e^{\frac{-2A-4B}{3}},\;\;\; e^{\Phi}=1,
\\[5pt]
& &G_{ij}d\xi_i d\xi_j=L^2 \left( e^{2A} dy_1^2 +e^{2B} ds^2_{\Sigma_2}+e^{2U}ds_{KE} ^2+e^{2V} \left( \eta + z A_1 \right)^2\right).
\end{aligned}
\end{equation}
We calculate 
\begin{equation}
\widehat{H}= {\cal N}^2 e^{2(2B+4U+V+A)},\;\;\; 
{\cal N}=\frac{(4\pi)^3  L^8}{108}.
\label{cccc}
\end{equation}
Then,  we obtain
\begin{equation}
c=\frac{ 27 {\cal N} e^{4U+V}}{8 \pi G_N^{(10)} (2B'+4U'+V'+A')^3}.
\label{c000}
\end{equation}
Focusing on the $H_2$ case, if we use the solution that 
describe this flow---see eq. \eqref{O19} we get an analytical expression,
\begin{equation}
c=\frac{\mathcal{N}}{\pi \ G_N^{(10)}} \left( \frac{1+e^{2r}}{1+2e^{2r}} \right)^3 \ .
\end{equation}
Notice that, by definition, this quantity gives the correct central charge 
in the UV (a constant, characterizing the 4-d fixed point). 
In the IR, the quantity turns out to be constant too, 
so it is capturing the presence of a 2-d fixed point. 
Nevertheless, it is probably not an appropriate candidate for 
a 'c-function between dimensions' 
as it is not necessarily increasing towards the UV. 
This is not in contradiction with the logic of 'a-theorems' and proofs like the
ones 
in \cite{Freedman:1999gp} or \cite{Komargodski:2011vj}, 
as the metric does not respect Lorentz invariance. 
Hence, it is not satisfying the assumptions of the theorems.
For the Donos-Gauntlett case analogous things happen. 
It would be very nice to try to apply the recent ideas of 
\cite{Gukov:2015qea}
to this flow. Notice that this feature of a 'wrong monotonicity' for the central
charge was also observed---for theories breaking Lorentz invariance in Higher Spin theories---see the papers \cite{Gutperle:2011kf}.

Let us move now to study other observables defining the 2-d and 4-d
QFTs.

\section{Entanglement entropy and Wilson loops}\label{sectionEEWilson}
In this section, we will complement the work done above,
by studying a couple of fundamental observables in the QFTs defined by 
the backgrounds in sections \ref{section1.1} to \ref{seccionlift}.

Whilst at the conformal points the functional dependence of
results is determined by the symmetries, the interest will be 
in the coefficients accompanying the dependence. Both observables
interpolate smoothly between the fixed points.

\subsection{Entanglement entropy}
\label{EE}

\def\reg{\mathrm{reg}}

The aim of this section is to compute the 
entanglement entropy on a strip, which extends along the 
direction $y_1 \in [-\frac{d}{2},\frac{d}{2}]$, and study 
how this observable transforms under NATD. 
The input backgrounds for our calculations will be the Donos-Gauntlett and the $H_2$ flow as well as their non-abelian T-duals. Since the procedure is the same in all of the cases and due to the similarity of the geometries we will present the results in a uniform way.

For the computation of the entanglement entropy one has to apply the 
Ryu-Takayanagi formula \cite{Ryu} for non-conformal metrics \cite{Nishioka:2006gr}. This prescription states that the holographic entanglement entropy between the strip and its complement is given by the minimal-area static surface that hangs inside the bulk, and whose boundary coincides with the boundary of the strip. The general form of the entanglement entropy for the non-conformal case is,

\begin{equation}
 \label{prescription} S = \frac{1}{4G_N^{(10)} } \int d^8 \sigma e^{-2
\Phi} \sqrt{G^{(8)}_{\rm ind}}~.
\end{equation}

For the strip, we chose the embedding functions to be $y_0 = \textrm{const}$ and $r=r(y_1)$ and then using the conservation of the 
Hamiltonian  we arrive at an expression for $r(y_1)$ that makes the area minimal under that embedding. With that we compute,

\begin{equation}
\label{Ryu-Takayanagi}
 S= \frac{\tilde{L}^8}{54}\frac{\pi}{G^{(10)}_N} vol (\Sigma_2)V_3 \int\limits_{r_*}^{\infty} dr \ e^{-A} \frac{G^2}{\sqrt{G^2 - G^2_*}} \ ,
\end{equation}
where   the form of the function $G$ depends on the geometry of the background 
that we consider, 
\begin{equation}
 G = \left\{ \begin{array}{ll}
                   e^{A+2B} & \textrm{for the twisted solutions}
                   \\[5pt]
                   e^{A+2B+4U+V} & \textrm{for the DG solution}
                  \end{array}
        \right. \ .
\end{equation}
Above, $r_*$ is the radial position of the hanging surface tip and we define $G_*=G(r_*)$. Also, with $vol(\Sigma_2)$ we denote the volume of 
the Riemann surface $\Sigma_2$. Notice that the form of the 
function $G$ is the same before and after NATD. 
Moreover we consider,
\begin{equation}
\label{Radii}
 \tilde{L} = \left\{ \begin{array}{ll}
                     L                     & \textrm{before NATD}
                   \\[5pt]
                   \widehat{L}& \textrm{after NATD}
                  \end{array}
        \right. \ ,
\end{equation}
%
The quantity $V_3$ is defined as,
\begin{equation}
\label{3dimVol}
 V_3 = \left\{ \begin{array}{ll}
                       16 \ \pi^2 & \textrm{before NATD}
                       \\[5pt]
                       \int d \chi \ d \xi \ d \r \ \r^2 \sin\chi & \textrm{after NATD}
                      \end{array}
            \right. \ .
\end{equation}
Before the NATD transformation $V_3$ comes from the 3-dimensional 
submanifold that is spanned by the coordinates 
$(\th_2,\phi_2,\psi)$, while after the NATD it comes 
from the submanifold that is spanned by the dual 
coordinates $(\r,\chi,\xi)$. This 
implies that the entropies before and after NATD are proportional all along the flow, 
for any strip length. 
A discussion on possible values of the  quantity 
$\int d \chi \ d \xi \ d \r \ \r^2 \sin\chi$ 
can be found in the quantized charges 
section \ref{chargessectionxx} and in the discussion 
on central charges in section \ref{centralchargessectionxx}.

When computed by the Ryu-Takayanagi formula eq. \eqref{Ryu-Takayanagi} the entanglement entropy 
is UV divergent. In order to solve this we compute the 
regularized entanglement entropy ($S^{\reg}$) by subtracting 
the divergent part of the integrand of  eq. \eqref{Ryu-Takayanagi}. 
The regularized entanglement entropy is given by,
\begin{equation}
\label{RegEE}
 S^{\reg} = \frac{\tilde{L}^8}{54}\frac{\pi}{G^{(10)}_N} vol (\Sigma_2)V_3 \Bigg\{ \int\limits_{r_*}^{\infty} dr \   \Big( \frac{e^{-A} G^2}{\sqrt{G^2 - G^2_*}} - F_{UV}  \Big)    - \int\limits^{r_*} dr \ F_{UV}  \Bigg\} \ ,
\end{equation}
where the last integral is an indefinite integral with the result being evaluated at $r=r_*$ and 
\begin{equation}
F_{UV}= \left\{ \begin{array}{ll}
                  \frac{1+e^{2r}}{3} & \textrm{for the $H_2$ twisted solution}
                   \\[5pt]
                   e^{2r}+\frac{\lambda^2}{3}& \textrm{for the Donos-Gauntlett solution}
                  \end{array}
        \right. \ .
\end{equation}
From the formulas \eqref{Radii}, \eqref{3dimVol} and \eqref{RegEE} it is obvious that the regularized entanglement entropies before and after the NATD transformation differ by the factor
\begin{equation}
\label{EEquotient}
 \frac{\widehat{S}^{\reg}}{S^{\reg}} = \frac{\widehat{L}^8}{L^8}\frac{\int d \chi \ d \xi \ d \r \ \r^2 \sin\chi}{16 \ \pi^2} \ .
\end{equation}
In the formula above we denoted by $\widehat{S}^{\reg}$ the value of the
entanglement entropy after the NATD transformation. 
 As discussed below eq. (\ref{vava}), the quantity $\big(e^{-2\Phi}\sqrt{\det[g]}\big)$ is invariant under NATD, and this explains why the ratio (\ref{EEquotient}) is constant along the flow.

At this point let us normalize the regularized 
entanglement entropy by defining the quantity,
\begin{equation}
\label{SPrimeDef}
 S' = \frac{54}{\tilde{L}^8} \frac{G^{(10)}_N}{\pi \ vol(\Sigma_2) V_3} \ S^{\reg} \ .
\end{equation}
In what follows we present the behavior of $S'$ in the UV and the IR for the geometries of interest. We express the results in terms of the width of the strip $d$,
%
\begin{equation}
 d = 2 G_* \int\limits_{r_*}^{\infty} dr \ \frac{e^{-A}}{\sqrt{G^2 - G_{*}^2}} \ .
\end{equation}
The UV/IR behavior written in 
eqs \eqref{TwistedUVEE}, \eqref{DGUVEE}, \eqref{TwistedIREE} 
and \eqref{DGIREE} below, 
are just consequences of the fact that in far UV and far 
IR the dual QFT is conformal. 
The functional forms are universal, so our main interest 
is the constant appearing in them, 
and also as a cross-check of numerical results.

\subsubsection*{Behavior in the UV}

\underline{Twisted geometries}

In the case of the twisted $H_2$ geometry  we find that the width of the strip is,
\begin{equation}
 d = e^{-r_*} \int\limits_{1}^{\infty} \frac{du}{u^2} \frac{2}{\sqrt{u^6 -1}} = \frac{2 \sqrt{\pi} \Gamma\big(  \frac{2}{3} \big)}{\Gamma\big(  \frac{1}{6} \big)} \ e^{-r_*} \ .
\end{equation}
Here in the integration we changed the variable $r$ by $u = \frac{e^r}{e^{r_*}}$. From the calculation of the normalized entropy $S'$ we observe that in the UV this behaves like $\frac{1}{d^2}$, namely
\begin{equation}
\label{TwistedUVEE}
 S' = - \frac{\pi^{3/2}}{6} \Bigg(  \frac{\Gamma\big(  \frac{2}{3} \big)}{\Gamma\big(  \frac{1}{6} \big)}  \Bigg)^3 \frac{1}{d^2} + \frac{1}{3} \ln d + \frac{1}{3} \ln \Bigg(  \frac{\Gamma\big(  \frac{1}{6}  \big)}{2 \sqrt{\pi}} \Gamma\Big(  \frac{2}{3}  \Big)  \Bigg) \ ,
\end{equation}
where we also included subleading and next-to subleading terms.

\noindent \underline{Donos-Gauntlett geometry}

Similarly in the case of the Donos-Gauntlett geometry we find that the width of the strip in terms of $r_*$ (considering also a subheading term) is,
\begin{equation}
 d = \frac{2\sqrt{\pi} \Gamma\big(  \frac{2}{3} \big)}{\Gamma\big(  \frac{1}{6}  \big)} e^{-r_*} + \l^2 \big(  \frac{5}{72} + \frac{11}{24} I_1  \big) e^{-3 r_*} \ ,
\end{equation}
where
\begin{equation}
 I_1 = \int\limits_{1}^{\infty} dz \ \frac{z^2}{(z^4 + z^2 + 1)\sqrt{z^6-1}} = 0.1896 \ldots
\end{equation}
Here as well we end up with a $\frac{1}{d^2}$ behavior, a logarithmic subleading contribution and a constant $c_1$ for the regularized entropy,
\begin{equation}
\label{DGUVEE}
 S' = - 2 \pi^{3/2} \Bigg(  \frac{\Gamma\big(  \frac{2}{3} \big)}{\Gamma\big(  \frac{1}{6} \big)}  \Bigg)^3 \frac{1}{d^2} + \frac{\l^2}{3} \ln d + c_1 \ ,
\end{equation}
where $c_1$ ,
\begin{equation}
 c_1 = \lambda^{2}\left(-\frac{1}{3}\ln\left(\frac{2\sqrt{\pi}\Gamma\left(\frac{2}{3}\right)}{\Gamma\left(\frac{1}{6}\right)}\right)+\left(\frac{\ln2}{9}+\frac{11}{48}I_{1}\right)+\frac{\Gamma\left(-\frac{1}{3}\right)}{6\Gamma\left(\frac{2}{3}\right)}\left(\frac{5}{72}+\frac{11}{24}I_{1}\right)\right) \ .
\end{equation}

\subsubsection*{Behavior in the IR}

\underline{Twisted geometries}

The calculation for the IR limit is more tricky. The origin of the subtlety is that the integrals we have to evaluate now run all along the flow and we do not know the analytical properties of the integrands. In order to address this issue we split the integration into the intervals $[r_*,a]$ and $[a,+\infty)$ choosing $a$ to be in the deep IR but always greater than $r_*$. 

Following this prescription in the calculation, for the 
width of the strip we find,
\begin{equation}
 d = 
 \frac{4}{3} \ e^{-\frac{3 r_{*}}{2}}.
\end{equation}
%
If we do the same analysis when we calculate the normalized entropy we find,
\begin{equation}
\label{TwistedIREE}
 S' = \frac{2}{9} \ln d + \frac{2}{9} \ln \frac{3}{2} \ .
\end{equation}
%
%
The logarithmic dependence  of the leading term on $d$ is the expected for a $1+1$ theory.

\noindent \underline{Donos-Gauntlett geometry}

We close this section presenting the corresponding results for the Donos-Gauntlett geometry. As in the case of the twisted geometries we split the integrations in the same way. Then for the width of the strip we find,
    \begin{equation}
 d = e^{-a_0} \frac{2\sqrt{2}}{3^\frac{3}{4}} e^{-\frac{3^\frac{3}{4}}{\sqrt{2}}r_*} \ .
    \end{equation}
The normalized entropy displays again a logarithmic behavior in terms of the width of the strip,
\begin{equation}
\label{DGIREE}
 S' = \frac{8}{9} \ln d + \frac{8}{9} \ln \Big(  e^{a_0} \frac{3^{\frac{3}{4}}}{\sqrt{2}}  \Big) + c_2 \ ,
\end{equation}
where the constant $c_2$ has the value,
\begin{equation}
 c_2= \int_{0}^{\infty} \left( e^{-A} G-  e^{2r}-\frac{4}{3} \right)dr + \int_{-\infty}^{0} \left( e^{-A} G- e^{2r}-\left(\frac{4}{3}\right)^{\frac{5}{4}} \right)dr = -0.0312 \ldots
\end{equation}

\begin{figure}
\centering
\label{fig: S'} 
\begin{subfigure}[b]{0.62\textwidth}
\centering
\includegraphics[width=\textwidth]{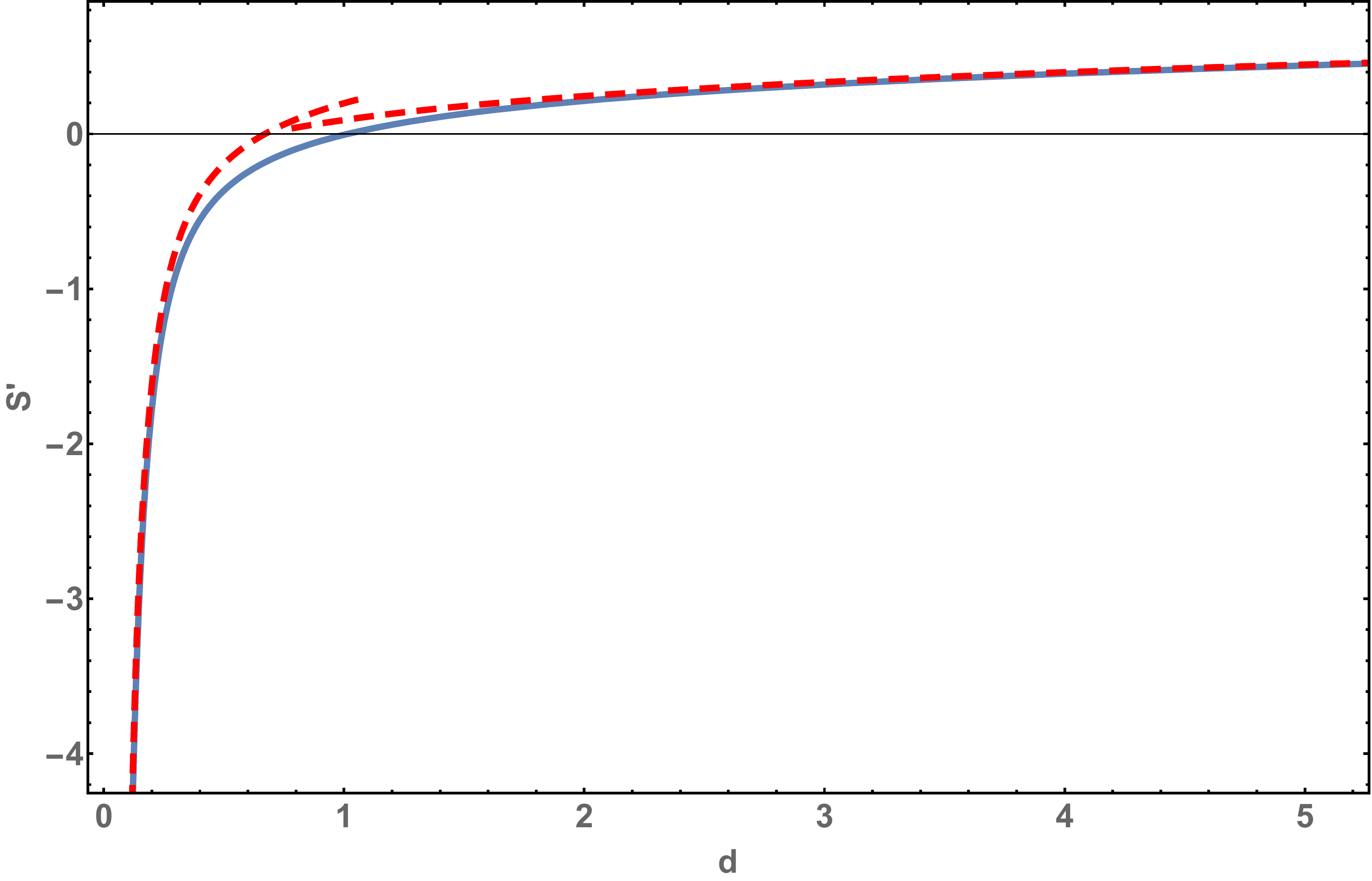}
\end{subfigure}
~
\begin{subfigure}[b]{0.62\textwidth}
\centering
\includegraphics[width=\textwidth]{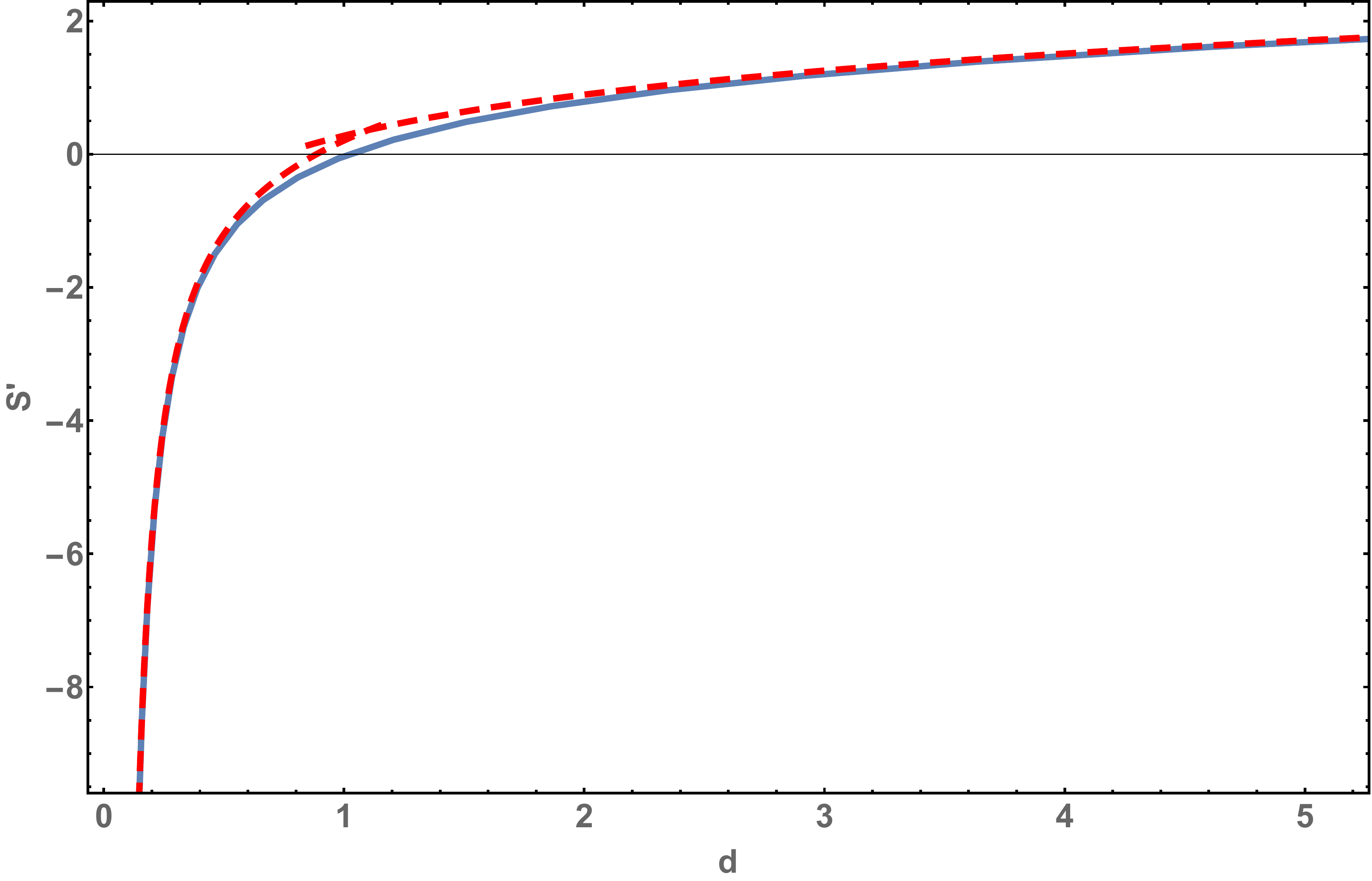}
\end{subfigure}
\caption{$S'$ as a function of $d$ for twisted $H_2$ (left) and Donos-Gauntlett solution (right).
The continuous curves correspond to the numerical value, while the dashed red ones to the UV and IR limits.
}
\label{EEnatduality0}
\end{figure}

The results for the entanglement entropy are shown in Fig. \ref{EEnatduality0}. We will now perform a similar analysis for Wilson loops.

\subsection{Wilson loop}

In this section we calculate the potential energy as a function 
of the separation in the $y_1$ direction (denoted as $d$)
for two non-dynamical sources added to the QFT \cite{Maldacena:1998im}. 
In holography this observable can be represented 
by a hanging string whose ends are separated by a 
distance $d$ along the $y_1$ direction. 
In our calculations we consider 
an embedding of the form $r = r(y_1)$ for the string. Such an embedding gives rise to the following induced metric on the string,
\begin{equation}
\label{InducedStringMetric}
 ds_{\textrm{st}}^2 = \tilde{L}^2 \Big(  e^{-2A} \ dy_0^2 + \big(  e^{2A} + r'^2  \big) \ dy_1^2  \Big) \ ,
\end{equation}
where $\tilde{L}$ is defined in \eqref{Radii}. It is obvious that the above induced metric is the same for 
all of the geometries that we have discussed in this chapter so far (even for the duals). For this reason we believe that it is not necessary to make any distinction with respect to these geometries for the moment. Moreover, this means that the observable is 'uncharged' under NATD and thus it has the same functional form when computed in the initial and dualized geometries. The interest will be 
in the numerical coefficients our calculation will give.

The Nambu-Goto lagrangian density for the string takes the form,
\begin{equation}
 \mathcal{L} = \frac{1}{2\pi\alpha'} \sqrt{-\det(g_{\textrm{ind}})} = \frac{\tilde{L}^2}{2\pi \alpha'} \ e^{A} \ \sqrt{e^{2A} + r'^2} \ ,
\end{equation}
where $g_{\textrm{ind}}$ stands for the induced metric \eqref{InducedStringMetric}. The conservation of the Hamiltonian implies that,
\begin{equation}
 \frac{e^{3A}}{\sqrt{e^{2A} + r'^2}} = e^{2 A_*} \ ,
\end{equation}
where $A_*$ is the value of the function $A(r)$ at the tip of the hanging string $r=r_*$. We can solve the last equation for $r'$ and use the result to calculate the distance between the endpoints of the string. If we do this we can express $d$ in terms of $r_*$
\begin{equation}
\label{QuarkDistance}
 d = e^{2 A_*} \int\limits_{r_*}^{\infty} dr \ \frac{e^{-A}}{\sqrt{e^{4A} - e^{4A_*}}} \ .
\end{equation}

The Nambu-Goto action now reads
\begin{equation}
\label{NambuGoto}
 S = \frac{T \tilde{L}^2}{\pi \alpha'} \int\limits_{r_*}^{\infty} dr \ \frac{e^{3A}}{\sqrt{e^{4A} - e^{4A_*}}} \ ,
\end{equation}
where $T = \int dt$. The integral in eq. \eqref{NambuGoto} is divergent since we are considering quarks of infinite mass sitting at the endpoints of the string. We can regularize  this integral by subtracting the mass of the two quarks and dividing by $T$ as it is shown below
\begin{equation}
\label{QuarkAntiquarkEnergy}
 \frac{E}{\tilde{L}^2} \alpha' = \frac{1}{\pi} \int\limits_{r_*}^{\infty} dr e^A \ \Big(  \frac{e^{2A}}{\sqrt{e^{4A} - e^{4A_*}}} - 1  \Big) - \frac{1}{\pi} \int\limits_{-\infty}^{r_*} dr \ e^{A} \ .
\end{equation}
This formula gives us the quark-antiquark energy. In order to calculate the same observable starting
with the NATD geometries one must take into account that the $\textrm{AdS}$ radius $L$ is different from that of the original geometries. In fact both results are related in the following way
\begin{equation}
\frac{\widehat{E}}{E} = \frac{\widehat{L}^2}{L^2} \ .
\end{equation}
In the last expression the hats refer to the dual quantities.

At this point we will explore the UV and IR limits of the quark-antiquark energy both for the twisted and the  Donos-Gauntlett geometries.

\subsubsection*{Behavior in the UV}

\underline{Twisted and Donos Gauntlett geometry}

First we focus on the twisted solution where the Riemann surface is the hyperbolic space, i.e. $\Sigma_2 = H_2$. In section \ref{section211xx} we saw that in this case the function $A(r)$ behaves like $A(r) \sim r$. Taking this into account we can compute the distance between the quarks from the formula \eqref{QuarkDistance}. The result of this is
\begin{equation}
 d = \frac{2 \sqrt{2} \ \pi^\frac{3}{2}}{\Gamma \Big(  \frac{1}{4} \Big)^2} e^{-r_*} \ .
\end{equation}
Solving this equation for $r_*$ we can substitute into the result coming from the formula \eqref{QuarkAntiquarkEnergy}. This will give the quark-antiquark energy in terms of $d$ which in our case is
\begin{equation}
\label{QQb1}
 E = - \frac{\tilde{L}^2}{\alpha'} \ \frac{4 \ \pi^2}{ \Gamma  \Big(  \frac{1}{4} \Big)^4} \frac{1}{d} \ ,
\end{equation}
as expected for a CFT. The main point of interest in the previous formula is in the numerical coefficient.


Similar considerations for the case of the Donos-Gauntlett geometry give the same results as in the twisted case above. This is because the asymptotic behavior of the function $A(r)$ in the UV is the same in both cases.

\subsubsection*{Behavior in the IR}

\underline{Twisted geometry}

Again in the IR region we address again the same difficulty that we found in the computation of the entanglement entropy. We use the same trick to overcome it, that is we split the integrations into the intervals $[r_*,a]$ and $[a, +\infty)$ where $a$ has value in the deep IR but always greater than $r_*$. 

In section \ref{section211xx} we saw that in the case where $\Sigma_2 = H_2$, the IR behavior of the function $A(r)$ is $A(r) \sim \frac{3}{2} \ r$. Applying this into the formula \eqref{QuarkDistance} we obtain the following result,
\begin{equation}
 d = \frac{4 \sqrt{2} \ \pi^\frac{3}{2}}{3 \ \Gamma \Big(  \frac{1}{4} \Big)^2} e^{-\frac{3}{2} r_*} \ .
\end{equation}
As before we solve the previous result for $r_*$ and we substitute it into the expression that we find from the calculation of the quark-antiquark potential. This way we express the energy as a function of the distance between the quarks,
\begin{equation}
\label{QQb2}
 E = - \frac{\tilde{L}^2}{\alpha'} \ \frac{16 \ \pi^2}{ 9 \ \Gamma \Big(\frac{1}{4} \Big)^4} \frac{1}{d} \ .
\end{equation}

\noindent \underline{Donos-Gauntlett geometry}

Repeating the same steps for the case of the Donos-Gauntlett geometry we find that the distance between the quarks is,
\begin{equation}
 d = \frac{4 \pi^\frac{3}{2}}{ 3^\frac{3}{2} \Gamma \Big( \frac{1}{4} \Big)^2} e^{-a_0 - \frac{3^{\frac{3}{4}}}{\sqrt{2}} r_*} \ .
\end{equation}
Then, expressing the energy in terms of the distance $d$ we find again a dependence proportional to $\frac{1}{d}$,
\begin{equation}
\label{QQb3}
 E = - \frac{\tilde{L}^2}{\alpha'} \frac{8 \pi^2}{3^{\frac{3}{2}} \Gamma \Big(  \frac{1}{4}  \Big)^4} \frac{1}{d} \ .
\end{equation}

Let us point out that the behavior in eqs \eqref{QQb1}, \eqref{QQb2} and \eqref{QQb3} are just consequences of the fact that far in the UV and far in the IR the QFT is conformal.

\begin{figure}[h]
\begin{subfigure}[h]{0.49\textwidth}
\centering
\includegraphics[width=\textwidth]{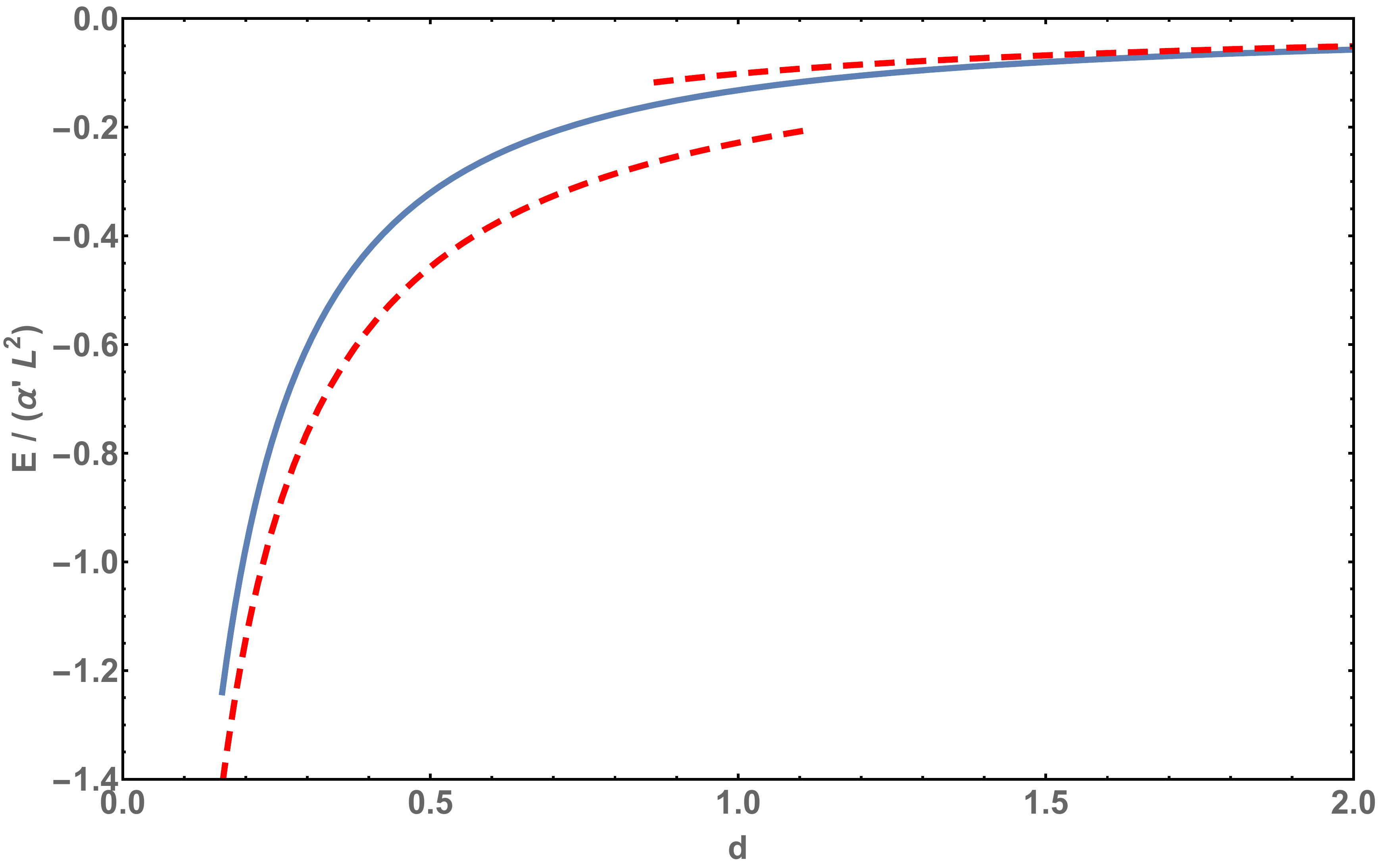}
\end{subfigure}
\begin{subfigure}[h]{0.49\textwidth}
\centering
\includegraphics[width=\textwidth]{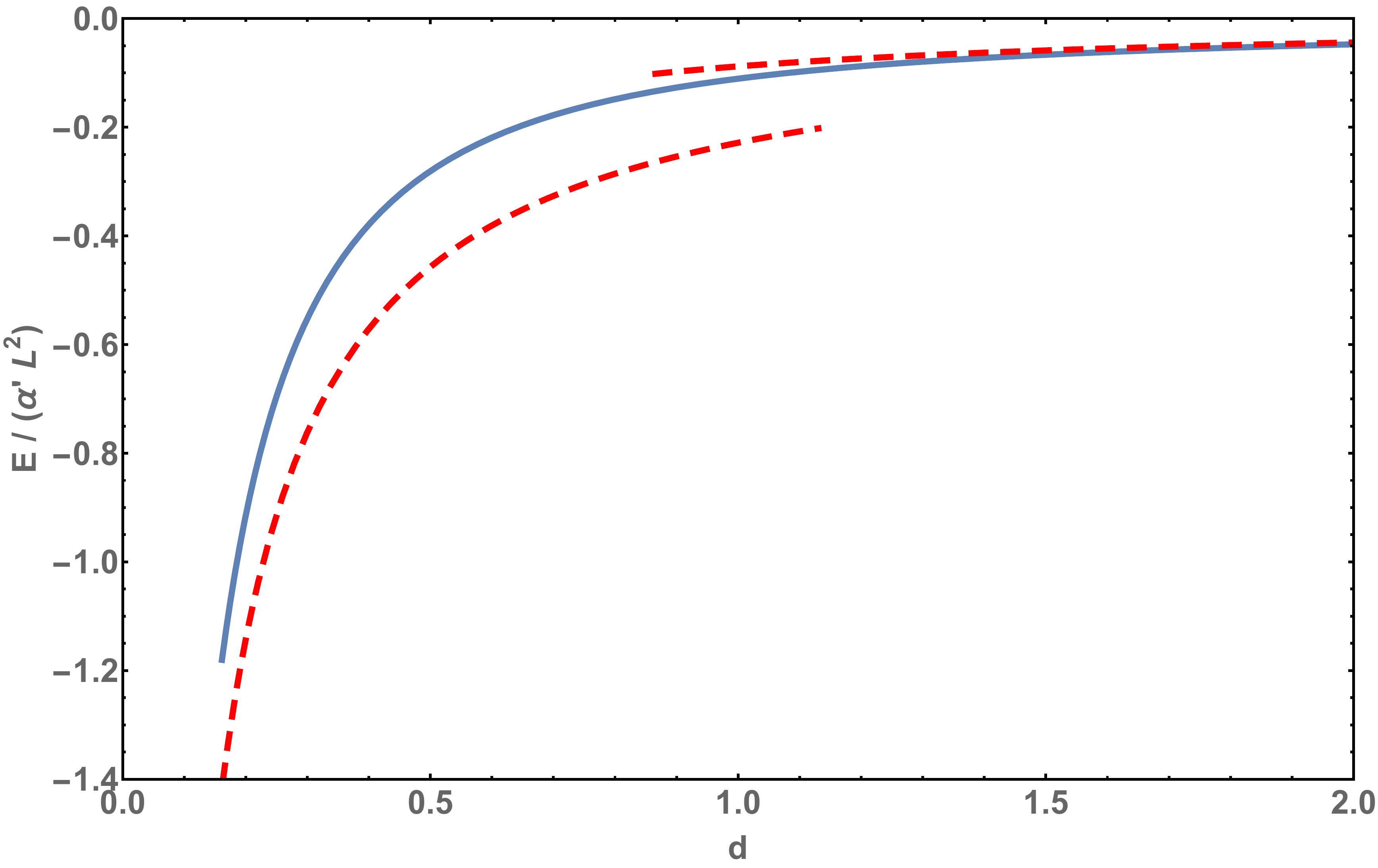}
\end{subfigure}

\caption{The quark-antiquark potential $\frac{E}{\alpha' L^{2}}$ as a function of the distance $d$ in the cases of the twisted $H_2$ (left) and Donos-Gauntlett (right) solutions. The continuous curves correspond to the numerical results and the dashed ones to the UV and IR limits.}
\label{WLnatduality00}
\end{figure}

The results for the quark-antiquark potential are shown in Fig. \ref{WLnatduality00}.

%
%
%
%
%
%
%
%

\section{Discussion}\label{conclusiones}

 We started by studying backgrounds dual to 
two-dimensional SUSY CFTs. The 2-d CFTs were obtained
by compactification of the four-dimensional Klebanov-Witten CFT
on a torus or on a compact hyperbolic plane. The 2-d CFT preserves (0,2)
SUSY.

On those type IIB backgrounds we performed a NATD transformation,
using an $SU(2)$-isometry of the 'internal space' (or, equivalently, 
a global symmetry of the dual CFT). As a result, we constructed 
{\it new, smooth} and SUSY preserving backgrounds in type IIA and 11 supergravity
with an $AdS_3$ fixed point. A further T-duality was used to construct new, smooth and SUSY
type IIB background whose IR is of the form $AdS_3\times M_7$ and all fluxes are
active. 

We analyzed the dual QFT by computing its observables, using
the smooth backgrounds mentioned above. By studying the Page charges, we observed that there is a correspondence between the branes of the starting type IIB solution and those of the type IIA solution after NATD.

The behavior of the central charge in the original CFT 
(the compactification of the Klebanov-Witten theory to 2-d) 
is $c\sim N_{D3}^2$, while
after the NATD goes like $c\sim N_{D6}^2 n^3$. This new (cubic)
dependence suggest a relation with long linear quivers, 
which would imply that $n$ is measuring the number of  D4- and NS5-branes.
The picture that emerges is that of a 2-d CFT living on the intersection
of D2- , D6- and NS5-branes, with induced D4 charge every time 
an NS-brane is crossed. Quantized charges support this interpretation. 

Entanglement entropy and Wilson loops
had the expected universal dependence on $d$ (the quark-antiquark separation or
the length of  strip separating the two regions) at the fixed points. The interest of the expressions
is on the coefficients, not determined by conformal invariance. Interestingly, along the flow the observables smoothly interpolate between the IR and the UV behaviors, which are fixed points of different dimensionality. Both for the entanglement entropy and the Wilson loops we found that the quotient of their values before and after NATD is constant along the flow, as it is expected. \newline

It would be interesting to connect these studies with previous calculations done for 
$AdS_3$, either at the sigma model or the supergravity level. We are presenting new backgrounds, hence new 2-d CFTs on which 
studies done in the past could be interesting to revisit. It would be also interesting to make more precise the QFT dual to these backgrounds. The key point needed for this purpose is to understand the global properties of the supergravity background after the NATD. Besides, it could be interesting to check whether our solutions fall within the existent classifications of, for example \cite{Gauntlett:2006ux},  \cite{Beck:2015hpa}. Probably, for this purpose the global properties of the solution are needed. Another interesting point would be to further study new observables that select (or explore) the values of $\rho=n\pi$
argued in this work to be special values of the $\rho$-coordinate, in the line of  \cite{Lozano:2016kum}\newline


\begin{subappendices}

\section{Appendix}

\setcounter{equation}{0}

\subsection{SUSY analysis}
\label{appendixsusy2}

\subsubsection{SUSY preserved by the twisted solutions}\label{susyvar}
In this appendix we write explicitly the variations of the dilatino and gravitino for the ansatz (2.1-2.4), for the 3 cases $H_2$, $S^2$ and $T^2$.  The SUSY transformations for the dilatino $\lambda$ and the gravitino $\psi_m$ for type IIB supergravity in string frame are \cite{3},
%
\begin{eqnarray}
 & & \delta_{\epsilon}\lambda = \left[ \frac{1}{2}\Gamma^m \partial_m \Phi + \frac{1}{4\cdot3!}H_{mnp}\Gamma^{mnp} \tau_3 - \frac{e^{\Phi} }{2}  F_m \Gamma^m(i\tau_2) -\frac{e^{\Phi}}{4\cdot3!} F_{mnp}\Gamma^{mnp} \tau_1 \right] \epsilon \ , 
\\[5pt]
 & & \delta_{\epsilon}\psi_m =\left[  \nabla_m+ \frac{1}{4\cdot2!}H_{mnp}\Gamma^{np} \tau_3 + \frac{e^{\Phi}}{8} \left( F_n \Gamma^n (i\tau_2)+\frac{1}{3!}F_{npq}\Gamma^{npq} \tau_1 + \frac{1}{2\cdot5!}F_{npqrt}\Gamma^{npqrt} (i\tau_2)  \right) \Gamma_m \right] \epsilon \ ,
 \nonumber
 \label{susy00}
\end{eqnarray}
%
where $\tau_i \ , \; i = 1,2,3$, are the Pauli matrices. Let us consider the $H_2$ case in detail (the $S^2$ case is obtained analogously) . Recall that the vielbein is written in (\ref{vielbein00}).

The dilatino variation vanishes identically, as the fields involved are vanishing. The $m=0$ component of the gravitino reads,
\begin{equation}
\delta_{\epsilon} \psi_0=\left[  \frac{A'}{2L} \Gamma_{04} - \frac{e^{-4U-V}}{2L} \Gamma_{04} \Gamma_{0123} i \tau_2 + \frac{e^{-2B-2U-V}}{16L} z \big(\Gamma_{014} - \Gamma_{239} \big) \big(   \Gamma_{78} - \Gamma_{56} \big) \Gamma_0 i\tau_2 \right] \epsilon \ .
\label{gravitino0}
\end{equation}
First, we use the chiral projection of type IIB,
\begin{equation}
\Gamma_{11}\epsilon=\epsilon \ ,
\label{gravitino00}
\end{equation}
where we define $\Gamma_{11}=\Gamma_{0123456789}$. We also impose the following projections (K\"ahler projections),
\begin{equation}
\Gamma_{56}\epsilon= - \Gamma_{78}\epsilon= - \Gamma_{49}\epsilon \ .
\label{gravitino01}
\end{equation}
Then, expression (\ref{gravitino0}) simplifies to,
\begin{equation}
\delta_{\epsilon} \psi_0= \Gamma_{04}\left[ \frac{A'}{2L}  - \frac{e^{-V-4U}}{2L}+\frac{e^{-2B-2U-V}}{4L}z \Gamma_{0178} i \tau_2 \right]\epsilon \ .
\label{gravitino02}
\end{equation}
We now impose the usual projection for the D3-brane,
\begin{equation}
\Gamma_{0123} \ i \tau_2 \epsilon= \epsilon \ ,
\label{gravitino03}
\end{equation}
and also a further projection related to the presence of the twisting,
\begin{equation}
\Gamma_{23}\epsilon= \Gamma_{78}\epsilon \ .
\label{gravitino04}
\end{equation}
Then, imposing that expression (\ref{gravitino02}) vanishes we obtain,
\begin{equation}
A'-e^{-V-4U}+\frac{z}{2} e^{-2B-2U-V}=0 \ .
\label{gravitino05}
\end{equation}
For the component $m=1$ of the gravitino equation, we obtain that it is zero when we impose the projections  and equation (\ref{gravitino05}). For the component  $m=2$ we have,
\begin{equation}
\delta_{\epsilon} \psi_2=\left[  \frac{B'}{2L} \Gamma_{24} +\frac{e^{-2B+V}}{4L} z \Gamma_{39} - \frac{e^{-V-4U}}{2L} \Gamma_{24}\Gamma_{0123} i \tau_2 - \frac{ e^{-2B-2U-V}}{4L} z \Gamma_{24} \Gamma_{0178} i\tau_2 \right] \epsilon \ .
\label{gravitino06}
\end{equation}
 Combining projections (\ref{gravitino01}) and (\ref{gravitino04}) we get  $\Gamma_{39} \epsilon=-\Gamma_{24} \epsilon$. Then, (\ref{gravitino06}) gives the condition,
 \begin{equation}
B'-e^{-V-4U}- \frac{z}{2}  e^{-2B-2U-V} -\frac{z}{2} e^{-2B+V}=0 \ .
\label{gravitino07}
\end{equation}
For $m=3$, after imposing the projections and equation \eqref{gravitino07} we arrive at,
\begin{equation}
\delta_{\epsilon} \psi_3= - \frac{e^{-B}}{2L} \cot \alpha \ (1+3z) \Gamma_{23} \epsilon \ .
\label{gravitino08}
\end{equation}
There are two contributions to this term, one coming from the curvature of the $H_2$ (through the spin connection) and another coming from the twisting $A_1$. That is, here we explicitly see that the twisting is introduced to compensate the presence of the curvature, in such  a way that some SUSY can be still preserved. Then, we impose,
\begin{equation}
z=-\frac{1}{3} \ .
\label{gravitino09}
\end{equation}
For $m=4$, the variation is,
\begin{equation}
\delta_{\epsilon} \psi_4= \frac{1}{L}\partial_r \epsilon - \frac{1}{2L} \left[ e^{-4U} + 2e^{-2B -2U-V} \right]  \epsilon \ .
\end{equation}
From the condition $\delta_{\epsilon} \psi_4=0$, we obtain a differential equation for $\epsilon$. Solving for it we arrive at the following form for the Killing spinor,
\begin{equation}
\epsilon= e^{1/2 \int (e^{-4U} + 2e^{-2B -2U-V} )  dr} \ \epsilon_0 \ ,
\end{equation}
where $\epsilon_0$ is spinor which is independent of the coordinate $r$. For $m=5,6,7,8$ the variations vanish as long as,
\begin{equation}
U'+e^{-V-4U}-e^{V-2U}=0 \ .
\label{gravitino011}
\end{equation}
 Finally, for $m=9$ the graviton variation vanishes if,
\begin{equation}
V'-3e^{-V}+2e^{V-2U}+e^{-V-4U}-\frac{z}{2} e^{-2B-2U-V} +\frac{z}{2}e^{-2B+V}=0 \ .
\label{gravitino012}
\end{equation}
Summarizing, the variations of the dilatino and gravitino vanish if we impose the following projections on the Killing spinor,
\begin{equation}
\Gamma_{11} \epsilon = \epsilon \ , ~~~~~~ \Gamma_{56} \epsilon = - \Gamma_{78} \epsilon= - \Gamma_{49} \epsilon \ ,  ~~~~~~ \Gamma_{0123} \ i \tau_2 \epsilon = \epsilon \ , ~~~~~~  \Gamma_{23} \epsilon =\Gamma_{49} \epsilon \ ,
\label{gravitino013}
\end{equation}
and the BPS equations (\ref{O17}), together with the condition $z=-1/3$.
%
%
For the case of the 2-torus, if we focus on the $m=3$ component,
\begin{equation}
\delta_{\epsilon} \psi_3= \left[   -\frac{z e^{-2B+V}}{4L}\Gamma_{29}  -\frac{3 z e^{-B}}{2L} \ \alpha \ \Gamma_{78} + \frac{B'}{2L} \Gamma_{34}  - \frac{e^{-V-4U}}{2L} \Gamma_{34} - \frac{z e^{-2B-2U-V}}{4L} \Gamma_{34} \right] \epsilon \ ,
\label{gravitino014}
\end{equation}
we see that there is one term depending on $\alpha$, due to the twisting. Contrary to the $H_2$ and $S^2$ cases, here there is no curvature term that could cancel it. This will force $z=0$, obtaining $A'=B'$, which does not permit an $AdS_3$ solution. 

Finally, after all this analysis we deduce that the Killing spinor does not depend explicitly on the coordinates $(\theta_2,\phi_2,\psi)$ on which we perform the NATD transformation. In fact it only has a dependence on the coordinate $r$.

\subsubsection{SUSY preserved by the NATD solutions}

In the above subsection we calculated the amount of SUSY that is preserved by the type IIB supergravity solutions of the section \ref{section211xx} by examining the dilatino and the gravitino variations. Here we compute the portion of SUSY that is preserved by a supergravity solution after a NATD transformation following the argument of \cite{Sfetsos:2010uq}, which has been proven in \cite{Kelekci:2014ima}. According to this, one just has to check the vanishing of the Lie-Lorentz (or Kosmann) derivative \cite{Kosmann} of the Killing spinor along the Killing vector that generates the isometry of the NATD transformation. More concretely, suppose that we want to transform a supergravity solution by performing a NATD transformation with respect to some isometry of the background that is generated by the Killing vector $k^\mu$. Then there is a simple criterion which states that if the Lie-Lorentz derivative of the Killing spinor along $k^\mu$ vanishes, then the transformed solution preserves the same amount of SUSY as the original solution. In the opposite scenario one has to impose more projection conditions on the Killing spinor in order to make the Lie-Lorentz derivative vanish. Thus in that case the dual background preserves less supersymmetry than the original one.

We recall that given a Killing vector $k^\mu$ the Lie-Lorentz derivative on a spinor $\epsilon$ along $k^\mu$ maps the spinor $\epsilon$ to another spinor and is defined as,
\begin{equation}
\label{KosmannDer}
 \mathcal{L}_{k} \epsilon =  k^\mu D_\mu \epsilon + \frac{1}{4} \big( \nabla_\mu k_\nu \big) \Gamma^{\mu\nu} \epsilon =        k^\m D_\m \epsilon + \frac{1}{8} (dk)_{\m\n} \Gamma^{\m\n} \epsilon \ ,
\end{equation}
where $D_\m \epsilon = \partial_\m \epsilon + \frac{1}{4} \omega_{\m\r\s} \Gamma^{\r\s} \epsilon$. For further details about the Lie-Lorentz derivative we urge the interested reader to consult \cite{Ortin:2002qb}.  

In this chapter we constructed type IIA supergravity solutions by applying a NATD transformation with respect to the $SU(2)$ isometry of the original backgrounds that corresponds to the directions $(\th_2,\phi_2,\psi)$. The non-vanishing components of the associated Killing vectors are,
\begin{equation}
\label{KillingV}
 \begin{array}{lll}
   k_{(1)}^{\th_2} = \sin\phi_2 \ , & k_{(1)}^{\phi_2} = \cot\th_2 \cos\phi_2 \ , & k_{(1)}^{\psi} = - \frac{\cos\phi_2}{\sin\th_2} \ ,
   \\[10pt]
   k_{(2)}^{\th_2} = \cos\phi_2 \ , & k_{(2)}^{\phi_2} = -\cot\th_2 \sin\phi_2 \ , & k_{(2)}^{\psi} = \frac{\sin\phi_2}{\sin\th_2} \ ,
   \\[10pt]
   k_{(3)}^{\phi_2} = 1 \ .  & \textrm{}   & \textrm{}
 \end{array}
\end{equation}
In what follows we will compute the Lie-Lorentz derivative along the three Killing vectors $(k_{(1)},k_{(2)},k_{(3)})$ using the geometries of the sections \ref{section211xx} and \ref{sectionDG}. It turns out that in all cases the Lie-Lorentz derivative vanishes without the requirement of imposing further projections on the Killing spinor. This means that the new solutions that we found using the technique of NATD preserve the same SUSY as the original solutions.
\newline

Let us now compute the Lie-Lorentz derivative along the Killing vector \eqref{KillingV} for the twisted geometries that are described by the formulas \eqref{NN02}-\eqref{vielbein00}. In the previous section, which deals with the supersymmetry of the starting solutions, we mentioned that the Killing spinor does not depend on the isometry coordinates $(\th_2,\phi_2,\psi)$. This means that the first term in \eqref{KosmannDer} reduces to,
\begin{equation}
 k_{(i)}^\m D_\m \epsilon = \frac{1}{4} \ \omega_{\m\r\s} \ k_{(i)}^\m \ \Gamma^{\r\s} \epsilon \ , \quad i=1,2,3 \ .
\end{equation}
Hence for each of the three Killing vectors we find,
\begin{equation}
 \label{KosmannTwisted1}
 \begin{aligned}
  &  k_{(1)}^\m D_\m \epsilon = \frac{z}{12} \ e^{2V-2B} \ \cos\phi_2 \sin\theta_2 \Gamma^{23} \epsilon - \frac{e^{2V-2U}}{6} \ \cos\phi_2 \sin\theta_2 \big(\Gamma^{56} - \Gamma'^{78} \big)  \epsilon
  \\[5pt] 
  & \qquad \quad - \frac{1}{2} \ \cos\phi_2 \sin\theta_2 \Gamma'^{78} \epsilon + \frac{e^{V-U}}{2 \sqrt{6}} \ \big(\cos\th_2 \cos\phi_2 \Gamma'^8 - \sin\phi_2 \Gamma'^7 \big) \Gamma^9 \epsilon 
  \\[5pt]
  & \qquad \quad + \frac{e^U \ U'}{2 \sqrt{6}} \big( \cos\theta_2 \cos\phi_2 \Gamma'^7 + \sin\phi_2 \Gamma'^8 \big) \Gamma^4 \epsilon + \frac{e^V \ V'}{6} \sin\theta_2 \cos\phi_2 \Gamma^{49} \epsilon \ ,
  \\[5pt]
 &  k_{(2)}^\m D_\m \epsilon = -\frac{z}{12} \ e^{2V-2B} \ \sin\phi_2 \sin\theta_2 \Gamma^{23} \epsilon + \frac{e^{2V-2U}}{6} \ \sin\phi_2 \sin\theta_2 \big( \Gamma^{56} - \Gamma'^{78} \big) \epsilon
  \\[5pt] 
  & \qquad \quad + \frac{1}{2} \ \sin\phi_2 \sin\theta_2 \Gamma'^{78} \epsilon - \frac{e^{V-U}}{2 \sqrt{6}} \ \big( \cos\th_2 \sin\phi_2 \Gamma'^8 + \cos\phi_2 \Gamma'^7 \big) \Gamma^9 \epsilon 
  \\[5pt]
  & \qquad \quad - \frac{e^U \ U'}{2 \sqrt{6}} \big( \cos\theta_2 \sin\phi_2 \Gamma'^7 - \cos\phi_2 \Gamma'^8 \big) \Gamma^4 \epsilon - \frac{e^V \ V'}{6} \sin\theta_2 \sin\phi_2 \Gamma^{49} \epsilon \ ,
  \\[5pt]
 &  k_{(3)}^\m D_\m \epsilon = -\frac{z}{12} \ e^{2V-2B} \ \cos\theta_2 \Gamma^{23} \epsilon + \frac{e^{2V-2U}}{6} \ \cos\theta_2 \big( \Gamma^{56} - \Gamma'^{78} \big) \epsilon + \frac{1}{2} \ \cos\theta_2 \Gamma'^{78} \epsilon
  \\[5pt] 
  & \qquad \quad + \frac{e^{V-U}}{2 \sqrt{6}} \ \sin\th_2 \Gamma'^8 \Gamma^9 \epsilon - \frac{e^U \ U'}{2 \sqrt{6}} \sin\theta_2 \Gamma^4 \Gamma'^7 \epsilon  - \frac{e^V \ V'}{6} \cos\theta_2 \Gamma^{49} \epsilon \ .
 \end{aligned} 
\end{equation}
For convenience we have defined the rotated $\Gamma$-matrices,
\begin{equation}
\label{GammaRot}
 \Gamma'^7 = \cos\psi \ \Gamma^7 + \sin\psi \ \Gamma^8 \ , \quad \Gamma'^8 = -\sin\psi \ \Gamma^7 + \cos\psi \ \Gamma^8 \ .
\end{equation}
Let us now compute the 1-forms that are dual to the Killing vectors. What one has to do is to lower the index of the Killing vectors \eqref{KillingV} using the metric \eqref{NN02} which gives the following result,
\begin{equation}
\label{KosmannTwisted2}
 \begin{aligned}
  & k_{(1)} = - \frac{L^2}{3} e^{2V} \sin\theta_2 \cos\phi_2 \ \big(\eta + z A_1\big) + \frac{L^2}{6} e^{2U} \big(  \sin\phi_2 d\theta_2 + \sin\theta_2 \cos\theta_2 \cos\phi_2 d\phi_2  \big) \ ,
  \\[5pt]
  & k_{(2)} = \frac{L^2}{3} e^{2V} \sin\theta_2 \sin\phi_2 \ \big(\eta + z A_1\big) + \frac{L^2}{6} e^{2U} \big(  \cos\phi_2 d\theta_2 - \sin\theta_2 \cos\theta_2 \sin\phi_2 d\phi_2  \big) \ ,
  \\[5pt]
  & k_{(3)} = \frac{L^2}{3} e^{2V} \cos\theta_2 \ \big(\eta + z A_1\big) + \frac{L^2}{6} e^{2U} \sin^2\theta_2 d\phi_2 \ .
 \end{aligned}
\end{equation}
The second term of \eqref{KosmannDer} can be computed by acting with the exterior derivative on the above 1-forms and contracting the result with $\Gamma$-matrices. Notice that in order to compare with \eqref{KosmannTwisted1} one has to express the components of $dk_{(i)}, \,\, i=1,2,3$ using the flat frame \eqref{vielbein00}. Finally for the second term of \eqref{KosmannDer} we find,
\begin{equation}
\label{KosmannCheck}
  \frac{1}{8} (dk_{(i)})_{\m\n} \Gamma^{\m\n} \epsilon = - k_{(i)}^\m D_\m \epsilon, \,\, i=1,2,3,
\end{equation}
which means that the Lie-Lorentz derivative along the Killing vectors $k_{(i)}, \,\, i=1,2,3$ vanishes.
\newline

In the case of the Donos-Gauntlett geometry \eqref{metric-bef} we notice that all the necessary expressions are quite similar to those computed in the previous subsection. This is because the only significant difference between the line element of the twisted geometries and that of the Donos-Gauntlett geometry is just a fiber term. As in the previous case, the Killing spinor does not depend on the isometry coordinates $(\theta_2,\phi_2,\psi)$. This implies that the derivative term $k^\m \partial_\m \epsilon$ in eq. \eqref{KosmannDer} has no contribution to the result. Then the first term of eq. \eqref{KosmannDer}, for each of the three Killing vectors, can be easily obtained from eq. \eqref{KosmannTwisted1} by setting $z=0$. Similarly, if we set $z=0$ into eq. \eqref{KosmannTwisted2} we find the 1-forms $k_{(1)}, k_{(2)}, k_{(3)}$ for the Donos-Gauntlett case. Once we know these 1-forms we can follow the same prescription as in the previous subsection and compute the second term of \eqref{KosmannDer} for each Killing vector. It happens again that this term, when computed for every Killing vector, is related to the first term by a minus sign and thus the Lie-Lorentz derivative vanishes without imposing further projections on the Killing spinor.
\newpage

\end{subappendices}


\chapter{Unquenched massive flavored ABJM}
\label{massiveABJM}

\section{Introduction}

 In section \ref{pureABJM} we introduced the ABJM model, which is an ${\cal N}=6$ supersymmetric $U(N)\times U(N)$  gauge theory with Chern-Simons levels $k$ and $-k$, coupled to matter fields which transform in the bifundamental representations $(N,\bar N)$ and $(\bar N, N)$ of the gauge group. When $N$ and $k$ are large the ABJM theory admits a gravity dual in type IIA supergravity in ten dimensions. The corresponding background is a geometry of the form $AdS_4\times {\mathbb C}{\mathbb P}^3$ with fluxes.

The ABJM model constitutes one of the most important examples of the AdS/CFT correspondence in which the high amount of supersymmetry allows to perform a large amount of checks and computations. Nevertheless, in its original formulation it only has fields in the bifundamental representation of the gauge group, and for condensed matter physics purposes it would be interesting to generalize it to a theory also with fundamental matter, \ie, with fields transforming in the $(N,1)$ and $(1,N)$ representations of the gauge group.  In section \ref{ABJMquenchedflavor} it was explained how to add fundamental matter to ABJM in the quenched approximation. Recall that in the supergravity description these flavors were added by considering D6-branes extended along the $AdS_4$ directions and wrapping a three-dimensional submanifold of ${\mathbb C}{\mathbb P}^3$.

In order to go beyond the quenched approximation, one must be able to solve the supergravity equations of motion  including the backreaction induced by  the source terms generated by  the flavor branes. The sources modify the Bianchi identities satisfied by the forms and the Einstein equations satisfied by the metric.  These equations with sources are, in general, very difficult to solve, since they contain Dirac $\delta$-functions whose support is the worldvolume of the branes. In order to bypass this difficulty we will follow the smearing approach, explained in section \ref{smearingtechnique} . When the branes are introduced in this way there are no $\delta$-function sources  in the equations of motion and  they become more tractable. This approach is valid in the so-called Veneziano limit \cite{Veneziano:1976wm}, in which both $N$ and $N_f$ are large and their ratio $N_f/N$ is fixed. 

A holographic dual to ABJM with unquenched {\it massless} flavors in the Veneziano limit was found in \cite{Conde:2011sw}. In this setup the flavor branes  fill the $AdS_4$ and are  smeared in the internal  ${\mathbb C}{\mathbb P}^3$  space in such a way that ${\cal N}=1$ supersymmetry is preserved. 
Notice, that since the flavor branes are not coincident, the flavor symmetry is $U(1)^{N_f}$ rather than $U(N_f)$. A remarkable feature  of the solution found in \cite{Conde:2011sw} is its simplicity and the fact that the ten-dimensional geometry is of the form $AdS_4\times {\cal M}_6$, where ${\cal M}_6$ is a compact six-dimensional manifold whose metric is a squashed version of the unflavored Fubini-Study metric of ${\mathbb C}{\mathbb P}^3$. The radii and squashing factors of this metric  depend non-linearly on the flavor deformation parameter 
${N_f\over N}\,\lambda$, where $\lambda=N/k$ is the 't Hooft coupling of the theory. Moreover, the dilaton is also constant and, since the metric contains an $AdS_4$ factor, the background is the gravity dual of a three-dimensional conformal field theory with flavor. Actually, it was checked in perturbation theory  in \cite{Bianchi} that the ABJM theory has conformal fixed points even after the addition of flavor. This solution captures rather well many of the effects due to loops of the fundamentals in  several observables \cite{Conde:2011sw}. Its generalization at non-zero temperature in \cite{Jokela:2012dw}
leads to  thermodynamics which passes several non-trivial tests required to a flavored black hole.

Contrary to other backgrounds with unquenched flavors, the supergravity solutions dual to ABJM with smeared sources are free of pathologies, both at the IR and the UV. This fact offers us a unique opportunity to study different flavor effects holographically in a well-controlled setup. In this chapter, we will study such  effects when {\it massive} flavors are considered. The addition of massive flavors breaks conformal invariance explicitly and, therefore, the corresponding dual geometry should not contain an Anti-de Sitter factor anymore.  Actually, for massive flavors the quark mass is an additional parameter at our disposal which we can vary and see what is the effect on the geometry and observables. Indeed, let $m_q$ denote the quark mass.  In the IR limit in which $m_q$ is very large we expect the quarks to be integrated out and their effects to disappear from the different observable quantities. Thus, in the IR limit we expect to find a geometry which reduces to the unflavored ABJM background. On the contrary, when $m_q\to 0$,  we are in the UV regime and we should recover the deformed Anti-de Sitter background of 
\cite{Conde:2011sw}. The important point to stress here is that the quark mass triggers a non-trivial renormalization group flow between two fixed points and that we can vary $m_q$ to enhance or suppress  the effects due to the loops of the fundamentals.

To find the supergravity solutions along the flow, we will adopt an ansatz with brane sources in which the metric and forms are squashed as in \cite{Conde:2011sw}. By imposing the preservation of ${\cal N}=1$ supersymmetry, the different functions of the ansatz must satisfy a system of first-order BPS equations, which reduce to a single second-order master equation. The full background can be reconstructed from the solution to the master equation. 

The flavor branes corresponding to massive flavors do not extend over the full range of the holographic coordinate. Indeed, their tip should lie at a finite  distance (related to the quark mass)  from the IR end of the geometry.  Moreover, in  the asymptotic UV region, the geometry we are looking for should reduce to the one in \cite{Conde:2011sw}, since the quarks should be effectively massless in that region. Therefore, we have to solve the BPS equations without sources at the IR and match this solution with another one in which the D6-brane charge is non-vanishing and such that it reduces to the massless flavored solution of \cite{Conde:2011sw} in the deep UV. Amazingly, we have been able to find an analytic solution in the region without sources which contains a free parameter which can be tuned in such a way that the background reduces to the massless flavored geometry in the asymptotic UV. This semi-analytic solution interpolates between two different  conformal $AdS$ geometries and contains the quark mass and the number of flavors as control parameters.

With the supergravity dual at our disposal, we can study the holographic flow for different observables. The general picture we get from this analysis is the following. Let $l$ be a length scale characterizing the observable. Then, the relevant parameter to explore the flow is the dimensionless quantity $m_q\,l$.  When $m_q\,l$ is very large (small) the observable is dominated by the IR unflavored (UV massless flavored) conformal geometry, whereas for intermediate values of $m_q\,l$ we move away from the fixed points. We will put a special emphasis on the study of the holographic entanglement entropy, following the prescription of \cite{Ryu}. In particular, we study the refined entanglement  entropy for a disk proposed in \cite{Liu:2012eea}, which can be used as a central function for the F-theorem \cite{Ftheorem}.  We check the monotonicity of the refined entropy along the flow (see \cite{Casini:2012ei} for a general proof of this monotonic character in three-dimensional theories).  Other observables we analyze are the Wilson loop and quark-antiquark potential, the two-point functions of high-dimension bulk operators, and the mass spectrum of quark-antiquark bound states. 

The rest of this chapter is divided into two parts. The first part starts in section \ref{review-ABJM} with a brief review of the ABJM solution. In section \ref{squashed_solutions} we introduce the squashed ansatz, write the master equation for the BPS geometries with sources, and classify its solutions according to their UV behavior. In section \ref{unflavored_running} we write the analytic solution of the unflavored system that was mentioned above while, in section \ref{interpolating} we construct solutions which interpolate between an unflavored IR region and a UV domain with D6-brane sources. The backgrounds corresponding to ABJM flavors with a given mass are studied in section \ref{massive_flavor}. 

In the second part of the chapter we study the different observables. In section \ref{Holographic_entanglement_entropy} we analyze the holographic entanglement entropy for a disk. section \ref{Wilson} is devoted to the calculation of the quark-antiquark potential from the Wilson loop. In section \ref{Two-point_section} we study the two-point functions of bulk operators with high mass, while the meson spectrum is obtained in section \ref{mesonsmassive}. Section \ref{conclu} contains a summary of our results and some conclusions. The chapter is completed with several appendices with detailed calculations and extensions of the results of the main text.

\section{Review of the ABJM solution}
\label{review-ABJM}

The ABJM theory was reviewed in detail in section \ref{pureABJM}, and in this section we will fix the notation for the expression of the metric, dilaton and forms in local coordinates. The ten-dimensional metric of the ABJM solution in string frame is given by:
\beq
ds^2\,=\,L_{ABJM}^2\,ds^2_{AdS_4}\,+\,4\,L_{ABJM}^2\,ds^2_{{\mathbb C}{\mathbb P}^3}\,\,,
\label{ABJM-metric3}
\eeq
where $ds^2_{AdS_4}$ and $ds^2_{{\mathbb C}{\mathbb P}^3}$ are respectively
 the  $AdS_4$  and ${\mathbb C}{\mathbb P}^3$ metrics. The former,  in Poincar\'e coordinates,  is given by:
\beq
ds^2_{AdS_4}\,=\,r^2\,d x_{1,2}^2\,+\,{dr^2\over r^2}\,\,,
\label{AdS4metric3}
\eeq
where $d x_{1,2}^2$ is the Minkowski metric in 2+1 dimensions.  In (\ref{ABJM-metric3})  $L_{ABJM}$ is the radius of the $AdS_4$ part of the metric and is given, in string units, by:
\beq
L_{ABJM}^4\,=\,2\pi^2\,{N\over k}\,\,,
\label{ABJM-AdSradius3}
\eeq
where $N$ and $k$  correspond, in the gauge theory dual, to the rank of the gauge groups and the Chern-Simons level, respectively. The ABJM background contains a constant dilaton, which can be written in terms of $N$ and $k$ as:
\beq
e^{\phi_{ABJM}}\,=\,{2L_{ABJM}\over k}\,\,=\,\,2\sqrt{\pi}\,\Big(\,{2N\over k^5}\,\Big)^{{1\over 4}}\,\,.
\label{ABJMdilaton3}
\eeq
Apart from the metric and the dilaton written above, the ABJM solution of type IIA supergravity contains a RR two-form $F_2$ and a RR four-form  $F_4$, whose expressions can be written as:
\beq
F_2\,=\,2k\,J\,\,,\qquad\qquad
F_4\,=\,{3\over 2}\,k\,L_{ABJM}^2\,\Omega_{AdS_4}\,=\,{3\pi\over \sqrt{2}}\,\,
\sqrt{kN}\,\Omega_{AdS_4}\,\,,
\label{F2-F4-ABJM3}
\eeq
where $J$ is the K\"ahler form of ${\mathbb C}{\mathbb P}^3$ and $\Omega_{AdS_4}$ is the volume element of the $AdS_4$ metric (\ref{AdS4metric3}).  It follows from (\ref{F2-F4-ABJM3}) that $F_2$ and $F_4$ are closed forms (\ie, $dF_2=dF_4=0$).

The metric of the ${\mathbb C}{\mathbb P}^3$ manifold in ({\ref{ABJM-metric3}) is the canonical Fubini-Study metric.  Following the approach of \cite{Conde:2011sw}, we will regard  ${\mathbb C}{\mathbb P}^3$ as an ${\mathbb S}^2$-bundle over ${\mathbb S}^4$, where the fibration is constructed by using the self-dual $SU(2)$ instanton on the four-sphere. This representation of  
${\mathbb C}{\mathbb P}^3$ is the one obtained when it is constructed as the twistor space of the four-sphere. As in \cite{Conde:2011sw}}, this ${\mathbb S}^4$-${\mathbb S}^2$ representation will allow us to deform the ABJM background by squashing appropriately the metric and forms, while keeping some amount of supersymmetry. More explicitly, we will write $ds^2_{{\mathbb C}{\mathbb P}^3}$  as:
\beq
ds^2_{{\mathbb C}{\mathbb P}^3}\,=\,{1\over 4}\,\,\Big[\,
ds^2_{{\mathbb S}^4}\,+\,\big(d x^i\,+\, \epsilon^{ijk}\,A^j\,x^k\,\big)^2\,\Big]\,\,,
\label{CP3=S4-S23}
\eeq
where $ds^2_{{\mathbb S}^4}$ is the standard metric for the unit four-sphere, $x^i$ ($i=1,2,3$) are Cartesian coordinates that parameterize  the unit two-sphere ($\sum_i (x^i)^2\,=\,1$) and $A^i$ are the components of the non-abelian one-form connection corresponding to the $SU(2)$ instanton. 
Let us now introduce  a specific system of  coordinates  to represent the metric (\ref{CP3=S4-S23}) and the two-form $F_2$. First of all,  let $\omega^i$ ($i=1,2,3$) be a set of  $SU(2)$ left-invariant one-forms satisfying $d\omega^i={1\over2}\,\epsilon_{ijk}\,\omega^j\wedge\omega^k$. Together with a new coordinate $\xi$, the $\omega^i$'s can be used to parameterize the metric of  the four-sphere ${\mathbb S}^4$ as:
\beq
ds^2_{{\mathbb S}^4}\,=\,
{4\over(1+\xi^2)^2}
\left[d\xi^2+{\xi^2\over4}\left((\omega^1)^2+(\omega^2)^2+(\omega^3)^2
\right)\right]\,\,,
\label{S4metric3}
\eeq
where $0\le \xi<+\infty$ is a non-compact coordinate. The $SU(2)$ instanton one-forms $A^i$ can be written in these coordinates as:
\beq
A^{i}\,=\,-{\xi^2\over 1+\xi^2}\,\,\omega^i\,\,. 
\label{A-instanton3}
\eeq
Let us next parameterize the $x^i$ coordinates of the unit ${\mathbb S}^2$ by two angles $\theta$ and $\varphi$ ($0\le\theta<\pi$, $0\le\varphi<2\pi$),
\beq
x^1\,=\,\sin\theta\,\cos\varphi\,\,,\qquad\qquad
x^2\,=\,\sin\theta\,\sin\varphi\,\,,\qquad\qquad
x^3\,=\,\cos\theta\,\,.
\label{cartesian_S23}
\eeq
Then, it is straightforward to demonstrate that the ${\mathbb S}^2$ part of the Fubini-Study metric can be written as:
\beq
\big(d x^i\,+\, \epsilon^{ijk}\,A^j\,x^k\,\big)^2\,=\,(E^1)^2\,+\,(E^2)^2\,\,,
\eeq
where  $E^1$ and $E^2$ are the following one-forms:
\bear
&&E^1=d\theta+{\xi^2\over1+\xi^2}\left(\sin\varphi\,\omega^1-\cos\varphi\,\omega^2\right)\,,
\rc\rc
&&E^2=\sin\theta\left(d\varphi-{\xi^2\over1+\xi^2}\,\omega^3\right)+{\xi^2\over1+\xi^2}\,
\cos\theta\left(\cos\varphi\,\omega^1+\sin\varphi\,\omega^2\right)\,.
\label{Es3}
\eear
Therefore,  the  $ {\mathbb C}{\mathbb P}^3$ metric can be written in terms of the one-forms defined above as:
\beq
ds^2_{{\mathbb C}\mathbb{P}^3}\,=\,{1\over 4}\,\Big[\,ds^2_{{\mathbb S}^4}\,+\,
(E^1)^2\,+\,(E^2)^2\,\Big]\,\,.
\label{CP3-metric3}
\eeq
We will now write the expression of $F_2$ in such a way that the ${\mathbb S}^4$-${\mathbb S}^2$ split structure is manifest. Accordingly, we define three new one-forms  $S^i$ $(i=1,2,3)$ as:
\bear
&&
S^1=\sin\varphi\,\omega^1-\cos\varphi\,\omega^2\,,\rc\rc
&&
S^2=\sin\theta\,\omega^3-\cos\theta\left(\cos\varphi\,\omega^1+
\sin\varphi\,\omega^2\right)\,,\rc\rc
&&
S^3=-\cos\theta\,\omega^3-\sin\theta\left(\cos\varphi\,\omega^1+
\sin\varphi\,\omega^2\right)\,.
\label{rotomega3}
\eear 
Notice that the $S^i$ are just the  $\omega^i$ rotated by the  two angles $\theta$ and $\varphi$.
In terms of the forms defined in (\ref{rotomega3})
 the line element  of the four-sphere is obtained by substituting $\omega^i\to S^i$ in (\ref{S4metric3}). Let us next define the one-forms ${\cal S}^{\xi}$   and ${\cal S}^{i}$ as:
\beq
{\cal S}^{\xi}\,=\,{2\over 1+\xi^2}\,d\xi\,\,,\qquad\qquad
{\cal S}^{i}\,=\,{\xi\over 1+\xi^2}\,S^i \,\,,\qquad(i=1,2,3)\,\,,
\label{calS3}
\eeq
in terms of which the metric of the four-sphere is  
$ds^2_{{\mathbb S}^4}=({\cal S}^{\xi})^2+\sum_i({\cal S}^{i})^2$.  Moreover, the RR two-form $F_2$  in (\ref{F2-F4-ABJM3}) can be written in terms of the one-forms defined in
(\ref{Es3}) and (\ref{calS3}) as:
\beq
F_2\,=\,{k\over 2}\,\Big(\,E^1\wedge E^2\,-\,\big(
{\cal S}^{\xi}\wedge {\cal S}^{3}\,+\,{\cal S}^1\wedge {\cal S}^{2}\big)\,\Big)\,\,.
\label{F2-ansatz3}
\eeq
The solution of type IIA supergravity reviewed above is a good gravity dual of the $U(N)_k\times U(N)_{-k}$ ABJM field theory when the $AdS$ radius $L_{ABJM}$  is large in string units and when the string coupling constant $e^{\phi}$ is small. From (\ref{ABJM-AdSradius3}) and (\ref{ABJMdilaton3}) it is straightforward to prove that these conditions are satisfied if $k$ and $N$ are in the range  $N^{{1\over 5}}\ll k\ll N$.

\section{Squashed solutions}
\label{squashed_solutions}

Let us consider the deformations of the ABJM background which preserve the ${\mathbb S}^4$-${\mathbb S}^2$ splitting. These deformed backgrounds will solve the equations of motion of type IIA supergravity (with sources) and will preserve at least two supercharges. We will argue below that some of these backgrounds are dual to Chern-Simons matter theories with fundamental massive flavors.

The general ansatz for the ten-dimensional metric of our solutions in string frame  takes the form:
\begin{equation}
d s^2_{10}=h^{-1/2}d x^2_{1,2}+h^{1/2}\left[d r^2+e^{2f}d s^2_{\mathbb{S}^4}+e^{2g}\left(\left(E^1\right)^2+\left(E^2\right)^2\right) \right]\,,
\label{metric_ansatz}
\end{equation}
where the warp factor $h$ and the functions $f$ and  $g$ depend on the holographic coordinate $r$. Notice that $f$ and $g$ determine the sizes of the ${\mathbb S}^4$ and ${\mathbb S}^2$ within the internal manifold.  Actually, their difference $f-g$ determines the squashing of the 
${\mathbb C}{\mathbb P}^3$  and will play an important role in characterizing our solutions. We will measure this squashing by means of the function $q$, defined as:
\beq
q\,\equiv\,e^{2f-2g}\,\,.
\eeq
Clearly, the ABJM solution has $q=1$. A departure from this value would signal a non-trivial deformation of the metric. Similarly, the RR two- and four-forms will be given by:
\begin{align}
F_4&=K\,d^3x\wedge d r\,, \label{eqn:F4}\\
F_2&=\frac{k}{2}\left(E^1\wedge E^2-\eta\left({\cal S}^{\xi}\wedge{\cal S}^3+{\cal S}^1\wedge{\cal S}^2\right)\right)\,,
\label{eqn:F2}
\end{align}
where $k$ is a constant and  $K=K(r)$, $\eta=\eta(r)$ are new functions. The background  is also endowed with  a dilaton $\phi=\phi(r)$. As compared with the ABJM value (\ref{F2-ansatz3}), the expression of $F_2$ in our ansatz contains the function $\eta(r)$ which generically introduces an asymmetry between the ${\mathbb S}^4$ and ${\mathbb S}^2$  terms.  Moreover, when $\eta\not=1$ the two-form $F_2$ is no longer closed and the corresponding Bianchi indentity is violated. Indeed, one can check that:
\bear
&&dF_2\,=\,-{k\over 2}\,\,(\eta-1)\,\,
\Big[\,
E^1\wedge ({\cal S}^{\xi}\wedge {\cal S}^{2}\,-\,{\cal S}^1\wedge {\cal S}^{3}\big)\,+\,
E^2\wedge ({\cal S}^{\xi}\wedge {\cal S}^{1}\,+\,{\cal S}^2\wedge {\cal S}^{3}\big)\,
\Big]\,\,-\rc\rc
&&\qquad\qquad\qquad\qquad
-{k\over 2}\,\,\eta'\,\,dr\wedge 
\Big({\cal S}^{\xi}\wedge {\cal S}^{3}\,+\,{\cal S}^1\wedge {\cal S}^{2}\Big)\,\,.
\label{massive-Omega}
\eear
The violation of the Bianchi identity of $F_2$ means that we have D6-brane sources in our model. 
Indeed, since $F_2=* \,F_8$, if  $dF_2\not= 0$ then  the Maxwell equation of $F_8$ contains a source term, which is due to the presence of D6-branes since the latter are electrically charged with respect to $F_8$. The charge distribution of the D6-brane sources is determined by the function $\eta$, which we  will call the profile function.

The function $K$ of the RR four-form can be related to the other functions of the ansatz by using 
its equation of motion $d*F_4=0$ and  the flux quantization condition for the integral of $*F_4$ over the internal manifold. The result is \cite{Conde:2011sw}:
\beq
K\,=\,3\pi^2\,N\,h^{-2}\,e^{-4f-2g}\,\,,
 \label{K-N}
 \eeq
where the integer $N$ is identified with the ranks of the gauge groups in the gauge theory dual (\ie, with the number of colors).

It is convenient  to introduce a new radial variable $x$, related to $r$ through the differential equation:
\beq
x\,{dr\over dx}\,=\,e^{g}\,\,.
\label{r-x-diff-eq}
\eeq
From now on, all functions of the holographic variable  are considered as functions of $x$, unless otherwise specified. 
The ten-dimensional metric in this new variable takes the form:
\beq
ds^2_{10}\,=\,h^{-{1\over 2}}\,dx^2_{1,2}\,+\,h^{{1\over 2}}\,
\Big[\,e^{2g}\,\,{dx^2\over x^2}\,+\,e^{2f}\,ds_{{\mathbb S}^4}^2\,+\,
e^{2g}\,\Big(\,\big(E^1\big)^2\,+\,\big(E^2\big)^2\Big)\,\Big]\,\,.
\eeq

It was shown in \cite{Conde:2011sw} that the background given by the ansatz written above preserves ${\cal N}=1$ supersymmetry in three dimensions if the functions satisfy a system of first-order differential equations.  It turns out that this BPS system can be reduced to a unique second-order differential equation for a particular combination of the functions of the ansatz. The details of this reduction are given in appendix \ref{BPS}. Here we will just present the final result of this analysis. First of all, let us define the function $W(x)$ as:
\beq
W(x)\,\equiv\,{4\over k}\,h^{{1\over 4}}\,e^{2f-g-\phi}\,x\,\,.
\label{W_definition}
\eeq
Then, the BPS system can be reduced to the following second-order non-linear differential equation for $W(x)$:
\beq
W''\,+\,4\eta'\,+\,(W'+4\eta)\,\Bigg[{W'+10\eta\over 3W}\,-\,
{W'+4\eta+6\over x (W'+4\eta)}\Bigg]\,=\,0\,\,.
\label{master_eq_W}
\eeq
We will refer to (\ref{master_eq_W}) as the master equation and to $W(x)$ as the master function. Interestingly, the BPS equations do not constrain the profile function $\eta$. Therefore,  we can choose $\eta(x)$ (which will fix the type of supersymmetric sources of our system) and afterwards we can solve (\ref{master_eq_W}) for $W(x)$. Given $\eta(x)$ and $W(x)$ one can obtain the other functions that appear in the metric. Indeed, as proved in appendix \ref{BPS}, $g(x)$ and $f(x)$ are given by:
\bear
&&e^{g(x)}\,=\,{x\over W^{{1\over 3}}}\,\exp\Big[{2\over 3}\int^x\,{\eta(\xi)d\xi\over W(\xi)}\Big]
\,\,,\rc\rc
&&e^{f(x)}\,=\,\sqrt{{3x\over W'+4\eta}}\,\,W^{{1\over 6}}\,
\exp\Big[{2\over 3}\int^x\,{\eta(\xi)d\xi\over W(\xi)}\Big]\,\,,
\label{g-f-W}
\eear
while the warp factor $h$ can be written as:
\beq
h(x)\,=\,4\pi^2\,{N\over k}\,e^{-g}\,
(W'+4\eta)\Bigg[\,\int_{x}^{\infty}\,
{\xi\,e^{-3 g(\xi)}\over W(\xi) ^2}
\,d\xi\,+\,\beta\,\Bigg]\,\,
\,\,,
\label{warp_factor_x}
\eeq
where $\beta$ is a constant that determines the behavior of $h$ as $x\to\infty$ ($\beta=0$ if we impose that  $h\to 0$ as $x\to\infty$). Finally,  the dilaton is given by:
\beq
e^{\phi(x)}\,=\,{12\over k}\,
{x \,h^{{1\over 4}}\over W^{{1\over 3}} (W'+4\eta)}\,\exp\Big[{2\over 3}\int^x\,{\eta(\xi)d\xi\over W(\xi)}\Big]\,\,.
\label{dilaton_x}
\eeq
From the expression of $f$ and $g$ in (\ref{g-f-W})
it follows   that the squashing function $q$ can be written in terms of the master function $W$ and its derivative as:
\beq
q\,=\,{3 W\over x(W'+4\eta)}\,\,.
\label{q_W_Wprime}
\eeq

\subsection{Classification of solutions}
Let us study the behavior of the solutions of the master equation in the UV region $x\to\infty$. This analysis will allow us to have a classification of the different solutions. We will assume that the profile function $\eta(x)$ reaches a constant value as $x\to\infty$, and we will denote:
\beq
\lim_{x\to\infty} \eta(x)\,=\,\eta_0\,\,.
\eeq
Let us restrict ourselves to the case in which $\eta_0\not=0$. We will  assume that $W(x)$ behaves for large $x$ as:
\beq
W(x)\,\approx A_0\,x^{\alpha}\,\,,
\qquad\qquad x\to\infty\,\,,
\label{Asym_W}
\eeq
where $A_0$ and $\alpha$ are constants. 
It is easy to check that this type of behavior is consistent only when  the exponent $\alpha\ge 1$ or, in other words, when $W(x)$ grows at least as a linear function of $x$ when  $x\to\infty$. 

We will also characterize the different solutions by the asymptotic value of the squashing function $q$, which determines the deformation of the internal manifold in the UV. Let us denote
\beq
q_0\,=\,\lim_{x\to\infty} q(x)\,\,. 
\eeq
It follows from (\ref{q_W_Wprime}) that the asymptotic value of the squashing function and that of the profile function are closely related.  Actually, this relation depends on whether the exponent $\alpha$ in (\ref{Asym_W}) is strictly greater or equal to one. Indeed, plugging  the asymptotic behavior (\ref{Asym_W}) in (\ref{q_W_Wprime}) one immediately proves that:
\beq
q_0\,=\,
\begin{cases}
{3\over \alpha}\,\,,\qquad\qquad
{\rm for\,\,} \alpha>1 \,\,,\cr\cr
{3A_0\over A_0+4\eta_0}\,\,,\qquad
{\rm for\,\,}\alpha=1\,\,.
\end{cases}
\label{q_0_alpha}
\eeq
This result indicates that we have to study separately the cases $\alpha>1$ and $\alpha=1$. As we show in the next two subsections these two different asymptotics correspond to two qualitatively different types of solutions.

\subsubsection{The asymptotic  $G_2$ cone}
\label{G2-cone}

Let us assume that the master function behaves as in (\ref{Asym_W}) for some $\alpha>1$.  By plugging this asymptotic form in the master equation (\ref{master_eq_W}) and keeping the leading terms  as $x\to\infty$, one readily verifies that the coefficient $A_0$ is  not constrained and that the exponent $\alpha$ takes the value:
\beq
\alpha\,=\,{3\over 2}\,\,.
\eeq
Therefore, it follows from (\ref{q_0_alpha}) that the asymptotic squashing is:
\beq
q_0=2\,\,.
\eeq
Let us evaluate the asymptotic form of all the functions of the metric. From  (\ref{g-f-W}), we get, at leading order:
\beq
e^{g}\approx C\,\sqrt{x}\,\,,
\eeq
where $C$ is a constant of integration. Moreover, since $q_0=2$, the asymptotic value of the function $f$ is:
\beq
e^{f}\approx \sqrt{2}\, C\,\sqrt{x}\,\,.
\eeq
Let us now evaluate the warp factor $h$ from  (\ref{warp_factor_x}). Clearly, we have to compute the integral:
\beq
\int_{x}^{\infty}\,
{\xi\,e^{-3 g(\xi)}\over W(\xi)^2}
\,d\xi\sim x^{-{5\over 2}}\,\,,
\label{asymp_integral_h}
\eeq
which vanishes when $x\to\infty$. Therefore, 
by choosing  the constant $\beta$ in (\ref{warp_factor_x}) to be non-vanishing we can neglect the integral (\ref{asymp_integral_h}) and, since $e^{g}\,W'\to{\rm constant}$, then the warp factor $h$ becomes also a constant when $x\to\infty$. To 
clarify the nature of the asymptotic metric, let us change variables, from $x$ to a new radial variable $\rho$, defined as $\rho=2C\sqrt{x}$. Then, after some constant rescalings of the coordinates the metric becomes:
\beq
ds^2_{10}\approx dx^2_{1,2}\,+\,ds^2_7\,\,,
\eeq
where $ds^2_7$ is:
\beq
ds^2_7= d\rho^2\,+\,{\rho^2\over 4}\,
\Big[2 ds^2_{{\mathbb S}^4}\,+\,
\left(E^1\right)^2+\left(E^2\right)^2\Big]\,\,.
\label{G2_cone_metric}
\eeq
The metric (\ref{G2_cone_metric}) is a Ricci flat cone with $G_2$ holonomy, whose principal orbits at fixed $\rho$ are ${\mathbb C}{\mathbb P}^3$ manifolds with a squashed Einstein metric. In the asymptotic region of large $\rho$ the line element (\ref{G2_cone_metric})  coincides with the metric of the resolved  Ricci flat cone found in \cite{G2cone}, which was constructed from the bundle of  self-dual two-forms over ${\mathbb S}^4$ and is topologically ${\mathbb S}^4\times {\mathbb R}^3$ (see \cite{Atiyah:2001qf} for applications of this manifold to the study of the dynamics of M-theory).

\subsubsection{The asymptotic  $AdS$ metric}
\label{AdS_UV}

Let us now explore the second possibility for the exponent $\alpha$ in (\ref{Asym_W}), namely $\alpha=1$. In this case the coefficient $A_0$ cannot be arbitrary. Indeed, by analyzing the master equation as $x\to\infty$ we find that $A_0$ and $\eta_0$ must be related as:
\beq
A_0^2\,+\,(9-\eta_0)\,A_0\,-\,20\,\eta_0^2\,=\,0\,\,.
\label{A_0_eq}
\eeq
On the other hand, $A_0$ should be related to the asymptotic squashing $q_0$ as in (\ref{q_0_alpha}), which we now write as:
\beq
A_0\,=\,{4q_0\over 3-q_0}\,\,\eta_0\,\,.
\label{A0_q0}
\eeq
By plugging (\ref{A0_q0}) into (\ref{A_0_eq}) we arrive at the following quadratic relation between $q_0$ and $\eta_0$:
\beq
q_0^2\,-3(1+\eta_0)\,q_0\,+\,5\eta_0\,=\,0\,\,.
\label{q0_eta0}
\eeq
Using this equation we can re-express $A_0$ as:
\beq
A_0\,=\,{q_0(\eta_0+q_0)\over 2-q_0}\,\,.
\eeq
Moreover, we can solve (\ref{q0_eta0}) for $q_0$ and obtain the following  two possible  asymptotic squashings in terms of $\eta_0$:
\beq
q_0^{\pm}\,=\,{1\over 2}\,\Big[\,3+3\eta_0\,\mp\,\sqrt{9\eta_0^2-2\eta_0+9}\,\Big]\,\,.
\label{q_0-pm}
\eeq
Thus, there are two possible branches in this case, corresponding to the two signs in (\ref{q_0-pm}). In this chapter we will only consider the $q_0^{+}$ case, since this is the one which has the same asymptotics as the ABJM solution  when there are no D6-brane sources. Indeed, (\ref{q_0-pm}) gives $q_0^{+}=1$ when $\eta_0=1$, which means that the internal manifold in the deep UV is just the un-squashed ${\mathbb C}{\mathbb P}^3$ (when $\eta_0=1$ there are no D6-brane sources in the UV, see (\ref{massive-Omega})).

Let us now study in detail the  asymptotic metric in the UV  corresponding to the $x\to\infty$ squashing  $q_0^{+}$ (which from now on we simply denote as $q_0$). By substituting  $\eta\to \eta_0$ and $W\to A_0\, x$ in (\ref{g-f-W}) and performing the integral, we get:
\beq
e^{g(x)}\,\approx C\,x^{{2\over 3}\,\big(1+{\eta_0\over A_0}\big)}\,\,,
\eeq
where $C$ is a constant of integration. Using (\ref{A0_q0}) this expression can be rewritten as:
\beq
e^{g(x)}\,\approx C\,x^{{1\over b}}\,\,,
\eeq
where  $b$ is given by:
\beq
b\,=\,{2\,q_0\over q_0+1}\,\,.
\label{b_new}
\eeq
The remaining functions of the metric can be found in a similar way. We get for $f$ and $h$ the following asymptotic expressions:
\beq
 e^{f(x)}\,\approx\,C\,\sqrt{q_0}\,x^{{1\over b}}\,\,,
 \qquad\qquad
 h(x)\,\approx\,4\pi^2\,{N\over k}\,{2-b\over 
 C^4\,A_0}\, \,{1\over x^{{4\over b}}}\,\,.
 \eeq
Let us write the above expressions in terms of the original $r$ variable, which can be related to $x$ by integrating the equation:
\beq
x\,{dr\over dx}\,=\,e^{g}\,\approx\,C\,x^{{1\over b}}\,\,.
\eeq
For large $x$ we get:
\beq
r\approx b\,C\,x^{{1\over b}}\,\,,
\eeq
and the functions $g$, $f$, and $h$ can be written  in terms of $r$ as:
\beq
e^{g}\approx {r\over b}\,\,,\qquad\qquad
e^{f}\approx {\sqrt{q_0}\over b}\,r\,\,,\qquad\qquad
h\approx {L_0^4\over r^4}\,\,,
\eeq
where $L_0$ is given by:
\beq
L_0^4\,=\,4\pi^2\,{N\over k}\,{(2-b)\,b^4\over A_0}\,\,.
\eeq
In terms of the asymptotic values $\eta_0$ and $q_0$,  $L_0$ can be written as:
\beq
L_0^4\,=\,128\pi^2\,{N\over k}\,
{(2-q_0)\,q_0^3\over (\eta_0+q_0)\,(q_0+1)^5}\,\,.
\label{L0_explicit}
\eeq
Using these results we find that the asymptotic metric takes the form:
\beq
ds^2\,\approx\,L_0^2\,\,ds^2_{AdS_4}\,+\,
{L_0^2\over b^2}\,\Big[\,q_0\,ds^2_{{\mathbb S}^4}\,+\,
(E^1)^2\,+\,(E^2)^2\,\Big]\,\,,
\label{metric-AdS-asymp}
\eeq
where we have rescaled the Minkowski  coordinates as $x^{\mu}\to L_0^2\,x^{\mu}$. The metric (\ref{metric-AdS-asymp}) corresponds to the product of $AdS_4$  space with radius $L_0$ and a squashed ${\mathbb C}{\mathbb P}^3$. The parameter $b$ will play an important role in the following. Its interpretation is rather clear from  (\ref{metric-AdS-asymp}): it represents the relative squashing of the 
${\mathbb C}{\mathbb P}^3$ part of the asymptotic metric with respect to the
$AdS_4$  part.

It is now straightforward to show that in the UV 
the dilaton reaches a constant value $\phi_0$, related to $q_0$ and $\eta_0$ as:
\beq
e^{\phi_0}\,\approx\,4\sqrt{2\pi}\,\,
\Bigg[\,
{(2-q_0)^{5}\over q_0\,(q_0+1)\,(\eta_0+q_0)^5}
\Bigg]^{{1\over 4}}\,\,
\Bigg({2N\over k^5}\Bigg)^{{1\over 4}}\,\,,
\label{dilaton-AdS_asymp}
\eeq
while the RR four-form approaches the value:
\beq
F_4\,\approx\,12\sqrt{2}\,\pi\,
\Bigg[\,
{q_0^5\,(\eta_0+q_0)\over 
(2-q_0)\,(q_0+1)^7}\Bigg]^{{1\over 2}}
\,\,
\sqrt{k\,N}\,\,\,\Omega_{AdS_4}\,\,,
\label{F4-AdS_asymp}
\eeq
where  $\Omega_{AdS_4}$ is the volume element of $AdS_4$. 

Interestingly, when the profile function $\eta$ is constant and equal to $\eta_0$, the metric, dilaton, and forms written above solve the BPS equations not only in the UV, but also in the full domain of the holographic coordinate. Equivalently, $W=A_0\,x$ is an exact solution to the master equation (\ref{master_eq_W}) if $\eta$ is constant and equal to $\eta_0$ and $A_0$ is given by (\ref{A0_q0}). Actually, when $\eta_0=1$ one can check that $q_0=b=1$ and the asymptotic background becomes the ABJM solution ($W=2x$ for this case). Moreover, when $\eta=\eta_0>1$ the background corresponds\footnote{
Notice that the  expression for $b$  written in (\ref{b_new}) is equivalent to the one obtained in \cite{Conde:2011sw}, namely:
\beq
b\,=\,{q_0(\eta_0+q_0)\over 2(q_0+\eta q_0-\eta_0)}\,\,.
\label{b-squashing}
\nonumber
\eeq
In order to check this equivalence it is convenient to use the following relation between $q_0$ and $\eta_0$:
$
q_0+\eta_0 q_0-\eta_0\,=\,(q_0+1)\,(\eta_0+q_0)/4\,\,.
$\
}  to the one found in \cite{Conde:2011sw} for the ABJM model  with unquenched massless flavors, if one identifies $\eta_0$ with $1+{3N_f\over 4k}$, where $N_f$ is the number of flavors.

The main objective of this chapter is the construction of solutions which interpolate between the $\eta=1$ ABJM background in the IR and the $AdS_4$ asymptotics with $\eta_0>1$ in the UV. Equivalently, we are looking for backgrounds such that the squashing function $q(x)$ varies from the value $q=1$ when $x\to 0$ to $q=q_0>1$ for $x\to\infty$. These backgrounds naturally correspond to gravity duals of Chern-Simons matter models with massive unquenched flavors.
In  the next section we present a one-parameter family of analytic unflavored solutions which coincide with the ABJM background in the deep IR and that have a squashing function $q$ which grows as we move towards the UV. In sections \ref{interpolating} and \ref{massive_flavor} we show that these running solutions can be used to construct the gravity duals to massive flavor that we are looking for.

\section{The unflavored system}
\label{unflavored_running}
In this section we will consider the particular case in which the profile is $\eta=1$. In this case $dF_2=0$ and there are no flavor sources. It turns out that one can find a particular analytic solution of the BPS system written in appendix \ref{BPS}. This solution was found in \cite{Conde:2011sw} in a power series expansion around the IR. Amazingly, this series can be summed exactly and a closed analytic form can be written for all functions. Let us first write them in the coordinate $r$. The functions $f$ and $g$ are given by:
\beq
e^{f}\,=\,r\,\sqrt{{1+c\,r\over 1+ 2c\,r}}\,\,,
\qquad\qquad
e^{g}\,=\,r\,{1+c\,r\over 1+ 2c\,r}\,\,,
\label{running_unflavor_in_r}
\eeq
where $c$ is a constant. For $c=0$ this solution is ABJM without flavor (\ie, $AdS_4\times {\mathbb C} {\mathbb P}^3$ with fluxes), while for $c\not=0$ it is a running background which reduces to ABJM in the IR, $r\to0$. The squashing function $q$ can be immediately obtained from  (\ref{running_unflavor_in_r}):
\beq
q\,=\,{1+2c\,r\over 1+ c\,r}\,\,.
\eeq
For $c\not=0$ the squashing function $q$ interpolates between the ABJM value $q=1$ in the IR and the UV value:
\beq 
q_0\,=\,2\,\,.
\eeq
The warp factor for this solution is:
\bear
&&h(r)\,=\,{2\pi^2\,N\over k}\,\,
{(1+ 2c\,r)^2\over r^4\,(1+c\,r)^2}\,\Bigg[1+2c\,r\Big(3cr(1+2c\,r)\,-\,1\Big)\,+\,\rc\rc
&&\qquad\qquad\qquad\qquad
+\,12 c^3\,(1+ c\,r)\,r^3\,\Big(
\log\Big[{cr\over 1+cr}\Big]\,+\,\alpha\Big)\,\Bigg]\,\,,
\eear
where $\alpha$ is a constant which has to be fixed by adjusting the behavior of the metric in the UV. Finally,  the dilaton can be related to the warp factor as:
\beq
e^{\phi}\,=\,{2\over k}\,{1+c\,r\over (1+ 2c\,r)^2}\,r\,h^{{1\over 4}}\,
\,.
\eeq

Let us now re-express this running analytic solution in terms of the variable $x$, related to $r$ by (\ref{r-x-diff-eq}), which in the present case  becomes:
\beq
{1\over x}\,{dx\over dr}\,=\,{1+2c\,r\over (1+ c\,r)r}\,\,.
\eeq
This equation can be easily integrated:
\beq
\gamma\,x\,=\,c\,r(1+c\,r)\,\,,
\label{x-r}
\eeq
where $\gamma$ is a constant of integration which parameterizes the freedom from passing to  the $x$ variable. By solving (\ref{x-r}) for $r$ we get:
\beq
r\,=\,{1\over  2c}\,\Big[\sqrt{1+4\,\gamma\, x}-1\,\Big]\,\,.
\label{rasfunctionofx}
\eeq
It is straightforward to write the functions $f$ and $g$ in terms of $x$:
\bear
&&e^{f}\,=\, {\gamma\over c}\,x\,\Big[{2\over \sqrt{1+4\,\gamma\, x}\,
(\sqrt{1+4\,\gamma\, x}+1)}\Big]^{{1\over 2}}\,\,,\rc\rc
&&e^{g}\,=\,  {\gamma\over c}\,{x\over \sqrt{1+4\,\gamma\, x}}\,\,,
\label{functions_running_x}
\eear
while the squashing function is:
\beq
q\,=\,2\,{\sqrt{1+4\,\gamma\, x}\over 1+\sqrt{1+4\,\gamma\, x}}\,\,.
\eeq

The warp factor $h$  in terms of the $x$ variable is:
\bear
&&h\,=\,{8\pi^2 N c^4\over k}\,\Big(1+{1\over 4\gamma \,x}\Big)\,
\Bigg[\Big({1\over 2}+ 6\gamma\,x\,+\,{1+
(1-6\gamma\,x)\sqrt{1+4\,\gamma\, x}\over 4\gamma\,x}\Big)
{\sqrt{1+4\,\gamma\, x}+1\over \gamma^2\,x^2}\,+\,\rc\rc
&&\qquad\qquad\qquad\qquad
+24\,\log\Big[{\sqrt{4\,\gamma\, x}\over \sqrt{1+4\,\gamma\, x}+1}\Big]\,+\,\alpha
\Bigg]\,\,.
\label{h_running-x}
\eear
By choosing appropriately the constant $\alpha$ in (\ref{h_running-x}) this running solution behaves as the $G_2$-cone in the UV region $x\to\infty$. 
The dilaton  as a function of $x$ is:
\beq
e^{\phi}\,=\,{2\over k}\,{\gamma\over c}\,{x\over 1+4\,\gamma\, x}\,h^{{1\over 4}}\,\,.
\eeq

Working in the variable $x$, it is very interesting to find the function $W(x)$. For the solution described above, $W$ can be found  by plugging the different  functions   in the definition (\ref{W_definition}). We find:
\beq
W(x)\,=\,{4\,(1+4\,\gamma\, x)\,x\over 1\,+\,\sqrt{1+4\,\gamma\, x}}\,\,.
\label{W_running_unflavored}
\eeq
One can readily check that the function written in (\ref{W_running_unflavored}) solves the master equation (\ref{master_eq_W}) for $\eta=1$.  For large $x$, the function $W(x)$ behaves as:
\beq
W\sim 8\sqrt{\gamma}\,\,x^{{3\over 2}}\,\,,
\eeq
which corresponds to an exponent $\alpha=3/2$ in (\ref{Asym_W}). This is consistent with the asymptotic value $q_0=2$ of the squashing found above. 

Let us finally point out that we have checked explicitly that the geometry discussed in this section is free of curvature singularities. 

\section{Interpolating solutions}
\label{interpolating}

Let us now construct solutions to the BPS equations which interpolate between an IR region in which there are no D6-brane sources (\ie, with $\eta=1$) and a UV region in which $\eta>1$ and, therefore, the Bianchi identity of $F_2$ is violated.  In the $r$ variable the profile $\eta(r)$ will be such that $\eta(r)=1$ for $r\le r_q$, while $\eta(r)>1$ for $r>r_q$. In the region $r\le r_q$ our interpolating solutions will reduce to  the unflavored running solution of section \ref{unflavored_running} for some value of the constant $c$. In order to match this solution with the one in the region $r>r_q$ it is convenient to work in the $x$ coordinate (\ref{x-r}). The point $r=r_q$ will correspond to some $x=x_q$. Notice, however, that we have some freedom  in performing the $r\to x$ change of variables. This freedom is parameterized by the constant $\gamma$ in (\ref{x-r}). We will fix this freedom by requiring that $x_q=1$, \ie, that the transition between the unflavored and flavored region takes place at the point $x=1$. Then,  (\ref{x-r}) immediately implies that $\gamma$ is given in terms of $c$ and $r_q$:
\beq
\gamma\,=\,c\,r_q\,(1+c\,r_q)\,\,.
\label{gamma_rq}
\eeq
We will use (\ref{gamma_rq}) to eliminate the constant $c$ in favor of $r_q$ and $\gamma$. Actually, if we define $\hat\gamma$ as:
\beq
\hat\gamma\,\equiv\,\sqrt{1+4\gamma}\,\,,
\label{hat_gamma}
\eeq
then $c$ is given by
\beq
c\,=\,{\hat\gamma-1\over 2\,r_q}\,\,.
\label{c_gamma}
\eeq
In this running solution the squashing factor $q$ is equal to one in the deep IR at 
$x=0$. When $x>0$ the function $q(x)$ grows monotonically until it reaches a certain value $\hat q$ at $x=1$, which is related to the parameter $\hat\gamma$ as:
\beq
\hat \gamma\,=\,{\hat q\over 2-\hat q}\,\,.
\eeq
In the region $x\ge 1$ we have to solve the master equation (\ref{master_eq_W}) with $\eta(x)>1$ and initial conditions given by the values of $W$ and $W'$ attained by the unflavored running solution at $x=1$. These values depend on the parameter $\hat\gamma$. They can be straightforwardly found by taking $x=1$ in the function (\ref{W_running_unflavored}) and in its derivative. We find:
\beq
W(x=1)\,=\,{4\,\hat\gamma^2\over 1+\hat \gamma}\,\,,
\qquad\qquad
W'(x=1)\,=\,6\,\hat\gamma-4\,\,.
\label{initial_W_Wprime}
\eeq
Let us now write the different functions of these interpolating solutions in the two regions 
$x\le 1$ and $x\ge 1$.  For $x\le 1$ we have to rewrite (\ref{functions_running_x}) after eliminating the constant $c$ by using (\ref{c_gamma}) (which implies that $\gamma/c=(\hat\gamma+1)\,r_q/2$). For the functions $f$, $g$, and the dilaton $\phi$ we get:
\bear
&&e^{f}\,=\,r_q \,{\hat \gamma+1\over \sqrt{2}}\,
 {x\over\Big[ \sqrt{1+4\,\gamma\, x}\,
(\sqrt{1+4\,\gamma\, x}+1)\Big]^{{1\over 2}}}
\,\,,\rc\rc
&&e^{g}\,=\,  r_q\,{\hat \gamma+1\over 2}\,{x\over \sqrt{1+4\,\gamma\, x}}\,\,,
\qquad\qquad\qquad\qquad (x\le 1)\,\,,\rc\rc
&&e^{\phi}\,=\, r_q\,{\hat \gamma+1\over k}\,{x\over 1+4\,\gamma\, x}\,
h^{{1\over 4}}\,\,,
\label{background_functions_xle1}
\eear
where $h$ is the function written in (\ref{h_running-x}) for $c=(\hat\gamma-1)/(2\,r_q)$. 
By using the general equations of section \ref{squashed_solutions}, the solution  for $x\ge 1$ can be written in terms of $W(x)$, which can be obtained by numerical integration of the master equation with initial conditions (\ref{initial_W_Wprime}). 
This defines a solution in the full range of $x$ for every $\gamma$ and $r_q$. 
Notice that (\ref{g-f-W}), (\ref{warp_factor_x}), and (\ref{dilaton_x}) contain arbitrary multiplicative constants, which we will fix by imposing continuity of $f$, $g$,  and $h$ at $x=1$. We get for $f$, $g$ and $\phi$:
\bear
&&e^{f}\,=\,r_q\,\Big[{(\hat \gamma+1)^2\over 2\,\hat \gamma}\Big]^{{1\over 3}}\,\,
\sqrt{{3x\over W'+4\eta}}\,\,W^{{1\over 6}}\,
\exp\Big[{2\over 3}\int^x_1\,{\eta(\xi)d\xi\over W(\xi)}\Big]
\,\,,\rc\rc
&&e^{g}\,=\, r_q\,\Big[{(\hat \gamma+1)^2\over 2\,\hat \gamma}\Big]^{{1\over 3}}\,\,
{x\over W^{{1\over 3}}}\,\exp\Big[{2\over 3}\int^x_1\,{\eta(\xi)d\xi\over W(\xi)}\Big]\,\,,
\qquad\qquad\qquad\qquad (x\ge 1)\,\,,\rc\rc
&&e^{\phi}\,=\, 
r_q\,\Big[{(\hat \gamma+1)^2\over 2\,\hat \gamma}\Big]^{{1\over 3}}\,\,{12\over k}\,
{x\,h^{{1\over 4}}\over W^{{1\over 3}} (W'+4\eta)}\,\exp\Big[{2\over 3}\int^x_1\,{\eta(\xi)d\xi\over W(\xi)}\Big]\,
\,\,.
\label{background_functions_xge1}
\eear
The warp factor $h(x)$ for $x\ge 1$ is given by (\ref{warp_factor_x}), where the integration constant $\beta$  is related to the constant $\alpha$ of (\ref{h_running-x}) by the following matching condition at $x=1$:
\beq
\lim_{x\to 1^-}\,h(x)\,=\,\lim_{x\to 1^+}\,h(x)\,\,.
\label{matching_h}
\eeq
For a given profile function $\eta(x)$, the solution described above depends on the parameter $\gamma$, which determines $W(x)$ for $x\le 1$ through (\ref{W_running_unflavored}) and sets the initial  conditions (\ref{initial_W_Wprime}) needed to integrate the master equation in the 
$x\ge 1$ region. The solution $W(x)$ obtained numerically in this way grows generically as $x^{3/2}$ for large $x$ which, according to our analysis in section \ref{G2-cone}, gives rise to the geometry of the $G_2$-cone in the UV. We are, actually, interested in obtaining solutions with the $AdS$ asymptotics discussed in section \ref{AdS_UV}, for a set of profiles that correspond to flavor D6-branes with a non-zero quark mass. In order to get these geometries we have to fine-tune the parameter $\gamma$ to some precise value which depends on the number of flavors. This analysis is presented in the next section.

\section{Massive flavor}
\label{massive_flavor}

We now apply the formalism developed so  far to find supergravity backgrounds representing massive flavors in ABJM. These solutions will depend on a deformation parameter $\hat \epsilon$, related to the total number of flavors $N_f$ and the Chern-Simons level $k$ as:
\beq
\hat\epsilon\,\equiv\,{3N_f\over 4k}\,\,,
\label{epsilonhat0}
\eeq
where the factor $3/4$ is introduced for convenience and $N_f/k$ is just ${N_f\over N}\,\lambda$ with $\lambda=N/k$ being the 't Hooft coupling. The profile $\eta$,  which corresponds to a set of smeared flavor D6-branes ending at $r=r_q$,  has been found in \cite{Conde:2011sw}. The main technique employed in \cite{Conde:2011sw} was the comparison between the smeared brane action for the distribution of flavor branes and the action for a fiducial embedding in a background of the type studied here.  This fiducial embedding was determined by using kappa symmetry. With our present conventions,\footnote{ The variable $\rho$ used in section 8 of 
\cite{Conde:2011sw} is related to $x$ by $x=e^{\rho}$.}  assuming as above that $r=r_q$ corresponds to $x=x_q=1$, the function $\eta(x)$ is given by:
\beq
\eta(x)\,=\,1\,+\,\hat \epsilon\,\Big(1-{1\over x^2}\Big)\,\Theta(x-1)\,\,,
\label{eta-flavor}
\eeq
where $\Theta(x)$ is the Heaviside step function. 
It follows from (\ref{eta-flavor}) that the asymptotic value $\eta_0$ of the profile is:
\beq
\eta_0\,=\,1+\hat \epsilon\,\,.
\label{eta_0-epsilon}
\eeq
We want to find interpolating solutions of the type studied in section \ref{interpolating} which have the AdS asymptotic behavior in the UV corresponding to the value of $\eta_0$ written in (\ref{eta_0-epsilon}). These solutions have an asymptotic squashing (corresponding to $q_0^{+}$ in (\ref{q_0-pm}))  given in terms of $\hat\epsilon$ as:
\beq
q_0\,=\,3\,+\,{3\over 2}\hat \epsilon\,-\,2\,
\sqrt{1\,+\,\hat\epsilon\,+\,{9\over 16}\,\hat\epsilon^{\,2}}\,\,.
\label{q0-epsilon}
\eeq
It follows from (\ref{q0-epsilon}) that the asymptotic squashing $q_0$ grows with the deformation parameter $\hat\epsilon$. Indeed, $q_0=1$ for $\hat\epsilon=0$, whereas for $\hat\epsilon\to\infty$ the squashing reaches its maximum  value: $q_0\to 5/3$. By using the relation between $b$ and $q_0$  (eq. (\ref{b_new})) we also conclude that $b\to 5/4$  when $\hat\epsilon\to\infty$. 
\begin{figure}[ht]
\center
\includegraphics[width=0.7\textwidth]{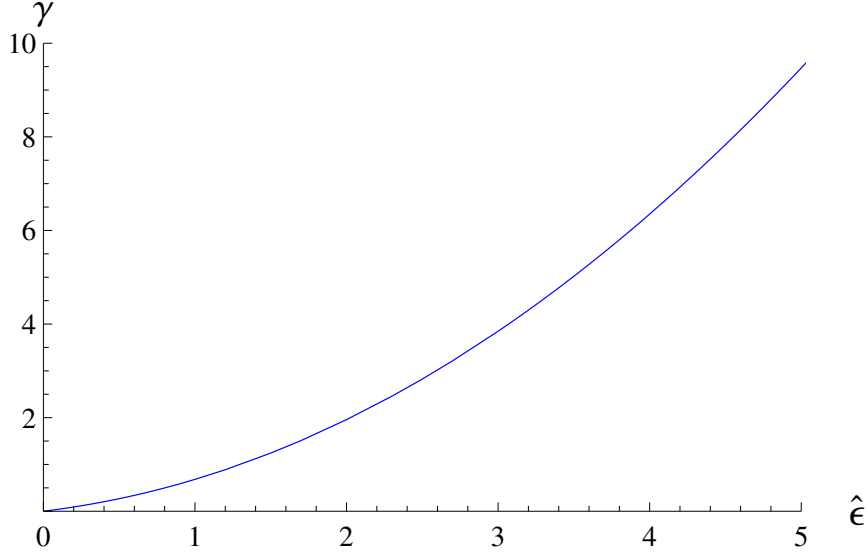}
\caption{Representation of $\gamma(\hat\epsilon)$.} 
\label{fig:gamma}
\end{figure}

\begin{figure}[ht]
\begin{center}
\includegraphics[width=0.44\textwidth]{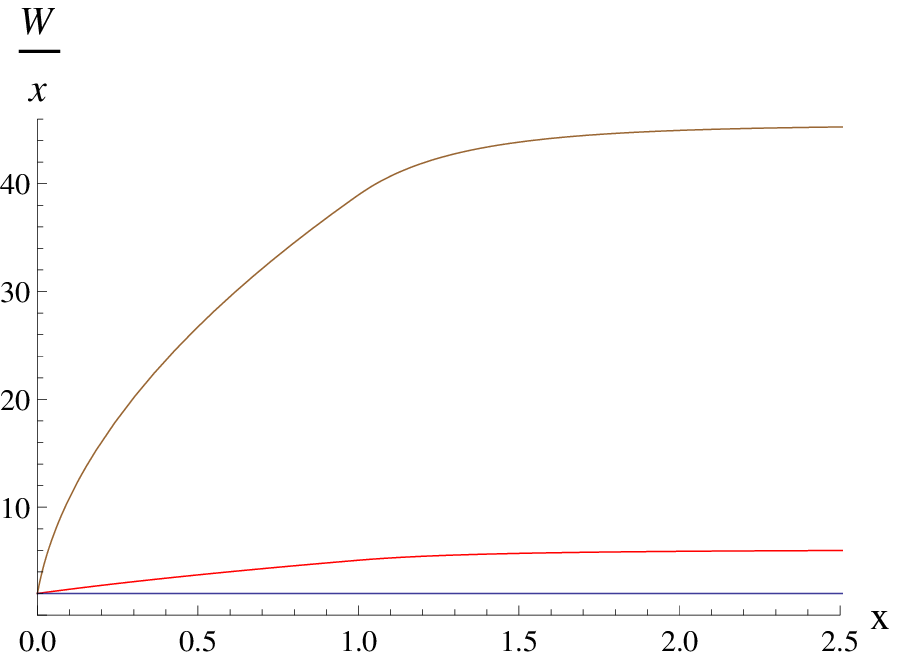}
\qquad\qquad
\includegraphics[width=0.44\textwidth]{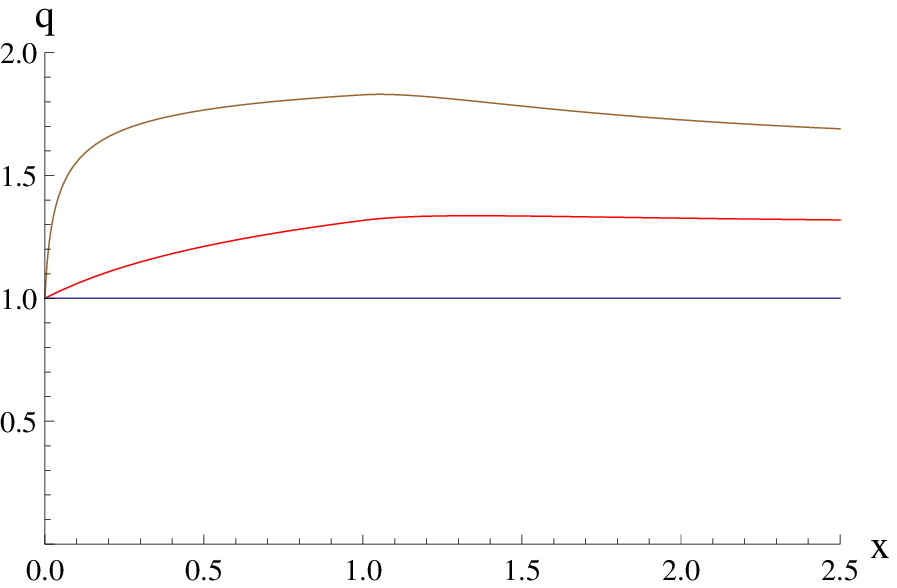}
\end{center}
\caption{Representation of $W(x)/x$ and $q(x)$ for different numbers of  flavors ($\hat\epsilon=0$ (bottom, blue), $\hat\epsilon=1$ (middle, red), and $\hat\epsilon=9$ (top, brown)). The plots on the left show that  $W(x)/x$ becomes constant as we approach the IR and UV conformal points at $x=0$ and $x=\infty$, respectively. Similarly, the squashing function $q(x)$ interpolates between $q=1$ at $x=0$ and $q=q_0$ at large $x$.} 
\label{fig:W_and_q}
\end{figure}

To find the solution for $x\ge 1$ we have to solve numerically the BPS system in this region.  The most efficient way to proceed is by looking at the master equation  for $W(x)$ with the initial conditions (\ref{initial_W_Wprime}). For a generic value of $\gamma$ the numerical solution either gives  rise to negative values of $W(x)$ (which is unphysical  for $k>0$, see  the definition (\ref{W_definition})) or behaves in the UV as $W(x)\sim x^{{3\over 2}}$, which corresponds to the $G_2$-cone asymptotics with $q_0=2$ discussed in section \ref{G2-cone}. Only when $\gamma$ is fine-tuned to some particular value (which depends on $\hat \epsilon$) we get in the UV that $W(x)\sim x$  and that $q_0$ is given by (\ref{q0-epsilon}).  To determine this critical value of $\gamma$ we have to perform a numerical shooting for every value of $\hat\epsilon$. In what follows we understand that $\gamma=\gamma(\hat\epsilon)$ is the function of the deformation parameter which results of this shooting. The  function $\gamma(\hat\epsilon)$ is plotted in Fig. \ref{fig:gamma}, where we notice that $\gamma(\hat\epsilon=0)=0$ and, therefore, we recover the unflavored ABJM background when the deformation parameter vanishes. In the opposite limit $\hat\epsilon\to\infty$, the function  $\gamma(\hat\epsilon)$ grows as 
$\hat\epsilon^2$. Actually,  $\gamma(\hat\epsilon)$ can be accurately represented by a function of the type:
\beq
 \gamma(\hat\epsilon)\,=\,\gamma_1\,\hat\epsilon\,+\,\gamma_2\,\hat\epsilon^{\,2}\,\,,
 \eeq
with $\gamma_1\,=\,0.351$ and $\gamma_2\,=\,0.309$. In Fig. \ref{fig:W_and_q} we plot  the function $W(x)$ and  the squashing function $q(x)$ for some selected values of $\hat\epsilon$.

Interestingly, we have found an analytical solution for small flavor parameter $\hat{\epsilon}$. We present the analytical solution up to order $\hat{\epsilon}^2$ in appendix \ref{Analytical_expansion}. Along this chapter, we will present the results using the numerical solution, which is valid for any value of $\hat{\epsilon}$, but a similar analysis could be done analytically for small flavor.

From the function $W(x)$ we can obtain $f$ and $g$ by performing the integrals in (\ref{background_functions_xge1}). The whole metric is determined if $h$ is known. We will compute $h$ from (\ref{warp_factor_x}) with $\beta=0$, which corresponds to requiring that $h\to 0$ in the UV. In the $x\le 1$ region the warp factor $h(x)$ is given by (\ref{h_running-x}), with the constant $\alpha$ determined by the matching condition (\ref{matching_h}).  The limit on the left-hand side of (\ref{matching_h}) can be determined  explicitly from (\ref{h_running-x}):
\bear
\lim_{x\to 1^-}h(x)=
{\pi^2\over 8 \,r_q^4}\,{N\over k}\,
{\big(\hat\gamma-1\big)^4\,(1+4\gamma)\over \gamma}
\Bigg[\Big({1\over 2}+6\gamma+{1+(1-6\gamma)\hat\gamma\over 4\gamma}\,\Big)
{\hat\gamma+1\over \gamma^2}+24\log {\sqrt{4\gamma}\over\hat\gamma+1}+
\alpha\,\Bigg],\rc
\eear
whereas $h(x\to 1^+)$ is given by:
\beq
\lim_{x\to 1^+}\,h(x)\,=\,
{48\,\pi^2\over r_q}\,{N\over k}\,
{\hat\gamma^2\over \hat\gamma+1}\,\,
\int_{1}^{\infty}\,{\xi\,e^{-3g(\xi)}\over  \big[ W(\xi)\big]^2}\,d\xi\,\,.
\eeq

Notice that $r_q$ (the value of the  $r$ coordinate at  the tip of the flavor branes) appears as a free parameter in eqs. (\ref{background_functions_xle1}) and (\ref{background_functions_xge1}). Actually, $r_q$ can be easily related to the mass $m_q$ of the quarks which deform the geometry. Indeed, by computing the Nambu-Goto action of a fundamental string stretching along the holographic direction between $r=0$ and $r=r_q$ at fixed value of all the other spacelike coordinates in the geometry (\ref{metric_ansatz}), we get that $m_q$ and $r_q$ are linearly related as:
\beq
m_q\,=\,{r_q\over 2\pi\alpha'}\,\,,
\label{rq_mq}
\eeq
where $\alpha'$ is the Regge slope (which we will take to be equal to one in most of our equations). Equivalently, we can relate $m_q$ to the constant $c$ appearing in the solution in the $x\le 1$ region:
\beq
m_q\,=\,{\sqrt{1+4\gamma(\hat\epsilon)}-1\over 4\pi}\,\,{1\over c}\,\,,
\label{mq-gamma-c}
\eeq
where $\gamma(\hat\epsilon)$ is the function obtained by the shooting and only depends on the deformation parameter $\hat\epsilon$.  

We have computed the curvature invariants for the flavored metric and we have checked that the geometry is regular both in the IR ($x\to 0$) and UV ($x\to\infty$). However, the curvature has a finite discontinuity at $x=1$, as can be directly concluded by inspecting Einstein's equations (see appendix \ref{BPS}). This ``threshold" singularity occurs at the point where the sources are added and could be avoided by smoothing the introduction of brane sources with an additional smearing (see the last article in \cite{CNP} for a similar analysis in other background). 

\subsection{UV asymptotics}
\label{UV_mass_corrections}

The full background in the $x\ge 1$ region must be found by numerical integration and shooting, as described above. However, in the UV region $x\to\infty$ one can solve the master equation (\ref{master_eq_W}) in power series for large $x$. Indeed, one can find 
a solution where $W(x)$ is represented as:
\beq
W(x)\,=\,x\,\,\sum_{i=0}^{\infty}\,{A_{2i}\over x^{2i}}\,\,,
\label{W_asymp_UV}
\eeq
where the coefficients $A_{2i}$ can be obtained recursively. The coefficient $A_0$ of the leading term was written in (\ref{A0_q0}). The next two coefficients are:
\beq
A_2\,=\,-{40\,\eta_0\,-\,11\,A_0\over 9+13\,\eta_0\,-\,2\,A_0}\,\hat\epsilon
\,\,,\qquad\qquad
A_4\,=\,{5A_2^2\,+\,20\,\hat\epsilon^2\,+\,25\,\hat\epsilon\,A_2\over 
9(1+3\eta_0-2A_0)}\,\,.
\eeq
Notice that a linear behavior of $W(x)$ with $x$ corresponds to a conformal $AdS$ background,  whereas the deviations from conformality are encoded in the non-linear corrections. 

From the result written above one can immediately  obtain the asymptotic behavior of the squashing function for large $x$. Indeed, let us use in the expression  of $q$ in terms of $W$ and $W'$  (eq. (\ref{q_W_Wprime})) the following large $x$ expansion: 
\beq
x\,(\,W'+4\eta\,)\,=\,(A_0+4\eta_0)\,x\,-\,{A_2+4(\eta_0-1)\over x}\,+\,\cdots\,\,.
\label{Wprime-eta-expansion}
\eeq
We get:
\beq
q(x)\,=\,q_0\,+\,{q_2\over x^2}\,+\,\cdots\,\,,
\eeq
where $q_0$ is the asymptotic value of the squashing (see (\ref{q0-epsilon}))  and $q_2$ is given by:
\beq
q_2\,=\,{2b\over 3(2-b)^2}\,\,\Big[\,(3-2b)\,{\eta_0-1\over \eta_0}\,+\,(3-b)\,{A_2\over A_0}\,\Big]\,\,,
\eeq
where $b$ is related to $q_0$ and $\hat\epsilon$ by  (\ref{b_new}) and (\ref{q0-epsilon}). Similarly, we can find analytically  the first corrections to the UV conformal behavior. The details of these calculations  are given in appendix \ref{UV_asymptotics}.  In this section we just present the final results.  First of all,  let us define the constant $\kappa$ (depending on the deformation $\hat\epsilon$) as:
\beq
\kappa\,\equiv\,b\,
\Big[{(\hat \gamma+1)^2\over 2\,\hat \gamma\,A_0}\Big]^{{1\over 3}}\,\,
\exp\Bigg[{2\over 3}\int^{\infty}_1\,\Big[{\eta(\xi)\over W(\xi)}\,-\,
{\eta_0\over A_0\,\xi}\Big]
d\xi\,\Bigg]\,\,.
\label{kappa_def}
\eeq
Then the functions $g$ and $f$ can be expanded for large $x$ as:
\beq
e^{g}\,=\, {\kappa\,r_q\over b}\,\,x^{{1\over b}}\,\,
\Big[\,1\,+\,{g_2\over x^2}\,+\,\cdots\Big]\,\,,
\qquad\qquad
e^{f}\,=\, \sqrt{q_0}\,\,{\kappa\,r_q\over b}\,\,x^{{1\over b}}\,\,
\Big[\,1\,+\,{f_2\over x^2}\,+\,\cdots\Big]\,\,,
\label{g_f_UVexpansions_x}
\eeq
where the coefficients  $g_2$  and $f_2$ are:
\bear
&&g_2\,=\,
{3-2b \over 6b}\,{\eta_0-1\over \eta_0}\,+\,{3-4b\over 6b}\,{A_2\over A_0}
\,\,,\rc\rc
&&f_2\,=\,{1\over 3}\,\Big(\,{3\over 2b}\,+\,{1\over 2-b}\,-1\,\Big)\,{A_2\over A_0}\,+\,
{(2+b)(3-2b)\over 6b(2-b)}{\eta_0-1\over \eta_0}\,\,.
\label{g2_f2}
\eear
Moreover, the  UV expansion of the warp factor $h$ and the dilaton is:
\beq
h(x)\,=\,{L_0^4\over \kappa^4\,r_q^4}\,
x^{-{4\over b}}
\,\Big[1\,+\,{h_2\over x^2}\,+\,\cdots\Big]\,\,,
\qquad\qquad
e^{\phi}\,=\, e^{\phi_0}\,\Big(1\,+\,{\phi_2\over x^2}\,+\,\cdots\Big)\,\,,
\label{UVexpansion_h_phi}
\eeq
with the coefficients $h_2$ and $\phi_2$ given by:
\bear
&&h_2\,=\,{3-2b\over 2b}\,
\Big(\,1+{8\over 3(b-2)}\,-\,{3\over 3+2b}\,\Big)\,{\eta_0-1\over \eta_0}\,+\,
\Big(\,1\,+\,{2\over 3(b-2)}\,-\,{2\over b}\,+\,{3\over 3+2b}\,\Big)\,{A_2\over A_0}
\,\,,\rc\rc
&&\phi_2\,=\,{3-2b\over 8b}\,\,\Big(\,1+{4b\over 2-b}-{3\over 3+2b}\,\Big)\,
{\eta_0-1\over \eta_0}-{3\over 4}\,\Big(\,1-{2\over 3(2-b)}-{1\over 3+2b}\,\Big)\,
{A_2\over A_0}\,\,.
\label{h_2_phi_2}
\eear
It is also interesting to write the previous expansions in terms of the $r$ variable. Again, the details are worked out in appendix \ref{UV_asymptotics} and the final result is:
\bear
&&e^{g(r)}= {r\over b}\,\big[1\,+\,\tilde g_2\,\Big({ r_q\over r}\Big)^{2b}\,+\,\cdots
\big]\,\,,
\qquad\qquad\,\,\,\,
e^{f(r)}= {\sqrt{q_0}\,\,r\over b}\,\big[1\,+\,\tilde f_2\,\Big({ r_q\over r}\Big)^{2b}\,+\,\cdots
\big]\,\,,\cr\cr
&&h(r)= \Big[{L_0\over r}\Big]^{4}\,\,
\Big[\,1\,+\,\tilde h_2\,\Big({r_q\over r}\Big)^{2b}\,+\,\cdots\Big]\,\,,
\qquad\,\,
e^{\phi}\,=\, e^{\phi_0}\,\Big(1\,+\,\tilde\phi_2\,\Big({ r_q\over r}\Big)^{2b}
\,+\,\cdots\Big)\,\,,\qquad\qquad
\label{All_UV_expansions_r}
\eear
where the coefficients $\tilde g_2$, $\tilde f_2$,  $\tilde h_2$,  and $\tilde \phi_2$ are related to the ones in (\ref{g2_f2}) and (\ref{h_2_phi_2}) as:
\bear
&& \tilde g_2\,=\,{2b\over 2b-1}\,\kappa^{2b}\,g_2\,\,,
\qquad\qquad\qquad\qquad
 \tilde f_2\,=\,  \kappa^{2b}\,\Big(\,f_2\,+\,{g_2\over 2b-1}\,\Big)\,\,,\rc\rc
&& \tilde h_2\,=\,  \kappa^{2b}\,\Big(\,h_2\,-\,{4g_2\over 2b-1}\,\Big)\,\,,
\qquad\qquad\,\,\,\,\,
\tilde \phi_2\,=\,\kappa^{2b}\,\phi_2\,\,.
\label{tilde_UV_coeff}
\eear
\begin{figure}[ht]
\center
\includegraphics[width=0.7\textwidth]{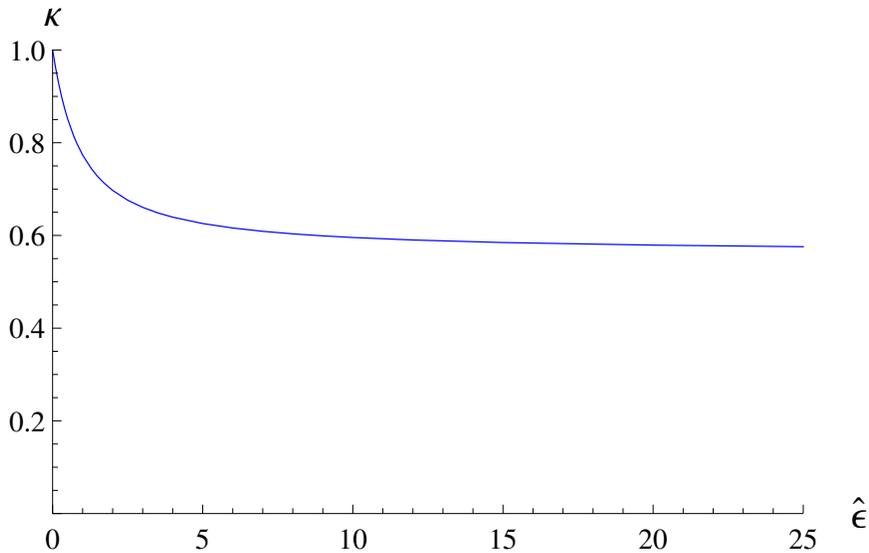}
\caption{Plot of  $\kappa$ as a function of the deformation parameter $\hat\epsilon$. The $\kappa$ asymptotes to some positive constant as $\hat\epsilon \to \infty$.} 
\label{fig: kappa}
\end{figure}
Recalling (see (\ref{rq_mq})) that $r_q=2\pi \,m_q$, it is clear from (\ref{All_UV_expansions_r}) that the deviation from conformality is controlled  by the quark mass and that the parameter $b$ determines the power of the first mass corrections. In our holographic context this is quite natural if one takes into account that $b$ determines the dimension $\Delta$ of the quark-antiquark bilinear operator in the theory with dynamical quarks ($\Delta=3-b$, see \cite{Conde:2011sw,Jokela:2012dw} and below).  The coefficients of these mass corrections depend on the constants $g_2$, $f_2$, $h_2$, and $\phi_2$ (whose analytic expressions we know from eqs. (\ref{g2_f2}) and (\ref{h_2_phi_2})), as well as on the constant $\kappa$, which must be determined numerically.   $\kappa$ as a function of the deformation parameter is plotted in Fig. \ref{fig: kappa}.  From this plot we notice that $\kappa(\hat\epsilon)$ interpolates continuously between $\kappa=1$ for $\hat\epsilon=0$ and some positive  constant value at large $\hat\epsilon$.

\section{Holographic entanglement entropy}
\label{Holographic_entanglement_entropy}

In a quantum theory the entanglement entropy $S_{A}$ between a spatial region $A$ and its complement is defined as the entropy seen by an observer in $A$ which has no access to the degrees of freedom living in the complement of $A$. It can be computed as the von Neumann entropy for the reduced density matrix obtained by taking the trace over the degrees of freedom of the complement of $A$. For quantum field theories admitting  a gravity dual, Ryu and Takayanagi proposed in \cite{Ryu} a simple prescription to compute $S_A$ from the corresponding supergravity background. The holographic entanglement entropy between $A$ and its complement in the proposal of \cite{Ryu}  is obtained by finding the eight-dimensional spatial surface $\Sigma$ whose boundary coincides with the boundary of $A$  and is such that it minimizes the functional:
\beq
S_{A}\,=\,{1\over 4\,G_{10}}\,\int_{\Sigma}\,
d^8\xi\,e^{-2\phi}\,\sqrt{\det g_8}\,\,,
\eeq
where  the $\xi$'s are a system of eight coordinates of $\Sigma$,  $G_{10}$ is the ten-dimensional Newton constant ($G_{10}=8\pi^6$ in  our units) and $g_8$ is the induced metric on $\Sigma$ in the string frame. The functional $S_{A}$ evaluated on the minimal surface $\Sigma$ is precisely the entanglement entropy between the region $A$ and its complement. 

\begin{figure}[ht]
\center
\includegraphics[width=0.5\textwidth]{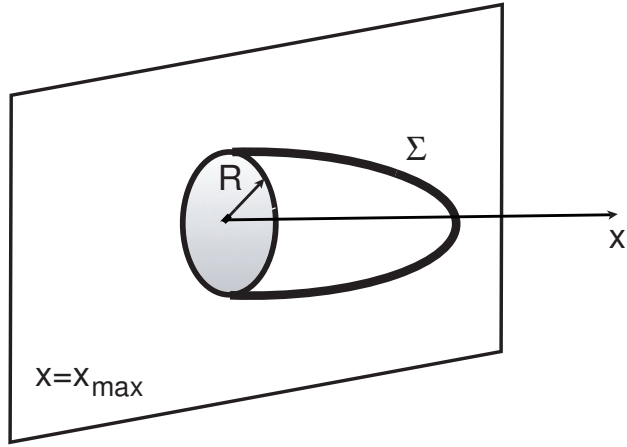}
\caption{The surface  $\Sigma$ ends on the disk of radius $R$ at the boundary $x=x_{max}\to \infty$.} 
\label{Sigma_disk}
\end{figure}

In our case $A$ is a region of the $(x^1,x^2)$-plane. In this section we will study in detail the case in which the region $A$  is a disk with radius $R$ as depicted in Fig. \ref{Sigma_disk} (see appendix \ref{entanglement_appendix} for the analysis of the entanglement entropy of a strip in the $(x^1,x^2)$-plane).  In order to deal with the disk case it is convenient to choose a system of polar coordinates for the plane:
\beq
(dx^1)^2\,+\,(dx^2)^2\,=\,d\rho^2\,+\,\rho^2\,d\Omega_1^2\,\,.
\eeq
We will describe the eight-dimensional fixed time surface $\Sigma$ by a function $\rho=\rho(x)$ with $\rho$ being the radial coordinate of the boundary plane and $x$ the holographic coordinate of the bulk. The eight-dimensional induced metric is:
\beq
ds^2_{8}\,=\,h^{-{1\over 2}}\,\big[\,\rho\,'^{\,2}+G(x)\,\big]\, dx^2\,+\,
h^{-{1\over 2}}\,\rho^2\,d\Omega_1^2\,
+h^{{1\over 2}}\,\Big[\,
e^{2f}\,ds_{{\mathbb S}^4}^2+
e^{2g}\,\Big(\,\big(E^1\big)^2\,+\,\big(E^2\big)^2\Big)\,\Big]\,\,,
\eeq
where $\rho\,'$ denotes the derivative with respect to  the holographic coordinate $x$  and   the function $G(x)$ is defined as:
\beq
G(x)\equiv {e^{2g}\,h\over x^2}\,\,.
\label{G_def}
\eeq
Let us next define a new function $H(x)$ as:
\beq
H(x)\,\equiv\,h^2\,e^{-4\phi}\,e^{8f+4g}\,\,.
\label{H_def}
\eeq
Then, the entanglement entropy as a function of $R$ is given by:
\beq
S(R)\,=\,{2\pi\,V_6\over 4 G_{10}}\,\int_{x_*}^{\infty} dx
\,\rho\,\sqrt{H(x)}\,\sqrt{\rho\,'^{\,2}+G(x)}\,\,,
\label{entropy_total_disk}
\eeq
where   $V_6=32\,\pi^3/3$ is the volume of the internal manifold
and $x_*$ is the $x$ coordinate of the turning point of $\Sigma$. 
The Euler-Lagrange equation of motion derived from the entropy  functional
 (\ref{entropy_total_disk})  is:
\beq
{d\over dx}\,\Bigg[\,
\sqrt{H(x)}\,{\rho\,\rho'\over 
\sqrt{\rho\,'^{\,2}+G(x)}}\,\Bigg]-\,\sqrt{H(x)}\,\,\sqrt{\rho\,'^{\,2}+G(x)}\,=\,0\,\,.
\label{Euler_Lagrange_disk}
\eeq
Notice that  the integrand in (\ref{entropy_total_disk}) depends on the independent variable $x$ and we therefore  cannot find a first-integral for the second-order  differential equation (\ref{Euler_Lagrange_disk}). Thus, we have to deal directly with 
(\ref{Euler_Lagrange_disk}), which must be solved with the following boundary conditions at the tip of $\Sigma$:
\beq
\rho(x=x_*)\,=\,0\,\,,
\qquad\qquad
\rho'(x=x_*)\,=\,+\infty\,\,.
\label{bc_x_tip}
\eeq
Notice also that the radius $R$ of the disk  at the boundary is just the UV limit of $\rho$:
\beq
\rho(x\to\infty)=R\,\,.
\eeq
The integral (\ref{entropy_total_disk}) for $S(R)$ diverges due to the 
contribution of the  UV region of  large $x$ . In order to characterize this divergence and to extract the finite part, let us study the behavior of the integrand in  (\ref{entropy_total_disk}) as $x\to\infty$. From the definitions of the functions $H(x)$ and $G(x)$  and the UV behavior written in  (\ref{g_f_UVexpansions_x}) and (\ref{UVexpansion_h_phi}), it follows that $H$ and $G$  display a power-like behavior as  $x\to\infty$, 
\beq
H(x)\approx H_{\infty}\,\,x^{{4\over b}}\,\,,
\qquad\qquad
G(x)\approx G_{\infty}\,\,x^{-2-{2\over b}}\,\,,
\qquad\qquad
(x\to\infty)\,\,,
\label{H_G_UV}
\eeq
where  the coefficients  $H_{\infty}$ and $G_{\infty}$ are
\beq
H_{\infty}\,=\,{L_0^8\,\kappa^4\,r_q^4\,q_0^4\,e^{-4\phi_0}\over b^{12}}\,\,,
\qquad\qquad
G_{\infty}\,=\,{L_0^4\over b^2\,r_q^2\,\kappa^2}\,\,.
\label{HG_infinity}
\eeq
 By taking the $x\to\infty$ values of $\rho$ and $\rho'$ ($\rho=R$ and  $\rho'=0$) inside the integral in (\ref{entropy_total_disk}), as well as the asymptotic form of $H(x)$ and $G(x)$ (eq. (\ref{H_G_UV})), we get:
\beq
S_{div}(R)\,=\,{\pi\,V_6\over 2 G_{10}}\,\,
R\,\sqrt{H_{\infty}\,G_{\infty}}\,
\int^{x_{max}}\,x^{{1\over b}-1}\,dx\,\,,
\eeq
where $x_{max}$ is the maximum value of the holographic coordinate $x$ (which acts as a UV regulator).  After performing the integral, we obtain:
\beq
S_{div}(R)\,=\,{\pi\,V_6\over 2 G_{10}}\,\,
R\,b\,\sqrt{H_{\infty}\,G_{\infty}}\,\,x_{max}^{{1\over b}}\,\,.
\label{S_div_first}
\eeq
Let us rewrite (\ref{S_div_first}) in terms of physically relevant quantities. First of all, we notice that:
\beq
b\,\sqrt{H_{\infty}\,G_{\infty}}\,=\,
{L_0^6\,q_0^2\,e^{-2\phi_0}\over b^6}\,\kappa\, r_q\,=\,
{3\pi^2\over 2}
{F_{UV} ({\mathbb S}^3)\over L_0^2}\,\kappa\, r_q\,\,,
\label{bHG_F}
\eeq
where $F_{UV} ({\mathbb S}^3)$ is the free energy\footnote{When the field theory is formulated on a three-sphere, its free energy is defined as:
\beq
F({\mathbb S}^3)\,=\,-\log |\,Z_{{\mathbb S}^3}\,|\,\,,
\eeq
where $Z_{{\mathbb S}^3}$ is the Euclidean path integral. For a CFT whose gravity dual is of the form $AdS_4\times {\cal M}_6$, the holographic calculation of $F({\mathbb S}^3)$  gives \cite{Emparan:1999pm}:
\beq
F({\mathbb S}^3)\,=\,{\pi L^2\over 2 G_N}\,\,,
\eeq
where $L$ is the $AdS_4$ radius and $G_N$ is the effective four-dimensional Newton's constant. } of the massless flavored theory on the three-sphere:
\beq
F_{UV} ({\mathbb S}^3)\,=\,{2\pi\over 3}\,{N^2\over \sqrt{2\lambda}}\,\,
\xi\big({N_f\over k}\big)\,\,,
\label{F_UV}
\eeq
where the function $\xi\Big({N_f\over k}\Big)$ is given by:
\beq
\xi\Big({N_f\over k}\Big)\equiv{1\over 16}\,\,
{q_0^{{5\over 2}}\,(\eta_0+q_0)^4\over (2-q_0)^{{1\over 2}}\,\,(q_0+\eta_0 q_0\,-\,\eta_0)^{{7\over 2}}}\,\,.
\label{xi-smeared}
\eeq
In (\ref{xi-smeared})  $\eta_0=1+\hat \epsilon$ and $q_0$ is written  in (\ref{q0-epsilon}) in  terms of the deformation parameter.  For the unflavored ABJM theory the free energy is given by (\ref{F_UV}) with $\xi=1$. This formula displays the famous $N^{{3\over 2}}$ behavior. The function 
$\xi(N_f/k)$ encodes the corrections to this behavior due to the smeared massless flavors. It was first computed in \cite{Conde:2011sw}, where it was shown that it is remarkably close to  the value found in \cite{Gaiotto:2009tk} for localized embeddings. 
The function $\xi$ is a monotonic function of the deformation parameter which grows as 
$\sqrt{\hat \epsilon}$ for large values of $\hat\epsilon$.

Using (\ref{bHG_F}) and the fact that, in the deep UV region of large $x$, $r_{max}=\kappa\, r_q\,x_{max}^{{1\over b}}$ (see (\ref{r-x_relation})), we can rewrite $S_{div}(R)$ as:
\beq
S_{div}(R)\,=\,{F_{UV} ({\mathbb S}^3)\over L_0^2}\,r_{max}\,R\,\,.
\label{Sdiv_disk}
\eeq
We notice in (\ref{Sdiv_disk}) that $S_{div}(R)$ diverges linearly with $r_{max}$. The coefficient of this divergent term is linear in the disk radius $R$ and in $F_{UV} ({\mathbb S}^3)$. The latter is a measure of the effective number of degrees of freedom of the flavored theory in the high-energy UV limit in which the flavors can be considered to be massless. The appearance of $F_{UV} ({\mathbb S}^3)$ in (\ref{Sdiv_disk}) is thus quite natural.

The separation between the divergent and finite parts of $S(R)$ has ambiguities. In order to solve these ambiguities, Liu and Mezei proposed in \cite{Liu:2012eea} to consider the function ${\cal F}(R)$, defined as:
\beq
{\cal F}(R)\,\equiv\,R\,{\partial S\over \partial R}\,-\,S\,\,.
\label{calF_definition}
\eeq
It was argued in \cite{Liu:2012eea} that ${\cal F}(R)$ is finite and a monotonic function of $R$ which provides a measure of the number of degrees of freedom of a system at a scale $R$. 

In a 3d CFT the entanglement entropy for a disk of radius $R$ has the form:
\beq
S_{CFT}(R)\,=\,\alpha \,R\,-\,\beta\,\,,
\label{CFT-entropy}
\eeq
where $\alpha$ is a UV divergent  non-universal part and $\beta$ is finite and independent of $R$.  It was shown in \cite{Casini:2011kv} that the finite part $\beta$ is equal to the free energy of the theory on ${\mathbb S}^3$. Notice that ${\cal F}=\beta$ when $S$ is of the form (\ref{CFT-entropy}). Therefore, for a conformal fixed point the function ${\cal F}$ is constant and equal to the free energy on the three-sphere of the corresponding CFT.  In the next subsection we will obtain the UV and IR values of  ${\cal F}$ and we will show that  they coincide  with the free energies on ${\mathbb S}^3$ of the massless flavored theory and of the unflavored ABJM model, respectively. 

It is interesting to point out that the entanglement entropy of a disk in a (2+1)-dimensional system at large $R$ can also be written in the form (\ref{CFT-entropy}), if one neglects terms  which vanish in the $R\to\infty$ limit.  In a gapped system, the $R$-independent part $\beta$ of the right-hand side of (\ref{CFT-entropy}) is the so-called topological entanglement entropy \cite{Kitaev:2005dm,Levin:2006zz} and serves to characterize topologically ordered many-body  states which contain non-local entanglement due to non-local correlations (examples of such states are the Laughlin states of the fractional quantum Hall effect or the ${\mathbb Z}_2$ fractionalized states). The  topological entanglement entropy $\beta$ is related to the so-called total quantum dimension 
${\cal D}$  of the system as  $\beta=\log {\cal D}$. In general ${\cal D}>1$ (or $\beta>0$) signals a topological order (for example ${\cal D}=\sqrt{q}$ for the quantum Hall system with filling fraction $\nu=1/q$, with $q$ an odd integer). 
\begin{figure}[ht]
\center
\includegraphics[width=0.7\textwidth]{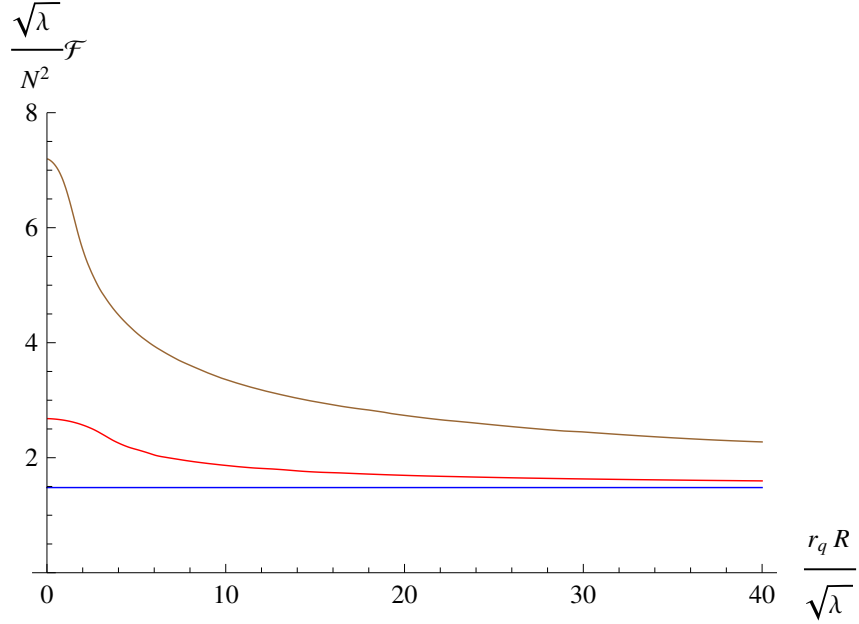}
\caption{Plot of the running free energy ${\cal F}$ as a function of $r_q\,R$ for $\hat\epsilon=0$ (bottom, blue), $1$ (middle, red), and $9$ (top, brown). } 
\label{curlyF}
\end{figure}

For our system, we can obtain the embedding function $\rho(x)$ by numerical integration of the differential equation (\ref{Euler_Lagrange_disk}) and then we can get the functions $S(R)$ and ${\cal F}(R)$ by using (\ref{entropy_total_disk}) and  the definition of ${\cal F}(R)$ in (\ref{calF_definition}). The results for the latter are plotted as a function of 
$R\,r_q\propto R\, m_q$ in Fig. \ref{curlyF}. We notice that  ${\cal F}(R)$ is a monotonically decreasing function that interpolates smoothly between  the two limiting values at $R\,m_q=0$ and $R\,m_q\to\infty$. The UV limit of ${\cal F}$ at small $R\,m_q$ equals the free energy (\ref{F_UV}) of the massless flavored theory, whereas for large $R\,m_q$ the function ${\cal F}$ approaches the free energy of the unflavored ABJM model (\ie,  the value in (\ref{F_UV}) with $\xi=1$). This behavior is in agreement with the general expectation in \cite{Liu:2012eea,Casini:2012ei}  and corresponds to a smooth decoupling of the massive flavors as their mass $m_q$ is increased continuously.  

We will study the UV and IR  limits of ${\cal F}$ analytically in the next two subsections.  Some details of these calculations are deferred to appendix \ref{entanglement_appendix}, where we also study the entanglement entropy for the strip geometry.

\subsection{UV limit}
\label{UV_limit_disk_entanglement}
In order to study the UV limit of the disk entanglement entropy, 
let us write the Euler-Lagrange equation (\ref{Euler_Lagrange_disk}) when $H(x)$ and $G(x)$ are given by their asymptotic values (\ref{H_G_UV}):
\beq
{d\over dx}\,\Bigg[\,
x^{{2\over b}}\,{\rho\,\rho'\over 
\sqrt{\rho\,'^{\,2}+G_{\infty}\,
x^{-2-{2\over b}}}}\,\Bigg]-\,x^{{2\over b}}\,\,\sqrt{\rho\,'^{\,2}+G_{\infty}\,
x^{-2-{2\over b}}}\,=\,0\,\,.
\label{profile_eq_UV}
\eeq
This equation can be solved exactly by the function:
\beq
\rho_{UV}(x)\,=\,\sqrt{R^2\,-\,b^2\,G_{\infty}\,x^{-{2\over b}}}\,\,,
\label{rho_exact}
\eeq
which clearly satisfies the initial conditions (\ref{bc_x_tip}), with the following value of  the turning point coordinate $x_*$:
\beq
x_{*}^{2/ b}\,=\,{b^2\,G_{\infty}\over R^2}\,\,.
\label{xstar}
\eeq
Since $G_{\infty}\propto r_q^{-2}$, it follows from (\ref{xstar}) that $x_{*}^{2/ b}\propto (r_q\,R)^{-2}$. Therefore, the turning point coordinate $x_*$ is large if $r_q$ or $R$ are small. In this case it would be justified to use the asymptotic UV values of the functions $G$ and $H$, since the minimal surface $\Sigma$ lies entirely in the large $x$ region.  Notice also that (\ref{rho_exact}) can be written  in terms of $x_*$ as:
\beq
\rho_{UV}\,=\,R\,\sqrt{1-\Big({x_*\over x}\Big)^{2/b}}\,\,.
\eeq
In order to calculate the entropy in this UV limit it is very useful to use the following relation satisfied by the function $\rho_{UV}(x)$ written in (\ref{rho_exact}):
\beq
\rho_{UV}\,\sqrt{\big(\rho\,'_{UV}\big)^{2}+G_{\infty}\,
x^{-2-{2\over b}}}\,=\,R\,\sqrt{G_{\infty}}\,\, x^{-1-{1\over b}}\,\,.
\label{rhoUV_relation}
\eeq
Making use of (\ref{rhoUV_relation})  in (\ref{entropy_total_disk}), we find the following expression for the entanglement entropy:
\beq
S_{UV}(R)\,=\,{\pi\,V_6\over 2 G_{10}}\,\,
R\,\sqrt{H_{\infty}\,G_{\infty}}\,
\int^{x_{max}}_{x_*}\,x^{{1\over b}-1}\,dx\,\,.
\eeq
The divergent part of this integral is due to its upper limit and is just given by (\ref{Sdiv_disk}). The finite part of $S$ is:
\beq
S_{finite,UV}\,=\,-{\pi\,V_6\over 2 G_{10}}\,\,b\,
\sqrt{H_{\infty}\,G_{\infty}}\,\,R\,x_*^{1\over b}\,=\,-{\pi\,V_6\over 2 G_{10}}\,\,
b^2\,G_{\infty}\,\sqrt{H_{\infty}}\,\,,
\label{Sfinite_UV_disk}
\eeq
where, in the second step, we used (\ref{xstar}) to eliminate $x_*$. Notice that  the right-hand side of (\ref{Sfinite_UV_disk}) is independent  of the disk radius $R$. Moreover, by using (\ref{HG_infinity})  and (\ref{bHG_F}) we find that:
\beq
S_{finite,UV}\,=\,-F_{UV} ({\mathbb S}^3)\,\,.
\eeq
Therefore, in this UV limit, the dependence on $R$ of the entanglement entropy takes the form (\ref{CFT-entropy}), where $\beta$ is just the free energy of the massless flavored theory on the three-sphere. It follows trivially from this form of $S_{UV}$ and  the definition (\ref{calF_definition}) that ${\cal F}_{UV}=\beta$ and therefore:
\beq
{\cal F}_{UV}\,\equiv\,
\lim_{r_qR\to 0}\,{\cal F}(R)\,=\,
F_{UV} ({\mathbb S}^3)\,\,.
\label{calF_UV}
\eeq
It is also possible to compute analytically the first correction to  (\ref{calF_UV}) for small values of  $r_q\,R$. The details of this calculation are given in appendix 
\ref{entanglement_appendix}. Here we will just present the final result, which can be written as:
\beq
{\cal F}(R)\,=\,F_{UV} ({\mathbb S}^3)\,+\,c_{UV}\,(r_q\,R)^{2(3-\Delta_{UV})}
\,+\,\cdots
\,\,,
\label{calF_nearUV}
\eeq
where $c_{UV}$ is a constant coefficient depending on the deformation parameter (see eq. (\ref{c_UV})) and $\Delta_{UV}\,=\,3-b$ is the dimension of the quark-antiquark bilinear operator in the UV flavored theory (this dimension was found  in section 7.3 of \cite{Conde:2011sw} from the analysis of the fluctuations of the flavor branes, see also \cite{Jokela:2012dw}). It is interesting to point out that (\ref{calF_nearUV}) is the  behavior expected \cite{Liu:2012eea} for a flow caused by a source deformation with a relevant operator of dimension $\Delta_{UV}$. Moreover, one can verify that $c_{UV}$ is negative for all values of the deformation parameter $\hat \epsilon$, which confirms that the UV fixed point is a local maximum of  ${\cal F}$.

\subsection{IR limit}
\label{IR_limit_disk_entanglement}

\begin{figure}[ht]
\center
\includegraphics[width=0.75\textwidth]{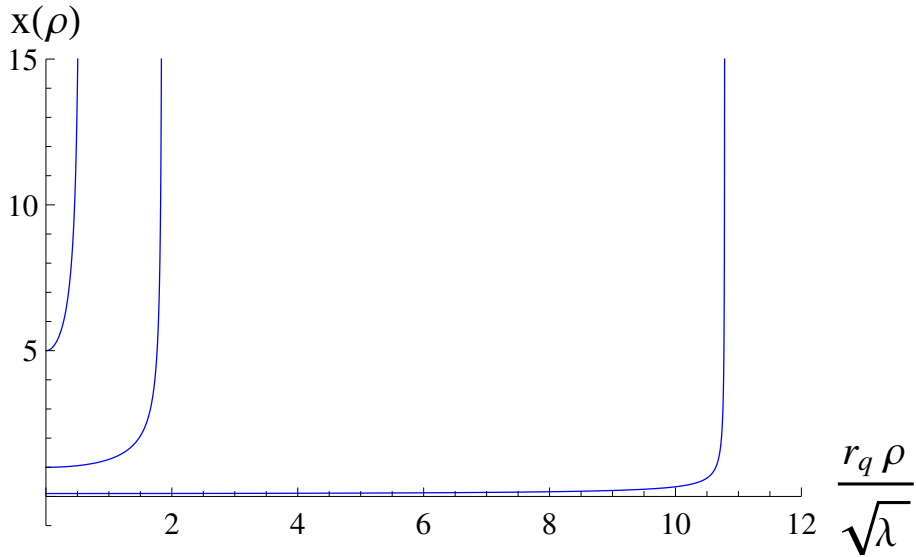}
\caption{Representation of the  embedding function $\rho(x)$  for different values of the turning point $x_*=0.1, 1, 5$ and $\hat\epsilon = 9$.} 
\label{profiles}
\end{figure}

Let us now analyze the IR limit of the entanglement entropy $S(R)$ and of the function ${\cal F}(R)$. This limit occurs when the 8d surface $\Sigma$ penetrates deeply into the geometry and, therefore, when the coordinate $x_*$ of the
turning point is small ($x_*\ll 1$). This happens when either the disk radius $R$  or the quark mass $m_q=r_q/2\pi$ are large.  Looking at the embedding function  $\rho(x)$ obtained by numerical integration of (\ref{Euler_Lagrange_disk}) one notices that, when $x_*$ is small, the function $\rho(x)$ is approximately constant and equal to its asymptotic value $\rho=R$ in the region $x\ge 1$ (see Fig. \ref{profiles}). Therefore, the dependence of $\rho$ on the holographic coordinate $x$ is determined by the integral of (\ref{Euler_Lagrange_disk})  in the region 
$x\le 1$, where the background is given by the unflavored running solution. Actually, when $x_*$ is small it is a good approximation to consider (\ref{Euler_Lagrange_disk}) for the unflavored background, \ie, when the constants $c,\gamma\to 0$, with $\gamma/c$ fixed and given by $\gamma/c=r_q$. In this limit the different functions of the background are:
\beq
e^{f}\approx e^{g}\approx r_q\,x\,\,,
\qquad\qquad
h\approx {1\over r_q^4}\,\,{2\pi^2\,N\over k}\,{1\over x^4}\,\,.
\label{functions_deep_IR}
\eeq
It follows that $G(x)$ and $H(x)$, as defined in (\ref{G_def}) and (\ref{H_def}), are then given by:
\beq
G_{IR}(x)\,=\,{1\over r_q^2}\,{2\pi^2\,N\over k}\,{1\over x^4}\,=\,G_{0}\,x^{-4}\,\,,
\qquad\qquad
H_{IR}(x)\,=\,r_q^4\,{4\pi^4\,N^2\over k^2}\,e^{-4\phi_{IR}}\,x^4\,=\,H_{0}\,x^{4}\,\,,
\label{G_H_IR}
\eeq
where, in the last step, we have defined the constants $G_0$ and $H_0$, and 
$\phi_{IR}$ is the constant dilaton corresponding to the unflavored background,
\beq
e^{\phi_{IR}}\,=\,2\sqrt{\pi}\,\Big({2N\over k^5}\Big)^{{1\over 4}}\,\,.
\eeq
Let us now elaborate  on the expression (\ref{entropy_total_disk}) for the entanglement entropy. We split the integration interval of the variable $x$ as 
$[x_*,\infty]=[x_*,1]\cup [1,\infty]$ and take  into account that one can put $\rho=R={\rm constant}$ in the region $x\ge 1$. We get:
\beq
S_{IR}(R)\,=\,
{\pi\,V_6\over 2 G_{10}}\,\int_{x_*}^{1} dx
\,\rho\,\sqrt{H_{IR}(x)}\,\sqrt{\rho\,'^{\,2}+G_{IR}(x)}\,+\,
{\pi\,V_6\over 2 G_{10}}\,R\,\int_{1}^{x_{max}} \,\,
\sqrt{H(x)\,G(x)}
\,\,.
\label{entropy_disk_IR}
\eeq
The second term in (\ref{entropy_disk_IR}) is linear in $R$ and will not contribute to ${\cal F}(R)$. To evaluate the first integral in (\ref{entropy_disk_IR}) we must determine the embedding function $\rho(x)$ by integrating  (\ref{Euler_Lagrange_disk}) when $G(x)$ and $H(x)$ are given by their IR values (\ref{G_H_IR}). The resulting equation is just the same as (\ref{profile_eq_UV}) with $G_{\infty}$ and $H_{\infty}$  substituted by $G_0$ and $H_0$ and $b=1$. Then, the function $\rho(x)$  can be written as in (\ref{rho_exact}), 
\beq
\rho_{IR}(x)\,=\,\sqrt{\hat R^{\,2}\,-\,G_0\,x^{-2}}\,\,,
\label{rho_IR_sol}
\eeq
where $\hat R$ is a constant. By requiring that $\rho_{IR}(x=1)=R$, we get:
\beq
\hat R^{\,2}\,=\,R^2\,+\,G_0\,\,.
\eeq
It follows from (\ref{rho_IR_sol})  that the coordinate $x_*$ of the turning point is given by:
\beq
x_{*}^{2}\,=\,{G_0\over \hat R^2}\,=\,{G_0\over G_0+R^2}\,\,.
\label{x_star_IR}
\eeq
Notice that when $r_q\to\infty$ (and $G_0\to 0$) or $R$ is large one can neglect the $G_0$ in the denominator of (\ref{x_star_IR}) and then 
$x_{*}^{2}\,\approx\, G_0\,R^{-2}\propto (r_q\,R)^{-2}$, which is a small number. Moreover, by using the explicit form (\ref{rho_IR_sol}) of $\rho(x)$ in this IR region, we get:
\beq
\rho_{IR}\,\sqrt{(\rho'_{IR})^2+G_{IR}}\,=\,\hat R\,\sqrt{G_{0}}\,\, x^{-2}\,\,,
\eeq
and the first integral in (\ref{entropy_disk_IR}) can be explicitly evaluated:
\beq
\int_{x_*}^{1} dx
\,\rho_{IR}\,\sqrt{H_{IR}(x)}\,\sqrt{(\rho'_{IR})^2+G_{IR}(x)}\,=\,
\hat R\,\sqrt{G_0\,H_0}\,-\,G_0\,\sqrt{H_0}\,\,.
\label{S_IR_integral}
\eeq
Notice that, at leading order $\hat R\approx R$ and, thus, the first term in (\ref{S_IR_integral}) does not contribute to ${\cal F}(R)$. Then, the IR limit of ${\cal F}$ is determined by the second contribution in (\ref{S_IR_integral}) and given by:
\beq
{\cal F}_{IR}\,=\,{\pi\,V_6\over 2 G_{10}}\,\,G_0\,\sqrt{H_0}\,\,.
\eeq
Moreover, from the values of $G_0$ and $H_0$ written in (\ref{G_H_IR}) we get:
\beq
G_0\,\sqrt{H_0}\,=\,{\pi^3\over \sqrt{2}}\,N^{{3\over 2}}\,k^{{1\over 2}}\,=\,
{3\pi^2\over 2}\,F_{IR} ({\mathbb S}^3)\,\,,
\eeq
where $F_{IR} ({\mathbb S}^3)$ is the free energy on the three-sphere of the unflavored ABJM theory:
\beq
F_{IR} ({\mathbb S}^3)\,=\,{\pi\sqrt{2}\over 3}\,k^{{1\over 2}}\,N^{{3\over 2}}\,\,.
\eeq
It follows that the IR limit of the ${\cal F}$ function is:
\beq
{\cal F}_{IR}\,
\equiv\,
\lim_{r_qR\to \infty}\,{\cal F}(R)\,=\,
F_{IR} ({\mathbb S}^3)\,\,,
\label{calF_IR}
\eeq
as expected in the deep IR limit in which the flavors become infinitely massive and can therefore be integrated out. The corrections to the result (\ref{calF_IR}) near the IR fixed point could be obtained by applying the techniques recently introduced in \cite{Liu:2013una}. We will not attempt to perform this calculation here.

\section{Wilson loops and the quark-antiquark potential}
\label{Wilson}
In this section we evaluate the expectation values of  the Wilson loop and the corresponding quark-antiquark potential for our model. We will employ the standard holographic prescription of refs. \cite{Maldacena:1998im,Rey:1998ik}, in which one considers a fundamental string hanging from the UV boundary. Then, one computes the regularized Nambu-Goto action for this configuration, from which the $q\bar q$ potential energy can be extracted. In a theory with dynamical flavors this potential energy contains information about the screening of external charges by the virtual quarks popping out from the vacuum. In our case we expect  having a non-trivial flow connecting two conformal behaviors as we move from the UV regime of small  $q\bar q$ separation (in units of the quark mass $m_q$) to the IR regime of large $q\bar q$ distance. We will verify below that this expectation is indeed fulfilled by our model.

Let us denote by $(t,x^1,x^2)$ the Minkowski coordinates and consider a fundamental string for which we take $(t,x^1)$ as its worldvolume coordinates. If the embedding is characterized by a function $x=x(x^1)$, with $x$ being the holographic coordinate, the induced metric is:
\beq
ds_2^2\,=\,-h^{-{1\over 2}}\,dt^2\,+\,h^{{1\over 2}}\,\Big[\,
{e^{2g}\over x^2}\,x'^{\,2}\,+\,h^{-1}\,\Big]\,(dx^1)^2\,\,,
\eeq
where $x'$ denotes the derivative of $x$ with respect to $x^1$. The Nambu-Goto lagrangian density takes the form:
\beq
{\cal L}_{NG}\,=\,{1\over 2\pi}\,\sqrt{-\det g_2}\,=\,
{1\over 2\pi}\,\sqrt{{e^{2g}\over x^2}\,x'^{\,2}\,+\,h^{-1}}\,\,.
\eeq
As ${\cal L}_{NG}$ does not depend on $x^1$, we have the following conservation law:
\beq
x'\,{\partial\,{\cal L}_{NG}\over \partial x'}\,-\,{\cal L}_{NG}\,=\,{\rm constant}\,\,.
\eeq
Therefore, if $x_*$ denotes the turning point of the string, we have the first integral of the equations of motion:
\beq
\sqrt{1+{e^{2g}\,h\over x^2}\,x'^{\,2}}\,=\,{\sqrt{h_*}\over \sqrt{h}}\,\,,
\label{first_integral}
\eeq
where $h_*\equiv h(x=x_*)$. Then $x'$ is given by:
\beq
x'\,=\,\pm {x \sqrt{h_*-h(x)}\over e^{g(x)}\,h(x)}\,\,,
\eeq
where the two signs correspond to the two branches of the hanging string. 
The $q\bar q$ separation $d$ in the $x^{1}$ direction is:
\beq
d\,=\,2\,\int_{x_*}^{\infty}\,
{e^{g(x)}\,h(x)\over x\,\sqrt{h_*-h(x)}}\,\,dx\,\,.
\label{d_x-integral}
\eeq
In order to compute the potential energy of the $q\bar q$ pair, let us evaluate the on-shell action. By using the first integral (\ref{first_integral}) it is straightforward to check that the on-shell value of the Nambu-Goto lagrangian density  is:
\beq
{\cal L}_{NG} ({\rm on-shell})\,=\,{1\over 2\pi}\,{\sqrt{h_*}\over h}\,\,.
\eeq
Therefore, the on-shell action becomes:
\beq
S_{{\rm on-shell}}\,=\,{T\over \pi}\,\,\int_{x_*}^{\infty}\,
{e^{g(x)}\over x}\,\,{\sqrt{h_*}\over \sqrt{h_*-h(x)}}\,dx\,\,,
\label{unregularized_NGaction}
\eeq
where $T=\int dt$. The integral (\ref{unregularized_NGaction}) is divergent and must be regularized as in \cite{Maldacena:1998im,Rey:1998ik} by subtracting the action of two straight strings stretched  between the origin and the UV boundary, which corresponds to subtracting the (infinite) quark masses in the static limit. After applying this procedure we arrive at the following  expression for the regulated on-shell action:
\beq
S_{{\rm on-shell}}^{{\rm reg}}\,=\,S_{{\rm on-shell}}\,-\,{T\over \pi}
\int_{0}^{\infty}\,{e^{g(x)}\over x}\,\,dx\,\,,
\eeq
from which we get the $q\bar q$ potential energy:
\beq
E_{q\bar q}\,=\,{1\over \pi}\,\int_{x_*}^{\infty}\,\,
{e^{g(x)}\over x}\,\,\Big[\,{\sqrt{h_*}\over \sqrt{h_*-h(x)}}\,-\,1\Big]
dx\,-\,{r_*\over \pi}\,\,,
\label{E_x-integral}
\eeq
where $r_*$ is the $r$ coordinate of the turning point:
\beq
r_*\,=\,\int^{x_*}_{0}\,{e^{g(x)}\over x}\,dx\,\,.
\eeq

\begin{figure}[ht]
\center
\includegraphics[width=0.75\textwidth]{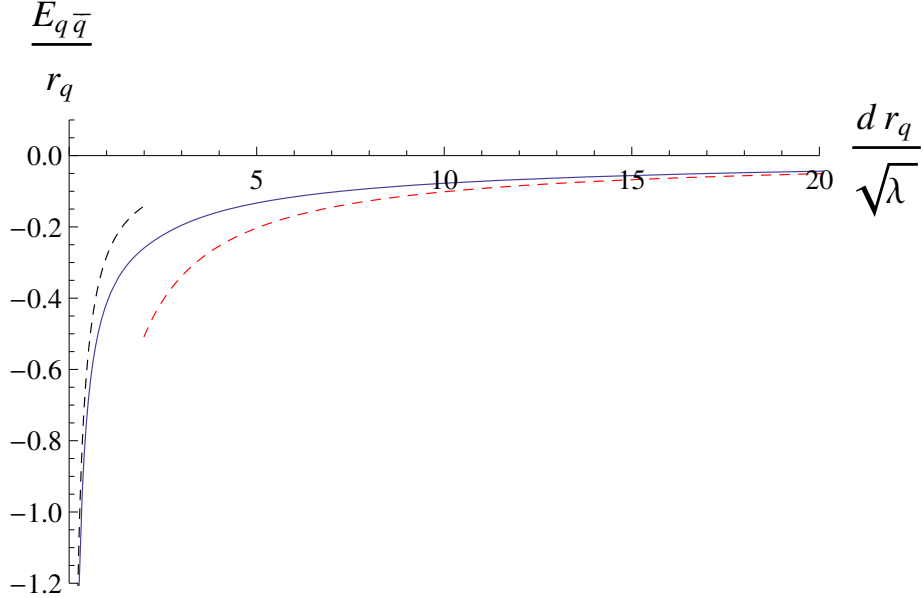}
\caption{Plot of the $q\bar q$ potential energy for $\hat\epsilon=9$. The numerical result is compared to the UV leading estimate (\ref{leading_UV_pot}) (black dashed curve  on the left) and to the leading IR potential (red dashed curve on the right). } 
\label{quark_potential}
\end{figure}

From (\ref{d_x-integral}) and (\ref{E_x-integral}) we have computed numerically the potential energy $E_{q\bar q}$ as a function of the $q\bar q$ distance $d$. The result of this numerical calculation is shown in Fig. \ref{quark_potential}.  As mentioned above, we expect to have a potential energy which interpolates between the  two conformal behaviors with $E_{q\bar q}\propto 1/d$ at the UV and IR. Actually, in the limiting cases in which  $r_q d$ is small or large the function $E_{q\bar q}(d)$ can be calculated analytically (see appendix \ref{asymp_qq_potential}). In both cases $E_{q\bar q}\propto 1/d$, but with different coefficients. Indeed, in the UV limit $r_q\,d\to 0$, the $q\bar q$ potential can be approximated as:\footnote{Also, the first correction to the UV conformal behavior (\ref{leading_UV_pot})  is computed in appendix \ref{asymp_qq_potential}.}
\beq
E_{q\bar q}^{UV}\approx-
{4\pi^3\,\sqrt{2\lambda}\over  \big[\Gamma\big({1\over 4}\big)\big]^4}\,\,\sigma\,\,
{1\over d}\,\,,
\qquad\qquad
( r_q\,d\to 0)\,\,,
\label{leading_UV_pot}
\eeq
where $\lambda=N/k$ is the 't Hooft coupling and $\sigma$ is the so-called screening function:
\beq
\sigma\,=\,\sqrt{
{2-q_0\over q_0(q_0+\eta_0 q_0-\eta_0)}}\,\,\,\,b^2\,=\,
{1\over 4}\,\,
{q_0^{{3\over 2}}\,\,(\eta_0+q_0)^2\,(2-q_0)^{{1\over 2}}
\over
(q_0+\eta_0 q_0-\eta_0)^{{5\over 2}}}\,\,.
\label{screening-sigma}
\eeq
Notice that $\sigma$ encodes all the dependence of the right-hand side of (\ref{leading_UV_pot}) on the number of flavors. Actually, the potential (\ref{leading_UV_pot}) is just the one corresponding to having massless flavors (which was first computed for this model in \cite{Conde:2011sw}), as expected in the high-energy UV regime in which all masses can be effectively neglected. The function $\sigma$ characterizes the corrections of the static $q\bar q$ potential due to the screening produced by the unquenched massless flavors ($\sigma\to 1$ for $N_f\to 0$, whereas $\sigma$ decreases as $\sigma\propto \sqrt{k/N_f}$ for $N_f$ large). In Fig.  \ref{quark_potential} we compare the leading UV result (\ref{leading_UV_pot}) with the numerical calculation in the small $r_q \,d$ region.

Similarly, one can compute analytically the $q\bar q$  potential in the region where $r_q\,d$ is large. At leading order the result is (see appendix \ref{asymp_qq_potential}):
\beq
E_{q\bar q}^{IR}\approx-
{4\pi^3\,\sqrt{2\lambda}\over  \big[\Gamma\big({1\over 4}\big)\big]^4}\,\,
{1\over d}\,\,,
\qquad\qquad
( r_q\,d\to \infty)\,\,.
\label{leading_IR_pot}
\eeq
In Fig.  \ref{quark_potential} we compare the analytic expression (\ref{leading_IR_pot}) to the numerical result in the large distance region. 
Notice that the difference between (\ref{leading_UV_pot}) and (\ref{leading_IR_pot}) is that the screening function $\sigma$ is absent in (\ref{leading_IR_pot}). Therefore, in the deep IR the flavor effects on the $q\bar q$ potential disappear, which is consistent with the intuition that massive flavors are integrated out at low energies.  

\section{Two-point functions of high dimension operators}
\label{Two-point_section}

In this section we study the two-point functions of bulk operators with high dimension. The form of these correlators can be obtained semiclassically by analyzing the geodesics of massive particles in the dual geometry \cite{Balasubramanian:1999zv,Louko:2000tp,Kraus:2002iv},
\beq
\langle {\cal O}(x)\,{\cal O}(y)\rangle\,\sim \,e^{-m\,{\cal L}_r(x,y)}\,\,,
\label{OO-vev}
\eeq
where $m$ is the mass of the bulk field dual to ${\cal O}$. We are assuming that $m$ is large in order to apply a saddle point approximation in the calculation of the correlator. In (\ref{OO-vev}) ${\cal L}_r(x,y)$ is a regularized length along a spacetime geodesic connecting the boundary points $x$ and $y$. To find these geodesics, let us write the Einstein frame metric of our geometry as:
\beq
ds^2_{10}\,=\,e^{-{\phi\over 2}}\,h^{-{1\over 2}}\,dx^2_{1,2}\,+\,
e^{-{\phi\over 2}}\,h^{{1\over 2}}\,
\Big[\,e^{2g}\,\,{dx^2\over x^2}\,+\,e^{2f}\,ds_{{\mathbb S}^4}^2\,+\,
e^{2g}\,\Big(\,\big(E^1\big)^2\,+\,\big(E^2\big)^2\Big)\,\Big]\,\,.
\eeq
Then, the induced metric for a curve parametrized as $x=x(x^1)$ is:
\beq
ds^2_{1}\,=\,e^{-{\phi\over 2}}\,h^{-{1\over 2}}\,\,
\Big(1+G(x)\,x'{\,^2}\Big)\,(dx^1)^2\,\,,
\eeq
with $x'=dx/dx^1$ and  $G(x)$ is the function  defined in (\ref{G_def}). Therefore, 
the action of a particle of mass $m$ whose worldline is the curve  $x=x(x^1)$ is:
\beq
S\,=\,m\int ds_1\,=\,m\int e^{-{\phi\over 4}}\,h^{-{1\over 4}}\,
\sqrt{1+G(x)\,x'{\,^2}}\,dx^1\,\,.
\label{particle_action}
\eeq
The geodesics we are looking for are solutions of the Euler-Lagrange equation derived from the action (\ref{particle_action}). This equation has a  first integral which is given by:
\beq
e^{{\phi(x)\over 4}}\,\big[h(x)\big]^{{1\over 4}}\,
\sqrt{1+G(x)\,x'{\,^2}}\,=\,
e^{{\phi_*\over 4}}\,h_*^{{1\over 4}}\,\,,
\label{first_integral_particle}
\eeq
where $\phi_*\equiv\phi(x=x_*)$ and $h_*\equiv h(x=x_*)$, with $x_*$ being the $x$ coordinate of the turning point, \ie, the minimum value of $x$ along the geodesic. It follows from (\ref{first_integral_particle}) that:
\beq
x\,'\,=\,\pm {1\over \sqrt{ G(x)}}\,
\sqrt{e^{{1\over 2}\,(\phi_*-\phi(x))}\,\,\Big({h_*\over h(x)}\Big)^{1\over 2}\,-\,1}\,\,.
\eeq
The spatial separation $l$ of the two points in the correlator can be obtained by integrating $1/x'$. We get:
\beq
l\,=\,2\,\int_{x_*}^{\infty}\,dx
{\sqrt{G(x)}\over
\sqrt{e^{{1\over 2}\,(\phi_*-\phi(x))}\,\,\big({h_*\over h(x)}\big)^{1\over 2}\,-\,1}
}\,\,.
\label{l_general}
\eeq
Moreover, the length of the geodesic can be obtained by integrating $ds_1$ over the worldline,
\beq
{\cal L}\,=\,2\,\int_{x_*}^{\infty}\,dx
{e^{-{\phi(x)\over 4}}\,\big[h(x)\big]^{-{1\over 4}}\,\,\sqrt{G(x)}
\over
\sqrt{1-e^{{1\over 2}\,(\phi(x))-\phi_*)}\,\,\big({h(x)\over h_*}\big)^{1\over 2}}
}\,\,.
\eeq
This integral is divergent. In order to regularize it, let us study the UV behavior of the integrand. For large $x$, the functions $h(x)$ and $G(x)$ behave as in (\ref{UVexpansion_h_phi}) and (\ref{H_G_UV}),  respectively. Thus, at leading order  for large $x$, 
\beq
\big[h(x)\big]^{-{1\over 4}}\,\,\sqrt{G(x)}\,\approx {L_0\over b}\,x^{-1}\,\,.
\eeq
In  the UV region $x\to\infty$, the integrand in ${\cal L}$ behaves approximately as
$2\,b^{-1}\,L_0\,e^{-\phi_0/4}\,x^{-1}$, which produces a logarithmic UV divergence when it is integrated. In order to tackle this divergence, let us regulate the integral by extending it up to some cutoff $x_{\max}$ and  renormalize the geodesic length by subtracting the divergent part. Accordingly, we define the renormalized geodesic length as:
\beq
{\cal L}_r\,=\,2\,\int_{x_*}^{x_{max}}\,dx
{e^{-{\phi(x)\over 4}}\,\big[h(x)\big]^{-{1\over 4}}\,\,\sqrt{G(x)}
\over
\sqrt{1-e^{{1\over 2}\,(\phi(x))-\phi_*)}\,\,\big({h(x)\over h_*}\big)^{1\over 2}}
}\,-\,{2\,L_0\,e^{-{\phi_0\over 4}}\over b}\,\log ({\cal C}\,x_{max})\,\,,
\label{renormalized_geodesic}
\eeq
where ${\cal C}$ is a constant to be fixed by  choosing a suitable normalization condition for the correlator. 

Our background interpolates between two limiting  $AdS_4$ geometries, at the UV and IR, with different radii. For an equal-time two-point function 
$\Big\langle {\cal O}(t, l)\,{\cal O}(t, 0)\Big\rangle$ the UV and IR limits should correspond to the cases in which $r_q\,l$ is small or large, respectively.  At the two endpoints  of the flow,  the theory is conformal invariant and the two-point correlator behaves as a power law in  $l$. We can use this fact to fix the normalization constant ${\cal C}$ in (\ref{renormalized_geodesic}). Actually, we will assume that 
 the field ${\cal O}$  is canonically normalized in the short-distance $r_q\,l\to 0$ limit and, therefore, the  UV limit of the two-point correlator is:
\beq
\Big\langle {\cal O}(t, l)\,{\cal O}(t, 0)\Big\rangle_{UV}\,=\,
{1\over  \big(r_q\,l/ \sqrt{\lambda} \big)^{2\Delta_{UV}}}\,\,,
\qquad\qquad
(r_q\,l\to 0)\,\,,
\label{UV_two-point}
\eeq
where the $\sqrt{\lambda}$ and $r_q$  factors have been introduced for convenience. 
In (\ref{UV_two-point}) $\Delta_{UV}$ is the conformal dimension of the operator ${\cal O}$ in the UV CFT, which for the dual of  a bulk field of mass $m$ is:
\beq
\Delta_{UV}\,=\,m\,L_0\,e^{-{\phi_0\over 4}}\,\,,
\label{DeltaUV}
\eeq
where we have taken into account that $m$ is large and that $L_0\,e^{-{\phi_0\over 4}}$ is the $AdS_4$ radius of the UV massless flavored geometry in the Einstein frame. It is shown in appendix \ref{asymp_two-point_functions} that, indeed,  the correlators derived from (\ref{renormalized_geodesic}) display the canonical form (\ref{UV_two-point}) if the constant ${\cal C}$ is chosen appropriately (see (\ref{calC_value})).  In appendix \ref{asymp_two-point_functions} we have also computed the first deviation from the conformal UV behavior. In this case the numerator on the right-hand side of (\ref{UV_two-point}) is not one but a function $f_{\Delta}(r_q\,l/\sqrt{\lambda})$ such that $f_{\Delta}(r_q\,l/\sqrt{\lambda}=0)=1$. We show in appendix 
\ref{asymp_two-point_functions} that  $f_{\Delta}(r_q\,l/\sqrt{\lambda})-1\propto (r_q\,l/\sqrt{\lambda})^{2b}$ for small 
$r_q\,l$. The explicit form of the first correction to the non-conformal behavior can be computed analytically  from the mass corrections of section \ref{UV_mass_corrections} and appendix \ref{UV_asymptotics} (see eqs. (\ref{nearUV_corr})-(\ref{cDelta})). 
\begin{figure}[ht]
\center
\includegraphics[width=0.75\textwidth]{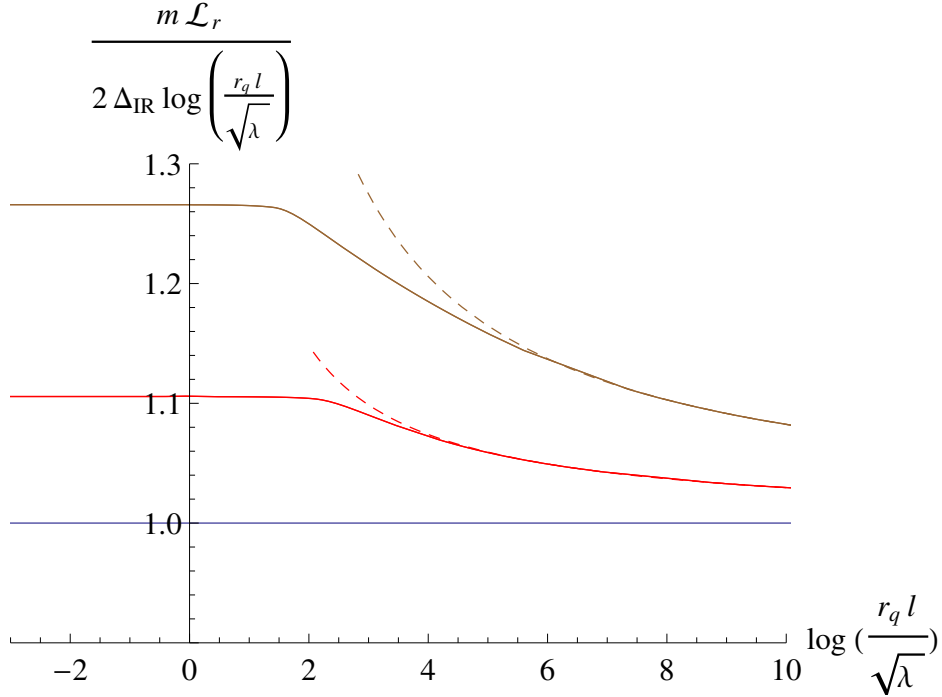}
\caption{Plot of $m\,{\cal L}_r/\big(2\Delta_{IR}\,\log (r_q\,l/\sqrt{\lambda})\big)$ versus the logarithm of 
$r_q\,l/\sqrt{\lambda}$. Notice that, according to (\ref{OO-vev}), 
$m\,{\cal L}_r=-\log \langle {\cal O}(t, l)\,{\cal O}(t, 0)\rangle$. The  three curves correspond to $\hat\epsilon=0$ (bottom, blue),  $\hat\epsilon=1$ (middle, red), and
$\hat\epsilon=9$ (top, brown). In the deep UV ($r_q\,l\to 0$) the curves approach the constant value $\Delta_{UV}/\Delta_{IR}$. The dashed curves correspond to the behavior 
(\ref{VEV_IR}), with the normalization constant ${\cal N}$ given in (\ref{Cal_N}).
}
\label{two-point_function}
\end{figure}

When the distance $r_q\,l$ is large the theory reaches a  new conformal point. Accordingly, the two-point function should behave again as a power law. Notice, however, that the conformal dimension $\Delta_{IR}$ in the IR of an operator dual to a particle of mass $m$ is different from the UV value (\ref{DeltaUV}). Indeed, in the IR the conformal dimension $\Delta_{IR}$ for an operator ${\cal O}$ of mass $m$ is the one corresponding to the unflavored ABJM theory,
\beq
\Delta_{IR}\,=\,m\,L_{ABJM}\,e^{-{\phi_{ABJM}\over 4}}\,\,,
\label{Delta_IR}
\eeq
where $L_{ABJM}$ and $\phi_{ABJM}$ are given, respectively, in (\ref{ABJM-AdSradius3}) and (\ref{ABJMdilaton3}). Actually, one can check that $\Delta_{UV}\ge \Delta_{IR}$ and that $\Delta_{UV}/ \Delta_{IR}\propto \hat\epsilon^{{1\over 16}}$ for large values of the deformation parameter $\hat\epsilon$. The calculation of the two-point function in the IR limit of large $r_q\,l$ is performed in detail in appendix  
\ref{asymp_two-point_functions}, with the result:
\beq
\Big\langle {\cal O}(t, l)\,{\cal O}(t, 0)\Big\rangle_{IR}\,=\,
{{\cal N}\over  \big(r_q\,l/ \sqrt{\lambda} \big)^{2\Delta_{IR}}}\,\,,
\qquad\qquad
(r_q\,l \to \infty)\,\,,
\label{VEV_IR}
\eeq
where ${\cal N}$ is a constant whose analytic expression is written  in (\ref{Cal_N}). Notice that 
${\cal N}\not=1$ due to our choice of the constant ${\cal C}$ in (\ref{renormalized_geodesic}), which corresponds to imposing the canonical normalization (\ref{UV_two-point}) to the two-point function in the UV regime. 

The results obtained by the numerical evaluation of the integral (\ref{renormalized_geodesic}) interpolate smoothly between the conformal behaviors (\ref{UV_two-point}) and (\ref{VEV_IR}). This is shown in Fig. \ref{two-point_function}, where we plot 
$-\log \langle {\cal O}(t, l)\,{\cal O}(t, 0)\rangle/(2\Delta_{IR}\,\log(r_q\,l/\sqrt{\lambda}))$ as a function of $\log(r_q\,l/\sqrt{\lambda})$. For small values of $r_q\,l/\sqrt{\lambda}$ the curve asymptotes to the ratio  $\Delta_{UV}/\Delta_{IR}$ of conformal  dimensions, in agreement with (\ref{UV_two-point}), whereas for large  $r_q\,l/\sqrt{\lambda}$ we recover the IR behavior (\ref{VEV_IR}).

\section{Meson spectrum}
\label{mesonsmassive}
Let us now test the flow encoded in our geometry by analyzing the mass spectrum of $q\bar q$ bound states. We will loosely refer to these bound states as mesons, although our background is not confining and quarkonia would be a more appropriate name for them. To carry out our analysis we will introduce additional external quarks, with a mass $\mu_q$ not necessarily equal to the mass $m_q$ of the quarks which backreact on the geometry. To distinguish between the two types of flavors we will call valence quarks to the additional ones, whereas the unquenched $N_f$ dynamical flavors of the geometry  will be referred to as sea quarks. The ratio $\mu_q/m_q$ of the masses of the two types of quarks will be an important quantity in what follows. Indeed, $\mu_q/m_q$ is the natural parameter for the holographic renormalization group trajectory.  When  $\mu_q/m_q$  is large (small) we expect to reach a UV (IR) conformal fixed point, whereas for intermediate values of this mass ratio the theory should flow in such a way that the  screening effects produced by the  sea quarks decrease as we move towards the IR. 

Within the context of the gauge/gravity  duality, 
the valence quarks can be introduced by adding an additional flavor D6-brane, which will be treated as a probe in the backreacted geometry. The mesonic mass spectrum can be obtained from the normalizable fluctuations of the D6-brane probe. The way in which the D6-brane probe is embedded in the ten-dimensional geometry preserving the supersymmetry of the background can be determined by using kappa symmetry. For fixed values of the Minkowski and holographic coordinates, the D6-brane extends over a cycle inside the ${\mathbb C}{\mathbb P}^3$ which has two directions along the 
${\mathbb S}^4$ base and one direction along the ${\mathbb S}^2$ fiber.  In order to specify further this configuration,  let us parameterize the $SU(2)$ left invariant one-forms $\omega_i$ of the four-sphere metric (\ref{S4metric3})  in terms of three angles $\hat\theta$, $\hat\varphi$ and $\hat \psi$,
\bear
\omega^1 & = & \cos\hat\psi\,d\,\hat\theta+\sin\hat\psi\,\sin\hat\theta\,d\hat\varphi\,\,, \rc
\omega^2 & = & \sin\hat\psi\,d\,\hat\theta-\cos\hat\psi\,\sin\hat\theta\,d\hat\varphi\,\,, \rc
\omega^3 & = & d\hat\psi+\cos\hat\theta \,d\hat\varphi\,\,,
\label{w1233}
\eear
with $0\le \hat\theta\le \pi$, $0\le\hat\varphi<2\pi$, $0\le\hat\psi \le 4\pi$.  Then, our D6-brane probe will be  extended along the Minkowski directions and embedded in the geometry in such a way that the angles $\hat\theta$ and $\hat\varphi$ are constant and that the angle $\theta$  of the ${\mathbb S}^2$ fiber depends on the holographic variable $x$. The pullbacks (denoted by a hat) of the left-invariant $SU(2)$ one-forms (\ref{w1233})  are $\hat \omega^{1}\,=\,\hat \omega^{2}\,=\,0$ and $\hat \omega^{3}\,=\,d\hat\psi$. 
The kappa symmetric configurations are those for which the function $\theta(x)$ satisfies the first order BPS equation \cite{Conde:2011sw}:
\beq
x\,{d\theta\over dx}\,=\,\cot\theta\,\,,
\eeq
which can be integrated as:
\beq
\cos\theta\,=\,{x_{*}\over x}\,\,.
\eeq
Here $x_*$ is the minimum value of the variable $x$ for the embedding, \ie, the value of $x$ for the tip of the brane. This minimum value of the coordinate $x$ for the embedding is related to the mass $\mu_q$ of the valence quarks introduced by the flavor probe. Indeed, by computing the Nambu-Goto action of a fundamental string stretched in the holographic direction between $x=0$ and $x=x_*$ we obtain $\mu_q$ as:
\beq
\mu_q\,=\,{1\over 2\pi\alpha'}\,\,\int_{0}^{x_*}\,\,
{e^{g(x)}\over x}\,dx\,\,.
\label{muq}
\eeq
In the following we will take the Regge slope $\alpha'=1$. Moreover,  to simplify the description of the embedding,  let us introduce the angular coordinate $\alpha$, defined as follows:
\beq
\xi\,=\,\tan \big({\alpha\over 2}\big)\,\,,
\eeq
and let us  define new angles  $\beta$ and $\psi$ as:
\beq
\beta\,=\,{\hat\psi\over 2}\,\,,\qquad\qquad
\psi\,=\,\varphi\,-\,{\hat\psi\over 2}\,\,,
\label{RP3-angles}
\eeq
where $\varphi$ is the angle in (\ref{cartesian_S23}). 
One can check that the ranges of the new angular variables are  $0\le\alpha <\pi$, $0\le \beta \,,\,\psi< 2\pi$. We will take the following set of worldvolume coordinates for the D6-brane:
\beq
\zeta^{a}\,=\,(x^{\mu}, x, \alpha, \beta, \psi)\,\,.
\eeq
Then, it is straightforward to verify that the  induced metric on the D6-brane worldvolume  takes the form:
\beq
ds^2_{7}\,=\,h^{-{1\over 2}}\,dx_{1,2}^2\,+\,{h^{{1\over 2}}\,e^{2g}\over x^2-x_*^2}\,
dx^2\,+\,h^{{1\over 2}}\,e^{2f}\,\Big[\,d\alpha^2\,+\,\sin^2\alpha\,d\beta^2\,\Big]\,+\,
(x^2-x_*^2)\,{h^{{1\over 2}}\,e^{2g}\over x^2}\,\big(d\psi\,+\,\cos\alpha\,d\beta\big)^2\,\,.
\label{induced_metric3}
\eeq
We will restrict ourselves to study a particular set of fluctuations of the D6-brane probe, namely the  fluctuations of the  worldvolume gauge field $A_{a}$. 
The equation for these fluctuations is:
\beq
\partial_{a}\,\Big[e^{-\phi}\,\sqrt{-\det g}\,\,g^{ac}\,g^{bd}\,F_{cd}\,\Big]\,=\,0\,\,,
\label{fluct_eqs}
\eeq
where $g_{ab}$ is the induced metric (\ref{induced_metric3}). More concretely,  we will study this equation for the following ansatz for $A_{a}$:
\beq
A_{\mu}\,=\,\xi_{\mu}\,e^{ik_{\nu} x^{\nu}}\,R(x)\,\,,\qquad
(\mu=0,1,2)\,\,,
\qquad\qquad
A_x\,=\,A_i\,=\,0\,\,,
\eeq
where  $\xi_{\mu}$ is a constant polarization vector and $A_i$ denote the components along the angular directions. These modes are dual to the vector mesons of the theory, with $k_{\mu}$ being  the momentum of the meson ($\eta^{\mu\nu}\,k_{\mu}\,k_{\nu}=-m^2$, with $m$ being the mass of the meson). The non-vanishing components of the field strength 
$F_{ab}$ are:
\beq
F_{\mu\nu}\,=\,i(k_{\mu}\,\xi_{\nu}\,-\,k_{\nu}\,\xi_{\mu})\,e^{ik_{\nu} x^{\nu}}\,R(x)\,\,,
\qquad\qquad
F_{x\mu}\,=\,\xi_{\mu}\,e^{ik_{\nu} x^{\nu}}\,R'(x)\,\,.
\eeq
The fluctuation equation (\ref{fluct_eqs}) is trivially satisfied when $b=i$, whereas for $b=x$ it is satisfied if the polarization is transverse:
\beq
\eta^{\mu\nu}\,k_{\mu}\,\xi_{\nu}\,=\,0\,\,.
\eeq
Moreover, by taking $b=\mu$ in (\ref{fluct_eqs}) we arrive at the following differential equation for the radial function $R$:
\beq
\partial_{x}\,\Big[\,{h^{{1\over 4}}\,e^{2f-\phi}\over x}\,(x^2-x_*^2)\,\partial_{x} R\,\Big]\,+\,
m^2\,{h^{{5\over 4}}\,e^{2f+2g-\phi}\over x}\,R\,=\,0\,\,.
\label{meson_fluct_eq}
\eeq
\begin{figure}[ht]
\center
\includegraphics[width=0.75\textwidth]{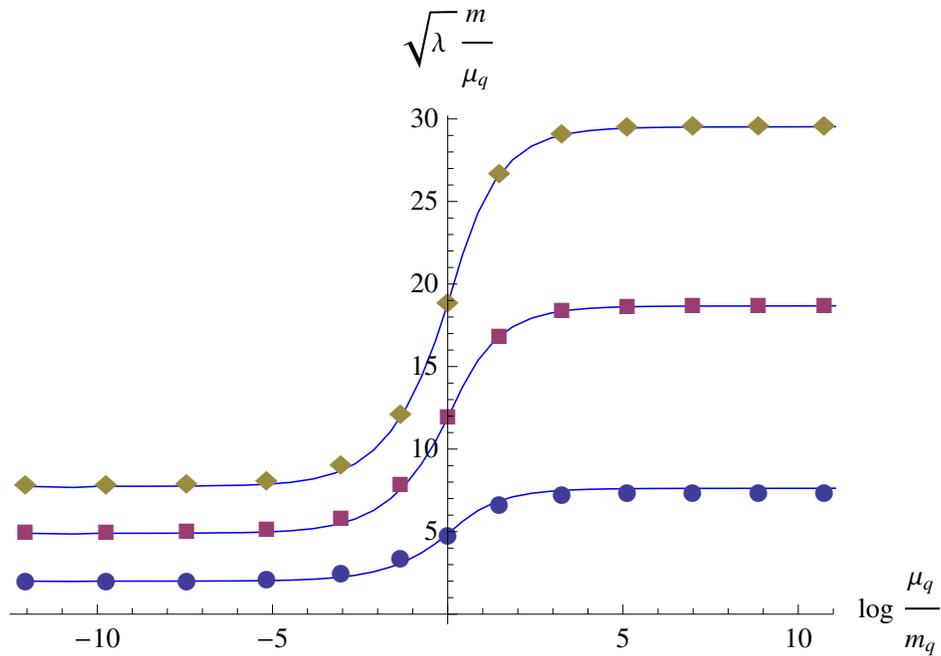}
\caption{Numerical values of the meson  masses for the first three levels  ($n=0,1,2$) as a function of the sea quark mass $m_q$ for deformation parameter $\hat\epsilon=9$. The solid curves depict  the WKB estimate (\ref{WKB_masses}).} 
\label{fig:masses}
\end{figure}

The mass levels correspond  to the   values of $m$ for which there are normalizable solutions of (\ref{meson_fluct_eq}). They can be obtained numerically by the shooting technique. One gets in this way a discrete spectrum depending on a quantization number $n$ ($n\in {\mathbb Z}$, $n\ge 0$). The numerical results for the first three levels are shown in Fig. \ref{fig:masses} as functions of the mass ratio
$\mu_q/m_q$. One notices in these results that the meson masses increase as we move from the IR ($\mu_q/m_q\to 0$) to the UV ($\mu_q/m_q\to \infty$). This non-trivial flow is due to the vacuum polarization effects of the sea quarks, which are enhanced as we move towards the UV and the sea quarks become effectively massless. This is the expected behavior of bound state masses for a theory in the Coulomb phase, since the screening effects reduce effectively the strength of the quark-antiquark force. 

One can get a very accurate description of the flow by applying the WKB approximation. The detailed calculation is presented in appendix \ref{appendix_WKB_masses}. The WKB formula for the mass spectrum is:
\beq
m_{WKB}\,=\,{\pi\over \sqrt{2}\, \xi(x_*)}\,\sqrt{(n+1)(2n+1)}\,\,,
\qquad\qquad (n=0,1,2,\cdots)\,\,,
\label{WKB_masses}
\eeq
where  $\xi(x_*)$ is the following integral:
\beq
\xi(x_*)\,=\,\int_{x_*}^{\infty}\,dx\,{e^{g(x)}\sqrt{h(x)}\over \sqrt{x^2-x_*^2}}\,\,.
\label{WKB_xi}
\eeq
The WKB mass levels (\ref{WKB_masses}) are compared with those obtained by the shooting technique in Fig. \ref{fig:masses}. We notice from these plots that the estimate  (\ref{WKB_masses}) describes rather well the numerical results along the flow. 
Moreover, we can use the UV and IR limits of the functions $g$ and $h$ to obtain the asymptotic form of the WKB spectrum at the endpoints of the flow. This analysis is performed in detail in appendix  \ref{appendix_WKB_masses} (see eqs. (\ref{m_WKB_UV}) and (\ref{m_WKB_IR})). As expected, in  the deep IR the mass levels coincide with those of the unflavored ABJM model. In this latter model the mass spectrum of vector mesons can be computed analytically since the fluctuation equation can be solved in terms of hypergeometric functions \cite{Hikida:2009tp,Jensen:2010vx}. When $\mu_q/m_q\to\infty$ the meson masses coincide with those obtained for the massless flavored model of \cite{Conde:2011sw}.

We can use the WKB formulas  (\ref{m_WKB_UV}) and (\ref{m_WKB_IR})  for the 
spectrum at the endpoints of the renormalization group flow to estimate the 
variation generated in the meson masses by changing the sea quark mass $m_q$ and switching on and off gradually the screening effects. It is interesting to point out that, within the WKB approximation, the ratio of these masses only depends on the number of flavors, and is given by:
\beq
{m_{_{WKB}}^{^{(UV)}}\over m_{_{WKB}}^{^{(IR)}}}\,=\,{\sqrt{\pi}\over 2}\,\,
{\Gamma\Big({b+1\over 2b}\Big)
\over 
\Gamma\Big({2b+1\over 2b}\Big)}\,\,{1\over \sigma}\,\,,
\label{UV-IR-mass_ratio}
\eeq
where $\sigma$ is the screening function defined in (\ref{screening-sigma}). As expected, when $N_f=0$ the right-hand side of (\ref{UV-IR-mass_ratio}) is equal to one, 
\ie,  there is no variation of the masses along the flow. On the contrary, when $N_f>0$ the UV/IR mass ratio in (\ref{UV-IR-mass_ratio}) is always greater than one, which means that the masses grow as we move towards the UV and the screening effects become more important. In appendix \ref{appendix_WKB_masses} we have expanded  
(\ref{UV-IR-mass_ratio}) for low values of the deformation parameter $\hat\epsilon$ (see (\ref{UV/IR_mass_ratio_lowNf})). Moreover, for large  $\hat\epsilon$ the UV/IR mass ratio grows as 
$\sqrt{\hat\epsilon}$ (see (\ref{UV/IR_mass_ratio_largeNf}) for the explicit formula).

\section{Discussion}
\label{conclu}

In this chapter we obtained a  holographic dual to Chern-Simons matter theory with unquenched flavor in the strongly-coupled Veneziano limit. The flavor degrees of freedom were added by a set of D6-branes smeared along the internal directions, which backreacted on the geometry by squashing it, while preserving ${\cal N}=1$ supersymmetry. We considered massive flavors and found a non-trivial holographic renormalization group flow connecting two scale-invariant fixed points: the unflavored ABJM theory at the IR and the massless flavored model at the UV. 

The quark mass $m_q$ played an important role as a control parameter of the solution. By increasing $m_q$ our solutions became closer to the unflavored ABJM model and we smoothly connected  the unquenched flavored model to the ABJM theory without fundamentals. After this soft introduction of flavor no pathological behavior was found. Indeed, our backgrounds had good IR and UV behaviors, contrary to what happens to other models with unquenched flavor \cite{Nunez:2010sf}. This made the ABJM model especially adequate to analyze the effects of unquenched fundamental matter in a holographic setup. 

We analyzed different flavor effects in our model. In general, the screening effects due to loops of fundamentals were controlled by the relative value of the quark mass $m_q$ with respect to the characteristic length scale $l$ of the observable. If $m_q\,l$ was small, which corresponds to the UV regime, the flavor effects were important, whereas they were suppressed if $m_q\,l$ is large, \ie, at the IR. Among the different observables that we analyzed, the holographic entanglement entropy for a disk was specially appropriate since it counts precisely the effective number of degrees of freedom which are relevant at the length scale  given by the radius of the disk. By using the refined entanglement entropy ${\cal F}$ introduced in \cite{Liu:2012eea}, we explicitly obtained the running of ${\cal F}$ and  verified the reduction of the number of degrees at the IR that was mentioned above. The other observables studied also supported this picture.  \newline

The results obtained in this chapter can be generalized in several directions. One possible generalization could be the construction of a black hole for the unquenched massive flavor. Such a background could serve to study the meson melting phase transition which occurs when the tip of the brane approaches the horizon. This system was studied in \cite{Jokela:2012dw}, in the case in which  the massive flavors are quenched and the corresponding flavor brane is a probe.  Another possibility would be trying to find a gravity dual of a theory in which the sum of the two Chern-Simons levels is non-vanishing. According to \cite{Gaiotto:2009mv} we should find a flavored solution of type  IIA supergravity with non-zero Romans mass. 

Moreover, to make contact with any condensed matter physics system, one is forced to consider non-vanishing components of the gauge field in the background or at the probe level.  As an initial step in this direction, in chapter \ref{chapterfour} we consider a probe D6-brane with electric field, magnetic field, charge density, currents and internal flux, in the ABJM background, and a gravity dual of a fractional quantum Hall system is constructed. Also, by adding only charge density to the probe D6-brane worldvolume, quantum phase transitions are studied in chapter \ref{chapterfive}. A possible generalization of the studies performed in chapters \ref{chapterfour} and  \ref{chapterfive} would be to consider the same probe D6-brane configuration with worldvolume gauge fields embedded in the background obtained in this chapter. A further generalization, and much more complicated, would be to consider the backreaction of the D6-branes with worldvolume gauge fields turned on.

\begin{subappendices}

\section{Appendices}
\label{appendixsusy}

\setcounter{equation}{0}

\subsection{BPS equations}
\label{BPS}

In this appendix we will derive the master equation (\ref{master_eq_W}), as well as the equations  that allow to construct the metric and dilaton from the master function $W(x)$ (\ie, (\ref{g-f-W}), (\ref{warp_factor_x}), and (\ref{dilaton_x})). 

Let us begin by writing the BPS equations that guarantee the preservation of ${\cal N}=1$ SUSY. They can be written in terms of the function $\Lambda$ introduced in \cite{Conde:2011sw}, which is  defined as the following combination  of the dilaton and the warp factor:
\beq
e^{\Lambda}\equiv e^{\phi}\,h^{-{1\over 4}}\,\,.
\label{Lambda-def}
\eeq
Then, it was proved in \cite{Conde:2011sw} that $\Lambda(r)$, $f(r)$, and $g(r)$ are solutions to the following system of first-order differential equations:
\bear
&&{d\Lambda\over dr}\,=\,k\,\eta\,e^{\Lambda-2f}\,-\,
{k\over 2}\,e^{\Lambda-2g}\,\,,\rc\rc
&&{d f\over dr}\,=\,{k\,\eta\over 4}\,e^{\Lambda-2f}\,-\,
{k\over 4}\,e^{\Lambda-2g}\,+\,e^{-2f+g}\,\,,\rc\rc
&&{d g\over dr}\,=\,{k\,\eta\over 2}\,e^{\Lambda-2f}\,+\,e^{-g}\,-\,e^{-2f+g}\,\,.
\label{3-system-flavored}
\eear
Moreover, the warp factor $h(r)$ can be recovered from $\Lambda$, $f$, and $g$ through:
\beq
h(r)\,=\,\,e^{-\Lambda(r)}\,\,\Big[\,\alpha\,-\,3\pi^2N\,
\int^r\,\,
e^{2\Lambda(z)-4f(z)-2g(z)}\,\,dz\,\,\Big]\,\,,
\label{warp-integral}
\eeq
where $\alpha$ is an integration constant. Given $h$ and $\Lambda$, the dilaton $\phi$ is obtained from (\ref{Lambda-def}).  The function $K$ of the RR four-form can be related to the other functions of the background by using (\ref{K-N}).  Alternatively, $K$ can be obtained from the BPS system as:
\beq
K\,=\,{d\over dr}\,\,\Big(\,e^{-\phi}\,\,h^{-{3\over 4}}\,\Big)\,\,.
\label{K-phi-h}
\eeq
In terms of the $x$ variable defined in (\ref{r-x-diff-eq}), the BPS system (\ref{3-system-flavored}) becomes:
\bear
&&x\,{d\Lambda\over dx}=k\,\eta\,e^{\Lambda-2f+g}-\frac{k}{2}\,e^{\Lambda-g}\,,\rc\rc
&&x\,{df\over dx}=\frac{k}{4}\,\eta\,e^{\Lambda-2f+g}-\frac{k}{4}\,e^{\Lambda-g}+e^{-2f+2g}\,,\rc\rc
&&x\,{dg\over dx}=\frac{k}{2}\,\eta\,e^{\Lambda-2f+g}+1-e^{-2f+2g}\,.
\label{BPS-system-x}
\eear
In order to reduce this system, let us define as in \cite{Conde:2011sw} the functions $\Sigma(x)$ and $\Delta(x)$,
\beq
\Sigma\,\equiv\, \Lambda-f\,\,,
\qquad\qquad
\Delta\,\equiv\, f-g\,\,.
\eeq
Then, one can easily show that $\Sigma(x)$ and $\Delta(x)$ satisfy the system:
\bear
&&x\,{d\Sigma\over dx}={k\over 4}\,e^{\Sigma}\left(3\eta\,e^{-\Delta}-e^{\Delta}\right)-e^{-2\Delta}\,,\rc\rc
&&x\,{d\Delta\over dx}=-{k\over 4}\,e^{\Sigma}\left(\eta\,e^{-\Delta}+e^{\Delta}\right)-1+2e^{-2\Delta}\,,
\label{system-Sigma-Delta}
\eear
whereas $g$ can be obtained from $\Sigma$ and $\Delta$ by integrating the equation:
\begin{equation}
x\,{d g\over dx}={k\over 2}\,\eta\,e^{\Sigma-\Delta}+1-e^{-2\Delta}\,.
\label{eqn:rec.g}
\end{equation}

Let us next define the master function  $W(x)$ as in (\ref{W_definition}). One immediately verifies that, in terms of the 
functions $\Delta$ and $\Sigma$,  this definition is equivalent to
\beq
W(x)= {4\over k}\,e^{\Delta-\Sigma}\,x\,\,.
\label{W-def}
\eeq
By computing the derivative of (\ref{W-def}) and using the BPS system (\ref{system-Sigma-Delta}), one can easily prove that:
\beq
{dW\over dx}\,=\,{12\over k}\,e^{-\Sigma-\Delta}\,-\,4\eta\,\,.
\label{W-prime}
\eeq
From (\ref{W-prime}) one immediately finds:
\beq
e^{\Sigma+\Delta}\,=\,{12\over k}\,{1\over W'+4\eta}\,\,,
\label{e-Sigma+Delta}
\eeq
where the prime denotes derivative with respect to $x$. Moreover, from the BPS system we can calculate the derivative of $\Sigma+\Delta$ and write the result as:
\beq
x\,{d\over dx}\,\Big(e^{\Sigma+\Delta}\Big)\,=\,
{2x\,\eta\, e^{\Sigma+\Delta}\over W}\,-\,
\Big[{k\over 2}\,e^{\Sigma+\Delta}\,+\,1\Big]\,e^{\Sigma+\Delta}\,+\,{4\over k}\,{x\over W}\,\,.
\label{d-Sigma+Delta}
\eeq
Plugging (\ref{e-Sigma+Delta}) into (\ref{d-Sigma+Delta}), we arrive at the following second-order equation for $W(x)$:
\beq
x\,{d\over dx}\,\Bigg(
{1\over W'+4\eta}\Bigg)\,+\,{W'+4\eta+6\over (W'+4\eta)^2}\,-\,
{x\over 3}\,{W'+10\eta\over W(W'+4\eta)}
\,=\,0\,\,,
\eeq
which can be straightforwardly shown to be equivalent to the master equation (\ref{master_eq_W}). 

Let us see now how one can reconstruct  the full solution from the knowledge of the function $W(x)$. First of all, we notice that from the expression of $W$ in (\ref{W-def}), we get:
\beq
e^{\Sigma-\Delta}\,=\,{4\over k}\,{x\over W}\,\,.
\eeq
By combining this expression with  (\ref{e-Sigma+Delta}) we obtain $\Sigma$ and $\Delta$:
\beq
e^{2\Sigma}\,=\,{48\over k^2}\,\,{x\over W(W'+4\eta)}\,\,,
\qquad\qquad
e^{2\Delta}\,=\,{3 W\over x(W'+4\eta)}\,\,.
\label{Sigma-Delta-q}
\eeq
By noticing that $e^{2\Delta}=q$ we arrive at the representation  of the squashing function $q$ written in 
(\ref{q_W_Wprime}). Moreover, by using this result in (\ref{eqn:rec.g}), we obtain the differential equation satisfied by $g$:
\beq
{dg\over dx}\,=\,{2\eta\over 3W}\,-\,{W'\over 3W}\,+\,{1\over x}\,\,,
\label{g_diff_eq_in_x}
\eeq
which allows to obtain $g(x)$ once $W(x)$ is known. The result of this integration is just the expression written in (\ref{g-f-W}). Moreover, taking into account the expression of the squashing factor $q$  we get precisely the expression of $f$ written in (\ref{g-f-W}).

Let us now compute $\Lambda$ by using $\Lambda=\Sigma+\Delta+g$ and (\ref{e-Sigma+Delta}). We get:
\beq
e^{\Lambda}\,=\,{12\over k}\,{e^{g}\over W'+4\eta}\,\,,
\eeq
and, after using (\ref{g-f-W}), we arrive at:
\beq
e^{\Lambda}\,=\,{12\over k}\,
{x\over W^{{1\over 3}} (W'+4\eta)}\,\exp\Big[{2\over 3}\int^x\,{\eta(\xi)d\xi\over W(\xi)}\Big]\,\,.
\label{Lambda-W}
\eeq
By using this result and (\ref{g-f-W}) in (\ref{warp-integral}), we get that the warp factor can be written as in (\ref{warp_factor_x}). The expression (\ref{dilaton_x}) for the dilaton is just a consequence of the definition  of $\Lambda$ in (\ref{Lambda-def}) and of (\ref{Lambda-W}).

\subsubsection{Equations of motion}

Let us now verify that the first-order BPS system (\ref{3-system-flavored}) implies the second-order equations of motion for the different fields. Let us work in Einstein frame and write the total action as:
\beq
S\,=\,S_{IIA}\,+\,S_{{\rm sources}}\,\,,
\label{total-action}
\eeq
where the action of type IIA supergravity is given by:
\beq
S_{IIA}\,=\,{1\over 2\kappa_{10}^2}\,\,\Bigg[\,
\int \sqrt{-g}\,\Big(R\,-\,{1\over 2}\,\partial_{\mu}\phi\,\partial^{\mu}\,\phi\,\Big)\,-\,
{1\over 2}\,\int\,\Big[\,e^{{3\phi\over 2}}\,\,{}^*F_2\wedge F_2+
e^{{\phi\over 2}}\,\,{}^*F_4\wedge F_4\Big]\Bigg]\,\,,
\label{IIA-action}
\eeq
and the source contribution is the DBI+WZ action for the set of smeared D6-branes. Let us write this last action as in \cite{Conde:2011sw}. First of all, we introduce a charge distribution three-form $\Omega$. Then, the DBI+WZ action 
is given by:
\beq
S_{{\rm sources}}\,=\,-T_{D_6}\,\int\,\Big(\,e^{{3\phi\over 4}}\,{\cal K}\,-\,C_7\,
\Big)\,\wedge\,\Omega\,\,,
\label{source-action}
\eeq
where the DBI term has been written in terms of the so-called calibration form (denoted by ${\cal  K}$), whose pullback to the worldvolume is equal to the induced volume form  for the supersymmetric embeddings. The expression of  ${\cal  K}$ has been written in \cite{Conde:2011sw}. Let us reproduce it here for completeness:
\beq
{\cal K}\,=\,-e^{012}\,\wedge\big(\,
e^{3458}\,-\,e^{3469}\,+\, e^{3579}\,+\,e^{3678}\,+\,e^{4567}\,+\,e^{4789}\,+\,e^{5689}
\,\big)\,\,,
\label{cal-K-explicit}
\eeq
where the $e^i$'s are the one-forms of the basis corresponding to the forms (\ref{Es3}) and (\ref{calS3}) (see \cite{Conde:2011sw} for further details). Notice that the equation of motion for $C_7$ derived from (\ref{total-action}) is just $dF_2=2\pi\,\Omega$. Therefore, the $\Omega$ for our ansatz can be read from the right-hand side of (\ref{massive-Omega}).

The Maxwell equations for the forms $F_2$ and $F_4$ derived from (\ref{total-action}) are:
\beq
d\,\Big(\,e^{{3\phi\over 2}}\,\,{}^*F_2\,\Big)\,=\,0\,\,,\qquad\qquad
d\,\Big(\,e^{{\phi\over 2}}\,\,{}^*F_4\,\Big)\,=\,0\,\,,
\label{Maxwell_F2_F4}
\eeq
while the equation for the dilaton is:
\beq
d\,{}^*d\phi\,=\,{3\over 4}\,\,e^{{3\phi\over 2}}\,\,{}^*F_2\wedge F_2+{1\over 4}\,
e^{{\phi\over 2}}\,\,{}^*F_4\wedge F_4\,+\,
{3\over 2}\,\kappa_{10}^2\,T_{D_6}\,\,e^{{3\phi\over 4}}\,{\cal K}\wedge\,\Omega\,\,.
\label{dilaton_eom}
\eeq
One can verify that, for our ansatz, (\ref{Maxwell_F2_F4}) and (\ref{dilaton_eom}) are a consequence of the BPS equations (\ref{3-system-flavored}). To carry out this verification we need to know the radial derivatives of $h(r)$ and $\phi(r)$ (which are not written in (\ref{3-system-flavored})). The derivative of $h$ can be related to the derivative of $\Lambda(r)$,
\beq
{dh\over dr}\,=\,-h\,{d\Lambda\over dr}\,-\,3\pi^2\,N\,e^{\Lambda\,-\,4f\,-\,2g}\,\,.
\eeq
The radial derivative of the dilaton can be put in terms of the derivative of $\Lambda$ and $h$ by using (\ref{Lambda-def}):
\beq
{d\phi\over dr}\,=\,{d\Lambda\over dr}\,+\,{1\over 4h}\,{dh\over dr}\,\,.
\eeq
It remains to check Einstein equations, which read:
\bear
&&R_{\mu\nu}\,-\,{1\over 2}\,g_{\mu\nu}\,R\,=\,
{1\over 2}\,\partial_{\mu}\phi\,\partial_{\nu}\phi\,-\,{1\over 4}\,g_{\mu\nu}\,
\partial_{\rho}\phi\,\partial^{\rho}\phi\,+\,
{1\over 4}\,e^{{3\phi\over 2}}\,\Big[\,2F_{\mu\rho}^{(2)}\,F_{\nu}^{(2)\,\,\rho}\,-\,{1\over 2}\,g_{\mu\nu}\,F_2^2\,\Big]\rc\rc
&&\qquad\qquad\qquad\qquad
+{1\over 48}\,
e^{{\phi\over 2}}\,\Big[\,4F_{\mu\rho\sigma\lambda}^{(4)}\,F_{\nu}^{(4)\,\,\rho\sigma\lambda}\,-\,
{1\over 2}\,g_{\mu\nu}\,F_4^2\,\Big]\,+\,T_{\mu\nu}^{{\rm sources}}\,\,,
\label{Einstein-eq}
\eear
where $T_{\mu\nu}^{{\rm sources}}$ is the stress-energy tensor for the flavor branes, which is defined as:
\beq
T_{\mu\nu}^{{\rm sources}}\,=\,-{2\kappa_{10}^2\over \sqrt{-g}}\,\,
{\delta S_{{\rm sources}}\over \delta g^{\mu\nu}}\,\,.
\label{Tmunu_def}
\eeq
In order to write the explicit expression for $T_{\mu\nu}^{{\rm sources}}$ derived from the definition (\ref{Tmunu_def}), let us introduce the following operation for any two $p$-forms $\omega_{(p)}$ and 
$\lambda_{(p)}$:
\beq
	\omega_{p} \lrcorner \lambda_{(p)} = \frac{1}{p!} \omega^{\mu_1 ... \mu_p}
	 \lambda_{\mu_1 ... \mu_p}\,\,.
\eeq
Then, by computing explicitly the derivative of the action (\ref{source-action}) with respect to the metric,  one can check that:
\beq
T_{\mu\nu}^{{\rm sources}}\,=\,\kappa_{10}^2\,T_{D_6}\,\,e^{{3\phi\over 4}}\,
\Big[\,\,g_{\mu\nu}\,
{}^*{\cal K} \lrcorner\,\Omega\,-\,
{1\over 2}\,\Omega_{\mu}^{\,\,\,\rho\sigma}\,\big({}^*{\cal K}\big)_{\nu\rho\sigma}
\Big]\,\,.
\eeq
It is now straightforward to compute explicitly the different components of this tensor. Written in  flat components in the basis in which the calibration form has the form (\ref{cal-K-explicit}), we get:\footnote{As compared to the  case studied in \cite{Conde:2011sw}, now we have terms proportional to $d\eta/dr$ that were absent for massless flavors.  
}
\begin{align}
&T_{00}=-T_{11}=-T_{22}=k\Big(\,\eta-1+\frac{e^g}{2}\,{d\eta\over d r}\,\Big)\,e^{-2f-g+{3\phi\over 2}}\,\,h^{-{3\over 4}}\,\,,\rc
&T_{33}=k(\eta-1)\,e^{-2f-g+{3\phi\over 2}}\,\,h^{-{3\over 4}}\,\,,\rc
&T_{ab}=-{k\over 2}\,\Big(\,\eta-1+\frac{e^g}{2}\,{d\eta\over d r}\,\Big)\,e^{-2f-g+{3\phi\over 2}}\,\,h^{-{3\over 4}}\delta_{ab}\,\,,
\qquad\qquad (a,b=4,\ldots, 7)\,,\rc
&T_{88}=T_{99}=-{k\over 2}\,\Big(\,\eta-1+e^g\,{d\eta\over d r}\,\Big)\,e^{-2f-g+{3\phi\over 2}}\,\,h^{-{3\over 4}}\,.
\end{align}
By using these values one can verify that,  the Einstein equations (\ref{Einstein-eq}) are indeed satisfied as a consequence of the first-order system (\ref{3-system-flavored}). Notice that, for the massive flavored background of section \ref{massive_flavor},  $d\eta/dr $ (and, therefore, $T_{\mu\nu}$) has a finite discontinuity at $r=r_q$. It follows from (\ref{Einstein-eq}) that the Ricci tensor $R_{\mu\nu}$  has also a finite jump at this point.


\subsection{Mass corrections in the UV }
\label{UV_asymptotics}

In this appendix we show how to obtain the first corrections to the conformal behavior of the metric and the dilaton from the UV asymptotic expansion of the master function $W(x)$ written in (\ref{W_asymp_UV}). First we notice that the ratio $\eta_0/ A_0$ can be written in terms of $b$ as:
\beq
{\eta_0\over A_0}\,=\,{3\over 2b}\,-\,1\,\,,
\label{eta0/A0}
\eeq
leading to  a useful identity:
\beq
x^{{1\over b}\,-\,{2\over 3}}\,\,\exp\Big[\,-{2\over 3}\,\int_{1}^{x}\,{\eta_0\over A_0}\,
{d\xi\over \xi}\,\Big]\,=\,1\,\,.
\eeq
Inserting the unit written in this way in the integral appearing in the expression of $e^{g}$ in (\ref{background_functions_xge1}), we get:
\beq
\exp\Big[{2\over 3}\int^x_1\,{\eta(\xi)d\xi\over W(\xi)}\Big]\,=\,x^{{1\over b}\,-\,{2\over 3}}\,\,
J\,\,{\cal G}(x)\,\,,
\eeq
where $J$  is the following constant integral (depending on $\hat\epsilon$):
\beq
J\,\equiv\,\exp\Bigg[{2\over 3}\int^{\infty}_1\,\Big[{\eta(\xi)\over W(\xi)}\,-\,
{\eta_0\over A_0\,\xi}\Big]
d\xi\,\Bigg]\,\,,
\eeq
and ${\cal G}(x)$ is the function
\beq
{\cal G}(x)\equiv 
\exp\Bigg[{2\over 3}\int^{x}_{\infty}\,\Big[{\eta(\xi)\over W(\xi)}\,-\,
{\eta_0\over A_0\,\xi}\Big]
d\xi\,\Bigg]\,\,.
\eeq
Using these results, we can immediately write:
\beq
e^{g}\,=\, r_q\,\Big[{(\hat \gamma+1)^2\over 2\,\hat \gamma}\Big]^{{1\over 3}}\,\,
J\,x^{{1\over b}}\,\,
\Big[{x\over W(x)}\Big]^{{1\over 3}}\,\,{\cal G}(x)
\,\,.
\eeq
The function ${\cal G}(x)$ can be easily expanded for large $x$. At first non-trivial order we get:
\beq
{\cal G}(x)\,=\, 1\,+\,{1\over 3 A_0}\,\Big[\,\eta_0\,-\,1\,+A_2\,{\eta_0\over A_0}\,\Big]\,\,
{1\over x^2}\,+\,\cdots\,\,.
\eeq
Using that, for large $x$,
\beq
\Big[{x\over W(x)}\Big]^{{1\over 3}}\,=\,
{1\over A_0^{{1\over 3}}}\,
\Big[\,1\,-\,{A_2\over 3\,A_0}\,{1\over x^2}\,+\,\cdots\Big]\,\,,
\eeq
we easily arrive at the expansion of $e^g$ written in (\ref{g_f_UVexpansions_x}). 
Let us next  find the large $x$ expansion for $f$, which can be obtained by using  the expansion of $g$ and $q$ in the relation $e^{f}=\sqrt{q}\,e^{g}$. We get:
\beq
e^{f}\,=\, \sqrt{q_0}\,\,
{\kappa\,\, r_q\over b}\,\,
J\,x^{{1\over b}}\,\,
\Big[\,1\,+\,{f_2\over x^2}\,+\,\cdots\Big]\,\,,
\eeq
where the coefficient $f_2$ is given by:
\beq
f_2\,=\,g_2\,+\,{1\over 2}\,{q_2\over q_0}\,=\,g_2\,+\,{2-b\over 2b}\,q_2\,\,.
\label{f2_g2_q2}
\eeq
It is straightforward to demonstrate that (\ref{f2_g2_q2}) coincides with the value of $f_2$ written in (\ref{g2_f2}).

Let us now analyze the UV expansion of $h$. First, we evaluate the integral appearing in 
(\ref{warp_factor_x}):
\beq
\int_{x}^{\infty}\,
{\xi\,e^{-3 g(\xi)}\over W(\xi)^2}
\,d\xi\,=\,
{2\hat \gamma\,b\over 3\,(\hat \gamma+1)^2\,A_0\,(Jr_q)^3}\,\,
x^{-{3\over b}}\,
\Big[\,1\,-{3\over 3+2b}\,\Big(3g_2+2 {A_2\over A_0}\Big)\,{1\over x^2}\,+\,\cdots
\Big]\,\,.
\eeq
By combining this result with the expansions (\ref{g_f_UVexpansions_x}) and (\ref{Wprime-eta-expansion}) we get:
\beq
h(x)\,=\,L_0^4\,
\Big[{2\,\hat \gamma\,A_0\over (\hat \gamma+1)^2 }\Big]^{{4\over 3}}\,\,
{x^{-{4\over b}}\over (J\,b\,r_q)^4}\,\Big[1\,+\,{h_2\over x^2}\,+\,\cdots\Big]\,\,,
\label{UVexpansion_h_x}
\eeq
where $L_0$ is the UV AdS  radius (\ref{L0_explicit}) and the coefficient $h_2$ is:
\beq
h_2\,=\,-{2(6+b)\over 3+2b}\,g_2\,-\,\Big({6\over 3+2b}\,+\,{b\over 3(2-b)}\Big)\,{A_2\over A_0}\,-\,{2\over 3}\,{3-2b\over 2-b}\,{\eta_0-1\over \eta_0}\,\,.
\label{h2_first}
\eeq
One can readily check that the prefactor in (\ref{UVexpansion_h_x}) coincides with the one in (\ref{UVexpansion_h_phi}) and that the coefficient $h_2$ written in  (\ref{h2_first}) is the same as the one in (\ref{h_2_phi_2}).

Let us now obtain the UV expansion of the function $\Lambda(x)$, defined in (\ref{Lambda-def}), which can be written as:
\beq
e^{\Lambda(x)}\,=\,{12\,r_q\,J\over k}\,\,
\Big[{(\hat \gamma+1)^2\over 2\,\hat \gamma}\Big]^{{1\over 3}}\,\,
{x^{{1\over b}}\over W'+4\eta}
\,\,\Big[{x\over W(x)}\Big]^{{1\over 3}}\,\,{\cal G}(x)\,\,.
\eeq
The UV expansion of this expression is:
\beq
e^{\Lambda}\,=\, {12\,r_q\,J\over k\,(A_0+4\eta_0)}\,\,
\Big[{(\hat \gamma+1)^2\over 2\,\hat \gamma\,A_0}\Big]^{{1\over 3}}\,\,
x^{{1\over b}}\,\,
\Big(1\,+\,{\Lambda_2\over x^2}\,+\,\cdots\Big)\,\,,
\eeq
where the coefficient $\Lambda_2$ can be written in terms of $g_2$ as:
\beq
\Lambda_2\,=\,g_2\,+\,{A_2+4(\eta_0-1)\over A_0+4\eta_0}\,=\,
g_2\,+\,{b\over 3(2-b)}\,\Big[{2(3-2b)\over b}\,{\eta_0-1\over \eta_0}\,+\,
{A_2\over A_0}\,\Big]\,\,.
\eeq
By using the value  of $g_2$ in (\ref{g2_f2}), we can find a more explicit expression for  $\Lambda_2$:
\beq
\Lambda_2\,=\,{3-2b\over 3}\,\Big[\,{1\over 2b}\,+\,{2\over 2-b}\,\Big]\,{\eta_0-1\over \eta_0}\,+\,{1\over 3}\Big[\,{3-4b\over 2b}\,+\,{b\over 2-b}\Big]\,\,{A_2\over A_0}\,\,.
\eeq
Let us next obtain the expansion of $e^{\phi}$ by using $e^{\phi}=e^{\Lambda}\,h^{{1\over 4}}$.  We get:
\beq
e^{\phi}\,=\, e^{\phi_0}\,\Big(1\,+\,{\phi_2\over x^2}\,+\,\cdots\Big)\,\,,
\eeq
where the prefactor is
\beq
e^{\phi_0}\,=\,{12\,L_0\over k(A_0+4\eta_0)\,b}\,\,,
\eeq
and can be shown to be the same as the asymptotic value of the dilaton written in (\ref{dilaton-AdS_asymp}). The coefficient $\phi_2$ is:
\beq
\phi_2\,=\,\Lambda_2\,+\,{1\over 4}\,h_2\,\,,
\eeq
which has been written  more explicitly in (\ref{h_2_phi_2}).

Finally,  let us write these results in terms of the $r$ variable. At  next-to-leading order, the relation between the coordinates $x$ and $r$ is obtained by integrating the differential equation:
\beq
{dr\over dx}\,=\,{e^{g}\over x}\,
=\,{\kappa\,r_q\over b}\,\big[\,
x^{{1\over b}-1}\,+\,g_2\,x^{{1\over b}-3}\,+\cdots\big]\,\,.
\eeq
We get:
\beq
r\, =\,\kappa\,r_q\,\big[\,x^{{1\over b}}\,-\,{g_2\over 2b-1}\,
x^{{1\over b}-2}\,+\cdots\big]\,\,.
\label{r-x_relation}
\eeq
This relation can be inverted as:
\beq
x= \Big({r\over \kappa\,r_q}\Big)^{b}\,\Big[\,
1\,+\,{b\over 2b-1}\,g_2\, \Big({ \kappa\,r_q\over r}\Big)^{2b}\,+\,\cdots
\,\Big]\,\,.
\label{x-r_relation}
\eeq
By plugging the expansion (\ref{x-r_relation}) into (\ref{g_f_UVexpansions_x}) and (\ref{UVexpansion_h_phi}) one easily arrives at (\ref{All_UV_expansions_r}) and (\ref{tilde_UV_coeff}).


\subsection{More on the entanglement entropy}
\label{entanglement_appendix}

Let us study analytically the entanglement entropy and the ${\cal F}$ function near the UV fixed point at $R=0$. We shall represent $S(R)$ in terms of a local functional 
${\cal L}={\cal L}(H(x), G(x), \rho(x))$ as:
\beq
S(R)\,=\,\int_{x_*}^{\infty}\,{\cal L}\,dx\,\,.
\label{entropy_calL}
\eeq
We will use this representation to compute the first-order variation of $S$ when the background functions $H$ and $G$ and the embedding function $\rho$ are varied around their UV values:
\beq
H(x)\,=\,H_{UV}(x)+\delta H(x)\,\,,
\qquad
G(x)\,=\,G_{UV}(x)+\delta G(x)\,\,,
\qquad
\rho(x)\,=\,\rho_{UV}(x)+\delta \rho(x)\,\,,
\label{H_G_rho_UVexpansion}
\eeq
where $H_{UV}=H_{\infty}\,x^{{4\over b}}$, $G_{UV}=G_{\infty}\,x^{-2-{2\over b}}$ and $\rho_{UV}(x)$ is the function written in (\ref{rho_exact}). At first order the corrections  $\delta H(x)$ and $\delta G(x)$ can be parameterized as:
\beq
\delta H(x)\,=\,H_{\infty}\,H_2\,x^{{4\over b}-2}\,\,,
\qquad
\delta G(x)\,=\,G_{\infty}\,G_2\,x^{-4-{2\over b}}\,\,.
\label{deltaHG_UV}
\eeq
The constants $H_2$ and $G_2$  in (\ref{deltaHG_UV}) can be related to the ones characterizing the behavior of the background at the UV:
\beq
H_2\,=\,2\,(h_2+4f_2+2g_2-2\phi_2)\,\,,
\qquad\qquad
G_2\,=\,h_2\,+\,2 g_2\,\,.
\eeq
It is useful  to define the following combination of $H_2$ and $G_2$:
\beq
{\cal H}_2\,=\, {H_2\over 2}+{G_2\over 2b+1}
\label{calH_2}\,\,.
\eeq
The perturbation of the profile $\delta \rho(x)$ is the  correction, at first-order,  of the solution of (\ref{Euler_Lagrange_disk}) which satisfies $\delta \rho(x\to\infty)=0$. 
By computing the first variation of  (\ref{Euler_Lagrange_disk})  we find that  $\delta \rho(x)$ is solution to the following second-order inhomogeneous  differential equation:
\bear
&&{d\over dx}\,\Bigg[{\sqrt{H_{UV}}\,\rho_{UV}\,\rho\,'_{UV}\over \sqrt{(\rho\,'_{UV})^2+G_{UV}}}\,\Big({\delta\rho'\over \rho\,'_{UV}}\,+\,{\delta\rho\over \rho_{UV}}\,+\,
{\delta H\over 2 H_{UV}}\,-\,{1\over 2}\,\,
{2\,\rho\,'_{UV}\,\delta\rho\,'\,+\,\delta G\over (\rho\,'_{UV})^2+G_{UV}}\Big)\Bigg]
\rc\rc\rc
&&\qquad
-\sqrt{H_{UV}}\,\sqrt{(\rho\,'_{UV})^2+G_{UV}}\,\Big({\delta H\over 2\,H_{UV}}\,+\,
{1\over 2}\,\,
{2\,\rho\,'_{UV}\,\delta\rho\,'\,+\,\delta G\over (\rho\,'_{UV})^2+G_{UV}}\Big)\,=\,0\,\,.
\label{ODE_deltarho}
\eear
More explicitly,  after using the UV values of $H$, $G$, and $\rho$, the differential  equation (\ref{ODE_deltarho}) can be written as:
\bear
&&{d\over dx}\,\Bigg[{2\,x^{{3\over b}+1}\,\rho^4_{UV}\over R^2\,G_{\infty}}\,\,
\delta\rho'\,+\,2b\,x^{{1\over b}}\,\delta\rho\,+\,b\,H_2\,x^{{1\over b}-2}\,\rho_{UV}\,-\,
{b\,G_2\over R^2}\,x^{{1\over b}-2}\,\rho^3_{UV}\Bigg]\rc\rc
&&\qquad\qquad\qquad\qquad
-2b\,x^{{1\over b}}\,\delta\rho'\,-\,{R^2\,H_2\,x^{{1\over b}-3}\over \rho_{UV}}\,-\,
G_2\,x^{{1\over b}-3}\, \rho_{UV}\,=\,0\,\,.
\label{ODE_deltarho_explicit}
\eear

Let us next calculate the first-order variation of the entanglement entropy. From (\ref{entropy_calL}) we get:
\beq
\delta S\,=\,\int_{x_*}^{\infty}\,dx\,\Bigg[
{\partial {\cal L}\over \partial H}\Big|_{UV}\delta H\,+\,
{\partial {\cal L}\over \partial G}\Big|_{UV}\delta G\Bigg]\,+\,
\Pi_{UV}(x)\,\delta\rho (x)\Bigg|^{x=\infty}_{x=x_*}\,\,,
\label{deltaS_UV}
\eeq
where $\Pi_{UV}(x)$ is defined as:
\beq
\Pi_{UV}(x)\equiv 
{\partial {\cal L}\over \partial \rho{\,'}}\Big|_{UV}\,=\,{2\over 3\pi^2}\,
\sqrt{H_{UV}(x)}\,\,{\rho_{UV}\,\rho\,'_{UV}\over 
\sqrt{(\rho\,'_{UV})^2+G_{UV}(x)}}\,\,.
\eeq
Using  the explicit form of $\rho_{UV}$ we get:
\beq
\Pi_{UV}(x)\,=\,{2b\over 3\pi^2}\,
\sqrt{H_{\infty}\,G_{\infty}}\,\,x^{{1\over b}}\,
\sqrt{1\,-\,{b^2\,G_{\infty}\over R^2}\,x^{-{2\over b}}}\,\,,
\eeq
and it follows that:
\beq
\Pi_{UV}(x=x_*)=0\,\,.
\eeq
Therefore, the lower limit contribution to the last term in (\ref{deltaS_UV}) vanishes. Let us now study the contribution to this term of the $x\to\infty$ limit. We will now check
that $\delta\rho(x)$ decreases as $x\to\infty$ in such a way that:
\beq
\lim_{x\to\infty}\,\Pi_{UV}(x)\,\,\delta\rho(x)\,=\,0\,\,.
\label{Pi_rho_xinfty}
\eeq
Thus, the upper limit of the last term in (\ref{deltaS_UV}) also vanishes. To prove (\ref{Pi_rho_xinfty}) we have to integrate the differential equation (\ref{ODE_deltarho_explicit}) and extract the large $x$ behavior of the solution. Amazingly, (\ref{ODE_deltarho_explicit}) can be integrated analytically. Its general solution can be written as the sum of two terms:
\beq
\delta\rho\,=\,\delta\rho_P\,+\delta\rho_G\,\,,
\eeq
where $\delta\rho_P$ is a particular solution of the equation and $\delta\rho_G$ is a general solution of the homogeneous part of (\ref{ODE_deltarho_explicit}). We have found a particular solution  $\delta\rho_P$, which can be written in terms of hypergeometric functions and is given by:
\bear
&&\delta\rho_P={R^{2b+2}{\cal H}_2\over 2b(b-1)(2b-1)G_{\infty}\,\rho_{UV}}\,\times\rc\rc
&&\qquad\qquad
\times
\Bigg[\Big(Rx^{{1\over b}}-b\,\sqrt{G_{\infty}}\Big)^{2-2b}
{}_2F_1\Big(2b-2, 2b-2;2b-1;{b\,\sqrt{G_{\infty}}\over b\sqrt{G_{\infty}}-R\,x^{{1\over b}}}\Big)\qquad\qquad
\rc\rc
&&\qquad\qquad
+\Big(R\,x^{{1\over b}}+b\,\sqrt{G_{\infty}}\Big)^{2-2b}\,
{}_2F_1\Big(2b-2, 2b-2;2b-1;{b\sqrt{G_{\infty}}\over b\sqrt{G_{\infty}}+R\,x^{{1\over b}}}\Big)\Bigg]
\rc\rc
&&\qquad\qquad
-\Big[{b^2\,G_2\,G_{\infty}\over 2\,(2b+1)}\,x^{-{2\over b}}\,+\,{R^2\,{\cal H}_2\over 2b-1}\,+\,{R^4\,{\cal H}_2\over b\,(b-1)\,(2b-1)\,G_{\infty}}\,x^{{2\over b}}\,\Big]\,
{1\over x^2\,\rho_{UV}}\,\,.
\label{delta_rho_particular}
\eear
We are only interested in the behavior of $\delta\rho_P$ for large $x$. It is straightforward to prove that, for large $x$,  $\delta\rho_P$ can be approximated as:
\beq
\delta\rho_P\approx {b^2\,G_{\infty}\,\big[bH_2\,-\,(b-1)\,G_2\big]\over 2\,(b+1)\,(2b-1)\,R}\,\,x^{-2-{2\over b}}\,\,.
\eeq
Thus, as $\Pi_{UV}\propto x^{{1\over b}}$ for large $x$,
\beq
\lim_{x\to\infty}\,\Pi_{UV}(x)\,\,\delta\rho_P(x)\,=\,0\,\,.
\eeq
The general solution of the homogeneous differential equation which vanishes as $x\to\infty$ is:
\beq
\delta\rho_G\,=\,{C\over x^{{1\over b}}\,\rho_{UV}}
\Big[{R\,x^{{1\over b}}\over 2b\,\sqrt{G_{\infty}}}\,
\log\,{R\,x^{{1\over b}}\,+\,b\,\sqrt{G_{\infty}}\over R\,x^{{1\over b}}\,-\,b\,\sqrt{G_{\infty}}}\,\,-\,1
\Big]\,\,,
\label{delta_rho_general}
\eeq
where $C$ is an arbitrary constant. For large $x$ the function in (\ref{delta_rho_general}) behaves as:
\beq
\delta\rho_G\,\approx  {C\,b^2\,G_{\infty}\over 3\,R^3}\,x^{-{3\over b}}\,\,.
\eeq
Therefore,
\beq
\lim_{x\to\infty}\,\Pi_{UV}(x)\,\,\delta\rho_G(x)\,=\,0\,\,.
\eeq
Then, eq. (\ref{Pi_rho_xinfty}) holds  and the last term in (\ref{deltaS_UV}) does not contribute to $\delta S$, as claimed above. Let us now calculate the other two contributions. First of all, the term due to the variation of $H$ is given by:
\bear
&&\int_{x_*}^{\infty}\,dx\,
{\partial {\cal L}\over \partial H}\Big|_{UV}\delta H\,=\,{1\over 3\pi^2}\,
\int_{x_*}^{\infty}\,dx\,\,
{\rho_{UV}\sqrt{(\rho\,'_{UV})^2+G_{UV}(x)}\over
\sqrt{H_{UV}(x)}}\,\,\delta H\rc\rc
&&\qquad\qquad\qquad\qquad
={1\over 3\pi^2}\,{bH_2\over 2b-1}\,\sqrt{H_{\infty}G_{\infty}}\,\,R\,x_*^{{1\over b}-2}\,\,,
\eear
which, after using (\ref{xstar})  to eliminate $x_*$, becomes:
\beq
\int_{x_*}^{\infty}\,dx\,
{\partial {\cal L}\over \partial H}\Big|_{UV}\delta H\,=\,{1\over 3\pi^2}\,
{H_2\,b^{2-2b}\over 2b-1}\,\sqrt{H_{\infty}}\,G_{\infty}^{1-b}\,R^{2b}\,\,.
\eeq
Similarly,
\bear
&&\int_{x_*}^{\infty}\,dx\,
{\partial {\cal L}\over \partial G}\Big|_{UV}\delta G\,=\,
{1\over 3\pi^2}\,\int_{x_*}^{\infty}\,dx\,
{\sqrt{H_{UV}(x)}\,\,\rho_{UV}\over
\sqrt{(\rho\,'_{UV})^2+G_{UV}(x)}}\,\,\delta G\rc\rc
&&\qquad\qquad\qquad\qquad
=\,
{2\over 3\pi^2}\,{b^2\over 4b^2-1}\,
{G_2\,\sqrt{H_{\infty}}\,G_{\infty}\over x_{*}^2}\,\,,
\eear
which, after eliminating $x_*$, gives:
\beq
\int_{x_*}^{\infty}\,dx\,
{\partial {\cal L}\over \partial G}\Big|_{UV}\delta G\,=\,
{2\over 3\pi^2}\,{G_2\,b^{2-2b}\over 4b^2-1}\,\sqrt{H_{\infty}}\,\,G_{\infty}^{1-b}\,\,R^{2b}\,\,.
\eeq
Thus,  $\delta S$ can be written as:
\beq
\delta S\,=\,{2\over 3\pi^2}\,{b^{2-2b}\over 2b-1}\,\,
\sqrt{H_{\infty}}\,\,G_{\infty}^{1-b}\,\,{\cal H}_2\,
R^{2b}\,\,,
\eeq
where ${\cal H}_2$ is the constant defined in (\ref{calH_2}). Using the definition (\ref{calF_definition}), it follows that the change in the function ${\cal F}$ is:
\beq
\delta {\cal F}\,=\,{2\over 3\pi^2}\,b^{2-2b}\,\sqrt{H_{\infty}}\,\,G_{\infty}^{1-b}
\,\,{\cal H}_2\,R^{2b}\,\,.
\eeq
Taking into account (\ref{HG_infinity})  and  (\ref{bHG_F}) this expression can be written as:
\beq
\delta {\cal F}\,=\,F_{UV} ({\mathbb S}^3)
\,\,{\cal H}_2\,
\Big({\kappa\over L_0^2}\Big)^{2b}\,(r_q\,R)^{2b}\,=\,F_{UV} ({\mathbb S}^3)\,
\,\,{\cal H}_2\,
x_*^{-2}\,\,.
\label{deltaF_R2b}
\eeq
Therefore, we can write  the  ${\cal F}$ function  near the UV fixed point as:
\beq
{\cal F}\,=\,F_{UV} ({\mathbb S}^3)\,+\,c_{UV}\,(r_q\,R)^{2b}\,\,,
\label{calF_nearUV_b}
\eeq
where the constant coefficient $c_{UV}$ can be read from (\ref{deltaF_R2b}),
\beq
c_{UV}\,=\,\Big({\kappa\over L_0^2}\Big)^{2b}\,\, F_{UV} ({\mathbb S}^3)
{\cal H}_2\,\,.
\label{c_UV}
\eeq

Eq. (\ref{calF_nearUV_b}) coincides with (\ref{calF_nearUV}) when the former is written in terms of the dimension  $\Delta_{UV}=3-b$ of the quark-antiquark bilinear operator in the massless flavored theory. Moreover, as $F_{UV} ({\mathbb S}^3)$ is always positive,  the sign of $c_{UV}$ depends on the sign of the coefficient ${\cal H}_2$. We have checked that ${\cal H}_2$ is negative for all values of the deformation parameter $\hat\epsilon$, in agreement with the expectation that ${\cal F}$  is maximized at the UV fixed point.

\subsubsection{Entanglement entropy on the strip}

Let us  now evaluate the entanglement entropy in   the case in which  the region $A$ is the strip $-{l\over 2}\,\le x^1\,\le +{l\over 2}$ of width $l$ in the $(x^1,x^2)$-plane (see Fig. \ref{Strip_surface}). In this case we consider a constant time surface $\Sigma$, whose embedding in the ten-dimensional space is determined by a function $x=x(x^1)$.  The induced metric on $\Sigma$ is,
\beq
ds^2_{8}\,=\,h^{-{1\over 2}}\,\Big[\,\Big(1+G(x)\,x'^{\,2}\,\Big)\, (dx^1)^2\,+\,
(dx^2)^2\,\Big]+h^{{1\over 2}}\,\Big[\,
e^{2f}\,ds_{{\mathbb S}^4}^2+
e^{2g}\,\Big(\,\big(E^1\big)^2\,+\,\big(E^2\big)^2\Big)\,\Big]\,\,,
\eeq
where $x'$ denotes the derivative of the holographic coordinate $x$ with respect to the cartesian coordinate $x^1$ and   the function $G(x)$ has been defined in (\ref{G_def}).
The entropy  functional  for the strip of width $l$ is given by:
\beq
S(l)\,=\,{V_6\,L_2\over 4 \,G_{10}}\,
\int_{-{l\over 2}}^{+{l\over 2}}\,dx^1\,\sqrt{H(x)}\,\sqrt{1+G(x)\,x'^{\,2}}\,\equiv\,
\int_{-{l\over 2}}^{+{l\over 2}}\,dx^1\,{\cal L}_{strip}
\,\,,
\eeq
where   $V_6$ is the volume of the internal manifold, whose value was given after (\ref{entropy_total_disk}),  $H(x)$ is the function defined in (\ref{H_def}), and $L_2=\int dx^2$ is the length of the strip.  As the integrand in $S(l)$ does not depend on the coordinate $x^1$, the corresponding  Euler-Lagrange equation admits the following first integral:\begin{figure}[ht]
\center
\includegraphics[width=0.35\textwidth]{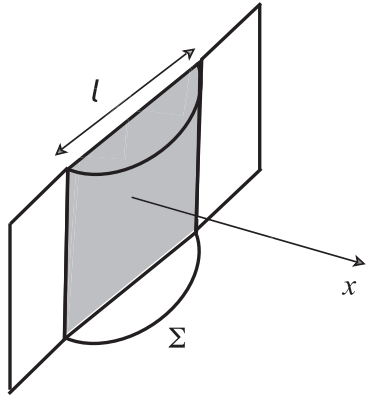}
\caption{The surface $\Sigma$ ends on a strip of length $l$ at the boundary.  } 
\label{Strip_surface}
\end{figure}

\beq
x'\,{{\cal L}_{strip}\over \partial x'}\,-\,{\cal L}_{strip}\,=\,{\rm constant}\,\,,
\eeq
or, more explicitly,
\beq
{\sqrt{H(x)}\over \sqrt{1\,+\,G(x)\,x'^{\,2}}}\,=\,\sqrt{H_*}\,\,,
\eeq
where $H_{*}=H(x=x_*)$ and $x_*$  is the holographic coordinate of the turning point. It follows from this last expression that $x'$ is given by:
\beq
x'\,=\,\pm {1\over \sqrt{G(x)}}\,\sqrt{{H(x)\over H_*}\,-\,1}\,\,.
\eeq
Therefore,  the width of the strip is given by the integral:
\beq
l\,=\,2 \sqrt{H_*}\,\int_{x_*}^{\infty}\,dx\,{\sqrt{G(x)}\over \sqrt{H(x)-H_*}}\,\,,
\label{width_strip}
\eeq
and the entropy $S(l)$ is:
\beq
S(l)\,=\,{V_6\,L_2\over 2 G_{10}}\,
\int_{x_*}^{\infty}\,dx\,{\sqrt{G(x)}\,H(x)\over \sqrt{H(x)-H_*}}\,\,.
\label{S_strip_total}
\eeq
The integral (\ref{S_strip_total}) for $S(l)$ is divergent  in the UV. Indeed, from the  behavior of the functions $H(x)$ and $G(x)$  at large $x$ (eqs. (\ref{H_G_UV}) and (\ref{HG_infinity})), it follows that, when $x\to\infty$,  the integrand in (\ref{S_strip_total})  behaves as: 
\beq
{\sqrt{G(x)}\,H(x)\over \sqrt{H(x)-H_*}}\,
\approx\,\sqrt{H_{\infty}\,G_{\infty}}\,\,x^{{1\over b}\,-\,1}\,\,,
\eeq
and $S(l)$ therefore diverges as $x_{\max}^{{1\over b}}$, where $ x_{\max}$ is the upper limit of the integral, which can be regarded as a UV cutoff. More explicitly, one can rearrange the integral in (\ref{S_strip_total}) as:
\bear
&&\int_{x_*}^{x_{\max}}\,dx\,{\sqrt{G(x)}\,H(x)\over \sqrt{H(x)-H_*}}\,=\,
\int_{x_*}^{x_{\max}}\,dx\,\Big[{\sqrt{G(x)}\,H(x)\over \sqrt{H(x)-H_*}}\,-\,
\sqrt{H_{\infty}\,G_{\infty}}\,\,x^{{1\over b}\,-\,1}\,\Big]\rc\rc
&&\qquad\qquad\qquad\qquad\qquad\qquad\qquad\qquad\qquad
+b\,\sqrt{H_{\infty}\,G_{\infty}}\,\big[\,x_{max}^{{1\over b}}\,-\,x_{*}^{{1\over b}}\,\big]\,\,,
\eear
where the divergent term is explicitly shown. We now define the rescaled functions $\hat G(x)$ and $\hat H(x)$ as:
\beq
\hat G(x)\equiv {G(x)\over G_{\infty}}\,\,,
\qquad\qquad
\hat H(x)\equiv {H(x)\over H_{\infty}}\,\,,
\qquad\qquad
\hat H_{*}\equiv {H_{*}\over H_{\infty}}
\,\,.
\eeq
Then, the finite term in $S(l)$ can be written as:
\beq
S_{finite}(l)\,=\,{V_6\,L_2\over 2 G_{10}}\,
\sqrt{H_{\infty}\,G_{\infty}}\Bigg[
\int_{x_*}^{\infty}\,dx\,\Bigg({\sqrt{\hat G(x)}\,\hat H(x)\over \sqrt{\hat H(x)-\hat H_*}}\,-\,
x^{{1\over b}-1}\,\Bigg)\,-\,b\,x_{*}^{{1\over b}}\,\Bigg]\,\,,
\label{Sfinite_strip}
\eeq
whereas the divergent term takes the form:
\beq
S_{div}\,=\,{V_6\,L_2\over 2 G_{10}}\,b\,
\sqrt{H_{\infty}\,G_{\infty}}\,\,x_{max}^{{1\over b}}\,\,.
\eeq
\begin{figure}[ht]
\center
\includegraphics[width=0.75\textwidth]{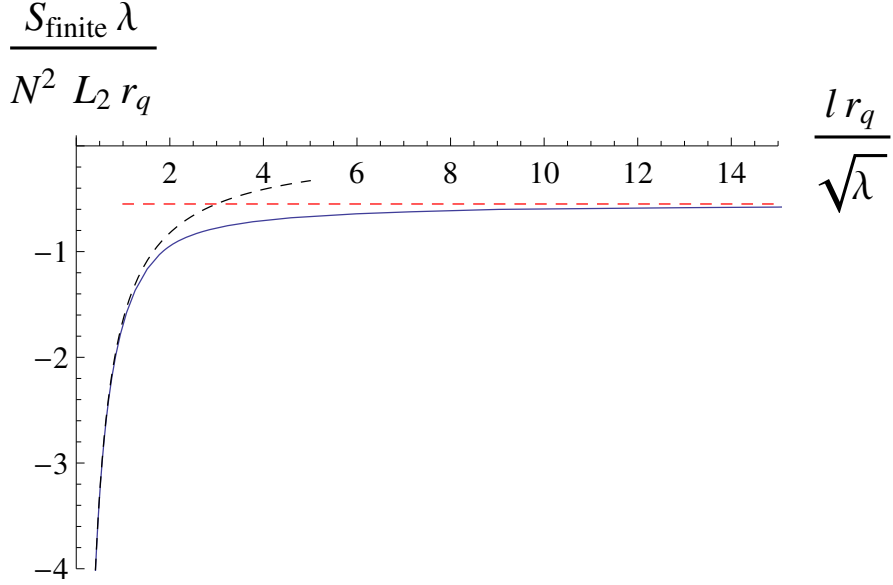}
\caption{Plot of the entanglement entropy versus the width of the strip. 
The solid curve corresponds to the numerical results for $\hat\epsilon=9$, while the 
the black dashed curve on the left  is the analytic UV result (\ref{Sfinite_strip_UV}) and the dashed red line  on the right corresponds to the IR value (\ref{S_infty}).} 
\label{Strip_entanglement}
\end{figure}

We have evaluated numerically the right-hand side of (\ref{Sfinite_strip}) as a function of the strip width $l$.  The result is displayed in Fig. \ref{Strip_entanglement}. One notices from these results that $S_{finite}(l)/L_2$ is negative and diverges as $-c/l$ for small $l$, where $c$ is a constant. Actually, for small $l$ the surface $\Sigma$ is in the UV region of the geometry and one can evaluate the entropy analytically in this limit. This is the subject of the next subsection, where we show that the constant $c$ is proportional to the free energy on the three-sphere of the massless flavored theory (see eq. (\ref{Sfinite_strip_UV}) below). The analytical UV calculation is compared to the numerical results in Fig. \ref{Strip_entanglement}.

\subsubsection{UV limit}

In order to study the UV limit it is convenient to change variables in the integrals (\ref{width_strip}) and (\ref{Sfinite_strip}). Let us introduce a new variable $z$, related to $x$ as $x=x_*\,z$. Then, $l$ can be represented as:
\beq
l\,=\,2\, x_{*}^{{2\over b}+1}\,\sqrt{G_{\infty}}\,\int_{1}^{\infty}\,dz\,{\sqrt{\hat G(x_*\,z)}\over \sqrt{\hat H(x_*\,z)-x_{*}^{4/ b}}}\,\,,
\eeq
while $S_{finite}(l)$ is given by:
\beq
S_{finite}(l)\,=\,-{V_6\,L_2\over 2 \,G_{10}}\,
\sqrt{H_{\infty}\,G_{\infty}}\,\,x_{*}^{{1\over b}}\,
\Bigg[b-\int_{1}^{\infty}\,dz\Bigg(x_{*}^{1-{1\over b}}\,
{\sqrt{\hat G(x_*\,z)}\,\hat H(x_*\,z)\over \sqrt{\hat H(x_*\,z)-x_{*}^{4/ b}}}\,-\,
z^{{1\over b}-1}\Bigg)\Bigg]\,\,.
\eeq
Let us now obtain the expressions of $l$ and $S_{finite}(l)$ in the limit in which $x_{*}\to\infty$. In this case the argument of the functions $\hat G$ and $\hat H$ in the integrals is always large and one can take $\hat G(x)\approx x^{-2-{2\over b}}$, $\hat H(x)\approx x^{{4\over b}}$. We get:
\beq
l\approx 2\,x_{*}^{-{1\over b}}\,\sqrt{G_{\infty}}\,\,I_1\,\,,
\label{l_UV_approx}
\eeq
where $I_1$ is the following integral:
\beq
I_1\,=\,\int_{1}^{\infty}\,{dz\over z^{1+{1\over b}}\,\sqrt{z^{{4\over b}}\,-\,1}}\,=\,
{b\,\sqrt{2}\,\pi^{{3\over 2}}\over  \big[\Gamma\big({1\over 4}\big)\big]^2}\,\,.
\label{I1}
\eeq
Using (\ref{I1}) in (\ref{l_UV_approx}), we get:
\beq
l\approx {b\,\sqrt{G_{\infty}}\over x_{*}^{{1\over b}}}\,\,
{2\,\sqrt{2}\,\pi^{{3\over 2}}\over \big[\Gamma\big({1\over 4}\big)\big]^2 }\,\,.
\eeq
Similarly, the finite part of the entropy is:
\beq
S_{finite}(l)\,\approx\,-{V_6\,L_2\over 2\, G_{10}}\,
\sqrt{H_{\infty}\,G_{\infty}}\,\,x_{*}^{{1\over b}}\,
\big(\,b\,-\,I_2\,\big)\,\,,
\label{S_finite_strip_I2}
\eeq
where $I_2$ is the integral:
\beq
I_2\,=\,\int_{1}^{\infty}\,z^{{1\over b}-1}\,\Bigg[
{z^{{2\over b}}\over \sqrt{z^{{4\over b}}\,-\,1}}\,-\,1\,\Bigg]\,dz\,=\,
b\,\Bigg[1\,-\,{\sqrt{2}\,\pi^{{3\over 2}}\over \big[\Gamma\big({1\over 4}\big)\big]^2}
\Bigg]\,\,.
\label{I2}
\eeq
Plugging (\ref{I2}) into (\ref{S_finite_strip_I2}), we arrive at:
 \beq
S_{finite}(l)\,\approx\,-{V_6\,L_2\over 2\, G_{10}}\,b\,
\sqrt{H_{\infty}\,G_{\infty}}\,\,x_{*}^{{1\over b}}\,
{\sqrt{2}\, \pi^{{3\over 2}}\over \big[\Gamma\big({1\over 4}\big)\big]^2 }\,\,.
\eeq
Eliminating $x_*$ in favor of $l$, we get:
\beq
S_{finite}(l)\,\approx\,-{2\pi^3\,V_6\,L_2\over G_{10}}\,\,
{b^2\,G_{\infty}\,\sqrt{H_{\infty}}\over 
\big[\Gamma\big({1\over 4}\big)\big]^4}\,\,
{1\over l}\,\,.
\eeq
However, from the definition of $G_{\infty}$ and  $H_{\infty}$ in (\ref{HG_infinity})  it follows that:
\beq
b^2\,G_{\infty}\,\sqrt{H_{\infty}}\,=\,
{L_0^8\,q_0^2\,e^{-2\phi_0}\over b^6}\,=\,{3\pi^2\over 2}
F_{UV} ({\mathbb S}^3)\,\,,
\label{GH-F-UV}
\eeq
where $F_{UV} ({\mathbb S}^3)$ is the free energy of the massless flavored theory on the three sphere (see (\ref{F_UV})). It follows that:
\beq
{S_{finite}(l)\over L_2}\,\approx\,-{4\pi^2\,F_{UV} ({\mathbb S}^3)\over
\big[\Gamma\big({1\over 4}\big)\big]^4}\,\,{1\over l}\,\,,
\label{Sfinite_strip_UV}
\eeq
which is the result we were looking for. From the comparison of Fig.  \ref{Strip_entanglement} to the numerical results we conclude that (\ref{Sfinite_strip_UV}) is a good description of the entropy in the small $l$ region. Notice that the coefficient of $1/l$ on the right-hand side of (\ref{Sfinite_strip_UV}) is determined by the free energy of the massless flavored theory. This is expected on general grounds in the UV region, since the masses of the quarks can be neglected in the high-energy regime.

\subsubsection{IR limit}
\label{Strip_entanglement_S_IR}

Let us now evaluate $S_{finite}(l)$ in the regime in which $l$ is large. In this case the surface $\Sigma$ penetrates deeply in the bulk and its tip is near the origin (\ie,  $x_*$ is small). We will proceed    as in section \ref{IR_limit_disk_entanglement} and split   the interval $[x_*,\infty]$ in the integrals (\ref{width_strip}) and (\ref{Sfinite_strip}) as $[x_*,\infty]=[x_*,x_a]\cup [x_a, \infty]$, where $x_a<1$ is considered  small enough so that one can use  the unflavored background functions in the interval $[x_*,x_a]$. Moreover, when 
$x\in [x_a, \infty]$ with $x_a\gg x_*$,  $H_*\propto x^4_{*}\ll H(x)\propto x^4$ and one can neglect the terms containing $H_*$ in the integrals. Then, the strip width $l$ can be approximated as:
\beq
l\,\approx {2c\over \gamma}\,\,{L^2_{ABJM}\over x_*}\,\,
\int_{1}^{{x_a\over x_*}}\,\,{dz\over z^2\,\sqrt{z^4-1}}\,+\,
2\sqrt{H_*}\,\int_{x_a}^{\infty}\,\,dx\,{\sqrt{G(x)}\over \sqrt{H(x)}}\,\,.
\label{strip_width_IR}
\eeq
When, $x_a\gg x_*$  and $x_*\to 0$  we can extend the first integral in (\ref{strip_width_IR}) up to $\infty$ and  neglect the contribution of the second integral (which is proportional to  $\sqrt{H_*}\propto x_*^2$). Then, $l$ can be approximately written as:
\beq
l\,\approx {c\over \gamma}\,L^2_{ABJM}\,{2\sqrt{2}\,\pi^{{3\over 2}}\over 
\big[\Gamma\big({1\over 4}\big)\big]^2}\,\,{1\over x_*}\,\,,
\label{strip_width_IR_simp}
\eeq
and, as expected, a small value of $x_*$ leads to a large value of $l$. Similarly, we can perform the same type of manipulations in the expression (\ref{Sfinite_strip}) of $S_{finite}$. We find:
\bear
&&{S_{finite}(l)\over L_2}\,\approx\,{2\over 3\pi^3}\,\,{\gamma\over c}\,\,
L^6_{ABJM}\,e^{-2\phi_{ABJM}}\,\,x_*\,\int_1^{{x_a\over x_*}}\,\,
{z^2\,dz\over \sqrt{z^4-1}}\rc\rc
&&\qquad\qquad\,\,+
{2\over 3\pi^3}\,\sqrt{G_{\infty}\,H_{\infty}}\,\Bigg[
\int_{x_a}^{\infty}\,\Big(\sqrt{\hat G(x)\,\hat H(x)}\,-\,x^{{1\over b}-1}\Big)\,dx\,-\,
b\,x_a^{{1\over b}}\,\Bigg]\,\,.\qquad
\label{strip_entropy_IR}
\eear
Performing the first integral in (\ref{strip_entropy_IR}) in the limit $x_*\to 0$ for fixed $x_a$, we arrive at:
\beq
{S_{finite}(l)\over L_2}\,\approx\,-
{2\sqrt{2}\over 3\pi^{{3\over 2}}\,\big[\Gamma\big({1\over 4}\big)\big]^2}\,\,
{\gamma\over c}\,\,L^6_{ABJM}\,e^{-2\phi_{ABJM}}\,x_*\,+\,{\cal S}_{\infty}\,\,,
\label{strip_entropy_IR_simp}
\eeq
where ${\cal S}_{\infty}$ is the constant:
\beq
{\cal S}_{\infty}={2\over 3\pi^3}\sqrt{G_{\infty}\,H_{\infty}}\Big[
\int_{x_a}^{\infty}\Big(\sqrt{\hat G(x)\hat H(x)}-x^{{1\over b}-1}\Big)dx-
bx_a^{{1\over b}}\Big]+{2\over 3\pi^3}\,{\gamma\over c}\,
L^6_{ABJM}e^{-2\phi_{ABJM}}x_a\,\,.
\label{calS_infty}
\eeq
We can now eliminate in (\ref{strip_entropy_IR_simp}) the turning point $x_*$ in favor of $l$ by using (\ref{strip_width_IR_simp}). We find:
\beq
{S_{finite}(l)\over L_2}\,\approx\,-
{4\pi^2\,F_{IR} ({\mathbb S}^3)\over
\big[\Gamma\big({1\over 4}\big)\big]^4}\,\,{1\over l}\,+\,{\cal S}_{\infty}\,\,,
\label{strip_entropy_IR_final}
\eeq
where we have written the result in terms of   
$F_{IR} ({\mathbb S}^3)$. The first term in (\ref{strip_entropy_IR_final}) is just the  strip entanglement  entropy for the unflavored theory. The constant ${\cal S}_{\infty}$ represents the asymptotic value of $S_{finite}(l)/ L_2$  as $l\to\infty$. One can approximate this constant by taking $x_a\to 0$ in (\ref{calS_infty}). After using (\ref{bHG_F}) to relate $G_{\infty}$ and $H_{\infty}$ to  
$F_{UV} ({\mathbb S}^3)$, we arrive at:
\beq
{\cal S}_{\infty}\,\approx\,r_q\,{\kappa\,F_{UV} ({\mathbb S}^3)\over \pi b\,L_0^2}\,\,
\int_{0}^{\infty}\Big(\sqrt{\hat G(x)\hat H(x)}-x^{{1\over b}-1}\Big)\,dx\,\,.
\label{S_infty}
\eeq


\subsection{Asymptotic quark-antiquark potential}
\label{asymp_qq_potential}

The purpose of this appendix is to obtain the analytic expressions for the $q\bar q$ potential energy in the UV and IR limit. We will start by calculating the leading and subleading UV potential. 

\subsubsection{UV potential}

Let us find the approximate value of the $q\bar q$ potential in the case in which the distance $d$ is small and the hanging string only explores the UV of the geometry. This is equivalent to considering the limit in which the turning point $x_*$ is large. It is then more convenient to perform a change of variables in the integrals (\ref{d_x-integral}) and (\ref{E_x-integral}) and write $d$ and $E_{q\bar q}$ as:
\bear
d&=&2\,\int_{1}^{\infty}\,
{e^{g(x_*\,z)}\,h(x_*\,z)\over z\,\sqrt{h_*-h(x_*\,z)}}\,\,dz\,\,,\rc\rc
E_{q\bar q}&=&{1\over \pi}\,\int_{1}^{\infty}\,\,
{e^{g(x_*\,z)}\over z}\,\,\Big[\,{\sqrt{h_*}\over \sqrt{h_*-h(x_*\,z)}}\,-\,1\Big]
dz\,-\,{r_*\over \pi}\,\,.
\label{d_E_z-integral}
\eear
We want to compute the leading value of $E_{q\bar q}$, as well as the first deviation from the conformal behavior. For this reason we will make use of the asymptotic expressions of the different functions of the metric derived in section \ref{UV_mass_corrections}.

Let us begin by computing the integrals in (\ref{d_E_z-integral}) in a power series expansion for large $x_*$. From (\ref{g_f_UVexpansions_x}) and (\ref{UVexpansion_h_phi}) we get:
\beq
e^{g(x_*\,z)}\,\,h(x_*\,z)\,=\,
{L_0^4\over b\,\kappa^3\,r_q^3}\,\,x^{-{3\over b}}_*\,z^{-{3\over b}}\,
\Big(1\,+\,{h_2+g_2\over x_*^2\,z^2}\,+\,\cdots\,\Big)\,\,,
\eeq
where  $g_2$ and $h_2$ are given in (\ref{g2_f2}) and (\ref{h_2_phi_2}) and $\kappa$ has been defined in (\ref{kappa_def}). Then, it follows that:
\beq
{1\over \sqrt{h_*-h(x_*\,z)}}\,=\,{\kappa^2\,r_q^2\over L_0^2}\,\,
x^{{2\over b}}_*\,\,{z^{{2\over b}}\over \sqrt{z^{{4\over b}}-1}}\,\,
\Big(1\,-\,{h_2\over 2\,x_*^2}\,{z^{{4\over b}+2}-1\over z^2\,
(z^{{4\over b}}-1)}
\,+\,\cdots\,\Big)\,.
\eeq
Using these results we obtain the following expansion of the $q\bar q$ separation $d$:
\beq
d\,=\,{2\,L_0^2\over b\,\kappa\,r_q\,x^{{1\over b}}_*}\,\,
\Big[\,I_1\,+\,{1\over x_*^2}\,\,\Big(\,(h_2+g_2)\,I_3\,-\,{h_2\over 2}\,I_4\,\Big)
+\,\cdots \Big]\,\,,
\eeq
where $I_1$ is the integral (\ref{I1}),  and $I_3$ and $I_4$ are:
\bear
&&I_3\,=\,\int_{1}^{\infty}\,{dz\over z^{3+{1\over b}}\,\sqrt{z^{{4\over b}}\,-\,1}}\,=\,
{b\sqrt{\pi}\over 4}\,\,G(b)\,\,,\rc\rc
&&I_4\,=\,\int_{1}^{\infty}\,
{z^{{4\over b}+2}\,-\,1\over  z^{3+{1\over b}}\,(z^{{4\over b}}\,-\,1)^{{3\over 2}}}\,\,dz\,=\,
{b\sqrt{\pi}\over 2}\,\Bigg[\,{3+2b\over 4}\,\,G(b)\,-\,
{\sqrt{2}\,\,\pi\over  \big[\Gamma\big({1\over 4}\big)\big]^2}\,
\Bigg]\,\,,
\label{I3_I4}
\eear
where $G(b)$ is the following ratio of Euler Gamma functions:
\beq
G(b)\equiv
{\Gamma\Big({3+2b\over 4}\Big)
\over 
\Gamma\Big({5+2b\over 4}\Big)}\,\,.
\eeq
Using these results we arrive at:
\beq
d\,\approx\,{2\,L_0^2\over \kappa\, r_q\,x_*^{{1\over b}}}\,
\Bigg[\,{\sqrt{2}\,\pi^{{3\over 2}}\over \big[\Gamma\big({1\over 4}\big)\big]^2}\,+\,
{\sqrt{\pi}\over 4\,x_*^2}\,\Big(g_2+(1-2b)\,{h_2\over 4}\Big)\,G(b)\,+\,
{\sqrt{2}\,\pi^{{3\over 2}}\over 4\, x_*^2}\,
{h_2\over \big[\Gamma\big({1\over 4}\big)\big]^2}\,\Bigg]\,\,.
\label{d-x_*}
\eeq

Let us now compute the $q\bar q$ energy at leading and next-to-leading order in the UV (large $x_*$ or small $d$). First, we need the expansion:
\beq
{\sqrt{h_*}\over \sqrt{h_*-h(x_*\,z)}}\,=\,
{z^{{2\over b}}\over \sqrt{z^{{4\over b}}-1}}\,+\,
{h_2\over 2\,x_*^2}\,\,{z^{{2\over b}-2}\,(1-z^2)\over (z^{{4\over b}}\,-\,1)^{{3\over 2}}}
\,+\,\cdots\,\,.
\eeq 
Then, it is easy to verify that $E_{q\bar q}$ can be expanded as:
\beq
E_{q\bar q}\approx
{\kappa\, r_q\over \pi\,b}\,x_*^{{1\over b}}\Bigg[\,
I_2\,+\,{1\over x_*^2}\,\Big(\,{h_2\over 2}\,I_5\,+\,g_2\,I_6\,\Big)\,\Bigg]\,-\,
{r_*\over \pi}\,\,,
\eeq
where  $I_2$ is the integral (\ref{I2})  and $I_5$ and $I_6$ are:
\bear
&&I_5\,=\,\int_{1}^{\infty}\,{z^{{3\over b}-3}(1-z^2)\over (z^{{4\over b}}-1)^{{3\over 2}}}
\,dz\,=\,{b\sqrt{\pi}\over 2}\,\,\Bigg[
{\sqrt{2}\,\pi\over \big[\Gamma\big({1\over 4}\big)\big]^2}\,-\,{2b+1\over 4}\,\,G(b)\,\Bigg]
\,\,,\rc\rc
&&I_6\,=\,\int_{1}^{\infty}\,z^{{1\over b}-3}\,\Bigg[
{z^{{2\over b}}\over \sqrt{z^{{4\over b}}\,-\,1}}\,-\,1\,\Bigg]\,dz\,=\,{b\over 2b-1}\,
\Big[\,{(2b+1)\sqrt{\pi}\over 4}\,\,G(b)\,-\,1\,\Big]\,\,.  ~~~~~~~~~~
\eear
To compute $E_{q\bar q}$ we also need $r_*$ as a function of $x_*$. It follows from (\ref{r-x_relation}) that:
\beq
r_*\,=\,\kappa\,r_q\,x_*^{{1\over b}}\Big[1\,-\,{g_2\over 2b-1}\,{1\over x_*^2}\,+\,\cdots\Big]\,\,.
\eeq
Then, one can check that:
\beq
E_{q\bar q}\approx\kappa\,r_q\,x_*^{{1\over b}}\Bigg[-{\sqrt{2\pi}\over 
\big[\Gamma\big({1\over 4}\big)\big]^2}+
{1\over x_*^2}\Bigg({\sqrt{2\pi}\over \big[\Gamma\big({1\over 4}\big)\big]^2}\,
{h_2\over 4}+{2b+1\over 4\sqrt{\pi}(2b-1)}
\Big(g_2+(1-2b)\,{h_2\over 4}\Big)\,G(b)\Bigg)\Bigg]\,\,.
\eeq
Let us write $E_{q\bar q}$ as a function of the $q\bar q$ separation $d$. For this purpose we have to eliminate $x_*$ in favor of $d$. By inverting (\ref{d-x_*}), we get:
\bear
&&\kappa\,r_q\,x_*^{{1\over b}}\approx {2\sqrt{2}\pi^{{3\over 2}}\over 
 \big[\Gamma\big({1\over 4}\big)\big]^2}\,\,{L_0^2\over d}\,+\,\rc\rc
 &&~
 +\,
 r_q\,{\sqrt{\pi}\over 2}\,
 \Bigg[{\kappa\,\big[\Gamma\big({1\over 4}\big)\big]^2
 \over 2\sqrt{2}\,\pi^{{3\over 2}}}\Bigg]^{2b}\,
 \Bigg({\sqrt{2}\,\pi\over \big[\Gamma\big({1\over 4}\big)\big]^2}\,
h_2+\Big(g_2+(1-2b)\,{h_2\over 4}\Big)\,G(b)\Bigg)
\Bigg({r_q \,d\over L_0^2}\Bigg)^{2b-1}.~~
\qquad\qquad
 \eear
Using this result, we get the following dependence of $E_{q\bar q}$ with the distance $d$
\beq
{E_{q\bar q}\over r_q}\approx-
 {4\pi^{2}\over 
 \big[\Gamma\big({1\over 4}\big)\big]^4}\,\,{L_0^2\over r_q\,d}\,+\,
 {\sqrt{2}\,\pi\over  \big[\Gamma\big({1\over 4}\big)\big]^2}\,
  \Bigg[{\kappa\,\big[\Gamma\big({1\over 4}\big)\big]^2
 \over 2\sqrt{2}\,\pi^{{3\over 2}}}\Bigg]^{2b}\,
 \Big({g_2\over 2b-1}-{h_2\over 4}\Big)\,G(b)
\,\Bigg({r_q\, d\over L_0^2}\Bigg)^{2b-1},
\label{potl_lead_and_sublead}
 \eeq
where we have represented this relation in terms of the rescaled quantities 
$E_{q\bar q}/ r_q$ and $r_q \,d/ L_0^2$. Notice that the leading term (the first term on the right-hand side of (\ref{potl_lead_and_sublead})) is given by the potential of the massless flavored background, as expected. 

\subsubsection{IR potential}
\label{qq_potential_IR}
We now estimate the $q\bar q$ potential for large separations. In this case we will content ourselves to compute the leading order contribution. For large $q\bar q$ separations the hanging fundamental string penetrates deeply in the geometry, which is equivalent to saying that $x_*$ is small.  We will follow an approach similar to the one in sections \ref{IR_limit_disk_entanglement}  and \ref{Strip_entanglement_S_IR} and we will split the $[x_*,\infty]$ interval of the integrals (\ref{d_x-integral}) and (\ref{E_x-integral}) as  $[x_*,\infty]=[x_*,x_a]\cup [x_a,\infty]$ with $x_a<1$ being  small. We will assume that $x_a$ is small enough so that the background functions are well approximated by (\ref{functions_deep_IR}) in the interval $[x_*,x_a]$ . Then, we can estimate the integral (\ref{d_x-integral}) for $d$ as:
\beq
d\,\approx\,{2\,L_{ABJM}^2\,x_*^2\over r_q}\,\,
\int_{x_*}^{x_a}\,{dx\over x^2\,\sqrt{x^4-x_*^4}}\,+\, {2\over \sqrt{h_*}}\,
\int_{x_a}^{\infty}\,{e^g\,h\over x}\,dx\,\,,
\label{d_deepIR}
\eeq
where $L_{ABJM}$ is the unflavored $AdS$ radius (\ref{ABJM-AdSradius3}) and we have used the fact that for $x_*\ll 1$ we have that $h_*\gg h(x)$  when $x\in  [x_a,\infty]$ and, therefore, in this interval we can neglect  $h(x)$ in the square root of the denominator of the integrand in (\ref{d_x-integral}). The first integral in (\ref{d_deepIR}) can be done analytically, yielding the result:
\beq
d\,\approx\,{2\,L_{ABJM}^2\over x_*\, r_q}\,\,\Bigg[
{\sqrt{2}\,\pi^{{3\over 2}}\over  \big[\Gamma\big({1\over 4}\big)\big]^2}\,-\,
{1\over 3}\,\Big({x_*\over x_a}\Big)^3\,
{}_2F_1\Big({1\over 2}, {3\over 4}; {7\over 4}; \Big({x_*\over x_a}\Big)^4\Big)\Bigg]\,+\,
{2\,x_*^2\, r_q^2\over L_{ABJM}^2}\,\int_{x_a}^{\infty}\,{e^g\,h\over x}\,dx\,\,.
\label{d_deepIR_explicit}
\eeq
For small $x_*$, at leading order, we get from (\ref{d_deepIR_explicit}):
\beq
d\,\approx\,{L_{ABJM}^2\over x_*\, r_q}\,\,
{2\sqrt{2}\,\pi^{{3\over 2}}\over  \big[\Gamma\big({1\over 4}\big)\big]^2}\,\,,
\label{d_deepIR_approx}
\eeq
which confirms that $r_q\,d$ is large when $x_*$ is small. We can make similar approximations in the integral (\ref{E_x-integral}) with the result:
\beq
E_{q\bar q}\approx {r_q\over \pi}\,\Bigg[\,
\int_{x_*}^{x_a}\,
\Big[\,{x^2\over \sqrt{x^4-x_*^4}}\,-\,1\Big]dx\,-\,x_*\,\Bigg]\,\,.
\eeq
Performing explicitly the integral, we get:
\beq
E_{q\bar q}\approx - {x_*\,r_q\over \pi}\,\Bigg[\,
{\sqrt{2}\,\pi^{{3\over 2}}\over  \big[\Gamma\big({1\over 4}\big)\big]^2}\,+\,
{x_a\over x_*}\,-\,{x_a\over x_*}\,
{}_2F_1\Big(-{1\over 4}, {1\over 2}; {3\over 4}; \Big({x_*\over x_a}\Big)^4\Big)\,\Bigg]\,\,,
\eeq
which, at leading order, becomes:
\beq
E_{q\bar q} \approx-
{\sqrt{2\pi}\over  \big[\Gamma\big({1\over 4}\big)\big]^2}\,\,
x_*\,r_q\,\,.
\eeq
Eliminating $x_*$ by using (\ref{d_deepIR_approx}) we arrive at the estimate (\ref{leading_IR_pot}).


\subsection{Asymptotics of the two-point functions}
\label{asymp_two-point_functions}
In this appendix we study the renormalized geodesic distance, and the corresponding two-point function of bulk operators with large conformal dimensions, in the UV limit of small separation $l$ and in the IR regime in which $r_q\,l\to \infty$. In the former case, our study will serve to fix the normalization constant ${\cal C}$ of (\ref{renormalized_geodesic}), as well as the analytic form of the correlator near the UV fixed point. This is the case we analyze first in the next subsection.

\subsubsection{UV behavior}
\label{UV_two-point_appendix}
We will now obtain the form of the correlator in the UV limit in which the turning point $x_*$ is large and the geodesic does not penetrate much into the bulk of the  geometry. In order to study this limiting case, it is convenient to perform a change of variables in the integral (\ref{l_general}) and write it as:
\beq
l\,=\,2\,x_{*}\,\int_{1}^{\infty}\,dz
{\sqrt{G(x_{*}\, z)}\over
\sqrt{e^{{1\over 2}\,(\phi_*-\phi(x_{*}\, z))}\,\,\big({h_*\over h(x_{*}\, z)}\big)^{1\over 2}\,-\,1}
}\,\,.
\label{l_xstar_z}
\eeq
Similarly, the renormalized geodesic length can be represented as:
\beq
{\cal L}_r\,=\,2\,x_{*}\,\int_{1}^{z_{max}}\,dz
{e^{-{\phi(x_{*}\, z)\over 4}}\,\big[h(x_{*}\, z)\big]^{-{1\over 4}}\,\,\sqrt{G(x_{*}\, z)}
\over
\sqrt{1-e^{{1\over 2}\,(\phi(x_{*}\, z))-\phi_*)}\,\,\big({h(x_{*}\, z)\over h_*}\big)^{1\over 2}}
}\,-\,{2\,L_0\,e^{-{\phi_0\over 4}}\over b}\,\log ({\cal C}\,x_{*}\, z_{max})\,\,,
\label{Lr_xstar_z}
\eeq
where $z_{max}=x_{max}/x_*$. When $x_*$ is large the argument of the functions in the integrals (\ref{l_xstar_z}) and (\ref{Lr_xstar_z}) is large and one can use the UV asymptotic expressions  (\ref{UVexpansion_h_phi}) and (\ref{H_G_UV}). Therefore, we can approximate $l$ as:
\beq
l\approx 2 \sqrt{G_{\infty}}\,x_{*}^{-{1\over b}}\,
\int_{1}^{\infty}\,{dz\over z^{1+{1\over b}}\,
\sqrt{z^{{2\over b}}-1}}\,\,.
\label{l_UV_integral}
\eeq
The integral  in (\ref{l_UV_integral})  just gives $b$ and, thus, we have:
\beq
l\approx 2\, b \,\sqrt{G_{\infty}}\,x_{*}^{-{1\over b}}\,=\,{2\over \kappa}\,{L_0^2\over r_q\,x_*^{{1\over b}}}\,\,.
\eeq
It follows that, when $x_*$ is large $r_q\,l$ is small. Thus the UV limit we are studying corresponds to $r_q\,l\to 0$. Similarly, the UV limit of ${\cal L}_r$ is:
\beq
{\cal L}_r\,\approx\,{2\,L_0\,e^{-{\phi_0\over 4}}\over b}\,
\Bigg[\int_{1}^{\infty}\,{dz\over z}\,\Big[{z^{{1\over b}}\over \sqrt{z^{{2\over b}}-1}}\,-\,1
\Big]\,-\,\log({\cal C}\,x_{*})\Bigg]\,\,.
\eeq
The integral in this last equation is:
\beq
\int_{1}^{\infty}\,{dz\over z}\,\Big[{z^{{1\over b}}\over \sqrt{z^{{2\over b}}-1}}\,-\,1
\Big]\,=\,b\,\log 2\,\,.
\eeq
On the other hand, the UV conformal dimension $\Delta_{UV}$  of a bulk field of mass $m$ has been written  in (\ref{DeltaUV}). Taking these facts into account, we can write:
\beq
e^{-m\,{\cal L}_r}\,\approx {\Big( b\,\sqrt{G_{\infty}}\,\,{\cal C}^{{1\over b}}\,
\Big)^{2\Delta_{UV}}\over l^{2\Delta_{UV}}}\,\,.
\label{emLr_UV}
\eeq
If the operator ${\cal O}$  is canonically normalized in the short-distance $r_q\,l\to 0$ limit, the coefficient in the numerator of (\ref{emLr_UV}) should be chosen to be one as in (\ref{UV_two-point}). Therefore, it follows that the constant ${\cal C}$ is fixed to
\beq
{\cal C}^{{1\over b}}={\sqrt{\lambda}\over r_q\, b\,\sqrt{G_{\infty}}}\,=\,{\kappa\sqrt{\lambda}\over L_0^2}\,\,.
\label{calC_value}
\eeq

Let us now evaluate the first correction of the two-point correlator around the UV fixed point. Let us expand the function $G(x)$ for large $x$ as in (\ref{H_G_rho_UVexpansion}) and (\ref{deltaHG_UV}). Then, for large $x_*$ we get:
\beq
\sqrt{G(x_{*}\, z)}\approx \sqrt{G_{\infty}}\,\,x_*^{-1-{1\over b}}\,
z^{-1-{1\over b}}\,\Big[1\,+\,{G_2\over 2}\,x_*^{-2}\,z^{-2}\Big]\,\,,
\eeq
where $G_2=h_2+2g_2$.  Similarly:
\beq
\sqrt{e^{{1\over 2}\,(\phi_*-\phi(x_{*}\, z))}\,\,\Big({h_*\over h(x_{*}\, z)}\Big)^{1\over 2}\,-\,1}\approx
\sqrt{z^{{2\over b}}-1}\,\Big[1+{\phi_2+h_2\over 4 x_*^2}\,\,{
z^{{2\over b}-2}\,(z^2-1)\over z^{{2\over b}}-1}\,\Big]\,\,.
\eeq
It follows that the separation $l$ of the two points of the correlator can be written as:
\beq
l\,\approx\,2\,b\,\sqrt{G_{\infty}}\,x_*^{-{1\over b}}
\Big[1\,+\,{1\over 2x_*^2}\,\big(G_2\,J_1\,-\,{\phi_2+h_2\over 2}\,J_2\big)\Big]\,\,,
\label{l_xstar}
\eeq
where $J_1$ and $J_2$ are the following integrals:
\bear
&&J_1\,\equiv\,{1\over b}\,\int_{1}^{\infty}\,{dz\over z^{3+{1\over b}}\,\sqrt{z^{{2\over b}}-1}}\,=\,
{\sqrt{\pi}\over 2}\,{\Gamma\big(1+b\big)
\over 
\Gamma\Big({3\over 2}+b\Big)}
\,\,,\rc\rc
&&J_2\,\equiv\,{1\over b}\,\int_{1}^{\infty}\,{z^2-1\over z^{3-{1\over b}}\,
\big(z^{{2\over b}}-1\big)^{{3\over 2}}}\,dz\,=\,
\sqrt{\pi}\,{\Gamma\big(1+b\big)\over 
\Gamma\Big({1\over 2}+b\Big)}\,-\,1\,\,.
\label{J1_J2}
\eear
Let us invert the relation (\ref{l_xstar}) at first order and write $x_*$ as a function of $l$. We get:
\beq
x_*\,\approx\, \Big[{2\sqrt{G_{\infty}}\,b\over l}\Big]^b\,\Big[1+c_{x_*}\,l^{2b}\Big]\,\,,
\label{xstar_l_fo}
\eeq
where $c_{x_*}$ is a constant given by:
\beq
c_{x_*}\,=\,{b\over \big(2\sqrt{G_{\infty}}\,b\big)^{2b}}\,\,
\Bigg[\Big({g_2\over 2b+1}\,-\,{2b-1\over 2b+1}\,{h_2\over 4}\,-\,{\phi_2\over 4}\Big)\,
\,{\sqrt{\pi}\,\Gamma\big(1+b\big)\over 
\Gamma\Big({1\over 2}+b\Big)}\,+\,{\phi_2+h_2\over 4}\,\Bigg]\,\,.
\eeq
Moreover,  the  renormalized geodesic length at first order takes the form:
\beq
{\cal L}_r\,\approx\,2\,L_0\,e^{-{\phi_0\over 4}}\,\Bigg[\,
\log\Big({2b \,r_q\,\,\sqrt{G_{\infty}}\over x_*^{{1\over b}}\sqrt{\lambda}}\Big)\,+\,
{1\over x_*^2}\,\Big[\,\Big(g_2\,-\,{\phi_2\over 4}\,+\,{h_2\over 4}\Big)\,J_3\,-\,
{\phi_2+h_2\over 4}\,J_2\,\Big]\Bigg]\,\,,
\eeq
where $J_2$ is the integral defined in (\ref{J1_J2}) and $J_3$ is:
\beq
J_3\,\equiv {1\over b}\,
\int_{1}^{\infty}\,{dz\over z^{3-{1\over b}}\,\sqrt{z^{{2\over b}}-1}}\,=\,
{\sqrt{\pi}\over 2}\,{\Gamma\big(b\big)
\over 
\Gamma\Big({1\over 2}+b\Big)}\,\,.
\eeq
Let us now write ${\cal L}_r$ in terms of $l$. By using (\ref{xstar_l_fo}) to eliminate $x_*$ in terms of $l$ we arrive at:
\beq
{\cal L}_r\,\approx\,2\,L_0\,e^{-{\phi_0\over 4}}\,\Big[\,\log \Big({r_q\, l\over \sqrt{\lambda}}\Big)\,+\,c_{{\cal L}}\,\Big({r_q\, l\over \sqrt{\lambda}}\Big)^{2b}\,\,\Big]\,\,,
\eeq
where the coefficient $c_{{\cal L}}$ is given by:
\beq
c_{{\cal L}}\,=\,\Bigg({\sqrt{\lambda}\over 2b\,r_q\,\sqrt{G_{\infty}}}\Bigg)^{2b}\,
\,{\sqrt{\pi}\,\Gamma\big(b\big)
\over 
\Gamma\Big({1\over 2}+b\Big)}\,\,
\Bigg[\,{g_2\over 2(2b+1)}\,-\,{\phi_2\over 8}\,-\,{1\over 8}\,{2b-1\over 2b+1}\,h_2\,\Bigg]\,\,.
\eeq
Therefore, the two-point correlator near the UV fixed point can be written as:
\beq
\Big\langle {\cal O}(t, l)\,{\cal O}(t, 0)\Big\rangle\,=\,
{f_{\Delta}(r_q\,l/\sqrt{\lambda})\over (r_q\, l/\sqrt{\lambda})^{2\Delta_{UV}}}\,\,,
\label{nearUV_corr}
\eeq
where the function $f_{\Delta}$ parameterizes the deviation from the conformal behavior near the UV fixed point and is given, at first order, by
\beq
f_{\Delta}(r_q\,l/\sqrt{\lambda})\,\approx\,1\,+\,c_{\Delta}\,\Big({r_q\,l\over \sqrt{\lambda}}\Big)^{2b}\,\,,
\label{fDelta_rewrite}
\eeq
where the new constant $c_{\Delta}$ is just:
\beq
c_{\Delta}\,=\,-{\Delta_{UV}\over 4}\,
{\sqrt{\pi}\,\Gamma\big(b\big)
\over \Gamma\Big({1\over 2}+b\Big)}\,
\Big({\kappa\over 2 \sqrt{2}\,\pi \,\sigma}\Big)^{2b}\,
\Big[\,{4\over 2b+1}\,g_2\,-\,{2b-1\over 2b+1}\,h_2\,-\,\phi_2\,\Big]\,\,.
\label{cDelta}
\eeq

\subsubsection{IR behavior}
\label{IR_two-point_appendix}

Let us now analyze the two-point functions in the IR limit, which corresponds to taking $x_*\to 0$. In this case the geodesic penetrates deeply in the bulk. 
We will proceed similarly to what we did in section \ref{qq_potential_IR} for the $q\bar q$ potential in the IR. Accordingly, we split the interval $[x_*, x_{max}]$ as
$[x_*, x_{max}]=[x_*, x_{a}]\cup [x_{a}, x_{max}]$ for some $x_a<1$ small. 
Near the turning point, \ie, when $x_*\le x\le x_a$, we can approximate the functions $h(x)$ and $g(x)$ as:
\beq
h(x)\approx  \Big({c\,L_{IR}\over \gamma}\Big)^4\,{1\over x^4}\,\,,
\qquad\qquad
e^{g}\approx {\gamma\over c}\,\,x\,\,,
\label{h_g_deepIR}
\eeq
where $L_{IR}=L_{ABJM}$ is the AdS radius of the unflavored ABJM solution and $\gamma$ and $c$ are the parameters of the unflavored running solution.  It follows that, in this IR region,  the function $G(x)$ behaves as:
\beq
G(x)\approx {c^2\over \gamma^2}\,{L_{IR}^4\over x^4}\,\,.
\eeq
Moreover, the dilaton $\phi_{IR}$ is constant and given by the ABJM value $\phi_{ABJM}$ written in (\ref{ABJMdilaton3}). 

Suppose that we have chosen some $x_a\gg x_*$.  Let us then split the integral (\ref{l_general}) for $l$ as:
\beq
l\,=\,2\,\Big[\int_{x_*}^{x_a}dx+\int_{x_a}^{x_{max}}dx\Big]\,\,
{\sqrt{G(x)}\over
\sqrt{e^{{1\over 2}\,(\phi_*-\phi(x))}\,\,\big({h_*\over h(x)}\big)^{1\over 2}\,-\,1}
}\,\equiv\,l_{IR}+l_{UV}\,\,,
\label{l_IRplusUV}
\eeq
where $x_{max}$ should be sent to $+\infty$ at the end of the calculation. Let us approximate the first integral in (\ref{l_IRplusUV}) by taking the functions in the deep IR, as in (\ref{h_g_deepIR}). We obtain:
\beq
l_{IR}\,\approx\,{2c\over \gamma}\,L_{IR}^2\,\int_{x_*}^{x_a}\,{dx\over x^2\,
\sqrt{{x^2\over x_*^2}-1}}\,=\,
{2c\over \gamma}\,{L_{IR}^2\over x_*}\,\sqrt{1-{x_*^2\over x_a^2}}\,\,.
\eeq
For small $x_*/x_a$ we can approximate this integral as:
\beq
l_{IR}\,\approx\,{2c\over \gamma}\,{L_{IR}^2\over x_*}\,\,.
\eeq
To evaluate $l_{UV}$ approximately we notice that $h_{*}\propto x_*^{-4}$ and, therefore, it is large for $x_*\to 0$. Then, we can neglect  the one inside the square root and approximate  $l_{UV}$ as:
\beq
l_{UV}\approx x_*\,{\gamma\,e^{-{\phi_{IR}\over 4}}\over c\,L_{IR}}\,\,
\int_{x_a}^{\infty} dx\,e^{{\phi(x)\over 4}}\,\big[h(x)\big]^{{1\over 4}}\,\sqrt{G(x)}\,\,.
\label{l_UV_estimate}
\eeq
Since the integral in (\ref{l_UV_estimate})  converges and is independent of $x_*$, it follows that $l_{UV}\sim x_*$ and, therefore, it can be neglected with respect to the large value of $l_{IR}\sim 1/x_*$.  Thus, we take
\beq
l\approx {\hat\gamma-1\over \gamma}\,{L_{IR}^2\over r_q\,x_*}\,\,.
\label{l_estimate_smallxs}
\eeq
Notice that it follows from this equation that $r_q\,l$ is large if $x_*$ is small, as it should be in the IR regime.  Let us next perform a similar analysis for the renormalized geodesic length ${\cal L}_r$. First, we consider the IR integral:
\bear
&&\int_{x_*}^{x_{a}}\,dx
{e^{-{\phi(x)\over 4}}\,\big[h(x)\big]^{-{1\over 4}}\,\,\sqrt{G(x)}
\over
\sqrt{1-e^{{1\over 2}\,(\phi(x))-\phi_*)}\,\,\big({h(x)\over h_*}\big)^{1\over 2}}}\,\approx\,
e^{-{\phi_{IR}\over 4}}\,L_{IR}\,\int_{x_*}^{x_{a}}\,
{dx\over x\sqrt{1-{x_*^2\over x^2}}}\,=\,\rc\rc
&&\qquad
=\,e^{-{\phi_{IR}\over 4}}\,L_{IR}\,\log\Big({x_a+\sqrt{x_a^2-x_*^2}\over x_*}\Big)
\,\approx\,e^{-{\phi_{IR}\over 4}}\,L_{IR}\,\log\big({2x_a\over x_*}\big)\,\,.
\eear
To evaluate the UV integral we proceed similarly to what we did for the integral for $l$ and, in this case,  we only keep the one inside the square root. Then:
\beq
\int_{x_{a}}^{x_{max}}\,dx
{e^{-{\phi(x)\over 4}}\,\big[h(x)\big]^{-{1\over 4}}\,\,\sqrt{G(x)}
\over
\sqrt{1-e^{{1\over 2}\,(\phi(x))-\phi_*)}\,\,\big({h(x)\over h_*}\big)^{1\over 2}}}\,\approx\,
\int_{x_{a}}^{x_{max}}\,dx\,
e^{-{\phi(x)\over 4}}\,\big[h(x)\big]^{-{1\over 4}}\,\,\sqrt{G(x)}\,\,.
\eeq
Therefore:
\beq
m\,{\cal L}_r\,\approx\,2\,\Delta_{IR}\log\big({2x_a\over x_*}\big)\,+\,2m\,
\int_{x_a}^{x_{max}}\,dx\,\,e^{-{\phi(x)\over 4}}\,\big[h(x)\big]^{-{1\over 4}}\,\,\sqrt{G(x)}\,
-\,2\,\Delta_{UV}\log\big({\kappa\,\sqrt{\lambda}\,x_{max}^{{1\over b}}\over L_0^2}\big)\,\,,\qquad
\eeq
where $\Delta_{IR}$ is the conformal dimension of the operator ${\cal O}$ in the IR conformal point, given by:
\beq
\Delta_{IR}\,=\,m\,L_{IR}\,e^{-{\phi_{IR}\over 4}}\,\,,
\eeq
which is just (\ref{Delta_IR}). 
Let us next define the following quantities:
\bear
&& {\cal I}_{IR}\,\equiv\,-2m\,\int_{x_a}^{1}\,dx\, 
e^{-{\phi(x)\over 4}}\,\big[h(x)\big]^{-{1\over 4}}\,\,\sqrt{G(x)}\,-\,2\, \Delta_{IR}\,\log x_a\,\,,\rc\rc
&& {\cal I}_{UV}\,\equiv\,-2m\,\int_{1}^{x_{max}}\,dx\, 
e^{-{\phi(x)\over 4}}\,\big[h(x)\big]^{-{1\over 4}}\,\,\sqrt{G(x)}\,+\,
{2\, \Delta_{UV}\over b}\,\log x_{max}\,\,.
\label{I_IR_UV_def}
\eear
Then, after using the relation (\ref{l_estimate_smallxs}) to eliminate $x_*$ in favor of $l$, we get:
\beq
m\,{\cal L}_r\,\approx\,\log (r_q\,l/\sqrt{\lambda})^{2\Delta_{IR}}\,+\,
\log\Big[\,\Big({2\gamma\over \hat\gamma-1}\,{\sqrt{\lambda}\over L^2_{IR}}\Big)^{2\Delta_{IR}}
\Big({\kappa\over L_0^2}\Big)^{-2\Delta_{UV}}\Big]\,-\, {\cal I}_{IR}\,-\, {\cal I}_{UV}\,\,.
\eeq
Then, it follows that the IR limit $r_q\, l\to \infty$ of the two-point correlator is as in (\ref{VEV_IR}),  where ${\cal N}$ is the normalization constant given by:
\beq
{\cal N}\,=\,{\Big({\kappa\,\sqrt{\lambda}\over L_0^2}\Big)^{2\Delta_{UV}}\over
\Big({2\gamma\over \hat\gamma-1}\,{\sqrt{\lambda}\over L^2_{IR}}\Big)^{2\Delta_{IR}}}\,\,
\exp\big[\, {\cal I}_{IR}\,+\,{\cal I}_{UV}\big]\,\,.
\label{Cal_N}
\eeq
It turns out that, in the expression of the integrals in (\ref{I_IR_UV_def}) we can take the limits $x_a\to 0$  (in ${\cal I}_{IR}$) and  $x_{max}\to\infty$ (in ${\cal I}_{UV}$). Actually, it can be easily proved that, after taking these limits,  ${\cal I}_{IR}$ and ${\cal I}_{UV}$ can be recast as:
\bear
&& {\cal I}_{IR}\,=\,2\Delta_{IR}\,\int_{0}^{1}\,
{dx\over x}\,\Bigg[\,1-\,e^{{\phi_{IR}-\phi(x)\over 4}}\,
\Big[ {h_{IR}(x)\over h(x)}\Big]^{{1\over 4}}\,
\sqrt{{G(x)\over G_{IR}(x)}}\,\Bigg]\,\,,\rc\rc
&& {\cal I}_{UV}\,=\,{2\Delta_{UV}\over b}\,\int_{1}^{\infty}\,
{dx\over x}\,\Bigg[\,1\,-\,e^{{\phi_{0}-\phi(x)\over 4}}\,
\Big[ {h_{UV}(x)\over h(x)}\Big]^{{1\over 4}}\,
\sqrt{{G(x)\over G_{UV}(x)}}\,\Bigg]\,\,.
\eear
Notice that the form of the correlator is consistent with the fact that the conformal symmetry is recovered in the IR limit $r_q\,l\to\infty$. The non-canonical normalization factor ${\cal N}$ is due to the fact that we chose to renormalize ${\cal L}$ in such a way that the correlator is canonically normalized in the opposite UV limit $r_q\,l\to 0$.


\subsection{WKB mass levels}
\label{appendix_WKB_masses}

Consider the following differential equation for the function $R(x)$:
\beq
\partial_{x}\,\Big[\,P(x)\,\partial_{x} R\,\Big]\,+\,
m^2\,Q(x)\,R\,=\,0\,\,,
\label{second_order_ODE}
\eeq
where $x$ takes values in the range  $x_*\le x\le \infty$, 
$ m$ is the mass parameter and $P(x)$ and $Q(x)$  are two
arbitrary functions that are independent of $ m$. We will assume that
near $x\approx x_*,\infty$ these functions behave as:
\bear
&&P\approx P_1(x-x_*)^{s_1}\,\,,
\,\,\,\,\,\,\,\,\,\,\,\,\,\,
Q\approx Q_1(x-x_*)^{s_2}\,\,,
\,\,\,\,\,\,\,\,\,\,\,\,\,\,{\rm as}\,\,x\to x_*\,\,,\rc\rc
&&P\approx P_2\,x^{r_1}\,\,,
\,\,\,\,\,\,\,\,\,\,\,\,\,\,
\qquad\,\,\,\,\,
Q\approx Q_2\, x^{r_2}\,\,,
\,\,\,\,\,\,\,\,\,\,\,\,\,\,{\rm as}\,\,x\to \infty\,\,,
\eear
where $P_i$, $Q_i$,  $s_i$, and $r_i$ are constants. By a of suitable change of variables, the differential equation (\ref{second_order_ODE}) can be converted into a Schr\" odinger equation, which only admits a discrete set of normalizable solutions for a set of values of $m$ parameterized by a quantum number $n\ge 0$. The mass levels for large values of  $n$ can be evaluated in the WKB approximation \cite{RS}, with the result:
\beq
m_{WKB}\,=\,{\pi\over \xi}\,
\sqrt{(n+1)\,\Big(\,n+{|s_1-1|\over s_2-s_1+2}+{|r_1-1|\over r_1-r_2-2}\,\Big)}\,\,,
\eeq
where $\xi=\xi(x_*)$ is the integral:
\beq
\xi\,=\,\int_{x_*}^{\infty}\,dx\,\sqrt{Q(x)\over P(x)}\,\,.
\eeq
In our case, the functions $P(x)$ and $Q(x)$ which correspond to the fluctuation equation (\ref{meson_fluct_eq}) are:
\beq
P(x)\,=\,{h^{{1\over 4}}\,e^{2f-\phi}\over x}\,(x^2-x_*^2)\,\,,
\qquad\qquad
Q(x)\,=\,{h^{{5\over 4}}\,e^{2f+2g-\phi}\over x}\,\,,
\eeq
and it is  immediate to extract the exponents $s_1$, $s_2$, $r_1$, and $r_2$:
\beq
s_1\,=\,1\,\,,\qquad\qquad
s_2\,=\,0\,\,,\qquad\qquad
r_1\,=\,1+{1\over b}\,\,,\qquad\qquad
r_2\,=-1-{1\over b}\,\,.
\eeq
Using these values one immediately finds that the WKB mass levels are given by (\ref{WKB_masses}).

\subsubsection{Asymptotic spectra}

Let us now evaluate analytically the meson spectrum in the two limiting cases in which the sea quark mass $m_q=r_q/2\pi$ is small or large compared to the valence quark mass $\mu_q$. Notice that $e^{g}\propto m_q$ (see eqs. (\ref{background_functions_xle1}) and (\ref{background_functions_xge1})). Therefore, (\ref{muq}) can be regarded as giving the relation $\mu_q/m_q$ as a function of $x_*$. 

Let us consider first the case in which $m_q$ is small. Clearly, when $m_q\to 0$ for fixed $\mu_q$ one necessarily   must have $x_*$  large. Actually, integrating (\ref{muq}) by using the asymptotic expression of $g$ written in (\ref{g_f_UVexpansions_x}) at leading order, we get:
\beq
\mu_q\,\approx\,\kappa\,\,
m_q\,\,x_*^{{1\over b}}\,\,,
\eeq
where the constant  $\kappa$ has been defined in (\ref{kappa_def}). 
Thus, when $m_q\to 0$ for fixed $\mu_q$ the $x$ coordinate of the tip of the flavor brane increases as:
\beq
x_*\,\sim\,m_q^{-b}\,\,\,,
\qquad\qquad
(m_q\to 0)\,\,.
\eeq
Let us now evaluate $\xi(x_*)$  when $x_*$ is large. First of all we perform a change of variables in the integral (\ref{WKB_xi}) and rewrite $\xi(x_*)$ as:
\beq
\xi(x_*)\,=\,\int_{1}^{\infty}\,dz\,{e^{g(x_*\,z)}\sqrt{h(x_*\,z)}\over \sqrt{z^2-1}}\,\,.
\label{WKB_xi_asymp}
\eeq
If $x_*$ is large,  the argument of the functions $g$ and $h$ in (\ref{WKB_xi_asymp}) is large and we can use their UV asymptotic expressions (\ref{g_f_UVexpansions_x}) and (\ref{UVexpansion_h_phi}) to evaluate them. At leading order, we get:
\beq
e^{g(x_*\,z)}\approx {r_*\over b}\,z^{{1\over b}}\,\,,
\qquad\qquad
h(x_*\,z)\approx {L_0^4\over r_*^4}\, z^{-{4\over b}}\,\,,
\eeq
where $r_*$ is the value of the $r$ coordinate corresponding to $x=x_*$. Therefore, for $m_q\to 0$ for fixed $\mu_q$ the $\xi$ is approximately
\beq
\xi\,\approx\,{L_0^2\over b\,r_*}\,\int_{1}^{\infty}\,{dz\over z^{{1\over b}}\,
\sqrt{z^2-1}}\,=\,{L_0^2\over \,r_*}\,\sqrt{\pi}\,\,
{\Gamma\Big({2b+1\over 2b}\Big)\over 
\Gamma\Big({b+1\over 2b}\Big)}\,\,.
\eeq
Using this result in (\ref{WKB_masses}) we get the following mass spectrum in the UV limit:
\beq
m_{_{WKB}}^{^{(UV)}}\,=\,{\sqrt{\pi}\over \sqrt{2}}\,{r_*\over L_0^2}\,
{\Gamma\Big({b+1\over 2b}\Big)\over 
\Gamma\Big({2b+1\over 2b}\Big)}\,\,
\sqrt{(n+1)(2n+1)}\,\,,
\qquad\qquad (\mu_q/m_q\,\to \infty)\,\,.
\eeq
Let  us write this result in terms of physical quantities. Recall that $L_0$ and $r_*$ are related to the 't Hooft coupling $\lambda$ and to the valence quark mass $\mu_q$ as:
\beq
L_0^2\,=\,\pi\sqrt{2\lambda}\,\sigma\,\,,
\qquad\qquad
r_*\,=\,2\pi\alpha'\,\mu_q\,\,,
\eeq
where $\sigma$ is the screening function written in (\ref{screening-sigma}). 
Then, for  large  $\mu_q/m_q$, we have:
\beq
m_{_{WKB}}^{^{(UV)}}\,=\,
{\sqrt{\pi}\,\mu_q\over \sigma\,\sqrt{\lambda}}\,\,
{\Gamma\Big({b+1\over 2b}\Big)
\over 
\Gamma\Big({2b+1\over 2b}\Big)}\,\,
\sqrt{(n+1)(2n+1)}\,\,,
\qquad\qquad (\mu_q/m_q\,\to \infty)\,\,,
\label{m_WKB_UV}
\eeq
where we have taken $\alpha'=1$. Eq. (\ref{m_WKB_UV}) is exactly the WKB mass spectrum one gets for vector mesons in the massless  flavored background of ref. 
\cite{Conde:2011sw}.

Let us next consider the limit in which $\mu_q/m_q$ is small.  Since $e^g\sim m_q$, it follows from (\ref{muq}) that $x_*$ must be small. Actually, we can estimate the relation  between $\mu_q/m_q$ and $x_*$ by extracting from (\ref{functions_running_x}) and (\ref{background_functions_xle1}) the approximate expression of $e^{g}$ near $x=0$. We get:
\beq
e^{g}\approx {\gamma\over c}\,x\,\approx\,\pi\, (\hat\gamma+1)\,m_q\,x\,\,.
\eeq
Then, we have:
\beq
\mu_q\approx {\hat\gamma+1\over 2}\,m_q\,x_*\,\,,
\eeq
and it follows that, for fixed $\mu_q$ and large $m_q$, the coordinate of the tip of the flavor brane behaves as:
\beq
x_*\,\sim\,{1\over m_q}\,\,,
\qquad\qquad
(m_q\to\infty)\,\,.
\eeq
Thus, for large sea quark mass $x_*\to 0$ and thus 
the dynamics of the fluctuating flavor brane is dominated by the IR, where the solution corresponds to the running solution of the unflavored system. In this case, we have at leading order near $x\sim 0$:
\beq
e^{g(x)}\approx {\gamma\over c}\,x\,\approx\,r_*\,{x\over x_*}\,\,,
\qquad\qquad
h(x)\,\approx\,{2\pi^2 \,N\over k}\,{c^4\over \gamma^4}\,{1\over x^4}\,\approx\,
{2\pi^2 \,N\over k}\,{1\over r_*^4}\,\,\Big({x_*\over x}\Big)^4\,\,,
\eeq
where, in the last step, we used that $c/\gamma\approx x_{*}/r_*$. Thus,
\beq
e^{g(x_*\,z)}\sqrt{h(x_*\,z)}\approx {\sqrt{2\pi^2\lambda}\over r_*}\,{1\over z}
\,\,,
\eeq
and, after evaluating the integral (\ref{WKB_xi_asymp}) and writing the result in terms of the 't Hooft coupling $\lambda$  and the valence quark mass $\mu_q$ ,  we have:
\beq
m_{_{WKB}}^{^{(IR)}}\,=\,
{2\mu_q\over \sqrt{\lambda}}\,
\sqrt{(n+1)(2n+1)}\,\,,
\qquad\qquad (\mu_q/m_q\,\to 0)\,\,,
\label{m_WKB_IR}
\eeq
which is exactly the mass spectrum of vector mesons in the unflavored ABJM model \cite{Jensen:2010vx}. By combining (\ref{m_WKB_UV}) and (\ref{m_WKB_IR}) we obtain the UV/IR mass relation written in (\ref{UV-IR-mass_ratio}). It is now straightforward to obtain  this ratio  for small and large values of the deformation parameter. Indeed, for small  $\hat\epsilon$ one can expand $m_{_{WKB}}^{^{(UV)}}/ m_{_{WKB}}^{^{(IR)}}$ as:
\beq
{m_{_{WKB}}^{^{(UV)}}\over m_{_{WKB}}^{^{(IR)}}}\,=\,1\,+\,
{3-\log 2\over 4}\,\,\hat\epsilon\,+\,{1\over 384}\,
\Big[12\Big(\log 2\big(\log 2+3\big)\,-\,3\Big)-\pi^2\Big]\,\hat\epsilon^2\,+\,\cdots\,\,.
\label{UV/IR_mass_ratio_lowNf}
\eeq
Moreover, in the opposite    limit in which $\hat\epsilon$ is very large, we have:
\beq
{m_{_{WKB}}^{^{(UV)}}\over m_{_{WKB}}^{^{(IR)}}}\,\approx\, 
{8\sqrt{2\pi}\over 5\sqrt{15}}\,
{\Gamma\Big({9\over 10}\Big)
\over 
\Gamma\Big({7\over 5}\Big)}\,\,\sqrt{\hat \epsilon}\,\,.
\label{UV/IR_mass_ratio_largeNf}
\eeq

\subsection{Analytical expansion in flavor}
\label{Analytical_expansion}

In this appendix we will construct an analytic solution to (\ref{master_eq_W}) in a power series
expansion in the flavor deformation parameter $\hat\epsilon$. 
We start by writing an ansatz for $W(x)$ as:
\begin{equation}
\label{eq:Wseries}
 W(x)=\sum_{n=0}W_n(x)\hat\epsilon^n=2x+  W_1(x) \hat{\epsilon}+  W_2(x)\hat{\epsilon}^2+  W_3(x)\hat{\epsilon}^3+ {\cal{O}}(\hat\epsilon^4) \ ,
\end{equation}
where we cut the series at level ${\cal{O}}(\hat\epsilon^3)$ up to which point we will write down explicit results in the following.
In (\ref{eq:Wseries}), we chose $W_0(x)=2x$, such that for $\hat{\epsilon}=0$ we recover the ABJM theory by construction.
Expanding the master equation  (\ref{master_eq_W}) in $\hat{\epsilon}$ yields, at first order,
\begin{equation}
36 x^2 - 84 x^4 + 36 x^3 W_1(x) - 12 x^4 W_1'(x) - 6 x^5 W_1''(x)=0
\end{equation}
A general solution is:
\begin{equation}
W_1(x)=\frac{7 x^2 - 2}{2 x} + a_1 x^2+\frac{b_1}{x^3} \ ,
\end{equation}
where $a_1$ and $b_1$ are integration constants. We wish to match with massless flavored ABJM in the UV, so we set $a_1=0$,
because the leading term in the UV should go as $x$. The second constant $b_1$ is fixed by demanding continuity at the boundary
of the cavity, that is, at $x=1$.  For that purpose, we will consider the running solution (\ref{W_running_unflavored}) inside the cavity and expand it in $\hat{\epsilon}$.
Recall that $\gamma$ depends on $\hat{\epsilon}$, for which we write an expansion:
\begin{equation}
\gamma=\gamma_1\hat{\epsilon} +\gamma_2\hat{\epsilon}^2 +\gamma_3\hat{\epsilon}^3 +\ldots \ .
\end{equation}
By expanding (\ref{W_running_unflavored}) we find
\begin{equation}
W_{running}(x)=2x \left\{1+3 \gamma_1 x  \hat{\epsilon} + (3 \gamma_2 x - 2 \gamma_1^2 x^2) \hat{\epsilon}^2+(3 \gamma_3 x - 4 \gamma_1 \gamma_2 x^2 +3 \gamma_1^3 x^3) \hat{\epsilon}^3\right\}+\ldots \ .
\end{equation}
Now, imposing continuity of $W_1(x)$ and its first derivative at $x=1$, we obtain:
\begin{equation}
\gamma_1 = \frac{2}{5} \quad ,\quad b_1 = - \frac{1}{10} \ .
\end{equation}
Thus, the exact solution to the master equation at first order in $\hat{\epsilon}$ reads:
\begin{equation}
 W_1(x)  = \left\{\begin{array}{ll}
                        \frac{12}{5} x^ 2~,  & \ x\leq 1\\
                        ~~ \\
                        \frac{7 x^2 - 2}{2 x} - \frac{1}{10 x^3}~, & \ x\geq 1 ~.
                      \end{array} \right. 
\end{equation}
It is straightforward to repeat the iterative procedure for higher $\hat{\epsilon}$ orders. The expression for $W_2 (x)$ reads:
\begin{equation}
 W_2(x)  = \left\{\begin{array}{ll}
                        -\frac{8 x^2}{875}  (70 x-171)~, & \ x\leq 1\\
                        ~~ \\
                        \frac{13125 x^8+3500 x^6-3962 x^4-3360 x^4 \log x+300 x^2-35}{14000 x^7}~, & \ x\geq 1 ~,
                      \end{array} \right. 
\end{equation}
and that for $W_3(x)$:
\begin{equation}
 W_3(x)  = \left\{\begin{array}{ll}
                        \frac{16 x^2 (135135 x (35 x-76)+2356888)}{197071875}~, & \ x\leq 1\\ 
                           ~~\\
                           ~~\\
                        -\frac{23}{208000 x^{11}}+\frac{89}{184800 x^9}-\frac{302}{39375 x^7}+\frac{8293}{98000 x^5}-\frac{2120717}{8400000 x^3} & \\
                           ~~\\
-\frac{\log (x) \left(10703 x^4+20160 x^4 \log (x)-3600 x^2+840\right)}{70000 x^7}-\frac{15 x}{32}+\frac{37}{96 x}~, & \ x\geq 1~.
                        \end{array} \right. 
\end{equation}
The constants $\gamma$, $\kappa$ and $b$ read:
\begin{eqnarray}
\gamma & = & \frac{2 \hat{\epsilon} }{5}+\frac{228 \hat{\epsilon}^2}{875}+{\cal{O}}(\hat\epsilon^3)~, \\ 
 \kappa & = &  1-\frac{33 \hat{\epsilon} }{80}+\frac{128881 \hat{\epsilon}^2}{448000} +{\cal{O}}(\hat\epsilon^3) \ , \\
b & = & 1+  \frac{\hat\epsilon}{4}-\frac{7\hat\epsilon^2}{42}+ {\cal{O}}(\hat\epsilon^3)\ .
\end{eqnarray}

The metric functions can then be explicitly solved for. We will write them down in the second order of the expansion parameter $\hat\epsilon$. The constants are fixed to make the functions
continuous across the boundary of the cavity $x=x_q=1$ corresponding to $r=r_q$. We only present the solutions in terms of $x$-variable, the transformation to the original coordinate
is readily obtained through the relationship $x\leftrightarrow r$ below (\ref{eq:rvsxsmalleps}).

The squashing functions read:
\begin{equation}
 \frac{e^f}{r_q}-x=  \left\{\begin{array}{ll}
                        \hat\epsilon  \left(\frac{2 x}{5}-\frac{3 x^2}{5}\right)  +  \hat\epsilon^2 \frac{1085 x^3-1104 x^2+176 x}{1750}+{\cal{O}}(\hat\epsilon^3)~, & \ x\leq 1\\
                           ~~\\
                           ~~\\
                        \hat\epsilon  \frac{-33 x^4-20 x^4 \log (x)+20 x^2-3}{80 x^3} &  \\
                           ~~\\
                        +\hat\epsilon^2 \frac{142881 x^8-130200 x^6+34426 x^4-6600 x^2+280 x^4 \log x \left(615 x^4+50 x^4 \log x-100 x^2-129\right)-315}{448000 x^7} \\
                           ~~\\
                        +{\cal{O}}(\hat\epsilon^3)~, &\ x\geq 1 ~.
                      \end{array} \right.
\end{equation}
and
\begin{equation}
 \frac{e^g}{r_q}-x=  \left\{\begin{array}{ll}
                        \hat\epsilon  \left(\frac{2 x}{5}-\frac{4 x^2}{5}\right)  +  \hat\epsilon^2 \frac{8}{875} \left(105 x^3-92 x^2+11 x\right)+{\cal{O}}(\hat\epsilon^3)~, & \ x\leq 1\\
                           ~~\\
                           ~~\\
                        \hat\epsilon  \frac{-53 x^4-20 x^4 \log x+20 x^2+1}{80 x^3} &  \\
                           ~~\\
                        +\hat\epsilon^2 \frac{903243 x^8-754600 x^6+133854 x^4+11400 x^2+840 x^4 \log x \left(715 x^4+50 x^4 \log x-100 x^2+43\right)+1015}{1344000 x^7} \\
                           ~~\\
                        +{\cal{O}}(\hat\epsilon^3)~, &\ x\geq 1 ~.
                      \end{array} \right. \label{egexpandedsecondorder}
\end{equation}
The relative squashing between the spheres reads:
\begin{equation}
 q-1=  \left\{\begin{array}{ll}
                        \hat\epsilon  \frac{2 x}{5}  +  \hat\epsilon^2 \frac{4}{875} x (57-70 x)+{\cal{O}}(\hat\epsilon^3)~, & \ x\leq 1\\
                           ~~\\
                        \hat\epsilon  \left(\frac{1}{2}-\frac{1}{10 x^4}\right)
                        +\hat\epsilon^2 \frac{-13125 x^8+17500 x^6-6006 x^4-10080 x^4 \log (x)-900 x^2+35}{42000 x^8} 
                        +{\cal{O}}(\hat\epsilon^3)~, &\ x\geq 1 ~.
                      \end{array} \right.
\end{equation}
In order to compute $h(x)$ outside the cavity we use expression (\ref{warp_factor_x}), and $\beta$ is fixed to zero. Furthermore, to
match it with the corresponding function inside the cavity (\ref{h_running-x}), we impose the matching condition
(\ref{matching_h}) . This condition fixes $\alpha$ to be:
\begin{equation}
\alpha =  \frac{25}{14\hat\epsilon^2}-\frac{4646}{539\hat\epsilon} +{\cal{O}}(\hat\epsilon^0)  \ ,  
\end{equation}
and the function $h(x)$ is:
\begin{eqnarray}
 &&\frac{kr_q^4}{4\pi^2N}h-\frac{1}{2 x^4} \nonumber\\
&& ~~\\
 &&=  \left\{\begin{array}{ll}
                        \hat\epsilon \frac{2 \left(x^3+14 x-14\right)}{35 x^4}+\hat\epsilon^2 \frac{2 \left(440 x^4-1230 x^3+1155 x^2-3652 x+2882\right)}{9625 x^4}  +{\cal{O}}(\hat\epsilon^3)~, & \ x\leq 1\\
                           ~~\\
                           ~~\\
                        \hat\epsilon \frac{91 x^4+140 x^4 \log x-84 x^2+9}{280 x^8} &  \\
                           ~~\\
                        +\hat\epsilon^2 \frac{1320 x^4 \log x \left(-665 x^4+700 x^4 \log x-840 x^2+306\right)+11 \left(-58659 x^6+14560 x^4+20424 x^2-5280\right) x^2+7465}{3696000 x^{12}} & \\
                           ~~\\
                        +{\cal{O}}(\hat\epsilon^3)~, &\ x\geq 1 ~.
                      \end{array} \right. ~.
\end{eqnarray}
The dilaton reads:
\begin{eqnarray}
 &&\left(\frac{k^5}{2\pi^2N}\right)^{1/4}x e^{\phi}-2 x \nonumber\\
&& ~~\\
 &&=  \left\{\begin{array}{ll}
                        \hat\epsilon \frac{2}{35} \left(x^2-14\right) x^2+\hat\epsilon^2 \frac{\left(165 x^5+1540 x^3+11060 x^2-70070 x+35112\right) x^2}{67375}  +{\cal{O}}(\hat\epsilon^3)~, & \ x\leq 1\\
                           ~~\\
                           ~~\\
                        -\hat\epsilon \frac{245 x^4-84 x^2+3}{70 x^3} &  \\
                           ~~\\
                        +\hat\epsilon^2 \frac{5322625 x^8-4117960 x^6+765534 x^4-110880 x^4 \log x-27720 x^2-575}{1078000 x^7}  & \\
                           ~~\\
                        +{\cal{O}}(\hat\epsilon^3)~, &\ x\geq 1 ~.
                      \end{array} \right.
\end{eqnarray}
and the expression for the function $e^{\Lambda}=e^{\phi}h^{-1/4}$ is:
\begin{eqnarray}
 &&\frac{k e^\Lambda}{r_q}-2x \nonumber \\
&& ~~\\
 &&=  \left\{\begin{array}{ll}
                        \hat\epsilon \frac{4}{5} x (1-4 x)   +  \hat\epsilon^2  \frac{16}{875} x \left(280 x^2-184 x+11\right)+{\cal{O}}(\hat\epsilon^3)~, & \ x\leq 1\\
                           ~~\\
                           ~~\\
                        -\hat\epsilon  \frac{153 x^4+20 x^4 \log x-60 x^2+3}{40 x^3} &  \\
                           ~~\\
                        +\hat\epsilon^2 \frac{3906243 x^8-3243800 x^6+692958 x^4-40200 x^2+840 x^4 \log x \left(1215 x^4+50 x^4 \log x-300 x^2-129\right)-385}{672000 x^7} \\
                           ~~\\
                        +{\cal{O}}(\hat\epsilon^3)~, &\ x\geq 1~.
                      \end{array} \right.
\end{eqnarray}

In order to transform the above formulas to the canonical radial coordinate $r$, let us write down the relation to $x$. Integrating equation (\ref{r-x-diff-eq})  by using (\ref{egexpandedsecondorder}) and fixing the integration constant by imposing the condition $r(x = 1) = r_q$, the relation between the holographic coordinates $r$ and $x$ outside the cavity can be computed. Expanding expression (\ref{rasfunctionofx}) in $\hat{\epsilon}$  we obtain the corresponding relation inside the cavity:
\begin{equation}
\label{eq:rvsxsmalleps}
 \frac{r}{r_q} -x =  \left\{\begin{array}{ll}
                        -\hat\epsilon  \frac{2}{5} x (x-1) + \hat\epsilon^2\frac{8}{875} x \left(35 x^2-46 x+11\right)   +{\cal{O}}(\hat\epsilon^3)~, & \ x\leq 1\\
                           ~~\\
                           ~~\\
                        +\hat\epsilon \left[ \frac{2}{3}+\frac{-99 x^4-60 x^4 \log x-60 x^2-1}{240 x^3}\right] &  \\
                           ~~\\
                        +\hat\epsilon^2 \frac{1159929 x^8+2515800 x^6-145894 x^4-6840 x^2+840 x^4 \log x \left(1845 x^4+150 x^4 \log x+300 x^2-43\right)-435}{4032000 x^7} \\
                           ~~\\
                        +{\cal{O}}(\hat\epsilon^3)~, &\ x\geq 1~.
                      \end{array} \right.
\end{equation}

Finally, introducing  (\ref{eta-flavor}) in expression (\ref{eqn:F2}) we obtain the expansion for  the $F_2$ form, and using (\ref{K-N}) and (\ref{r-x-diff-eq}) in expression (\ref{eqn:F4}), and expanding to second order in $\hat{\epsilon}$, we obtain the expression for the $F_4$ form.




\end{subappendices}

\newpage


\chapter{Holographic fractional Hall effect}
\label{chapterfour}

\section{Introduction}\label{intro}

Since its discovery more than thirty years ago, the QHE has been the subject of intense research.  Nevertheless, some aspects of the FQHE involve strongly-coupled  dynamics and are still not fully understood. The holographic AdS/CFT duality has proven to be a powerful tool in the study of quantum matter in the strongly-coupled regime, since it provides answers to difficult field theory questions by using classical gravitational theories in higher dimensions.  Therefore, it is quite natural to explore the possibility of constructing holographic models of the (F)QHE and to extract properties that are very difficult to obtain via weakly-coupled many-body field theory.

In recent years, two types of holographic models of the QHE have been proposed. The  first class consists of bottom-up models in Einstein-Maxwell-axio-dilaton theories \cite{KeskiVakkuri:2008eb, Goldstein:2009cv,Goldstein:2010aw, Bayntun:2010nx,Gubankova:2010rc}. These models are endowed with an $SL(2,{\mathbb Z})$ duality and, as a consequence, they capture some observed features of QH physics. However, it is very difficult to engineer these types of models to have a mass gap; \cite{Lippert:2014jma} is so far the only example of a gapped model in this class. 

The second approach to holographically realize the QHE makes use of top-down D-brane constructions \cite{Bergman:2010gm,Jokela:2011eb,Kristjansen:2012ny}, in which a (2+1)-dimensional gauge theory  with fermions in the fundamental representation is modeled  by a suitable D$p$-D$q$ brane intersection. The limit in which the D$q$-brane is treated as a probe in the D$p$-brane background corresponds in the field theory dual to the so-called quenched approximation in which loops of fundamental fermions are neglected. In this approach, the worldvolume theory of the probe brane encodes the physics of the fermions.  Generically, the probe brane crosses the horizon, yielding a black hole embedding, which is dual to a gapless metallic state.  The quantum Hall state is realized holographically as a Minkowski embedding, in which the brane ends smoothly above the black hole horizon. The distance from the horizon at which the probe caps off determines the mass gap. \newline

In this chapter, we want to engineer quantum Hall states in the flavored ABJM theory.  Such Hall states are only possible if parity is broken, which can be accomplished by turning on an appropriate internal flux on the D6-brane worldvolume. However, treating the backreaction of this internal flux is quite challenging.  For now, we will start with a single quenched massive quark in the background of $N_f$ unquenched massless quarks, a system analyzed in \cite{Jokela:2013qya, Jokela:2012dw}.  We then will turn on a parity-breaking internal flux on the worldvolume of this probe D6-brane.

In the presence of this internal flux, the Wess-Zumino term of the probe action contains the term $\int \hat C_1\wedge F^3$, where $\hat C_1$ is the pullback of the RR potential one-form. In the ABJM background $C_1$ has only internal components. Therefore, after integrating over the internal directions, we are left with an axionic term $F\wedge F$ along $AdS_4$, which indeed breaks parity and corresponds to a Chern-Simons term on the boundary. 

Even in the probe limit, choosing a consistent ansatz for this internal flux, which must also be quantized appropriately, is not obvious.   We can, however, take a cue from the ABJ model \cite{Aharony:2008gk}, \ie,  the $U(N+M)_{k}\times U(N)_{-k}$ Chern-Simons matter theory, which can be engineered in string theory by adding fractional D2-branes to the ABJM setup.  The corresponding gravity dual can be obtained from the ABJM solution by turning on a flat Neveu-Schwarz $B_2$ field proportional to the K\"ahler form of ${\mathbb C}{\mathbb P}^3$.  The pullback of this parity-breaking $B_2$ on a probe D6-brane can alternately be viewed as a worldvolume gauge field flux.  Inspired by this example, we will generalize this ABJ solution into an ansatz for the case with no background $B_2$ field and only a probe worldvolume flux, but with backreacted massless flavors.

Equipped with this ansatz for the internal gauge flux, we will show that, indeed, there are quantum Hall states in this setup.  From the QH perspective, one can regard the effects of the massless, backreacted quarks as representing intrinsic disorder due to the quantum fluctuations of the massive quark. We will compute the contribution of these fluctuations to the conductivities in the form of an integral extended in the holographic direction, from the tip of the brane to the $AdS$ boundary. 

Surprisingly, we will find a very special family of explicit, supersymmetric, gapped QH solutions at zero temperature.  These BPS solutions have nonzero charge density and equal electric and magnetic fields, and we can compute the Hall conductivity, including the effects of quark loops, analytically.\newline

The rest of this chapter is organized as follows. In section \ref{Background} we review the ABJM background with flavor at finite temperature.  Then, in section \ref{probes_with_flux}, we consider the embedding of a probe D6-brane with internal flux.  We first present in section \ref{internal_flux_quantization} the ansatz for the internal components of the worldvolume gauge field that will be used throughout the chapter and discuss the corresponding flux quantization condition. In section \ref{Full_ansatz} we generalize these results to nonvanishing background electric and magnetic fields, as well as to nonzero charge density and currents. We compute the corresponding longitudinal and transverse conductivities in section \ref{conductivities}. In section \ref{em_symmetry} we analyze the residual $SO_+(1,1)$ boost invariance of our system at zero temperature.  An analytic supersymmetric solution of the equations of motion at zero temperature is presented in section \ref{BPS-sol}.  Section \ref{mesons} is devoted to the analysis of quark-antiquark bound states, \ie, mesons.  In particular, we study the effect of the broken parity on the mass spectrum. In section \ref{discussion} we discuss our results.

The chapter is completed  with several appendices  \ref{appendixsusy4} where we provide details of our background geometry and discuss the quantization condition of the worldvolume flux obtained by comparison with the ABJ solution. We also include a detailed analysis of the equations of motion of the probe and kappa symmetry. Besides, we include the equations governing the fluctuations of the probe , where we also estimate the meson masses using a WKB approximation.


\section{The flavored ABJM black hole}
\label{Background}
In this section we will review, following \cite{Conde:2011sw, Jokela:2013qya, Jokela:2012dw}, the background geometry corresponding to the ABJM model with unquenched massless flavors in the smeared approximation at finite temperature. Additional details of this supergravity solution are given in appendix \ref{Background_details}. The ten-dimensional metric, in string frame,  has the form
\beq
ds^2\,=\,L^2\,\,ds^2_{BH_4}\,+\,ds^2_{6}\,\,,
\label{flavoredBH-metric}
\eeq
where $L$ is the radius of curvature, $ds^2_{BH_4}$ is the metric of a planar black hole in the four-dimensional Anti-de Sitter space, given by
\beq\label{BH4-metric}
 ds^2_{BH_4} = -r^2h(r) dt^2+\frac{dr^2}{r^2h(r)}+r^2\big[dx^2+dy^2\big] \ ,
\eeq
and  $ds^2_{6}$ is the metric of the compact internal six-dimensional manifold. The blackening factor $h(r)$ is given by
\beq
h(r)\,=\,1\,-\,\frac{r_h^3}{r^3} \ ,
\label{blackening-factor}
\eeq
where the horizon radius $r_h$ is related to the temperature $T$ by
$T={3\,r_h\over 4\pi}$. The internal metric $ds^2_{6}$ in (\ref{flavoredBH-metric}) is a deformation of the Fubini-Study metric of ${\mathbb C}{\mathbb P}^3$, realized as an ${\mathbb S}^2$-bundle over ${\mathbb S}^4$. Let 
$ds^2_{{\mathbb S}^4}$ be the standard metric for the unit round four-sphere and let 
$z^i$ ($i=1,2,3$) be  three Cartesian coordinates parameterizing the unit two-sphere ($\sum_i (z^i)^2\,=\,1$). Then, $ds^2_{6}$ can be written as:
\beq
ds^2_{6}\,=\,{L^2\over b^2}\,\,\Big[\,
q\,ds^2_{{\mathbb S}^4}\,+\,\big(d z^i\,+\, \epsilon^{ijk}\,A^j\,z^k\,\big)^2\,\Big] \ ,
\label{internal-metric-flavored}
\eeq
where $A^i$ are the components of the non-abelian one-form connection corresponding to an $SU(2)$ instanton. In appendix \ref{Background_details} we give a more explicit representation of the $ds^2_{6}$  line element  in terms of alternative coordinates. 

The parameters $b$ and $q$   in (\ref{internal-metric-flavored})  are constant squashing factors which  encode the effect of the massless flavors in the backreacted metric. Indeed, when $q=b=1$ the metric (\ref{internal-metric-flavored}) is just the canonical Fubini-Study metric of  the ${\mathbb C}{\mathbb P}^3$ manifold with radius $2L$ in  the so-called twistor representation. In this case 
 (\ref{flavoredBH-metric}) is the metric of the unflavored ABJM model at nonzero temperature. When the effect of the delocalized D6-brane sources is taken into account, the resulting metric is deformed as in (\ref{internal-metric-flavored}). It was shown in \cite{Conde:2011sw} that at zero temperature the particular deformation written in (\ref{internal-metric-flavored}) preserves ${\cal N}=1$ SUSY.

 The parameter $b$  in (\ref{internal-metric-flavored}) represents the relative squashing of the ${\mathbb C}{\mathbb P}^3$ part of the metric with respect to the $AdS_4$ part due to the flavor, while $q$ parameterizes an internal deformation which preserves the ${\mathbb S}^4$-${\mathbb S}^2$ split of the twistor representation of ${\mathbb C}{\mathbb P}^3$.  The explicit expressions for the coefficients $q$ and $b$ found in  \cite{Conde:2011sw} are given in  (\ref{q0-epsilon}) and (\ref{b_new}) respectively, where we drop the subindex $0$ for simplicity in the rest of the thesis. Recall that $\hat{\epsilon}$ is given by (\ref{epsilonhat0}) and $\eta$ is given by (\ref{eta_0-epsilon}).
As functions of $\hat \epsilon$, the squashing parameters $q$ and $b$ are monotonically  increasing functions, which approach the values $q\approx 5/3$ and $b\approx 5/4$ as 
$\hat\epsilon\to\infty$. 
Recall that another way to encode the loop effects of the massless sea quarks is the screening factor $\sigma$, defined in (\ref{screening-sigma}).
Without flavors, $\sigma = 1$, and as $\hat\epsilon\to\infty$, $\sigma \to 0$.
 The $AdS$ radius $L$ can then expressed in terms of $\lambda$  and the screening factor:
\beq
L^2\,=\,\pi\sqrt{2\lambda}\,\sigma \ .
\label{AdS_radius}
\eeq

The complete solution of type IIA supergravity with sources is endowed with RR two- and four-forms $F_2$ and $F_4$, as well as with a constant dilaton $\phi$ (whose value depends on $N$, $N_f$, and $k$). Their explicit expressions are given in the appendix \ref{Background_details}.


\section{D6-brane probes with flux}
\label{probes_with_flux}

We are interested in the dynamics of a massive quark holographically dual to a probe D6-brane with internal flux in the flavored ABJM background. The D6-brane extends along $r$ and the three Minkowski directions and, wraps on the internal manifold a three-cycle topologically equivalent to ${\mathbb R}{\mathbb P}^3={\mathbb S}^3/{\mathbb Z}_2$. This three-cycle will be parameterized by three angles $\alpha$, $\beta$, and $\psi$, and will be characterized by an embedding function $\theta(r)$. With this embedding, the D6-brane then has an induced metric given by (for details see appendix \ref{Background_details}):
\bear
 {ds^2_{7}\over L^2} &=& r^2\,\left[-h(r)\,dt^2+
 dx^2+dy^2\right]+
  {1\over r^2 }\,
\left({1\over h(r)}+{r^2\,\theta'^{\,2}\over b^2}\,\right)\,dr^2 \rc\rc
&&\qquad +\,{1\over b^2}\,\Big[
q\,d\alpha^2+q\,\sin^2\alpha \,d\beta^2+ \sin^2\,\theta\,\left(\,d\psi\,+\,\cos\alpha\,d\beta\,\right)^2\,\Big] \ ,
\label{induced_metric_Hall}
\eear
where $0\le \alpha < \pi$, $0\le \beta, \psi<2\pi$, and $\theta=\theta(r)$ determines the profile of the probe brane. Notice that the second line in (\ref{induced_metric_Hall}) is the line element of a squashed ${\mathbb R}{\mathbb P}^3$.

For a supersymmetric configuration at zero temperature, it is possible to use kappa symmetry to find an explicit solution for $\theta(r)$ (see the analysis in \cite{Conde:2011sw} and in appendix \ref{kappa}).  But, in general we will have to numerically solve the equations of motion to find $\theta(r)$.

The thermodynamic properties of D6-branes embedded in this way were studied in detail in \cite{Jokela:2012dw}. Here we will generalize some of these results by including worldvolume gauge fields.  In particular, we will turn on a nontrivial flux on the internal cycle. In the rest of this section we will determine the form of this internal worldvolume flux which gives rise to a consistent solution of the brane equations of motion.

\subsection{Internal flux}
\label{internal_flux_quantization}

Since we are primarily interested in gapped, QH states, let us focus on Minkowski (MN) embeddings of the probe, in which the brane ends smoothly at a radial position $r_*$ above the horizon, \ie,  $r_* > r_h$.  The D6-brane can cap off smoothly if, at the tip of the brane $r=r_*$, the angle $\theta$ reaches its minimal value $\theta=0$ where an ${\mathbb S}^{1} \subset {\mathbb R}{\mathbb P}^3$ shrinks to zero.  At the tip, the last term of (\ref{induced_metric_Hall}) vanishes and the induced metric takes the form:
\beq
\label{induced_metric_tip}
 {ds^2_{7}\over L^2}\Big|_{r=r_*}=r^2\,\big[-h_* dt^2+ dx^2+dy^2\big]+
  {q\over b^2}\,\,
 \Big[d\alpha^2\,+\,\sin^2\alpha \ d\beta^2\Big] \ ,
 \eeq
 where $h_*=h(r=r_*)$. From (\ref{induced_metric_tip}), we see that at the tip of the brane the coordinates $\alpha$ and $\beta$ span a non-collapsing ${\mathbb S}^{2}_{*}$.  As in other probe-brane QH models \cite{Bergman:2010gm, Jokela:2011eb}, we want to turn on a flux of the worldvolume gauge field $F$ on this non-shrinking sphere.

 Of course, this flux must be quantized appropriately. We will adopt the following quantization condition:
\beq
{1\over 2\pi\alpha'}\,\int_{{\mathbb S}^{2}_{*}}\,F\,=\,{2\pi M\over k} \ ,
\qquad\qquad M\in {\mathbb Z}\,\,.
\label{wv_quantization}
\eeq
 Notice that, compared with the ordinary flux quantization condition of the worldvolume gauge field, we are considering in (\ref{wv_quantization}) $M/k$ fractional units of flux. In appendix \ref{Background_details} we verify that (\ref{wv_quantization}) is the correct prescription for the flux quantization by studying the background without massless flavors, \ie, $N_f = 0$. In this case one can induce an internal $F$ flux through 
${\mathbb S}^{2}_{*}$ by switching on a flat Neveu-Schwarz  $B_2$ field proportional to the K\"ahler form of ${\mathbb C}{\mathbb P}^3$. Then, the quantization condition 
(\ref{wv_quantization}) follows from the fractional holonomy of $B_2$ along the ${\mathbb C}{\mathbb P}^1$ cycle of ${\mathbb C}{\mathbb P}^3$. In this setup the integer $M$ is the number of fractional D2-branes and this configuration is dual to the ABJ model \cite{Aharony:2008gk} with gauge group $U(N+M)_{k}\times U(N)_{-k}$. We also check in appendix \ref{Background_details} that $M$ can be identified with the Page charge for fractional D2-branes.

Let us now write a concrete ansatz for the internal gauge field $F$. We will represent $F$ in terms of a potential one-form  $A$ given by:
\beq
A\,=\,L^2\,a(r)\,(d\psi+\cos\alpha\,d\beta)\ ,
\label{A_internal}
\eeq
where the $L^2$ factor is introduced for convenience and $a=a(r)$ is a function of the radial coordinate which determines the varying flux on the $(\alpha,\beta)$ two-sphere. The field strength $F=dA$ corresponding to (\ref{A_internal}) is simply:
\beq
F\,=\,L^2\Big[a'(r)\,dr\wedge (d\psi+\cos\alpha d\beta)\,-\,
a(r)\,\sin\alpha\,d\alpha\wedge d\beta\Big] \ ,
\label{F_internal}
\eeq
which restricted to ${\mathbb S}^{2}_{*}$ becomes:
\beq
F\big|_{{\mathbb S}^{2}_{*}}\,=\,-L^2\,a_*\,\sin\alpha\,d\alpha\wedge d\beta \ ,
\eeq
where $a_*\equiv a(r=r_*)$ is the value of the flux function at the tip. It follows that
\beq
\int_{{\mathbb S}^{2}_{*}}\,F\,=\,-4\pi\,L^2\,a_* \ ,
\eeq
and the condition  (\ref{wv_quantization}) quantizes the values of   $a_*$ in the following way:
\beq
a_*\,=\,-{\pi M\over k L^2}\,\,,
\qquad\qquad M\in {\mathbb Z} \ .
\label{a*_quantization}
\eeq
Let us denote the value of the flux function at the tip as:
\beq
a_*\,=\,-Q \ .
\label{a_*_Q}
\eeq
To write the quantization condition (\ref{a*_quantization}) in terms of physical quantities, recall that the $AdS$ radius $L$ can be written as in (\ref{AdS_radius}). Plugging this into (\ref{a*_quantization}), we find the following quantization condition for $Q$:
\beq
Q\,=\,{\sqrt{\lambda}\over \sqrt{2}\,\sigma}\,{M\over N} \ ,
\qquad\qquad M\in{\mathbb Z} \ .
\eeq

Using the ansatz (\ref{F_internal}) for the internal flux, we can try to find a solution for a MN embedding of the probe D6-brane.  In appendix \ref{EOMs} we check that (\ref{F_internal}), together with embedding ansatz corresponding to the induced metric (\ref{induced_metric_Hall}), is a consistent truncation of the equations of motion of the probe. 

At zero temperature, we have found an analytic solution for $\theta(r)$ and the flux function $a=a(r)$ which preserves two of the four supercharges of the ${\cal N}=1$ superconformal background. The explicit calculations are performed in appendix \ref{kappa} with the use of kappa symmetry. Here we just quote the result for $\theta(r)$ and $a(r)$:
\bear
\cos\theta(r) &=&\Big({r_*\over r}\Big)^{b} 
\label{theta_SUSY_EBzero} \\
a(r) &=&-Q\,(\cos\theta(r))^{{1\over q}}\,=\,-Q\,\Big({r_*\over r}\Big)^{{b\over q}} \ .
\label{a_SUSY_EBzero}
\eear

However, to realize the quantum Hall states we are interested in, we need to generalize our ansatz 
for the gauge field to include electric and magnetic fields, as well as the components dual to the charge density and current.  We analyze this more general set up in the next subsection.

\subsection{Background fields and currents}
\label{Full_ansatz}

If we want a more general ansatz that includes background electric and magnetic fields and the associated charged current, we need to consider other components of the worldvolume gauge field.  In the standard way, a  magnetic field $B$ and an electric field $E$ are added by turning on the radial zero modes of $F_{xy}$ and $F_{0x}$.  The charge density is holographically related to $F_{r0}$, the longitudinal and Hall currents come from $F_{rx}$ and $F_{ry}$.  We therefore take the worldvolume gauge field to have the form:
\beq
A\,=\,L^2\,\Big[\,a_0(r)\,dt\,+(Et+\,a_x(r))\,dx
\,+\,(B\,x+a_y(r))\,dy\,+\,a(r)\,(d\psi+\cos\alpha\,d\beta)\,\Big] \ .
\label{A_full_ansatz}
\eeq
We can continue to use the induced metric ansatz given by (\ref{induced_metric_Hall}), characterized by the embedding function $\theta=\theta(r)$. 

Interestingly, due to our choice in (\ref{A_full_ansatz}) of the internal components of the gauge field, the dependence of the action 
 on the internal angles of the ${\mathbb R}{\mathbb P}^3$ cycle factorizes and consequently, we can consistently take the functions $\theta$, $a$, $a_0$, $a_x$, and $a_y$ to depend only on the radial variable. After integrating over the internal angles $\alpha$, $\beta$, and $\psi$, the DBI action of the D6-brane for our ansatz can be written as:
\beq
S_{DBI}\,=\int d^3x\,dt\,{\cal L}_{DBI} \ ,
\eeq
where the DBI lagrangian density ${\cal L}_{DBI}$ can be compactly written as:
\beq
{\cal L}_{DBI}\,=\,-{8\pi^2\,L^7\,T_{D6}\,e^{-\phi}\over b^4}\,
{\sqrt{(B^2+r^4)h-E^2}\,\sqrt{q^2+b^4\,a^2}\over \sqrt{h}}\,\sqrt{\Delta} \ ,
\eeq
where $T_{D6}$ is the D6-brane tension and the quantity $\Delta$ is defined to be
\bear
\Delta &=& b^4r^2 h a'^{\,2}\,+\,\sin^2\theta\,\Bigg[b^2+r^2 h\,\theta'^{\,2}\, \rc\rc
&&\qquad\qquad\qquad+
{b^2h\over E^2-(B^2+r^4)h}\,\Big[(B a_0'+E a_y')^2\,+\,r^4 (a_0'^{\,2}-h a_x'^{\,2}+ ha_y'^{\,2})\Big]\Bigg]\,\,.\qquad\qquad
\label{Delta_def_general}
\eear
The Wess-Zumino term of the action is:
\beq
S_{WZ}\,=\,T_{D6}\int_{{\cal M}_7}\,\left(\hat C_7\,+\,\hat C_5\wedge F\,+\,{1\over 2}\,
\hat C_3\wedge F\wedge F\,+\,{1\over 6}\hat C_1\wedge F\wedge F\wedge F \right) \ ,
\label{Wess_Zumino}
\eeq
where, $\hat C_7$,  $\hat C_5$,  $\hat C_3$,  and $\hat C_1$ are the pullbacks to the D6-brane of the RR gauge fields.  All of these terms, except for $\hat C_5\wedge F$, give non-vanishing contributions to the equations of motion.\footnote{One subtlety is that when the backreaction of the flavors is included, the RR field strength $F_2$ is not closed, implying that there is no well-defined RR potential $C_1$.  However, the equations of motion derived from (3.17) only contain $F_2$ and therefore can be generalized to the unquenched case; see appendix \ref{EOMs} for details.}

In the holographic setup, the charge density is encoded in the bulk by the radial electric displacement field $\tilde D(r)$, which is given by the derivative of the DBI lagrangian density with respect to the radial component of the physical electric field. From the ansatz (\ref{A_full_ansatz}), and taking into account the physical gauge field $A_{phys}=A/(2\pi\alpha')$, we find:
\beq
\tilde D\,=\,{\partial {\cal L}_{DBI}\over \partial A_{0,phys}'}\,=\,{2\pi \alpha'\over L^2}\,
{\partial {\cal L}_{DBI}\over \partial a_{0}'} \ .
\label{tilde_D_def}
\eeq
We will set $\alpha'=1$ from now on. In order to write  $\tilde D(r)$ in a compact fashion, let us define a function $g(r)$ as:
\beq
g(r)\,=\,{q+\eta\over 2b(2-q)}\,{r^4\,h^{{3\over 2}}\,\sin^2\theta\,\sqrt{q^2+b^4\,a^2}
\over \sqrt{(B^2+r^4)\,h\,-E^2}\,\sqrt{\Delta}} \ .
\eeq
Then, one can show that:
\beq
\tilde D(r)\,=\,{N\sigma^2\over 4\pi}\,\tilde d(r)\ ,
\eeq
where $\sigma$ is the screening factor defined in (\ref{screening-sigma}) and $\tilde d(r)$ is the function:
\beq
\tilde d(r)\,\equiv \,{g\over h}\,\left[\left(1+{B^2\over r^4}\right)\,a_0'\,+\, {BE\over r^4}\,a_y'\,\right]\,\,.
\label{tilded_def}
\eeq
The total charge density is obtained by taking the boundary value of $\tilde D(r)$, which is proportional to:
\beq
d\,=\,\lim_{r\to\infty}\,\tilde d(r) \ .
\eeq
Similarly, the physical currents along the $x$ and $y$ directions are given by:
\beq
J_x\,=\,{2\pi \alpha'\over L^2}\,
{\partial {\cal L}_{DBI}\over \partial a_{x}'} \ ,
\qquad\qquad
\tilde J_y\,=\,{2\pi \alpha'\over L^2}\,
{\partial {\cal L}_{DBI}\over \partial a_{y}'} \ .
\label{Jxy_def}
\eeq
One can readily prove that:
\beq
J_x\,=\,{N\sigma^2\over 4\pi}\,j_x\,\,,
\qquad\qquad\qquad
\tilde J_y\,=\,{N\sigma^2\over 4\pi}\,\tilde j_y \ ,
\eeq
where $j_x$ turns out to be:
\beq
j_x\,=\,- g\,a_x' \ ,
\label{eom_ax_jx}
\eeq
and $\tilde j_y(r)$ is:
\beq
\tilde j_y(r)\,\equiv \,g\,\left[-\left(1-{E^2\over r^4 h}\right)\,a_y'\,+\,
{BE\over r^4 h}\,a_0'\,\right] \ .
\label{tildejy_def}
\eeq

The equations of motion for the probe are worked out in detail in appendix \ref{EOMs}.
In particular, $J_x$ is constant in $r$  (see  (\ref{eom_ax_general_case})) and represents the longitudinal current parallel to the electric field. 
On the other hand, $\tilde J_y(r)$ depends on the holographic variable. The transverse current $J_y$ is obtained as the value of $\tilde J_y(r)$  at the UV boundary $r\to\infty$ which, according to (\ref{Jxy_def}),  is determined from the limit:
\beq
 j_y\,=\,\lim_{r\to\infty}\,\tilde j_y(r) \ .
\eeq

The radial dependence of $\tilde d$ and $\tilde j_y$ is determined by the $a_0$ and $a_y$ equations of motion, (\ref{eom_a0_general_case}) and (\ref{eom_ax2_general_case}). With the definitions introduced above, they can be simply written as:
\bear
\label{eom_dtilde}
\partial_r\,\tilde d &=& B(\eta\,\cos\theta\,a'-a\,\sin\theta\,\theta')~, \\
\label{eom_jytilde}
\partial_r\,\tilde  j_y &=& E(\eta\,\cos\theta\,a'-a\,\sin\theta\,\theta')\,\,.
\eear
In the unflavored case $\eta=1$, these two equations (\ref{eom_dtilde}) and (\ref{eom_jytilde}) can be integrated once because their right-hand side is proportional to $\partial_r (a\cos\theta)$. Indeed, for the unflavored background $a_0(r)$ and $a_y(r)$ are cyclic and can be eliminated by performing the appropriate Legendre transformation. 

This is not the case, however, when $\eta\not=1$. We can formally integrate (\ref{eom_dtilde}) and (\ref{eom_jytilde}), defining the integral $I(r)$ as:
\bear
\label{I_integral_def}
I(r) &\equiv& \,\int_r^{\infty}\,\Big(\eta\,\cos\theta(\bar r)\,a'(\bar r)\,-\,a(\bar r)\,\sin\theta(\bar r)\,\theta'(\bar r)\Big)\,d\bar r  \\
\label{I_partial_integral}
&=& \,-\cos\theta(r)\,a(r)\,+\,(\eta-1)\,\int_r^{\infty} \cos \theta(\bar r)\,a'(\bar r)\,d\bar r  \ ,
\eear
where we have integrated by parts to obtain the second line.  Clearly, 
\beq
\lim_{r\to \infty}\,I(r)\,=\,0 \ ,
\eeq
and equations (\ref{eom_dtilde}) and (\ref{eom_jytilde}) can be written as
\beq
\tilde d(r)\,=\,d\,-\,B\,I(r)\ ,
\qquad\qquad
\tilde j_y(r)\,=\, j_y\,-\,E\,I(r)\ .
\label{tilde_d_jy_integrated}
\eeq

Since $a_0$ and $a_y$ are no longer cyclic, we need a new strategy to solve the equations of motion. Interestingly, there is still one conserved quantity associated with the equations of motion for $A_0$ and $A_y$.  Eq.~({\ref{constant_of_motion}) can be recast as the radial independence of the quantity:
\beq
\Pi \,\equiv\,E\,\tilde d(r)\,-\,B\,\tilde j_y(r)\ .
\label{Pi_def}
\eeq
Indeed, the equation $\partial_r\,\Pi=0$ follows immediately from (\ref{eom_dtilde}) and (\ref{eom_jytilde}). This implies that $\Pi$ can be written in terms of quantities evaluated at the boundary:
\beq
\Pi\, = \, E \, d\,-\,B\, j_y\ .
\label{Pi_UV}
\eeq

One can now try to write $a_0'$, $a_y'$, and $a_x'$ in terms of the embedding function $\theta(r)$ and the flux function $a(r)$. Let us work this out in detail. First, we notice that one can invert Eqs.~(\ref{tilded_def}) and (\ref{tildejy_def}) and write $a_0'$ and $a_y'$ in terms of $\tilde d$ and $\tilde j_y$:
\beq
a_0'\,=\,{h\Big(1-{E^2\over r^4\,h}\Big)\,\tilde d\,+\,{EB\over r^4}\,\tilde j_y\over
g\Big(1+{B^2\over r^4}\,-\,{E^2\over r^4\,h}\Big)}\ ,
\qquad\qquad
a_y'\,=\,{{EB\over r^4}\,\tilde d\,-\,\Big(1+{B^2\over r^4}\Big)\,\tilde j_y
\over
g\Big(1+{B^2\over r^4}\,-\,{E^2\over r^4\,h}\Big)}\ .
\label{a_0_y_disp_quenched}
\eeq
Notice that (\ref{a_0_y_disp_quenched}) are not actually solutions for $a_0'$ and $a_y'$ since $g$ on the right-hand side is written in terms of these same fields. However, one can write an expression of $g$ in terms of $\theta$ and $a$. Let us define $X$ as:
\beq
X\equiv h\left(1+{B^2\over r^4}\,-\,{E^2\over r^4\,h}\right)
\Bigg[\left({q+\eta\over 2 b^2 (2-q)}\right)^2
r^4 h (q^2+b^4\,a^2)\,\sin^2\theta +h\tilde d^{\,2}-j_x^2-\tilde j_y^{\,2}\Bigg]- \,
{\left(hB\tilde d-E\tilde j_y\right)^2\over r^4}\ .
\eeq
Then, after some calculation, we obtain:
\beq
\label{gdef}
g\,=\,{\sin\theta\,\sqrt{X}\over
\Big(1+{B^2\over r^4}\,-\,{E^2\over r^4\,h}\Big)\,
\sqrt{b^2\,r^2\,h\,a'^{\,2}\,+\,\sin^2\theta\,(1+{r^2\over b^2}\,h\,\theta'^{\,2})}}\,\,.
\eeq
Therefore, we have for $a_0'$, $a_x'$, and $a_y'$:
\bear
&&a_0'\,=\,{
\sqrt{b^2\,r^2\,h\,a'^{\,2}\,+\,\sin^2\theta\,(1+{r^2\over b^2}\,h\,\theta'^{\,2})}\over \sin\theta\sqrt{X}}\,
\Bigg[h\left(1-{E^2\over r^4\,h}\right)\,\tilde d\,+\,{EB\over r^4}\,\tilde j_y\Bigg]\,\,,\rc\rc
&&a_x'\,=\,-{
\sqrt{b^2\,r^2\,h\,a'^{\,2}\,+\,\sin^2\theta\,(1+{r^2\over b^2}\,h\,\theta'^{\,2})}\over 
\sin\theta\sqrt{X}}\,
\left(1+{B^2\over r^4}\,-\,{E^2\over r^4\,h}\right)\,j_x\,\,,\rc\rc
&&a_y'\,=\,{
\sqrt{b^2\,r^2\,h\,a'^{\,2}\,+\,\sin^2\theta\,(1+{r^2\over b^2}\,h\,\theta'^{\,2})}\over
 \sin\theta\sqrt{X}}\,
\Bigg[
{EB\over r^4}\,\tilde d\,-\,\left(1+{B^2\over r^4}\right)\,\tilde j_y
\Bigg]\,\,.
\label{aprimes_tilded_jy}
\eear
The right-hand sides of (\ref{aprimes_tilded_jy}) contain the radial functions $\tilde d$ and $\tilde j_y$, which in turn can be written as in (\ref{tilde_d_jy_integrated}) in terms of the constants $d$ and $j_y$, and the integral $I(r)$ defined in (\ref{I_integral_def}). 

In principle, we could  use (\ref{aprimes_tilded_jy}) to eliminate $a_0'$, $a_x'$, and $a_y'$ from the equations of motion and to reduce the system to an effective problem for the functions 
$\theta(r)$ and $a(r)$. However, when $\eta\not=1$, the functions $\tilde d(r)$ and $\tilde j_y$ depend non-locally on $\theta(r)$ and $a(r)$ and the corresponding reduced equations of motion would be a system of integro-differential equations for $\theta(r)$ and $a(r)$, which does not seem to be very easy to solve in practice.  In the case in which there are no flavors backreacting on the geometry, \ie, when $\eta=q=b=1$, the integral $I(r)$ is just 
$I=-\cos\theta\, a$ and we can write $\tilde d$ and $\tilde j_y$ simply as
$\tilde d\,=\,d\,+\,B\cos\theta\,a$ and 
$\tilde j_y\,=\,j_y+E\cos\theta \,a$. Thus, 
in this quenched case one can eliminate the gauge fields $a_0$, $a_x$, and $a_y$ and reduce the problem to a system of two coupled, second-order differential equations for  $\theta(r)$ and $a(r)$.

\subsection{Minkowski embeddings}

Having obtained the equations of motion for the D6-brane probe, the next step is to try to solve them.  Although, as we will discuss in section \ref{BPS-sol}, there are special analytic BPS solutions, in general we will have to resort to numerics. 

Probe brane solutions are categorized into two classes by their IR behavior.  The generic solution is a black hole embedding, in which the brane falls into the horizon; these correspond holographically to gapless, compressible states.  In certain special circumstances, the brane can end smoothly at some $r=r_*$ when a wrapped cycle shrinks to zero size; these are Minkowski (MN) embeddings.  MN solutions with broken parity correspond to gapped, quantum Hall states.

As discussed above in section \ref{internal_flux_quantization}, for a D6-brane probe in the flavored ABJM background, MN embeddings occur when $\theta(r_*) = 0$ for some $r_*$.  In order to have a physical, finite-energy solution, the embedding $\theta(r)$ and the worldvolume gauge field $F$ must be regular at the tip; that is, the induced metric (\ref{induced_metric_Hall}) must be smooth, and $a'$, $a_0'$, $a_x'$, $a_y'$ must all be finite. Given that the function $g$ (\ref{gdef}) vanishes at the tip of the brane, the regularity of $a_0'$ and $a_y'$ at the tip, combined with (\ref{tilded_def}) and (\ref{tildejy_def}), implies that
\beq
\label{tip_conditions}
\tilde d(r_*)\,=\, j_x \, = \,\tilde j_y(r_*)\,=\,0\,\,.
\eeq
We can interpret this condition to mean that there are no sources at the tip, which is physically sensible as the D6-brane could not support such a source.  Suppose that $\tilde d(r_*) \not= 0$; this radial displacement field would have to be sourced, for example, by fundamental strings stretching from the horizon.  Due to the shrinking cycle, the effective radial tension of the D6-brane vanishes at the tip, so these strings would then pull the D6-brane into the horizon, resulting in a black hole embedding.

The filling fraction $\nu$ is defined by
\beq
\nu \, = \, 2\pi \, \frac{D_{phys}}{B_{phys}} \ ,
\eeq
where the physical magnetic field $B_{phys}$ is related to $B$ by
\beq
B_{phys}\,=\,{L^2\over 2\pi}\,B\,=\,\sqrt{{\lambda\over 2}}\,\sigma\,B\ .
\eeq
Combining (\ref{tip_conditions}) with (\ref{tilde_d_jy_integrated}) gives $d = BI(r_*)$, and the filling fraction for MN solutions is therefore
\beq
\label{nu_MN}
\nu\,=\,{N\sigma\over \sqrt{2\lambda}}\,\frac{d}{B}\,=\,{N\sigma\over \sqrt{2\lambda}}\,I(r_*) \ ,
\eeq
or, more explicitly using (\ref{I_partial_integral}),
\beq
\nu\,=\,{M\over 2}\,\left[
1\,+\,(\eta-1)\,\int_{r_*}^{\infty}\,
\cos\theta(r)\,{a'(r)\over Q}\,dr\,\right]\ ,
\label{nu_flux}
\eeq
where $M$ is the quantization integer and $Q$ is minus the flow function at the tip (see (\ref{a_*_Q})).  Note that,  (\ref{nu_flux}) shows explicitly that, for a QH state with nonzero charge density, a nonzero flux is required. Moreover, $\nu$ is the sum of two contributions. The first term in (\ref{nu_flux}) is proportional to the flux at the tip. The second term is only nonzero in the unquenched case $\eta\not=1$ and contains an integral from the tip to the boundary. In terms of $N_f$ and $k$, $\nu$ takes the form:
\beq
\nu\,=\,{M\over 2}\,\left[1\,+\,{3N_f\over 4 k}\,
\int_{r_*}^{\infty}\,
\cos\theta(r)\,{a'(r)\over Q}\,dr\,\right]\ .
\label{filling_flux}
\eeq
It follows that $\nu$ is a half-integer in the quenched case but gets corrections due to the massless sea quark loops in the unquenched Veneziano limit.  

Numerically integrating the equations of motion, we have verified that there are MN solutions obeying the tip regularity conditions (\ref{tip_conditions}).  At this point, we will be content with evidence for MN solutions with nonzero charge density $d$ and magnetic field $B$.  We will defer a more thorough study of the possible MN solutions to the future.

\section{Conductivities}
\label{conductivities}

We are interested in analyzing the longitudinal and transverse conductivity of our configurations. In order to relate these quantities with the variables we have employed, let us point out that the physical electric field $E_{phys}$ is related to the quantity $E$  used above  as:
\beq
E_{phys}\,=\,{L^2\over 2\pi}\,E\,=\,\sqrt{{\lambda\over 2}}\,\sigma\,E\ .
\eeq
The longitudinal  and transverse conductivities $\sigma_{xx}$ and $\sigma_{xy}$  are defined in terms of $J_x$, $J_y\equiv\tilde J_y(r\to \infty)$ and  $E_{phys}$ as:
\beq
\sigma_{xx}\,=\,{J_x\over E_{phys}}\ ,
\qquad\qquad
\sigma_{xy}\,=\,{J_y\over  E_{phys}}\ .
\eeq
The conductivities can be written as
\beq
\sigma_{xx}\,=\,{N\sigma\over 2\pi\sqrt{2\lambda}}\,{j_x\over E}\ ,
\qquad\qquad
\sigma_{xy}\,=\,{N\sigma\over 2\pi\sqrt{2\lambda}}\,{j_y\over E}\ .
\eeq
In the next two subsections we obtain formulas for $\sigma_{xx}$ and $\sigma_{xy}$ for the two types of embeddings (Minkowski and black hole) of the D6-brane probe.

\subsection{Quantum Hall states}
\label{QH_conductivity}

Let us now suppose that we have a Minkowski (MN) embedding. To compute the conductivities, we will adapt the method of \cite{Bergman:2010gm, Jokela:2011eb} but with some new twists.  In particular, we use the invariance of $\Pi$ under the the holographic flow.  The conductivity comes directly from the condition (\ref{tip_conditions}) that there are no charge sources at the tip $r=r_*$. Since $j_x$ in (\ref{eom_ax_jx}) must be equal to zero, 
\beq
\sigma_{xx}\,=\,0 \ .
\eeq
Furthermore, (\ref{tip_conditions}) implies that $\Pi$ vanishes at $r=r_*$ and, since it is radially invariant, $\Pi=0$  at all values of $r$. From (\ref{Pi_UV}), we see that this is equivalent to $E\, d = B\, j_y$; the Hall conductivity is then:
\beq
\sigma_{xy}\,=\,{N\sigma\over 2\pi\sqrt{2\lambda}}\,{j_y\over E}\,
=\,{N\sigma\over 2\pi\sqrt{2\lambda}}\,{d\over B}\ .
\eeq
From (\ref{nu_MN}), we find that
\beq
\sigma_{xy}\,=\,{\nu\over 2\pi} \ , 
\label{sigma_xy_flux}
\eeq
which is exactly what one would expect for a QH state.

\subsection{Gapless states}

Let us now consider black hole embeddings, in which the D6-brane crosses the horizon at $r=r_h$. These embeddings correspond to gapless states. To compute the conductivity, we employ the pseudohorizon argument of \cite{Karch:2007pd} to Eq.~(\ref{aprimes_tilded_jy}).  Let $r=r_p$ be the position of the pseudohorizon, which is determined by the conditions:
\bear
h_p\,(r_p^4+B^2) &=& E^2~, \rc\rc 
j_x^2\,+\,\tilde j_y^2(r_p) &=& \left({q+\eta\over 2 b^2 (2-q)}\right)^2\,h_p\,r_p^4\,\sin^2\theta_p\,\left(q^2+b^4\,a_p^2\right)\,+\,h_p\,\tilde d(r_p)^2~, \rc\rc
E\,\tilde j_y(r_p) &=& B\,h_p\,\tilde d(r_p) \ ,
\eear
where $h_p\equiv h(r_p)$, $\theta_p\equiv \theta(r_p)$, and $a_p\equiv a(r_p)$. 
It follows that the currents in $x$- and $y$-directions are given by:
\bear
j_x&=&\sqrt{h_p}\,
\Bigg[\left(1-{B^2 h_p\over E^2}\right)\,\tilde d^{\,2}(r_p)\,+\,
\left({q+\eta\over 2 b^2 (2-q)}\right)^2\,r_p^4\,(q^2+b^4\,a_p^2)\,\sin^2\theta_p
\Bigg]^{{1\over 2}}~, \rc\rc
j_y&=&{B\,h_p\over E}\,d\,+\,E\,\left[1-{B^2 h_p\over E^2}\right]\,I(r_p) \ .
\eear
Notice that the previous expression involves the value of the integral $I$ extended between the $r_p$ and the boundary. Therefore,  the conductivities are:
\bear
\sigma_{xx}&=&{N\sigma\over 2\pi\sqrt{2\lambda}}\,
{\sqrt{h_p}\over E}\,
\Bigg[\left(1-{B^2 h_p\over E^2}\right)\,\tilde d^{\,2}(r_p)\,+\,
\left({q+\eta\over 2 b^2 (2-q)}\right)^2\,r_p^4\,(q^2+b^4\,a_p^2)\,\sin^2\theta_p
\Bigg]^{{1\over 2}} ~, \rc\rc
\sigma_{xy}&=&{N\sigma\over 2\pi\sqrt{2\lambda}}\,\Bigg[
{B\,h_p\over E^2}\,d\,+\,\left[1-{B^2 h_p\over E^2}\right]\,I(r_p)\Bigg]
\ .
\label{metallic_conductivities}
\eear
For small electric field, $r_p$ is close to the horizon radius $r_h$. At first order in $E^2$ we can estimate $r_p$ as:
\beq
r_p\,\approx \,r_h\,\left(1\,+\,{E^2\over 3(r_h^4+B^2)}\right) \ .
\eeq
With this result we can write $h_p$ approximately as:
\beq
h_p\approx {E^2\over r_h^4+B^2}\ .
\eeq
Applying these results to (\ref{metallic_conductivities}), we obtain the linear conductivities:
\bear
 \label{longitudentalconductivity}
\sigma_{xx} &\approx& {N\sigma\over 2\pi\sqrt{2\lambda}}\,
{r_h^2\over  r_h^4+B^2}\,
\Bigg[\tilde d_h^2\,+\,\left({q+\eta\over 2 b^2 (2-q)}\right)^2\,
(r_h^4+B^2)\,(q^2+b^4\,a_h^2)\,\sin^2\theta_h\Bigg]^{{1\over 2}} ~, ~~~~~~~~~~~~ \\
\label{Hallconductivity}
\sigma_{xy} &\approx& {N\sigma\over 2\pi\sqrt{2\lambda}}\,\Bigg[
{B\,\tilde d_h\over r_h^4+B^2}\,+\, I_h\Bigg] \ ,
\eear
where $I_h\equiv I(r_h)$ is defined in (\ref{I_partial_integral}).

These conductivities are analogous to the conductivities found in the metallic phases of other similar probe brane models, for example \cite{Lifschytz:2009si, Bergman:2010gm, Jokela:2011eb}.  One important difference is that here, the unquenched sea quarks reduce the conductivity by the screening factor $\sigma$.

The longitudinal conductivity (\ref{longitudentalconductivity}) receives contributions from two sources:  the first term under the square root is due to the charge density at the horizon $\tilde d_h$, and the other term can be interpreted as being due to thermal pair production.  At vanishing magnetic field and nonzero charge density, $\sigma_{xx}$ diverges as $r_h^{-2}$.  Charge carriers can only scatter off the thermal bath, and at zero temperature, momentum conservation implies an infinite DC conductivity.  For nonzero $B$, $\sigma_{xx}$ vanishes in the zero-temperature limit, as implied by Lorentz invariance.

The Hall conductivity  (\ref{Hallconductivity}) is the sum of two terms, which appear to correspond to the contributions of two types of charge carriers:  the charges at the horizon $\tilde d_h$, which are sensitive to the heat bath and contribute to $\sigma_{xx}$, and the charges $B I_h = d-\tilde d_h$, which are smeared radially along the D6-brane and do not interact with the dissipative heat bath at all.  In the limit where $d/B \to I_h$, \ie, $\tilde d \to 0$, the Hall conductivity smoothly approach the results found above for the MN embedding (\ref{sigma_xy_flux}). Varying the $d/B$ from zero to $I_h$ and plotting the conductivity on the $(\sigma_{xy},\sigma_{xx})$-plane is expected to reproduce the behavior, seen also in \cite{Jokela:2011eb}, qualitatively similar to the semi-circle law experimentally observed in QH systems \cite{Dykhne}.

\section{Boost invariance at zero temperature}
\label{em_symmetry}

At zero temperature and before adding an electric field, the system is Lorentz invariant.  
In probe brane constructions, the zero-temperature limit of a black hole embedding is often problematic.  However, Minkowski embeddings are perfectly well defined in the zero-temperature limit since the brane never reaches the black hole horizon.  One important feature of this model and others in its class \cite{Bergman:2010gm, Jokela:2011eb, Kristjansen:2012ny}, is that MN embeddings can occur at nonzero charge density.

Turning on an electric field in the $x$-direction breaks rotation invariance, and the full Lorentz symmetry is reduced to a (1+1)-dimensional subgroup: boosts in the $y$-direction form a set of $SO_+(1,1)$ transformations which rotate the electromagnetic field and the currents. When $h=1$, the equations of motion studied in section \ref{Full_ansatz}  and appendix \ref{EOMs} are not modified under these transformations.

In terms of the boundary variables, a boost with rapidity $\gamma$ acts as 
\beq
\begin{pmatrix}
E\\  B 
\end{pmatrix}
\to\,{\cal M}_{\gamma}\,\begin{pmatrix}
E\\  B 
\end{pmatrix}
\,\,,\qquad\qquad
\begin{pmatrix}
d\\  j_y
\end{pmatrix}
\to\,{\cal M}_{\gamma}\,\begin{pmatrix}
d\\  j_y
\end{pmatrix}\ .
\label{proper_o11}
\eeq
where ${\cal M}_{\gamma}$ is the symmetric matrix:
\beq
{\cal M}_{\gamma}\equiv 
\begin{pmatrix}
  \cosh \gamma &&\sinh \gamma \\
  {} && {} \\
 \sinh \gamma && \cosh \gamma
 \end{pmatrix}\ .
 \label{M_alpha}
\eeq
The transverse conductivity $\sigma_{xy}$ is invariant under the boost because it is determined only by the flux (\ref{sigma_xy_flux}). 

The boundary electromagnetic fields and currents are packaged holographically in the bulk worldvolume field strength $F$. From the transformation properties of $F$ due to a boost in the bulk, one can reproduce the transformation (\ref{proper_o11}) of $E$ and $B$ and see that the radial components of $F$ transform as
\beq
\begin{pmatrix}
F_{r0}\\  F_{ry}
\end{pmatrix}
=
\begin{pmatrix}
a_0'\\  a_y'
\end{pmatrix}
\to\,{\cal M}_{-\gamma}\,\begin{pmatrix}
a_0'\\ a_y'
\end{pmatrix}\ ;
\eeq
therefore, the symmetry acts contravariantly on $(a_0',  a_y')$. Using Eqs.~(\ref{tilded_def}) and (\ref{tildejy_def}), one can demonstrate that the functions $\tilde d$ and $\tilde j_y$ also rotate via ${\cal M}_{\gamma}$, namely:
\beq
\begin{pmatrix}
\tilde d\\  \tilde j_y
\end{pmatrix}
\to\,{\cal M}_{\gamma}\,\begin{pmatrix}
\tilde d\\  \tilde j_y
\end{pmatrix}~,
\eeq
which matches, for $r = \infty$, the transformation (\ref{proper_o11}) of $d$ and $j_y$. One can also check that  the quantity $\Pi$ defined in (\ref{Pi_def}) is invariant.

Apart from the boosts, the equations of motion are also invariant under the two types of discrete operations, which are the elements of $O(1,1)$ not connected to the identity. The first of these operations is the 
electric field inversion ${\cal P}_{E}$, which acts as:
\beq
{\cal P}_{E}: \ 
\begin{pmatrix}
E\\  B 
\end{pmatrix}
\to\,
 \begin{pmatrix}
-E\\  \ B 
\end{pmatrix}\  , \ \ 
\begin{pmatrix}
a_0'\\  a_y'
\end{pmatrix}\to
\begin{pmatrix}
\ a_0'\\  -a_y'
\end{pmatrix}\ , \ \
\begin{pmatrix}
\tilde d\\  \tilde j_y
\end{pmatrix}\,\to\,
\begin{pmatrix}
\ \tilde d\\ - \tilde j_y
\end{pmatrix}\ .
\label{PE_def}
\eeq
 Under ${\cal P}_E$, the function $\Pi$ changes its sign, \ie, $\Pi$ behaves as a pseudoscalar.  However, the conductivity $\sigma_{xy}$ is left invariant.
 Similarly, the equations of motion are invariant under a magnetic field inversion 
 ${\cal P}_{B}$, defined as:
\beq
{\cal P}_{B}:\
\begin{pmatrix}
E\\  B 
\end{pmatrix}
\to\,
 \begin{pmatrix}
\ E\\  -B 
\end{pmatrix}\ , \ \
\begin{pmatrix}
a_0'\\  a_y'
\end{pmatrix}\to
\begin{pmatrix}
-a_0'\\  \,\,\,a_y'
\end{pmatrix}\ , \ \
\begin{pmatrix}
\tilde d\\  \tilde j_y
\end{pmatrix}\,\to\,
\begin{pmatrix}
-\tilde d\\ \ \ \tilde j_y
\end{pmatrix}\ .
\eeq
Under this transformation, $\Pi$ again changes its sign and $\sigma_{xy}$ is again invariant. 

We can use the $O(1,1)$ symmetry to classify the different configurations in terms of the sign of the following quadratic forms 
\beq
{\cal Q}_1\equiv E^2\,-\,B^2 \ ,
\qquad\qquad
{\cal Q}_2\equiv d^{\,2}\,-\,j_y^{\,2}\ ,
\qquad\qquad
{\cal Q}_3\,\equiv B\,d\,-\,E\,j_y \ ,
\eeq
which are  left invariant by the $O(1,1)$ transformations.  We will call solutions with ${\cal Q}_1 >0 $ ``electric-like'' , and those with ${\cal Q}_1 <0$ ``magnetic-like''.  We can also have solutions with  ${\cal Q}_1=0$, which we call ``null" solutions. Notice that an electric-like (magnetic-like) solution can be connected continuously to a solution with $B=0$ ($E=0$) and, an electric-like solution cannot be transformed into a magnetic-like one.  We could similarly classify the solutions according to the sign of ${\cal Q}_2$ and  ${\cal Q}_3$. 

For MN embeddings, however, ${\cal Q}_2$ and ${\cal Q}_3$ are not independent but are rather proportional to ${\cal Q}_1$.  The regularity at the tip (\ref{tip_conditions}) implies $d/B = j_y/E = I(r_*)$.  From these relations, we find
\beq
{\cal Q}_2 = {\cal Q}_3 I(r_*) =  - {\cal Q}_1 I(r_*)^2 \ .
\eeq

In the next section we will find an analytic solution to the equations of motion for which the three ${\cal Q}_i$ invariants vanish. Intuitively, one would think that these null solutions have a large degree of symmetry. In particular, they are related to the $E=B=a_0'=a_y'=0$ solution by the infinite boost ${\cal M}_{-\infty}$.  Indeed, we will prove that these are BPS solutions preserving a fraction of the supersymmetry of the background.

\section{The BPS solution}
\label{BPS-sol}

In this section we find a simple, exact MN solution of the zero-temperature equations of motion (\ref{eom_a0_general_case})-(\ref{eom_theta_general_case}).  This solution preserves one supercharge, or one quarter of the supersymmetry of the background.  Accordingly, we will refer to this solution as the BPS solution.

Let us first consider the probe D6-brane in the absence of electric and magnetic fields and with $a_0' = a_y' = 0$.  At zero temperature, we found a SUSY solution in section \ref{internal_flux_quantization} for which the embedding function $\theta(r)$ and the flux function $a(r)$ satisfy the system of first-order BPS equations (\ref{BPSeq_theta}) and (\ref{BPSeq_a}) derived in appendix \ref{kappa}:
\beq
r\,\theta'\,=\,b\,\cot\theta\ ,
\qquad\qquad\qquad
{a'\over a}\,=\,-{b\over q\,r}\ .
\label{BPSeq_theta_a}
\eeq

We now generalize this supersymmetric solution to include electric and magnetic fields, as well as charge density and current, provided they satisfy a BPS condition:
\beq
E\,=\,B \ ,
\qquad\qquad
a_0'\,=\,-a_y' \ .
\label{EB_BPS}
\eeq
In addition, we take $a_x'=0$. 
Notice that, since $h=1$, (\ref{EB_BPS}) implies (\ref{constant_of_motion}) is trivially satisfied with the constant on the right-hand side equal to zero.  
Moreover, the equations of motion for $a_0$ (\ref{eom_a0_general_case}) and $a_y$  (\ref{eom_ax2_general_case}) become equivalent. The quantity $\Delta$ defined in (\ref{Delta_def_general}) greatly simplifies and satisfies:
\beq
{\sqrt{q^2+b^4\,a^2}\over \sqrt{\Delta}}\Bigg|_{BPS}\,=\,{q\over b} \ .
\eeq
As we saw in section \ref{QH_conductivity}, for MN embeddings $\Pi = 0$.  Combining this fact with (\ref{EB_BPS}), yields $\tilde d(r)=\tilde j_y(r)$, and in particular, $d=j_y$. 

Using these results it is straightforward to verify that the equations (\ref{eom_theta_general_case}) and (\ref{eom_apsi_general_case}) for 
$\theta(r)$ and $a(r)$ become:
\bear
&&\partial_r\,\big[r^4\,\sin^2\theta\,\theta'\big]\,-\,b^2\,r^2\,  
\left( 1+\frac{b^4}{q^2}a^2-b^2 r^2 a'^2  \right) 
 \cot\theta\,-(3-2b)\,b\, 
r^2\sin\theta\cos\theta\,=\,0\ ,\rc\rc
&&\partial_r\,\big[r^4\,a'\big]\,+\,{b\over q}\,\Big(3-{b\over q}\Big)\,r^2\,a\,=\,0 \ .
\label{eom_theta_a}
\eear
Note that these equations are just the same as in the $E=B=a_0'=a_y'=0$ case. One can also check that the second-order equations (\ref{eom_theta_a}) are satisfied if the first-order ones in (\ref{BPSeq_theta_a}) are fulfilled.  So, the solutions for $\theta(r)$ and $a(r)$ are just as in section \ref{internal_flux_quantization} (see (\ref{theta_SUSY_EBzero}) and (\ref{a_SUSY_EBzero})):
\bear
\cos\theta &=& \left({r_*\over r}\right)^{b} ~,
\label{theta_SUSY} \\ 
a &=& -Q\,\Big({r_*\over r}\Big)^{{b\over q}}\,=\,-Q\,(\cos\theta)^{{1\over q}} \ .
\label{a_SUSY}
\eear

It remains to solve for $a_0$. Its equation of motion (\ref{eom_a0_general_case}) simplifies to:
\beq
\partial_r\,\left[r^2\,\sin^2\theta\,a_0'\right]\,=\,-2B\,{(2-q)\,b^3\over q^2}\,
{a\cos\theta\over r} \ .
\label{eom_a0_BPS}
\eeq
Plugging in the solutions (\ref{theta_SUSY}) for $\theta(r)$ and (\ref{a_SUSY}) for $a(r)$, it is now straightforward to integrate (\ref{eom_a0_BPS}). Using the relation $q=b/(2-b)$, we get:
\beq
a_0'\,=\,{1\over r^2}\,\,{1\over 1-\Big({r_*\over r}\Big)^{2b}}\,\Bigg[\frac{2b^2(2-q)}{q(q+\eta)} d\,-\,
(4-3b)(2-b)b\,Q\,B\,\Big({r_*\over r}\Big)^{2}\Bigg] \ ,
\eeq
The regularity condition of $a_0'$ at the tip of the brane fixes a relation between $d$, $Q$, and $B$:
\beq
d \,=\,\frac{(q+\eta)(2-b)}{2}\,Q\,B \ .
\label{a_0_infty}
\eeq
From the first equality of (\ref{nu_MN}), the filling fraction $\nu$ for this SUSY solution is then:
\beq
\label{nu_SUSY_a}
\nu\,=\,{N\sigma\over \sqrt{2\lambda}}\,\frac{d}{B}\,= (q+\eta)(2-b) \, \frac{M}{4} \  .
\eeq
As we found in (\ref{nu_flux}) for general MN solutions, the filling fraction is proportional to the internal flux.
In addition, (\ref{nu_SUSY_a}) can be rewritten as
\beq
\nu\,=\,\left[1+{3N_f\over 8k}\,\big(1-\gamma_{m}\big)\,\right]\,{M\over 2} \ ,
\label{nu_SUSY_b}
\eeq
where $\gamma_m\,=\,b-1$ is anomalous dimension of the quark mass (see \cite{Conde:2011sw, Jokela:2013qya}). Notice that $\gamma_m$ depends on $N_f/k$ and controls the  coefficient of the contribution of the quarks loops to $\nu$. 

For the BPS solution, the integral $I(r)$ can be explicitly performed using the expressions (\ref{theta_SUSY}) and (\ref{a_SUSY}) for $\theta(r)$ and $a(r)$.  In particular,
\beq
I(r_*) = Q + (\eta - 1) \int_{r_*}^{\infty}\,\cos\theta a'\,dr\,=\, {q+\eta \over q+1} \, Q \ .
\eeq
As a cross-check, we can compute the filling fraction using the second equality in (\ref{nu_MN}):
\beq
\nu \,=\,{N\sigma\over \sqrt{2\lambda}}\,I(r_*)  \,=\, {q+\eta\over q+1} \,\frac{M}{2} \ ,
\label{nu_SUSY_c}
\eeq
which using the relation  $q=b/(2-b)$ (see (\ref{b_new})) matches (\ref{nu_SUSY_a}).

We can also use the integrated formula for $I(r)$ to write explicit expressions for $\tilde d$ and $\tilde j_y$:
\beq
\tilde d(r)\,= \tilde j_y(r) \, = \,d\,-\,\frac{(q+\eta)(2-b)}{2}\, Q \, B \left({r_*\over  r}\right)^2 \ .
\eeq
Note that, in particular, $d=j_y$.  Notice that the non-constant terms in $\tilde d$ and  $\tilde j_y$ behave as $r^{-2}$, with no flavor corrections, which is probably a consequence of the non-anomalous dimensions of these currents.

\subsection{Nonzero temperature generalization}

Let us now consider the system at $T>0$ (\ie, when $h\not= 1$). It is possible to find a truncation of the general system of equations which defines a solution that  can be regarded as the $T>0$  generalization of the BPS system studied above. This truncation occurs when the following relations are satisfied:
\beq
E\,=\,B\,\,,
\qquad\qquad
a_0'=-h\,a_y'\ .
\label{BPS_T}
\eeq
Notice that Eq.~(\ref{constant_of_motion}) continues to be trivially satisfied.  
Moreover, Eqs.~(\ref{eom_a0_general_case}) and (\ref{eom_ax2_general_case}) still reduce to a single equation which is now:
\beq
{q+\eta\over 2b(2-q)}\partial_r
\left[
{\sqrt{q^2+b^4\,a^2}\sqrt{\Big(1-{1\over h}\Big)B^2+r^4}\over \sqrt{\Delta}}\,
\sin^2\theta\,
a_0'\right]-
B(\eta\,\cos\theta\,a'-a\,\sin\theta\,\theta')=0~,
\label{eom_a0_BPS_T}
\eeq
and where $\Delta$ takes the value:
\beq
\Delta\,=\,b^4r^2 h a'^{\,2}\,+\,\sin^2\theta\,\left[b^2+r^2 h\,\theta'^{\,2}\,+\,
b^2\,\Big({1\over h}-1\Big)
a_0'^{\,2}\right]\ .
\label{Delta_BPS_T}
\eeq
The equations for the flux function $a(r)$ and the embedding function $\theta(r)$ can likewise be straightforwardly  derived from the probe action.

\section{Spectrum of mesons}
\label{mesons}

The addition of flux in the internal directions induces the breaking of parity symmetry in the Minkowski worldvolume directions. In this section we explore the effect of this parity violation on the mass spectrum of quark-antiquark  bound states which, in an abuse of language, we will refer to as mesons. The standard method to find the meson spectrum in the holographic correspondence is to analyze the normalizable fluctuations of the worldvolume gauge and scalar fields of the flavor brane.\footnote{See \cite{Hikida:2009tp,Jensen:2010vx,Zafrir:2012yg} for the calculation of the meson spectrum in the unflavored ABJM model.} Here we will restrict ourselves to analyzing the fluctuations of the gauge field $A$ around the zero-temperature supersymmetric configuration with only the internal components of $A$ are switched on. Accordingly, let us take the worldvolume gauge field  as:
\beq
A\,=\,A^{(0)}\,+\,\delta A\ ,
\eeq
where $\delta A$  is assumed to be small and the unperturbed gauge potential is given by:
\beq
A^{(0)}\,=\,L^2\,a(r)\,(d\psi+\cos\alpha\, d\beta)\ ,
\eeq
with $a(r)$ being the flux function for the SUSY embedding (\ref{a_SUSY}). We will also assume that the embedding function $\theta(r)$ does not fluctuate and is given by (\ref{theta_SUSY}).  We will denote by $f$ the first-order correction to the worldvolume field strength (\ie, $f=d\delta A$). We will assume that $f$ has only components along the $AdS$ directions. Its  components are:
\beq
f_{mn}\,=\,\partial_m \delta A_n-\partial_n\delta A_m\ ,
\eeq
where the indices $m$ and $n$ run over the $AdS$ directions. One can verify that these modes are a consistent truncation of the full set of fluctuations of the probe. In appendix 
\ref{Fluctuations} we obtain the equations of motion for $\delta A$ by computing the first variation of the equations of motion for the probe. These equations can be written as the Euler-Lagrange equations for the following second-order  effective lagrangian density:
\beq
{\cal L}^{(2)}=-{1\over 4}
{r^2\over (2-b)b^2}\,\left(1+(2-b)^2\,b^2\,Q^2(\cos\theta)^{2\over q}\right)
f^{mn}f_{mn}
-{Q\over 4 b\,L^4}(4-3b)
\big(\cos\theta)^{{2\over b}}
\tilde f^{mn}f_{mn}\ ,
\label{axion_QED}
\eeq
where the indices in $f^{mn}$ are raised with  the inverse of the open string metric ${\cal G}^{mn}$:
\beq
{\cal G}^{x^{\mu}\,x^{\nu}}\,=\,{\eta^{\mu\nu}\over r^2\,L^2}\,\,,
\qquad\qquad
{\cal G}^{rr}\,=\,{r^2\over L^2}\,\,{\sin^2\theta\over 1+(2-b)^2\,b^2\,Q^2\,(\cos\theta)^{2\over q}}\,\,,
\eeq
and the dual field $\tilde f^{mn}$ is defined as:
\beq
\tilde f^{\,mn}\,=\,{1\over 2}\, 
\epsilon^{mnpq}\,f_{pq}\,\,.
\label{tildef_def}
\eeq
Notice that the lagrangian density (\ref{axion_QED}) is that of axion electrodynamics in $AdS_4$, with the axion depending on the holographic direction, showing explicitly the breaking of parity in $AdS$  when the flux is turned on.  The equation of motion derived from ${\cal L}^{(2)}$ is:
\beq
\partial_m\,\Big[r^2\Big(1+(2-b)^2\,b^2\,Q^2\,(\cos\theta)^{2\over q}\Big)\,
{\cal G}^{mp}\,{\cal G}^{nq}\,f_{pq}\,\Big]\,-\,
{\Lambda\over L^4}\,\tilde f^{rn}\,=\,0 \ ,
\label{fluct_eq_flavored}
\eeq
where $\Lambda=\Lambda(r)$ is  the function written in (\ref{calO_Lambda_explicit}). 
To solve (\ref{fluct_eq_flavored}) let us first choose the gauge in which $\delta A_r=0$, and let us separate variables in the remaining components of $\delta A$ as follows:
\beq
\delta A_{\mu}\,=\,\xi_{\mu}\,e^{ik_{\nu}\,x^{\nu}}\,R(r)\ ,
\qquad (\mu=0,1,2)\ ,
\label{ansatz_a_fluct}
\eeq
where $\xi_{\mu}$ is a constant polarization vector. The gauge condition $\delta A_r=0$, together with (\ref{fluct_eq_flavored}), imposes the following transversality condition on $\xi_{\mu}$:
\beq
k\cdot \xi\,=\,\eta^{\mu\nu}\,k_{\mu}\,\xi_{\nu}\,=\,0\ .
\label{transversality}
\eeq
When they are normalizable,  these fluctuations  are dual to vector mesons, whose mass  $m$ is given by:
\beq
m^2\,=\,-\eta^{\mu\nu}\,k_{\mu}\,k_{\nu}\ .
\label{mass_mesons}
\eeq
In order to find the equation for $R(r)$, let us choose, without loss of generality,  the Minkowski momentun $k^{\mu}$ as:
\beq
k^{\mu}\,=\,(\omega, k, 0)\ ,
\label{momentun}
\eeq
\ie, we choose our coordinates in such a way that the momentum is oriented along the $x$-direction. Notice that the mass is just $m=\sqrt{\omega^2-k^2}$. The polarization transverse to (\ref{momentun}) is:
\beq
\xi_{\mu}\,=\,\left(-{k\over \omega}\,\xi_1\,,\,\xi_1\,,\xi_2\right)\ ,
\label{polarization}
\eeq
where $\xi_1$ and $\xi_2$ are undetermined.  Let us next consider the following complex combinations of  $\xi_1$ and $\xi_2$:
\beq
\chi_{\pm}\,=\,\sqrt{1\,-\,{k^2\over \omega^2}}\,\xi_1\,\pm\,i\xi_2\ .
\label{chi_pm}
\eeq
Then, as shown in appendix \ref{Fluctuations}, we can solve the fluctuation equation  (\ref{fluct_eq_flavored}) by taking $\chi_+\not=0$, $\chi_-=0$ or 
$\chi_+=0$, $\chi_-\not=0$ provided the radial function in the ansatz (\ref{ansatz_a_fluct}) satisfies the following ordinary differential equation:
\beq
{\cal O}\,R_{\pm}\,\pm\,m\,\Lambda\,R_{\pm}\,=\,0\ ,
\label{R_pm_eq}
\eeq
where ${\cal O}$ is the second-order differential operator defined in (\ref{calO_Lambda_explicit}) and 
$R_{+}=R_{+}(r)$ ($R_{-}=R_{-}(r)$) is the radial function for the solution with $\chi_+\not=0$ ($\chi_-\not=0$). Notice that the $\chi_+$ and $\chi_-$ modes are two helicity states which correspond to two different circular polarizations of the vector meson in the $x-y$ plane. They are exchanged by the parity transformation $\xi_2\to-\xi_2$, as  is obvious from  their definition (\ref{chi_pm}).

\begin{figure}
\centering
\label{fig: S'} 
\begin{subfigure}[b]{0.62\textwidth}
\centering
\includegraphics[width=\textwidth]{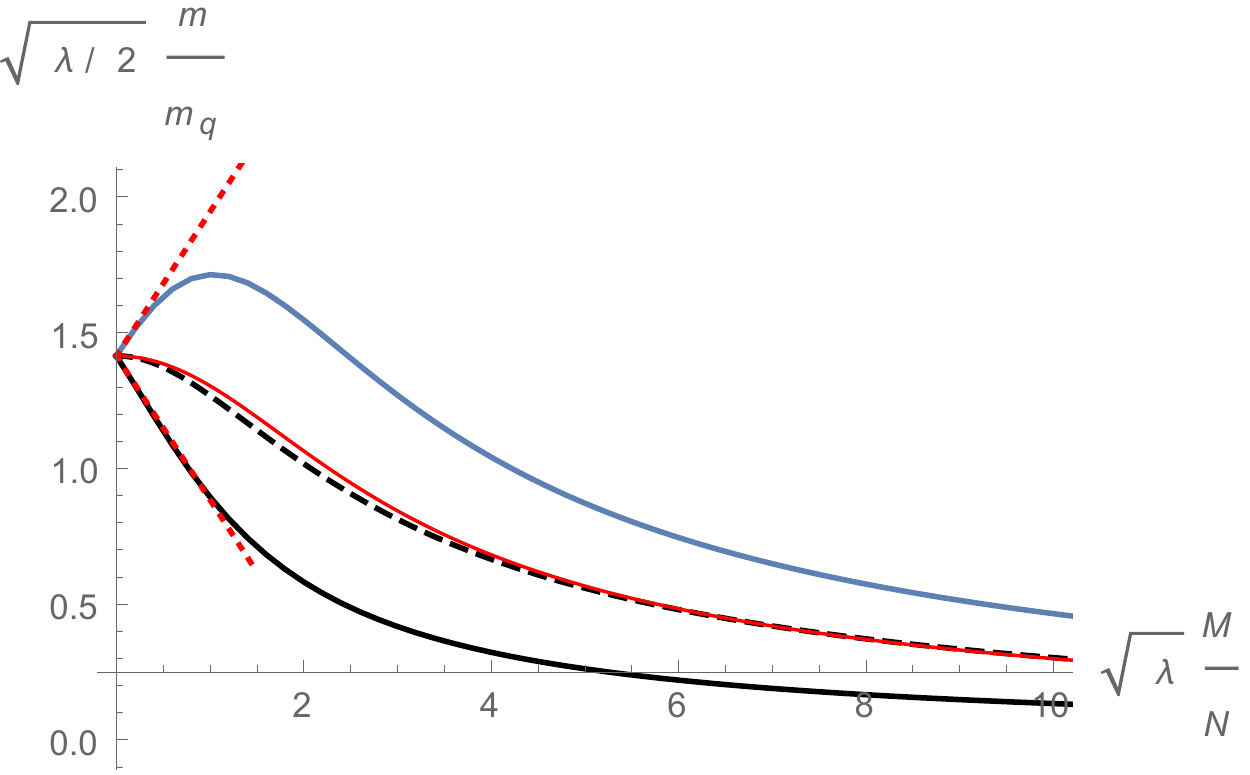}
\end{subfigure}
~~

~~

\begin{subfigure}[b]{0.62\textwidth}
\centering
\includegraphics[width=\textwidth]{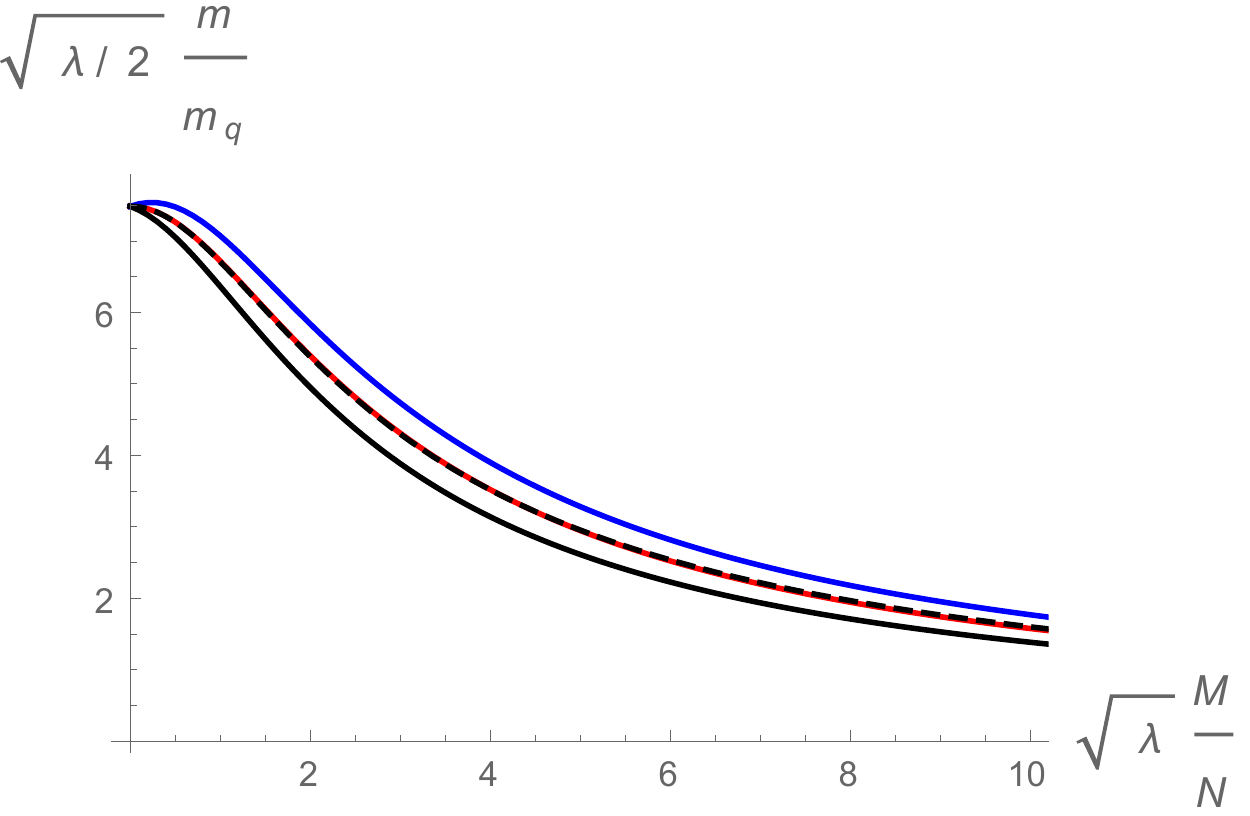}
\end{subfigure}
 \caption{Meson masses  in the quenched background as a function of the flux integer $M$ for the lightest meson with excitation integer $n=0$  (up) and for $n=3$ (down). The upper blue (lower black) curves corresponds to the mode $\chi_-$ ($\chi_+$). The intermediate red curve is the average of the two curves, and the dashed black curve is the WKB estimate (\ref{WKB_averaged}). The two straight lines on the left panel are the first-order results written in  (\ref{n=0_flux_split}). }
 \label{mass_spectrum}
\end{figure}


To find the mass spectrum of the mesons we must determine the values of $m$ for which there exists a normalizable solution to (\ref{R_pm_eq}). In general this must be done numerically by using the shooting technique, although, we can make analytic estimates using the WKB approximation, which we describe in detail in appendix \ref{Fluctuations}. The result is a tower of solutions with increasing masses which depend on the flux.  These masses depend on the location of the tip $r_*$, which can be related to the mass of the quarks $m_q$ as:
\beq
m_q\,=\,\sqrt{{\lambda\over 2}}\,\,\sigma\,r_*\ .
\eeq

In Fig.~\ref{mass_spectrum} we illustrate the dependence on the flux of the mass of the lightest state, for the quenched background with $N_f=0$. The spectra of the two helicity states are different. The mass splitting obviously vanishes at $Q=0$ and changes with the amount of  flux. 

For $N_f=0$  it is possible to compute analytically the meson mass splitting at first order in $Q$. Indeed, the fluctuation equation at first order in $Q$ for the unflavored background is:
\beq
\partial_r\,\big[\,(r^2-r_*^2)\partial_r\,R_{\pm}\,\big]\,+\,{m^2\over r^2}\,R_{\pm}\,\pm\,
2mQ\,{r_*^2\over r^3}\,R_{\pm}\,=\,0 .
\label{fluct_unflavored_orderQ}
\eeq
For $Q=0$,  Eq.~(\ref{fluct_unflavored_orderQ}) can be solved analytically in terms of hypergeometric functions. Let $R_{n}(r)$ be the normalizable regular solutions of 
(\ref{fluct_unflavored_orderQ}) for $Q=0$; they are given by:
\beq
R_{n}(r)\,=\,\Big({r_*\over r}\Big)^{2n+1}\,\,
F\left(-n-{1\over 2}, -n; 1;1-{r^2\over r_*^2}\,\right) \ ,
\qquad\qquad
n\,=\,0,1,\ldots\ .
\eeq
The corresponding  mass levels are:
\beq
{m_n\over r_*}\,=\, \sqrt{2(n+1)(2n+1)} \ ,
\qquad\qquad
n\,=\,0,1,\ldots\ .
\eeq
Let us now solve (\ref{fluct_unflavored_orderQ}) at first order in $Q$. We write:
\beq
R_{\pm,n}(r)\,=\,R_n(r)\pm Q\,\delta R_{n}(r)\,+\,{\mathcal O}(Q^2)\ ,
\eeq
where the two signs are in correspondence with the ones in (\ref{fluct_unflavored_orderQ}). The mass levels associated to $R_{\pm,n}$ will be denoted by $m_{\pm,n}$. The lightest regular normalizable modes at first order in $Q$ are given by:
\beq
R_{\pm,0}(r)\,=\,{r_*\over r}\pm Q\,
\left[{c\over r}+{1\over \sqrt{2}}\,\Big({r_*\over r}\Big)^{2}
\left(1\,-\,{r\over r_*}\,\log\left(1+{r_*\over r}\right)\right)\right]\,+\,{\mathcal O}(Q^2)\,\,,
\eeq
with $c$ being an integration constant. The masses for these modes are:
\beq
{m_{\pm,0}\over r_*}\,=\,\sqrt{2}\mp\,{3\over 4}\,Q\,+\,{\mathcal O}(Q^2)\ .
\label{n=0_flux_split}
\eeq
In Fig.~\ref{mass_spectrum} we plot these first-order results and we compare them with the numerical calculations. The first-order correction in the flux can similarly be obtained for the modes with $n\ge 1$, and the general form for the mass splitting is: 
\beq
{m_{-,n}-m_{+,n}\over r_*}\,=\,
{(2n+1)(3+4n)\over 2\,(n+1)\,\pi}\,
\left[\,{\Gamma\Big(n+{1\over 2}\Big)\over 
\Gamma\big(n+1\big)}\right]^2\,\,Q\,+\,{\mathcal O}(Q^2)\ .
\eeq
It follows from this expression that the first-order mass splitting becomes $4Q/\pi\,+\,{\mathcal O}(Q^2)$ as $n\to\infty$.

As shown in Fig.~\ref{mass_spectrum}, 
the mass averaged over the two helicities is well approximated by the WKB method:
\beq
{m_{+,n}+m_{-,n}\over 2}\approx m_{WKB}\ .
\eeq
Let us write the WKB estimate of this helicity average in the unflavored background. By
using the values of the WKB masses written  in appendix \ref{Fluctuations}, we have:
\beq
{m_{+,n}+m_{-,n}\over 2}\approx 
{r_*\over F\Big(-{1\over 2}, {1\over 2}; 1;-Q^2\,\Big) }\,
\sqrt{2(n+1)(2n+1)}\,\,.
\label{WKB_averaged}
\eeq

\section{Discussion}
\label{discussion}

In this chapter, we initiated a study of D6-brane probes with
parity-breaking flux in the ABJM background with unquenched massless
flavors. Minkowski embeddings of these probe branes holographically
described massive fermions in quantum Hall states. The filling fraction
was a half-integer in the quenched case, but received corrections when
the dynamics of the sea of massless flavors is included. The
conductivities, both in the gapped Minkowski embeddings and in the
metallic black hole ones, depended on the parity-breaking flux but also
contained a contribution from the dynamical flavors. This was interpreted
as an effect of the intrinsic disorder due to quantum fluctuations of
the fundamental degrees of freedom.

Despite the complexity of the equations of motion we managed to obtain
an explicit, analytic family of supersymmetric solutions with nonzero
charge density, electric, and magnetic fields. For these gapped QH
solutions, we obtained an analytic expression for the Hall
conductivity, which includes the effects of quark loops. We also
analyzed the residual $SO_+(1,1)$ boost invariance of the system at zero
temperature; this is a powerful tool for generating non-supersymmetric
solutions with general electric and magnetic fields starting from
solutions with either $E$=0 or $B$=0. We also explored the effect of
the parity violation on the computation of the meson spectrum. We
restricted our analysis to the fluctuations of the gauge field around
the zero-temperature supersymmetric configuration in which only the internal gauge
field components were switched on.

There remain many open topics for further investigation.  Although, we presented here a set of analytic, BPS solutions for a very specific set of parameters, more general solutions with nonzero temperature and arbitrary $d$, $Q$, $E$, and $B$ still need to be studied, probably numerically.  A thorough analysis of the thermodynamics and phase structure is needed to provide a complete understanding of this model. For example, we anticipate a phase transition as the temperature is increased from the MN to the black hole embeddings.\footnote{For a massless probe brane embedding, this phase transition is related to the breaking of chiral symmetry.  See {\it e.g.}~\cite{Preis:2010cq, Preis:2012fh, Filev:2007gb, Filev:2010pm} for the holographic realization of (inverse) magnetic catalysis and \cite{Filev:2011mt, Erdmenger:2011bw, Jokela:2013qya} for the study of the effect of the dynamical flavors on magnetic catalysis.}  

The effects of internal flux and the sea of massless quarks are particularly interesting.  And, we would also like to understand how the BPS solutions fit in to the complete picture. 

Because there are so many possible parameters to vary, it makes sense to start by isolating one or two.  A good first step could be to analyze, along the lines of \cite{Jokela:2012dw}, the thermodynamics of the D6-brane probe with only the internal flux, presented in section \ref{Full_ansatz}.  In the absence of massless flavors, this system is essentially a probe D6-brane in the ABJ background, but with zero worldvolume gauge field.  This is another system which deserves a detailed thermodynamic study.

Another interesting open problem is the study of flavor branes with internal flux in the ABJM background with unquenched massive quarks presented in chapter \ref{massiveABJM}. Recall  that the geometry found in chapter \ref{massiveABJM} is a running solution flowing between two $AdS$ spaces,  in  which the control parameter is the mass of the sea quarks. One expects to find conductivities depending on the mass of the dynamical quarks, which interpolate between the values found here (for massless sea quarks) and the unflavored values (for infinitely massive quarks).

In this chapter we considered brane probes with electric and magnetic fields in their worldvolume and we have neglected the backreaction of these electromagnetic fields. Computing this backreaction is a very complicated task. One possible intermediate step could be considering the geometry dual to the non-commutative ABJM model, which was found in \cite{Imeroni:2008cr} by applying a TsT duality transformation. By  adding internal flux to the probe branes we should be able to find Hall states similar to those found here \cite{work_in_progress}.

We initiated a very limited fluctuation analysis in section \ref{mesons}, and a more thorough study is needed.  One goal would be to compute the full meson dispersion about the BPS solution of section \ref{BPS-sol}.  The lightest neutral excitations of QH fluids are magneto-rotons, collective excitations whose minimum energy is at nonzero momentum.  They have been detected in experiments, for example \cite{rotonexp}, and they have also been found in other holographic probe-brane QH models \cite{Jokela:2010nu, Jokela:2011sw}.  Naturally, we would like to know if the spectrum of the model presented here also includes rotons.

Homogeneity-breaking instabilities seem to be a general feature of black hole embeddings in related brane models \cite{BallonBayona:2012wx, Bergman:2011rf, Jokela:2012se, Jokela:2012vn}, which are examples of the general type of instability described in \cite{Nakamura:2009tf}.  In some cases, examples of spatially-modulated ground states have been found explicitly; for example, see \cite{Rozali:2012es, Withers:2013loa, Withers:2014sja, Jokela:2014dba}.  A thorough analysis of the quasi-normal mode spectrum is needed to determine whether such instabilities exists in this model.  If so, the ABJM system, due to its symmetries and other special properties, might afford an ideal laboratory to study inhomogeneous phases.

Another interesting area to explore is alternative quantization of the D6-brane worldvolume gauge field.  In a four-dimensional bulk, the gauge field can take Dirichlet, Neumann, or mixed boundary conditions in the UV, and these choices correspond to different boundary CFTs dual to strongly coupled anyon fluids in dynamical gauge fields. In particular, by changing the quantization as in \cite{Jokela:2013hta, Brattan:2013wya, Jokela:2014wsa}, this ABJM system could be turned from a quantum Hall fluid into an anyon superfluid.

In the next chapter, we will focus on a D6-brane probe with density turned on, and all the rest of worldvolume gauge fields turned off. Of course, the equations of motion are a particular case of those presented in this chapter, but we will give a detailed analysis of the solutions. Besides, we give a detailed study of the fluctuations. The main feature that we have found is the interesting behavior of the quantum phase transitions involved.

\begin{subappendices}

\section{Appendices}
\label{appendixsusy4}

\setcounter{equation}{0}

\subsection{Details of the background geometry}
\label{Background_details}

In this appendix we specify the coordinate system we employ to represent the metric and forms of the background. Let us begin with the four-sphere part of the internal metric (\ref{internal-metric-flavored}). Most of the following expressions are basically the ones presented in section (\ref{review-ABJM}) where we rewrite $\xi$ in terms of a new angle $\alpha$ related by $\frac{2\xi}{1+\xi^2}=\sin \alpha$. Let $\omega^i$ ($i=1,2,3$) be the $SU(2)$ left-invariant one-forms which satisfy $d\omega^i={1\over2}\,\epsilon_{ijk}\,\omega^j\wedge\omega^k$. Together with a new coordinate $\alpha$, the $\omega^i$'s can be used to parameterize the metric of  the four-sphere ${\mathbb S}^4$ as:
\beq
ds^2_{{\mathbb S}^4}\,=\,d\alpha^2\,+\,
{\sin^2\alpha\over 4}
\left[(\omega^1)^2+(\omega^2)^2+(\omega^3)^2
\right]\ ,
\label{S4metric}
\eeq
where $0\le \alpha<\pi$. The $SU(2)$ instanton one-forms $A^i$  which fiber the ${\mathbb S}^2$ over the $ {\mathbb S}^4$ in (\ref{internal-metric-flavored}) can be written in these coordinates as:
\beq
A^{i}\,=\,-\sin^2\left({\alpha\over 2}\right)
\,\,\omega^i\,\,. 
\label{A-instanton}
\eeq
Let us next parametrize the $z^i$ coordinates of the ${\mathbb S}^2$ by means of two angles $\theta$ and $\varphi$ ($0\le\theta<\pi$, $0\le\varphi<2\pi$), namely:
\beq
z^1\,=\,\sin\theta\,\cos\varphi\,\,,\qquad\qquad
z^2\,=\,\sin\theta\,\sin\varphi\,\,,\qquad\qquad
z^3\,=\,\cos\theta\,\,.
\label{cartesian-S2}
\eeq
Then, one can easily prove that the  $ {\mathbb S}^2$ part of the metric 
(\ref{internal-metric-flavored}) can be written as:
\beq
\left(d x^i\,+\, \epsilon^{ijk}\,A^j\,z^k\,\right)^2\,=\,(E^1)^2\,+\,(E^2)^2\ ,
\eeq
where  $E^1$ and $E^2$ are the following one-forms:
\bear
E^1&=&d\theta+\sin^2\big({\alpha\over 2}\big)\,
\left[\sin\varphi\,\omega^1-\cos\varphi\,\omega^2\right]~,
\rc\rc
E^2&=&\sin\theta\left[d\varphi-\sin^2\big({\alpha\over 2}\big)
\,\omega^3\right]+\sin^2\big({\alpha\over 2}\big)\,
\cos\theta\left[\cos\varphi\,\omega^1+\sin\varphi\,\omega^2\right]\ .
\label{Es}
\eear

In order to write the explicit expression for $F_2$,  we  first define three new one-forms  $S^i$ $(i=1,2,3)$ as the following rotated version of the $\omega^i$'s:
\bear
S^1&=&\sin\varphi\,\omega^1-\cos\varphi\,\omega^2~, \rc\rc
S^2&=&\sin\theta\,\omega^3-\cos\theta\left(\cos\varphi\,\omega^1+
\sin\varphi\,\omega^2\right)~, \rc\rc
S^3&=&-\cos\theta\,\omega^3-\sin\theta\left(\cos\varphi\,\omega^1+
\sin\varphi\,\omega^2\right)\ .
\label{rotomega}
\eear 
Next, we define the one-forms ${\cal S}^{\alpha}$   and ${\cal S}^{i}$ as:
\beq
{\cal S}^{\alpha}\,=\,d\alpha\,\,,\qquad\qquad
{\cal S}^{i}\,=\,{\sin\alpha\over 2}\,S^i \,\,,\qquad(i=1,2,3)\ ,
\label{calS}
\eeq
in terms of which the metric of the four-sphere is just 
$ds^2_{{\mathbb S}^4}=({\cal S}^{\alpha})^2+\sum_i({\cal S}^{i})^2$.  

With these definitions,  the ansatz for the RR two-form $F_2$ for the flavored background  is
\beq
F_2\,=\,{k\over 2}\,\,\Big[\,\,
E^1\wedge E^2\,-\,\eta\,
\big({\cal S}^{\alpha}\wedge {\cal S}^{3}\,+\,{\cal S}^1\wedge {\cal S}^{2}\big)
\,\,\Big]\ .
\label{F2-flavored}
\eeq
Note that the two-form $F_2$ is not closed when $\eta\not=1$; $dF_2$ is proportional to the charge distribution three-form of the flavor D6-branes. 
The RR four-form $F_4$ is:
\beq
F_4\,=\,{3k\over 4}\,\,\,{(\eta+q)b\over 2-q}\,\,L^2\,\,\Omega_{BH_4}\ ,
\eeq
where $\Omega_{BH_4}$ is the volume-form of the four-dimensional black hole (\ref{BH4-metric}). The solution is completed by a constant dilaton $\phi$ given by
\beq
e^{-\phi}\,=\,{b\over 4}\,{\eta+q\over 2-q}\,{k\over L}\ . 
\label{dilaton-flavored-squashings}
\eeq

Let us now spell out the embedding of the D6-brane probe in the background geometry. We first represent the $SU(2)$ left-invariant one-forms $\omega^i$ in terms of three angles $\hat\theta$, $\hat\varphi$, and $\hat\psi$ as:
\bear
\omega^1 & = & \cos\hat\psi\,d\hat\theta+\sin\hat\psi\,\sin\hat\theta\,d\hat\varphi~, \rc
\omega^2 & = & \sin\hat\psi\,d\hat\theta-\cos\hat\psi\,\sin\hat\theta\,d\hat\varphi~, \rc
\omega^3 & = & d\hat\psi+\cos\hat\theta \,d\hat\varphi\,\,.
\label{w123}
\eear
In these coordinates our embedding is defined by the conditions:
\beq
\hat\theta\, ,\ \hat\varphi \,=\,{\rm constant}\ ,
\label{RP3-cycle}
\eeq
with the coordinate $\theta$ defined in (\ref{cartesian-S2}) being a function of the radial coordinate $r$.  The relation of the coordinates defined here with those used in (\ref{induced_metric_Hall}) to write the internal part of the induced metric is as follows: The angle $\alpha$ here is equal to the one introduced in (\ref{S4metric}), while $\beta$ and $\psi$ are given by
\beq
\beta\,=\,{\hat\psi\over 2}\,\,,\qquad\qquad
\psi\,=\,\varphi\,-\,{\hat\psi\over 2}\ .
\label{RP3-angles}
\eeq
It is now easy to check that the pullback of the metric (\ref{flavoredBH-metric}) to the worldvolume is, indeed, the line element written in (\ref{induced_metric_Hall}).

\subsubsection{Matching the unflavored ABJ model}
\label{matching_ABJ}

Let us now explore our prescription (\ref{wv_quantization}) in the case of the unflavored model. The main point is that, when $N_f=0$, the worldvolume gauge field for the supersymmetric configurations can be understood as induced by a flat NSNS $B_2$ field of the bulk, which is proportional to the K\"ahler form $J$ of ${\mathbb C}{\mathbb P}^3$. When the coefficient multiplying 
$J$ is appropriately quantized, the corresponding supergravity solution is the dual of the ABJ theory with gauge group $U(N+M)_{k}\times U(N)_{-k}$. We will see that the rank difference $M$ can be identified with the quantization integer in (\ref{wv_quantization}).  

Let us begin our analysis by writing the K\"ahler form of  ${\mathbb C}{\mathbb P}^3$ in our variables:
\beq
J\,=\,{1\over 4}\,\,\Big(\,E^1\wedge E^2\,-\,\big(
{\cal S}^{\alpha}\wedge {\cal S}^{3}\,+\,{\cal S}^1\wedge {\cal S}^{2}\big)\,\Big)\ .
\label{Kahler}
\eeq
The pullback of $J$  to the probe brane worldvolume is:
\beq
\hat J\,=\,{\theta'\,\sin\theta\over 4}\,dr\wedge \big[\,
d\psi+\cos\alpha\,d\beta\,\big]\,+\,{\cos\theta\sin\alpha\over 4}\,d\alpha\wedge d\beta\ ,
\eeq
and, as claimed, it has the form written in  (\ref{F_internal}) if we identify the flux function $a(r)$ with:
\beq
a(r)\,=-Q\,\cos\theta(r)\ ,
\label{a_theta_unflavored}
\eeq
where $Q$ is a constant. Actually, one can check that the worldvolume gauge field $F$ (\ref{F_internal}) for this unflavored case can be written in terms of the pullback of $J$ as:
\beq
F\,=\,4\,L^2\,Q\,\hat J\,\,.
\label{internal_flux}
\eeq
We will see below  in appendix \ref{kappa} that the relation (\ref{a_theta_unflavored}) between the flux and embedding functions is dictated by supersymmetry when $N_f=0$. Notice also that the flux function at the tip is just $Q$, as in (\ref{a_*_Q}).
 
In the DBI+WZ action for D-branes, the worldvolume gauge field $F$ is always combined additively with the pullback of $B_2$. It follows that,  in this case, the worldvolume flux can alternatively be thought of as induced by the following NSNS $B_2$ field:
\beq
B_2\,=\,4\,Q\,L^2\,J\,\,.
\label{B2-para}
\eeq
Notice that $B_2$ is a closed two-form, and it has the same form as in the proposed gravity dual of the ABJ model \cite{Aharony:2008gk} with gauge group $U(N+M)_{k}\times U(N)_{-k}$, where $Q$ is related to $M$. Actually, the integer $M$ is determined by the discrete holonomy of $B_2$ on the ${\mathbb C}{\mathbb P}^1$ cycle of the ${\mathbb C}{\mathbb P}^3$ space, which is inherited from the holonomy of the three-dimensional three-form potential of the eleven-dimensional supergravity along the torsion cycle of the ${\mathbb S}^7/{\mathbb Z_k}$. Let us compute explicitly the integral of the 
two-form (\ref{B2-para}) along the ${\mathbb C}{\mathbb P}^1$.  In our coordinates (see \cite{Conde:2011sw}) the  ${\mathbb C}{\mathbb P}^1$ is obtained by keeping the coordinates of the ${\mathbb S}^4$ cycle fixed. Therefore, the pullback of $J$ is just:
\beq
J_{ |_{{\mathbb C}{\mathbb P}^1}}\,=\,{1\over 4}\,\sin\theta\,d\theta\wedge d\varphi\,\,,
\eeq
and thus the integral of $J$ along the ${\mathbb C}{\mathbb P}^1$ is:
\beq
\int_{{\mathbb C}{\mathbb P}^1}\,J\,=\,\pi\,\,.
\eeq
It follows from (\ref{B2-para})  that:
\beq
\int_{{\mathbb C}{\mathbb P}^1}\,B_2\,=\,4\pi\,L^2\,Q\,\,.
\eeq
Let us now use our quantization condition (\ref{a*_quantization}) and the identification 
(\ref{a_*_Q}) to write the period of $B_2$ in terms of $k$ and  the quantization integer $M$. We get:
\beq
\int_{{\mathbb C}{\mathbb P}^1}\,B_2\,=\,(2\pi)^2\,{M\over k}\,\,,
\eeq
which is the fractional holonomy proposed in \cite{Aharony:2008gk} for the gravity dual of the $U(N+M)_{k}\times U(N)_{-k}$ theory. 

The coefficient $Q$ can also be fixed by looking at the Page charge  $Q_4$ for fractional D2-branes (D4-branes wrapped on a ${\mathbb C}{\mathbb P}^1$ two-cycle), which is given by the following integral over the ${\mathbb C}{\mathbb P}^2$ dual to the   ${\mathbb C}{\mathbb P}^1$  where the D4-branes are wrapped:
\beq
Q_4\,=\,{1\over (2\pi)^3}\,\int_{{\mathbb C}{\mathbb P}^2}\,\Big[F_4+B_2\wedge F_2]
\,\,. 
\label{Q_4}
\eeq
We require that $Q_4$ is equal to our quantization integer $M$, which can then be interpreted as the number of fractional D2-branes. Taking into account (\ref{B2-para}) and that  $F_2=2k\,J$ for this unflavored case, we get:
\beq
Q_4\,=\,{k\,L^2\,Q\over  \pi^3}\,\int_{{\mathbb C}{\mathbb P}^2}\,
J\wedge J\,\,.
\eeq
To compute this integral we use the fact that the equation of the
${\mathbb C}{\mathbb P}^2$ cycle in our coordinates is $\varphi=\theta=\pi/2$ (see appendix A in \cite{Conde:2011sw}), which implies:
\beq
J\wedge J_{ |_{{\mathbb C}{\mathbb P}^2}}\,=\,{1\over 16}\,
\sin^2{\alpha\over 2}\,\sin\alpha\,
\,d\alpha\wedge \omega^{1}\wedge \omega^{2}\,\wedge \omega^{3}\,\,.
\eeq
Then, it follows that:
\beq
\int_{{\mathbb C}{\mathbb P}^2}\,J\wedge J\,=\,\pi^2\,\,.
\eeq
Therefore,
\beq
Q_4\,=\,{k\,L^2\,Q\over  \pi}\,\,,
\eeq
and the quantization condition $Q_4=M$ coincides with the one obtained in (\ref{a*_quantization}) for $a_*\,=\,-Q$.

\subsection{Probe brane equation of motion}
\label{EOMs}

Let us consider a D$p$-brane probe propagating in a background of type II supergravity.  Let $g_{ij}$ denote the components of the induced metric on the worldvolume:
\beq
g_{ij}\,=\,g_{mn}\,\partial_i\,X^{m}\,\partial_j\,X^n\,\,,
\eeq
where the $X^{n}$ are coordinates of the ten-dimensional space and  $g_{mn}$ is the target space metric of the background. 
In what follows $m,n,\ldots$ will denote indices of the target space, whereas $i,j,\ldots$ will represent worldvolume indices. Let us denote by $M$ the following matrix:
\beq
M\,=\,g+F\,\,,
\eeq
where $F=dA$ is the worldvolume gauge field. Then, the action of a D$p$-brane probe can be written as:
\beq
S_{D_p}\,=\,S_{DBI}+S_{WZ}\,\,,
\eeq
where the DBI and WZ terms are:
\beq
S_{DBI}\,=\,-T_{Dp}\,\int_{{\cal M}_{p+1}}\,d^{p+1}\xi\,\,e^{-\phi}\,
\sqrt{-\det M}\,\,,
\qquad\qquad
S_{WZ}\,=\,T_{Dp}\,\int_{{\cal M}_{p+1}}\,e^{F}\wedge C\,\,,
\eeq
with $T_{Dp}$ being the D$p$-brane tension (from now on  in this appendix we will take $T_{Dp}=1$) and $C=\sum_r C_r$ is the sum of RR potentials. 
In order to write the equations of motion derived from this action following the analysis of section 2 of \cite{Skenderis:2002vf}, let us consider the inverse $M^{-1}=[M^{ij}]$  of the matrix $M=[M_{ij}]$ and let us decompose $M^{-1}$ in its symmetric and antisymmetric parts as:
\beq
M^{-1}\,=\,{\cal G}^{-1}\,+\,{\cal J}\,\,,
\eeq
where ${\cal J}=[{\cal J}^{ij}]$ is the antisymmetric component of $M^{-1}$ and ${\cal G}^{-1}=[{\cal G}^{ij}]$ is the inverse open string metric. Then,  the equation of motion of the gauge field component $A_j$ is \cite{Skenderis:2002vf}:
\beq
\partial_j\Big(e^{-\phi}\,\sqrt{-\det M}\,{\cal J}^{ji}\Big)\,=\,j^{i}\,\,,
\label{eom_gauge_general}
\eeq
where the source current for the gauge field $j^{i}$ is given by:
\beq
j^{i}\equiv {\delta S_{WZ}\over \delta A_i}\,\,.
\eeq
Moreover, the equation for the scalar field $X^m$ becomes \cite{Skenderis:2002vf}:
\bear
&&-\partial_i\Big(e^{-\phi}\,\sqrt{-\det M}\,{\cal G}^{ij}\,\partial_j\,
X^n\,g_{nm}\Big)\, \,\rc\rc
&&\qquad\qquad
+\sqrt{-\det M}\,\Big({e^{-\phi}\over 2}\,
{\cal G}^{ji}\,\partial_i\,X^n\,\partial_j\,X^p\,g_{np,m}\,-\,e^{-\phi}\,\partial_m\,\phi\Big)\,=\,
j_m\,\,,\qquad\qquad
\label{eom_scalars_general}
\eear
where the source for the scalar $X^m$ is:
\beq
j_m\equiv {\delta S_{WZ}\over \delta X^m}\,\,.
\eeq

\subsubsection{Currents for the D6-brane}

Let us  write the form of the currents for the case of a D6-brane probe. In this case,
the WZ term of the action is:
\beq
S_{WZ}\,=\,\int_{{\cal M}_7}\,\left(\hat C_7\,+\,\hat C_5\wedge F\,+\,{1\over 2}\,
\hat C_3\wedge F\wedge F\,+\,{1\over 6}\hat C_1\wedge F\wedge F\wedge F
\right)\ .
\label{WZ-D6}
\eeq
Let us perform a general variation of the worldvolume gauge field $F\to F+d(\delta A)$, under which $S_{WZ}$ varies as:
\beq
\delta S_{WZ}\,=\,\int_{{\cal M}_7}\,
\left( \hat C_5\,+\,\hat C_3\wedge F\,+\,{1\over 2}\,\hat C_1\wedge F\wedge F\right)
\wedge d(\delta A)\ .
\eeq
In order to compute the current associated to the worldvolume gauge field, we use the fact that,  for any odd-dimensional form ${\cal O}$, one has
\beq
{\cal O}\wedge d(\delta A)\,=\,d{\cal O}\wedge \delta A\,-\,d({\cal O}\wedge \delta A)\ .
\eeq
The total derivative generates a boundary term which vanishes since we are assuming that $A$ is fixed at the boundary in the variational process.\footnote{Note that, although we have chosen Dirichlet boundary conditions for $A$ here, $A$ can, in fact, have arbitrary mixed boundary conditions, corresponding to alternative quantization, as discussed in \cite{Jokela:2013hta}.} Taking into account that, with our notation $F_4=-dC_3$, we get:
\beq
\delta S_{WZ}\,=\,\int_{{\cal M}_7}\,
\left( \hat F_6\,-\,\hat F_4\wedge F\,+\,{1\over 2}\,\hat F_2\wedge F\wedge F\right)
\wedge \delta A\ .
\eeq
Then, the gauge current along the worldvolume direction $i$ is given by the expression:
\beq
j^i\,d^7\xi\,=\,\left(\hat F_6\,-\,\hat F_4\wedge F\,+\,{1\over 2}\,\hat F_2\wedge F\wedge F\right)\wedge d\xi^i\ .
\eeq

In order to compute the source current $j_m$ for $X^m$, we should vary in (\ref{WZ-D6}) the  scalars which enter the pullback of the RR potentials. It turns out that the final expression can be written in a rather simple form, which we will now spell out. 
Let $V=V^{m}\,{\partial \over \partial X^m}$ be any vector field in target space. The interior product of $V$ with a $p$-form $\omega$ is a $(p-1)$-form $ \iota_V\omega$ defined as follows. Let $\omega$ be:
\beq
\omega\,=\,{1\over p!}\,\omega_{n_1,\ldots ,n_p}\,
dX^{n_1}\wedge\cdots\wedge dX^{n_p}\,\,.
\eeq
Then, $ \iota_V\omega$  is given by:
\beq
\iota_V\omega\,=\,{1\over (p-1)!}\,V^{m}\,
\omega_{m,m_1,\ldots, m_{p-1}}\,dX^{m_1}\wedge \cdots\wedge dX^{m_{p-1}}\,\,.
\eeq
Let $\iota_m\omega$ denote the interior product of $\omega$ and the vector $\partial/\partial X^m$:
\beq
\iota_m\omega\,\equiv\,\iota_{{\partial\over \partial X^m}}\omega\,\,. 
\eeq
Then, the current $j_m$, corresponding to the scalar $X^{m}$,  can be written as:
\beq
j_m\,d^7\xi\,=\,\widehat{\iota_m F_8}\,+\,\widehat{\iota_m F_6}\wedge F\,-\,{1\over 2}\,
\widehat{\iota_m F_4}\wedge F\wedge F\,+\,{1\over 6}\,
\widehat{\iota_m F_2}\wedge F\wedge F\wedge F\,\,,
\label{scalar_current}
\eeq
where the hat denotes the pullback of the different $\iota_m F_r$ to the worldvolume. In (\ref{scalar_current}) $F_8$ and $F_6$ are defined as  Hodge duals of $F_2$ and $F_4$, respectively, \ie, $F_8=-*F_2$ and $F_6=-*F_4$.

Notice that we have derived the expressions of $j^i$ and $j_m$ from the action (\ref{WZ-D6}), where we have assumed the existence of the RR potentials $C_r$.  In the case of backreacting flavor some Bianchi identities are  violated and, as a consequence, some of the RR potentials do not exist. However, the currents  $j^i$ and $j_m$ (and the corresponding equations of motion) only depend on the RR field strengths and their pullbacks, and then they can be generalized to the case in which we include the backreaction. This is the point of view we will adopt in what follows.

\subsubsection{The equations of motion for our ansatz}

We now write explicitly the equations of motion for the D6-brane with a gauge potential $A$ as the one written in (\ref{A_full_ansatz}). We will also assume that the embedding is defined by the conditions (\ref{RP3-cycle}) with $\theta=\theta(r)$ being a function of $r$ to be determined. The set of worldvolume coordinates we will employ is:
\beq
\xi^{i}\,=\,(t, x, y , r, \alpha,\beta,\psi)\,\,,
\eeq
where $\alpha$, $\beta$, and $\psi$ are the angles defined in (\ref{S4metric}) and (\ref{RP3-angles}).  First of all, let us write the non-zero components of the worldvolume gauge field strength $F$ corresponding to the potential  (\ref{A_full_ansatz}):
\bear
&&F_{t\,x}\,=\,L^2\,E\,\,,\qquad\qquad
F_{x\,y}\,=\,L^2\,B\,\,,\rc\rc
&&F_{r\,t}\,=\,L^2\,a_0'\,\,,\qquad\qquad
F_{r\,x}\,=\,L^2\,a_x'\,\,,\qquad\qquad
F_{r\,y}\,=\,L^2\,a_y'\,\,,\rc\rc
&&F_{r\,\psi}\,=\,L^2\,a'\,\,,\qquad\qquad
F_{r\,\beta}\,=\,L^2\,a'\cos\alpha\,\,,\qquad\qquad
F_{\alpha\,\beta}\,=\,-L^2\,a\,\sin\alpha\,\,,
\qquad\qquad\qquad
\eear
where the prime denotes the derivative with respect to the radial variable. 
Notice that, in our ansatz, isotropy in the $x-y$ plane is explicitly broken by the electric field in the $x$-direction.

We will start by computing the different components of the currents appearing in (\ref{eom_gauge_general}) and (\ref{eom_scalars_general}). It is straightforward  to prove that $\hat F_6=0$, and it therefore does not contribute to 
$j^i$ and $j_m$.  The non-vanishing components of the gauge current $j^i$ are:
\bear
&&j^t\,=\,{kL^4\over 2}\,B\,\sin\alpha\,(\eta\,\cos\theta\,a'\,-\,a\,\sin\theta\,\theta')\,\,,\rc\rc
&&j^{y}\,=\,{kL^4\over 2}\,E\,\sin\alpha\,(\eta\,\cos\theta\,a'\,-\,a\,\sin\theta\,\theta')\,\,,\rc\rc
&&j^{\psi}\,=\,{kL^4\over 2}\,\sin\alpha\,\Bigg({3b\over 2}\,{\eta+q\over 2-q}\,r^2\,a\,-\,
\eta\,(B\,a_0'+E\,a_y')\cos\theta \Bigg)\,\,.
\label{gauge_current_comps_general}
\eear
We now work out the current for the three transverse scalars.  First we compute the interior products of $F_8$ with the tangent vectors along the three scalar directions $m=\theta, \hat\theta, \hat\varphi$. We find:
\beq
\widehat{\iota_{\theta} F_8}\,=\,-{(3-2b)(q+\eta)q\over 8 b^3 (2-q)}\,kL^6\,\sin
\alpha\,r^2\,\sin(2\theta)\,\,d^7\xi\,\,.
\label{iF8_scalars}
\eeq
Moreover,  the product of $F_8$  with the other two tangent vectors gives a result proportional  to $\widehat{\iota_{\theta} F_8}$:
\beq
\widehat{\iota_{\hat\theta} F_8}\,=\,-\sin^2\Big({\alpha\over 2}\Big)\,\sin(\beta-\psi)\,
\widehat{\iota_{\theta} F_8}\,\,,
\qquad
\widehat{\iota_{\hat\varphi} F_8}\,=\,\sin\hat\theta\sin^2\Big({\alpha\over 2}\Big)\,\cos(\beta-\psi)\,
\widehat{\iota_{\theta} F_8}\,\,.
\label{iF8_hat_scalars}
\eeq
We already mentioned that $F_6$ does not contribute since its pullback is zero. It is also immediate to check that $F_4$  does not have components along the transverse scalars and will not contribute to $j_m$. The contribution of $F_2$ to $j_{\theta}$ is determined by:
\beq
{1\over 6}\,
\widehat{\iota_{\theta} F_2}\wedge F\wedge F\wedge F\,=\,{kL^6\over 2}\,
\sin\alpha\,a\,\sin\theta\,(B\,a_0'\,+\,E\,a_y')
\,\,d^7\xi\,\,,
\eeq
while the result for the other scalars are:
\bear
&&\widehat{\iota_{\hat\theta} F_2}\wedge F\wedge F\wedge F\,=\,
-\sin^2\Big({\alpha\over 2}\Big)\,\sin(\beta-\psi)\,
\widehat{\iota_{\theta} F_2}\wedge F\wedge F\wedge F\,\,,\rc\rc
&&\widehat{\iota_{\hat\varphi} F_2}\wedge F\wedge F\wedge F\,=\,
\sin\hat\theta\sin^2\Big({\alpha\over 2}\Big)\,\cos(\beta-\psi)\,
\widehat{\iota_{\theta} F_2}\wedge F\wedge F\wedge F\,\,.
\label{iF2_hat_scalars}
\eear
Notice that the proportionality factors in (\ref{iF8_hat_scalars}) and (\ref{iF2_hat_scalars}) are the same. Thus, the current for the scalar $\theta$ becomes:
\beq
j_{\theta}\,=\,-{kL^6\over 2}\,\sin\alpha\,\sin\theta\,
\Bigg({(3-2b)(q+\eta)q\over 2\, b^3 (2-q)}\,r^2\cos\theta\,-\,a\,(B\,a_0'\,+\,E\,a_y')\,\Bigg)\,\,.
\label{j_theta_general_case}
\eeq
Moreover, the other two components of $j_m$ are:
\beq
j_{\hat\theta}\,=\,-\sin^2\Big({\alpha\over 2}\Big)\,\sin(\beta-\psi)\,j_{\theta}\,\,,
\qquad
j_{\hat\varphi}\,=\,\sin\hat\theta\,\sin^2\Big({\alpha\over 2}\Big)\,\cos(\beta-\psi)\,
\,j_{\theta}\,\,.
\label{j_hat_scalars_exp}
\eeq

Let us now  finally write the equations of motion  for the different gauge field components and scalars. We have to compute the different components of the antisymmetric tensor ${\cal J}^{ij}$, as well as the elements of the inverse open string metric ${\cal G}^{ij}$.  This calculation is straightforward (although rather tedious in some cases) and we limit ourselves to give the final result for the equations. The equation of $A_t$ is:
\bear
&&{q+\eta\over 2b(2-q)}\,\partial_r
\Bigg[{\sqrt{h}\,\sqrt{q^2+b^4\,a^2}\over \sqrt{\Delta}
\sqrt{(B^2+r^4)h-E^2}}\,\sin^2\theta\,
\big[B(B\,a_0'\,+\,E\,a_y')\,+r^4\,a_0'\big]
\Bigg]
\,\,\rc\rc
&&\qquad\qquad\qquad\qquad
-B(\eta\,\cos\theta\,a'-a\,\sin\theta\,\theta')\,=\,0\,\,,
\label{eom_a0_general_case}
\eear
where $\Delta$ is the quantity defined in (\ref{Delta_def_general}).  The equation for 
 $A_{x}$ can be integrated once ($a_x$ is a cyclic variable) to give the following equation for $a_x'$:
\beq
r^4\,h^{{3\over 2}}\,\sin^2\theta\,
{\sqrt{q^2+b^4\,a^2}\,a_x'\over 
\sqrt{\Delta}\,\sqrt{(B^2+r^4)h-E^2}}\,=\,
{\rm constant}\,\,.
\label{eom_ax_general_case}
\eeq
The equation for $A_{y}$ is also non-trivial and given by:
\bear
&&{q+\eta\over 2b(2-q)}\,\partial_r
\Bigg[{\sqrt{h}\,\sqrt{q^2+b^4\,a^2}\over \sqrt{\Delta}
\sqrt{(B^2+r^4)h-E^2}}\,\sin^2\theta\,
\big[E(B\,a_0'\,+\,E\,a_y')\,-r^4\,h\,a_y'\big]
\Bigg]
\,\,\rc\rc
&&\qquad\qquad\qquad\qquad
-E(\eta\,\cos\theta\,a'-a\,\sin\theta\,\theta')\,=\,0\,\,.
\label{eom_ax2_general_case}
\eear
It is easy to demonstrate that the equations for $A_r$, $A_{\alpha}$, and $A_{\beta}$ are trivially satisfied by our ansatz. The only non-trivial equation for the gauge field that remains to write is the one corresponding to $A_{\psi}$, which is:
\bear
&&\partial_r\Bigg[{r^2\,\sqrt{h}\,\sqrt{q^2+b^4\,a^2}\,\sqrt{(B^2+r^4)h-E^2}
\over \sqrt{\Delta}}\,a'\Bigg]-
{\sqrt{\Delta}\,\sqrt{(B^2+r^4)h-E^2}\over\sqrt{h}\, \sqrt{q^2+b^4\,a^2}}\,a\, \,\rc\rc
&&\qquad\qquad\qquad\qquad
+3r^2 a\,-\,{2(2-q)\,\eta\over b(q+\eta)}\,(B\,a_0'\,+\,E\,a_y')\cos\theta=0\,\,.
\label{eom_apsi_general_case}
\eear
Finally, one can prove that the three equations for the transverse scalars $\theta$, $\hat \theta$, and $\hat\varphi$ are the same, namely:
\bear
&&\partial_r\Bigg[{r^2\,\sin^2\theta\,\sqrt{h}\,\sqrt{q^2+b^4\,a^2}\,\sqrt{(B^2+r^4)h-E^2}
\over \sqrt{\Delta}}\,\theta'\Bigg] \rc\rc
&&\qquad\qquad
-\,{\sqrt{q^2+b^4\,a^2}\,\sqrt{(B^2+r^4)h-E^2}\over \sqrt{h}\,\sqrt{\Delta}}
\big[\Delta\,-\,b^4\,r^2\,h\,a'^{\,2}\big]
\cot{\theta} \qquad\rc\rc
&&\qquad
-(3-2b)\,q\,r^2\,\sin\theta\cos\theta\,+\,{2b^3(2-q)\over q+\eta}\,a\,\sin\theta\,
(B\,a_0'\,+\,E\,a_y')\,=\,0
\,\,.\qquad\qquad
\label{eom_theta_general_case}
\eear

Eq. (\ref{eom_ax_general_case}) allows us to eliminate $a_x'$, after which
 we have four second-order, coupled differential equations (\ref{eom_a0_general_case})-(\ref{eom_theta_general_case}) for four radial functions of $a_0$, $a_y$, $a$, and $\theta$. Solving this system in general is a quite formidable task. For this reason it is worth to look for simplifications and partial integrations. Notice that (\ref{eom_a0_general_case}) and (\ref{eom_ax2_general_case}) present some electric-magnetic symmetry. 
Actually, by combining these equations one easily finds the following constant of motion:
\beq
{r^4\,\sqrt{h}\,\sqrt{q^2+b^4\,a^2}\over \sqrt{\Delta}
\sqrt{(B^2+r^4)h-E^2}}\,\sin^2\theta\,
\big[E\,a_0'\,+h\,B\,a_y'\big]\,=\,{\rm constant}\,\,,
\label{constant_of_motion}
\eeq
which  could be used to eliminate $a_0'$ or $a_y'$ from the system of equations. Moreover, in the unflavored case ($\eta=b=q=1$), the last two terms in (\ref{eom_a0_general_case}) and (\ref{eom_ax2_general_case}) can be combined to construct the radial derivative of $a\cos\theta$, which leads to two constants of motion. In this unflavored case, $a_0$ and $a_y$  are cyclic variables and can be eliminated.

\subsection{Kappa symmetry analysis}
\label{kappa}

The kappa symmetry matrix for a D$p$-brane in the type IIA theory is given by:
\beq
d^{p+1}\,\zeta\,\Gamma_{\kappa}\,=\,
{1\over \sqrt{- \det (g+F)}}\,e^{F}\wedge X\,\,,
\eeq
where $g$ is  the induced metric,  $\zeta^{\alpha}$ ($\alpha=0,\ldots ,p)$ are a set of worldvolume coordinates and $X$ is the polyform matrix:
\beq
X\,=\,\sum_{n}\,\gamma_{2n+1}\,\big(\Gamma_{11}\big)^{n+1}\,\,,
\eeq	
with $\gamma_{2n+1}$ being the $(2n+1)$-form whose components are the  antisymmetrized products of the induced Dirac matrices $\gamma_{\mu}$:
\beq
\gamma_{2n+1}\,=\,{1\over (2n+1)!}\,\gamma_{\mu_1\cdots \mu_{2n+1}}\,\,
d \zeta^{\mu_1}\,\wedge\cdots\wedge d \zeta^{\mu_{2n+1}}\,\,.
\eeq
In particular, we are interested in the case of a D6-brane with a flux across a (non-compact) four-cycle. The corresponding kappa symmetry matrix takes the form:
\beq
d^{7}\, \zeta\,\Gamma_{\kappa}\,=\,
{1\over \sqrt{ -\det (g+F)}}\,\Big[\,\gamma_{(7)}\,+\,F\wedge \gamma_{(5)}\, \Gamma_{11}\,+\,
{1\over 2}\,F\wedge F\wedge  \gamma_{(3)}\,+\,{1\over 6}\,
F\wedge F\wedge  F\wedge \gamma_{(1)}\,\Gamma_{11}
\Big]\,\,.
\label{Gamma_kappa_FF}
\eeq

Let us now study the conditions imposed by kappa symmetry in the case in which
the embedding is determined by the conditions (\ref{RP3-cycle}),  the worldvolume gauge field takes the form (\ref{A_full_ansatz}) with $a_x=0$, and the background is the zero-temperature supergravity solution  of \cite{Conde:2011sw}. We begin by computing  the pullbacks  of the left-invariant $SU(2)$ one-forms  $\omega^i$ of (\ref{w123}) in the $\alpha$, $\beta$, and $\psi$  variables:
\beq
\hat \omega^{1}\,=\,\hat \omega^{2}\,=\,0\,\,,\qquad\qquad
\hat \omega^{3}\,=\,2d\beta\,\,,
\eeq
whereas those of ${\cal S}^i$ and $E^i$ are:
\bear
&&\hat {\cal S}^{\alpha}\,=\,d\alpha\,\,,
\qquad
\hat {\cal S}^{1}\,=\,0\,\,,
\qquad
\hat {\cal S}^{2}\,=\,\sin\alpha\sin\theta\,d\beta\,\,,
\qquad
\hat {\cal S}^{3}\,=\,-\sin\alpha\cos\theta\,d\beta\,\,,\qquad\rc\rc
&&\hat E^{1}\,=\,\theta'\,dr\,\,,\qquad\qquad
\hat E^{2}\,=\,\sin\theta\,(d\psi+\cos\alpha\,d\beta)\,\,.
\label{pullbacks_S_E_newangles}
\eear
The pullbacks of the frame one-forms  used in appendix B of \cite{Conde:2011sw} are:
\bear
&& \hat e^{\mu}\,=\,L\,r\,dx^{\mu}\,\,,\qquad\qquad
\hat e^{3}\,=\,{L\over r}\,dr\,\,,\qquad\qquad
\hat e^{4}\,=\,{\sqrt{q}\over b}\,L\,d\alpha\,\,,
\qquad\qquad\qquad\qquad\rc\rc
&&\hat e^{5}\,=\,0\,\,,\qquad\qquad
\hat e^{6}\,=\,L\,{\sqrt{q}\over b}\,\sin\alpha\sin\theta\,d\beta\,\,,
\qquad\qquad
\hat e^{7}\,=\,-L\,{\sqrt{q}\over b}\,\sin\alpha\cos\theta\,d\beta
\,\,,\rc\rc
&&\hat e^{8}\,=\,{L\over b}\,\theta'\,dr\,\,,\qquad
\hat e^{9}\,=\,{L\over b}\,\sin\theta\,
(d\psi+\cos\alpha\,d\beta)\,\,.
\label{pullback-es_newangles}
\eear
The corresponding  induced gamma matrices become:
\bear
&&\gamma_{x^{\mu}}\,=\,L\,r\,\Gamma_{\mu}\,\,,\qquad\qquad
\gamma_{r}\,=\,{L\over r}\,\Big(\,\Gamma_3\,+\, {r\over b}\,\theta'\,\Gamma_8\,\Big)\,\,,
\qquad\qquad
\gamma_{\alpha}\,=\,L\,{\sqrt{q}\over b}\,\,\,\Gamma_4\,\,,
\qquad\qquad
\rc\rc
&&\gamma_{\beta}\,=\,L\,{\sqrt{q}\over b}\,\sin\alpha\sin\theta\,\,
\Big[\,\Gamma_{6}-\cot\theta\,\Gamma_7+{\cot\alpha\over \sqrt{q}}\,\Gamma_9\,\Big]\,\,,
\qquad\qquad
\gamma_{\psi}\,=\,{L\,\sin\theta\over b}\,\Gamma_9\,\,.
\label{induced_gammas_newangles}
\eear

Let us next compute the different contributions on the right-hand side of (\ref{Gamma_kappa_FF}). First of all we notice that:
\beq
\gamma_{(7)}\,=\,d^7\,\zeta\,\,\gamma_*\,\,,
\eeq
where $\gamma_*$ is the antisymmetrized product of all induced gamma matrices, namely:
\beq
\gamma_*\,=\,\gamma_{t \,x\,y\, r\,\alpha\,\beta\,\psi}\,\,.
\eeq
In terms of flat 10d matrices, $\gamma_*$ can be written as:
\beq
\gamma_*\,=\,{q\over b^3}\,L^7\,r^2\,
\sin\alpha\,\sin^2\theta\,\,
\Gamma_{012}\,\,\Big(\,\Gamma_3\,+\,{r\,\theta'\over b}\,\Gamma_8\,\Big)\,
\Gamma_4\,\Big(\,\Gamma_6\,-\,\cot\theta\,\Gamma_7\,\Big)\,\Gamma_9\,\,.
\eeq
With our notation, the supersymmetric embeddings are those that satisfy $\Gamma_{\kappa}\,\epsilon\,=\,-\epsilon$, where $\epsilon$ is a Killing spinor of the background. To implement this relation we impose that $\epsilon$ satisfies the projection corresponding to a D2-brane, \ie,
\beq
\Gamma_{012}\,\epsilon\,=\,-\epsilon\,\,.
\label{D2-brane_projection}
\eeq
We also impose that $\epsilon$ satisfies the generic projections found in appendix B of \cite{Conde:2011sw} for a generic ABJM-like geometry (Eqs.~(B.4) and (B.14) in \cite{Conde:2011sw}):
\beq
\Gamma_{47}\,\epsilon\,=\,\Gamma_{56}\,\epsilon\,=\,\Gamma_{89}\,\epsilon\,\,,
\qquad\qquad
\Gamma_{3458}\,\epsilon\,=\,-\,\epsilon\,\,.
\label{internal_projections}
\eeq
From these projections  it follows that:
\beq
\Gamma_{3469}\,\epsilon\,=\,-\Gamma_{8479}\,\epsilon\,=\,\epsilon\,\,,
\qquad\qquad
\Gamma_{3479}\,\epsilon\,=\,\Gamma_{8469}\epsilon\,=\,-\Gamma_{38}\,\epsilon\,\,.
\eeq
Using  (\ref{D2-brane_projection}) and (\ref{internal_projections}), $\gamma_*\,\epsilon$ reduces to:
\beq
\gamma_*\,\epsilon\,=\,-{q\over b^3}\,L^7\,r^2\,\sin\alpha\, \sin^2\theta \,\,
\left[\,1\,+\,{r\,\theta'\over b}\,\cot\theta\,+\,\Big(\cot\theta\,-\,{r\,\theta'\over b}\,\Big)\,
\Gamma_{38}\,\right]\epsilon\ .
\label{gamma_star_epsilon_unquenched}
\eeq

From the condition that $ \gamma_*$ acts diagonally  on $\epsilon$ (\ie, $ \gamma_* $ acts on $\epsilon$ as a matrix proportional to the unit matrix), we get the following equation for the embedding angle:
\beq
r\,\theta'\,=\,b\,\cot\theta\,\,.
\label{BPSeq_theta}
\eeq
Moreover, when (\ref{BPSeq_theta}) and the projections (\ref{D2-brane_projection}) and (\ref{internal_projections})  hold, $\gamma_{(7)}$ acts on $\epsilon$ as:
\beq
\gamma_{(7)}\,\epsilon\,=\,-d^7\,\zeta\,{q\over b^3}\,L^7\,r^2\,\sin\alpha\,\epsilon\,\,.
\eeq
Let us now study the terms in (\ref{Gamma_kappa_FF}) that are linear in the worldvolume gauge field $F$. Let us write these terms as:
\beq
F\wedge \gamma_{(5)} \Gamma_{11}\,=\,d^7\zeta\,\big[\,
\Gamma^{flux}+\Gamma^{Min}\,\big]\,\,,
\label{F_gamma5}
\eeq
where $\Gamma^{flux}$ contains the contributions of the components of $F$ along the internal directions and $\Gamma^{Min}$ is the contribution of the components of $F$ with legs along the Minkowski spacetime. It is readily verified that:
\beq
\Gamma^{flux}\,=\,
\gamma_{t\,x\,y}\,\Gamma_{11}\,\Big[\,\gamma_{\alpha\,\beta}\,F_{r\,\psi}\,-\,\gamma_{\alpha\psi}\,F_{r\,\beta}
\,+\,\gamma_{r\psi}\,F_{\alpha\,\beta}\,\Big]\,\,.
\label{Gamma_flux}
\eeq
The antisymmetric products of induced gamma matrices appearing on (\ref{Gamma_flux}) can be straightforwardly computed from (\ref{induced_gammas_newangles}):
\bear
&&\gamma_{\alpha\,\beta}\,=\,{q\over b^2}\,L^{2}\,\,\sin\alpha\,\sin\theta\,
\Big[\Gamma_{46}\,-\,\cot\theta\,\Gamma_{47}\,+\,
{\cot\alpha\over \sqrt{q}}\,\,\Gamma_{49}\,\Big]\,\,,\rc\rc
&&\gamma_{\alpha\psi}\,=\,{\sqrt{q}\over b^2}\,
L^{2}\,\sin\theta\,\Gamma_{49}\,\,,\rc\rc
&&\gamma_{r\psi}\,=\,{L^{2}\over b}\,{\sin\theta\over r}\,
\Big[\,\Gamma_{39}\,+\,{r\,\theta'\over b}\,\Gamma_{89}\,\Big]\,\,,
\label{Gamma_products_flux}
\eear
On the other hand, $\Gamma^{Min}$ is given by:
\beq
\Gamma^{Min}\,=\,L^2\,\Big(
E\,\gamma_{y r}\,-\,a_0'\,\gamma_{x y}\,+\,B\,\gamma_{t r}\,-\,a_y'\,
\gamma_{t x}\Big)\,\gamma_{\alpha\beta\psi}\,\Gamma_{11}\,\,.
\label{Gamma_Min}
\eeq
The products of the  induced Dirac matrices needed to compute $\Gamma^{Min}$ are:
\bear
&&\gamma_{y r}\,=\,L^2\,\Big[\Gamma_{23}\,+\,{r\over b}\,\theta'\,\Gamma_{28}\Big]
\,\,,\qquad\qquad
\gamma_{x y}\,=\,L^2\,r^2\,\Gamma_{12}\,\,,\rc\rc
&&\gamma_{t r}\,=\,L^2\,\Big[\Gamma_{03}\,+\,{r\over b}\,\theta'\,\Gamma_{08}\Big]
\,\,,\qquad\qquad
\gamma_{t x}\,=\,L^2\,r^2\,\Gamma_{01}\,\,,\rc\rc
&&\gamma_{\alpha\beta\gamma}\,=\,L^3\,{q\over b^3}\,\sin\alpha\sin^2\theta\,
\Big(\Gamma_{469}\,-\,\cot\theta\,\Gamma_{479}\Big)\,\,.
\label{Gamma_products_Min}
\eear
A quick inspection of the different terms appearing in $\Gamma^{flux}\epsilon$ and $\Gamma^{Min}\epsilon$  reveals that, after using the projections (\ref{D2-brane_projection}) and (\ref{internal_projections}), all terms contain products of $\Gamma$ matrices and there are no terms containing the unit matrix. Therefore, to implement the condition $\Gamma_{\kappa}\,\epsilon=-\epsilon$ we should require that
 $\Gamma^{flux}\epsilon=\Gamma^{Min}\epsilon=0$. By combining (\ref{Gamma_flux}) and (\ref{Gamma_products_flux}) we find that the product of $\Gamma$'s contained in  $\Gamma^{flux}\,\epsilon$ is:
\bear
&&{1\over L^4}\,\,\Big[\,\gamma_{\alpha\beta}\,F_{r\,\psi}\,-\,\gamma_{\alpha\psi}\,
F_{r\,\beta}
\,+\,\gamma_{r\psi}\,F_{\alpha\beta}\,\Big]\,\epsilon\,=\,
{\sin\alpha\,\sin\theta\over b}\,
\Big({q\over b}\,a'\,+\,{a\over r}\Big)
\,\Gamma_{46}\,\epsilon\,\rc\rc
&&\qquad\qquad\qquad\qquad\qquad
-{\sin\alpha\over b^2}\,
\Big(q\cos\theta\,a'\,+\,\sin\theta\,\theta'\,a\Big)\,
\Gamma_{47}\,\epsilon\,\,.
\eear
After using the equation (\ref{BPSeq_theta}) satisfied by the angle $\theta(r)$, we find that  $\Gamma^{flux}\epsilon=0$ if the flux function $a(r)$ satisfies the following first-order equation:
\beq
{a'\over a}\,=\,-{b\over q\,r}\,\,.
\label{BPSeq_a}
\eeq
When $E=B=0$ and $a_0'=a_y'=0$, Eqs.~(\ref{BPSeq_theta}) and (\ref{BPSeq_a}) guarantee that the embedding preserves two of the four supersymmetries of the background. If this is not the case, we should continue analyzing the remaining terms in $\Gamma_{\kappa}$. From (\ref{Gamma_Min}) and (\ref{Gamma_products_Min}) we get:
\bear
{1\over L^7}\,\Gamma^{Min}\,\epsilon&=&{q\over b^3}\,\sin\alpha\,\sin^2\theta\,
\left[E\left(\Gamma_{23}+{r\over b}\theta'\,\Gamma_{28} \right)+
B \left(\Gamma_{03}+{r\over b}\,\theta'\,\Gamma_{08}\right)-r^2\,a_0'\,\Gamma_{12}-
r^2\,a_y'\,\Gamma_{01}\right]\,\rc\rc
&&\qquad\qquad\qquad\qquad\qquad\qquad
\times \Big(\Gamma_{469}\,-\,\cot\theta\,\Gamma_{479}
\Big)\,\Gamma_{11}\,\epsilon\,\,.
\eear
After using the projections (\ref{internal_projections}) we can write the action of $\Gamma^{Min}$ on the Killing spinor $\epsilon$ as:
\bear
{1\over L^7}\,\Gamma^{Min}\,\epsilon &=& {q\over b^3}\,\sin\alpha\,\sin^2\theta\,
\Bigg[(E\Gamma_{2}+B\Gamma_{0})\left[1+{r\theta'\over b}\cot\theta+
\left(\cot\theta-{r\theta'\over b}\right)\,\Gamma_{38}\right]\Gamma_{11}\,\rc\rc
&&\qquad
+\Big(a_0'\,\Gamma_{2}\,-\,a_y'\,\Gamma_{0}\Big)\,
r^2\, (1+\cot\theta\,\Gamma_{38})\,\Gamma_{13}\,\Gamma_{11}\,\Bigg]\epsilon\,\,.
\eear
Using the BPS equation for $\theta$ (\ref{BPSeq_theta}), we can rewrite this last expression as:
\bear
{1\over L^7}\Gamma^{Min}\,\epsilon&=&{q\over b^3}\sin\alpha\,
\left[(E\Gamma_{2}+B\Gamma_{0})\Gamma_{11}+
\big(a_0'\Gamma_{2}-a_y'\Gamma_{0}\big)
r^2\sin^2\theta (1+\cot\theta\, \Gamma_{38})\,\Gamma_{13}\Gamma_{11}\right]\epsilon\,\,. \rc
\eear
To ensure that $\Gamma^{Min}\,\epsilon=0$ we first impose one of the following two extra projections on $\epsilon$:
\beq
\Gamma_{02}\,\epsilon\,=\,\pm\epsilon\,\,.
\label{extra_projection}
\eeq
Notice that the conditions (\ref{extra_projection}) are compatible with the projections (\ref{D2-brane_projection}) and (\ref{internal_projections}) that we have imposed so far. We get
\beq
{1\over L^7}\Gamma^{Min}\,\epsilon={q\over b^3}\sin\alpha\,
\big[(E\mp B)\Gamma_{2}\Gamma_{11}+\big(a_0'\pm a_y'\big)
r^2\sin^2\theta (1+\cot\theta\Gamma_{38})\,\Gamma_{213}\Gamma_{11}\big]\,\epsilon\,\,,
\eeq
and we have that  $\Gamma^{Min}\,\epsilon=0$ if $E$, $B$, $a_0'$, and $a_y'$   satisfy the following conditions:
\beq
E\,=\,\pm B\,\,,
\qquad\qquad
a_0'\,=\,\mp a_y'\,\,.
\label{BPS_EB}
\eeq
The two signs correspond to the two projections in (\ref{extra_projection}) (in section \ref{BPS-sol} we have chosen the upper signs). Therefore, after imposing these conditions, we have
\beq
F\wedge \gamma_{(5)}\,\epsilon\,=\,0\,\,.
\eeq
Notice that the extra projection (\ref{extra_projection}) is only needed if the worldvolume gauge field has components along the Minkowski directions. Furthermore, one can check that the BPS equations (\ref{BPSeq_theta}), (\ref{BPSeq_a}), and (\ref{BPS_EB}) and the projections (\ref{D2-brane_projection}), (\ref{internal_projections}), and (\ref{extra_projection}) imply that the remaning terms  in $\Gamma_{\kappa}$
act on $\epsilon$ as:
\bear
&&{1\over 2}\,F\wedge F\wedge  \gamma_{(3)}\,\epsilon\,=\,-
d^7\,\zeta\,\,{b\,L^7\over q}\,r^2\,a^2\,\sin\alpha\,\epsilon\,\,,\rc\rc
&&{1\over 6}\,
F\wedge F\wedge  F\wedge \gamma_{(1)}\,\Gamma_{11}
\,\epsilon\,=\,0\,\,.
\eear
It follows that:
\beq
d^7\,\zeta\,\Gamma_{\kappa}\,\epsilon_{|_{BPS}}\,=\,-
{d^7\,\zeta\over \sqrt{-\det(g+F)}{|_{BPS}}}\,{b\,L^7\over q}\,r^2\,
\left({q^2\over b^4}\,+\,a^2\right)\,\sin\alpha\,\epsilon_{|_{BPS}}\ ,
\eeq
and one can verify by computing the DBI determinant for the BPS configuration that, indeed, $\Gamma_{\kappa}\,\epsilon_{|_{BPS}}=-\epsilon_{|_{BPS}}$. 


\subsection{Fluctuations}
\label{Fluctuations}

To find the equations satisfied by the fluctuations at first order, we just compute the variation of the gauge field equations (\ref{eom_gauge_general}). One can check that the variation of $\det M$ is zero at first order and, as a consequence, the equations for the fluctuations are:
\beq
\partial_j\Big(e^{-\phi}\,\sqrt{-\det M}\,\delta{\cal J}^{ji}\Big)\,=\,\delta j^{i}\,\,.
\label{eom_fluct_gauge}
\eeq
We will restrict our attention to the case in which the only non-zero components of $\delta A$ are those along the Minkowski directions, 
\beq
\delta A\,=\,c_{\mu}(x^{\nu},r)\,dx^{\mu}\,\,.
\label{fluct_ansatz}
\eeq
Notice that in (\ref{fluct_ansatz}) we are assuming that the $c_{\mu}$'s do not depend on the internal angles. It is then easy to verify that, when the index $i$ corresponds to one of those internal directions,  the equation of motion (\ref{eom_fluct_gauge}) is satisfied automatically by the ansatz (\ref{fluct_ansatz}). Moreover, when $i=r$ this equation reduces to the following Lorentz condition:
\beq
-\partial_0\,c_0\,+\,\partial_1 c_1\,+\,\partial_2 c_2\,=\,0\,\,.
\label{Lorentz_trasn}
\eeq
Finally, when $i=\mu=0,1,2$, Eq.~(\ref{eom_fluct_gauge}) becomes:
\bear
&&{b\over q}\,\partial_r\,\Bigg(
{r^2\sin^2\theta\,\sqrt{q^2+b^4a^2}\over 
\sqrt{b^2\sin^2\theta+r^2(b^4\,a'^2\,+\,\sin^2\theta\,\theta'^{\,2})}}
\partial_r\,c^{\mu}\,\Bigg)\rc\rc
&&\qquad\qquad\qquad\qquad
+{1\over bq}\,{\sqrt{q^2+b^4a^2}\over r^2}\,
\sqrt{b^2\sin^2\theta+r^2(b^4\,a'^2\,+\,\sin^2\theta\,\theta'^{\,2})}\,
\partial^{\nu}\partial_{\nu}\,c^{\mu}\rc\rc
&&\qquad\qquad\qquad\qquad\
+{2b^2\over q}{2-q\over \eta+q}\,\big(\eta\cos\theta\,a'\,-\,a\sin\theta\theta')\,
\epsilon^{\mu\alpha\beta}\,\partial_{\alpha}\,c_{\beta}\,=\,0\,\,,
\label{eom_fluct_explicit}
\eear
where $c^{\mu}\,=\,\eta^{\mu\nu}\,c_{\nu}$ and, in our conventions, $\epsilon^{012}=1$. 
To solve these equations, let us separate variables in  $c_{\mu}(x^{\nu},r)$ as:
\beq
c_{\mu}(x^{\nu},r)\,=\,\xi_{\mu}\,e^{ik_{\nu}\,x^{\nu}}\,R(r)\,\,,
\qquad (\mu=0,1,2)\,\,,
\label{ansatz_a_fluct_c}
\eeq
where $\xi_{\mu}$ is a constant polarization vector. It follows immediately  from (\ref{Lorentz_trasn}) that this vector satisfies the transversality condition (\ref{transversality}).

In order to write the fluctuation equation for the radial function $R$ in a compact form, let us define the differential  operator ${\cal O}$, which acts on any function of the radial coordinate $R(r)$ as:
\bear
&&{\cal O}\,R\,\equiv\,{b\over q}\,
\partial_r\left[\,
{r^2\sin^2\theta\,\sqrt{q^2+b^4a^2}\over 
\sqrt{b^2\sin^2\theta+r^2(b^4\,a'^2\,+\,\sin^2\theta\,\theta'^{\,2})}}
\partial_r\,R
\right] \rc\rc
&&
\qquad\qquad\qquad\qquad\qquad
+{m^2\over bq}\,{\sqrt{q^2+b^4a^2}\over r^2}\,
\sqrt{b^2\sin^2\theta+r^2(b^4\,a'^2\,+\,\sin^2\theta\,\theta'^{\,2})}\,\,R \ ,
\qquad\qquad
\eear
where $m$ is the mass of the dual meson (see (\ref{mass_mesons})). 
We also define  the function $\Lambda(r)$ as:
\beq
\Lambda(r)\,\equiv\,
{2b^2\over q}{2-q\over \eta+q}\,\big(\eta\cos\theta\,a'\,-\,a\sin\theta\theta')\,\,.
\eeq
Then, the fluctuation equation can be written as:
\beq
\xi^{\mu}\,{\cal O}\,R\,+\,i\epsilon^{\mu\alpha\beta}\,
k_{\alpha}\,\xi_{\beta}\,\Lambda R\,=\,0\,\,.
\label{Minkowski_fluct_eqs}
\eeq
Moreover, by substituting the values of the functions $\theta(r)$ and $a(r)$ which correspond to a SUSY embedding (\ref{theta_SUSY}) and (\ref{a_SUSY}), we can greatly simplify the operator ${\cal O}$ and the function $\Lambda$. We get:
\bear
&&{\cal O}\,R\,=\,\partial_r\left[r^{2-2b}\big(r^{2b}-r_*^{2b}\big)\partial_r\,R\right]\,+\,
{m^2\over r^{2(3-b)}}\,\big(r^{2(2-b)}+(2-b)^2\,b^2\,Q^2\,r_*^{2(2-b)}\big)\,R\,\,,\rc\rc
&&\Lambda\,=\,2b\,(2-b)\,(4-3b)\,Q\,{r_*^2\over r^3}\,\,.
\label{calO_Lambda_explicit}
\eear
The three equations in  (\ref{Minkowski_fluct_eqs}) are coupled to each other. Let us see how they can be decoupled and reduced to a single ordinary differential equation. First of all, without loss of generality we pick the Minkowski momentum as $k^{\mu}\,=\,(\omega, k, 0)$ with  the  meson mass being  $m=\sqrt{\omega^2-k^2}$. The transverse polarization has been written in (\ref{polarization}) in terms of two unknown constants $\xi_1$ and $\xi_2$. For this parametrization  of $\xi_{\mu}$ one can
show that the equations for $\mu=0$ and $\mu=1$ in   (\ref{Minkowski_fluct_eqs})  are equivalent and that the remaining two equations are just:
\bear
&& \xi_1\,{\cal O}\,R\,+\,i\omega \xi_2\,\Lambda R\,=\,0\,\,,\rc\rc
&& \xi_2\,{\cal O}\,R\,-\,i\omega\left[1\,-\,{k^2\over \omega^2}\right]\,\xi_1\,
\Lambda R\,=\,0\,\,.
\eear
To decouple these equations, let us consider the complex combinations $\chi_{\pm}$ defined in (\ref{chi_pm}). Then, one can  straightforwardly show that the system (\ref{Minkowski_fluct_eqs}) can be reduced to the equations:
\beq
\chi_{+}\,\Big({\cal O}\,R\,+\,m\,\Lambda\,R\Big)\,=\,0\,\,,
\qquad\qquad
\chi_{-}\,\Big({\cal O}\,R\,-\,m\,\Lambda\,R\Big)\,=\,0\,\,.
\label{chi-equation}
\eeq
Obviously, $\chi_{\pm}$ can be eliminated from (\ref{chi-equation})  when they are non-vanishing  and the system can be reduced to two ordinary differential equations for the radial functions $R_{\pm}$, which can be written as:
\beq
\partial_r\,\big[r^{2-2b}(r^{2b}-r_*^{2b})\partial_r R_{\pm}\big]+
\Big[{m^2\over r^{2(3-b)}}\,\Big(r^{2(2-b)}+(2-b)^2\,b^2\,Q^2\,r_*^{2(2-b)}\Big)\,\pm\,
m\,\Lambda(r)\Big]R_{\pm}=0\,\,.
\label{R-eq}
\eeq
To find the mass spectrum we must compute the values of $m$ leading to a normalizable solution. This can be done numerically by the shooting technique.  We present these numerical results for the two types of modes  in Figs.~\ref{mass_spectrum} and \ref{mass_spectrum_unquenched}.

\begin{figure}
\centering
\label{fig: S'} 
\begin{subfigure}[b]{0.62\textwidth}
\centering
\includegraphics[width=\textwidth]{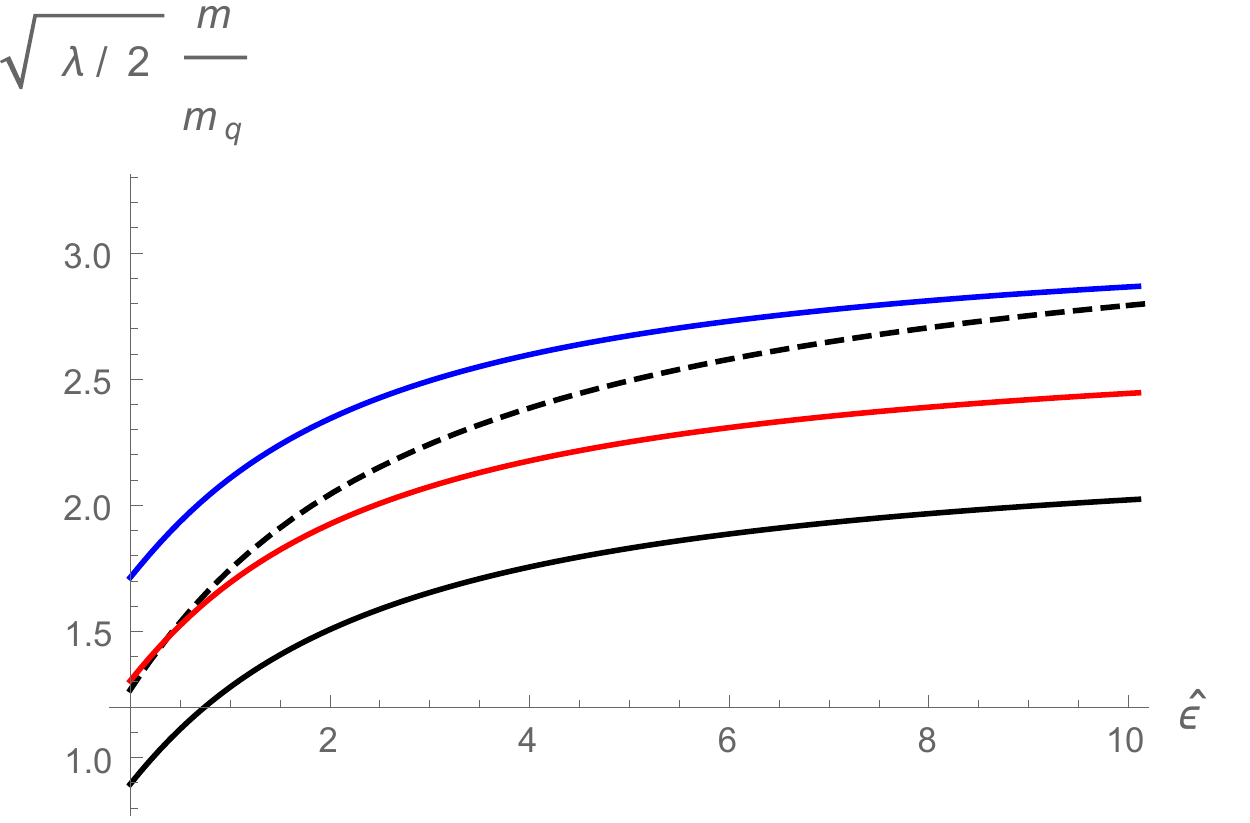}
\end{subfigure}
~~

~~

\begin{subfigure}[b]{0.62\textwidth}
\centering
\includegraphics[width=\textwidth]{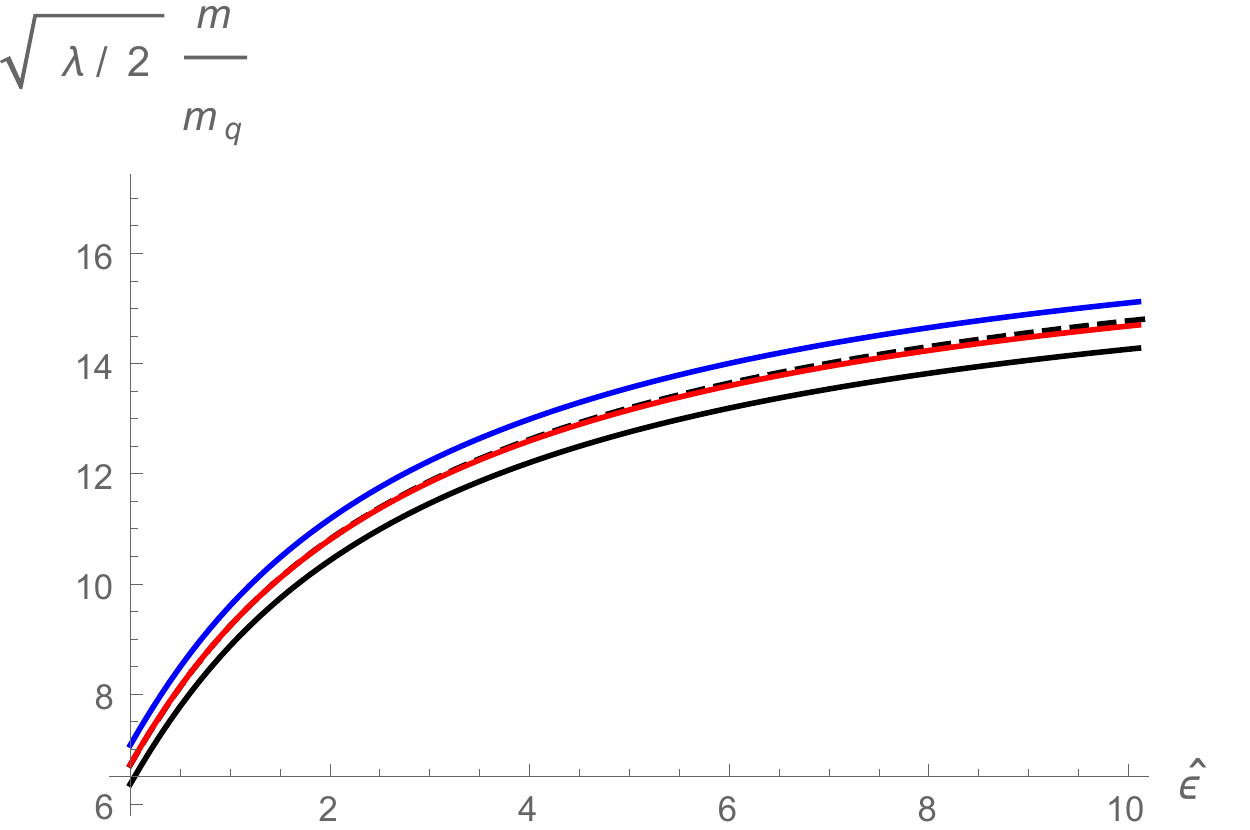}
\end{subfigure}
 \caption{Meson masses in the unquenched background as a function of the flavor deformation parameter $\hat \epsilon$ for $\sqrt\lambda M/N=1$ and two values of the excitation integer: $n=0$ (left) and $n=3$ (right).  The upper blue (lower black) curve corresponds to the mode $\chi_-$ ($\chi_+$). The intermediate red curve is the average of the two curves and the dashed black curve is the WKB estimate (\ref{WKB_masses_general}).}
 \label{mass_spectrum_unquenched}
\end{figure}


\subsubsection{WKB mass spectrum}

When the mass $m$ is large we can neglect the term containing the function $\Lambda$ in the fluctuation equation (\ref{R-eq}), and we can estimate the mass levels by using the WKB method developed in \cite{RS}. Indeed,  let us consider a differential equation of the form
\beq
\partial_{r}\,\big(\,f(r)\,\partial_{r}\,R\,\big)\,+\,
 m^2\,h(r)\,R\,=\,0\,\,,
\label{ODE}
\eeq
where $ m$ is the mass parameter and $f(r)$ and $h(r)$  are two
arbitrary functions that are independent of $ m$. We will assume that
near $r\approx r_*$ and $r\approx \infty$ these functions behave as:
\bear
&&f\approx f_1(r-r_*)^{s_1}\,\,,
\,\,\,\,\,\,\,\,\,\,\,\,\,\,
h\approx h_1(r-r_*)^{s_2}\,\,,
\,\,\,\,\,\,\,\,\,\,\,\,\,\,{\rm as}\,\,r\to r_*\,\,,\rc\rc
&&f\approx f_2\,r^{r_1}\,\,,
\,\,\,\,\,\,\,\,\,\,\,\,\,\,
h\approx h_2\, r^{r_2}\,\,,
\,\,\,\,\,\,\,\,\,\,\,\,\,\,{\rm as}\,\,r\to \infty\,\,,
\label{coeff_exp_wkb_def}
\eear
where $f_i$, $h_i$, $s_i$, and $r_i$ are constants. Then, the mass levels for large quantum number $n$ can be approximately written in terms of these constants as \cite{RS}:
\beq
 m^2_{WKB}\,=\,{\pi^2\over 
\xi^2}\,(n+1)\,\bigg(n\,+\,{|s_1-1|\over s_2-s_1+2}+{|r_1-1|\over r_1-r_2-2}
\bigg)\,\,,\quad\quad \quad (n\ge 0)\,\,,
\label{generallWKB}
\eeq
where $\xi$ is the following integral:
\beq
\xi\,=\,\int_{r_*}^{\infty} dr \,\,\sqrt{{h(r)\over f(r)}}\,\,.
\label{xi-WKBintegral}
\eeq
In our case  $f$ and $h$ are the functions:
\bear
&&f(r)\,=\,r^{2-2b}(r^{2b}-r_*^{2b})\,\,,\rc\rc
&&h(r)\,=\,{1\over r^{2(3-b)}}\,\Big(r^{2(2-b)}+
(2-b)^2\,b^2\,Q^2\,r_*^{2(2-b)}\Big)\,\,.
\eear
The behavior of these functions at $r=r_*$ is characterized by the following values of the coefficients and exponents defined in (\ref{coeff_exp_wkb_def}):
\bear
&&f_1\,=\,2\,b\,r_*\,\,,\qquad\qquad\qquad\qquad\qquad
 s_1\,=\,1\,\,,\rc\rc
&&h_1\,=\,{1+(2-b)^2\,b^2\,Q^2\over r_*^2}\,\,,\qquad\qquad\,\,
 s_2\,=\,0\,\,.
\eear
Similarly, for the behavior at large $r$ we obtain:
\bear
&&f_2\,=\,1\,\,,\qquad\qquad \qquad\qquad \qquad\qquad\qquad
r_1\,=\,2\,\,,\rc\rc
&&h_2\,=\,1\,\,,\qquad\qquad \qquad\qquad \qquad\qquad\qquad
  r_2\,=-2\,\,.
\eear
Therefore, the WKB mass spectrum is:
\beq
m_{WKB}\,=\,{\pi\over \sqrt{2}\,\xi(b,Q)}\,\sqrt{(n+1)(2n+1)}\,\,,
\label{WKB_masses_general}
\eeq
where $ \xi(b,Q)$ is the following integral:
\beq
\xi(b,Q)\,\equiv\,{1\over r_*}\,
\int_1^{\infty}\,{dz\over z^{2(2-b)}}\,
{\sqrt{z^{2(2-b)}\,+\,(2-b)^2\,b^2\,Q^2}\over
\sqrt{z^{2b}-1}}\,\,.
\eeq
By expanding in series the square root in the numerator and integrating term by term, we can express $ \xi(b,Q)$ as the following series:
\beq
\xi(b,Q)\,=\,-{1\over 4\,b\,r_*}\,
\sum_{p=0}^{\infty}\,(-1)^p\,
\big[(2-b)\,b\,Q\big]^{2p}\,
{\Gamma\Big(p-{1\over 2}\Big)\over  p!}\,\,
{\Gamma\Big({1+2p(2-b)\over 2b}\Big)\over 
\Gamma\Big({1+2p(2-b)\over 2b}+{1\over 2}\Big)}\,\,.
\eeq
Some particular values of the integral  $ \xi(b,Q)$ are:
\bear
&& \xi(b=1,Q)\,=\,{\pi\over 2 r_*}\,\,F\Big(-{1\over 2}, {1\over 2}; 1;-Q^2\,\Big)
\,\,,\rc\rc
&& \xi(b,Q=0)\,=\,{\sqrt{\pi}\over r_*}\,\,
{\Gamma\Big({2b+1\over 2b}\Big)\over 
\Gamma\Big({b+1\over 2b}\Big)}\,\,.
\eear
Interestingly, for $b=1$ and $Q=0$ (the unflavored model without internal flux) the WKB formula for the mass levels is exact. Moreover, for large $Q$ we can approximate $\xi(b,Q)$ as:
\beq
\xi(b,Q)\,\approx\,{(2-b)\,b\,Q\over r_*}\,
\int_1^{\infty}\,{dz\over z^{2(2-b)}
\sqrt{z^{2b}-1}
}\,=\,{\sqrt{\pi}\,Q\over r_*}\,{(2-b)\,b\over 3-b}\,
{\Gamma\Big({3+b\over 2b}\Big)\over 
\Gamma\Big({3\over 2b}\Big)}
\,\,.
\eeq
It follows that, for fixed quantum number $n$,  the WKB mass levels  for large $Q$ decrease as $1/Q$ according to the equation:
\beq
m_{WKB}\,\approx\,{\sqrt{\pi}\,r_*\over \sqrt{2}\,Q}\,{3-b\over (2-b)\,b}\,
{\Gamma\Big({3\over 2b}\Big)\over \Gamma\Big({3+b\over 2b}\Big)}\,
\sqrt{(n+1)(2n+1)}\,\,.
\eeq
In Fig.~\ref{mass_spectrum_unquenched} we compare the WKB estimates using (\ref{WKB_masses_general})  and the numerical results. The WKB method, however, is not valid at large values of $Q$, as it falls off the validity regime of \cite{RS}. Our numerical studies verified this expectation.

\end{subappendices}

\newpage


\chapter{Quantum phase transitions in the ABJM model}
\label{chapterfive}

\section{Introduction}

Quantum phase transitions are transitions that happen at zero temperature and that are induced by quantum fluctuations. They occur when some control parameters are varied and tuned to critical values, at which the ground state of the system undergoes a macroscopic rearrangement and the energy levels develop a non-analytic behavior on these parameters. Although, the quantum phase transitions occur at zero temperature, they determine the behavior of the system at low temperature in the so-called quantum critical regime, which is a region of the phase diagram surrounding the quantum critical point (see, for example \cite{Sachdev} \cite{Vojta} for reviews).

Strong coupling is a natal environment, where one expects quantum phase transitions. Therefore, a natural question is whether holography could be useful to search and characterize new types of quantum critical matter. Indeed, it is extremely important to develop new theoretical models which could shed light on the nature of quantum criticality and could serve to establish new paradigms to describe these phenomena. 

In the recent years different holographic models displaying quantum phase transitions have been studied in the literature (see, for example \cite{Karch:2007br,Jensen:2010vd,Jensen:2010ga,Evans:2010iy,Iqbal:2010eh,Iqbal:2011aj,D'Hoker:2012ih,Filev:2014mwa}). We are especially interested in top-down models, for which the field theory dual is clearly identified. In particular, we will deal with probe flavor D-branes in a gravitational background,  that corresponds, in the field theory side, to adding fields on the fundamental representation of the gauge group which act as charge carriers. When $N_f$ flavor D-branes are added to a geometry generated by $N_c$ color branes with $N_f\ll N_c$, we can use the probe approximation and neglect the backreaction of the flavor branes on the geometry. 
This precludes the fundamentals being dynamical and they are treated as quenched in the field theory.

The worldvolume dynamics of the flavor branes is governed by an action which has two pieces. The first one is the standard Dirac-Born-Infeld (DBI) action, which contains a gauge field. The other one is the Wess-Zumino (WZ) action which couples the brane to the Ramond-Ramond potentials of the background. The effects from the latter typically lead to far reaching consequences. In this probe brane setup it is rather simple to generate a configuration dual to a compressible state with non-zero charge density \cite{Apreda:2005yz,Kim:2006gp,Horigome:2006xu,Parnachev:2006ev,Kobayashi:2006sb}. Indeed, the charge density is dual to a radial electric field on the worldvolume. When the density is non-vanishing all consistent embeddings reach the horizon, \ie\  are black hole embeddings, whereas at zero  density there could also be Minkowski embeddings which always stay outside the horizon.\footnote{Suitable WZ terms would allow regular Minkowski embeddings even at non-zero charge density \cite{Bergman:2010gm,Jokela:2011eb}.}  It was shown in \cite{Karch:2007br} for the D3-D7 and D3-D5 systems that a quantum phase transition takes place at zero temperature at the point where the charge density vanishes, which corresponds to the chemical potential being equal to the quark mass. This phase transition is of second order and is realized in the holographic dual as a topology change of the embedding (from the black hole to Minkowski).  In \cite{Ammon:2012je} the critical exponents of the transition were found, corresponding to a non-relativistic scale invariant field theory with hyperscaling violation. These results were generalized in \cite{Itsios:2016ffv} to generic D$p$-D$(p+4)$ and D$p$-D$(p+2)$ intersections. 

The aim of this chapter is to study the quantum phase transitions of brane probes in the gravity dual of the ABJM Chern-Simons matter theory, especially in the Veneziano limit. Recall that the ABJM theory (explained in detail in section \ref{pureABJM}) is an $U(N)\times U(N)$ Chern-Simons gauge theory in $2+1$ dimensions with levels $(k,-k)$ and bifundamental fields transforming in the $(N, \bar N )$ and $(\bar N,  N )$ representations of the gauge group. When $N$ and $k$ are large the theory admits a supergravity description in the ten-dimensional type IIA theory. The corresponding geometry is of the form $AdS_4\times {\mathbb C}{\mathbb P}^3$ with fluxes.

The flavors in the ABJM theory are fields transforming in the fundamental representations $(N,1)$ and $(1,N)$ of the gauge group. In the holographic dual these flavors are introduced by means of D6-branes extended in $AdS_4$ and wrapping an ${\mathbb R}{\mathbb P}^3$ cycle inside the ${\mathbb C}{\mathbb P}^3$ internal manifold. In the probe approximation these holographic quarks have been reviewed in section \ref{ABJMquenchedflavor}. Moreover, by using the smearing technique (reviewed in section \ref{smearingtechnique}), when $N_f$ is large,  one can obtain simple analytic geometries encoding the effects of dynamical quarks in holography.  The geometry generated by the backreaction of massless flavors in ABJM has been obtained in \cite{Conde:2011sw} at zero temperature and generalized in \cite{Jokela:2012dw} to non-vanishing temperature. A detailed review on this solution was given in section \ref{Background}, together with some details in appendix \ref{appendixsusy4}. The backreaction affects the ABJM geometry  rather mildly since the metric differs from the unflavored one by constant squashing factors which depend on $N_f$. For massive quarks this construction was carried out in chapter \ref{massiveABJM}.

In this chapter we probe the ABJM background (with and without massless dynamical quarks included) with a flavor D6-brane corresponding to a massive quark. We study the dynamics of this probe at zero temperature and non-vanishing  charge density. This dynamics is governed by the DBI action, with the WZ term playing a fundamental role. 
We are interested in the phase structure of the system as the charge density is varied and, in particular, in analyzing the phase transition that occurs when the charge density is small.

The D6-brane embedding ansatz is a particular case of the one studied in section \ref{probes_with_flux}. Consequently, the equations of motion obtained in this chapter are a particular case of the ones of section \ref{probes_with_flux}. In chapter \ref{chapterfour} we only solve the equations for a concrete case. In this chapter, we solve the equations of motion for a different particular case.

We first study the probe in the unflavored ABJM background. Working at zero temperature, we find a continuous quantum phase transition at the point where the charge density vanishes. This transition is similar to the one that happens in the D$p$-D$q$ systems  in \cite{Ammon:2012je,Itsios:2016ffv} and corresponds to passing from a black hole to a Minkowski embedding. However, the scaling behavior of the probe near the critical point differs from the ones found in \cite{Ammon:2012je,Itsios:2016ffv}. Indeed, we find that the corresponding critical exponents are different and, in addition, our system displays multiplicative logarithmic corrections to the scaling behavior.

\begin{figure}[ht]
\center
 \includegraphics[width=0.50\textwidth]{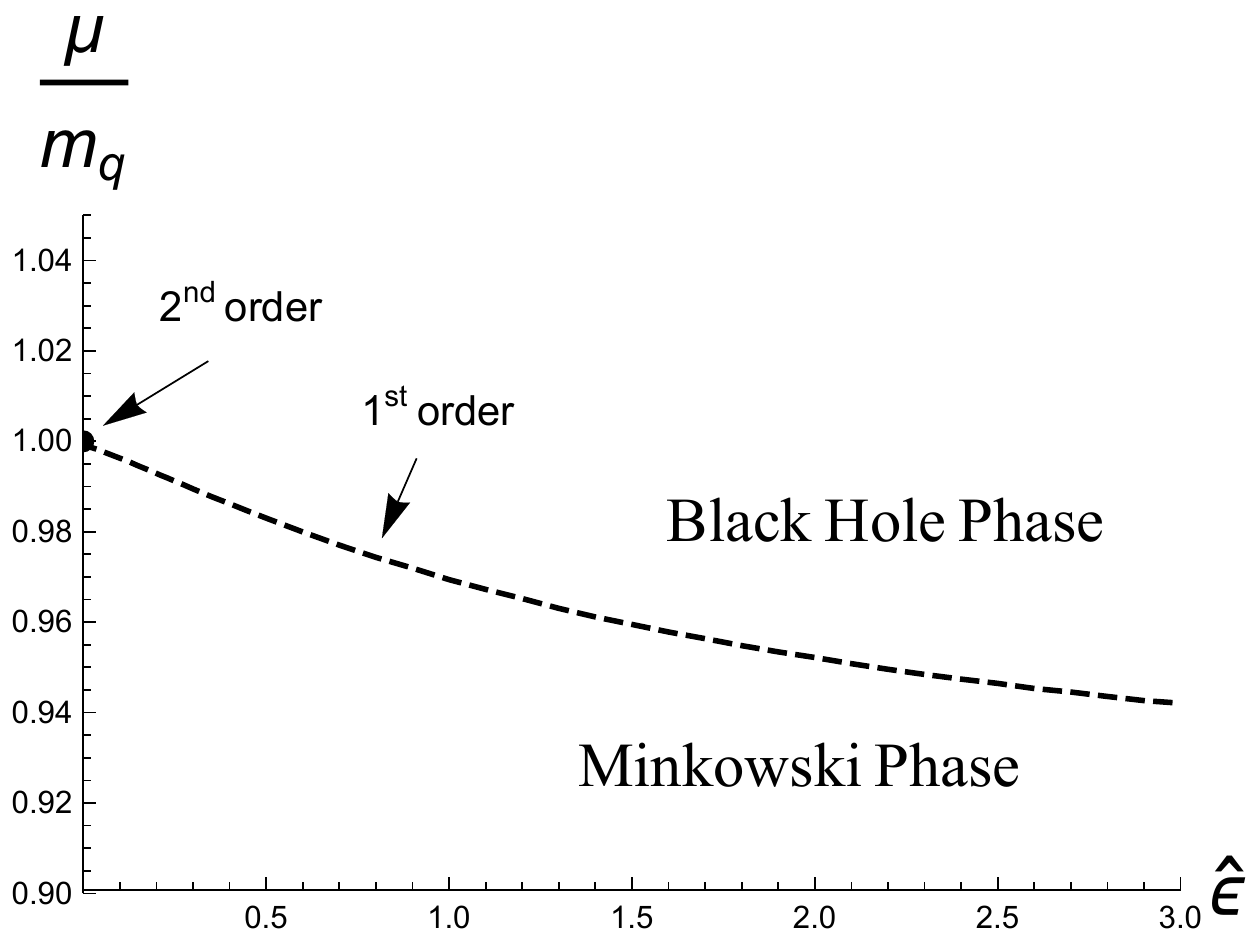}
  \caption{The phase diagram of the unquenched  ABJM model at zero temperature separates two different domains. At high enough chemical potential to quark mass ratios, the system is in the so-called black hole phase, which corresponds to a metallic behavior. The lower domain stands for Minkowski phase, where the system is gapped to charged excitations and resembles an insulating phase. The two domains are separated by a curve of first order phase transitions, whose location depends on the amount of flavor in the background $\hat \epsilon\propto N_f$ (see (\ref{epsilonhat0})) for a given chemical potential. The curve ends at the second order critical point in the quenched limit $N_f\to 0$. Interestingly, the corresponding critical exponents characterizing the continuous phase transition exhibit multiplicative logarithmic corrections.} 
  \label{fig:phasediagram}
\end{figure}

We also study the effects due to the presence of unquenched dynamical quarks in the background. In general, the inclusion of the flavor backreaction in holography is quite challenging. However, in the ABJM model the deformation of the geometry due to massless flavors seems quite mild and this gives us a unique opportunity to explore the different flavor effects. What we found below is that the influence on the phase transition of the unquenched case is not so moderate as their effects of the geometry could suggest. Indeed, we show below that the flavored black hole to Minkowski phase transition occurs at non-zero density and, moreover, it is of first order.  We have been able to compute  several quantities characterizing this discontinuous transition, such as its latent heat and the speed of sound close to the transition point.  \newline

The rest of this chapter is organized as follows. In section \ref{Probes} we study the embeddings of flavor D6-branes in the unquenched massless flavored ABJM background, both at zero and non-zero temperature. Section \ref{zeroT_thermo} is devoted to the analysis of the zero temperature thermodynamics and to explore  the quantum phase transitions in the unflavored and flavored cases. In section \ref{diffusion} we determine the charge susceptibility and diffusion constants at non-zero temperature. In section \ref{fluctuations} we analyze the fluctuations of the probe  and, in particular, we calculate the speed of its zero sound mode. In section \ref{summary} we summarize our results and discuss possible future research directions. The chapter is completed with appendix \ref{fluctuation_apendix} in which we carry out in detail the analysis of the fluctuations of the D6-brane.

\section{Probes on the flavored ABJM }
\label{Probes}

We are interested in analyzing the behavior of a flavor D6-brane probe in the background described in section \ref{Background} of chapter \ref{chapterfour} (for details see also appendix \ref{appendixsusy4}).  This flavor brane is extended along the $AdS_4$ coordinates $(x^{\mu}, r)$ and wraps a compact three-dimensional submanifold of the internal space.  Let us recall the expression (\ref{induced_metric_Hall}) of the induced metric on the worldvolume of the  flavor D6-brane:
\bear
 {ds^2_{7}\over L^2} &=& r^2\,\left[-h(r)\,dt^2+
 dx^2+dy^2\right]+
  {1\over r^2 }\,
\left({1\over h(r)}+{r^2\,\theta'^{\,2}\over b^2}\,\right)\,dr^2 \rc\rc
&&\qquad +\,{1\over b^2}\,\Big[
q\,d\alpha^2+q\,\sin^2\alpha \,d\beta^2+ \sin^2\,\theta\,\left(\,d\psi\,+\,\cos\alpha\,d\beta\,\right)^2\,\Big] \ ,
\label{induced_metric}
\eear
where $\alpha$, $\beta$, and $\psi$ are angles taking values in the range 
$0\le \alpha < \pi$, $0\le \beta, \psi<2\pi$, and $\theta=\theta(r)$ is an angle which determines the profile of the probe brane.  We want to deal with a system with non-zero baryonic charge density. Therefore, we should have a non-zero value of the $tr$ component of the worldvolume gauge field strength $F=dA$. Accordingly, we will adopt the  following  ansatz:
\beq
\theta\,=\,\theta(r)\,\,,
\qquad\qquad
A\,=\,L^2\,A_t(r)\,dt\,\,.
\label{unperturbed_theta_A}
\eeq
Recall that the D6-brane probe is governed by the standard DBI+WZ action:
\beq
S\,=\,S_{DBI}\,+\,S_{WZ}\,\,,
\eeq
where $S_{DBI}$ and $S_{WZ}$ are given by:
\bear
S_{DBI} & = & -T_{D6}\,\int_{{\cal M}_7}\,d^7\zeta\,e^{-\phi}\,
\sqrt{-\det (g+F)} ~, \rc
S_{WZ} & = & T_{D6}\int_{{\cal M}_7}\,\left(\hat C_7\,+\,\hat C_5\wedge F\,+\,{1\over 2}\,
\hat C_3\wedge F\wedge F\,+\,{1\over 6}\hat C_1\wedge F\wedge F\wedge F \right). ~~
\label{DBI_Wess_Zumino}
\eear
In (\ref{DBI_Wess_Zumino})   $g$ is the  induced metric on the worldvolume and the $\hat C_p$'s are the pullbacks of the different RR potentials of the background. In the flavored ABJM background $dF_2\not=0$ and, therefore, the RR potential $C_1$ is not well-defined. In this unquenched case one should work directly with the equations of motion of the probe derived from $S$, which contain the RR field strengths $F_p$  and do not contain the potentials (see appendix \ref{EOMs}). Nevertheless, to determine the embedding corresponding to the ansatz (\ref{unperturbed_theta_A}), only the term with $C_7$ in (\ref{DBI_Wess_Zumino}) is relevant (the explicit expression of $C_7$ can be found in \cite{Conde:2011sw,Jokela:2012dw}).

We will use the following system of worldvolume coordinates 
$\zeta^{a}\,=\,(x^{\mu}, r, \alpha, \beta, \psi)$. After integrating over the internal coordinates, we can write the action in the form:
\beq
S\,=\,\int d^3x\,dr\,{\cal L}\,\,,
\eeq
where ${\cal L}$ is the lagrangian density of the probe, given by:
\beq
{\cal L}\,=\,-{\cal N}\,r^2\,\sin\theta\,
\Big[\sqrt{b^2(1-A_t'^{\,2})+r^2\,h\,\theta'^{\,2}}-b\sin\theta\,-\,r\,\cos\theta\,\theta'\Big] \ .
\label{total_lag_density}
\eeq
Here and in the following, the prime denotes differentiation with respect to $r$. In (\ref{total_lag_density})   ${\cal N}$ is  a constant given by:
\beq
{\cal N}\,=\,{8\pi^2\,L^7\,T_{D6}\,e^{-\phi}\over b^4}\,q\,\,.
\label{calN_def}
\eeq

Notice that the ansatz \ref{unperturbed_theta_A} is a particular case of the ansatz for the embedding of the D6-brane considered in section \ref{probes_with_flux} of chapter \ref{chapterfour}, for $E=B=a_x(r)=a_y(r)=a(r)=0$. Consequently, the equations of motion obtained from the lagrangian \ref{total_lag_density} are a particular case of the equations of motion in section \ref{probes_with_flux}.

In the lagrangian (\ref{total_lag_density})
the variable $A_t$ is cyclic and its equation of motion can be integrated once as:
\beq
r^2\,\sin\theta\,{A_t'\over 
\sqrt{1-A_t'^{\,2}+{r^2\over b^2}\,h\,\theta'^{\,2}}}\,=\,d\,\,,
\eeq
where $d$ is a constant, which is proportional to the charge density. This equation can be inverted to give:
\beq
A_t'\,=\, {d\over b}{\sqrt{b^2+r^2\,h\,\theta'^{\,2}}\over \,\sqrt{d^2+r^4\sin^2\theta}}\,\,.
\label{At_prime}
\eeq
According to the standard AdS/CFT dictionary  the chemical potential $\mu$ is identified with the value of $A_t$ at the UV:
\beq
\mu\,=\,A_t(r\to\infty)\,\,.
\label{mu_At_UV}
\eeq
For a black hole embedding one can write an expression for $A_t$ as an integral over the radial variable $r$. Indeed, in this case we integrate (\ref{At_prime}) with the condition $A_t(r=r_h)=0$, namely:
\beq
A_t(r)\,=\,{d\over b}\,\int_{r_h}^{r}\,{\sqrt{b^2+\tilde r^2\,h\,\theta'^{\,2}}\over \,\sqrt{d^2+\tilde r^4\sin^2\theta}}
\,\,d\tilde r\,\,.
\eeq
Then, it follows that the chemical potential $\mu$ for a black hole embedding is:
\beq
\mu\,=\,{d\over b}\,\int_{r_h}^{\infty}\,{\sqrt{b^2+r^2\,h\,\theta'^{\,2}}\over \,\sqrt{d^2+r^4\sin^2\theta}}
\,dr\,\,.
\label{chemical_pot}
\eeq

Let us now write the equation of motion for $\theta(r)$:
\beq
\partial_r\,\Bigg[
{r^4\,h\,\sin\theta\over \sqrt{1-A_t'^{\,2}+{r^2\over b^2}\,h\,\theta'^{\,2}}}\,\,\theta'\Bigg]\,-\,
b\,r^2\,\cos\theta\Bigg[(3-2b)\sin\theta\,+\,b\,\sqrt{1-A_t'^{\,2}+{r^2\over b^2}\,h\,\theta'^{\,2}}
\,\Bigg]\,=\,0\,\,.
\label{eom_theta_At}
\eeq
Using (\ref{At_prime}) to eliminate $A_t'$, we can rewrite (\ref{eom_theta_At}) as:
\beq
\partial_r\,\Bigg[ 
{r^2\,h\,\sqrt{d^2+r^4\sin^2\theta}\over 
\sqrt{b^2+r^2\,h\,\theta'^{\,2}}}\,\theta'\Bigg]\,-\,\,r^2\cos\theta\sin\theta\,
\Bigg[3-2b+{r^2\,\sqrt{b^2+r^2\,h\,\theta'^{\,2}}\over
\sqrt{d^2+r^4\sin^2\theta}}\Bigg]\,=\,0\,\,.
\label{eom_theta_d}
\eeq

Eq. (\ref{eom_theta_d}) must be solved numerically, except in the case of vanishing temperature and density, where an analytic supersymmetric solution is available \cite{Conde:2011sw}.   All solutions of (\ref{eom_theta_d}) reach the UV with an angle which approaches asymptotically the value $\theta=\pi/2$. Actually, for large $r$ the deviation of $\theta$ with respect to this asymptotic value can be represented as:
\beq
{\pi\over 2}-\theta(r)\sim {m\over r^b}\,+\,{c\over r^{3-b}}\,+\,\cdots\,\,,
\label{asymptotic}
\eeq
where $b$ is the constant (depending of the flavor deformation parameter $\hat\epsilon$) defined in (\ref{b_new}) and $m$ and $c$ are constants related to the quark mass and the condensate, respectively. The precise holographic dictionary for our probes has been worked out in \cite{Jokela:2012dw}. For our purposes it is sufficient to recall that the physical quark mass $m_q$ is proportional to $m^{{1\over b}}$. This non-trivial exponent is related to the anomalous mass dimension $\gamma_m=b-1$, which enters in the (holographic) Callan-Zymanzik equation \cite{Jokela:2013qya}.

\begin{figure}[ht]
\center
 \includegraphics[width=0.60\textwidth]{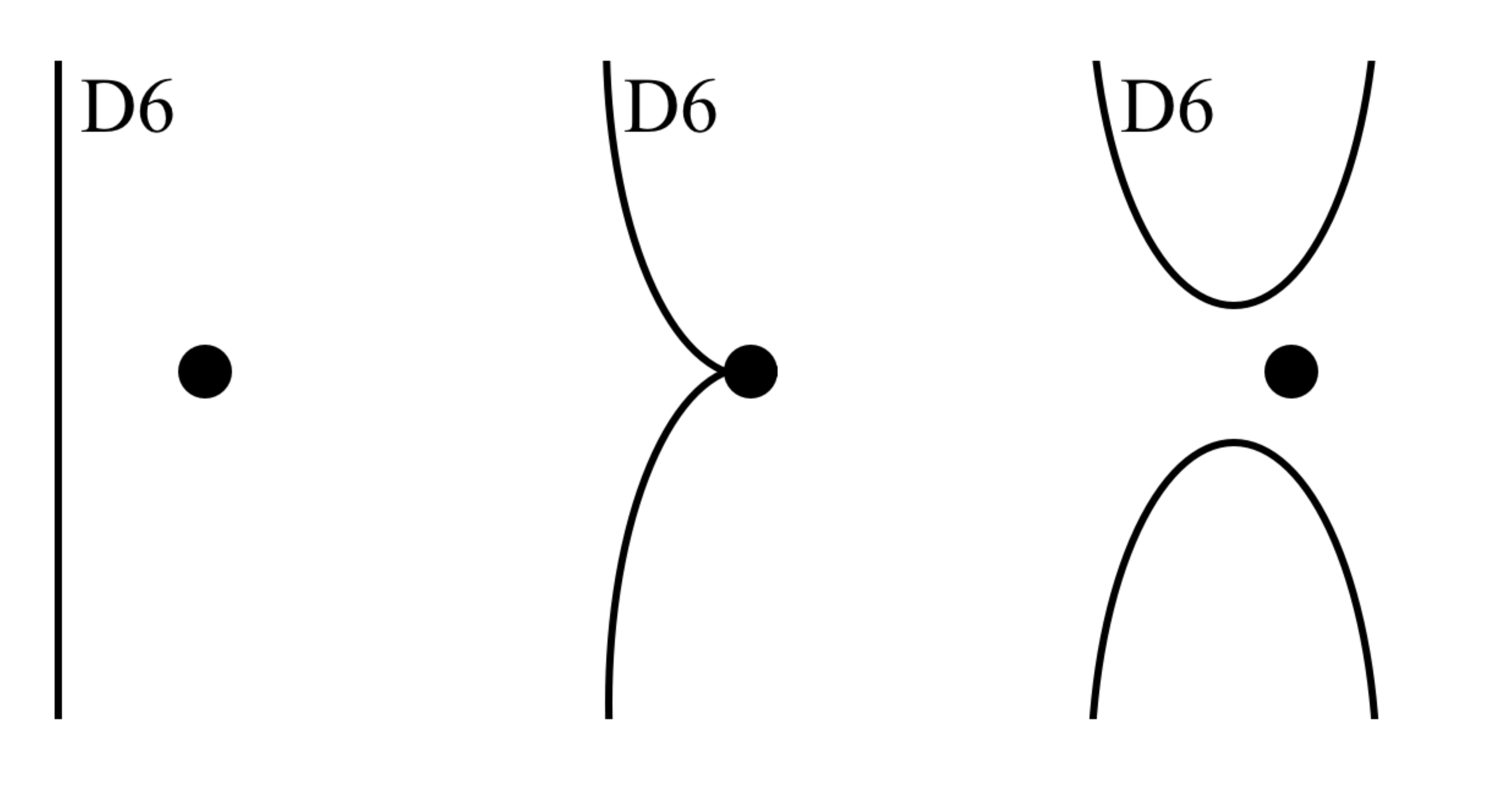}
  \caption{We sketch the three possible embeddings available in the model at non-zero chemical potentials at non-vanishing mass parameter at zero temperature. The left-most profile corresponds to Minkowski embeddings, where the D6-brane does not enter the Poincar\'e horizon, displayed as the black dot. The middle profile corresponds to that of a black hole embedding penetrating the horizon, while the right-most profile stands for D6-anti-D6-brane embeddings. This figure is adapted from the one in \cite{Karch:2007br}, in the context of a D3-D7 model, where a clean flat space interpretation can be given.} 
  \label{embeddings2}
\end{figure}

The different solutions of (\ref{eom_theta_d}) are obtained by imposing suitable boundary conditions  at the  IR. We will study them in the next two subsections, starting with the embeddings at zero temperature.  
There are three different kinds of embeddings, sketched in Fig.~\ref{embeddings2}. They are introduced one-by-one in the following subsection.

\subsection{Embeddings at zero temperature}
\label{zeroT_embeddings}

\begin{figure}[ht]
\center
 \includegraphics[width=0.50\textwidth]{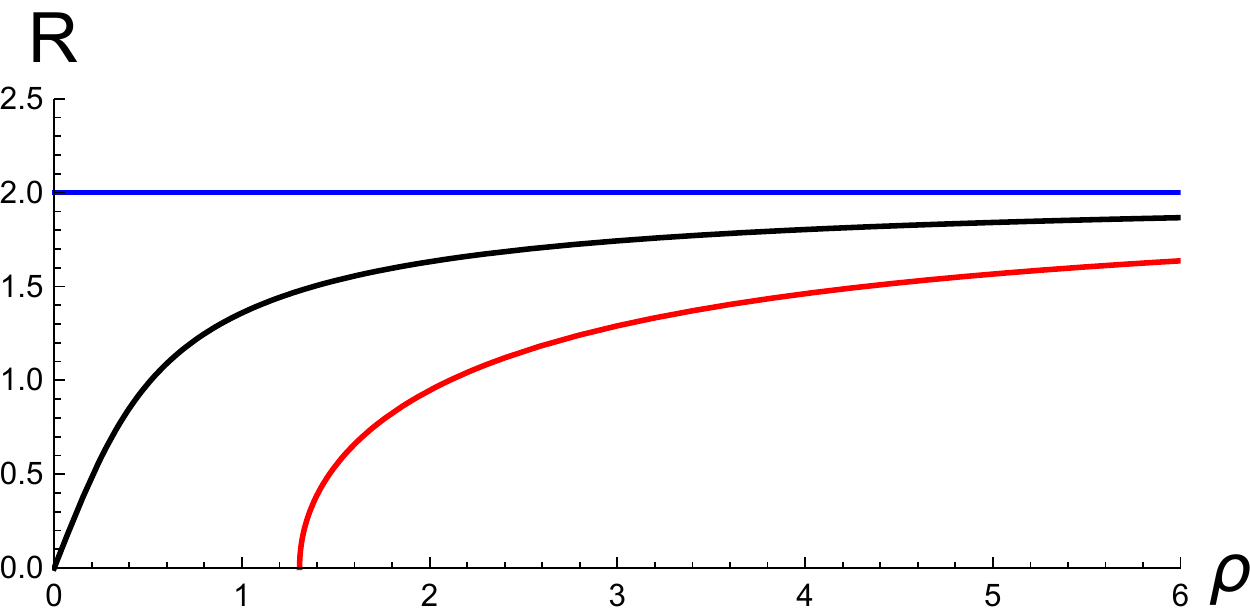}
  \caption{We depict the profiles of the three possible types of embeddings in the $(\rho, R)$ coordinates defined in
  (\ref{R_rho_def}) at zero temperature. The top-most curve corresponds to the Minkowski embedding, while the middle curve entering in the Poincar\'e horizon corresponds to the black hole embedding. The bottom curve stands for one of branches of the brane-antibrane embeddings.} 
  \label{embeddings}
\end{figure}


Let us now consider eq. (\ref{eom_theta_d}) for $T=0$ (\ie\ for $h=1$). One can verify by numerical integration that  (\ref{eom_theta_d}) admits a family of solutions in which the embeddings reach the origin $r=0$ at any given value of $\theta_0=\theta(r=0)$, quantities which we shall denote as initial angles.  These solutions are called black hole embeddings as they are continuously connected with their $T\ne 0$ counterparts.  Actually, one can solve (\ref{eom_theta_d}) for $h=1$ in a power series expansion near $r=0$ as:
\beq
\theta(r)=\theta_0\,+\,b(3-2b)\,{\sin\theta_0\cos\theta_0\over 6d}\,r^2\,+\,\cdots\,\,.
\eeq
These solutions can be found numerically by imposing the initial conditions $\theta(r=0)=\theta_0$ and $\theta'(r=0)=0$. The mass parameter $m$ of the embedding (determined by the value of $r^b\cos\theta$ at $r\to\infty$) is related to the initial angle $\theta_0$.  Given the embedding, the chemical potential  can be obtained  by evaluating  the integral (\ref{chemical_pot}). When $\theta_0\to \pi/2$ the mass approaches zero. In fact, the whole embedding becomes trivial with constant angle. When $\theta_0\to 0$, on the other hand, the embedding becomes increasingly spiky and the corresponding chemical potential approaches the value:
\beq
\lim_{\theta_0\to 0}\,\mu\,=\,m^{{1\over b}} \ .
\label{mu_for_zero_theta}
\eeq

\begin{figure}[ht]
\center
 \includegraphics[width=0.50\textwidth]{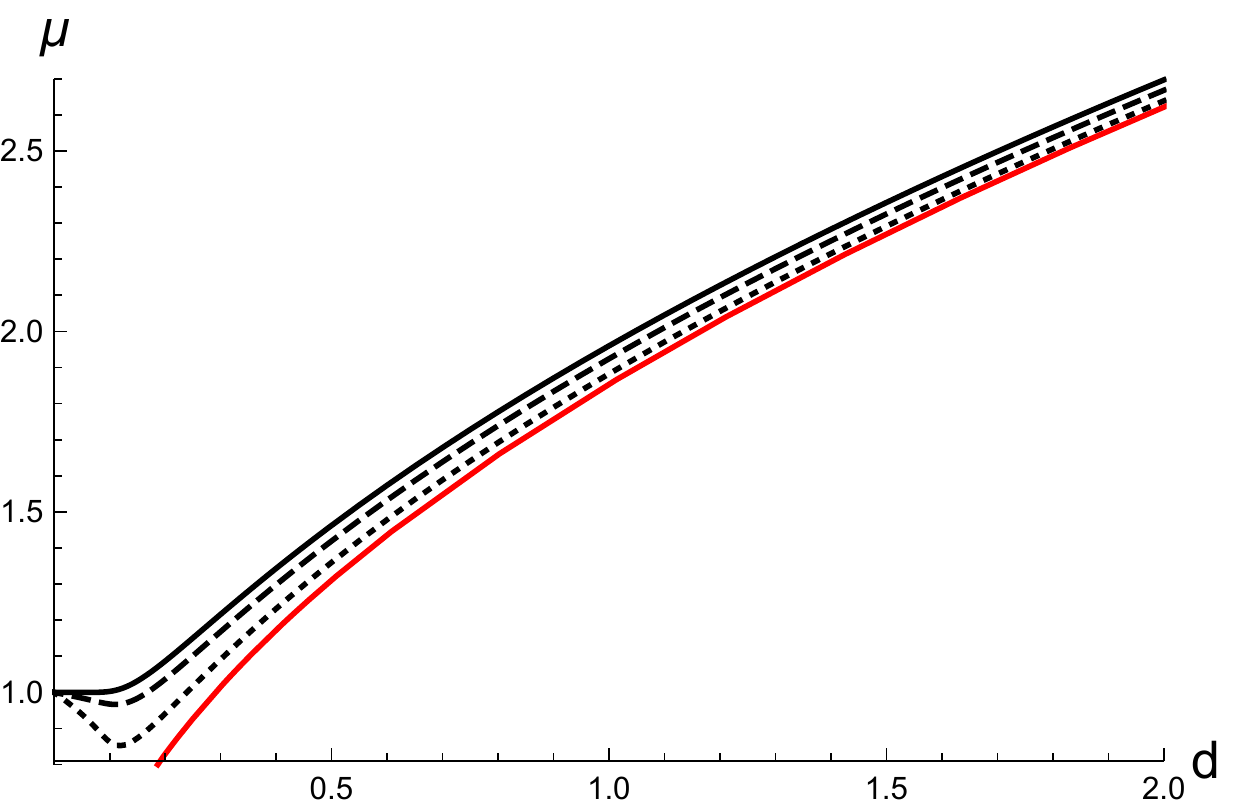}
  \caption{We plot the chemical potential $\mu$ as a function of the density $d$ for fixed quark mass. The continuous black curve corresponds to the unflavored case ($b=1)$, the dashed curve is for $b=1.1$, while the dotted curve is for $b=1.25$ (corresponding to $\hat\epsilon\to\infty$).  All curves are for $m=1$. The  continuous red curve corresponds to the conformal D3-D5 system with massless quarks, for which $\mu=\gamma\, d^{{1\over 2}}$, with 
 $\gamma={1\over 4\sqrt{\pi}}\,\Gamma(1/4)^2$. }
\label{plot_chemical_pot}
\end{figure}

We have verified (\ref{mu_for_zero_theta}) numerically. This result can also be easily demonstrated analytically as follows. Let us first introduce the Cartesian-like coordinates $(\rho, R)$, related to $(\theta, r)$ as:
\beq
R\,=\,r^{b}\,\cos\theta\,\,,
\qquad\qquad
\rho\,=\,r^{b}\,\sin\theta\,\,.
\label{R_rho_def}
\eeq
In these coordinates the black hole embeddings start in the IR at the origin $R=\rho=0$ with a certain angle $\theta_0$ with respect to the $R$-axis and they end at the UV at $R=m$ with $\rho\to\infty$ (see Fig.~\ref{embeddings}).  If the initial angle $\theta_0$ is very small, the embeddings are very spiky and approach the maximal value $R=m$ very fast  for very small values of the coordinate $\rho$. Instead of parameterizing the embedding as $\theta=\theta(r)$, it is more convenient in this situation to represent it as $\rho=\rho(R)$. It is then straightforward to demonstrate that $\mu$ is given by the integral:
\beq
\mu\,=\,{d\over b}\,\int_0^{m}\,
{\sqrt{1+({d\rho\over dR})^2}\over 
\sqrt{d^2(\rho^2+R^2)^{1-{1\over b}}+\rho^2 (\rho^2+R^2)^{{1\over b}}} }\,\,
dR\,\,.
\label{mu_R_rho}
\eeq
For $\theta_0\to 0$ the coordinate $\rho$ is very close to zero except when $R\approx m$ and we can approximate the integral (\ref{mu_R_rho}) by taking $\rho\approx 0$ in the integrand. We get:
\beq
\mu\,\approx\,{1\over b}\,\int_{0}^{m} R^{{1\over b}-1}\,\,dR\,=\,m^{{1\over b}}\,\,,
\eeq
in agreement with (\ref{mu_for_zero_theta}). For a fixed value of the mass parameter $m$, the limiting value (\ref{mu_for_zero_theta})  corresponds to sending $d\to 0$. Actually, the dependence of $\mu$ on $d$ for fixed $m$ can be obtained numerically by performing the integral (\ref{chemical_pot}). The result is shown in Fig.~\ref{plot_chemical_pot}, where we notice an important difference between the unflavored and flavored cases. Indeed, when $N_f=0$ the chemical potential $\mu$ grows monotonically with $d$, starting from its minimal value $\mu=m$ at $d=0$.  When $d$ is large the chemical potential grows as $\mu\propto d^{{1\over 2}}$, which is the behavior expected in a conformal theory in 2+1 dimensions.  On the contrary, when the backreaction  of the flavors is added, $\mu$ decreases for small values of $d$ until it reaches a minimum at a non-zero value of $d$ and then it grows and converges eventually to the unflavored case. The presence of minimum in the $\mu=\mu(d)$ curve means that the charge susceptibility $\chi=\partial d/\partial \mu$ diverges at $d\not=0$, signaling a discontinuous phase transition at a non-zero density. We will confirm this fact below.

The black hole embeddings considered above are not the only possible ones. Indeed, there are also two other configurations in which the brane does not reach the $r=0$ origin. The so-called brane-antibrane embeddings are characterized by the initial boundary conditions:
\beq
\theta(r_0)\,=\,{\pi\over 2}\,\,,
\qquad\qquad
\theta'(r_0)\,=\,\infty\,\,,
\eeq
where $r_0$ is the minimal value of $r$. In terms of the $(\rho, R)$ variables the brane is orthogonal to the $\rho$-axis in the IR (at $\rho=\rho_0=r_0^b$, $R=0$) and becomes parallel to the $\rho$-axis as $\rho$ becomes large (see Fig.~\ref{embeddings}). Notice that $dR/d\rho$ diverges at $\rho=\rho_0$, which indicates that the brane has a turn-around point where the brane jumps to a second branch.

A third class of configurations are the so-called Minkowski embeddings, in which the brane reaches the $R$-axis at some non-zero value of $R$, as shown in Fig.~\ref{embeddings}. Due to charge conservation these embeddings are  only consistent if the density $d$ is zero. When this is the case there are analytic solutions which preserve some amount of supersymmetry \cite{Conde:2011sw}.  In terms of the $(r, \theta)$ variables, these embedding are:
\beq
\cos\theta(r)\,=\,{m\over r^b}\,\,,
\qquad\qquad
(d=0)\,\,.
\label{SUSY_emb}
\eeq
Equivalently $R=m$. Notice that in this case the minimal value of $r$ is $r_0=m^{{1\over b}}$. Moreover, when $d$ vanishes it follows from (\ref{At_prime}) that $A_t'=0$ and, therefore, the gauge field $A_t$ is an arbitrary constant, which equals the chemical potential $\mu$.  Thus, the SUSY embeddings (\ref{SUSY_emb})  correspond to $d=0$, with $\mu$ being a free parameter. 

Notice that, in this zero temperature case, the mass parameter $m$ can be scaled out by a suitable change of the radial variable followed by some redefinitions. Indeed, from (\ref{asymptotic}) we conclude that $m$ can be taken to be one if one changes variables from $r$ to $\tilde r=r/m^{{1\over b}}$. Then, it follows from (\ref{chemical_pot}) that $m$ can be eliminated from this  last equation if $d$ and $\mu$ are written in terms of the rescaled quantities $\tilde d$ and $\tilde \mu$, defined as $\tilde d =d/m^{{2\over b}}$ and  $\tilde\mu=\mu/m^{{1\over b}}$.

In section \ref{zeroT_thermo} we will determine which of these three types of embeddings at zero temperature is thermodynamically favored. We will carry out this analysis by comparing their thermodynamic potentials $\Omega$ in the grand canonical ensemble.

\subsection{Embeddings at finite temperature}\label{finiteT_embeddings}

As will become clear later, we need to extend some of our analysis to small and non-zero temperature. All three types of embeddings, as discussed in the preceding section, extend continuously to $T\ne 0$. However, as our main motivation in this work are the quantum critical phenomena, we will restrict our attention in the black hole phase.
Let us thus only consider the black hole embeddings at non-zero temperature. These embeddings reach the horizon $r=r_h$ with some angle $\theta=\theta_0$. Near $r=r_h$ we can solve (\ref{eom_theta_d}) in powers of $r-r_h$. The first two terms in this expansion are:
\beq
\theta(r)\,=\,\theta_0\,+\,\theta_1\,(r-r_h)\,+\,\cdots\,\,,
\label{expansion_theta_nh}
\eeq
where the constant $\theta_1$ is given by:
\beq
\theta_1\,=\,b\,
{\cos\theta_0\,\sin\theta_0\,\big[b\,r_h^2\,+\,(3-2b)\,\sqrt{d^2+r_h^4\,\sin^2\theta_0}\big]\over
3(d^2\,+\,r_h^4\,\sin^2\theta_0)}\,\,r_h\,\,.
\label{theta_nh}
\eeq
To get the full $\theta(r)$ function we need to integrate numerically (\ref{eom_theta_d}) with the initial condition at $r=r_h$ given by (\ref{theta_nh}). Notice that (\ref{eom_theta_d}) depends explicitly on $r_h$ through the blackening factor $h$. It turns out that the horizon radius $r_h$ can be scaled out by an appropriate change of variables followed by a redefinition of the density $d$. Indeed, let us define the reduced variable $\hat r$ and  density $\hat d$ as:
\beq
\hat r\,=\,{r\over r_h}\,\,,
\qquad\qquad
\hat d\,=\,{d\over r_h^2}\,\,.
\label{hat_r_d}
\eeq
Then, it is readily verified that the embedding equation in terms of $\hat r$ is just (\ref{eom_theta_d}) with $r_h=1$ and $d$ substituted by $\hat d$. Other quantities can be similarly rescaled. Indeed, let us define $\hat \mu$ and $\hat m$ as:
\beq
\hat\mu\,=\,{\mu\over r_h}\,\,,
\qquad\qquad
\hat m\,=\,{m\over r_h^{b}}\,\,.
\eeq
It is straightforward to find an expression of $\hat\mu$ in terms of the rescaled quantities:
\beq
\hat \mu\,=\,{\hat d\over b}\,\int_{1}^{\infty}\,{\sqrt{b^2+\hat r^2\,h (\hat r)\,\big({d\theta\over d\hat r}\big)^{\,2}}
\over \,\sqrt{\hat d^2+\hat r^4\sin^2\theta}}
\,d\hat r\,\,.
\label{hat_chemical_pot}
\eeq
Notice also that the ratio ${\hat m}^{1\over b}/\hat \mu$ does not depend on $r_h$:
\beq
{{\hat m}^{1\over b}\over \hat \mu}\,=\,
{{ m}^{1\over b}\over  \mu}\,\,.
\eeq

\section{Zero temperature thermodynamics}
\label{zeroT_thermo}

The zero-temperature grand canonical potential $\Omega$ is given by minus the on-shell action of the probe brane:
\beq
\Omega\,=\,-S_{on-shell}\,\,.
\eeq
Notice that, as pointed out in \cite{Jokela:2012dw}, the on-shell action of our ABJM system is finite and does not need to be regulated. Indeed, the WZ term of the action serves as a regulator of the DBI term, giving rise to consistent thermodynamics. The explicit expression of $\Omega$ at zero temperature is given by:
\beq
\Omega\,=\,{\cal N}\,\int_{r_0}^{\infty}\,r^2\,\sin\theta\,\
\Bigg[{r^2\sin\theta\,\sqrt{b^2+r^2\,\theta'^2}\over \sqrt{d^2+r^4\sin^2\theta}}\,-\,
b\sin\theta\,-\,r\cos\theta\,\theta'\Bigg]dr\,\,.
\label{Omega_integral}
\eeq
Other thermodynamic properties at $T=0$ can be obtained from (\ref{Omega_integral}). For example, the pressure $P$ is just:
\beq
P\,=\,-\Omega\,\,.
\label{P_Omega}
\eeq
Moreover,  we can evaluate $\Omega$ for the different embeddings and determine the one that is favored at different values of the chemical potential. One can verify by plugging (\ref{SUSY_emb}) in (\ref{Omega_integral})  that $\Omega=0$ for the SUSY embeddings (\ref{SUSY_emb})  which have  zero density $d$ and arbitrary $\mu$. In the case of the black hole embeddings the situation varies greatly when the backreaction is included. Indeed, for the unflavored background with $b=1$ the grand canonical potential of the black hole embeddings is always negative and grows monotonically as $\mu$ decreases towards its minimal value $\mu\searrow m$, where $\Omega=0$ and $d=0$ (see Fig.~\ref{Omega_mu}, left). On the contrary, in the flavored backgrounds with $b>1$,  the grand canonical potential is negative for large values of $\mu$ and vanishes for some $\mu=\mu_c$ which corresponds to a non-zero density $d=d_c$ (see Fig.~\ref{Omega_mu}, right). From this point on, $\Omega\ge 0$, reaching a maximum  positive value, which corresponds to the minimum value of the chemical potential $\mu$. It is at this point where the black hole embedding ceases to exist as it annihilates with another (unstable) black hole embedding. This latter black hole branch is the one which connects with the Minkowski embeddings at larger $mu$, \ie\ until the grand potential reaches the value $\Omega=0$ when $\mu=m^{{1\over b}}$ and $d=0$.  The grand canonical potential for the brane-antibrane embeddings is always non-negative and decreases monotonically as $\mu$ grows ($\mu\le m^{{1\over b}}$ for these embeddings). This structure in the $(\mu,\Omega)$ plane is the well-known swallow-tail shape, typical of first-order phase transitions.

\begin{figure}[ht]
\center
 \includegraphics[width=0.40\textwidth]{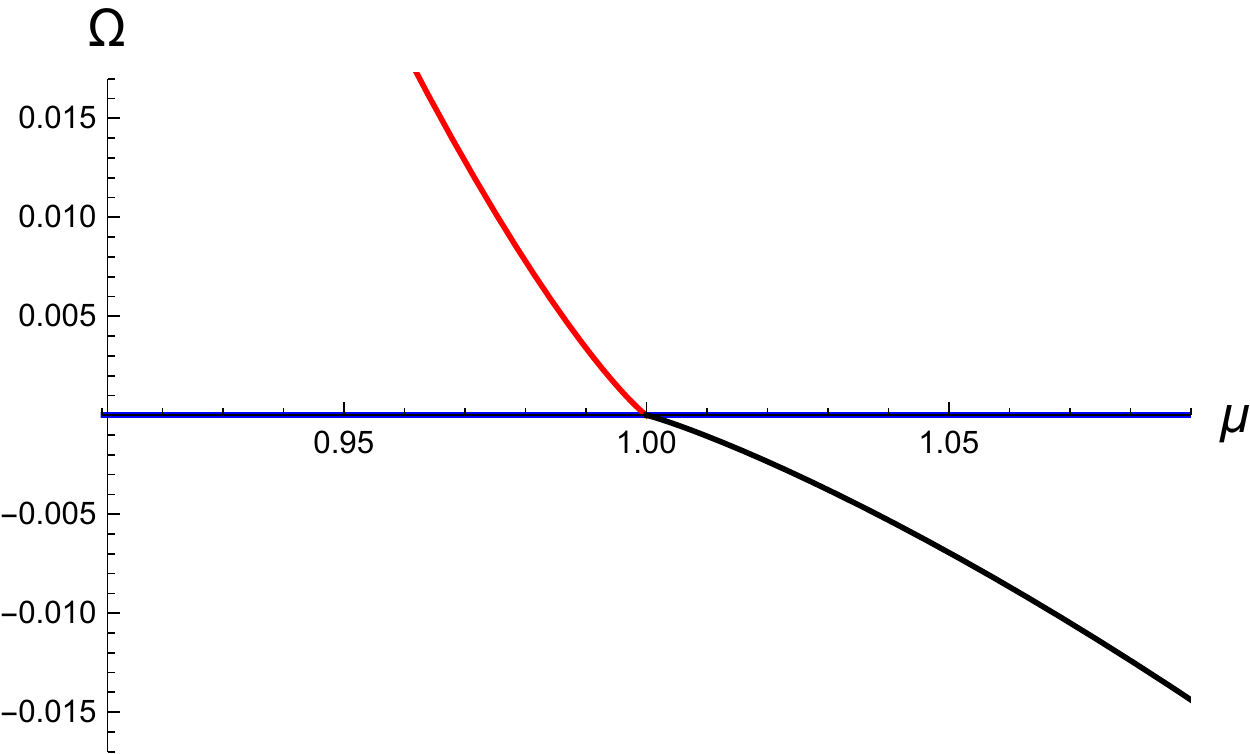}
 \qquad\qquad
  \includegraphics[width=0.40\textwidth]{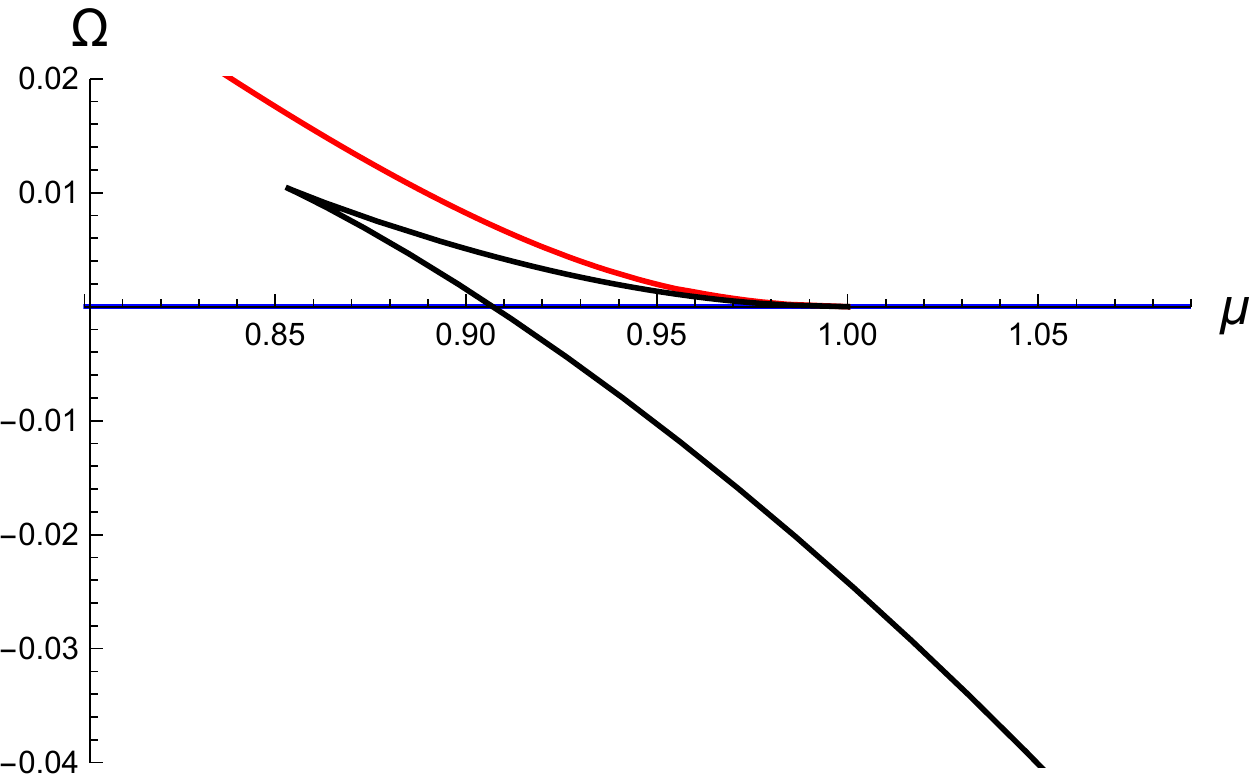}
  \caption{We plot the grand canonical potential $\Omega$ as a function of the chemical potential $\mu$ for the unflavored (left) and flavored (right) models. The black (red) curve corresponds to the black hole (brane-antibrane) embedding. The supersymmetric Minkowski embeddings $\Omega(\mu)=0$ we have represented with a blue curve on the horizontal axis. The curves for the flavored model on the right have been obtained for $b=1.25$. All curves are with $m=1$.}
\label{Omega_mu}
\end{figure}

From the numerical results displayed in Fig.~\ref{Omega_mu} it is clear that the black hole embeddings are thermodynamically preferred for values of $\mu$ such that their grand canonical potential $\Omega_{bh}$ is negative. Moreover, when $\mu$ is such that $\Omega_{bh}>0$,  the Minkowski embeddings (with $d=\Omega=0$) are preferred. Notice also that the brane-antibrane configurations are always thermodynamically disfavored.  Therefore, at $\mu=\mu_c$ such that $\Omega_{bh}(\mu_c)=0$  there is a black hole-Minkowski embedding phase transition. In Fig.~\ref{Omega_mu} we see that the nature of this quantum phase transition  for the unflavored model is very different from that of the backreacted background. Indeed, in the quenched unflavored case we have a continuous second order phase transition in which the density $d$ vanishes in both phases at the transition point 
$\mu_c=m$. In section \ref{unflavored_transition} we will study in detail this quantum critical point and we will characterize the  scaling of the different physical quantities near the transition. 

In the unquenched flavored model the phase transition at $\mu=\mu_c$ is discontinuous since $d$ jumps from a non-zero value in the black hole phase to $d=0$ in the Minkowski phase. Therefore, we have a first-order phase transition, for which we will determine the latent heat and other quantities in section \ref{flavored_transition}.

\begin{figure}[ht]
\center
 \includegraphics[width=0.40\textwidth]{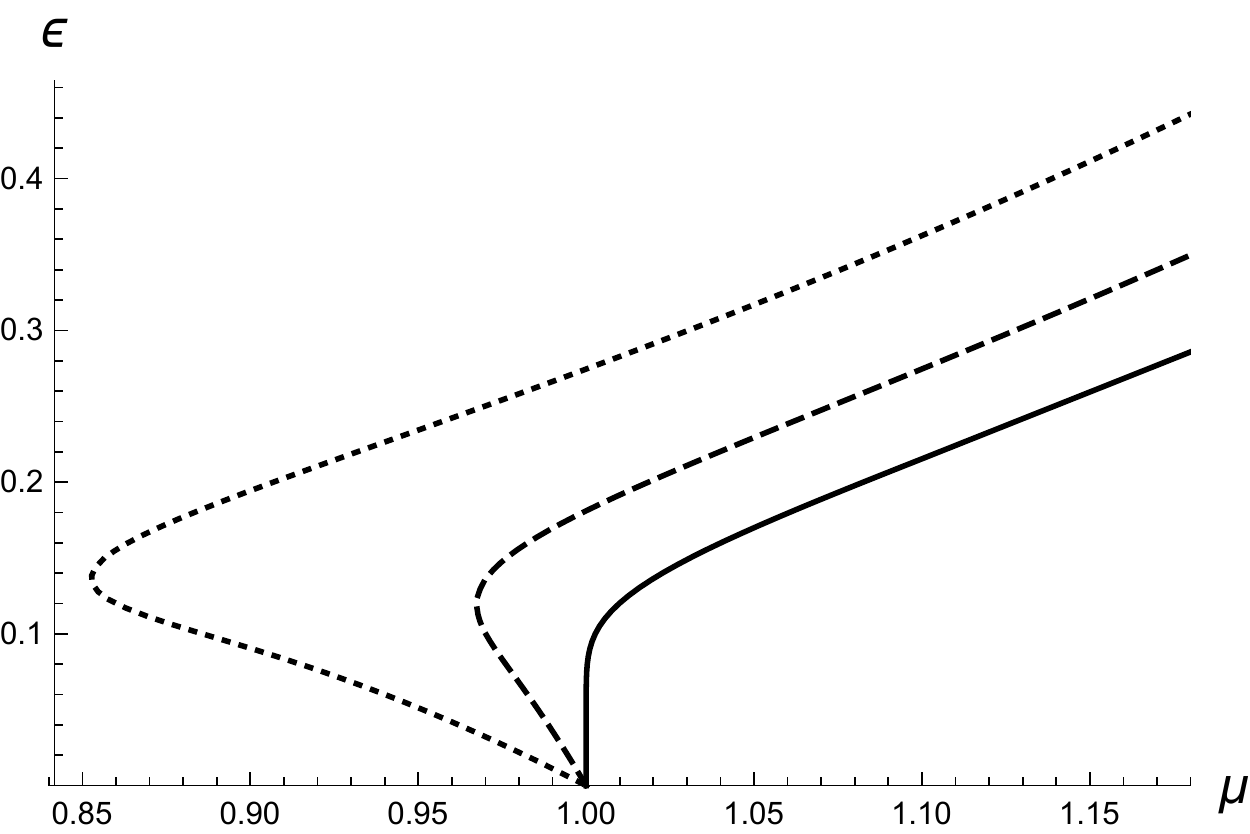}
  \caption{We depict the internal energy $\epsilon$ as a function of the chemical potential $\mu$ for the unflavored model (continuous curve) and for the flavored model with $b=1.1$ (dashed curve) and $b=1.25$ (dotted curve). In all the cases we have used $m=1$.}
\label{energy_density}
\end{figure}

Once the grand canonical potential $\Omega$ is known, we can determine other thermodynamic functions. Indeed, the charge density $\rho_{ch}$ is given by:
\beq
\rho_{ch}\,=\,-{\partial \Omega\over \partial\mu}\,\,.
\label{rho_ch_def}
\eeq
By computing numerically the derivative in (\ref{rho_ch_def}) at fixed mass $m$,  we have checked that 
$\rho_{ch}$ is related to $d$ as:
\beq
\rho_{ch}\,=\,{\cal N}\,b\,d\,\,,
\label{rho_ch_d}
\eeq
where ${\cal N}$ is the normalization constant (\ref{calN_def}).  Eq. (\ref{rho_ch_d}) confirms our identification of the constant $d$. The energy density $\epsilon$ can be obtained as:
\beq
\epsilon\,=\,\Omega\,+\,\mu\,\rho_{ch} \ .
\eeq
More explicitly, after using (\ref{Omega_integral}), (\ref{chemical_pot}),  and (\ref{rho_ch_d}), we have the following integral expression for $\epsilon$,
\beq
\epsilon\,=\,{\cal N}\,\int_{r_0}^{\infty}\,
\Big[\sqrt{b^2+r^2\,\theta'^2}\, \sqrt{d^2+r^4\sin^2\theta}\,-\,b\,r^2\,\sin^2\theta\,-\,r^3\,\sin\theta\,\cos\theta\,\theta'
\Big]dr\,\,,
\eeq
where $r_0$ is the minimal value of $r$ for the embedding. In Fig.~\ref{energy_density} we plot $\epsilon$  for black hole embeddings as a function of $\mu$, both for the quenched and unquenched model.  We notice that the energy density in the quenched theory grows monotonically with the chemical potential, starting from the value $\epsilon=0$ at the transition point at $\mu=m$. On the contrary, when dynamical quarks are added to the background, the function $\epsilon_{bh}$ is not monotonic and becomes double-valued, with a point where $\partial\epsilon/\partial\mu= \mu \partial \rho_{ch}/\partial \mu$ blows up. This is, of course, consistent with the results plotted in Fig.~\ref{plot_chemical_pot}.

The speed of the first sound is defined as:
\beq
u_s^2\,=\,{\partial P\over \partial \epsilon}\,\,.
\label{speed_of_first_sound}
\eeq
We evaluated   numerically the derivative in (\ref{speed_of_first_sound}) for black hole embeddings  by using 
(\ref{P_Omega}) and (\ref{Omega_integral}). The results are represented in  Fig.~
\ref{speed_of_sound}, both for the quenched and unquenched cases.  Again, they are very different in these two cases. In the quenched model $u_s^2$ is always non-negative and decreases monotonically when $m/\mu$ varies in the physical interval $[0,1]$. In Fig.~\ref{speed_of_sound} (left) we compare  $u_s^2$  for our quenched system with the corresponding values for the D3-D5 model \cite{Ammon:2012je, Itsios:2016ffv}. In the unquenched case  $u_s^2$ is not monotonic and becomes negative for small $\mu$, which again signals a discontinuous phase transition. 

\begin{figure}[ht]
\center
 \includegraphics[width=0.40\textwidth]{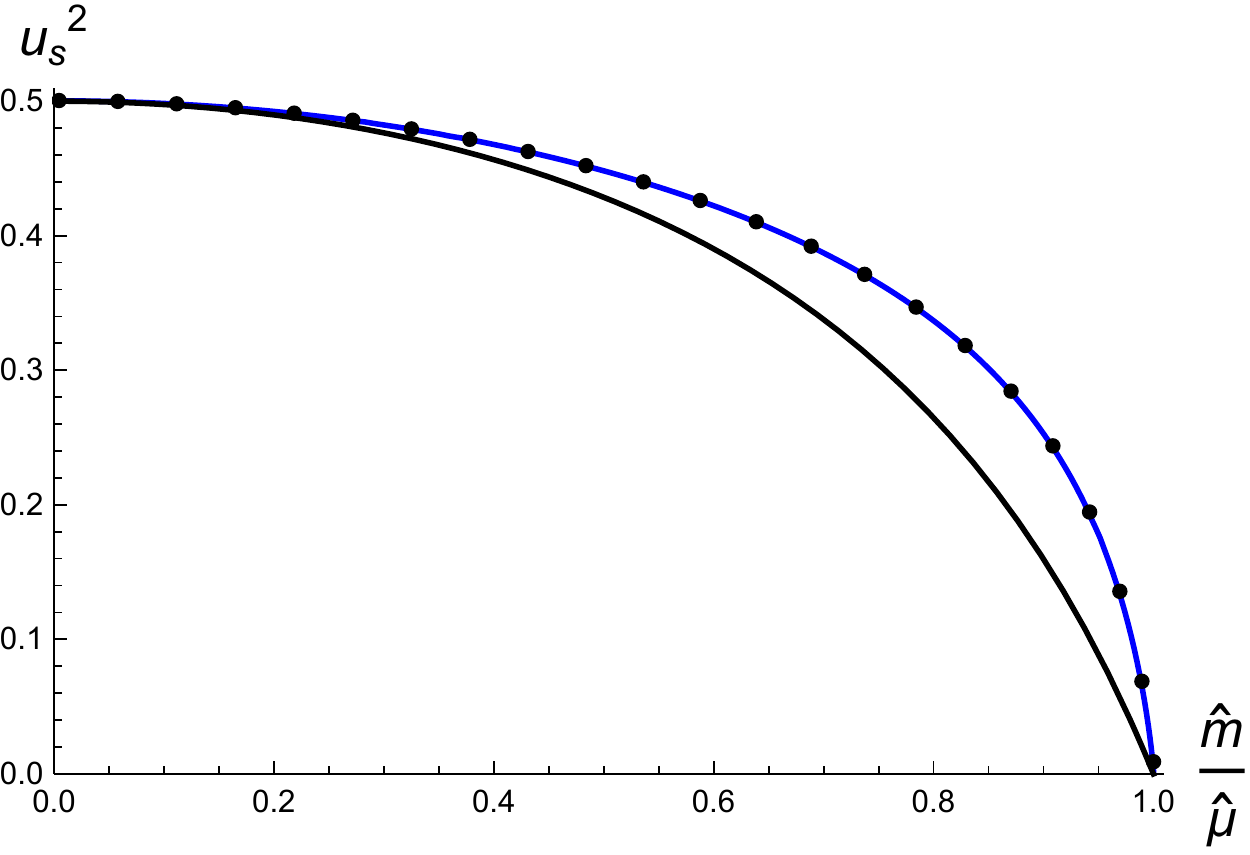}
 \qquad\qquad
  \includegraphics[width=0.40\textwidth]{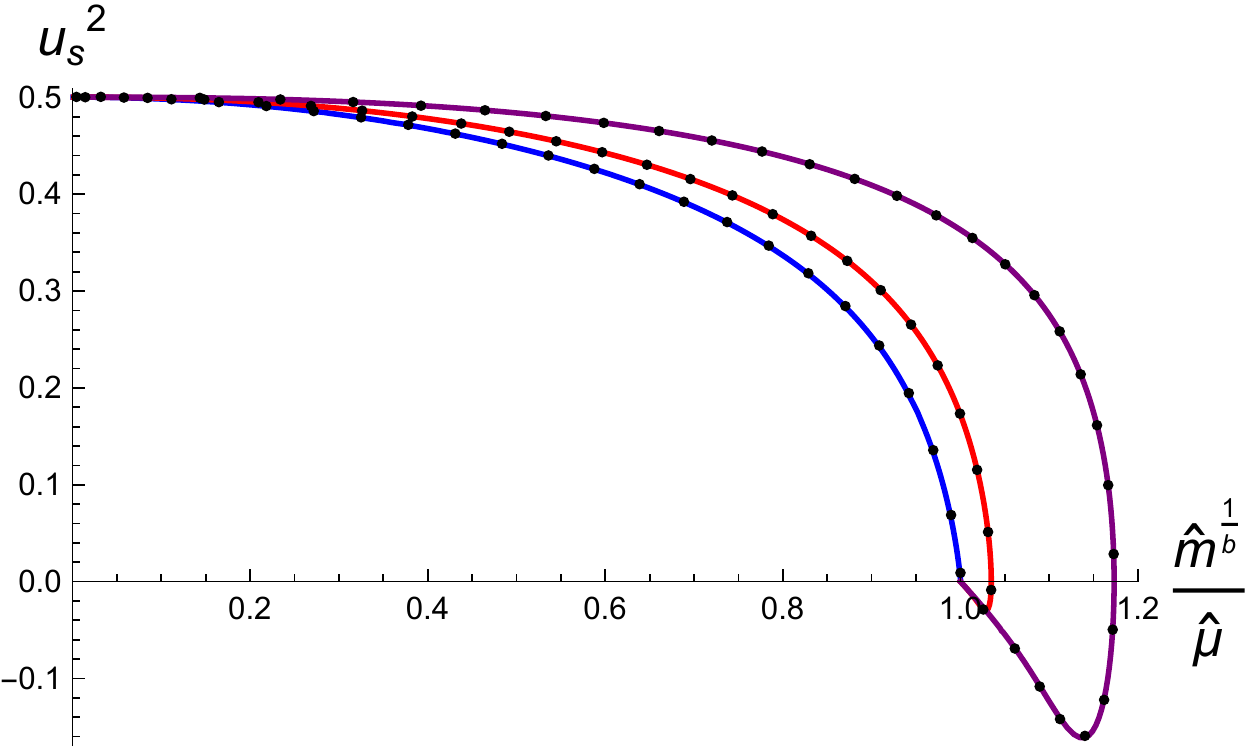}
  \caption{Left: We plot $u_s^2$ as a function of $m/\mu$ for the unflavored background (blue curve). We compare with the same quantity for the D3-D5 model (black curve).  Right: We plot $u_s^2$ for different numbers of flavors: $b=1$ (blue), $b=1.1$ (red), and $b=1.25$ (purple). In both plots the points are the values of the  square of the speed of zero sound obtained by integrating the fluctuation equations of section \ref{fluctuations}; we conclude that the speeds of first and zero sounds agree in this model.}
\label{speed_of_sound}
\end{figure}

\subsection{The unflavored transition}
\label{unflavored_transition}

We have shown above that the unflavored system experiences a continuous phase transition  at $\mu=m$ and $T=0$. In this section we look in more detail the behavior of the system near this quantum critical point. Accordingly, let us define $\bar\mu$ as:
\beq
\bar\mu\,=\,\mu\,-\,m\,\,.
\label{mu_bar_def}
\eeq
Clearly $\bar\mu=0$ is the location of the phase transition. Therefore,  we expect that the grand canonical potential $\Omega$  behaves in a non-analytic form  near $\bar\mu=0$. We assume that the system displays a scaling behavior near the critical point. The goal of this section is to characterize this behavior  in terms of a set of critical exponents. 

Let us consider a system with hyperscaling violation exponent $\theta$ and dynamical exponent $z$ in $n$ spatial dimensions ($n=2$ in our case). Recall that in such a system $n-\theta$ is the effective number of spatial dimensions near the critical point and $z$ is the effective dimension of the energy.  Therefore $[\bar\mu]=z$ and the energy densities (such as our grand canonical potential $\Omega$) should have a dimension equal to $n-\theta+z$. These dimension assignments allow us to write $\Omega$ near $\bar\mu=0$ as:
\beq
\Omega\,\approx\, -C\,\bar\mu^{{n+z-\theta\over z}}\,\Big(\big|\log {\bar\mu\over m}\big|\Big)^{-\zeta}\,\,,
\label{Critical_Omega}
\eeq
where $C>0$ is a constant.  Eq. (\ref{Critical_Omega})  is a generalization of the expression written in \cite{Ammon:2012je}  by including a logarithmic multiplicative term with some new exponent $\zeta$. We show below that $\zeta$ cannot be zero in our ABJM case.  This is to be compared with the D$p$-D$q$ systems studied in \cite{Ammon:2012je, Itsios:2016ffv}, where  $\zeta=0$. Similar multiplicative logarithmic corrections to the scaling has been studied in general in \cite{Kenna} for thermal phase transitions. 

The charge density $\rho_{ch}={\cal N}\,d$  is obtained by computing the derivative of $\Omega$ with respect to $\bar\mu$. We get:
\beq
{\cal N}\,d\,\approx\,C\,\bar\mu^{{n-\theta\over z}}\,
\,\Big(\big|\log {\bar\mu\over m}\big|\Big)^{-\zeta}\,
\Big[1+{n-\theta\over z}\,+\,{\zeta\over \big|\log {\bar\mu\over m}\big|}\Big]\,\,.
\label{critical_d}
\eeq
Let us  next consider, following \cite{Ammon:2012je}, the non-relativistic energy density $e$,  defined  as:
\beq
e\,=\,\epsilon\,-\,\rho_{ch}\,m\,=\,\Omega+\,\rho_{ch}\,\bar\mu\,\,.
\eeq
Near the critical point, $e$ behaves as:
\beq
e\,\approx\, 
C\,\bar\mu^{{n+z-\theta\over z}}\,\Big(\big|\log {\bar\mu\over m}\big|\Big)^{-\zeta}\,
\Big[
{n-\theta\over z}\,+\,{\zeta\over  \big|\log {\bar\mu\over m}\big|}
\Big]\,\,,
\eeq
and it is very convenient to consider the ratio $e/P$, which is given by:
\beq
{e\over P}\,\approx\,{n-\theta\over z}\,+\,{\zeta\over  \big|\log {\bar\mu\over m}\big|}\,\,.
\eeq
If $\theta\not=n$ the ratio $e/P$ reaches a constant non-vanishing value as $\bar\mu\to0$. This is clearly not the case  for our system, as illustrated in Fig.~\ref{e/p_cs}. Therefore,  our system should have $\theta=2$. Moreover, the  logarithmic exponent $\zeta$ should be non-zero and positive.\footnote{Indeed, if we had $\theta=2$ and $\zeta=0$ the charge density $d$ in (\ref{critical_d})  would be non-zero at the critical point, which is not the case for our ABJM system.} Therefore we get the following leading behavior for our system:
\beq
\rho_{ch}\,=\,{\cal N}\,d\,\approx \,{C\over \Big(\big|\log {\bar\mu\over m}\big|\Big)^{\zeta}}\,\,,
\qquad\qquad\qquad\qquad
{e\over P}\,\approx {\zeta\over \big|\log {\bar\mu\over m}\big|}\,\,.
\label{rho_e_over_P}
\eeq
We can also compute the speed of sound $u_s$ near the critical point by using (\ref{speed_of_first_sound}), with the result:
\beq
u_s^2\,\approx\, {1\over \zeta}\,{\bar\mu \big|\log {\bar\mu\over m}\big|\over m+\bar\mu}\,
\Big[\,1-{1\over 1+\zeta+\big|\log {\bar\mu\over m}\big|}\,\Big]\,\,,
\eeq
which, at leading order for $\bar\mu\to 0$ becomes simply:
\beq
u_s^2\,\approx\, {1\over \zeta}\,{\bar\mu\over m}\,\big|\log {\bar\mu\over m}\big|\,\,.
\label{us_scaling}
\eeq
\begin{figure}[ht]
\center
 \includegraphics[width=0.40\textwidth]{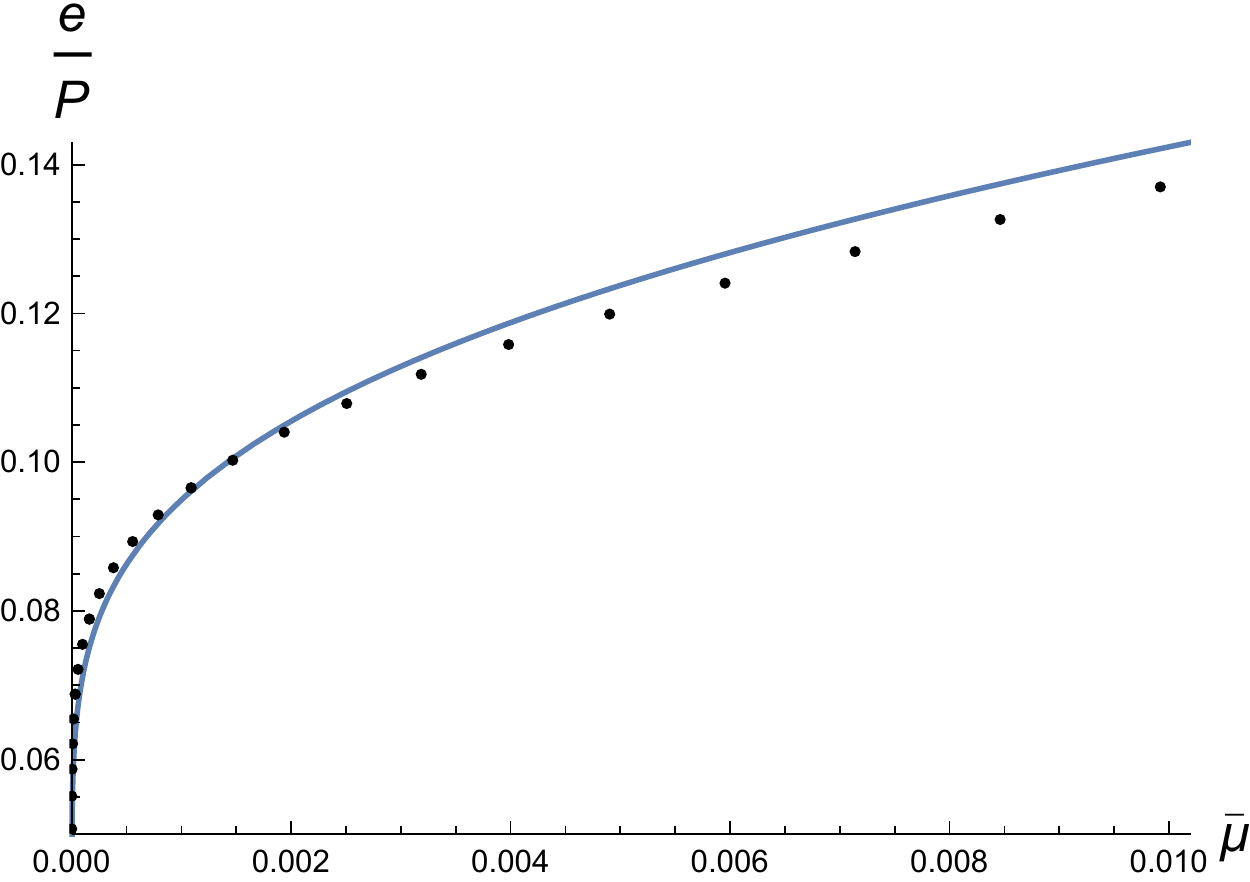}
 \qquad\qquad
  \includegraphics[width=0.40\textwidth]{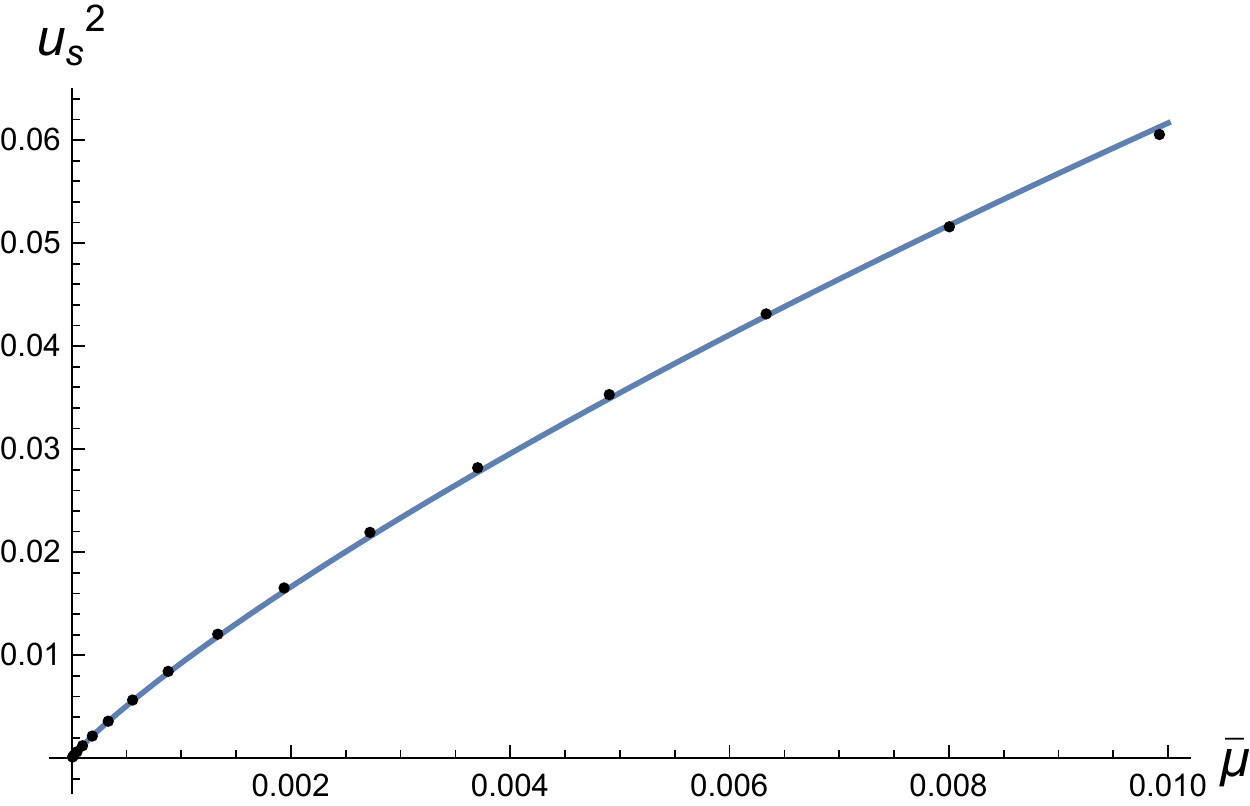}
  \caption{On the left we plot the numerical values of the  ratio $e/P$ as a function of $\bar\mu$. The continuous line 
  is a fit to the expression written in (\ref{rho_e_over_P}). The value of $\zeta$ obtained in this fit is $\zeta=0.65575$.  On the right we plot the values of $u_s^2$, together with the scaling expression (\ref{us_scaling}) for  $\zeta=0.74689$.
  }
\label{e/p_cs}
\end{figure}

To determine the value of the exponent $\zeta$ we can fit the numerical values of $e/P$ and $u_s^2$ near $\bar\mu=0$ to our scaling expressions (\ref{rho_e_over_P}) and (\ref{us_scaling}). Due to the logarithmic behavior of these quantities we must explore very small values of $\bar\mu$. The results of these fits is shown in Fig.~\ref{e/p_cs}. The values of $\zeta$ obtained are in the range $\zeta=0.65-0.75$.

Let us determine, following the reasoning in \cite{Ammon:2012je}, the dynamical critical exponent $z$ by dimensional analysis of   the dispersion relation of the sound mode, which is of the form $\omega\,=\,u_s\, k$, where 
$u_s$ is given by (\ref{us_scaling}) near the critical point $\bar\mu=0$. Actually, we will see that the speed of the zero sound, obtained by numerical integration of the fluctuation equations of the probe brane, is exactly the same as the one determined by (\ref{speed_of_first_sound}). Near $\bar\mu=0$ eq. (\ref{us_scaling}) tells us that $u_s\sim \sqrt{\bar\mu}$  (times a logarithmic correction) and, since $[\omega]=[\bar\mu]=z$ and $[k]=1$,  the dimensional consistency of the dispersion relation $\omega\,=\,u_s\, k$ implies that $z=2$. Therefore, the values of $\theta$ and $z$ for our system are:
\beq
\theta=2\,\,,
\qquad\qquad
z\,=\,2\,\,.
\label{theta_z}
\eeq
Notice that the value of $\theta$ just found differs from the value   $\theta=1$ obtained in \cite{Ammon:2012je} for the conformal systems D3-D7 and D3-D5.

Let us now consider the system  at small non-zero temperature $T\ll \bar \mu$. We can evaluate the free energy 
$f$ at first order in $T$ by using the results of \cite{Karch:2009eb}. Notice that at $T=0$, $f=\epsilon$. Indeed, according to the analysis of
\cite{Karch:2009eb}, when $T$ is small the free energy density can be approximated as:
\beq
f(\mu,m,T)\,=\,f(\mu,m,T=0)\,+\,\pi\, \rho_{ch}\,T\,+\,{\mathcal O}(T^2)\,\,.
\eeq
Then, the non-relativistic free energy  density is given by:
\beq
f_{non-rel}(\mu,m,T)\,=\,f(\mu,m, T)\,-\, \rho_{ch}\,m\,=\,e\,+\,
\pi\, \rho_{ch}\,T\,+\,{\mathcal O}(T^2)\,\,.
\label{f_non_rel_def}
\eeq
Evaluating  the right-hand side of  (\ref{f_non_rel_def}) for our system, we get the following expression of $f_{non-rel}$ for small 
$\bar\mu$ and $T/\bar\mu$:
\beq
f_{non-rel}(\mu,m,T) = C\,{\bar\mu\over 
\Big(\big|\log {\bar\mu\over m}\big|\Big)^{\zeta+1}}\,\Big[\zeta\,+\,
\pi\,\big|\log {\bar\mu\over m}\big|\,{T\over \bar\mu}\,+\,\cdots\Big]\,\,.
\label{f_non_rel_ABJM}
\eeq
On general grounds, near a quantum phase transition the free energy density should behave as a homogeneous function when the control parameter $\bar\mu$ and the temperature $T$ are scaled as $\bar\mu\to \Lambda^{{1\over \nu}}\,\bar\mu$,  $T\to \Lambda^{z}\, T$, where $\nu$ is the critical exponent that characterizes the divergence of the correlation length $\xi\sim (T-0)^{-\nu}$ \cite{Vojta}.  Eq. (\ref{f_non_rel_ABJM}) is the first order term of a Taylor expansion of the scaling function of $f_{non-rel}$. If we disregard the logarithmic terms in (\ref{f_non_rel_ABJM}) (which give rise to  subleading terms when $\bar\mu\to 0$), it follows that $T$ and $\bar\mu$ should be scaled by the same power of the scale factor $\Lambda$. Since $z=2$ for our system, we must have $\nu=1/2$. Eq.  (\ref{f_non_rel_ABJM}) also determines the value of the exponent $\alpha$ which characterizes the scaling of the heat capacity $c_V\sim (T-0)^{-\alpha}$. Indeed, according to the analysis of \cite{Ammon:2012je} the global power of $\bar\mu$  in $f_{non-rel}(\mu,m,T)$  should be $2-\alpha$. If we ignore again the logarithmic correction, this prescription gives $\alpha=1$. Therefore, we have obtained  that the critical exponents $\alpha$ and 
$\nu$ are given by
\beq
\alpha\,=\,1\,\,,
\qquad\qquad
\nu\,=\,{1\over 2}\,\,.
\label{alpha_nu}
\eeq
Notice that the values of $\theta$, $z$, $\alpha$, and $\nu$ listed in (\ref{theta_z})  and (\ref{alpha_nu}) satisfy the hyperscaling relation 
\beq
(n+z-\theta)\,\nu\,=\,2-\alpha\,\,,
\eeq
with $n=2$.

\subsection{The flavored transition}
\label{flavored_transition}

We already pointed out above that the black hole-Minkowski phase transition with dynamical quarks in the background is of first order. At the transition point  the density jumps from being $d=d_c\not=0$ in the black hole phase to $d=0$ in the Minkowski phase. We have investigated numerically the dependence of $d_c$ on $\hat\epsilon$ and $m$ and we found that, with big accuracy, this dependence can be written as:
\beq
d_c(\hat\epsilon, m)\,=\,\tilde d_{c}(\hat \epsilon)\,\,
m^{{2\over b}}\,=\,\tilde d_{c}(\hat \epsilon)\,\,m_q^2\,\,,
\label{quadratic_law}
\eeq
where $m_q=m^{{1\over b}}$ is proportional to the physical mass of the quarks. Notice that the dependence on $m$ written in (\ref{quadratic_law}) is the one expected by the rescaling argument given at the end of section \ref{zeroT_embeddings}.

The flavor dependent coefficient of the quadratic law (\ref{quadratic_law}) grows monotonically  with $\hat\epsilon$, as shown in Fig.~\ref{flavored_transition_plots} (left).  For small $\hat\epsilon$ this growth is very fast and saturates very quickly for larger values of the deformation parameter.  

\begin{figure}[ht]
\center
 \includegraphics[width=0.40\textwidth]{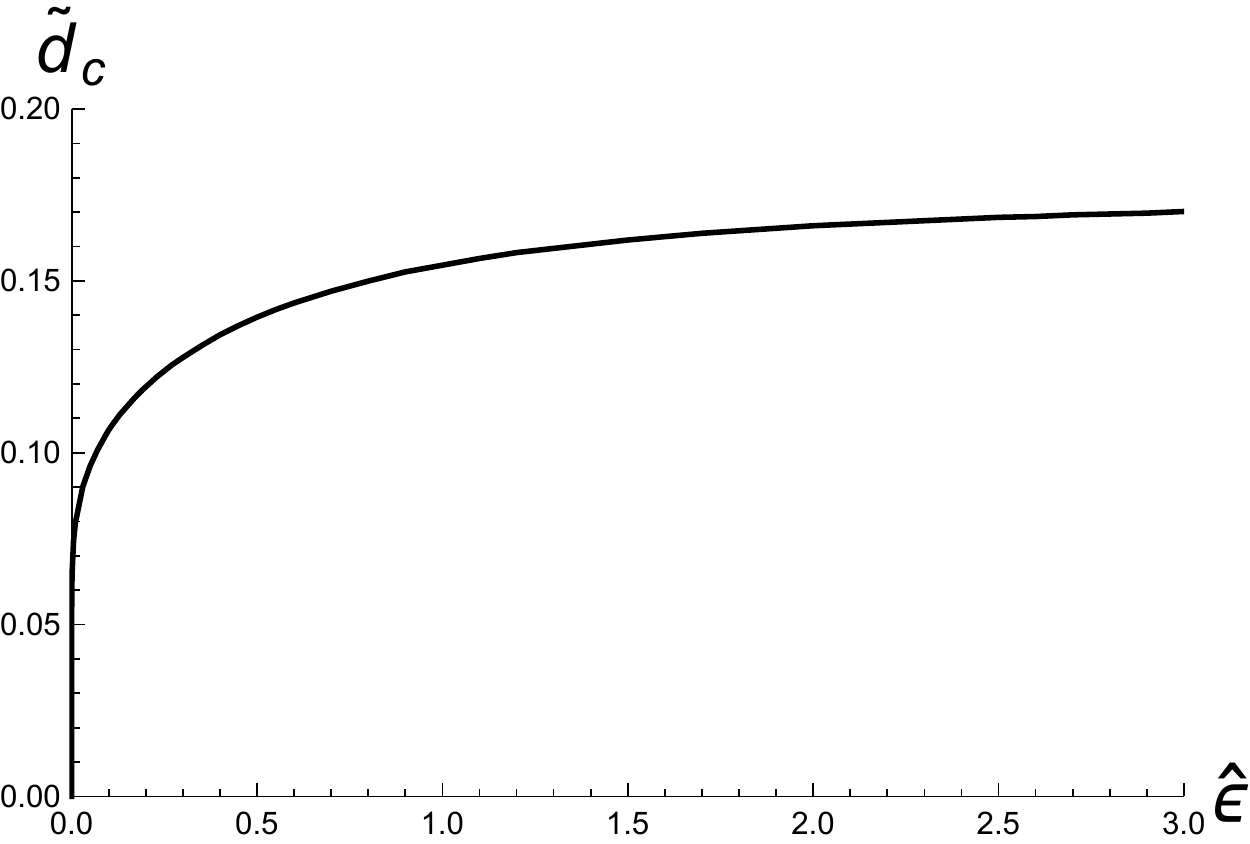}
 \qquad\qquad
  \includegraphics[width=0.40\textwidth]{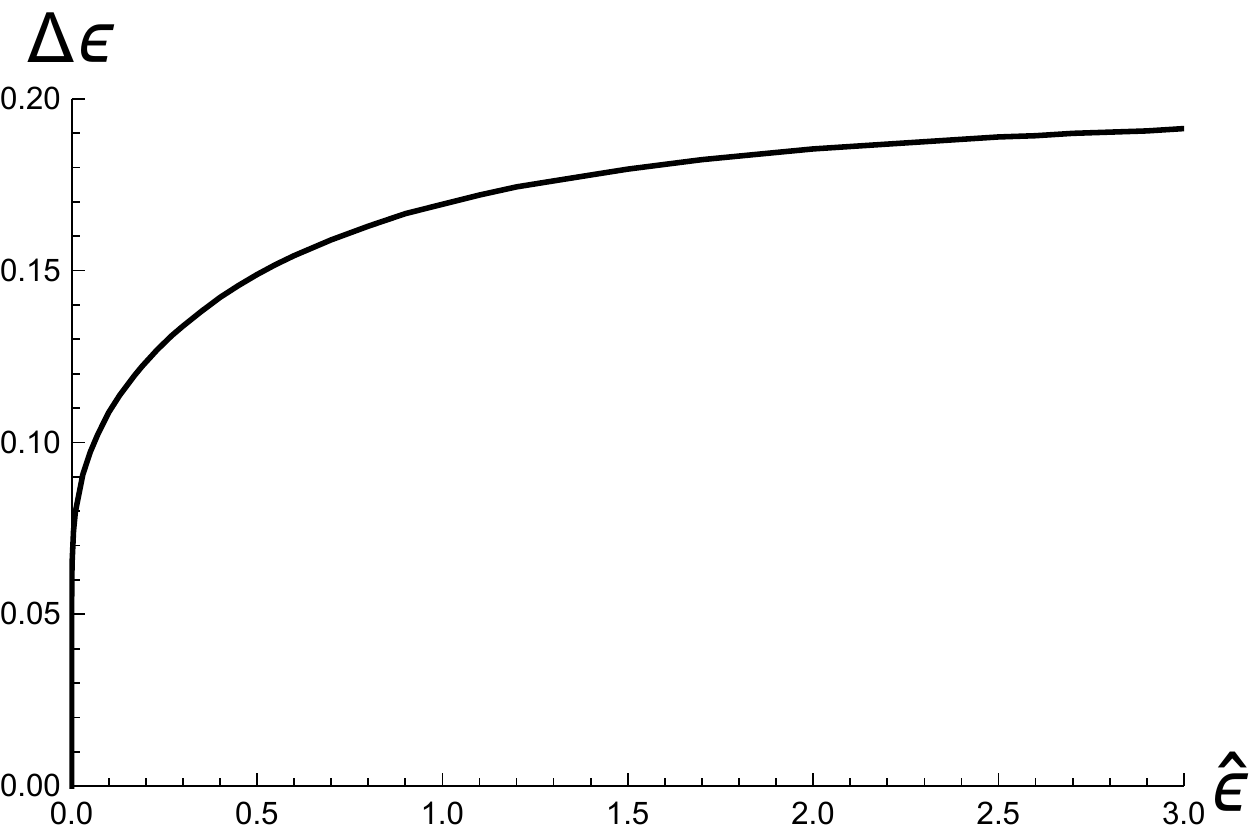}
  \caption{Left: We plot the function $\tilde d_{c}(\hat \epsilon)$ introduced in (\ref{quadratic_law}) at $m=1$. Right: We depict the latent heat $\Delta \epsilon$ for $m=1$ as a function of $\hat\epsilon$.}
\label{flavored_transition_plots}
\end{figure}

The phase transition occurs at a critical  chemical potential $\mu_c< m^{{1\over b}}=m_q$. Actually, 
the ratio $\mu_c/m_q$ is a decreasing function of $\hat \epsilon$ which approaches the value $\mu_c/m_q\approx 0.9$ when $\hat\epsilon\to\infty$. It is also interesting to point out that the value of $\mu$ where the speed of sound 
vanishes (see Fig.~\ref{speed_of_sound})) corresponds to the turning point of $\Omega$ as a function of $\mu$ for a black hole embedding, \ie\ to the minimum value of $\mu$ for such embeddings. The phase transition occurs for  a value of $\mu$ close to its lowest value where $u_s^2$ is still positive. Moreover, it follows from the above discussion that $\mu_c\sim d_c^{{1\over 2}}$.

We also studied the latent heat of the phase transition, \ie\ the difference $\Delta\epsilon$ of the internal energy of the two phases. Notice that, as $\Omega=0$ in both phases at the transition point and $\rho_{ch}=0$ in the Minkowski phase,  $\Delta\epsilon$ is simply obtained by evaluating $\mu\,\rho_{ch}$ at the black hole side of the transition:
\beq
\Delta \epsilon\,=\,(\mu\,\rho_{ch})_{bh}\,\,.
\eeq
The behavior of this quantity with the number of flavors  when $m=1$ is displayed in Fig.~\ref{flavored_transition_plots} (right).  We notice that the latent heat resembles the behavior of the critical density. 
We have also verified that $\Delta \epsilon$ grows with the quark mass as $\Delta \epsilon\sim m_q^3\,=\,m^{{3\over b}}$.

Most of the figures that we have presented above have been produced using $m=1$. It is however simple to obtain the results for any value of $m$ by using the rescaling argument presented above. Indeed, one can readily show that the different quantities scale with $m_q=m^{1\over b}$ as:
\beq
 \epsilon\sim \Omega \sim m_q^3 \ , \ \mu \sim m_q \ , \ d\sim m_q^2 \ .
\eeq
We have checked that this behavior is confirmed by our numerical results.

\section{Charge susceptibility and diffusion constant}
\label{diffusion}

Let us now consider the system at non-zero temperature and compute the charge susceptibility, which is defined as:
\beq
\chi\,=\,{\partial\rho_{ch}\over \partial\mu}\,\,.
\eeq
Taking into account that the charge density $\rho_{ch}$ is related to $d$ as $\rho_{ch}\,=\,{\cal N}\,b\,d$ (\ref{rho_ch_d}), we can rewrite this last expression as:
\beq
\chi^{-1}\,=\,{1\over {\cal N}\,b}\,{\partial \mu\over \partial d}\,\,.
\label{charge_sus_def}
\eeq
We now evaluate  explicitly the derivative in (\ref{charge_sus_def}) as:
\beq
{\partial \mu\over \partial d}\,=\,\int_{r_h}^{\infty}\,
{\partial  A_t'\over \partial d} \,dr\,\,.
\label{d_mu_d}
\eeq
The derivative inside the  integral in (\ref{d_mu_d})  can be computed directly. We get:
\beq
{\partial  A_t'\over \partial d}\,=\,{\sqrt{\Delta}\over b}\,
{r^2\sin\theta\over d^2+r^4\sin^2\theta}\,
\Bigg[1\,-\,d\Bigg(\cot\theta\,{\partial\theta\over \partial d}\,-\,{r^2\,h\,\theta'\over \Delta}\,
{\partial\theta'\over \partial d}\Bigg)\Bigg]\,\,.
\eeq
where $\Delta$ is defined as:
\beq
\Delta\,\equiv\,b^2(1-A_t'^{\,2})\,+\,r^2\,h\,\theta'^{\,2}\,=\,
{r^4\sin^2\theta(b^2+r^2\,h\,\theta'^{\,2})\over
d^2+r^4\sin^2\theta}\,\,.
\label{Delta_definition}
\eeq
Thus, the charge susceptibility can be written in the form:
\beq
\chi^{-1}\,=\,{1\over {\cal N}}\,\int_{r_h}^{\infty}\,dr
{\sqrt{\Delta}\over b^2}\,
{r^2\sin\theta\over d^2+r^4\sin^2\theta}\,
\Bigg[1\,-\,d\Bigg(\cot\theta\,{\partial\theta\over \partial d}\,-\,{r^2\,h\,\theta'\over \Delta}\,
{\partial\theta'\over \partial d}\Bigg)\Bigg]\,\,.
\label{susceptibility}
\eeq

\begin{figure}[ht]
\center
 \includegraphics[width=0.40\textwidth]{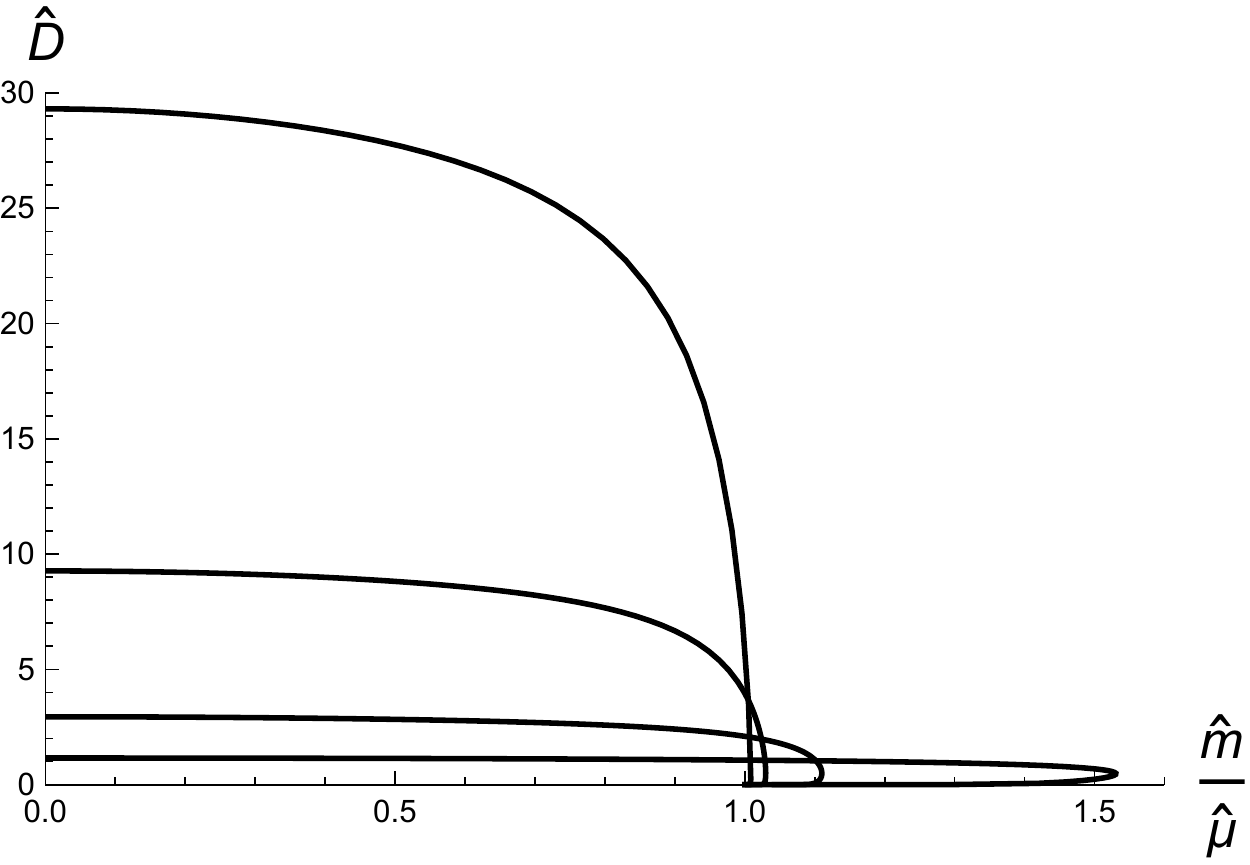}
 \qquad\qquad
  \includegraphics[width=0.40\textwidth]{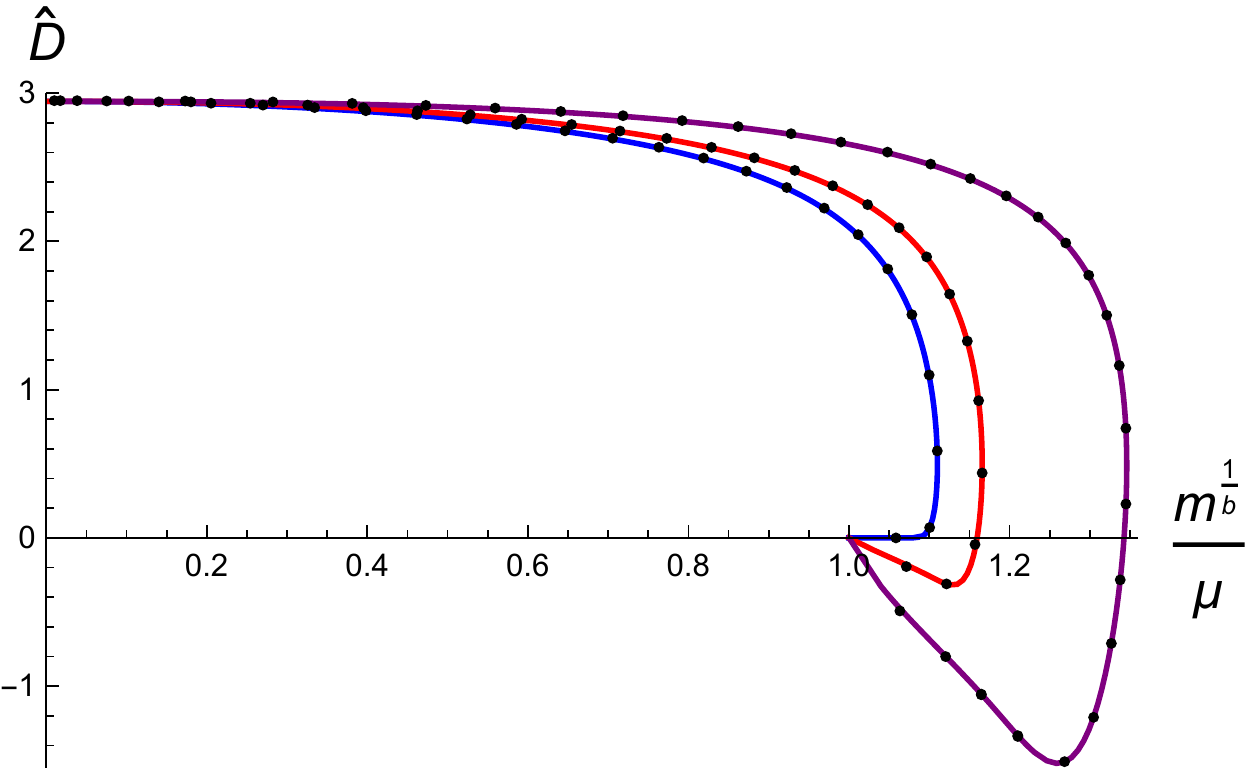}
  \caption{We plot the rescaled diffusion constant $\hat D=r_h\,D$ as a function of the ratio of the mass and the chemical potential. Left: We plot the values of $\hat D$ for the unflavored theory for different values of the rescaled density $\hat d=d/r_h^2$. The values of $\hat d$ in this plot  are $\hat d=1,10,100,1000$ (bottom-up). Right: We plot the values of $\hat D$ for $\hat d=10$ and for different number of flavors: $b=1$ (blue), $b=1.1$ (red), and $b=1.25$ (purple), inside-out. The continuous curves are obtained by the Einstein relation and the points correspond to the diffusive fluctuation modes of the probe.}
\label{Diffusion_constant_plots}
\end{figure}

The charge diffusion constant $D$ can be related to the charge susceptibility and to the DC conductivity $\sigma$ by the Einstein relation:
\beq
D\,=\,\sigma\,\chi^{-1}\,\,.
\label{Einstein_rel}
\eeq
The value of $\sigma$ can be obtained from the two-point correlators  of the transverse currents. This calculation is performed in detail in appendix \ref{fluctuation_apendix}. Alternatively, $\sigma$ can be computed by employing the Karch-O'Bannon method \cite{Karch:2007pd}, as was done for the ABJM model in chapter \ref{chapterfour}.  The result obtained by these two methods agree and is given by:
\beq
\sigma\,=\,{\cal N}\,{b\over r_h^2}\,\sqrt{d^2+r_h^4\,\sin^2\theta_h}\,\,.
\label{dc_conductivity}
\eeq
We can now plug  (\ref{susceptibility}) and (\ref{dc_conductivity}) into the right-hand side of (\ref{Einstein_rel}) to get the diffusion constant $D$. The final result is:
\beq
D\,=\,{\sqrt{d^2+r_h^4\,\sin^2\theta_h}\over b\,r_h^2}\,
\int_{r_h}^{\infty}\,dr
\,{r^2\sin\theta\sqrt{\Delta}\over d^2+r^4\sin^2\theta}\,
\Bigg[1\,-\,d\Bigg(\cot\theta\,{\partial\theta\over \partial d}\,-\,{r^2\,h\,\theta'\over \Delta}\,
{\partial\theta'\over \partial d}\Bigg)\Bigg]\,\,.
\label{Diffusion_Einstein}
\eeq
In the case of massless quarks, the embedding is just $\theta={\rm constant}=\pi/2$ and the integral (\ref{Diffusion_Einstein}) can be evaluated in analytic form. We get:
\beq
D_{m=0}\,=\,{\sqrt{d^2+r_h^4}\over r_h^3}\,
F\Big({3\over 2}, {1\over 2}; {5\over 4};-{d^2\over r_h^4}\Big)\,\,.
\label{massless_diffusion_constant}
\eeq
In the general case of massive quarks we have evaluated (\ref{Diffusion_Einstein}) numerically for the unflavored and  flavored backgrounds as a function of the chemical potential. The results of these calculations are displayed in Fig.~\ref{Diffusion_constant_plots}. In the unflavored background $D$ is always non-negative and vanishes when 
$\mu=m$ (see Fig.~\ref{Diffusion_constant_plots}, left).  On the contrary, when $N_f\not= 0$ the diffusion constant  
is maximal for large chemical potential (and given by the massless value (\ref{massless_diffusion_constant})) and 
becomes negative after $\mu$ reaches its minimal value, which means that the system becomes unstable and that the first-order phase transition at $T=0$ survives at non-zero temperature. In the next section  we obtain the diffusion constant by looking at the fluctuation modes of the probe in the hydrodynamical regime. The corresponding values of $D$ are also plotted in Fig.~\ref{Diffusion_constant_plots}, where we notice that they agree perfectly with the values found above by using Einstein relation.

\section{Fluctuations}
\label{fluctuations}

We now want to carry out a dynamic (\ie\ time-dependent) study of our system,  to complement the static analysis performed so far. Accordingly, let us consider the generic $T\not=0$ background and let us allow the probe brane to fluctuate around the black hole embeddings described in section \ref{Probes}. In general, the equations of motion of these fluctuations are very complicated since the different fluctuation modes are coupled. However, there are certain modes that can be decoupled from the rest and, therefore, they constitute a consistent truncation of the general system of equations. In this section we will study one of these restricted sets of fluctuations, which involves the gauge   field  $A$ and the transverse scalar $\theta$.  These fields take the form:
\beq
A\,=\,L^2\,\big[A_t(r)\,dt\,+\,a_t(t,x,r)\,dt+a_x(t,x,r)\,dx+a_r(t,x,r)\,dr]\,\,,
\quad
\theta\,=\,\theta(r)+\lambda(t,x,r)\,\,\,,
\label{longitudinal_fluct_ansatz}
\eeq
where $a_t$, $a_x$, $a_r$, and $\lambda$ are the first-order perturbations. One can check that the ansatz (\ref{longitudinal_fluct_ansatz}) is indeed a consistent truncation of the equations of motion. These truncated equations  can be derived from a second order lagrangian density ${\cal L}^{(2)}$, which is derived in detail in appendix \ref{fluctuation_apendix}.  The expression for  ${\cal L}^{(2)}$ is:
\bear
&&{\cal L}^{(2)}=-{\cal N}\,r^2\,\sin\theta\sqrt{\Delta}\Bigg[{1\over 4}\,{\cal G}^{nm}\,{\cal G}^{pq}\,f_{mq}\,f_{np}\,+\,{L^2\over 2b^2}\,
\Big(1\,-\,{r^2\,h\,\theta'^{\,2}\over \Delta}\Big)\,{\cal G}^{mn}
\partial_m\lambda\,\partial_n\lambda \rc
&&
\qquad\qquad\qquad\qquad
+\Big(\big(b-{3\over 2}\big)\,{\sin\theta\over \sqrt{\Delta}}\,+\,
{r^2\,h\,\theta'^{\,2}-\Delta\over 2\sin^2\theta\Delta}\Big)\lambda^2- 
{d^2\theta'^{\,2}\over 2\, b^2\, r^4\sin^2\theta\Delta}\,(\partial_t\lambda)^2
\rc
&&
\qquad\qquad\qquad\qquad
+{d\theta'\over b r^2\sin\theta\sqrt{\Delta}}\,{\cal G}^{mn}\partial_m \lambda\, f_{nt}
+{b\,d\cot\theta\over L^2 r^2\sin\theta\sqrt{\Delta}}\,\lambda f_{tr}\Bigg]\,\,,
\label{total_lag_fuct}
\eear
where ${\cal G}^{mn}$ is the open string metric defined in (\ref{open_string_metric_def}), $f_{mn}$ is the field strength for $a_m$ ($f=L^2 da$)  and $\Delta$ is given by (\ref{Delta_definition}). Let us now write the different equations of motion which can be derived from the total lagrangian (\ref{total_lag_fuct}). The non-zero values of ${\cal G}^{mn}$ are written in (\ref{open_string_metric_components}).  First of all, we write the equation of motion for $a_r$ (in the $a_r=0$ gauge):
\beq
{b^2+r^2\,h\,\theta'^{\,2}\over h\,\Delta}\,\partial_t\,a_t'\,-\,\partial_x\,a_x'\,-\,
{d\over b\sin\theta\sqrt{\Delta}}\Big(\theta'\,\partial_t\,\lambda'\,-\,{\Delta\over r^2\,h}\,\cot\theta\,
\partial_t\lambda\Big)\,=\,0\,\,.
\label{eom_ar}
\eeq
The equation of motion for $a_t$ is:
\bear
&&\partial_r\,\Bigg[
{b^2\,r^2\,\sin\theta\over \Delta^{{3\over 2}}}\,(b^2+r^2\,h\,\theta'^2)\,a_t'\,
+\,d\,b\Big(\cot\theta\,\lambda\,-\,{r^2\,h\,\theta'\over \Delta}\,\lambda'\Big)\Bigg]  \rc
&&
\qquad\qquad
+{\sin\theta(b^2+r^2\,h\,\theta'^{\,2})\over r^2\, h\, \sqrt{\Delta}}\,
\partial_x(\partial_x a_t\,-\,\partial_t a_x)\,-\,
{d\,\theta'\over b\,r^2}\,\partial_x^2\,\lambda\,=\,0\,\,, 
\label{eom_at}
\eear
while the equation of motion for $a_x$ becomes:
\beq
\partial_r\,\Big({b^2\,r^2\,h\,\sin\theta\over \sqrt{\Delta}}\,a_x'\Big)\,+\,
{\sin\theta(b^2+r^2\,h\,\theta'^2)\over r^2 \,h\,\sqrt{\Delta}}\,
\partial_t(\partial_x\,a_t-\partial_t a_x)\,-\,{d\,\theta'\over b\,r^2}\,\partial_t\partial_x\lambda\,=\,0\,\,.
\label{eom_ax}
\eeq
Finally, the equation of motion for the scalar $\lambda$ is:
\bear
&&\partial_r\,\Bigg[{r^2\sin\theta h\over \sqrt{\Delta}}\,\Big[r^2\,
\Big(1\,-\,{r^2\,h\,\theta'^{\,2}\over \Delta}\Big)\lambda'+
{b\,d\,\theta'\over \sin\theta\sqrt{\Delta}}\,a_t'\Big]\Bigg] \\
&&
\qquad
+b\,d\cot\theta\,a_t'\,
+{d\theta'\over b\,r^2}\,\partial_x(\partial_x a_t-\partial_t a_x) +r^2\sin\theta\,\sqrt{\Delta}\,\Big[\big(3-2b){\sin\theta\over \sqrt{\Delta}}\,+\,
{\Delta-r^2\, h\,\theta'^2\over \sin^2\theta\Delta}\,\Big]\,\lambda \rc
&&
+ {\sin\theta\over b^2\sqrt{\Delta}}\Big[
{(b^2+r^2\,h\,\theta'^2)(r^2\,h\,\theta'^{\,2}-\Delta)\over h\,\Delta}\,-\,
{d^2\,\theta'^{\,2}\over r^2\sin^2\theta}\,\Big]\partial_t^2\lambda+
{\sin\theta\sqrt{\Delta}\over b^2}\,
\Big(1\,-\,{r^2\,h\,\theta'^{\,2}\over \Delta}\Big)\partial_x^2\lambda\,=\,0\,\,.\nonumber
\label{eom_lambda}
\eear
Let us  next Fourier transform the gauge field  and the scalar to momentum space as:
\bear
a_\nu(r, t, x) & = & \int {d\omega\,dk\over (2\pi)^2}\,
a_\nu(r, \omega, k)\,e^{-i\omega\,t\,+\,i k x}~,  \rc
\lambda(r, t, x) & = & \int {d\omega\,dk\over (2\pi)^2}\,
\lambda(r, \omega, k)\,e^{-i\omega\,t\,+\,i k x}\,\,,
\eear
and  define the electric field $E$ as the following  gauge-invariant combination:
\beq
E\,=\,k\,a_t\,+\,\omega\,a_x\,\,.
\label{E_at_ax}
\eeq
Then, the  equation of motion for $a_r$ in momentum space is:
\beq
{b^2+r^2\,h\,\theta'^{\,2}\over h\,\Delta}\,\omega\,a_t'\,+\,k\,a_x'\,-\,
{\omega\,d\over b\sin\theta\sqrt{\Delta}}\Big(\theta'\,\lambda'\,-\,{\Delta\over r^2\,h}\,\cot\theta\,
\lambda\Big)\,=\,0\,\,.
\eeq
We now combine this last equation with the definition of $E$. We get $a_t'$ and $a_x'$ as functions of $E$ and $\lambda$:
\bear
a_t' & = & {k\,h\,\Delta\over \Delta h k^2-(b^2+r^2 h\theta'^2)\omega^2}\,E'-
{\omega^2\,h\,d\,\sqrt{\Delta}\over b\big[\Delta h k^2-(b^2+r^2 h \theta'^2)\omega^2\big]\sin\theta}\,
\big(\theta'\lambda'-{\Delta\over r^2 h}\cot\theta\lambda\big)~, \rc
a_x' & = & {-(b^2+r^2 h \theta'^2)\omega\over \Delta h k^2-(b^2+r^2 h \theta'^2)\omega^2}\,E'+{\omega\,k\,h\,d\,\sqrt{\Delta}\over b\big[\Delta h k^2-(b^2+r^2 h \theta'^2)\omega^2\big]\sin\theta}\big(\theta'\lambda'-{\Delta\over r^2h}\cot\theta\lambda\big) \ .\rc
\label{at_ax_E}
\eear
After using these equations, it is easy to check that (\ref{eom_at}) and (\ref{eom_ax}) become equivalent and equal to the following differential equation for the electric field $E$:
\bear
&&
\partial_r\Bigg[{b^2\,r^2\,h\over 
(b^2+r^2\,h\,\theta'^{\,2})\omega^2-\Delta\,h\,k^2}\Bigg({\sin\theta\over \sqrt{\Delta}} (b^2+r^2\,h\,\theta'^2)\,E'\,-\,{k\,d\,h\over b}\,\big(\theta'\,\lambda'-{\Delta\over r^2\,h}\cot\theta\lambda\big)
\Bigg)\Bigg] \rc
&&
\qquad\qquad\qquad\qquad\qquad\qquad\qquad\qquad
+{\sin\theta (b^2+r^2\,h\,\theta'^{\,2})\over r^2\,h\,\sqrt{\Delta}}\,E\,-\,
{d\theta'\over b\,r^2}\,k\,\lambda\,=\,0 \ .
\label{E_fluc_eq}
\eear
Let us now write the equation of motion for the scalar field $\lambda$ in momentum space as:
\bear
\label{lambda_fluc_eq}
&& 0 = \partial_r\,\Bigg[{r^2\sin\theta h\over \sqrt{\Delta}}\,\Big[r^2\,
\Big(1\,-\,{r^2\,h\,\theta'^{\,2}\over \Delta}\Big)\lambda'+
{b\,d\,\theta'\over \sin\theta\sqrt{\Delta}}\,a_t'\Big]\Bigg] \\
&&
\quad
+b\,d\cot\theta\,a_t'\,
-k\,{d\theta'\over b\,r^2}\,E +r^2\sin\theta\,\sqrt{\Delta}\,\Big[\big(3-2b){\sin\theta\over \sqrt{\Delta}}\,+\,
{\Delta-r^2\, h\,\theta'^2\over \sin^2\theta\Delta}\,\Big]\,\lambda\rc
&&
\quad
-\omega^2\, {\sin\theta\over b^2\sqrt{\Delta}}\Big[
{(b^2+r^2\,h\,\theta'^2)(r^2\,h\,\theta'^{\,2}-\Delta)\over h\,\Delta}\,-\,
{d^2\,\theta'^{\,2}\over r^2\sin^2\theta}\,\Big]\lambda-k^2\,
{\sin\theta\sqrt{\Delta}\over b^2}\,
\Big(1-{r^2\,h\,\theta'^{\,2}\over \Delta}\Big)\lambda \ ,\nonumber
\eear
where it should be understood that $a_t'$ is given by the first equation in (\ref{at_ax_E}). The fluctuation equations (\ref{E_fluc_eq}) and (\ref{lambda_fluc_eq}) depend explicitly on the horizon radius $r_h$ through the blackening factor $h$.  This dependence can be eliminated by performing the familiar rescaling of the  radial variable 
and of the different quantities appearing in the equations. Indeed, let us rescale the radial variable $r$ and the density $d$ as in (\ref{hat_r_d}).  Moreover, we also define the rescaled frequency and momentum as:
\beq
\hat \omega\,=\,{\omega\over r_h}\,\,,
\qquad\qquad
\hat k\,=\,{k\over r_h}\,\,.
\label{hat_omega_k}
\eeq
Then, one can easily verify that the resulting equations of motion are independent of $r_h$ if the fields $E$ and $\lambda$  are rescaled appropriately.  Actually, since only the relative power of $r_h$ in these two fields matters, we can decide not to rescale the electric field $E$.  The rescaling of the scalar $\lambda$ that allows to eliminate $r_h$ is:
\beq
\hat\lambda\,=\,r_h^{2}\,\lambda\,\,.
\label{hat_lambda}
\eeq
The resulting equations of motion are just (\ref{E_fluc_eq}) and (\ref{lambda_fluc_eq}) with $r_h=1$ and with all quantities replaced by their hatted counterparts. 

The collective excitations of the brane system are dual to the quasinormal modes of the probe. The latter can be obtained by solving (\ref{E_fluc_eq}) and (\ref{lambda_fluc_eq}) for low $\omega$ and $k$ by imposing infalling boundary conditions at the horizon and the vanishing of the source terms at the UV.  At low temperature, in the so-called collisionless  quantum regime, the dominant excitation is the holographic zero sound \cite{Karch:2008fa,Karch:2009zz,Bergman:2011rf} (see also \cite{Jokela:2012se,Jokela:2012vn,Brattan:2012nb,Davison:2011ek,Kulaxizi:2008jx,Kulaxizi:2008kv}), whose dispersion relation has the form:
\beq
\hat \omega\,=\,\pm c_s\,\hat k\,-\,i\Gamma(\hat k, \hat d)\,\,.
\label{zero_sound_disp_rel}
\eeq
In (\ref{zero_sound_disp_rel}) $c_s$ is the speed of zero sound and $\Gamma$ is the attenuation. We have integrated numerically the fluctuation equations when  $\hat d$ is  large (\ie\ at low temperature) 
and we have found the value of $c_s$, both for the unflavored and the flavored backgrounds. The main conclusion from this calculation is that $c_s$ is equal to the speed of the  first sound $u_s$ (given by (\ref{speed_of_first_sound})).  As shown in Fig.~\ref{speed_of_sound}, $c_s$ reaches its maximal value ($c_2=1/\sqrt{2}$) when $m/\mu=0$, where the system is conformally invariant. In the unflavored case $c_s$ is always positive and vanishes at the quantum critical point at $\mu=m$ (see Fig.~\ref{speed_of_sound}, left). When dynamical quarks are included $c_s$ becomes imaginary when $\mu$ reaches its minimal value, which occurs when the Minkowski embeddings are thermodynamically favored.  

At higher temperature (\ie\ with small $\hat d$) the system enters into the hydrodynamic diffusive regime. The dominant mode in this case has purely imaginary frequency and a spectrum of the form:
\beq
\hat\omega\,=\,-i\hat D\,\hat k^2\,\,,
\label{hat_diffusion}
\eeq
where $\hat D$ is the rescaled diffusion constant,
\beq
\hat D\,=\,r_h\,D\,\,.
\eeq
As in the zero sound case, this dynamic calculation of the diffusion constant yields the same result as the static one. Indeed, the results obtained by numerical integration of  (\ref{E_fluc_eq}) and (\ref{lambda_fluc_eq}) coincide with the ones obtained from the Einstein relation (\ref{Diffusion_Einstein}), as shown in Fig.~\ref{Diffusion_constant_plots}.

Let us finish this section with the following observation. A careful reader would had expected some discussion on the possible instability as the WZ action has a term $C_1\wedge F\wedge F\wedge F$ which is the source of 
striping via a generic mechanism introduced in \cite{Nakamura:2009tf}. Indeed, the occurrence of tachyonic fluctuations have been confirmed in similar brane models \cite{Bergman:2011rf,Jokela:2012se}, with the subsequent construction of the striped ground state \cite{Jokela:2014dba}. In the current work, we analyzed the fluctuations of the transverse gauge field, where such an instability is expected. In this sector, one needs to analyze the coupled fluctuations of the internal gauge field $a$ and the transverse Minkowski gauge field $a_y$ at non-vanishing momentum. The corresponding equations of motion are presented in appendix \ref{app:transverse}. While we did see the precursor of the instability, a purely imaginary mode first ascending towards the upper half of the complex $\omega$ plane and then descending as a function of $k$, we were unsuccessful to finding parameter values for which case the mode would had actually become unstable. We expect that in the case in which an internal flux is turned on at the unperturbed level,  where the  contribution of the pullback of $\hat C_1$  at the background level  is non-vanishing,  the relevant WZ term can become sizable and thus implies striping in some range of parameters.

\section{Discussion}
\label{summary}

In this chapter we studied the phase diagram of a D6-brane probe with non-vanishing charge density in a background dual to the ABJM Chern-Simons matter theory with dynamical massless flavors at zero temperature. We analyzed the phase transition between black hole and Minkowski embeddings at zero temperature and non-vanishing chemical potential. This transition is a holographic model of a conductor-insulator phase transition between a gapless (black hole) phase and a gapped (Minkowski) phase. 

In the unflavored background we found that this transition occurs when the charge density vanishes and is of second order. Moreover, we were able to characterize the scaling behavior of the probe near the critical point. Interestingly, we found logarithmic multiplicative corrections. In the background with dynamical quarks the transition of the probe is of first order and takes place when the density is non-zero. Therefore, we have shown that, even if the change of the metric due to the backreaction to the flavor is seemingly mild, the physical effects are very important. 

It is interesting to compare our results with the one corresponding to the (2+1)-dimensional  D3-D5 intersection \cite{Karch:2007br,Ammon:2012je}. When the mass $m$ of the quarks is zero, the gravitational descriptions of both systems are equivalent and have the same thermodynamic quantities. However, for non-conformal embeddings with $m\not=0$, the ABJM probe action gets a non-trivial contribution from the Wess-Zumino term. This term is responsible for the different critical behaviors of the systems even in the absence of backreaction. \newline

Let us now discuss some possible extensions of the work performed in this chapter. First of all, it would be interesting to extend our study of the Minkowski-black hole embedding phase transition  to non-zero temperature, in order to completely determine the phase diagram of the model. In the absence of the chemical potential $\mu=0$, this analysis was performed in \cite{Jokela:2012dw}. Another possible generalization would be to consider the case of massive dynamical quarks presented in chapter \ref{massiveABJM}. recall that this solution contains a scale (the mass of the sea quarks) and it would be very interesting to explore how it affects the results found here.

Turning on a suitable NSNS flat $B$ field in the ABJM supergravity solution we get the so-called ABJ background, which is dual to a Chern-Simons matter theory with gauge group $U(N+M)\times U(N)$ \cite{Aharony:2008gk}. The $B$ field breaks parity in 2+1 dimensions. The embedding of flavor brane probes in the ABJ background has been analyzed in chapter \ref{chapterfour} and the relation to the quantum Hall effect was doped out. It would be interesting to analyze possible quantum phase transitions in this ABJ system.

One of the main motivations of this work was the analysis of the effects of the dynamical quarks in the phase diagram of holographic compressible matter. We achieved this objective only partially since our backreacted background did not include the effect of the charge density on the flavor brane.  On general grounds, one would expect having a Lifshitz geometry in the IR of such a background. Indeed, this is precisely what happens in the geometry recently found in \cite{Faedo:2015urf}, corresponding to an intersection of color D2-branes and flavor D6-branes. The study of the quantum phase transitions, as well as the collective excitations of the flavor brane, in this background is of great interest.

\begin{subappendices}

\section{Appendix}

\setcounter{equation}{0}

\subsection{Fluctuation analysis}
\label{fluctuation_apendix}


Let us consider fluctuations of the gauge field $A$ and the embedding function $\theta$ as in (\ref{longitudinal_fluct_ansatz}).  We expand the induced metric $g$ and the gauge field strength as:
\beq
g\,=\,g^{(0)}\,+\,g^{(1)}\,+\,g^{(2)}\,\,,
\qquad\qquad
F\,=\,F^{(0)}\,+\,f\,\,,
\eeq
where $g^{(0)}$ is the metric written in (\ref{induced_metric}) and $F^{(0)}$ is the field strength of the unperturbed gauge connection (\ref{unperturbed_theta_A}), while $f=L^2 da$ and 
the first and second order  induced metrics 
$g^{(1)}$ and $g^{(2)}$ are given by:
\bear
&&g^{(1)}_{ij}\,d\zeta^i\,d\zeta^j\,=\,{L^2\over b^2}\,\Big[
2\,\theta' \,(\lambda'\,dr+\partial_t\lambda\,dt+\partial_x\lambda\,dx) dr\,+\,
\lambda\,\sin(2\theta)\,(d\psi+\cos\alpha \,d\beta)^2\Big] \rc
&&g^{(2)}_{ij}\,d\zeta^i\,d\zeta^j\,=\,{L^2\over b^2}\,\Big[
(\lambda'\,dr+\partial_t\lambda\,dt+\partial_x\lambda\,dx)^2+\lambda^2\,\cos (2\theta)\,
(d\psi+\cos\alpha\, d\beta)^2\Big]\,\,.
\eear
Let us next write the inverse of the zeroth-order DBI matrix $g^{(0)}+F^{(0)}$  as:
\beq
\Big(g^{(0)}+F^{(0)}\Big)^{-1}\,=\,{\cal G}^{-1}+{\cal J}\,\,,
\label{open_string_metric_def}
\eeq	
where ${\cal G}^{-1}$ is the symmetric part (the inverse open string metric) and ${\cal J}$ is the antisymmetric part. In order to write the different elements of 	${\cal G}$  and ${\cal J}$ it is quite convenient to introduce 
 the quantity $\Delta$ defined in (\ref{Delta_definition}). 
 In terms of $\Delta$, the equation for the embedding takes the form:
\beq
\partial_r\Big[{r^4\,h\,\sin\theta\over \sqrt{\Delta}}\,\theta'\Big]\,-\,r^2\,\sin\theta\cos\theta\,
\left[3-2b+{\sqrt{\Delta}\over \sin\theta}\right]\,=\,0\,\,.
\eeq
Then, the non-vanishing components of the open string metric are:
\bear
&&{\cal G}^{tt}\,=\,-{b^2+r^2\,h\,\theta'^{\,2}\over L^2\,r^2\,h\,\Delta}\,\,,
\qquad\qquad
{\cal G}^{xx}\,=\,{\cal G}^{yy}\,=\,{1\over L^2\,r^2}\,\,,\rc\rc
&&{\cal G}^{rr}\,=\,{b^2\,r^2\,h\over L^2\Delta}\,\,,
\qquad\qquad
{\cal G}^{\alpha\alpha}\,=\,{b^2\over L^2\,q}\,\,,
\qquad\qquad
{\cal G}^{\beta\beta}\,=\,{b^2\over L^2\,q\,\sin^2\alpha}\,\,,\rc\rc
&&{\cal G}^{\beta\psi}\,=\,-{b^2\cos\alpha\over L^2\,q\,\sin^2\alpha}\,\,,
\qquad\qquad
{\cal G}^{\psi\psi}\,=\,{b^2\over L^2 q}\,\Big(\cot^2\alpha\,+\,{q\over \sin^2\theta}\Big)\,\,.
\label{open_string_metric_components}
\eear
The only non-vanishing components of the antisymmetric tensor are:
\beq
{\cal J}^{tr}\,=\,-{\cal J}^{rt}\,=\,-{d\,b\over L^2\,r^2\,\sin\theta\sqrt{\Delta}}\,\,.
\eeq

At second order in the fluctuations, the DBI action is:
\bear
S_{DBI}^{(2)} & = &-T_{D6}\int d^7 \zeta e^{-\phi}\sqrt{-\det (g^{(0)}+F^{(0)})}\Bigg[{1\over 2} Tr\big({\cal G}^{-1}g^{(2)}\big)
+{1\over  8}\Big( Tr\big({\cal G}^{-1}g^{(1)}\big)+ Tr\big({\cal J}f\big)\Big)^2\rc
&&-{1\over 4}
Tr\Big[({\cal G}^{-1}\,g^{(1)}\big)^2+({\cal J}\,g^{(1)}\big)^2+
4\,{\cal G}^{-1}\,g^{(1)}\,{\cal J}\,f\,+\,({\cal G}^{-1}\,f\big)^2\,+\,({\cal J}\,f\big)^2\Big]\Bigg]\,\,.
\eear
To evaluate this expression we use:
\bear
&& Tr\big({\cal G}^{-1}\,g^{(2)}\big)\,=\,
{L^2\over b^2}\,{\cal G}^{mn}\partial_m\lambda\,\partial_n\lambda\,+\,
\big(\cot^2\theta-1)\,\lambda^2 ~, \rc
&& Tr\big({\cal G}^{-1}\,g^{(1)}\big)\,=\,{2L^2\over b^2}\,\theta'\,{\cal G}^{rr}\,\lambda'\,+\,2\cot\theta\,\lambda ~, \rc
&& Tr\Big[\big({\cal G}^{-1}\,g^{(1)}\big)^2\Big]\,=\,{4 L^4\over b^4}\,\theta'^{\,2}\,({\cal G}^{rr})^2\,
(\lambda')^2\,+\,4\,\cot^2\theta\,\lambda^2+{2L^4\over b^4}\,\theta'^{\,2}\,{\cal G}^{rr}\,
{\cal G}^{mn}\,\partial_m\lambda\partial_n\lambda\,\,,\qquad\qquad
\eear
where the indices $n$ and $m$ run over the Minkowski and radial directions. After integrating over the internal angles, we get the following second-order DBI lagrangian:
\bear
{\cal L}^{(2)}_{DBI} & = & -{\cal N}r^2\sin\theta\sqrt{\Delta}\Bigg[{1\over 4}{\cal G}^{nm}{\cal G}^{pq}f_{mq}f_{np}+{L^2\over 2b^2}\Big(1-{L^2\over b^2}\theta'^2{\cal G}^{rr}\Big){\cal G}^{mn}\partial_m\lambda\partial_n\lambda-{1\over 2}\lambda^2 \\
&&+{L^2\over b^2}\theta'\cot\theta{\cal G}^{rr}\lambda\partial_r\lambda-{1\over 2}\Big({A_t'\theta'\over \Delta}\Big)^2(\partial_t\lambda)^2+{A_t'\theta'\over \Delta}{\cal G}^{mn}\partial_m\lambda f_{nt}+{b^2\over L^2}{A_t'\over \Delta}\cot\theta\lambda f_{tr}\Bigg] \nonumber\ .
\eear
The WZ term at second order yields the following lagrangian density:
\beq
{\cal L}^{(2)}_{WZ}\,=\,{\cal N}\,r^2\,b\,\Big[
{r\over b}\,\cos(2\theta)\,\lambda\partial_r\lambda\,+\,
\Big(\cos(2\theta)-{r\over b}\sin(2\theta)\,\theta'\Big)\lambda^2\Big]\,\,.
\eeq
Let us now simplify these expressions. First of all, we should eliminate $A_t'$. With this purpose we notice that:
\beq
{A_t'\over \Delta}\,=\,{d\over b\,r^2\,\sin\theta\sqrt{\Delta}}\,\,.
\eeq
Secondly, we rewrite the terms with $\lambda\partial_r \lambda$ by integrating by parts and neglecting  the total derivative generated in this process. In the WZ lagrangian we use
\beq
{r^3\over b}\,\cos(2\theta)\,\lambda\partial_r\lambda\,=\,
\Big({r^3\over b}\sin(2\theta)\,\theta'\,-\,
{3\over 2}\,{r^2\over b}\cos(2\theta)\,\Big)\,\lambda^2\,+\,
\partial_r\Big[{1\over 2}\,{r^3\over b}\,\cos(2\theta)\,\lambda^2\Big]\,\,.
\eeq
The resulting WZ lagrangian takes the form:
\beq
{\cal L}^{(2)}_{WZ}\,=\,{\cal N}\,r^2\,b\,\big(1-{3\over 2b}\big)\,\cos(2\theta)\,\lambda^2\,\,.
\eeq
In the DBI part, we first write:
\beq
{r^4\sin\theta h\over \sqrt{\Delta}}\,\theta'\,\cot\theta\,\lambda\partial_r\lambda\,=\,
-{1\over 2}\,\partial_r\Big({r^4\sin\theta h\over \sqrt{\Delta}}\,\theta'\,\cot\theta\Big)\,\lambda^2\,+\,
\partial_r\Big({r^4\sin\theta h\over 2 \sqrt{\Delta}}\,\theta'\,\cot\theta\,\lambda^2\Big)\,\,.
\eeq
It follows that we can make the following substitution in ${\cal L}_{DBI}$:
\bear
{r^4\sin\theta h\over \sqrt{\Delta}}\,\theta'\,\cot\theta\,\lambda\partial_r\lambda
\to
-\partial_r\Big({r^4\sin\theta h\over 2 \sqrt{\Delta}}\,\theta'\cot\theta\Big)\lambda^2=
-\partial_r\Big({r^4\sin\theta h\over 2 \sqrt{\Delta}}\,\theta'\Big)\cot\theta\lambda^2+
{r^4 h\theta'^{\,2}\over 2\sin\theta\sqrt{\Delta}}\lambda^2\,\,,\rc
\eear
which, after using eq. (\ref{eom_theta_At}) for $\theta(r)$, can be written as:
\beq
{r^4\sin\theta h\over \sqrt{\Delta}}\,\theta'\,\cot\theta\,\lambda\partial_r\lambda
\to
\Bigg(\big(b-{3\over 2}\big)\,r^2\,\cos^2\theta\,+\,r^2\,
{r^2 h \theta'^{\,2}-\cos^2\theta\Delta\over 2\sin\theta\sqrt{\Delta}}\Bigg)\lambda^2\,\,.
\eeq
Taking these results into account, it is straightforward to verify that the total lagrangian density 
${\cal L}^{(2)}={\cal L}^{(2)}_{DBI}+{\cal L}^{(2)}_{WZ}$  can be written as in (\ref{total_lag_fuct}).

\subsubsection{Transverse fluctuations}\label{app:transverse}
We now consider fluctuations of the gauge field along the transverse direction $y$. It turns out that these fluctuations 
are coupled to those along the internal directions. Actually, we can write the following consistent ansatz:
\beq
A\,=\,L^2\,\Big[A_t(r)\,dt\,+\,e^{-i\omega t+i kx}\,
a_y(r)\,dy\,+\,e^{-i\omega t+i kx}\,a(r)\,\big(\cos\alpha\,d\beta\,+\,d\psi\big)\Big]\,\,,
\eeq
where $a_y$ and $a$ are first-order fluctuations. The equation of motion for $a_y$ is given by:
\bear
&&\partial_r\,\Big({b^2\,r^2\,h\,\sin\theta\over \sqrt{\Delta}}\,a_y'\Big)\,+\,{\sin\theta\over r^2\,h\sqrt{\Delta}}\,
\Big[\omega^2(b^2+r^2\,h\theta'^2)-k^2\,h\,\Delta\Big]\,a_y \rc
&&\qquad\qquad
+{2i\,k\,d\,\cot\theta\over r^2}\,{b^2(2-q)\eta\over q(q+\eta)}\,\sqrt{\Delta}\,
\,a\,=\,0\,\,,
\eear
whereas that for $a$ is:
\bear
&&\partial_r\,\Big({b\,q\,r^4\,h\,\over \sin\theta\sqrt{\Delta}}\,a'\Big)\,+\,3b\,r^2\,a\,-\,{b\over q}\,r^2\sin\theta\,\sqrt{\Delta}\,a\,+
\omega^2\,{q\over b}\,{b^2+r^2\,h\,\theta'^2\over \sin\theta\,h\,\sqrt{\Delta}}\,a \rc
&&\qquad\qquad
-k^2\,{q\over b}\,{\sqrt{\Delta}\over \sin\theta}\,a\,-\,{2ikd\cot\theta \over r^2}\,{(2-q)\eta\over (q+\eta)\,b}\,\,\sqrt{\Delta}\,
a_y\,=\,0\,\,.
\eear
For our purposes, it is enough to consider the fluctuations at zero momentum ($k=0$). In this case the equation for $a_y$ is decoupled from the internal fluctuation $a$ and becomes:
\beq
\partial_r\,\Big({b^2\,r^2\,h\,\sin\theta\over \sqrt{\Delta}}\,a_y'\Big)\,+\,{\sin\theta\over r^2\,h\sqrt{\Delta}}\,
\omega^2(b^2+r^2\,h\,\theta'^2)\,a_y\,=\,0\,\,.
\label{a_y_eq_zerok}
\eeq
Explicitly, this equation for $a_y$ can be written as:
\beq
a_y''\,+\,\partial_r\,\log\Bigg[
{\sqrt{d^2+r^4\sin^2\theta}\over \sqrt{b^2+r^2\,h\,\theta'^{\,2}}}\,h\Bigg] a_y'\,+\,\omega^2\,
{b^2+r^2\,h\,\theta'^{\,2}\over b^2\,r^4\,h^2}\,a_y\,=\,0\,\,.
\label{ay_eq_k0}
\eeq
Let us expand this equation near $r=r_h$. First, we expand the embedding as in (\ref{expansion_theta_nh}):
\beq
\theta(r)\,=\,\theta_h+\theta_h'\,(r-r_h)\,+\,\cdots\,\,,
\eeq
where $\theta_h'$ is given by (see (\ref{theta_nh})):
\beq
\theta_h'\,=\,{b\,r_h\over 3}\,{\sin\theta_h\,\cos\theta_h\over d^2+r_h^4\,\sin^2\theta_h}\,
\Big[b\,r_h^2\,+\,(3-2b)\,\sqrt{d^2+r_h^4\,\sin^2\theta_h}\Big]\,\,.
\eeq
The coefficients of (\ref{ay_eq_k0})  will be expanded as:
\bear
\partial_r\,\log\Bigg[{\sqrt{d^2+r^4\sin^2\theta}\over \sqrt{b^2+r^2\,h\,\theta'^{\,2}}}\,h\Bigg] & = & {1\over r-r_h}\,+\,d_1\,\cdots \rc
\omega^2{b^2+r^2\,h\,\theta'^{\,2}\over b^2\,r^4\,h^2} & = & {A\over (r-r_h)^2}+{c_2\over r-r_h}\,+\,\cdots\,\,,
\eear
where $A$, $d_1$, and $c_2$ are:
\bear
A & = & {\omega^2\over 9\,r_h^2}~, \rc
d_1 & = & -{2\over r_h}\,{d^2\over d^2+r_h^4\sin^2\theta_h}\,+\,{r_h^4\,\sin\theta_h\,\cos\theta_h\over d^2+r_h^4\sin^2\theta_h}\,\theta_h'\,-\,
{3 r_h\over 2 b^2}\,\big(\theta_h')^2~, \rc
c_2 & = & {\big(\theta_h')^2\over 3\, b^2\,r_h}\,\omega^2\,\,.
\eear
We now solve the equation of motion for $a_y$ in a Frobenius series around $r=r_h$ as:
\beq
a_y(r)\,=\,(r-r_h)^{\alpha}\,\big(1\,+\,\beta (r-r_h)\,+\,\cdots\big)\,\,,
\label{Frobenius_ay}
\eeq
where, for infalling boundary conditions,  the  exponent $\alpha$ is given by:
\beq
\alpha\,=\,-\,{i\omega\over 3 r_h}\,\,.
\eeq
We will also perform a low frequency expansion by considering $k\sim {\cal O} (\epsilon)$ and 
$\omega\sim { \cal O} (\epsilon^2)$. Then one can show that  $\beta\sim{ \cal O} (\epsilon^2)$ and is given by:
\beq 
\beta\,\approx\,-\alpha\,d_1\,\,.
\eeq
Let us now take the limits in opposite order. First, we consider the low frequency limit. At leading order, we can neglect the last term in (\ref{a_y_eq_zerok}) and write the equation for $a_y$ as:
\beq
\partial_r\,\Big({b^2\,r^2\,h\,\sin\theta\over \sqrt{\Delta}}\,a_y'\Big)\,=\,0\,\,.
\eeq
This equation can be immediately integrated:
\beq
a_y'\,=\,{\cal C}\,{\sqrt{\Delta}\over b^2\,r^2\,h\,\sin\theta}\,\equiv \,{{\cal C}\over G(r)}\,\,,
\eeq
where ${\cal C}$ is a constant of integration and, in the last step, we have defined the function $G(r)$. This solution can be expanded near the horizon $r=r_h$ as:
\beq
a_y'\,=\,{\cal C}\,{r_h\over 3 b\,\sqrt{d^2+r_h^4\,\sin^2\theta_h}}\,{1\over r-r_h}\,+\,\cdots\,\,.
\label{ayprime_low_nh}
\eeq
Let us now compare this near-horizon expansion with the one written in (\ref{Frobenius_ay}) for low frequency. First, we compute $a_y'$ by direct differentiation of  the expansion (\ref{Frobenius_ay}):
\beq
a_y'\,=\,\alpha (r-r_h)^{\alpha-1}\,\Big(1+\beta (r-r_h)\,+\,\cdots\Big)\,+\,(r-r_h)^{\alpha}(\beta+\cdots) \ .
\eeq
Taking into account that $\alpha\sim{\cal O}(\epsilon^2)$ and $\beta\sim{\cal O}(\epsilon^2)$, we get, at leading order in 
$\epsilon$, that:
\beq
a_y'\,=\,{\alpha \over r-r_h}\,+\,\cdots\,\,.
\label{ayprime_nh_low}
\eeq
Thus, matching (\ref{ayprime_nh_low}) and (\ref{ayprime_low_nh}), we get that the constant ${\cal C}$  is given by:
\beq
{\cal C}\,=\,{3b\over r_h}\,\sqrt{d^2+r_h^4\,\sin^2\theta_h}\,\,\alpha\,=\,
-i\,{b\over r_h^2}\,\sqrt{d^2+r_h^4\,\sin^2\theta_h}\,\omega\,\,.
\eeq
Therefore, we can write 
\beq
a_y'\,=\,-{i\over G(r)}\,{b\over r_h^2}\,\sqrt{d^2+r_h^4\,\sin^2\theta_h}\,\,\omega\,\,.
\eeq
Let us now obtain the $\left\langle J_y\,J_y\right\rangle $ correlator from these results. The term in the lagrangian density depending on $a_y$ is given by:
\beq
{\cal L}(a_y)\,=\,-{\cal N}\,r^2\,\sin\theta\,\sqrt{\Delta}\,{\cal G}^{yy}\,{\cal G}^{rr}\,\big(f_{yr}\big)^2\,\,.
\eeq
Taking into account that $f_{ry}=L^2\,a_y'$ and that:
\beq
r^2\,\sin\theta\,\sqrt{\Delta}\,{\cal G}^{yy}\,{\cal G}^{rr}\,=\,{G(r)\over L^4}\,\,,
\eeq
we arrive at:
\beq
{\cal L}(a_y)\,=\,-{\cal F}\,\big(a_y'\big)^2\,\,,
\eeq
where ${\cal F}$ is given by:
\beq
{\cal F}\,=\,-{\cal N}\,G(r)\,\,.
\eeq
Therefore, the on-shell boundary action of $a_y$ is:
\beq
S_{{\rm on-shell}}(a_y)\,=\,\int\,d^3\,x\,\Big( {\cal F}\,a_y\,a_y'\Big)_{r\to\infty}\,\,.
\eeq
The two-point function of the transverse currents, at zero momentum, is given by:
\beq
\left\langle J_y(k)\,J_y(-k)\right\rangle\Big|_{k=0} =\Big( {\cal F}\,a_y'\Big)_{r\to\infty}\,\equiv
{\cal N}\,\Gamma_{\omega}\,i\,\omega\,\,,
\label{correlator_formula}
\eeq
where we have defined the quantity $\Gamma_{\omega}$. 
From the explicit expressions of ${\cal F}$ and $a_y$, we get:
\beq
{\cal F}\,a_y'\,=\,{\cal N}\,\,{b\over r_h^2}\,\sqrt{d^2+r_h^4\,\sin^2\theta_h}\,\,i\omega\,\,.
\eeq
Thus $\Gamma_{\omega}$ is given by:
\beq
\Gamma_{\omega}\,=\,{b\over r_h^2}\,\sqrt{d^2+r_h^4\,\sin^2\theta_h}\,\,.
\eeq
From this result we get the DC conductivity, namely:
\beq
\sigma\,=\,{\cal N}\,\Gamma_{\omega}\,=\,{\cal N}\,{b\over r_h^2}\,\sqrt{d^2+r_h^4\,\sin^2\theta_h}\,\,,
\label{dc_conductivity_ABJM}
\eeq
which is just the result written in (\ref{dc_conductivity}).

\end{subappendices}


\chapter{Conclusions}

In this thesis we have studied new cases of the AdS/CFT correspondence by using non-abelian T-duality and by including fundamental matter. Furthermore, we have constructed new holographic models of condensed matter physics systems.

\begin{itemize}

\item In the first chapter we gave a general introduction to the gauge/gravity duality. We reviewed the basic aspects of supergravity and T-duality. Moreover, we have introduced the AdS/CFT correspondence and how to add flavor to it. Finally, we commented on some applications of the AdS/CFT duality.

\item In the second chapter we have reviewed some applications of non-abelian T-duality. In the context of type IIA/B supergravity, the non-abelian T-duality is used to obtain new solutions from previously know solutions. We have obtained new interesting cases of the gauge/gravity duality, in particular, new $AdS_3$ fixed points. Furthermore, we have analyzed the field theory duals of the new supergravity solutions, through the computation of some observables. 

\item In the third chapter we have constructed a new type IIA supergravity background with D6-branes sources, whose field theory dual corresponds to a generalization of the ABJM theory that includes unquenched massive flavors. 
We have analyzed several observables, and the results confirmed a RG flow between two fixed points at the IR and the UV

\item In the fourth chapter we have constructed a supergravity solution based on the ABJM model that presents quantum Hall states. We have introduced D6-flavor branes with non-trivial worldvolume gauge fields. This set up allowed to have Hall states, and the filling fraction was computed. We have found supersymmetric Hall states for particular values of the parameters.

\item In the fifth chapter we have obtained a supergravity solution based on the ABJM model, dual to a field theory exhibiting quantum phase transitions. The fundamental matter was introduced via D6-branes, and charge density was turned on. In the ABJM background, the fundamental matter undergoes a second order phase transition at zero charge density. In the partially backreacted ABJM solution, the phase transition becomes first order, and takes place at a finite value of the charge density.

\end{itemize}



\end{document}